\documentclass[12pt,twoside]{article}

\usepackage{amsmath}
\usepackage{graphicx}
\usepackage{epsfig}
\usepackage[figuresright]{rotating}
\usepackage{pstricks}
\usepackage{mciteplus}
\usepackage{authblk}

\usepackage[dvips]{hyperref}                              
\usepackage{relsize}                                      
\def\hfurl#1{http://hfag.phys.ntu.edu.tw/b2charm/#1} 
\newcommand{\mysection}[1]{\section{\boldmath #1}}
\newcommand{\mysubsection}[1]{\subsection[#1]{\boldmath #1}}
\newcommand{\mysubsubsection}[1]{\subsubsection[#1]{\boldmath #1}}
\newcommand{\mysubsubsubsection}[1]{\subsubsubsection{\boldmath #1}}

\newcommand{\lesssim}{\ensuremath{\raise-.5ex\hbox{$\buildrel<\over\sim$}\,}} 

\def\dof{{\rm dof}}

\newcommand\VCKM{{V}}
\newcommand\etacpf{{\eta_f}}
\newcommand\etacp{{\eta}}

\renewcommand\Im{{\rm Im}} 
\renewcommand\Re{{\rm Re}}

\newcommand\Abar{\kern 0.18em\overline{\kern -0.18em A}{}}
\newcommand\Af{A_f}
\newcommand\Abarf{\Abar_f}
\newcommand\Afbar{A_{\bar f}}
\newcommand\Abarfbar{\Abar_{\bar f}}
\newcommand\Acp{{\cal A}}
\newcommand\Adirnoncp{\ensuremath{\langle{\cal A}_{f\bar f}\rangle}\xspace}
\newcommand\mc{\multicolumn}

\newcommand {\cbf}{\ensuremath{{\cal B}}}

\newcommand {\vcb}{\ensuremath{|V_{cb}|}}
\newcommand {\vub}{\ensuremath{|V_{ub}|}}

\def\Bp      {\ensuremath{B^{+}}}
\def\Bm      {\ensuremath{B^{-}}}
\def\Bz      {\ensuremath{B^{0}}}
\def\Bs      {\ensuremath{B_{s}}}

\newcommand{\BzbDplnu}    {\ensuremath{\Bzb \to D^{+}\ell^{-}\nub}}
\newcommand{\BzbDstarlnu} {\ensuremath{\Bzb \to D^{*+}\ell^{-}\nub}}

\newcommand {\rhoz} {\ensuremath{\rho^0}\hbox{ }}

\usepackage{relsize}

\def\beq{\begin{equation}}
\def\eeq#1{\label{#1}\end{equation}}
\def\eeqn{\end{equation}}

\def\beqa{\begin{eqnarray}}
\def\eeqa#1{\label{#1}\end{eqnarray}}
\def\eeqan{\end{eqnarray}}

\let\bar=\overbar

\def\etal{{\it et al.}}
\def\ie{{\it i.e.}}
\def\eg{{\it e.g.}}
\def\etc{{\it etc.}}
\def\cf{{\it cf.}}

\def\Dslash{\ensuremath{\not{\hbox{\kern-4pt $D$}}}\xspace}
\def\dslash{\not{\hbox{\kern-2pt $\del$}}}

\def\BR{\mbox{\rm BR}}
\def\ee{e^+e^-}

\def\alphas{\alpha_s}
\def\msb{{\bar{\ssstyle M \kern -1pt S}}}

\RequirePackage{xspace}

\usepackage{relsize}
\def\babar{\mbox{\slshape B\kern-0.1em{\smaller A}\kern-0.1em
    B\kern-0.1em{\smaller A\kern-0.2em R}}\xspace}
\def\belle{\mbox{\normalfont Belle}\xspace}
\def\cdf{\mbox{\normalfont CDF}\xspace}
\newcommand{\dzero}{D\O\xspace}

\def\ee         {\ensuremath{e^-e^-}\xspace}

\def\mtau       {\ensuremath{\tau}\xspace}

\def\ellp       {\ensuremath{\ell^+}\xspace}

\def\nub        {\ensuremath{\overline{\nu}}\xspace}

\def\nub        {\ensuremath{\overline{\nu}}\xspace}

\def\nul        {\ensuremath{\nu_\ell}\xspace}

\def\Z      {\ensuremath{Z^0}\xspace}

\def\ubar  {\ensuremath{\overline u}\xspace}

\def\dbar  {\ensuremath{\overline d}\xspace}
\def\ddbar {\ensuremath{d\overline d}\xspace}

\def\sbar  {\ensuremath{\overline s}\xspace}

\def\b  {\ensuremath{b}\xspace}
\def\bbar  {\ensuremath{\overline b}\xspace}

\def\piz   {\ensuremath{\pi^0}\xspace}

\def\pip   {\ensuremath{\pi^+}\xspace}
\def\pim   {\ensuremath{\pi^-}\xspace}
\def\pipi  {\ensuremath{\pi^+\pi^-}\xspace}

\def\etapr {\ensuremath{\eta^{\prime}}\xspace}

\def\Kbar  {\kern 0.2em\overline{\kern -0.2em K}{}\xspace}

\def\Kpm   {\ensuremath{K^\pm}\xspace}
\def\Kmp   {\ensuremath{K^\mp}\xspace}
\def\Kp    {\ensuremath{K^+}\xspace}
\def\Km    {\ensuremath{K^-}\xspace}
\def\KS    {\ensuremath{K^0_{\scriptscriptstyle S}}\xspace} 
\def\KL    {\ensuremath{K^0_{\scriptscriptstyle L}}\xspace}

\def\Kstar   {\ensuremath{K^*}\xspace}

\def\Kstarpm   {\ensuremath{K^{*\pm}}\xspace}
\def\Kstarmp   {\ensuremath{K^{*\mp}}\xspace}
\def\Kz   {\ensuremath{K^0}\xspace}
\def\Kzb   {\ensuremath{\Kbar^0}\xspace}
\def\KzKzb {\ensuremath{K^0 \kern -0.16em \Kzb}\xspace}

\def\KorKstarpm {\ensuremath{K^{(*)\pm}}\xspace}

\def\Dz    {\ensuremath{D^0}\xspace}
\def\Dbar  {\kern 0.2em\overline{\kern -0.2em D}{}\xspace}

\def\Dzb   {\ensuremath{\Dbar^0}\xspace}
\def\DzDzb {\ensuremath{D^0 {\kern -0.16em \Dzb}}\xspace}

\def\Dstar   {\ensuremath{D^*}\xspace}

\def\Dstarp  {\ensuremath{D^{*+}}}
\def\Dstarm  {\ensuremath{D^{*-}}}
\def\DorDstar   {\ensuremath{D^{(*)}}\xspace}
\def\DorDstarz  {\ensuremath{D^{(*)0}}\xspace}
\def\DorDstarzb {\ensuremath{\Dbar^{(*)0}}\xspace}

\def\Ds    {\ensuremath{D^+_s}\xspace}

\def\Bz    {\ensuremath{B^0}\xspace}
\def\B     {\ensuremath{B}\xspace}
\def\Bbar  {\kern 0.18em\overline{\kern -0.18em B}{}\xspace}
\def\Bb    {\ensuremath{\Bbar}\xspace}
\def\Bzb   {\ensuremath{\Bbar^0}\xspace}
\def\Bu    {\ensuremath{B^+}\xspace}

\def\Bpm   {\ensuremath{B^\pm}\xspace}
\def\Bmp   {\ensuremath{B^\mp}\xspace}
\def\Bs    {\ensuremath{B_s}\xspace}
\def\Bsb   {\ensuremath{\Bbar_s}\xspace}
\def\BB    {\ensuremath{B\Bbar}\xspace} 
\def\BzBzb {\ensuremath{B^0 {\kern -0.16em \Bzb}}\xspace}

\def\jpsi  {\ensuremath{{J\mskip -3mu/\mskip -2mu\psi\mskip 2mu}}\xspace}

\mathchardef\Upsilon="7107
\def\Y#1S{\ensuremath{\Upsilon{(#1S)}}\xspace}

\mathchardef\Deltares="7101
\mathchardef\Xi="7104
\mathchardef\Lambda="7103
\mathchardef\Sigma="7106
\mathchardef\Omega="710A
\def\Deltabar   {\kern 0.25em\overline{\kern -0.25em \Deltares}{}\xspace}
\def\Lbar {\kern 0.2em\overline{\kern -0.2em\Lambda\kern 0.05em}\kern-0.05em{}\xspace}
\def\Sigbar{\kern 0.2em\overline{\kern -0.2em \Sigma}{}\xspace}
\def\Xibar{\kern 0.2em\overline{\kern -0.2em \Xi}{}\xspace}
\def\Obar{\kern 0.2em\overline{\kern -0.2em \Omega}{}\xspace}
\def\Nbar{\kern 0.2em\overline{\kern -0.2em N}{}\xspace}
\def\Xb{\kern 0.2em\overline{\kern -0.2em X}{}}

\def\BR{{\ensuremath{\cal B}}}

\newcommand{\tev}{\ensuremath{\mathrm{Te\kern -0.1em V}}\xspace}
\newcommand{\gev}{\ensuremath{\mathrm{Ge\kern -0.1em V}}\xspace}
\newcommand{\mev}{\ensuremath{\mathrm{Me\kern -0.1em V}}\xspace}
\newcommand{\kev}{\ensuremath{\mathrm{ke\kern -0.1em V}}\xspace}
\newcommand{\ev}{\ensuremath{\mathrm{e\kern -0.1em V}}\xspace}
\newcommand{\gevc}{\ensuremath{{\mathrm{Ge\kern -0.1em V\!/}c}}\xspace}
\newcommand{\mevc}{\ensuremath{{\mathrm{Me\kern -0.1em V\!/}c}}\xspace}
\newcommand{\gevcc}{\ensuremath{{\mathrm{Ge\kern -0.1em V\!/}c^2}}\xspace}
\newcommand{\mevcc}{\ensuremath{{\mathrm{Me\kern -0.1em V\!/}c^2}}\xspace}

\def\invfb   {\ensuremath{\mbox{\,fb}^{-1}}\xspace}
\def\mus  {\ensuremath{\rm \,\mus}\xspace}

\def\ps   {\ensuremath{\rm \,ps}\xspace}

\def\mus        {\ensuremath{\,\mu{\rm s}}\xspace}    
\def\ps         {\ensuremath{{\rm \,ps}}\xspace}  
\def\gsim{{~\raise.15em\hbox{$>$}\kern-.85em
          \lower.35em\hbox{$\sim$}~}\xspace}
\def\lsim{{~\raise.15em\hbox{$<$}\kern-.85em
          \lower.35em\hbox{$\sim$}~}\xspace}

\def\CP                 {\ensuremath{C\!P}\xspace}
\def\CPT                {\ensuremath{C\!PT}\xspace}
\def\ra                 {\ensuremath{\to}\xspace}

\def\pep2{PEP-II}

\def\rhobar {\ensuremath{\overline{\rho}}\xspace}
\def\etabar {\ensuremath{\overline{\eta}}\xspace}

\def\Vud  {\ensuremath{|V_{ud}|}\xspace}

\def\Vus  {\ensuremath{|V_{us}|}\xspace}

\def\Vub  {\ensuremath{|V_{ub}|}\xspace}

\def\stwob{\ensuremath{\sin\! 2 \beta   }\xspace}

\def\deltamd{\ensuremath{{\rm \Delta}m_d}\xspace}

\xspace
\newcommand{\fds}{\ensuremath{f_{D_s}}\xspace}

\def\jetset74   {\mbox{\tt Jetset \hspace{-0.5em}7.\hspace{-0.2em}4}}

\newcommand{\aerr}[4]   {\mbox{${{#1}^{+ #2}_{- #3}\pm #4}$}}
\newcommand{\berr}[4]   {\mbox{${{#1}\pm #2^{+ #3}_{- #4}}$}}
\newcommand{\cerr}[3]   {\mbox{${{#1}^{+ #2}_{- #3}}$}}
\newcommand{\aerrsy}[5] {\mbox{${{#1}^{+ #2 + #4}_{- #3 - #5}}$}}

\newcommand{\err}[3]   {\mbox{${{#1}\pm{#2}\pm{#3}}$}}

\newcommand{\nodata}{$$}
\newcommand{\vs}{\mbox{$vs.$}}
\def\etapr{{\eta^{\prime}}}

\def\sgline{\noalign{\vskip 0.10truecm\hrule\vskip 0.10truecm}}
\def\sglinespt{\noalign{\vskip 0.05truecm\hrule}}
\def\sglinespb{\noalign{\hrule\vskip 0.05truecm}}

\newcommand{\kz}    {\mbox{$K^0$}}

\newcommand{\RPP}{}

\renewcommand{\mysection}[1]{\section[#1]{#1}} 

\begin{document}

\setcounter{page}{1}
\thispagestyle{empty}
\renewcommand\Affilfont{\itshape\small} 

\title{Averages of $b$-hadron, $c$-hadron, and $\tau$-lepton Properties \\ 
\vskip0.20in
\large{\it Heavy Flavor Averaging Group (HFAG):}
\vspace*{-0.20in}}
\author[1]{D.~Asner}\affil[1]{Pacific Northwest National Laboratory, USA}
\author[2]{Sw.~Banerjee} \affil[2]{University of Victoria, Canada}
\author[3]{R.~Bernhard}\affil[3]{University of Z\"{u}rich, Switzerland}
\author[4]{S.~Blyth}\affil[4]{National United University, Taiwan}
\author[5]{A.~Bozek}\affil[5]{University of Krakow, Poland}
\author[6]{C.~Bozzi}\affil[6]{INFN Ferrara, Italy} 
\author[7]{D.~G.~Cassel} \affil[7]{Cornell University, USA}
\author[8]{G.~Cavoto}\affil[8]{INFN Rome, Italy}
\author[6]{G.~Cibinetto}
\author[9]{J.~Coleman} \affil[9]{University of Liverpool, UK}
\author[10]{W.~Dungel}\affil[10]{Austrian Academy of Sciences, Austria}
\author[11]{T.~J.~Gershon}\affil[11]{University of Warwick, UK}
\author[7]{L.~Gibbons}
\author[12]{B.~Golob}\affil[12]{University of Ljubljana, Slovenia}
\author[13]{R.~Harr}\affil[13]{Wayne State University, USA}
\author[14]{K.~Hayasaka} \affil[14]{Nagoya University, Japan}
\author[15]{H.~Hayashii} \affil[15]{Nara Womana's University, Japan}
\author[16]{C.-J.~ Lin}\affil[16]{Lawrence Berkeley National Laboratory, USA}
\author[17]{D.~Lopes Pegna}\affil[17]{Princeton University, USA}
\author[18]{R.~Louvot}\affil[18]{Ecole Polytechnique F\'{e}d\'{e}rale de Lausanne (EPFL), Switzerland}
\author[19]{A.~Lusiani} \affil[19]{INFN Pisa, Italy}
\author[20]{V.~L\"{u}th}\affil[20]{SLAC National Accelerator Laboratory, USA}
\author[21]{B.~Meadows} \affil[21]{University of Cincinnati, USA}
\author[22]{S.~Nishida}\affil[22]{KEK, Tsukuba, Japan}
\author[23]{D.~Pedrini} \affil[23]{INFN Milano-Bicocca, Italy}
\author[24]{M.~Purohit}\affil[24]{University of South Carolina, USA}
\author[25]{M.~Rama}\affil[25]{INFN Frascati, Italy}
\author[2]{M.~Roney}
\author[18]{O.~Schneider}
\author[10]{C.~Schwanda}
\author[21]{A.~J.~Schwartz}
\author[26]{B.~Shwartz}\affil[26]{Budker Institute of Nuclear Physics, Russia}
\author[27]{J.~G.~Smith}\affil[27]{University of Colorado, USA}
\author[28]{R.~Tesarek}\affil[28]{Fermilab, USA}
\author[28]{D.~Tonelli}
\author[22]{K.~Trabelsi}
\author[29]{P.~Urquijo}\affil[29]{Syracuse University, USA}
\author[30]{R.~Van Kooten}\affil[30]{Indiana University, USA}

\date{6 September 2011 \\
(original version: 8 October 2010)}
\maketitle

\begin{abstract}
This article reports world averages for measurements of $b$-hadron, $c$-hadron, 
and $\tau$ lepton properties obtained by the Heavy Flavor Averaging Group (HFAG) 
using results available at least through the end of 2009. Some of the world 
averages presented use data available through the spring of 2010.
For the averaging, common input parameters used in the various analyses 
are adjusted (rescaled) to common values, and known correlations are taken 
into account. The averages include branching fractions, lifetimes, neutral 
meson mixing parameters, \CP~violation parameters, and parameters of 
semileptonic decays.
\end{abstract}

\newpage
\tableofcontents
\newpage


\mysection{Introduction}
\label{sec:intro}

Flavor dynamics is an important element in understanding the nature of
particle physics.  The accurate knowledge of properties of heavy flavor
hadrons, especially $b$ hadrons, plays an essential role for
determining the elements of the Cabibbo-Kobayashi-Maskawa (CKM)
weak-mixing matrix~\cite{Cabibbo:1963yz,Kobayashi:1973fv}. 
Since the \belle\ and \babar\ $e^+e^-$ $B$ factory 
experiments began collecting data, the size of available 
$B$ meson samples has dramatically increased, and the 
accuracies of measurements have greatly improved. 
The CDF and \dzero\ experiments at the Fermilab Tevatron 
have also provided important results on $B$ and $D$ meson
decays, most notably the discovery of $B^0_s$-$\Bsb^0$ mixing,
and confirmation of $D^0$-$\Dzb$ mixing.
 
The Heavy Flavor Averaging Group (HFAG) was formed in 2002 to 
continue the activities of the LEP Heavy Flavor Steering 
group~\cite{Abbaneo:2000ej_mod,*Abbaneo:2001bv_mod_cont}. 
This group was responsible for calculating averages of 
measurements of $b$-flavor related quantities. HFAG has evolved 
since its inception and currently consists of seven subgroups:
\begin{itemize}
\item the ``$B$ Lifetime and Oscillations'' subgroup provides 
averages for $b$-hadron lifetimes, $b$-hadron fractions in 
$\Upsilon(4S)$ decay and $p\bar{p}$ collisions, and various 
parameters governing $B^0$-$\Bzb$ and $B_s^0$-$\Bsb^0$ mixing;

\item the ``Unitarity Triangle Parameters'' subgroup provides
averages for time-dependent $\CP$ asymmetry parameters and 
resulting determinations of the angles of the CKM unitarity triangle;

\item the ``Semileptonic $B$ Decays'' subgroup provides averages
for inclusive and exclusive $B$-decay branching fractions, and
subsequent determinations of the CKM matrix elements 
$|V_{cb}|$ and $|V_{ub}|$;

\item the ``$B$ to Charm Decays'' subgroup provides averages of 
branching fractions for $B$ decays to final states involving open 
charm or charmonium mesons;

\item the ``Rare Decays'' subgroup provides averages of branching 
fractions and $\CP$ asymmetries for charmless, radiative, 
leptonic, and baryonic $B$ meson decays;

\item the ``Charm Physics'' subgroup provides averages of branching 
fractions for $D$ meson hadronic and semileptonic decays, properties 
of excited $D^{**}$ and $D^{}_{sJ}$ mesons, averages of $D^0$-$\Dzb$ 
mixing and $\CP$ and $T$ violation parameters, and an average value 
for the $D^{}_s$ decay constant~$f^{}_{D_s}$.

\item the ``Tau Physics'' subgroup provides documentation and
averages for the $\tau$ lepton mass and branching fractions,
and documents upper limits for $\tau$ lepton-flavor-violating decays.
\end{itemize}
The ``Lifetime and Oscillations'' and ``Semileptonic'' subgroups continue the 
activities of the LEP working groups with some reorganization, i.e., merging 
four groups into two. 
The ``Unitary Triangle,'' ``$B$ to Charm Decays,'' and ``Rare Decays''
subgroups were formed to provide averages for new results obtained
from the $B$ factory experiments (and now also from the Fermilab 
Tevatron experiments).
The ``Charm'' and ``Tau''  subgroups were formed more recently in 
response to the wealth of new data concerning $D$ and $\tau$ decays. 
All subgroups include representatives from \belle\ and \babar\ and, 
when relevant, CLEO, CDF, and \dzero.

This article is an update of the ``End of 2007'' HFAG 
preprint~\cite{Barberio:2008fa}. Here we report world 
averages using results available at least through the end 
of 2009. Averages reported in Chapters~\ref{sec:life_mix} and
\ref{sec:charm_physics} incorporate results available through 
the spring of 2010. In general, we use all publicly available 
results that have written documentation. These include preliminary 
results presented at conferences or workshops.
However, we do not use preliminary results that remain unpublished 
for an extended period of time, or for which no publication is planned. 
Close contacts have been established between representatives from
the experiments and members of subgroups that perform averaging 
to ensure that the data are prepared in a form suitable for 
combinations.  

In the case of obtaining a world average for which $\chi^2/\dof > 1$,
where $\dof$ is the number of degrees of freedom in the average
calculation, we do not scale the resulting error, as is presently 
done by the Particle Data Group~\cite{PDG_2010}. Rather, 
we examine the systematics of each measurement to better understand them. 
Unless we find possible systematic discrepancies between the measurements, 
we do not apply any additional correction to the calculated error. 
We provide the confidence level of the fit as an indicator for the 
consistency of the measurements included in the average. In case some
special treatment was necessary to calculate an average, or if an
approximation used in the average calculation may not be good enough 
(\eg, assuming Gaussian errors when the likelihood function indicates 
non-Gaussian behavior), we include a warning message.

Chapter~\ref{sec:method} describes the methodology used for calculating
averages. In the averaging procedure, common input parameters used in 
the various analyses are adjusted (rescaled) to common values, and, 
where possible, known correlations are taken into account. 
Chapters~\ref{sec:life_mix}--\ref{sec:tau} present world 
average values from each of the subgroups listed above. A brief 
summary of the averages presented is given in Chapter~\ref{sec:summary}.   
A complete listing of the averages and plots are also available on the 
HFAG web site:
\vskip0.15in\hskip0.75in
\vbox{
{\tt http://www.slac.stanford.edu/xorg/hfag } and 
\vskip0.15in
{\tt http://belle.kek.jp/mirror/hfag } (KEK mirror site).
}

\section{Methodology } 
\label{sec:method} 

The general averaging problem that HFAG faces is to combine 
information provided by different measurements of the same parameter
to obtain our best estimate of the parameter's value and
uncertainty. The methodology described here focuses on the problems of
combining measurements performed with different systematic assumptions
and with potentially-correlated systematic uncertainties. Our methodology
relies on the close involvement of the people performing the
measurements in the averaging process.

Consider two hypothetical measurements of a parameter $x$, which might
be summarized as
\begin{align*}
x &= x_1 \pm \delta x_1 \pm \Delta x_{1,1} \pm \Delta x_{2,1} \ldots \\
x &= x_2 \pm \delta x_2 \pm \Delta x_{1,2} \pm \Delta x_{2,2} \ldots
\; ,
\end{align*}
where the $\delta x_k$ are statistical uncertainties, and
the $\Delta x_{i,k}$ are contributions to the systematic
uncertainty. One popular approach is to combine statistical and
systematic uncertainties in quadrature
\begin{align*}
x &= x_1 \pm \left(\delta x_1 \oplus \Delta x_{1,1} \oplus \Delta
x_{2,1} \oplus \ldots\right) \\
x &= x_2 \pm \left(\delta x_2 \oplus \Delta x_{1,2} \oplus \Delta
x_{2,2} \oplus \ldots\right)
\end{align*}
and then perform a weighted average of $x_1$ and $x_2$, using their
combined uncertainties, as if they were independent. This approach
suffers from two potential problems that we attempt to address. First,
the values of the $x_k$ may have been obtained using different
systematic assumptions. For example, different values of the \Bz
lifetime may have been assumed in separate measurements of the
oscillation frequency $\deltamd$. The second potential problem is that
some contributions of the systematic uncertainty may be correlated
between experiments. For example, separate measurements of $\deltamd$
may both depend on an assumed Monte-Carlo branching fraction used to
model a common background.

The problems mentioned above are related since, ideally, any quantity $y_i$
that $x_k$ depends on has a corresponding contribution $\Delta x_{i,k}$ to the
systematic error which reflects the uncertainty $\Delta y_i$ on $y_i$
itself. We assume that this is the case and use the values of $y_i$ and
$\Delta y_i$ assumed by each measurement explicitly in our
averaging (we refer to these values as $y_{i,k}$ and $\Delta y_{i,k}$
below). Furthermore, since we do not lump all the systematics
together,
we require that each measurement used in an average have a consistent
definition of the various contributions to the systematic uncertainty.
Different analyses often use different decompositions of their systematic
uncertainties, so achieving consistent definitions for any potentially
correlated contributions requires close coordination between HFAG and
the experiments. In some cases, a group of
systematic uncertainties must be lumped to obtain a coarser
description that is consistent between measurements. Systematic uncertainties
that are uncorrelated with any other sources of uncertainty appearing
in an average are lumped with the statistical error, so that the only
systematic uncertainties treated explicitly are those that are
correlated with at least one other measurement via a consistently-defined
external parameter $y_i$. When asymmetric statistical or systematic
uncertainties are quoted, we symmetrize them since our combination
method implicitly assumes parabolic likelihoods for each measurement.

\begin{figure}
\begin{center}
\includegraphics[width=6.0in]{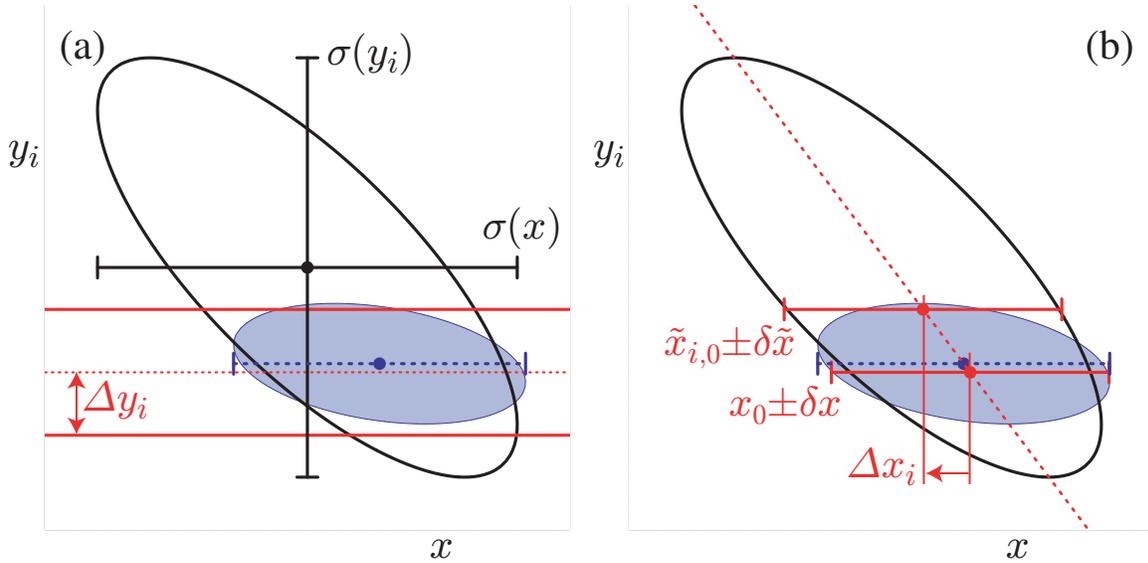}
\end{center}
\caption{The left-hand plot (a) compares the 68\% confidence-level
  contours of a
  hypothetical measurement's unconstrained (large ellipse) and
  constrained (filled ellipse) likelihoods, using the Gaussian
  constraint on $y_i$ represented by the horizontal band. The solid
  error bars represent the statistical uncertainties $\sigma(x)$ and
  $\sigma(y_i)$ of the unconstrained likelihood. The dashed
  error bar shows the statistical error on $x$ from a
  constrained simultaneous fit to $x$ and $y_i$. The right-hand plot
  (b) illustrates the method described in the text of performing fits
  to $x$ with $y_i$ fixed at different values. The dashed
  diagonal line between these fit results has the slope
  $\rho(x,y_i)\sigma(y_i)/\sigma(x)$ in the limit of a parabolic
  unconstrained likelihood. The result of the constrained simultaneous
  fit from (a) is shown as a dashed error bar on $x$.}
\label{fig:singlefit}
\end{figure}

The fact that a measurement of $x$ is sensitive to the value of $y_i$
indicates that, in principle, the data used to measure $x$ could
equally-well be used for a simultaneous measurement of $x$ and $y_i$, as
illustrated by the large contour in Fig.~\ref{fig:singlefit}(a) for a hypothetical
measurement. However, we often have an external constraint $\Delta
y_i$ on the value of $y_i$ (represented by the horizontal band in
Fig.~\ref{fig:singlefit}(a)) that is more precise than the constraint
$\sigma(y_i)$ from
our data alone. Ideally, in such cases we would perform a simultaneous
fit to $x$ and $y_i$, including the external constraint, obtaining the
filled $(x,y)$ contour and corresponding dashed one-dimensional estimate of
$x$ shown in Fig.~\ref{fig:singlefit}(a). Throughout, we assume that
the external constraint $\Delta y_i$ on $y_i$ is Gaussian.

In practice, the added technical complexity of a constrained fit with
extra free parameters is not justified by the small increase in
sensitivity, as long as the external constraints $\Delta y_i$ are
sufficiently precise when compared with the sensitivities $\sigma(y_i)$
to each $y_i$ of the data alone. Instead, the usual procedure adopted
by the experiments is to perform a baseline fit with all $y_i$ fixed
to nominal values $y_{i,0}$, obtaining $x = x_0 \pm \delta
x$. This baseline fit neglects the uncertainty due to $\Delta y_i$, but
this error can be mostly recovered by repeating the fit separately for
each external parameter $y_i$ with its value fixed at $y_i = y_{i,0} +
\Delta y_i$ to obtain $x = \tilde{x}_{i,0} \pm \delta\tilde{x}$, as
illustrated in Fig.~\ref{fig:singlefit}(b). The absolute shift,
$|\tilde{x}_{i,0} - x_0|$, in the central value of $x$ is what the
experiments usually quote as their systematic uncertainty $\Delta x_i$
on $x$ due to the unknown value of $y_i$. Our procedure requires that
we know not only the magnitude of this shift but also its sign. In the
limit that the unconstrained data is represented by a parabolic
likelihood, the signed shift is given by
\begin{equation}
\Delta x_i = \rho(x,y_i)\frac{\sigma(x)}{\sigma(y_i)}\,\Delta y_i \;,
\end{equation}
where $\sigma(x)$ and $\rho(x,y_i)$ are the statistical uncertainty on
$x$ and the correlation between $x$ and
$y_i$ in the unconstrained data.
While our procedure is not
equivalent to the constrained fit with extra parameters, it yields (in
the limit of a parabolic unconstrained likelihood) a central value
$x_0$ that agrees 
to ${\cal O}(\Delta y_i/\sigma(y_i))^2$ and an uncertainty $\delta x
\oplus \Delta x_i$ that agrees to ${\cal O}(\Delta y_i/\sigma(y_i))^4$.

In order to combine two or more measurements that share systematics
due to the same external parameters $y_i$, we would ideally perform a
constrained simultaneous fit of all data samples to obtain values of
$x$ and each $y_i$, being careful to only apply the constraint on each
$y_i$ once. This is not practical since we generally do not have
sufficient information to reconstruct the unconstrained likelihoods
corresponding to each measurement. Instead, we perform the two-step
approximate procedure described below.

\begin{figure}
\begin{center}
\includegraphics[width=6.0in]{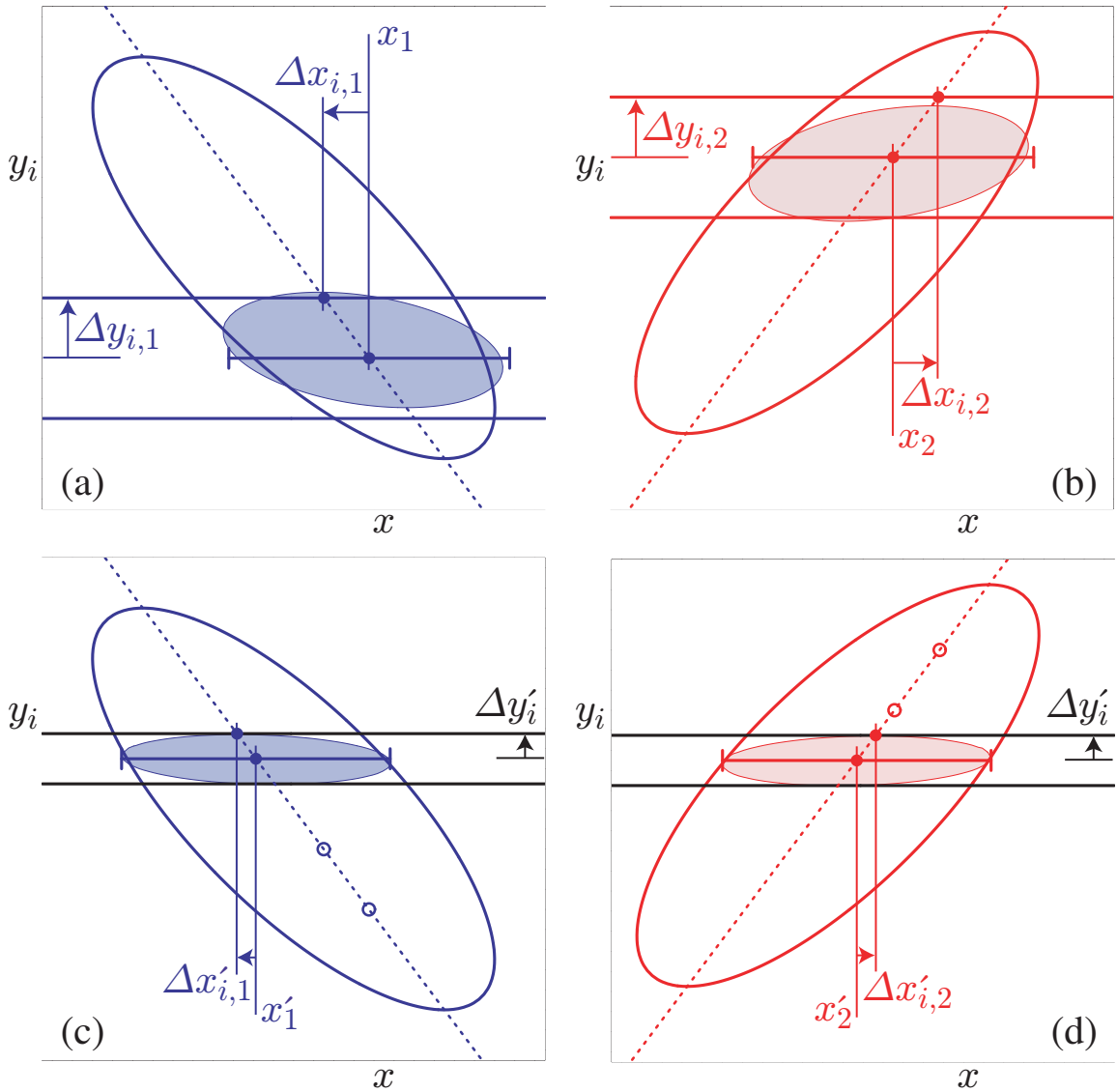}
\end{center}
\caption{The upper plots (a) and (b) show examples of two individual
  measurements to be combined. The large ellipses represent their
  unconstrained likelihoods, and the filled ellipses represent their
  constrained likelihoods. Horizontal bands indicate the different
  assumptions about the value and uncertainty of $y_i$ used by each
  measurement. The error bars show the results of the approximate
  method described in the text for obtaining $x$ by performing fits
  with $y_i$ fixed to different values. The lower plots (c) and (d)
  illustrate the adjustments to accommodate updated and consistent
  knowledge of $y_i$ as described in the text. Open circles mark the
  central values of the unadjusted fits to $x$ with $y$ fixed; these
  determine the dashed line used to obtain the adjusted values. }
\label{fig:multifit}
\end{figure}

Figs.~\ref{fig:multifit}(a,b) illustrate two
statistically-independent measurements, $x_1 \pm (\delta x_1 \oplus
\Delta x_{i,1})$ and $x_2\pm(\delta x_i\oplus \Delta x_{i,2})$, of the same
hypothetical quantity $x$ (for simplicity, we only show the
contribution of a single correlated systematic due to an external
parameter $y_i$). As our knowledge of the external parameters $y_i$
evolves, it is natural that the different measurements of $x$ will
assume different nominal values and ranges for each $y_i$. The first
step of our procedure is to adjust the values of each measurement to
reflect the current best knowledge of the values $y_i'$ and ranges
$\Delta y_i'$ of the external parameters $y_i$, as illustrated in
Figs.~\ref{fig:multifit}(c,b). We adjust the
central values $x_k$ and correlated systematic uncertainties $\Delta
x_{i,k}$ linearly for each measurement (indexed by $k$) and each
external parameter (indexed by $i$):
\begin{align}
x_k' &= x_k + \sum_i\,\frac{\Delta x_{i,k}}{\Delta y_{i,k}}
\left(y_i'-y_{i,k}\right)\\
\Delta x_{i,k}'&= \Delta x_{i,k}\cdot \frac{\Delta y_i'}{\Delta
  y_{i,k}} \; .
\end{align}
This procedure is exact in the limit that the unconstrained
likelihoods of each measurement is parabolic.

\begin{figure}
\begin{center}
\includegraphics[width=3.5in]{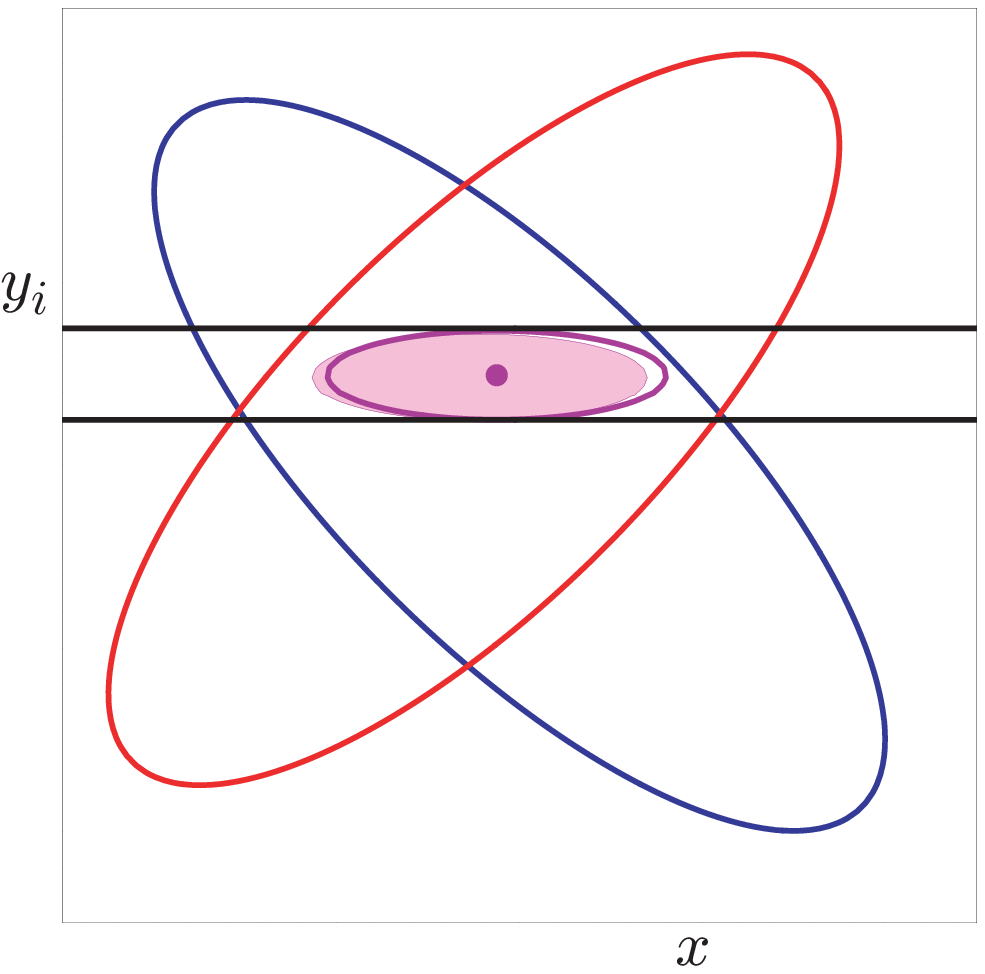}
\end{center}
\caption{An illustration of the combination of two hypothetical
  measurements of $x$ using the method described in the text. The
  ellipses represent the unconstrained likelihoods of each measurement,
  and the horizontal band represents the latest knowledge about $y_i$ 
  that is used to adjust the individual measurements. The filled small
  ellipse shows the result of the exact method using 
  ${\cal L}_{\text{comb}}$, and the hollow small ellipse and dot show 
  the result of the approximate method using $\chi^2_{\text{comb}}$.}
\label{fig:fit12}
\end{figure}

The second step of our procedure is to combine the adjusted
measurements, $x_k'\pm (\delta x_k\oplus \Delta x_{k,1}'\oplus \Delta
x_{k,2}'\oplus\ldots)$ using the chi-square 
\begin{equation}
\chi^2_{\text{comb}}(x,y_1,y_2,\ldots) \equiv \sum_k\,
\frac{1}{\delta x_k^2}\left[
x_k' - \left(x + \sum_i\,(y_i-y_i')\frac{\Delta x_{i,k}'}{\Delta y_i'}\right)
\right]^2 + \sum_i\,
\left(\frac{y_i - y_i'}{\Delta y_i'}\right)^2 \; ,
\end{equation}
and then minimize this $\chi^2$ to obtain the best values of $x$ and
$y_i$ and their uncertainties, as illustrated in
Fig.~\ref{fig:fit12}. Although this method determines new values for
the $y_i$, we do not report them since the $\Delta x_{i,k}$ reported
by each experiment are generally not intended for this purpose (for
example, they may represent a conservative upper limit rather than a
true reflection of a 68\% confidence level).

For comparison, the exact method we would
perform if we had the unconstrained likelihoods ${\cal L}_k(x,y_1,y_2,\ldots)$
available for each
measurement is to minimize the simultaneous constrained likelihood
\begin{equation}
{\cal L}_{\text{comb}}(x,y_1,y_2,\ldots) \equiv \prod_k\,{\cal
  L}_k(x,y_1,y_2,\ldots)\,\prod_{i}\,{\cal 
  L}_i(y_i) \; ,
\end{equation}
with an independent Gaussian external constraint on each $y_i$
\begin{equation}
{\cal L}_i(y_i) \equiv \exp\left[-\frac{1}{2}\,\left(\frac{y_i-y_i'}{\Delta
 y_i'}\right)^2\right] \; .
\end{equation}
The results of this exact method are illustrated by the filled ellipses
in Figs.~\ref{fig:fit12}(a,b) and agree with our method in the limit that
each ${\cal L}_k$ is parabolic and that each $\Delta
y_i' \ll \sigma(y_i)$. In the case of a non-parabolic unconstrained
likelihood, experiments would have to provide a description of ${\cal
  L}_k$ itself to allow an improved combination. In the case of
$\sigma(y_i)\simeq \Delta y_i'$, experiments are advised to perform a
simultaneous measurement of both $x$ and $y$ so that their data will
improve the world knowledge about $y$. 

 The algorithm described above is used as a default in the averages
reported in the following sections.  For some cases, somewhat simplified
or more complex algorithms are used and noted in the corresponding 
sections. Some examples for extensions of the standard method for extracting
averages are given here. These include the case where measurement errors
depend on the measured value, i.e. are relative errors, unknown
correlation coefficients and the breakdown of error sources.

For measurements with Gaussian errors, the usual estimator for the
average of a set of measurements is obtained by minimizing the following
$\chi^2$:
\begin{equation}
\chi^2(t) = \sum_i^N \frac{\left(y_i-t\right)^2}{\sigma^2_i} ,
\label{eq:chi2t}
\end{equation}
where $y_i$ is the measured value for input $i$ and $\sigma_i^2$ is the
variance of the distribution from which $y_i$ was drawn.  The value $\hat{t}$
of $t$ at minimum $\chi^2$ is our estimator for the average.  (This
discussion is given for independent measurements for the sake of
simplicity; the generalization to correlated measurements is
straightforward, and has been used when averaging results.) 
The true $\sigma_i$ are unknown but typically the error as assigned by the
experiment $\sigma_i^{\mathrm{raw}}$ is used as an estimator for it.
Caution is advised,
however, in the case where $\sigma_i^{\mathrm{raw}}$
depends on the value measured for $y_i$. Examples of this include
an uncertainty in any multiplicative factor (like
an acceptance) that enters the determination of $y_i$, i.e. the $\sqrt{N}$
dependence of Poisson statistics, where $y_i \propto N$
and $\sigma_i \propto \sqrt{N}$.
Failing to account for this type of
dependence when averaging leads to a biased average.
Biases in the average can be avoided (or at least reduced)
by minimizing the following
$\chi^2$:
\begin{equation}
\chi^2(t) = \sum_i^N \frac{\left(y_i-t\right)^2}{\sigma^2_i(\hat{t})} .
\label{eq:chi2that}
\end{equation}
In the above $\sigma_i(\hat{t})$ is the uncertainty
assigned to input $i$ that includes the assumed dependence of the
stated error on the value measured.  As an example, consider 
a pure acceptance error, for which
$\sigma_i(\hat{t}) = (\hat{t} / y_i)\times\sigma_i^{\mathrm{raw}}$ .
It is easily verified that solving Eq.~\ref{eq:chi2that} 
leads to the correct behavior, namely
$$ 
\hat{t} = \frac{\sum_i^N y_i^3/(\sigma_i^{\mathrm{raw}})^2}{\sum_i^N y_i^2/(\sigma_i^{\mathrm{raw}})^2},
$$
i.e. weighting by the inverse square of the 
fractional uncertainty, $\sigma_i^{\mathrm{raw}}/y_i$.
It is sometimes difficult to assess the dependence of $\sigma_i^{\mathrm{raw}}$ on
$\hat{t}$ from the errors quoted by experiments.  


Another issue that needs careful treatment is the question of correlation
among different measurements, e.g. due to using the same theory for
calculating acceptances.  A common practice is to set the correlation
coefficient to unity to indicate full correlation.  However, this is
not a ``conservative'' thing to do, and can in fact lead to a significantly
underestimated uncertainty on the average.  In the absence of
better information, the most conservative choice of correlation coefficient
between two measurements $i$ and $j$
is the one that maximizes the uncertainty on $\hat{t}$
due to that pair of measurements:
\begin{equation}
\sigma_{\hat{t}(i,j)}^2 = \frac{\sigma_i^2\,\sigma_j^2\,(1-\rho_{ij}^2)}
   {\sigma_i^2 + \sigma_j^2 - 2\,\rho_{ij}\,\sigma_i\,\sigma_j} ,
\label{eq:correlij}
\end{equation}
namely
\begin{equation}
\rho_{ij} = \mathrm{min}\left(\frac{\sigma_i}{\sigma_j},\frac{\sigma_j}{\sigma_i}\right) ,
\label{eq:correlrho}
\end{equation}
which corresponds to setting $\sigma_{\hat{t}(i,j)}^2=\mathrm{min}(\sigma_i^2,\sigma_j^2)$.
Setting $\rho_{ij}=1$ when $\sigma_i\ne\sigma_j$ can lead to a significant
underestimate of the uncertainty on $\hat{t}$, as can be seen
from Eq.~\ref{eq:correlij}.

Finally, we carefully consider the various sources of error
contributing to the overall uncertainty of an average.
The overall covariance matrix is constructed from a number of
individual sources, e.g.
$\mathbf{V} = \mathbf{V_{stat}+V_{sys}+V_{th}}$.
The variance on the average $\hat{t}$ can be written
\begin{eqnarray}
\sigma^2_{\hat{t}} 
 &=& 
\frac{ \sum_{i,j}\left(\mathbf{V^{-1}}\, 
\mathbf{[V_{stat}+V_{sys}+V_{th}]}\, \mathbf{V^{-1}}\right)_{ij}}
{\left(\sum_{i,j} V^{-1}_{ij}\right)^2}
= \sigma^2_{stat} + \sigma^2_{sys} + \sigma^2_{th} .
\end{eqnarray}
Written in this form, one can readily determine the 
contribution of each source of uncertainty to the overall uncertainty
on the average.  This breakdown of the uncertainties is used 
in the following sections.

Following the prescription described above, the central values and
errors are rescaled to a common set of input parameters in the averaging
procedures according to the dependency on any of these input parameters.
We try to use the most up-to-date values for these common inputs and 
the same values among the HFAG subgroups. For the parameters whose
averages are produced by HFAG, we use the values in the current 
update cycle.  For other external parameters, we use the most
recent PDG values available (usually Ref.~\cite{PDG_2010}). 
The parameters and values used are listed in each subgroup section.

\clearpage
%
%
%
%

%

%
%
%
%

%

\renewcommand{\floatpagefraction}{0.8}
\renewcommand{\topfraction}{0.9}

\newcommand{\comment}[1]{}

\newcommand{\auth}[1]{#1,}
\newcommand{\coll}[1]{#1 Collaboration,}
\newcommand{\authcoll}[2]{#1 \etal\ (#2 Collaboration),}
\newcommand{\authgrp}[2]{#1 \etal\ (#2),}
\newcommand{\titl}[1]{``#1'',} 
\newcommand{\J}[4]{{#1} {\bf #2}, #3 (#4)}
\newcommand{\subJ}[1]{submitted to #1}
\newcommand{\PRL}[3]{\J{Phys.\ Rev.\ Lett.}{#1}{#2}{#3}}
\newcommand{\subPRL}{\subJ{Phys.\ Rev.\ Lett.}}
\newcommand{\PRD}[3]{\J{Phys.\ Rev.\ D}{#1}{#2}{#3}}
\newcommand{\subPRD}{\subJ{Phys.\ Rev.\ D}}
\newcommand{\PREP}[3]{\J{Phys.\ Reports}{#1}{#2}{#3}}
\newcommand{\ZPC}[3]{\J{Z.\ Phys.\ C}{#1}{#2}{#3}}
\newcommand{\PLB}[3]{\J{Phys.\ Lett.\ B}{#1}{#2}{#3}}
\newcommand{\subPLB}{\subJ{Phys.\ Lett.\ B}}
\newcommand{\EPJC}[3]{\J{Eur.\ Phys.\ J.\ C}{#1}{#2}{#3}}
\newcommand{\NPB}[3]{\J{Nucl.\ Phys.\ B}{#1}{#2}{#3}}
\newcommand{\subNPB}{\subJ{Nucl.\ Phys.\ B}}
\newcommand{\NIMA}[3]{\J{Nucl.\ Instrum.\ Methods A}{#1}{#2}{#3}}
\newcommand{\subNIMA}{\subJ{Nucl.\ Instrum.\ Methods A}}
\newcommand{\JHEP}[3]{\J{J.\ of High Energy Physics }{#1}{#2}{#3}}
\newcommand{\JPG}[3]{\J{J.\ of Physics G}{#1}{#2}{#3}}
\newcommand{\ARNS}[3]{\J{Ann.\ Rev.\ Nucl.\ Sci.}{#1}{#2}{#3}}
\newcommand{\newref}{\\}

\newcommand{\particle}[1]{\ensuremath{#1}\xspace}
\renewcommand{\ee}{\particle{e^+e^-}}
\newcommand{\Ups}{\particle{\Upsilon(4S)}}
\newcommand{\Upsfive}{\particle{\Upsilon(5S)}}
\renewcommand{\b}{\particle{b}}
\renewcommand{\B}{\particle{B}}
\newcommand{\Bd}{\particle{B^0}}
\renewcommand{\Bs}{\particle{B^0_s}}
\renewcommand{\Bu}{\particle{B^+}}
\newcommand{\Bc}{\particle{B^+_c}}
\newcommand{\Bdbar}{\particle{\bar{B}^0}}
\newcommand{\Bsbar}{\particle{\bar{B}^0_s}}
\newcommand{\Lb}{\particle{\Lambda_b^0}}
\newcommand{\Xib}{\particle{\Xi_b}}
\newcommand{\Xibd}{\particle{\Xi_b^-}}
\newcommand{\Omegab}{\particle{\Omega_b^-}}
\newcommand{\Lc}{\particle{\Lambda_c^+}}

\newcommand{\fBs}{\ensuremath{f_{\particle{s}}}\xspace}
\newcommand{\fBd}{\ensuremath{f_{\particle{d}}}\xspace}
\newcommand{\fBu}{\ensuremath{f_{\particle{u}}}\xspace}
\newcommand{\fbb}{\ensuremath{f_{\rm baryon}}\xspace}

\newcommand{\dmd}{\ensuremath{\Delta m_{\particle{d}}}\xspace}
\newcommand{\dms}{\ensuremath{\Delta m_{\particle{s}}}\xspace}
\newcommand{\xd}{\ensuremath{x_{\particle{d}}}\xspace}
\newcommand{\xs}{\ensuremath{x_{\particle{s}}}\xspace}
\newcommand{\yd}{\ensuremath{y_{\particle{d}}}\xspace}
\newcommand{\ys}{\ensuremath{y_{\particle{s}}}\xspace}
\newcommand{\chibar}{\ensuremath{\overline{\chi}}\xspace}
\newcommand{\chid}{\ensuremath{\chi_{\particle{d}}}\xspace}
\newcommand{\chis}{\ensuremath{\chi_{\particle{s}}}\xspace}
\newcommand{\Gd}{\ensuremath{\Gamma_{\particle{d}}}\xspace}
\newcommand{\DGd}{\ensuremath{\Delta\Gd}\xspace}
\newcommand{\DGGd}{\ensuremath{\DGd/\Gd}\xspace}
\newcommand{\Gs}{\ensuremath{\Gamma_{\particle{s}}}\xspace}
\newcommand{\DGs}{\ensuremath{\Delta\Gs}\xspace}
\newcommand{\DGGs}{\ensuremath{\Delta\Gs/\Gs}\xspace}
\newcommand{\ASLd}{\ensuremath{{\cal A}_{\rm SL}^\particle{d}}\xspace}
\newcommand{\ASLs}{\ensuremath{{\cal A}_{\rm SL}^\particle{s}}\xspace}
\newcommand{\ASLb}{\ensuremath{{\cal A}_{\rm SL}^\particle{b}}\xspace}

\renewcommand{\BR}[1]{\particle{{\cal B}(#1)}}
\newcommand{\CL}[1]{#1\%~\mbox{CL}}
\newcommand{\Qjet}{\ensuremath{Q_{\rm jet}}\xspace}

\newcommand{\labe}[1]{\label{equ:#1}}
\newcommand{\labs}[1]{\label{sec:#1}}
\newcommand{\labf}[1]{\label{fig:#1}}
\newcommand{\labt}[1]{\label{tab:#1}}
\newcommand{\refe}[1]{\ref{equ:#1}}
\newcommand{\refs}[1]{\ref{sec:#1}}
\newcommand{\reff}[1]{\ref{fig:#1}}
\newcommand{\reft}[1]{\ref{tab:#1}}
\newcommand{\Ref}[1]{Ref.~\cite{#1}}
\newcommand{\Refs}[1]{Refs.~\cite{#1}}
\newcommand{\Refss}[2]{Refs.~\cite{#1} and \cite{#2}}
\newcommand{\Refsss}[3]{Refs.~\cite{#1}, \cite{#2} and \cite{#3}}
\newcommand{\eq}[1]{(\refe{#1})}
\newcommand{\Eq}[1]{Eq.~(\refe{#1})}
\newcommand{\Eqs}[1]{Eqs.~(\refe{#1})}
\newcommand{\Eqss}[2]{Eqs.~(\refe{#1}) and (\refe{#2})}
\newcommand{\Eqssor}[2]{Eqs.~(\refe{#1}) or (\refe{#2})}
\newcommand{\Eqsss}[3]{Eqs.~(\refe{#1}), (\refe{#2}), and (\refe{#3})}
\newcommand{\Figure}[1]{Figure~\reff{#1}}
\newcommand{\Figuress}[2]{Figures~\reff{#1} and \reff{#2}}
\newcommand{\Fig}[1]{Fig.~\reff{#1}}
\newcommand{\Figs}[1]{Figs.~\reff{#1}}
\newcommand{\Figss}[2]{Figs.~\reff{#1} and \reff{#2}}
\newcommand{\Figsss}[3]{Figs.~\reff{#1}, \reff{#2}, and \reff{#3}}
\newcommand{\Section}[1]{Section~\refs{#1}}
\newcommand{\Sec}[1]{Sec.~\refs{#1}}
\newcommand{\Secs}[1]{Secs.~\refs{#1}}
\newcommand{\Secss}[2]{Secs.~\refs{#1} and \refs{#2}}
\newcommand{\Secsss}[3]{Secs.~\refs{#1}, \refs{#2}, and \refs{#3}}
\newcommand{\Table}[1]{Table~\reft{#1}}
\newcommand{\Tables}[1]{Tables~\reft{#1}}
\newcommand{\Tabless}[2]{Tables~\reft{#1} and \reft{#2}}
\newcommand{\Tablesss}[3]{Tables~\reft{#1}, \reft{#2}, and \reft{#3}}

\newcommand{\subsubsubsection}[1]{\vspace{2ex}\par\noindent {\bf\boldmath\em #1} \vspace{2ex}\par}


\newcommand{\definemath}[2]{\newcommand{#1}{\ensuremath{#2}\xspace}}

\definemath{\hfagCHIBARLEPval}{0.1259}
\definemath{\hfagCHIBARLEPerr}{\pm0.0042}
\definemath{\hfagTAUBDval}{1.518}
\definemath{\hfagTAUBDerr}{\pm0.007}
\definemath{\hfagTAUBUval}{1.641}
\definemath{\hfagTAUBUerr}{\pm0.008}
\definemath{\hfagRTAUBUval}{1.081}
\definemath{\hfagRTAUBUerr}{\pm0.006}
\definemath{\hfagTAUBSval}{1.458}
\definemath{\hfagTAUBSerr}{\pm0.030}
\definemath{\hfagRTAUBSval}{0.960}
\definemath{\hfagRTAUBSerr}{\pm0.020}
\definemath{\hfagTAULBval}{1.425}
\definemath{\hfagTAULBerr}{\pm0.032}
\definemath{\hfagTAUBBval}{1.382}
\definemath{\hfagTAUBBerr}{\pm0.029}
\definemath{\hfagRTAUBBval}{0.910}
\definemath{\hfagRTAUBBerr}{\pm0.020}
\definemath{\hfagTAUXBval}{1.49}
\definemath{\hfagTAUXBerp}{^{+0.19}}
\definemath{\hfagTAUXBern}{_{-0.18}}
\definemath{\hfagTAUBval}{1.568}
\definemath{\hfagTAUBerr}{\pm0.009}
\definemath{\hfagTAUXBDval}{1.56}
\definemath{\hfagTAUXBDerp}{^{+0.27}}
\definemath{\hfagTAUXBDern}{_{-0.25}}
\definemath{\hfagTAUOBval}{1.13}
\definemath{\hfagTAUOBerp}{^{+0.53}}
\definemath{\hfagTAUOBern}{_{-0.40}}
\definemath{\hfagTAUBCval}{0.461}
\definemath{\hfagTAUBCerr}{\pm0.036}
\definemath{\hfagTAUBSSLval}{1.455}
\definemath{\hfagTAUBSSLerr}{\pm0.030}
\definemath{\hfagTAUBSSLXval}{1.456}
\definemath{\hfagTAUBSSLXerr}{\pm0.031}
\definemath{\hfagTAUBSMEANCONval}{1.477}
\definemath{\hfagTAUBSMEANCONerp}{^{+0.021}}
\definemath{\hfagTAUBSMEANCONern}{_{-0.022}}
\definemath{\hfagTAUBSJFval}{1.477}
\definemath{\hfagTAUBSJFerr}{\pm0.046}
\definemath{\hfagRTAUBSSLval}{0.958}
\definemath{\hfagRTAUBSSLerr}{\pm0.020}
\definemath{\hfagRTAUBSMEANCONval}{0.973}
\definemath{\hfagRTAUBSMEANCONerr}{\pm0.015}
\definemath{\hfagRTAUBSMEANCONsig}{1.8}
\definemath{\hfagONEMINUSRTAUBSMEANCONpercent}{(2.7\pm1.5)\%}
\definemath{\hfagRTAULBval}{0.939}
\definemath{\hfagRTAULBerr}{\pm0.022}
\definemath{\hfagTAUBVTXval}{1.572}
\definemath{\hfagTAUBVTXerr}{\pm0.009}
\definemath{\hfagTAUBLEPval}{1.537}
\definemath{\hfagTAUBLEPerr}{\pm0.020}
\definemath{\hfagTAUBJPval}{1.533}
\definemath{\hfagTAUBJPerp}{^{+0.038}}
\definemath{\hfagTAUBJPern}{_{-0.034}}
\definemath{\hfagNSIGMATAULBCDFTWO}{3.3}
\definemath{\hfagSDGDGDval}{0.011}
\definemath{\hfagSDGDGDerr}{\pm0.037}
\definemath{\hfagTAUBSSHORTval}{1.47}
\definemath{\hfagTAUBSSHORTerr}{\pm0.16}
\definemath{\hfagBRDSDSval}{0.049}
\definemath{\hfagBRDSDSerr}{\pm0.014}
\definemath{\hfagDGSGSBRDSDSval}{+0.103}
\definemath{\hfagDGSGSBRDSDSerr}{\pm0.032}
\definemath{\hfagTAUBSMEANval}{1.506}
\definemath{\hfagTAUBSMEANerr}{\pm0.032}
\definemath{\hfagDGSGSval}{+0.105}
\definemath{\hfagDGSGSerr}{\pm+0.060}
\definemath{\hfagDGSGSlow}{-0.011}
\definemath{\hfagDGSGSupp}{+0.224}
\definemath{\hfagRHODGSGSTAUBSMEAN}{+9.999999}
\definemath{\hfagDGSval}{+0.070}
\definemath{\hfagDGSerr}{\pm0.039}
\definemath{\hfagDGSlow}{-0.006}
\definemath{\hfagDGSupp}{+0.146}
\definemath{\hfagRHODGSTAUBSMEAN}{+9.999999}
\definemath{\hfagTAUBSLval}{1.431}
\definemath{\hfagTAUBSLerp}{^{+0.038}}
\definemath{\hfagTAUBSLern}{_{-0.037}}
\definemath{\hfagTAUBSHval}{1.590}
\definemath{\hfagTAUBSHerp}{^{+0.075}}
\definemath{\hfagTAUBSHern}{_{+0.071}}
\definemath{\hfagDGSGSCONBDval}{N/A}
\definemath{\hfagDGSGSCONBDerp}{^{N/A}}
\definemath{\hfagDGSGSCONBDern}{_{N/A}}
\definemath{\hfagTAUBSMEANCONXval}{1.477}
\definemath{\hfagTAUBSMEANCONXerp}{^{+0.021}}
\definemath{\hfagTAUBSMEANCONXern}{_{-0.022}}
\definemath{\hfagDGSGSCONval}{+0.072}
\definemath{\hfagDGSGSCONerp}{^{+0.049}}
\definemath{\hfagDGSGSCONern}{_{-0.051}}
\definemath{\hfagDGSGSCONlow}{-0.028}
\definemath{\hfagDGSGSCONupp}{+0.167}
\definemath{\hfagRHODGSGSTAUBSMEANCON}{+9.999999}
\definemath{\hfagDGSCONval}{+0.049}
\definemath{\hfagDGSCONerp}{^{+0.033}}
\definemath{\hfagDGSCONern}{_{-0.034}}
\definemath{\hfagDGSCONlow}{-0.019}
\definemath{\hfagDGSCONupp}{+0.112}
\definemath{\hfagRHODGSTAUBSMEANCON}{+9.999999}
\definemath{\hfagTAUBSLCONval}{1.425}
\definemath{\hfagTAUBSLCONerp}{^{+0.037}}
\definemath{\hfagTAUBSLCONern}{_{-0.035}}
\definemath{\hfagTAUBSHCONval}{1.532}
\definemath{\hfagTAUBSHCONerr}{\pm0.049}
\definemath{\hfagDGSGSCONBDCONval}{N/A}
\definemath{\hfagDGSGSCONBDCONerp}{^{N/A}}
\definemath{\hfagDGSGSCONBDCONern}{_{N/A}}
\definemath{\hfagTAUBSMEANCONXXval}{1.479}
\definemath{\hfagTAUBSMEANCONXXerr}{\pm0.021}
\definemath{\hfagDGSGSCONXval}{+0.089}
\definemath{\hfagDGSGSCONXerr}{\pm0.032}
\definemath{\hfagDGSGSCONXlow}{+0.025}
\definemath{\hfagDGSGSCONXupp}{+0.150}
\definemath{\hfagRHODGSGSTAUBSMEANCONX}{+9.999999}
\definemath{\hfagDGSCONXval}{+0.060}
\definemath{\hfagDGSCONXerr}{\pm0.021}
\definemath{\hfagDGSCONXlow}{+0.017}
\definemath{\hfagDGSCONXupp}{+0.101}
\definemath{\hfagRHODGSTAUBSMEANCONX}{+9.999999}
\definemath{\hfagTAUBSLCONXval}{1.416}
\definemath{\hfagTAUBSLCONXerr}{\pm0.027}
\definemath{\hfagTAUBSHCONXval}{1.548}
\definemath{\hfagTAUBSHCONXerp}{^{+0.036}}
\definemath{\hfagTAUBSHCONXern}{_{-0.037}}
\definemath{\hfagDGSGSCONBDCONXval}{N/A}
\definemath{\hfagDGSGSCONBDCONXerp}{^{N/A}}
\definemath{\hfagDGSGSCONBDCONXern}{_{N/A}}
\definemath{\hfagFCWval}{0.513}
\definemath{\hfagFCWerr}{\pm0.006}
\definemath{\hfagFNWval}{0.487}
\definemath{\hfagFNWerr}{\pm0.006}
\definemath{\hfagFFWval}{1.052}
\definemath{\hfagFFWerr}{\pm0.025}
\definemath{\hfagNSIGMAFFW}{2.1}
\definemath{\hfagFCNval}{0.513}
\definemath{\hfagFCNerr}{\pm0.013}
\definemath{\hfagFNNval}{0.487}
\definemath{\hfagFNNerr}{\pm0.013}
\definemath{\hfagFFNval}{1.053}
\definemath{\hfagFFNerr}{\pm0.054}
\definemath{\hfagFCval}{0.513}
\definemath{\hfagFCerr}{\pm0.007}
\definemath{\hfagFNval}{0.487}
\definemath{\hfagFNerr}{\pm0.007}
\definemath{\hfagFFval}{1.052}
\definemath{\hfagFFerr}{\pm0.028}
\definemath{\hfagNSIGMAFF}{1.9}
\definemath{\hfagFPRODval}{0.512}
\definemath{\hfagFPRODerr}{\pm0.019}
\definemath{\hfagFSUMval}{0.999}
\definemath{\hfagFSUMerr}{\pm0.030}
\definemath{\hfagFSFIVEOSval}{0.206}
\definemath{\hfagFSFIVEOSsta}{\pm0.010}
\definemath{\hfagFSFIVEOSsys}{\pm0.024}
\definemath{\hfagFSFIVEOSerr}{\pm0.027}
\definemath{\hfagFSFIVERLval}{0.215}
\definemath{\hfagFSFIVERLerr}{\pm0.031}
\definemath{\hfagFUDFIVEval}{0.763}
\definemath{\hfagFUDFIVEerr}{\pm0.046}
\definemath{\hfagFSFIVEval}{0.202}
\definemath{\hfagFSFIVEerr}{\pm0.036}
\definemath{\hfagFNBFIVEval}{0.035}
\definemath{\hfagFNBFIVEerr}{\pm0.057}
\definemath{\hfagLFSFACTOR}{}
\definemath{\hfagLFBSNOMIXval}{0.087}
\definemath{\hfagLFBSNOMIXerr}{\pm0.014}
\definemath{\hfagLFBBNOMIXval}{0.099}
\definemath{\hfagLFBBNOMIXerr}{\pm0.016}
\definemath{\hfagLFBDNOMIXval}{0.407}
\definemath{\hfagLFBDNOMIXerr}{\pm0.009}
\definemath{\hfagWFSFACTOR}{1.4}
\definemath{\hfagWFBSNOMIXval}{0.100}
\definemath{\hfagWFBSNOMIXerr}{\pm0.017}
\definemath{\hfagWFBBNOMIXval}{0.089}
\definemath{\hfagWFBBNOMIXerr}{\pm0.022}
\definemath{\hfagWFBDNOMIXval}{0.405}
\definemath{\hfagWFBDNOMIXerr}{\pm0.012}
\definemath{\hfagTFSFACTOR}{}
\definemath{\hfagTFBSNOMIXval}{0.094}
\definemath{\hfagTFBSNOMIXerr}{\pm0.016}
\definemath{\hfagTFBBNOMIXval}{0.262}
\definemath{\hfagTFBBNOMIXerr}{\pm0.073}
\definemath{\hfagTFBDNOMIXval}{0.322}
\definemath{\hfagTFBDNOMIXerr}{\pm0.032}
\definemath{\hfagCHIBARTEVval}{0.147}
\definemath{\hfagCHIBARTEVerr}{\pm0.011}
\definemath{\hfagCHIBARSFACTOR}{1.8}
\definemath{\hfagCHIBARval}{0.1284}
\definemath{\hfagCHIBARerr}{\pm0.0069}
\definemath{\hfagWFBSMIXval}{0.120}
\definemath{\hfagWFBSMIXerr}{\pm0.019}
\definemath{\hfagTFBSMIXval}{0.172}
\definemath{\hfagTFBSMIXerr}{\pm0.031}
\definemath{\hfagLFBSMIXval}{0.116}
\definemath{\hfagLFBSMIXerr}{\pm0.012}
\definemath{\hfagCHIDUval}{0.182}
\definemath{\hfagCHIDUerr}{\pm0.015}
\definemath{\hfagCHIDWUval}{0.1864}
\definemath{\hfagCHIDWUerr}{\pm0.0022}
\definemath{\hfagXDWval}{0.771}
\definemath{\hfagXDWerr}{\pm0.007}
\definemath{\hfagXDWUval}{0.771}
\definemath{\hfagXDWUerr}{\pm0.007}
\definemath{\hfagDMDWval}{0.508}
\definemath{\hfagDMDWsta}{\pm0.003}
\definemath{\hfagDMDWsys}{\pm0.003}
\definemath{\hfagDMDWerr}{\pm0.004}
\definemath{\hfagDMDWUval}{0.508}
\definemath{\hfagDMDWUerr}{\pm0.004}
\definemath{\hfagLFBSval}{0.103}
\definemath{\hfagLFBSerr}{\pm0.009}
\definemath{\hfagLFBBval}{0.090}
\definemath{\hfagLFBBerr}{\pm0.015}
\definemath{\hfagLFBDval}{0.403}
\definemath{\hfagLFBDerr}{\pm0.009}
\definemath{\hfagLRHOFBBFBS}{+0.035}
\definemath{\hfagLRHOFBDFBS}{-0.523}
\definemath{\hfagLRHOFBDFBB}{-0.870}
\definemath{\hfagWFBSval}{0.109}
\definemath{\hfagWFBSerr}{\pm0.012}
\definemath{\hfagWFBBval}{0.083}
\definemath{\hfagWFBBerr}{\pm0.020}
\definemath{\hfagWFBDval}{0.404}
\definemath{\hfagWFBDerr}{\pm0.012}
\definemath{\hfagWRHOFBBFBS}{-0.053}
\definemath{\hfagWRHOFBDFBS}{-0.475}
\definemath{\hfagWRHOFBDFBB}{-0.854}
\definemath{\hfagTFBSval}{0.111}
\definemath{\hfagTFBSerr}{\pm0.014}
\definemath{\hfagTFBBval}{0.211}
\definemath{\hfagTFBBerr}{\pm0.069}
\definemath{\hfagTFBDval}{0.339}
\definemath{\hfagTFBDerr}{\pm0.031}
\definemath{\hfagTRHOFBBFBS}{-0.582}
\definemath{\hfagTRHOFBDFBS}{+0.426}
\definemath{\hfagTRHOFBDFBB}{-0.984}
\definemath{\hfagDMDHval}{0.496}
\definemath{\hfagDMDHsta}{\pm0.010}
\definemath{\hfagDMDHsys}{\pm0.009}
\definemath{\hfagDMDHerr}{\pm0.013}
\definemath{\hfagDMDBval}{0.508}
\definemath{\hfagDMDBsta}{\pm0.003}
\definemath{\hfagDMDBsys}{\pm0.003}
\definemath{\hfagDMDBerr}{\pm0.005}
\definemath{\hfagDMDTWODval}{0.509}
\definemath{\hfagDMDTWODsta}{\pm0.004}
\definemath{\hfagDMDTWODsys}{\pm0.004}
\definemath{\hfagDMDTWODerr}{\pm0.006}
\definemath{\hfagTAUBDTWODval}{1.527}
\definemath{\hfagTAUBDTWODsta}{\pm0.006}
\definemath{\hfagTAUBDTWODsys}{\pm0.008}
\definemath{\hfagTAUBDTWODerr}{\pm0.010}
\definemath{\hfagRHOstaDMDTAUBD}{-0.19}
\definemath{\hfagRHOsysDMDTAUBD}{-0.26}
\definemath{\hfagRHODMDTAUBD}{-0.23}
\definemath{\hfagQPDBval}{1.0024}
\definemath{\hfagQPDBerr}{\pm0.0023}
\definemath{\hfagQPDAval}{1.0030}
\definemath{\hfagQPDAerr}{\pm0.0017}
\definemath{\hfagASLDBval}{-0.0047}
\definemath{\hfagASLDBerr}{\pm0.0046}
\definemath{\hfagASLDAval}{-0.0058}
\definemath{\hfagASLDAerr}{\pm0.0034}
\definemath{\hfagREBDBval}{-0.0012}
\definemath{\hfagREBDBerr}{\pm0.0011}
\definemath{\hfagREBDAval}{-0.0015}
\definemath{\hfagREBDAerr}{\pm0.0008}
\definemath{\hfagASLSval}{-0.0088}
\definemath{\hfagASLSsta}{\pm0.0043}
\definemath{\hfagASLSsys}{\pm0.0039}
\definemath{\hfagASLSerr}{\pm0.0058}
\definemath{\hfagQPSval}{1.0044}
\definemath{\hfagQPSsta}{\pm0.0022}
\definemath{\hfagQPSsys}{\pm0.0019}
\definemath{\hfagQPSerr}{\pm0.0029}
\definemath{\hfagDMSval}{17.78}
\definemath{\hfagDMSerr}{\pm0.12}
\definemath{\hfagXSval}{26.3}
\definemath{\hfagXSerr}{\pm0.4}
\definemath{\hfagCHISval}{0.49928}
\definemath{\hfagCHISerr}{\pm0.00002}
\definemath{\hfagRATIODMDDMSval}{0.0286}
\definemath{\hfagRATIODMDDMSerr}{\pm0.0003}
\definemath{\hfagVTDVTSval}{0.2062}
\definemath{\hfagVTDVTSexx}{\pm0.0011}
\definemath{\hfagVTDVTSthp}{^{+0.0080}}
\definemath{\hfagVTDVTSthn}{_{-0.0060}}
\definemath{\hfagVTDVTSerp}{^{+0.0081}}
\definemath{\hfagVTDVTSern}{_{-0.0061}}
\definemath{\hfagXIval}{1.210}
\definemath{\hfagXIerp}{^{+0.047}}
\definemath{\hfagXIern}{_{-0.035}}
\definemath{\hfagBETASCOMBAval}{+0.41}
\definemath{\hfagBETASCOMBAerp}{^{+0.18}}
\definemath{\hfagBETASCOMBAern}{_{-0.15}}
\definemath{\hfagBETASCOMBAlow}{+0.16}
\definemath{\hfagBETASCOMBAupp}{+0.75}
\definemath{\hfagBETASCOMBBval}{+1.16}
\definemath{\hfagBETASCOMBBerp}{^{+0.15}}
\definemath{\hfagBETASCOMBBern}{_{-0.18}}
\definemath{\hfagBETASCOMBBlow}{+0.82}
\definemath{\hfagBETASCOMBBupp}{+1.41}
\definemath{\hfagPHISCOMBAval}{-0.83}
\definemath{\hfagPHISCOMBAerp}{^{+0.30}}
\definemath{\hfagPHISCOMBAern}{_{-0.36}}
\definemath{\hfagPHISCOMBAlow}{-1.50}
\definemath{\hfagPHISCOMBAupp}{-0.32}
\definemath{\hfagPHISCOMBBval}{-2.31}
\definemath{\hfagPHISCOMBBerp}{^{+0.36}}
\definemath{\hfagPHISCOMBBern}{_{-0.30}}
\definemath{\hfagPHISCOMBBlow}{-2.82}
\definemath{\hfagPHISCOMBBupp}{-1.64}
\definemath{\hfagDGSCOMBAval}{+0.150}
\definemath{\hfagDGSCOMBAerp}{^{+0.055}}
\definemath{\hfagDGSCOMBAern}{_{-0.056}}
\definemath{\hfagDGSCOMBAlow}{+0.060}
\definemath{\hfagDGSCOMBAupp}{+0.297}
\definemath{\hfagDGSCOMBBval}{-0.150}
\definemath{\hfagDGSCOMBBerp}{^{+0.056}}
\definemath{\hfagDGSCOMBBern}{_{-0.055}}
\definemath{\hfagDGSCOMBBlow}{-0.297}
\definemath{\hfagDGSCOMBBupp}{-0.060}
\definemath{\hfagNSIGMASM}{2.3}
\definemath{\hfagBETASCOMBACONval}{+0.41}
\definemath{\hfagBETASCOMBACONerp}{^{+0.10}}
\definemath{\hfagBETASCOMBACONern}{_{-0.08}}
\definemath{\hfagBETASCOMBACONlow}{+0.22}
\definemath{\hfagBETASCOMBACONupp}{+0.60}
\definemath{\hfagBETASCOMBBCONval}{+1.18}
\definemath{\hfagBETASCOMBBCONerp}{^{+0.07}}
\definemath{\hfagBETASCOMBBCONern}{_{-0.10}}
\definemath{\hfagBETASCOMBBCONlow}{+1.00}
\definemath{\hfagBETASCOMBBCONupp}{+1.36}
\definemath{\hfagPHISCOMBACONval}{-0.82}
\definemath{\hfagPHISCOMBACONerp}{^{+0.16}}
\definemath{\hfagPHISCOMBACONern}{_{-0.21}}
\definemath{\hfagPHISCOMBACONlow}{-1.20}
\definemath{\hfagPHISCOMBACONupp}{-0.45}
\definemath{\hfagPHISCOMBBCONval}{-2.36}
\definemath{\hfagPHISCOMBBCONerp}{^{+0.20}}
\definemath{\hfagPHISCOMBBCONern}{_{-0.14}}
\definemath{\hfagPHISCOMBBCONlow}{-2.72}
\definemath{\hfagPHISCOMBBCONupp}{-1.99}
\definemath{\hfagDGSCOMBACONval}{+0.150}
\definemath{\hfagDGSCOMBACONerp}{^{+0.045}}
\definemath{\hfagDGSCOMBACONern}{_{-0.049}}
\definemath{\hfagDGSCOMBACONlow}{+0.075}
\definemath{\hfagDGSCOMBACONupp}{+0.228}
\definemath{\hfagDGSCOMBBCONval}{-0.150}
\definemath{\hfagDGSCOMBBCONerp}{^{+0.042}}
\definemath{\hfagDGSCOMBBCONern}{_{-0.049}}
\definemath{\hfagDGSCOMBBCONlow}{-0.220}
\definemath{\hfagDGSCOMBBCONupp}{-0.071}
\definemath{\hfagNSIGMASMCON}{2.8}
\definemath{\hfagBETASCOMBACONXval}{+0.37}
\definemath{\hfagBETASCOMBACONXerp}{^{+0.11}}
\definemath{\hfagBETASCOMBACONXern}{_{-0.16}}
\definemath{\hfagBETASCOMBACONXlow}{+0.10}
\definemath{\hfagBETASCOMBACONXupp}{+0.59}
\definemath{\hfagBETASCOMBBCONXval}{+1.19}
\definemath{\hfagBETASCOMBBCONXerp}{^{+0.17}}
\definemath{\hfagBETASCOMBBCONXern}{_{-0.13}}
\definemath{\hfagBETASCOMBBCONXlow}{+0.97}
\definemath{\hfagBETASCOMBBCONXupp}{+1.47}
\definemath{\hfagPHISCOMBACONXval}{-0.75}
\definemath{\hfagPHISCOMBACONXerp}{^{+0.32}}
\definemath{\hfagPHISCOMBACONXern}{_{-0.21}}
\definemath{\hfagPHISCOMBACONXlow}{-1.19}
\definemath{\hfagPHISCOMBACONXupp}{-0.21}
\definemath{\hfagPHISCOMBBCONXval}{-2.38}
\definemath{\hfagPHISCOMBBCONXerp}{^{+0.25}}
\definemath{\hfagPHISCOMBBCONXern}{_{-0.34}}
\definemath{\hfagPHISCOMBBCONXlow}{-2.94}
\definemath{\hfagPHISCOMBBCONXupp}{-1.93}
\definemath{\hfagDGSCOMBACONXval}{+0.054}
\definemath{\hfagDGSCOMBACONXerp}{^{+0.026}}
\definemath{\hfagDGSCOMBACONXern}{_{-0.015}}
\definemath{\hfagDGSCOMBACONXlow}{+0.025}
\definemath{\hfagDGSCOMBACONXupp}{+0.097}
\definemath{\hfagDGSCOMBBCONXval}{-0.054}
\definemath{\hfagDGSCOMBBCONXerp}{^{+0.016}}
\definemath{\hfagDGSCOMBBCONXern}{_{-0.026}}
\definemath{\hfagDGSCOMBBCONXlow}{-0.099}
\definemath{\hfagDGSCOMBBCONXupp}{-0.024}
\definemath{\hfagNSIGMASMCONX}{2.7}

\newcommand{\unit}[1]{~\ensuremath{\rm #1}\xspace}
\renewcommand{\ps}{\unit{ps}}
\newcommand{\invps}{\unit{ps^{-1}}}
\newcommand{\TeV}{\unit{TeV}}
\newcommand{\MeVcc}{\unit{MeV/\mbox{$c$}^2}}
\newcommand{\MeV}{\unit{MeV}}

\definemath{\hfagCHIBARLEP}{\hfagCHIBARLEPval\hfagCHIBARLEPerr}
\definemath{\hfagTAUBD}{\hfagTAUBDval\hfagTAUBDerr\ps}
\definemath{\hfagTAUBDnounit}{\hfagTAUBDval\hfagTAUBDerr}
\definemath{\hfagTAUBU}{\hfagTAUBUval\hfagTAUBUerr\ps}
\definemath{\hfagTAUBUnounit}{\hfagTAUBUval\hfagTAUBUerr}
\definemath{\hfagRTAUBU}{\hfagRTAUBUval\hfagRTAUBUerr}
\definemath{\hfagTAUBS}{\hfagTAUBSval\hfagTAUBSerr\ps}
\definemath{\hfagTAUBSnounit}{\hfagTAUBSval\hfagTAUBSerr}
\definemath{\hfagRTAUBS}{\hfagRTAUBSval\hfagRTAUBSerr}
\definemath{\hfagTAULB}{\hfagTAULBval\hfagTAULBerr\ps}
\definemath{\hfagTAULBnounit}{\hfagTAULBval\hfagTAULBerr}
\definemath{\hfagTAUBB}{\hfagTAUBBval\hfagTAUBBerr\ps}
\definemath{\hfagTAUBBnounit}{\hfagTAUBBval\hfagTAUBBerr}
\definemath{\hfagRTAUBB}{\hfagRTAUBBval\hfagRTAUBBerr}
\definemath{\hfagTAUXBerr}{\hfagTAUXBerp\hfagTAUXBern}
\definemath{\hfagTAUXB}{\hfagTAUXBval\hfagTAUXBerr\ps}
\definemath{\hfagTAUXBnounit}{\hfagTAUXBval\hfagTAUXBerr}
\definemath{\hfagTAUB}{\hfagTAUBval\hfagTAUBerr\ps}
\definemath{\hfagTAUBnounit}{\hfagTAUBval\hfagTAUBerr}
\definemath{\hfagTAUXBDerr}{\hfagTAUXBDerp\hfagTAUXBDern}
\definemath{\hfagTAUXBD}{\hfagTAUXBDval\hfagTAUXBDerr\ps}
\definemath{\hfagTAUXBDnounit}{\hfagTAUXBDval\hfagTAUXBDerr}
\definemath{\hfagTAUOBerr}{\hfagTAUOBerp\hfagTAUOBern}
\definemath{\hfagTAUOB}{\hfagTAUOBval\hfagTAUOBerr\ps}
\definemath{\hfagTAUOBnounit}{\hfagTAUOBval\hfagTAUOBerr}
\definemath{\hfagTAUBC}{\hfagTAUBCval\hfagTAUBCerr\ps}
\definemath{\hfagTAUBCnounit}{\hfagTAUBCval\hfagTAUBCerr}
\definemath{\hfagTAUBSSL}{\hfagTAUBSSLval\hfagTAUBSSLerr\ps}
\definemath{\hfagTAUBSSLnounit}{\hfagTAUBSSLval\hfagTAUBSSLerr}
\definemath{\hfagTAUBSSLX}{\hfagTAUBSSLXval\hfagTAUBSSLXerr\ps}
\definemath{\hfagTAUBSSLXnounit}{\hfagTAUBSSLXval\hfagTAUBSSLXerr}
\definemath{\hfagTAUBSMEANCONerr}{\hfagTAUBSMEANCONerp\hfagTAUBSMEANCONern}
\definemath{\hfagTAUBSMEANCON}{\hfagTAUBSMEANCONval\hfagTAUBSMEANCONerr\ps}
\definemath{\hfagTAUBSMEANCONnounit}{\hfagTAUBSMEANCONval\hfagTAUBSMEANCONerr}
\definemath{\hfagTAUBSJF}{\hfagTAUBSJFval\hfagTAUBSJFerr\ps}
\definemath{\hfagTAUBSJFnounit}{\hfagTAUBSJFval\hfagTAUBSJFerr}
\definemath{\hfagRTAUBSSL}{\hfagRTAUBSSLval\hfagRTAUBSSLerr}
\definemath{\hfagRTAUBSMEANCON}{\hfagRTAUBSMEANCONval\hfagRTAUBSMEANCONerr}
\definemath{\hfagRTAULB}{\hfagRTAULBval\hfagRTAULBerr}
\definemath{\hfagTAUBVTX}{\hfagTAUBVTXval\hfagTAUBVTXerr\ps}
\definemath{\hfagTAUBVTXnounit}{\hfagTAUBVTXval\hfagTAUBVTXerr}
\definemath{\hfagTAUBLEP}{\hfagTAUBLEPval\hfagTAUBLEPerr\ps}
\definemath{\hfagTAUBLEPnounit}{\hfagTAUBLEPval\hfagTAUBLEPerr}
\definemath{\hfagTAUBJPerr}{\hfagTAUBJPerp\hfagTAUBJPern}
\definemath{\hfagTAUBJP}{\hfagTAUBJPval\hfagTAUBJPerr\ps}
\definemath{\hfagTAUBJPnounit}{\hfagTAUBJPval\hfagTAUBJPerr}
\definemath{\hfagSDGDGD}{\hfagSDGDGDval\hfagSDGDGDerr}
\definemath{\hfagTAUBSSHORT}{\hfagTAUBSSHORTval\hfagTAUBSSHORTerr\ps}
\definemath{\hfagTAUBSSHORTnounit}{\hfagTAUBSSHORTval\hfagTAUBSSHORTerr}
\definemath{\hfagBRDSDS}{\hfagBRDSDSval\hfagBRDSDSerr}
\definemath{\hfagDGSGSBRDSDS}{\hfagDGSGSBRDSDSval\hfagDGSGSBRDSDSerr}
\definemath{\hfagTAUBSMEAN}{\hfagTAUBSMEANval\hfagTAUBSMEANerr\ps}
\definemath{\hfagTAUBSMEANnounit}{\hfagTAUBSMEANval\hfagTAUBSMEANerr}
\definemath{\hfagDGSGS}{\hfagDGSGSval\hfagDGSGSerr}
\definemath{\hfagDGS}{\hfagDGSval\hfagDGSerr\invps}
\definemath{\hfagDGSnounit}{\hfagDGSval\hfagDGSerr}
\definemath{\hfagTAUBSLerr}{\hfagTAUBSLerp\hfagTAUBSLern}
\definemath{\hfagTAUBSL}{\hfagTAUBSLval\hfagTAUBSLerr\ps}
\definemath{\hfagTAUBSLnounit}{\hfagTAUBSLval\hfagTAUBSLerr}
\definemath{\hfagTAUBSHerr}{\hfagTAUBSHerp\hfagTAUBSHern}
\definemath{\hfagTAUBSH}{\hfagTAUBSHval\hfagTAUBSHerr\ps}
\definemath{\hfagTAUBSHnounit}{\hfagTAUBSHval\hfagTAUBSHerr}
\definemath{\hfagDGSGSCONBDerr}{\hfagDGSGSCONBDerp\hfagDGSGSCONBDern}
\definemath{\hfagDGSGSCONBD}{\hfagDGSGSCONBDval\hfagDGSGSCONBDerr}
\definemath{\hfagTAUBSMEANCONXerr}{\hfagTAUBSMEANCONXerp\hfagTAUBSMEANCONXern}
\definemath{\hfagTAUBSMEANCONX}{\hfagTAUBSMEANCONXval\hfagTAUBSMEANCONXerr\ps}
\definemath{\hfagTAUBSMEANCONXnounit}{\hfagTAUBSMEANCONXval\hfagTAUBSMEANCONXerr}
\definemath{\hfagDGSGSCONerr}{\hfagDGSGSCONerp\hfagDGSGSCONern}
\definemath{\hfagDGSGSCON}{\hfagDGSGSCONval\hfagDGSGSCONerr}
\definemath{\hfagDGSCONerr}{\hfagDGSCONerp\hfagDGSCONern}
\definemath{\hfagDGSCON}{\hfagDGSCONval\hfagDGSCONerr\invps}
\definemath{\hfagDGSCONnounit}{\hfagDGSCONval\hfagDGSCONerr}
\definemath{\hfagTAUBSLCONerr}{\hfagTAUBSLCONerp\hfagTAUBSLCONern}
\definemath{\hfagTAUBSLCON}{\hfagTAUBSLCONval\hfagTAUBSLCONerr\ps}
\definemath{\hfagTAUBSLCONnounit}{\hfagTAUBSLCONval\hfagTAUBSLCONerr}
\definemath{\hfagTAUBSHCON}{\hfagTAUBSHCONval\hfagTAUBSHCONerr\ps}
\definemath{\hfagTAUBSHCONnounit}{\hfagTAUBSHCONval\hfagTAUBSHCONerr}
\definemath{\hfagDGSGSCONBDCONerr}{\hfagDGSGSCONBDCONerp\hfagDGSGSCONBDCONern}
\definemath{\hfagDGSGSCONBDCON}{\hfagDGSGSCONBDCONval\hfagDGSGSCONBDCONerr}
\definemath{\hfagTAUBSMEANCONXX}{\hfagTAUBSMEANCONXXval\hfagTAUBSMEANCONXXerr\ps}
\definemath{\hfagTAUBSMEANCONXXnounit}{\hfagTAUBSMEANCONXXval\hfagTAUBSMEANCONXXerr}
\definemath{\hfagDGSGSCONX}{\hfagDGSGSCONXval\hfagDGSGSCONXerr}
\definemath{\hfagDGSCONX}{\hfagDGSCONXval\hfagDGSCONXerr\invps}
\definemath{\hfagDGSCONXnounit}{\hfagDGSCONXval\hfagDGSCONXerr}
\definemath{\hfagTAUBSLCONX}{\hfagTAUBSLCONXval\hfagTAUBSLCONXerr\ps}
\definemath{\hfagTAUBSLCONXnounit}{\hfagTAUBSLCONXval\hfagTAUBSLCONXerr}
\definemath{\hfagTAUBSHCONXerr}{\hfagTAUBSHCONXerp\hfagTAUBSHCONXern}
\definemath{\hfagTAUBSHCONX}{\hfagTAUBSHCONXval\hfagTAUBSHCONXerr\ps}
\definemath{\hfagTAUBSHCONXnounit}{\hfagTAUBSHCONXval\hfagTAUBSHCONXerr}
\definemath{\hfagDGSGSCONBDCONXerr}{\hfagDGSGSCONBDCONXerp\hfagDGSGSCONBDCONXern}
\definemath{\hfagDGSGSCONBDCONX}{\hfagDGSGSCONBDCONXval\hfagDGSGSCONBDCONXerr}
\definemath{\hfagFCW}{\hfagFCWval\hfagFCWerr}
\definemath{\hfagFNW}{\hfagFNWval\hfagFNWerr}
\definemath{\hfagFFW}{\hfagFFWval\hfagFFWerr}
\definemath{\hfagFCN}{\hfagFCNval\hfagFCNerr}
\definemath{\hfagFNN}{\hfagFNNval\hfagFNNerr}
\definemath{\hfagFFN}{\hfagFFNval\hfagFFNerr}
\definemath{\hfagFC}{\hfagFCval\hfagFCerr}
\definemath{\hfagFN}{\hfagFNval\hfagFNerr}
\definemath{\hfagFF}{\hfagFFval\hfagFFerr}
\definemath{\hfagFPROD}{\hfagFPRODval\hfagFPRODerr}
\definemath{\hfagFSUM}{\hfagFSUMval\hfagFSUMerr}
\definemath{\hfagFSFIVEOS}{\hfagFSFIVEOSval\hfagFSFIVEOSerr}
\definemath{\hfagFSFIVEOSfull}{\hfagFSFIVEOSval\hfagFSFIVEOSsta\hfagFSFIVEOSsys}
\definemath{\hfagFSFIVERL}{\hfagFSFIVERLval\hfagFSFIVERLerr}
\definemath{\hfagFUDFIVE}{\hfagFUDFIVEval\hfagFUDFIVEerr}
\definemath{\hfagFSFIVE}{\hfagFSFIVEval\hfagFSFIVEerr}
\definemath{\hfagFNBFIVE}{\hfagFNBFIVEval\hfagFNBFIVEerr}
\definemath{\hfagLFBSNOMIX}{\hfagLFBSNOMIXval\hfagLFBSNOMIXerr}
\definemath{\hfagLFBBNOMIX}{\hfagLFBBNOMIXval\hfagLFBBNOMIXerr}
\definemath{\hfagLFBDNOMIX}{\hfagLFBDNOMIXval\hfagLFBDNOMIXerr}
\definemath{\hfagWFBSNOMIX}{\hfagWFBSNOMIXval\hfagWFBSNOMIXerr}
\definemath{\hfagWFBBNOMIX}{\hfagWFBBNOMIXval\hfagWFBBNOMIXerr}
\definemath{\hfagWFBDNOMIX}{\hfagWFBDNOMIXval\hfagWFBDNOMIXerr}
\definemath{\hfagTFBSNOMIX}{\hfagTFBSNOMIXval\hfagTFBSNOMIXerr}
\definemath{\hfagTFBBNOMIX}{\hfagTFBBNOMIXval\hfagTFBBNOMIXerr}
\definemath{\hfagTFBDNOMIX}{\hfagTFBDNOMIXval\hfagTFBDNOMIXerr}
\definemath{\hfagCHIBARTEV}{\hfagCHIBARTEVval\hfagCHIBARTEVerr}
\definemath{\hfagCHIBAR}{\hfagCHIBARval\hfagCHIBARerr}
\definemath{\hfagWFBSMIX}{\hfagWFBSMIXval\hfagWFBSMIXerr}
\definemath{\hfagTFBSMIX}{\hfagTFBSMIXval\hfagTFBSMIXerr}
\definemath{\hfagLFBSMIX}{\hfagLFBSMIXval\hfagLFBSMIXerr}
\definemath{\hfagCHIDU}{\hfagCHIDUval\hfagCHIDUerr}
\definemath{\hfagCHIDWU}{\hfagCHIDWUval\hfagCHIDWUerr}
\definemath{\hfagXDW}{\hfagXDWval\hfagXDWerr}
\definemath{\hfagXDWU}{\hfagXDWUval\hfagXDWUerr}
\definemath{\hfagDMDW}{\hfagDMDWval\hfagDMDWerr\invps}
\definemath{\hfagDMDWnounit}{\hfagDMDWval\hfagDMDWerr}
\definemath{\hfagDMDWfull}{\hfagDMDWval\hfagDMDWsta\hfagDMDWsys\invps}
\definemath{\hfagDMDWnounitfull}{\hfagDMDWval\hfagDMDWsta\hfagDMDWsys}
\definemath{\hfagDMDWU}{\hfagDMDWUval\hfagDMDWUerr\invps}
\definemath{\hfagDMDWUnounit}{\hfagDMDWUval\hfagDMDWUerr}
\definemath{\hfagLFBS}{\hfagLFBSval\hfagLFBSerr}
\definemath{\hfagLFBB}{\hfagLFBBval\hfagLFBBerr}
\definemath{\hfagLFBD}{\hfagLFBDval\hfagLFBDerr}
\definemath{\hfagWFBS}{\hfagWFBSval\hfagWFBSerr}
\definemath{\hfagWFBB}{\hfagWFBBval\hfagWFBBerr}
\definemath{\hfagWFBD}{\hfagWFBDval\hfagWFBDerr}
\definemath{\hfagTFBS}{\hfagTFBSval\hfagTFBSerr}
\definemath{\hfagTFBB}{\hfagTFBBval\hfagTFBBerr}
\definemath{\hfagTFBD}{\hfagTFBDval\hfagTFBDerr}
\definemath{\hfagDMDH}{\hfagDMDHval\hfagDMDHerr\invps}
\definemath{\hfagDMDHnounit}{\hfagDMDHval\hfagDMDHerr}
\definemath{\hfagDMDHfull}{\hfagDMDHval\hfagDMDHsta\hfagDMDHsys\invps}
\definemath{\hfagDMDHnounitfull}{\hfagDMDHval\hfagDMDHsta\hfagDMDHsys}
\definemath{\hfagDMDB}{\hfagDMDBval\hfagDMDBerr\invps}
\definemath{\hfagDMDBnounit}{\hfagDMDBval\hfagDMDBerr}
\definemath{\hfagDMDBfull}{\hfagDMDBval\hfagDMDBsta\hfagDMDBsys\invps}
\definemath{\hfagDMDBnounitfull}{\hfagDMDBval\hfagDMDBsta\hfagDMDBsys}
\definemath{\hfagDMDTWOD}{\hfagDMDTWODval\hfagDMDTWODerr\invps}
\definemath{\hfagDMDTWODnounit}{\hfagDMDTWODval\hfagDMDTWODerr}
\definemath{\hfagDMDTWODfull}{\hfagDMDTWODval\hfagDMDTWODsta\hfagDMDTWODsys\invps}
\definemath{\hfagDMDTWODnounitfull}{\hfagDMDTWODval\hfagDMDTWODsta\hfagDMDTWODsys}
\definemath{\hfagTAUBDTWOD}{\hfagTAUBDTWODval\hfagTAUBDTWODerr\ps}
\definemath{\hfagTAUBDTWODnounit}{\hfagTAUBDTWODval\hfagTAUBDTWODerr}
\definemath{\hfagTAUBDTWODfull}{\hfagTAUBDTWODval\hfagTAUBDTWODsta\hfagTAUBDTWODsys\ps}
\definemath{\hfagTAUBDTWODnounitfull}{\hfagTAUBDTWODval\hfagTAUBDTWODsta\hfagTAUBDTWODsys}
\definemath{\hfagQPDB}{\hfagQPDBval\hfagQPDBerr}
\definemath{\hfagQPDA}{\hfagQPDAval\hfagQPDAerr}
\definemath{\hfagASLDB}{\hfagASLDBval\hfagASLDBerr}
\definemath{\hfagASLDA}{\hfagASLDAval\hfagASLDAerr}
\definemath{\hfagREBDB}{\hfagREBDBval\hfagREBDBerr}
\definemath{\hfagREBDA}{\hfagREBDAval\hfagREBDAerr}
\definemath{\hfagASLS}{\hfagASLSval\hfagASLSerr}
\definemath{\hfagASLSfull}{\hfagASLSval\hfagASLSsta\hfagASLSsys}
\definemath{\hfagQPS}{\hfagQPSval\hfagQPSerr}
\definemath{\hfagQPSfull}{\hfagQPSval\hfagQPSsta\hfagQPSsys}
\definemath{\hfagDMS}{\hfagDMSval\hfagDMSerr\invps}
\definemath{\hfagDMSnounit}{\hfagDMSval\hfagDMSerr}
\definemath{\hfagXS}{\hfagXSval\hfagXSerr}
\definemath{\hfagCHIS}{\hfagCHISval\hfagCHISerr}
\definemath{\hfagRATIODMDDMS}{\hfagRATIODMDDMSval\hfagRATIODMDDMSerr}
\definemath{\hfagVTDVTSerr}{\hfagVTDVTSerp\hfagVTDVTSern}
\definemath{\hfagVTDVTSthe}{\hfagVTDVTSthp\hfagVTDVTSthn}
\definemath{\hfagVTDVTS}{\hfagVTDVTSval\hfagVTDVTSerr}
\definemath{\hfagVTDVTSfull}{\hfagVTDVTSval\hfagVTDVTSexx\hfagVTDVTSthe}
\definemath{\hfagXIerr}{\hfagXIerp\hfagXIern}
\definemath{\hfagXI}{\hfagXIval\hfagXIerr}
\definemath{\hfagBETASCOMBAerr}{\hfagBETASCOMBAerp\hfagBETASCOMBAern}
\definemath{\hfagBETASCOMBA}{\hfagBETASCOMBAval\hfagBETASCOMBAerr}
\definemath{\hfagBETASCOMBBerr}{\hfagBETASCOMBBerp\hfagBETASCOMBBern}
\definemath{\hfagBETASCOMBB}{\hfagBETASCOMBBval\hfagBETASCOMBBerr}
\definemath{\hfagPHISCOMBAerr}{\hfagPHISCOMBAerp\hfagPHISCOMBAern}
\definemath{\hfagPHISCOMBA}{\hfagPHISCOMBAval\hfagPHISCOMBAerr}
\definemath{\hfagPHISCOMBBerr}{\hfagPHISCOMBBerp\hfagPHISCOMBBern}
\definemath{\hfagPHISCOMBB}{\hfagPHISCOMBBval\hfagPHISCOMBBerr}
\definemath{\hfagDGSCOMBAerr}{\hfagDGSCOMBAerp\hfagDGSCOMBAern}
\definemath{\hfagDGSCOMBA}{\hfagDGSCOMBAval\hfagDGSCOMBAerr\invps}
\definemath{\hfagDGSCOMBAnounit}{\hfagDGSCOMBAval\hfagDGSCOMBAerr}
\definemath{\hfagDGSCOMBBerr}{\hfagDGSCOMBBerp\hfagDGSCOMBBern}
\definemath{\hfagDGSCOMBB}{\hfagDGSCOMBBval\hfagDGSCOMBBerr\invps}
\definemath{\hfagDGSCOMBBnounit}{\hfagDGSCOMBBval\hfagDGSCOMBBerr}
\definemath{\hfagBETASCOMBACONerr}{\hfagBETASCOMBACONerp\hfagBETASCOMBACONern}
\definemath{\hfagBETASCOMBACON}{\hfagBETASCOMBACONval\hfagBETASCOMBACONerr}
\definemath{\hfagBETASCOMBBCONerr}{\hfagBETASCOMBBCONerp\hfagBETASCOMBBCONern}
\definemath{\hfagBETASCOMBBCON}{\hfagBETASCOMBBCONval\hfagBETASCOMBBCONerr}
\definemath{\hfagPHISCOMBACONerr}{\hfagPHISCOMBACONerp\hfagPHISCOMBACONern}
\definemath{\hfagPHISCOMBACON}{\hfagPHISCOMBACONval\hfagPHISCOMBACONerr}
\definemath{\hfagPHISCOMBBCONerr}{\hfagPHISCOMBBCONerp\hfagPHISCOMBBCONern}
\definemath{\hfagPHISCOMBBCON}{\hfagPHISCOMBBCONval\hfagPHISCOMBBCONerr}
\definemath{\hfagDGSCOMBACONerr}{\hfagDGSCOMBACONerp\hfagDGSCOMBACONern}
\definemath{\hfagDGSCOMBACON}{\hfagDGSCOMBACONval\hfagDGSCOMBACONerr\invps}
\definemath{\hfagDGSCOMBACONnounit}{\hfagDGSCOMBACONval\hfagDGSCOMBACONerr}
\definemath{\hfagDGSCOMBBCONerr}{\hfagDGSCOMBBCONerp\hfagDGSCOMBBCONern}
\definemath{\hfagDGSCOMBBCON}{\hfagDGSCOMBBCONval\hfagDGSCOMBBCONerr\invps}
\definemath{\hfagDGSCOMBBCONnounit}{\hfagDGSCOMBBCONval\hfagDGSCOMBBCONerr}
\definemath{\hfagBETASCOMBACONXerr}{\hfagBETASCOMBACONXerp\hfagBETASCOMBACONXern}
\definemath{\hfagBETASCOMBACONX}{\hfagBETASCOMBACONXval\hfagBETASCOMBACONXerr}
\definemath{\hfagBETASCOMBBCONXerr}{\hfagBETASCOMBBCONXerp\hfagBETASCOMBBCONXern}
\definemath{\hfagBETASCOMBBCONX}{\hfagBETASCOMBBCONXval\hfagBETASCOMBBCONXerr}
\definemath{\hfagPHISCOMBACONXerr}{\hfagPHISCOMBACONXerp\hfagPHISCOMBACONXern}
\definemath{\hfagPHISCOMBACONX}{\hfagPHISCOMBACONXval\hfagPHISCOMBACONXerr}
\definemath{\hfagPHISCOMBBCONXerr}{\hfagPHISCOMBBCONXerp\hfagPHISCOMBBCONXern}
\definemath{\hfagPHISCOMBBCONX}{\hfagPHISCOMBBCONXval\hfagPHISCOMBBCONXerr}
\definemath{\hfagDGSCOMBACONXerr}{\hfagDGSCOMBACONXerp\hfagDGSCOMBACONXern}
\definemath{\hfagDGSCOMBACONX}{\hfagDGSCOMBACONXval\hfagDGSCOMBACONXerr\invps}
\definemath{\hfagDGSCOMBACONXnounit}{\hfagDGSCOMBACONXval\hfagDGSCOMBACONXerr}
\definemath{\hfagDGSCOMBBCONXerr}{\hfagDGSCOMBBCONXerp\hfagDGSCOMBBCONXern}
\definemath{\hfagDGSCOMBBCONX}{\hfagDGSCOMBBCONXval\hfagDGSCOMBBCONXerr\invps}
\definemath{\hfagDGSCOMBBCONXnounit}{\hfagDGSCOMBBCONXval\hfagDGSCOMBBCONXerr}


\mysection{\b-hadron production fractions, lifetimes and mixing parameters}
\labs{life_mix}


Quantities such as \b-hadron production fractions, \b-hadron lifetimes, 
and neutral \B-meson oscillation frequencies have been studied
in the nineties at LEP and SLC 
(\ee colliders at $\sqrt{s}=m_{\particle{Z}}$) 
as well as at the 
first version of the Tevatron
(\particle{p\bar{p}} collider at $\sqrt{s}=1.8\TeV$). 
Since then 
precise measurements of the \Bd and \Bu lifetimes, as well as of the 
\Bd oscillation frequency, have also been performed at the 
asymmetric \B factories, KEKB and PEPII
(\ee colliders at $\sqrt{s}=m_{\Ups}$) while measurements related 
to the other \b-hadrons, in particular \Bs, \Bc and \Lb, 
are being performed at the upgraded Tevatron ($\sqrt{s}=1.96\TeV$).
In most cases, these basic quantities, although interesting by themselves,
became 
necessary ingredients for the more complicated and 
refined analyses 
at the asymmetric \B factories
and at the Tevatron, 
in particular the time-dependent \CP asymmetry measurements.
It is therefore important that the best experimental
values of these quantities continue to be kept up-to-date and improved. 

In several cases, the averages presented in this chapter are 
needed and used as input for the results given in the subsequent chapters. 
Within this chapter, some averages need the knowledge of other 
averages in a circular way. This coupling, which appears through the 
\b-hadron fractions whenever inclusive or semi-exclusive measurements 
have to be considered, has been reduced significantly in the last several years 
with increasingly precise exclusive measurements becoming available. 

In addition to \b-hadron fractions, lifetimes and 
mixing parameters, this chapter also deals with the 
CP-violating phase $\beta_s$, which is the phase 
difference between the \Bs mixing amplitude and the 
$b\to c\bar{c}s$ decay amplitude. The angle $\beta$, 
which is the equivalent of $\beta_s$ for the \Bd 
system, is discussed in Chapter~\ref{sec:cp_uta}. 

\mysubsection{\b-hadron production fractions}
\labs{fractions}
 
We consider here the relative fractions of the different \b-hadron 
species found in an unbiased sample of weakly-decaying \b hadrons 
produced under some specific conditions. The knowledge of these fractions
is useful to characterize the signal composition in inclusive \b-hadron 
analyses, or to predict the background composition in exclusive analyses.
Many analyses in \B physics need these fractions as input. We distinguish 
here the following three conditions: \Ups decays, \Upsfive decays, and 
high-energy collisions (including \Z decays). 

\mysubsubsection{\b-hadron production fractions in \Ups decays}
\labs{fraction_Ups4S}

Only pairs of the two lightest (charged and neutral) \B mesons 
can be produced in \Ups decays, 
and it is enough to determine the following branching 
fractions:
\begin{eqnarray}
f^{+-} & = & \Gamma(\Ups \to \particle{B^+B^-})/
             \Gamma_{\rm tot}(\Ups)  \,, \\
f^{00} & = & \Gamma(\Ups \to \particle{B^0\bar{B}^0})/
             \Gamma_{\rm tot}(\Ups) \,.
\end{eqnarray}
In practice, most analyses measure their ratio
\begin{equation}
R^{+-/00} = f^{+-}/f^{00} = \Gamma(\Ups \to \particle{B^+B^-})/
             \Gamma(\Ups \to \particle{B^0\bar{B}^0}) \,,
\end{equation}
which is easier to access experimentally.
Since an inclusive (but separate) reconstruction of 
\Bu and \Bd is difficult, specific exclusive decay modes, 
${\Bu} \to x^+$ and ${\Bd} \to x^0$, are usually considered to perform 
a measurement of $R^{+-/00}$, whenever they can be related by 
isospin symmetry (for example \particle{\Bu \to J/\psi K^+} and 
\particle{\Bd \to J/\psi K^0}).
Under the assumption that $\Gamma(\Bu \to x^+) = \Gamma(\Bd \to x^0)$, 
\ie\ that isospin invariance holds in these \B decays,
the ratio of the number of reconstructed
$\Bu \to x^+$ and $\Bd \to x^0$ mesons is proportional to
\begin{equation}
\frac{f^{+-}\, \BR{\Bu\to x^+}}{f^{00}\, \BR{\Bd\to x^0}}
= \frac{f^{+-}\, \Gamma({\Bu}\to x^+)\, \tau(\Bu)}%
{f^{00}\, \Gamma({\Bd}\to x^0)\,\tau(\Bd)}
= \frac{f^{+-}}{f^{00}} \, \frac{\tau(\Bu)}{\tau(\Bd)}  \,, 
\end{equation} 
where $\tau(\Bu)$ and $\tau(\Bd)$ are the \Bu and \Bd 
lifetimes respectively.
Hence the primary quantity measured in these analyses 
is $R^{+-/00} \, \tau(\Bu)/\tau(\Bd)$, 
and the extraction of $R^{+-/00}$ with this method therefore 
requires the knowledge of the $\tau(\Bu)/\tau(\Bd)$ lifetime ratio. 

\begin{table}
\caption{Published measurements of the $\Bu/\Bd$ production ratio
in \Ups decays, together with their average (see text).
Systematic uncertainties due to the imperfect knowledge of 
$\tau(\Bu)/\tau(\Bd)$ are included. The latest \babar result\cite{Aubert:2004rz}
supersedes  the earlier \babar measurements \cite{Aubert:2001xs,Aubert:2004ur}.}
\labt{R_data}
\begin{center}
\begin{tabular}{lccll}
\hline
Experiment & Ref. & Decay modes & Published value of & Assumed value \\
and year & & or method & $R^{+-/00}=f^{+-}/f^{00}$ & of $\tau(\Bu)/\tau(\Bd)$ \\
\hline
CLEO,   2001 & \cite{Alexander:2000tb}  & \particle{J/\psi K^{(*)}} 
             & $1.04 \pm0.07 \pm0.04$ & $1.066 \pm0.024$ \\
\babar, 2002 & \cite{Aubert:2001xs} & \particle{(c\bar{c})K^{(*)}}
             & $1.10 \pm0.06 \pm0.05$ & $1.062 \pm0.029$\\ 
CLEO,   2002 & \cite{Athar:2002mr}  & \particle{D^*\ell\nu}
             & $1.058 \pm0.084 \pm0.136$ & $1.074 \pm0.028$\\
\belle, 2003 & \cite{Hastings:2002ff} & dilepton events 
             & $1.01 \pm0.03 \pm0.09$ & $1.083 \pm0.017$\\
\babar, 2004 & \cite{Aubert:2004ur} & \particle{J/\psi K}
             & $1.006 \pm0.036 \pm0.031$ & $1.083 \pm0.017$ \\
\babar, 2005 & \cite{Aubert:2004rz} & \particle{(c\bar{c})K^{(*)}}
             & $1.06 \pm0.02 \pm0.03$ & $1.086 \pm0.017$\\ 
\hline
Average      & & & \hfagFF~(tot) & \hfagRTAUBU \\
\hline
\end{tabular}
\end{center}
\end{table}

The published measurements of $R^{+-/00}$ are listed 
in \Table{R_data} together with the corresponding assumed values of 
$\tau(\Bu)/\tau(\Bd)$.
All measurements are based on the above-mentioned method, 
except the one from \belle, which is a by-product of the 
\Bd mixing frequency analysis using dilepton events
(but note that it also assumes isospin invariance, 
namely $\Gamma(\Bu \to \ell^+{\rm X}) = \Gamma(\Bd \to \ell^+{\rm X})$).
The latter is therefore treated in a slightly different 
manner in the following procedure used to combine 
these measurements:
\begin{itemize} 
\item each published value of $R^{+-/00}$ from CLEO and \babar
      is first converted back to the original measurement of 
      $R^{+-/00} \, \tau(\Bu)/\tau(\Bd)$, using the value of the 
      lifetime ratio assumed in the corresponding analysis;
\item a simple weighted average of these original
      measurements of $R^{+-/00} \, \tau(\Bu)/\tau(\Bd)$ from 
      CLEO and \babar (which do not depend on the assumed value 
      of the lifetime ratio) is then computed, assuming no 
      statistical or systematic correlations between them;


\item the weighted average of $R^{+-/00} \, \tau(\Bu)/\tau(\Bd)$ 
      is converted into a value of $R^{+-/00}$, using the latest 
      average of the lifetime ratios, $\tau(\Bu)/\tau(\Bd)=\hfagRTAUBU$ 
      (see \Sec{lifetime_ratio});
\item the \belle measurement of $R^{+-/00}$ is adjusted to the 
      current values of $\tau(\Bd)=\hfagTAUBD$ and 
      $\tau(\Bu)/\tau(\Bd)=\hfagRTAUBU$ (see \Sec{lifetime_ratio}),
      using the quoted systematic uncertainties due to these parameters;
\item the combined value of $R^{+-/00}$ from CLEO and \babar is averaged 
      with the adjusted value of $R^{+-/00}$ from \belle, assuming a 100\% 
      correlation of the systematic uncertainty due to the limited 
      knowledge on $\tau(\Bu)/\tau(\Bd)$; no other correlation is considered. 
\end{itemize} 
The resulting global average, 
\begin{equation}
R^{+-/00} = \frac{f^{+-}}{f^{00}} =  \hfagFF \,,
\labe{Rplusminus}
\end{equation}
is consistent with an equal production of charged and neutral \B mesons, 
although only at the $\hfagNSIGMAFF \sigma$ level.

On the other hand, the \babar collaboration has 
performed a direct measurement of the $f^{00}$ fraction 
using a novel method, which does not rely on isospin symmetry nor requires 
the knowledge of $\tau(\Bu)/\tau(\Bd)$. Its analysis, 
based on a comparison between the number of events where a single 
$B^0 \to D^{*-} \ell^+ \nu$ decay could be reconstructed and the number 
of events where two such decays could be reconstructed, yields~\cite{Aubert:2005bq}
\begin{equation}
f^{00}= 0.487 \pm 0.010\,\mbox{(stat)} \pm 0.008\,\mbox{(syst)} \,.
\labe{fzerozero}
\end{equation}

The two results of \Eqss{Rplusminus}{fzerozero} are of very different natures 
and completely independent of each other. 
Their product is equal to $f^{+-} = \hfagFPROD$, 
while another combination of them gives $f^{+-} + f^{00}= \hfagFSUM$, 
compatible with unity.
Assuming\footnote{A few non-$\B\bar{B}$
decay modes of the $\Upsilon(4S)$ 
($\Upsilon(1S)\pi^+\pi^-$,
$\Upsilon(2S)\pi^+\pi^-$, $\Upsilon(1S)\eta$) 
have been observed with branching fractions
of the order of $10^{-4}$~\cite{Aubert:2006bm,*Sokolov:2006sd,*Aubert_mod:2008bv},
corresponding to a partial
width several times larger than that in the \ee channel.
However, this can still be
neglected and the assumption $f^{+-}+f^{00}=1$ remains valid
in the present context of the determination of $f^{+-}$ and $f^{00}$.}
 $f^{+-}+f^{00}= 1$, also consistent with 
CLEO's observation that the fraction of \Ups decays 
to \BB pairs is larger than 0.96 at \CL{95}~\cite{Barish:1995cx},
the results of \Eqss{Rplusminus}{fzerozero}
can be averaged (first converting \Eq{Rplusminus} 
into a value of $f^{00}=1/(R^{+-/00}+1)$) 
to yield the following more precise estimates:
\begin{equation}
f^{00} = \hfagFNW  \,,~~~ f^{+-} = 1 -f^{00} =  \hfagFCW \,,~~~
\frac{f^{+-}}{f^{00}} =  \hfagFFW \,.
\end{equation}
The latter ratio differs from one by $\hfagNSIGMAFFW \sigma$.

\mysubsubsection{\b-hadron production fractions in \Upsfive decays}
\labs{fraction_Ups5S}

\newcommand{\fsfive}{\ensuremath{f_s^{\Upsfive}}}
\newcommand{\fudfive}{\ensuremath{f_{u,d}^{\Upsfive}}}
\newcommand{\fnBfive}{\ensuremath{f_{B\!\!\!\!/}^{\Upsfive}}}

Hadronic events produced in $e^+e^-$ collisions at the \Upsfive energy
can be classified into three categories: 
light-quark ($u$, $d$, $s$, $c$) continuum events, $b\bar{b}$ continuum events,
and \Upsfive events. The latter two cannot be distinguished and will be called
$b\bar{b}$ events in the following. These $b\bar{b}$ events, which also include 
$b\bar{b}\gamma$ events because of possible initial-state radiation, 
can hadronize in different final states.
We define \fudfive\ as
the fraction of $b\bar{b}$ events with a pair of non-strange 
bottom mesons 
(final states $B\bar{B}$,
$B\bar{B}^*$, $B^*\bar{B}$, $B^*\bar{B}^*$,
$B\bar{B}\pi$, $B\bar{B}^*\pi$, $B^*\bar{B}\pi$,
$B^*\bar{B}^*\pi$, and $B\bar{B}\pi\pi$, 
where
$B$ denotes a $B^0$ or $B^+$ meson and 
$\bar{B}$ denotes a $\bar{B}^0$ or $B^-$ meson), 
\fsfive\ as
the fraction of $b\bar{b}$ events with a pair of strange 
bottom mesons (final states 
$B_s^0\bar{B}_s^0$, $B_s^0\bar{B}_s^{0*}$, $B_s^{0*}\bar{B}_s^0$, and
$B_s^{0*}\bar{B}_s^{0*}$), and 
\fnBfive\ as the fraction of $b\bar{b}$ events without 
bottom meson in the final state.
Note that the excited bottom-meson states decay via $B^* \to B \gamma$ and
$B_s^{0*} \to B_s^0 \gamma$.
These fractions satisfy
\begin{equation}
\fudfive + \fsfive + \fnBfive = 1 \,.
\labe{sum_frac_five}
\end{equation} 

\begin{table}
\caption{Published measurements of \fsfive.
All values have been obtained assuming $\fnBfive=0$. 
They are quoted 
as in the original publication, except for the most
recent measurement which is quoted as 
$1-\fudfive$, with \fudfive\ from Ref.~\cite{Drutskoy:2010an}.
The last line gives our average of \fsfive\ assuming $\fnBfive=0$.}
\labt{fsFiveS}
\begin{center}
\begin{tabular}{lll}
\hline
Experiment, year, dataset                 & Decay mode or method    & Value of \fsfive\  \\
CLEO, 2006, 0.42\invfb~\cite{Huang:2006em_mod}     & $\Upsfive\to D_{s}X$     & $0.168 \pm 0.026^{+0.067}_{-0.034}$  \\
             & $\Upsfive \to \phi X$    & $0.246 \pm 0.029^{+0.110}_{-0.053}$ \\
             & $\Upsfive \to B\bar{B}X$ & $0.411 \pm 0.100 \pm 0.092$ \\  
             & CLEO average of above 3  & $0.21^{+0.06}_{-0.03}$      \\  \hline
Belle, 2006, 1.86\invfb~\cite{Drutskoy:2006fg} & $\Upsfive \to D_s X$     & $0.179 \pm 0.014 \pm 0.041$ \\
             & $\Upsfive \to D^0 X$     & $0.181 \pm 0.036 \pm 0.075$ \\  
             & Belle average of above 2 & $0.180 \pm 0.013 \pm 0.032$ \\  \hline 
Belle, 2010, 23.6\invfb~\cite{Drutskoy:2010an} & $\Upsfive \to B\bar{B}X$ & $0.263 \pm 0.032 \pm 0.051$ 
\\ \hline
\multicolumn{2}{l}{Average of all above 
after adjustments to inputs of \Table{fsFiveS_external}} & 
\hfagFSFIVERL             \\  \hline 
\end{tabular}
\end{center}
\end{table}

\begin{table}
\caption{External inputs on which the \fsfive\ averages are based.}
\labt{fsFiveS_external}
\begin{center}
\begin{tabular}{lcl}
\hline
Branching fraction   & Value     & Explanation and reference \\
\hline
${\cal B}(B\to D_s X)\times {\cal B}(D_s \to \phi\pi)$ & 
$0.00374\pm 0.00014$ & derived from~\cite{PDG_2010} \\
${\cal B}(B^0_s \to D_s X)$ & 
$0.92\pm0.11$ & model-dependent estimate~\cite{Artuso:2005xw} \\
${\cal B}(D_s \to \phi\pi)$ & 
$0.045\pm0.004$ & \cite{PDG_2010} \\
${\cal B}(B\to D^0 X)\times {\cal B}(D^0 \to K\pi)$ & 
$0.0243\pm0.0011$ & derived from~\cite{PDG_2010} \\
${\cal B}(B^0_s \to D^0 X)$ & 
$0.08\pm0.07$ & model-dependent estimate~\cite{Drutskoy:2006fg,Artuso:2005xw} \\
${\cal B}(D^0 \to K\pi)$ & 
$0.0389\pm0.0005$ & \cite{PDG_2010} \\
${\cal B}(B \to \phi X)$ & 
$0.0343\pm0.0012$ & world average~\cite{PDG_2010,Huang:2006em_mod} \\
${\cal B}(B^0_s \to \phi X)$ &
$0.161\pm0.024$ & model-dependent estimate~\cite{Huang:2006em_mod} \\
\hline
\end{tabular}
\end{center}
\end{table}

The CLEO and Belle collaborations have published in 2006
measurements of several inclusive \Upsfive branching fractions, 
${\cal B}(\Upsfive\to D_s X)$, 
${\cal B}(\Upsfive\to \phi X)$ and 
${\cal B}(\Upsfive\to D^0 X)$, 
from which they extracted the
model-dependent estimates of \fsfive\
reported in \Table{fsFiveS}. This extraction was performed 
under the implicit assumption  
$\fnBfive=0$, using the relation 
\begin{equation}
\frac12{\cal B}(\Upsfive\to D_s X)=\fsfive\times{\cal B}(B_s^0\to D_s X) + 
\left(1-\fsfive-\fnBfive\right)\times{\cal B}(B\to D_s X) \,,
\labe{Ds_correct}
\end{equation}
and similar relations for
${\cal B}(\Upsfive\to D^0 X)$ and ${\cal B}(\Upsfive\to \phi X)$.
We list also in \Table{fsFiveS} 
the values of \fsfive\ derived from measurements of
$\fudfive={\cal B}(\Upsfive\to B\bar BX)$~\cite{Huang:2006em_mod,Drutskoy:2010an}, as well as our average value of  \fsfive,
all obtained under the assumption $\fnBfive=0$.

Since the observation of \Upsfive\ decays to final states without 
bottom hadrons~\cite{Abe:2007tk},
the assumption $\fnBfive=0$ is no longer valid. 
We therefore perform a $\chi^2$ fit of the original 
measurements of the \Upsfive\ branching fractions of
Refs.~\cite{Huang:2006em_mod,Drutskoy:2006fg,Drutskoy:2010an},
using the inputs of \Table{fsFiveS_external}
and the constraints of \Eqss{sum_frac_five}{Ds_correct},
to simultaneously extract \fudfive, \fsfive\ and \fnBfive. Taking all known 
correlations into account, the best fit values are
\begin{eqnarray}
\fudfive &=& \hfagFUDFIVE \,, \labe{fudfive} \\
\fsfive  &=& \hfagFSFIVE  \,, \labe{fsfive}  \\
\fnBfive &=& \hfagFNBFIVE \,. \labe{fnBfive}
\end{eqnarray}


The 
\Upsfive\ resonance has been observed to decay to 
$\Upsilon(1S)\pi^+\pi^-$,
$\Upsilon(2S)\pi^+\pi^-$,
$\Upsilon(3S)\pi^+\pi^-$ and
$\Upsilon(1S)K^+K^-$ final states~\cite{Abe:2007tk}.
The sum of these measured branching fractions, adding also the 
contribution of the 
$\Upsilon(1S)\pi^0\pi^0$,
$\Upsilon(2S)\pi^0\pi^0$,
$\Upsilon(3S)\pi^0\pi^0$ and
$\Upsilon(1S)K^0\bar{K}^0$ final states
assuming isospin conservation,
amounts to
$$
{\cal B}(\Upsfive\to \Upsilon(nS)hh) = 0.028\pm0.003 \,,
~~~~\mbox{for $n=1,2,3$ and $h=\pi,K$}\,,
$$
%
which represents a lower bound for \fnBfive. 
Our central value of \Eq{fnBfive} is indeed larger than this bound. 


The production of $B^0_s$ mesons at the \Upsfive
is observed to be dominated by the $B_s^{0*}\bar{B}_s^{0*}$
channel, 
with $\sigma(e^+e^- \to B_s^{0*}\bar{B}_s^{0*})/%
\sigma(e^+e^- \to B_s^{0(*)}\bar{B}_s^{0(*)})
= (90.1^{+3.8}_{-4.0}\pm 0.2)\%$~\cite{Louvot:2008sc}.
The proportion of the various production channels 
for non-strange $B$ mesons have also been recently measured~\cite{Drutskoy:2010an}.

\mysubsubsection{\b-hadron production fractions at high energy}
\labs{fractions_high_energy}
\labs{chibar}

At high energy, all species of weakly-decaying \b hadrons 
can be produced, either directly or in strong and electromagnetic 
decays of excited \b hadrons.
It is often assumed 
that the fractions of these different species 
are the same in unbiased samples of high-$p_{\rm T}$ \b jets 
originating from \particle{Z^0} decays or from \particle{p\bar{p}} 
collisions at the Tevatron.
This hypothesis is plausible considering that, in both cases, 
the last step of the jet hadronization is a non-perturbative
QCD process occurring at the scale of $\Lambda_{\rm QCD}$.
On the other hand, there is no strong argument to claim that these 
fractions should be strictly equal, so this assumption 
should be checked experimentally.
Although the available data is not sufficient at 
this time to perform a significant check, 
it is expected that more data from 
Tevatron Run~II may improve this situation and 
allow one to confirm or disprove this assumption with reasonable 
confidence. Meanwhile, the attitude adopted here is that these 
fractions are assumed to be equal at all high-energy colliders
until demonstrated otherwise by experiment.\footnote{It is likely
that the \b-hadron fractions in low-$p_{\rm T}$ jets at a hadronic machine 
be different; in particular, beam-remnant effects may enhance the \b-baryon production.}
However, as explained below, the measurements performed at LEP and at 
the Tevatron show discrepancies. Therefore we present three sets 
of averages: one set including only measurements performed at LEP, 
a second set including only measurements performed at the Tevatron, 
and a third set including measurements performed at both LEP and Tevatron. 

Contrary to what happens in the charm sector where the fractions of \particle{D^+} 
and \particle{D^0} are different, the relative amount of \Bu and \Bd is not affected by the 
electromagnetic decays of excited ${\Bu}^*$ and ${\Bd}^*$ states and strong decays of excited
${\Bu}^{**}$ and ${\Bd}^{**}$ states. Decays of the type \particle{{\Bs}^{**} \to B^{(*)}K}
also contribute to the \Bu and \Bd rates, but with the same magnitude if mass effects
can be neglected.  
We therefore assume equal production of \Bu and \Bd. We also  
neglect the production of weakly-decaying states
made of several heavy quarks (like \Bc and other heavy baryons) 
which is known to be very small. Hence, for the purpose of determining 
the \b-hadron fractions, we use the constraints
\begin{equation}
\fBu = \fBd ~~~~\mbox{and}~~~ \fBu + \fBd + \fBs + \fbb = 1 \,,
\labe{constraints}
\end{equation}
where \fBu, \fBd, \fBs and \fbb
are the unbiased fractions of \Bu, \Bd, \Bs and \b baryons, respectively.

The LEP experiments have measured
$\fBs \times \BR{\Bs\to\particle{D_s^-} \ell^+ \nu_\ell \mbox{$X$}}$~\cite{Abreu:1992rv,*Acton:1992zq,*Buskulic:1995bd}, 
$\BR{\b\to\Lb} \times \BR{\Lb\to\Lc\ell^-\bar{\nu}_\ell \mbox{$X$}}$~\cite{Abreu:1995me,Barate:1997if}
and $\BR{\b\to\Xib^-} \times \BR{\Xi_b^- \to \Xi^-\ell^-\overline\nu_\ell 
\mbox{$X$}}$~\cite{Buskulic:1996sm,Abdallah:2005cw}\footnote{The DELPHI result 
of \Ref{Abdallah:2005cw} is considered to supersede an older one~\cite{Abreu:1995kt}.}
from partially reconstructed final states 
including a lepton, \fbb
from protons identified in \b events~\cite{Barate:1997ty}, and the 
production rate of charged \b hadrons~\cite{Abdallah:2003xp}. 
The various \b-hadron fractions 
have also been measured at CDF using lepton-charm final 
states~\cite{Affolder:1999iq,Aaltonen:2008zd,Aaltonen:2008eu}\footnote{CDF updated their measurement of 
\fbb/\fBd~\cite{Affolder:1999iq} to account for a measured 
$p_{\rm T}$ dependence between exclusively reconstructed $\Lambda_b$ and $B^0$~\cite{Aaltonen:2008eu}.}
and double semileptonic decays with
\particle{K^*\mu\mu} and \particle{\phi\mu\mu}
final states~\cite{Abe:1999ta}.  Recent measurements of heavy flavor baryon production at 
the Tevatron are included in the determination of 
\fbb~\cite{Abazov_mod:2007ub,Abazov:2008qm,Aaltonen:2009ny}\footnote{\dzero reports $f_{\Omega_b^-}/f_{\Xi_b^-}$.
We use the CDF+\dzero average of $f_{\Xi_b^-}/f_{\Lambda_b}$ to obtain 
$f_{\Omega_b^-}/f_{\Lambda_b}$ and then combine with the CDF result.}
using the constraint
\begin{eqnarray}
\fbb & = & f_{\Lambda_b} + f_{\Xi_b^0} + f_{\Xi_b^-} + f_{\Omega_b^-} \nonumber \\
     & = & f_{\Lambda_b}\left(1 + 2\frac{f_{\Xi_b^-}}{f_{\Lambda_b}} 
           + \frac{f_{\Omega_b^-}}{f_{\Lambda_b}}\right),
\end{eqnarray}
where isospin invariance is assumed in the production of $\Xi_b^0$ and $\Xi_b^-$. 
Other \b baryons are expected to decay strongly or electromagnetically to those baryons listed.
For the production measurements, both CDF and \dzero\ reconstruct their \b baryons exclusively
to final states which include a $J/\psi$ and a hyperon ($\Lambda_b\rightarrow J/\psi \Lambda$, 
$\Xi_b^- \rightarrow J/\psi \Xi^-$ and $\Omega_b^- \rightarrow J/\psi \Omega^-$).  
We assume that the partial decay width of a \b baryon to a $J/\psi$ and the corresponding hyperon
is equal to the partial width of any other \b baryon to a $J/\psi$ and the corresponding hyperon.

All these published results have been combined
following the procedure and assumptions described in~\cite{Abbaneo:2000ej_mod,*Abbaneo:2001bv_mod_cont},
to yield $\fBu=\fBd=\hfagWFBDNOMIX$, 
$\fBs=\hfagWFBSNOMIX$ and $\fbb=\hfagWFBBNOMIX$
under the constraints of \Eq{constraints}. 
Following the PDG prescription, we have scaled the
combined uncertainties on these fractions 
by \hfagWFSFACTOR to account for slight 
discrepancies in the input data.
Repeating the combinations,
we obtain $\fBu=\fBd=\hfagLFBDNOMIX$,
$\fBs=\hfagLFBSNOMIX$ and $\fbb=\hfagLFBBNOMIX$ when using the LEP data only,
and $\fBu=\fBd=\hfagTFBDNOMIX$, $\fBs=\hfagTFBSNOMIX$ 
$\fbb = \hfagTFBBNOMIX$ when using the Tevatron data only.  When the Tevatron and LEP
data are separated, we find no need to scale the uncertainties of either combination.
For these combinations other external inputs are used, \eg\ the branching 
ratios of \B mesons to final states with a \particle{D}, \particle{D^*} or 
\particle{D^{**}} in semileptonic decays, which are needed to evaluate the 
fraction of semileptonic \Bs decays with a \particle{D_s^-} in the final state.

Time-integrated mixing analyses performed with lepton pairs 
from \particle{b\bar{b}} 
events produced at high-energy colliders measure the quantity 
\begin{equation}
\chibar = f'_{\particle{d}} \,\chid + f'_{\particle{s}} \,\chis \,,
\labe{chibar}
\end{equation}
where $f'_{\particle{d}}$ and $f'_{\particle{s}}$ are 
the fractions of \Bd and \Bs hadrons 
in a sample of semileptonic \b-hadron decays, and where \chid and \chis 
are the \Bd and \Bs time-integrated mixing probabilities.
Assuming that all \b hadrons have the same semileptonic decay width implies 
$f'_i = f_i R_i$, where $R_i = \tau_i/\tau_{\b}$ is the ratio of the lifetime 
$\tau_i$ of species $i$ to the average \b-hadron lifetime 
$\tau_{\b} = \sum_i f_i \tau_i$.
Hence measurements of the mixing probabilities
\chibar, \chid and \chis can be used to improve our 
knowledge of \fBu, \fBd, \fBs and \fbb.
In practice, the above relations yield another determination of 
\fBs obtained from \fbb and mixing information, 
\begin{equation}
\fBs = \frac{1}{R_{\particle{s}}}
\frac{(1+r)\overline{\chi}-(1-\fbb R_{\rm baryon}) \chid}{(1+r)\chis - \chid} \,,
\labe{fBs-mixing}
\end{equation}
where $r=R_{\particle{u}}/R_{\particle{d}} = \tau(\Bu)/\tau(\Bd)$.

The published measurements of \chibar performed by the LEP
experiments have been combined by the LEP Electroweak Working Group to yield 
$\chibar = \hfagCHIBARLEP$~\cite{LEPEWWG:2005ema_mod}.
This can be compared with the Tevatron average, 
$\chibar = \hfagCHIBARTEV$,
obtained from a CDF measurement with Run~I data~\cite{Acosta:2003ie_mod}
and from a recent \dzero measurement with Run~II data~\cite{Abazov:2006qw}.
The two averages deviate
from each other by $\hfagCHIBARSFACTOR\,\sigma$; 
this could be an indication that the production fractions of \b hadrons 
at the \particle{Z} peak or at the Tevatron are not the same. 
Although this discrepancy 
is not very significant it should be carefully monitored in the future. 
We choose to combine these two results in a simple weighted average,
assuming no correlations, and, following the PDG prescription, we 
multiply the combined uncertainty by \hfagCHIBARSFACTOR to account 
for the discrepancy. Our world average is then
$\chibar = \hfagCHIBAR$.

\begin{table}
\caption{Time-integrated mixing probability \chibar (defined in \Eq{chibar}), 
and fractions of the different \b-hadron species in an unbiased sample of 
weakly-decaying \b hadrons, obtained from both direct
and mixing measurements.
The last column includes measurements performed at both LEP and Tevatron.}
\labt{fractions}
\begin{center}
\begin{tabular}{llccc}
\hline
Quantity            &                      & in $Z$ decays   & at Tevatron    & combined    \\
\hline
Mixing probability  & $\overline{\chi}$    & \hfagCHIBARLEP  & \hfagCHIBARTEV & \hfagCHIBAR \\
\Bu or \Bd fraction & $\fBu = \fBd$        & \hfagLFBD       & \hfagTFBD      & \hfagWFBD   \\
\Bs fraction        & $\fBs$               & \hfagLFBS       & \hfagTFBS      & \hfagWFBS   \\
\b-baryon fraction  & $\fbb$               & \hfagLFBB       & \hfagTFBB      & \hfagWFBB   \\
\multicolumn{2}{l}{Correlation between \fBs and $\fBu = \fBd$}            & \hfagLRHOFBDFBS & \hfagTRHOFBDFBS & \hfagWRHOFBDFBS \\
\multicolumn{2}{l}{Correlation between \fbb and $\fBu = \fBd$} & \hfagLRHOFBDFBB & \hfagTRHOFBDFBB & \hfagWRHOFBDFBB \\
\multicolumn{2}{l}{Correlation between \fbb and \fBs}       & \hfagLRHOFBBFBS & \hfagTRHOFBBFBS & \hfagWRHOFBBFBS \\
\hline
\end{tabular}
\end{center}
\end{table}

Introducing the \chibar average in \Eq{fBs-mixing}, together with our world average 
$\chid = \hfagCHIDWU$ (see \Eq{chid} of \Sec{dmd}), the assumption $\chis= 1/2$ 
(justified by \Eq{chis} in \Sec{dms}), the 
best knowledge of the lifetimes (see \Sec{lifetimes}) and the estimate of \fbb given above, 
yields $\fBs = \hfagWFBSMIX$ 
(or $\fBs = \hfagLFBSMIX$ using only LEP data, 
or $\fBs = \hfagTFBSMIX$ using only Tevatron data),
an estimate dominated by the mixing information. 
Taking into account all known correlations (including the one introduced by \fbb), 
this result is then combined with the set of fractions obtained from direct measurements 
(given above), to yield the 
improved estimates of \Table{fractions}, 
still under the constraints of \Eq{constraints}.\footnote{%
The combined value of \fbb is smaller than the  
results from either LEP or Tevatron separately.
This seemingly surprising result  
arises from the smaller uncertainties on the other fractions 
and the application of the unitarity constraint of \Eq{constraints}.}
As can be seen, our knowledge on the mixing parameters 
substantially reduces the uncertainty on \fBs, and this 
even in the case of the world averages where a rather strong 
deweighting was introduced in the computation of \chibar.
It should be noted that the results 
are correlated, as indicated in \Table{fractions}.


%
%

\mysubsection{\b-hadron lifetimes}
\labs{lifetimes}

In the spectator model the decay of \b-flavored hadrons $H_b$ is
governed entirely by the flavor changing \particle{b\to Wq} transition
($\particle{q}=\particle{c,u}$).  For this very reason, lifetimes of all
\b-flavored hadrons are the same in the spectator approximation
regardless of the (spectator) quark content of the $H_b$.  In the early
1990's experiments became sophisticated enough to start seeing the
differences of the lifetimes among various $H_b$ species.  The first
theoretical calculations of the spectator quark effects on $H_b$
lifetime emerged only few years earlier.

Currently, most of such calculations are performed in the framework of
the Heavy Quark Expansion, HQE.  In the HQE, under certain assumptions
(most important of which is that of quark-hadron duality), the decay
rate of an $H_b$ to an inclusive final state $f$ is expressed as the sum
of a series of expectation values of operators of increasing dimension,
multiplied by the correspondingly higher powers of $\Lambda_{\rm
QCD}/m_b$:
\begin{equation}
\Gamma_{H_b\to f} = |CKM|^2\sum_n c_n^{(f)}
\Bigl(\frac{\Lambda_{\rm QCD}}{m_b}\Bigr)^n\langle H_b|O_n|H_b\rangle,
\labe{hqe}
\end{equation}
where $|CKM|^2$ is the relevant combination of the CKM matrix elements.
Coefficients $c_n^{(f)}$ of this expansion, known as Operator Product
Expansion~\cite{Shifman:1986mx,*Chay:1990da,*Bigi:1992su,*Bigi:1992su_erratum},
can be calculated perturbatively.  Hence, the HQE
predicts $\Gamma_{H_b\to f}$ in the form of an expansion in both
$\Lambda_{\rm QCD}/m_{\b}$ and $\alpha_s(m_{\b})$.  The precision of
current experiments makes it mandatory to go to the next-to-leading
order in QCD, {\em i.e.}\ to include correction of the order of
$\alpha_s(m_{\b})$ to the $c_n^{(f)}$'s.  All non-perturbative physics
is shifted into the expectation values $\langle H_b|O_n|H_b\rangle$ of
operators $O_n$.  These can be calculated using lattice QCD or QCD sum
rules, or can be related to other observables via the
HQE~\cite{Bigi:1995jr,*Bellini:1996ra}.  One may reasonably expect that powers of
$\Lambda_{\rm QCD}/m_{\b}$ provide enough suppression that only the
first few terms of the sum in \Eq{hqe} matter.

Theoretical predictions are usually made for the ratios of the lifetimes
(with $\tau(\Bd)$ chosen as the common denominator) rather than for the
individual lifetimes, for this allows several uncertainties to cancel.
The precision of the current HQE calculations (see
\Refs{Ciuchini:2001vx,*Beneke:2002rj,*Franco:2002fc,Tarantino:2003qw,*Gabbiani:2003pq,Gabbiani:2004tp} for the latest updates)
is in some instances already surpassed by the measurements,
\eg\ in the case of $\tau(\Bu)/\tau(\Bd)$.  Also, HQE calculations are
not assumption-free.  More accurate predictions are a matter of progress
in the evaluation of the non-perturbative hadronic matrix elements and
verifying the assumptions that the calculations are based upon.
However, the HQE, even in its present shape, draws a number of important
conclusions, which are in agreement with experimental observations:
\begin{itemize}
\item The heavier the mass of the heavy quark the smaller is the
  variation in the lifetimes among different hadrons containing this
  quark, which is to say that as $m_{\b}\to\infty$ we retrieve the
  spectator picture in which the lifetimes of all $H_b$'s are the same.
   This is well illustrated by the fact that lifetimes are rather
   similar in the \b sector, while they differ by large factors
   in the \particle{c} sector ($m_{\particle{c}}<m_{\b}$).
\item The non-perturbative corrections arise only at the order of
  $\Lambda_{\rm QCD}^2/m_{\b}^2$, which translates into 
  differences among $H_b$ lifetimes of only a few percent.
\item It is only the difference between meson and baryon lifetimes that
  appears at the $\Lambda_{\rm QCD}^2/m_{\b}^2$ level.  The splitting of the
  meson lifetimes occurs at the $\Lambda_{\rm QCD}^3/m_{\b}^3$ level, yet it is
  enhanced by a phase space factor $16\pi^2$ with respect to the leading
  free \b decay.
\end{itemize}

To ensure that certain sources of systematic uncertainty cancel, 
lifetime analyses are sometimes designed to measure a 
ratio of lifetimes.  However, because of the differences in decay
topologies, abundance (or lack thereof) of decays of a certain kind,
{\em etc.}, measurements of the individual lifetimes are more 
common.  In the following section we review the most common
types of the lifetime measurements.  This discussion is followed by the
presentation of the averaging of the various lifetime measurements, each
with a brief description of its particularities.



\mysubsubsection{Lifetime measurements, uncertainties and correlations}

In most cases lifetime of an $H_b$ is estimated from a flight distance
and a $\beta\gamma$ factor which is used to convert the geometrical
distance into the proper decay time.  Methods of accessing lifetime
information can roughly be divided in the following five categories:
\begin{enumerate}
\item {\bf\em Inclusive (flavor-blind) measurements}.  These
  measurements are aimed at extracting the lifetime from a mixture of
  \b-hadron decays, without distinguishing the decaying species.  Often
  the knowledge of the mixture composition is limited, which makes these
  measurements experiment-specific.  Also, these
  measurements have to rely on Monte Carlo for estimating the
  $\beta\gamma$ factor, because the decaying hadrons are not fully
  reconstructed.  On the bright side, these usually are the largest
  statistics \b-hadron lifetime measurements that are accessible to a
  given experiment, and can, therefore, serve as an important
  performance benchmark.
\item {\bf\em Measurements in semileptonic decays of a specific
  {\boldmath $H_b$\unboldmath}}.  \particle{W}from \particle{\b\to Wc}
  produces $\ell\nu_l$ pair (\particle{\ell=e,\mu}) in about 21\% of the
  cases.  Electron or muon from such decays is usually a well-detected
  signature, which provides for clean and efficient trigger.
  \particle{c} quark from \particle{b\to Wc} transition and the other
  quark(s) making up the decaying $H_b$ combine into a charm hadron,
  which is reconstructed in one or more exclusive decay channels.
  Knowing what this charmed hadron is allows one to separate, at least
  statistically, different $H_b$ species.  The advantage of these
  measurements is in statistics, which usually is superior to that of the
  exclusively reconstructed $H_b$ decays.  Some of the main
  disadvantages are related to the difficulty of estimating lepton+charm
  sample composition and Monte Carlo reliance for the $\beta\gamma$
  factor estimate.
\item {\bf\em Measurements in exclusively reconstructed hadronic decays}.
  These
  have the advantage of complete reconstruction of decaying $H_b$, which
  allows one to infer the decaying species as well as to perform precise
  measurement of the $\beta\gamma$ factor.  Both lead to generally
  smaller systematic uncertainties than in the above two categories.
  The downsides are smaller branching ratios, larger combinatoric
  backgrounds, especially in $H_b\rightarrow H_c\pi(\pi\pi)$ and
  multi-body $H_c$ decays, or in a hadron collider environment with
  non-trivial underlying event.  $H_b\to J/\psi H_s$ are relatively
  clean and easy to trigger on $J/\psi\to \ell^+\ell^-$, but their
  branching fraction is only about 1\%.
\item {\bf\em Measurements at asymmetric B factories}. 

In the $\Ups\rightarrow B \bar{B}$ decay, the \B mesons (\Bu or \Bd) are
essentially at rest in the \Ups frame.  This makes direct lifetime
measurements impossible in experiments at symmetric colliders producing 
\Ups at rest. 
At asymmetric \B factories the \Ups meson is boosted
resulting in \B and \particle{\bar{B}} moving nearly parallel to each 
other with the same boost. The lifetime is inferred from the distance $\Delta z$        
separating the \B and \particle{\bar{B}} decay vertices along the beam axis 
and from the \Ups boost known from the beam energies. This boost is equal to 
$\beta \gamma \approx 0.55$ (0.43) in the \babar (\belle) experiment,
resulting in an average \B decay length of approximately 250~(190)~$\mu$m. 
In order to determine the charge of the \B mesons in each event, one of the them is
fully reconstructed in a semileptonic or hadronic decay mode.
The other \B is typically not fully reconstructed, only the position
of its decay vertex is determined from the remaining tracks in the event.
These measurements benefit from large statistics, but suffer from poor proper time 
resolution, comparable to the \B lifetime itself. This resolution is dominated by the 
uncertainty on the decay vertices, which is typically 50~(100)~$\mu$m for a
fully (partially) reconstructed \B meson. 
With very large future statistics,
the resolution and purity could be improved (and hence the systematics reduced)
by fully reconstructing both \B mesons in the event. 
 
\item {\bf\em Direct measurement of lifetime ratios}.  This method has
  so far been only applied in the measurement of $\tau(\Bu)/\tau(\Bd)$.
  The ratio of the lifetimes is extracted from the dependence of the
  observed relative number of \Bu and \Bd candidates (both reconstructed
  in semileptonic decays) on the proper decay time.
\end{enumerate}

In some of the latest analyses, measurements of two (\eg\ $\tau(\Bu)$ and
$\tau(\Bu)/\tau(\Bd)$) or three (\eg\ $\tau(\Bu)$,
$\tau(\Bu)/\tau(\Bd)$, and \dmd) quantities are combined.  This
introduces correlations among measurements.  Another source of
correlations among the measurements are the systematic effects, which
could be common to an experiment or to an analysis technique across the
experiments.  When calculating the averages, such correlations are taken
into account per general procedure, described in
\Ref{LEPBOSC:1996}.

\mysubsubsection{Inclusive \b-hadron lifetimes}

The inclusive \b hadron lifetime is defined as $\tau_{\b} = \sum_i f_i
\tau_i$ where $\tau_i$ are the individual species lifetimes and $f_i$ are
the fractions of the various species present in an unbiased sample of
weakly-decaying \b hadrons produced at a high-energy
collider.\footnote{In principle such a quantity could be slightly
different in \particle{Z} decays and at the Tevatron, in case the
fractions of \b-hadron species are not exactly the same; see the
discussion in \Sec{fractions_high_energy}.}  This quantity is certainly
less fundamental than the lifetimes of the individual species, the
latter being much more useful in comparisons of the measurements with
the theoretical predictions.  Nonetheless, we perform the averaging of
the inclusive lifetime measurements for completeness as well as for the
reason that they might be of interest as ``technical numbers.''

\begin{table}[tp]
\caption{Measurements of average \b-hadron lifetimes.}
\labt{lifeincl}
\begin{center}
\begin{tabular}{lcccl} \hline
Experiment &Method           &Data set & $\tau_{\b}$ (ps)       &Ref.\\
\hline
ALEPH  &Dipole               &91     &$1.511\pm 0.022\pm 0.078$ &\cite{Buskulic:1993gj}\\
DELPHI &All track i.p.\ (2D) &91--92 &$1.542\pm 0.021\pm 0.045$ &\cite{Abreu:1994dr}$^a$\\
DELPHI &Sec.\ vtx            &91--93 &$1.582\pm 0.011\pm 0.027$ &\cite{Abreu:1996hv}$^a$\\
DELPHI &Sec.\ vtx            &94--95 &$1.570\pm 0.005\pm 0.008$ &\cite{Abdallah:2003sb}\\
L3     &Sec.\ vtx + i.p.     &91--94 &$1.556\pm 0.010\pm 0.017$ &\cite{Acciarri:1997tt}$^b$\\
OPAL   &Sec.\ vtx            &91--94 &$1.611\pm 0.010\pm 0.027$ &\cite{Ackerstaff:1996as}\\
SLD    &Sec.\ vtx            &93     &$1.564\pm 0.030\pm 0.036$ &\cite{Abe:1995rm}\\ 
\hline
\multicolumn{2}{l}{Average set 1 (\b vertex)} && \hfagTAUBVTXnounit &\\
\hline\hline
ALEPH  &Lepton i.p.\ (3D)    &91--93 &$1.533\pm 0.013\pm 0.022$ &\cite{Buskulic:1995rw}\\
L3     &Lepton i.p.\ (2D)    &91--94 &$1.544\pm 0.016\pm 0.021$ &\cite{Acciarri:1997tt}$^b$\\
OPAL   &Lepton i.p.\ (2D)    &90--91 &$1.523\pm 0.034\pm 0.038$ &\cite{Acton:1993xk}\\ 
\hline
\multicolumn{2}{l}{Average set 2 ($\b\to\ell$)} && \hfagTAUBLEPnounit &\\
\hline\hline
CDF1   &\particle{J/\psi} vtx&92--95 &$1.533\pm 0.015^{+0.035}_{-0.031}$ &\cite{Abe:1997bd} \\ 
\hline\hline
\multicolumn{2}{l}{Average of all above} && \hfagTAUBnounit & \\
\hline
\multicolumn{5}{l}{$^a$ \footnotesize The combined DELPHI result quoted in
\cite{Abreu:1996hv} is 1.575 $\pm$ 0.010 $\pm$ 0.026 ps.} \\[-0.5ex]
\multicolumn{5}{l}{$^b$ \footnotesize The combined L3 result quoted in \cite{Acciarri:1997tt} 
is 1.549 $\pm$ 0.009 $\pm$ 0.015 ps.}
\end{tabular}
\end{center}
\end{table}

In practice, an unbiased measurement of the inclusive lifetime is
difficult to achieve, because it would imply an efficiency which is
guaranteed to be the same across species.  So most of the measurements
are biased.  In an attempt to group analyses which are expected to
select the same mixture of \b hadrons, the available results (given in
\Table{lifeincl}) are divided into the following three sets:
\begin{enumerate}
\item measurements at LEP and SLD that accept any \b-hadron decay, based 
      on topological reconstruction (secondary vertex or track impact
      parameters);
\item measurements at LEP based on the identification
      of a lepton from a \b decay; and
\item measurements at the Tevatron based on inclusive 
      \particle{H_b\to J/\psi X} reconstruction, where the
      \particle{J/\psi} is fully reconstructed.
\end{enumerate}

The measurements of the first set are generally considered as estimates
of $\tau_{\b}$, although the efficiency to reconstruct a secondary
vertex most probably depends, in an analysis-specific way, on the number
of tracks coming from the vertex, thereby depending on the type of the
$H_b$.  Even though these efficiency variations can in principle be
accounted for using Monte Carlo simulations (which inevitably contain
assumptions on branching fractions), the $H_b$ mixture in that case can
remain somewhat ill-defined and could be slightly different among
analyses in this set.

On the contrary, the mixtures corresponding to the other two sets of
measurements are better defined in the limit where the reconstruction
and selection efficiency of a lepton or a \particle{J/\psi} from an
$H_b$ does not depend on the decaying hadron type.  These mixtures are
given by the production fractions and the inclusive branching fractions
for each $H_b$ species to give a lepton or a \particle{J/\psi}.  In
particular, under the assumption that all \b hadrons have the same
semileptonic decay width, the analyses of the second set should measure
$\tau(\b\to\ell) = (\sum_i f_i \tau_i^2) /(\sum_i f_i \tau_i)$ which is
necessarily larger than $\tau_{\b}$ if lifetime differences exist.
Given the present knowledge on $\tau_i$ and $f_i$,
$\tau(\b\to\ell)-\tau_{\b}$ is expected to be of the order of 0.01\ps.

Measurements by SLC and LEP experiments are subject to a number of
common systematic uncertainties, such as those due to (lack of knowledge
of) \b and \particle{c} fragmentation, \b and \particle{c} decay models,
\BR{B\to\ell}, \BR{B\to c\to\ell}, \BR{c\to\ell}, $\tau_{\particle{c}}$,
and $H_b$ decay multiplicity.  In the averaging, these systematic
uncertainties are assumed to be 100\% correlated.  The averages for the
sets defined above (also given in \Table{lifeincl}) are
\begin{eqnarray}
\tau(\b~\mbox{vertex}) &=& \hfagTAUBVTX \,,\\
\tau(\b\to\ell) &=& \hfagTAUBLEP  \,, \\
\tau(\b\to\particle{J/\psi}) &=& \hfagTAUBJP\,,
\end{eqnarray}
whereas an average of all measurements, ignoring mixture differences, 
yields \hfagTAUB.

\mysubsubsection{\Bd and \Bu lifetimes and their ratio}
\labs{taubd}
\labs{taubu}
\labs{lifetime_ratio}

\begin{table}[tp]
\caption{Measurements of the \Bd lifetime.}
\labt{lifebd}
\begin{center}
\begin{tabular}{lcccl} \hline
Experiment &Method                    &Data set &$\tau(\Bd)$ (ps)                  &Ref.\\
\hline
ALEPH  &\particle{D^{(*)} \ell}       &91--95 &$1.518\pm 0.053\pm 0.034$          &\cite{Barate:2000bs}\\
ALEPH  &Exclusive                     &91--94 &$1.25^{+0.15}_{-0.13}\pm 0.05$     &\cite{Buskulic:1996hy}\\
ALEPH  &Partial rec.\ $\pi^+\pi^-$    &91--94 &$1.49^{+0.17+0.08}_{-0.15-0.06}$   &\cite{Buskulic:1996hy}\\
DELPHI &\particle{D^{(*)} \ell}       &91--93 &$1.61^{+0.14}_{-0.13}\pm 0.08$     &\cite{Abreu:1995mc}\\
DELPHI &Charge sec.\ vtx              &91--93 &$1.63 \pm 0.14 \pm 0.13$           &\cite{Adam:1995mb}\\
DELPHI &Inclusive \particle{D^* \ell} &91--93 &$1.532\pm 0.041\pm 0.040$          &\cite{Abreu:1996gb}\\
DELPHI &Charge sec.\ vtx              &94--95 &$1.531 \pm 0.021\pm0.031$          &\cite{Abdallah:2003sb}\\
\end{tabular}
\end{center}
\end{table}

After a number of years of dominating these averages the LEP experiments
yielded the scene to the asymmetric \B~factories and
the Tevatron experiments.  The \B~factories have been very successful in
utilizing their potential -- in only a few years of running, \babar and,
to a greater extent, \belle, have struck a balance between the
statistical and the systematic uncertainties, with both being close to
(or even better than) the impressive 1\%.  In the meanwhile, CDF and
\dzero have emerged as significant contributors to the field as the
Tevatron Run~II data flowed in.  Both appear to enjoy relatively small
systematic effects, and while current statistical uncertainties of their
measurements are factors of 2 to 4 larger than those of their \B-factory
counterparts, both Tevatron experiments stand to increase their samples
by almost an order of magnitude.

\begin{table}[tbp]
\caption{Measurements of the \Bu lifetime.}
\labt{lifebu}
\begin{center}
\begin{tabular}{lcccl} \hline
Experiment &Method                 &Data set &$\tau(\Bu)$ (ps)                 &Ref.\\
\hline
ALEPH  &\particle{D^{(*)} \ell}    &91--95 &$1.648\pm 0.049\pm 0.035$          &\cite{Barate:2000bs}\\
ALEPH  &Exclusive                  &91--94 &$1.58^{+0.21+0.04}_{-0.18-0.03}$   &\cite{Buskulic:1996hy}\\
DELPHI &\particle{D^{(*)} \ell}    &91--93 &$1.61\pm 0.16\pm 0.12$             &\cite{Abreu:1995mc}$^a$\\
DELPHI &Charge sec.\ vtx           &91--93 &$1.72\pm 0.08\pm 0.06$             &\cite{Adam:1995mb}$^a$\\
DELPHI &Charge sec.\ vtx           &94--95 &$1.624\pm 0.014\pm 0.018$          &\cite{Abdallah:2003sb}\\
L3     &Charge sec.\ vtx           &94--95 &$1.66\pm  0.06\pm 0.03$            &\cite{Acciarri:1998uv}\\
OPAL   &\particle{D^{(*)} \ell}    &91--93 &$1.52 \pm 0.14\pm 0.09$            &\cite{Akers:1995pa}\\
\end{tabular}
\end{center}
\end{table}

At present time we are in an interesting position of having three sets
of measurements (from LEP/SLC, \B factories and the Tevatron) that
originate from different environments, obtained using substantially
different techniques and are precise enough for incisive comparison.


\begin{table}[tb]
\caption{Measurements of the ratio $\tau(\Bu)/\tau(\Bd)$.}
\labt{liferatioBuBd}
\begin{center}
\begin{tabular}{lcccl} 
\hline
Experiment &Method                 &Data set &Ratio $\tau(\Bu)/\tau(\Bd)$      &Ref.\\
\hline
ALEPH  &\particle{D^{(*)} \ell}    &91--95 &$1.085\pm 0.059\pm 0.018$          &\cite{Barate:2000bs}\\
ALEPH  &Exclusive                  &91--94 &$1.27^{+0.23+0.03}_{-0.19-0.02}$   &\cite{Buskulic:1996hy}\\
DELPHI &\particle{D^{(*)} \ell}    &91--93 &$1.00^{+0.17}_{-0.15}\pm 0.10$     &\cite{Abreu:1995mc}\\
DELPHI &Charge sec.\ vtx           &91--93 &$1.06^{+0.13}_{-0.11}\pm 0.10$     &\cite{Adam:1995mb}\\
DELPHI &Charge sec.\ vtx           &94--95 &$1.060\pm 0.021 \pm 0.024$         &\cite{Abdallah:2003sb}\\
L3     &Charge sec.\ vtx           &94--95 &$1.09\pm 0.07  \pm 0.03$           &\cite{Acciarri:1998uv}\\
OPAL   &\particle{D^{(*)} \ell}    &91--93 &$0.99\pm 0.14^{+0.05}_{-0.04}$     &\cite{Akers:1995pa}\\
\end{tabular}
\end{center}
\end{table}

The averaging of $\tau(\Bu)$, $\tau(\Bd)$ and $\tau(\Bu)/\tau(\Bd)$
measurements is summarized in \Tablesss{lifebd}{lifebu}{liferatioBuBd}.
For $\tau(\Bu)/\tau(\Bd)$ we averaged only the measurements of this
quantity provided by experiments rather than using all available
knowledge, which would have included, for example, $\tau(\Bu)$ and
$\tau(\Bd)$ measurements which did not contribute to any of the ratio
measurements.

The following sources of correlated (within experiment/machine)
systematic uncertainties have been considered:
\begin{itemize}
\item for SLC/LEP measurements -- \particle{D^{**}} branching ratio uncertainties~\cite{Abbaneo:2000ej_mod,*Abbaneo:2001bv_mod_cont},
momentum estimation of \b mesons from \particle{Z^0} decays
(\b-quark fragmentation parameter $\langle X_E \rangle = 0.702 \pm 0.008$~\cite{Abbaneo:2000ej_mod,*Abbaneo:2001bv_mod_cont}),
\Bs and \b baryon lifetimes (see \Secss{taubs}{taulb}),
and \b-hadron fractions at high energy (see \Table{fractions}); 
\item for \babar measurements -- alignment, $z$ scale, PEP-II boost,
sample composition (where applicable);
\item for \dzero and CDF Run~II measurements -- alignment (separately
within each experiment).
\end{itemize}
The resultant averages are:
\begin{eqnarray}
\tau(\Bd) & = & \hfagTAUBD \,, \\
\tau(\Bu) & = & \hfagTAUBU \,, \\
\tau(\Bu)/\tau(\Bd) & = & \hfagRTAUBU \,.
\end{eqnarray}
%
%
%

\mysubsubsection{\Bs lifetime}
\labs{taubs}

Similar to the kaon system, neutral \B mesons contain
short- and long-lived components, since the
light (L) and heavy  (H)
eigenstates, $\B_{\rm L}$ and $\B_{\rm H}$, differ not only
in their masses, but also in their widths 
with $\Delta\Gamma = \Gamma_{\rm L} - \Gamma_{\rm H}$. 
In the case of the \Bs system, $\DGs$ can
be particularly large. The current theoretical
prediction in the Standard Model for
the fractional width difference is
$\DGs = 0.096 \pm 0.039$~\cite{Lenz:2006hd,Beneke:1998sy},
where $\Gs = (\Gamma_{\rm L} + \Gamma_{\rm H})/2$.
Specific measurements of \DGs and \Gs are explained
in \Sec{DGs}, but the result for
\Gs is quoted here.

Neglecting \CP violation in $\Bs-\Bsbar$ mixing, 
which is expected to be small~\cite{Lenz:2006hd,Beneke:1998sy}, the
\Bs mass eigenstates are also \CP eigenstates. In
the Standard Model assuming no \CP violation in
the \Bs system,
$\Gamma_{\rm L}$ is the width of
the \CP-even state and
$\Gamma_{\rm H}$ the width of
the \CP-odd state.
Final states can be decomposed into
\CP-even and \CP-odd components, each with a different
lifetime.

In view of a possibly substantial width difference,
and the fact that various
decay channels will have different proportions of 
the $\B_{\rm L}$ and $\B_{\rm H}$ eigenstates,
the straight average of all available 
\Bs lifetime measurements
is rather ill-defined.  Therefore,
the \Bs lifetime measurements are broken down into
four categories and averaged separately.

\begin{itemize}
\item 
{\bf\em Flavor-specific decays}, such as semileptonic
$\particle{B_s} \to \particle{D_s \ell \nu}$
or $\particle{B_s} \to \particle {D_s \pi}$, will
have equal 
fractions of $\B_{\rm L}$ and $\B_{\rm H}$ at time
zero, where
$\tau_{\rm L} = 1/\Gamma_{\rm L}$ 
is expected to be the shorter-lived component and
$\tau_{\rm H} = 1/\Gamma_{\rm H}$ 
expected
to be the longer-lived component.  A superposition
of two exponentials thus results with decay
widths $\Gs \pm \DGs /2$.
Fitting to a single exponential one obtains a
measure of the flavor-specific 
lifetime~\cite{Hartkorn:1999ga}:
\begin{equation}
\tau(\Bs)_{\rm fs} = \frac{1}{\Gs}
\frac{{1+\left(\frac{\DGs}{2\Gs}\right)^2}}{{1-\left(\frac{\DGs}{2\Gs}\right)^2}
}.
\end{equation}
As given in \Table{lifebs}, the flavor-specific 
\Bs lifetime world average is:
\begin{equation}
\tau(\Bs)_{\rm fs} = \hfagTAUBSSL \,.
\labe{fslife_const2}
\end{equation}
This world average will be used later in \Sec{DGs} in combination
with other measurements to find
$\bar{\tau}(\Bs) = 1/\Gs$ and $\DGs$.

The following correlated systematic errors were considered:
average \B lifetime used in backgrounds,
\Bs decay multiplicity, and branching ratios used to determine 
backgrounds (\eg\ \BR{B\to D_s D}).
A knowledge of the multiplicity of \Bs decays is important for
measurements that partially reconstruct the final state such as 
\particle{\B\to D_s \mbox{$X$}} (where $X$ is not a lepton). 
The boost deduced from Monte Carlo simulation depends on the multiplicity used.
Since this is not well known, the multiplicity in the simulation is
varied and this range of values observed is taken to be a systematic.
Similarly not all the branching ratios for the potential background
processes are measured. Where they are available, the PDG values are
used for the error estimate. Where no measurements are available
estimates can usually be made by using measured branching ratios of
related processes and using some reasonable extrapolation.
\end{itemize}



\begin{table}[tb]
\caption{Measurements of the \Bs lifetime obtained from simple exponential fits,
without attempting to separate the \CP-even and \CP-odd components.}
\labt{lifebs}
\begin{center}
\begin{tabular}{lcccl} \hline
Experiment & Method           & Data set & $\tau(\Bs)$ (ps)               & Ref. \\
\hline
ALEPH  & \particle{D_s \ell}  & 91--95 & $1.54^{+0.14}_{-0.13}\pm 0.04$   & \cite{Buskulic:1996ei}          \\
CDF1   & \particle{D_s \ell}  & 92--96 & $1.36\pm 0.09 ^{+0.06}_{-0.05}$  & \cite{Abe:1998cj}           \\
DELPHI & \particle{D_s \ell}  & 91--95 & $1.42^{+0.14}_{-0.13}\pm 0.03$   & \cite{Abreu:2000sh}          \\
OPAL   & \particle{D_s \ell}  & 90--95 & $1.50^{+0.16}_{-0.15}\pm 0.04$   & \cite{Ackerstaff:1997qi}  \\
\dzero & \particle{D_s \mu}  & 02--04 & $1.398 \pm 0.044 ^{+0.028}_{-0.025}   $   & \cite{Abazov:2006cb}       \\ 
CDF2   & \particle{D_s \pi (X)} 
                              & 02--06 & $1.518 \pm 0.041 \pm 0.027     $   & \cite{CDFnote9203:2008,*CDFnote9203:2008_cont}$^p$      \\
CDF2   & \particle{D_s \ell}  & 02--04 & $1.381 \pm 0.055 ^{+0.052}_{-0.046} $ &
\cite{CDFnote7757:2005}$^p$ \\
\end{tabular}
\end{center}
\end{table}

\begin{itemize}
\item
{\bf\em \boldmath $\Bs\to\Ds X$ decays}.
Included in \Table{lifebs} are measurements
of lifetimes using samples of \particle{\Bs} decays to
\particle{D_s} plus
hadrons, and hence into a less known mixture
of \CP-states.  A lifetime
weighted this way can still be a useful input
for analyses examining such an inclusive sample.
These are separated in \Table{lifebs} and combined
with the semileptonic lifetime to obtain:
\begin{equation}
\tau(\Bs)_{\particle{D_s {\rm X}}} = \hfagTAUBS \,.
\end{equation}

\item
{\bf\em Fully exclusive 
{\boldmath \Bs $\to J/\psi \phi$ \unboldmath}decays}
are expected to be
dominated by the \CP-even state and its lifetime.
First measurements of the \CP mix for this decay mode
are outlined in \Sec{DGs}.
CDF and \dzero measurements 
based on simple exponential fits of the \particle{\Bs\to J/\psi\phi} 
lifetime distribution 
are combined into an average
given in \Table{lifebs}.  There are no correlations
between the measurements for this fully exclusive
channel, and the world average for this 
specific decay is:
\begin{equation}
\tau(\Bs)_{\particle{J/\psi \phi}} = \hfagTAUBSJF \,.
\end{equation}
A caveat is that different experimental acceptances
will likely lead to different admixtures of the 
\CP-even and \CP-odd states, and fits to a single
exponential may result in inherently different 
measurements of these quantities.

\item
{\bf\em Decays to (almost) pure \boldmath\CP-even eigenstates}, such as 
$\Bs \to K^+ K^-$ and $\Bs \to D_s^{(*)+}D_s^{(*)+}$ decays
which are expected to be \CP even to within 5\%, and hence 
allow the measurement of the lifetime of the ``light''
mass eigenstate $\tau_L = 1/\Gamma_L$. 
ALEPH has measured $1.27 \pm 0.33 \pm 0.08$~ps with $\Bs \to D_s^{(*)+}D_s^{(*)+}$ decays~\cite{Barate:2000kd},
while CDF has measured $1.53 \pm 0.18 \pm 0.02$~ps with $\Bs \to K^+ K^-$ in Run~II~\cite{Tonelli:2006np}.
The average of these two measurements is:
\begin{equation}
\tau_L = 1/\Gamma_L = \tau(\Bs\to\mbox{\CP even}) = \hfagTAUBSSHORT \,.
\labe{tau_CPeven}
\end{equation}
\end{itemize}

Finally, as will be shown in \Sec{DGs}, measurements
of $\DGs$, including separation into
\CP-even and \CP-odd components, give\footnote{%
A recent CDF result, 
$1/\Gs = 1.530 \pm 0.025 \pm 0.012$~\cite{CDFnote10206:2010},
has not yet been included in this average.
}
\begin{equation}
\bar{\tau}(\Bs) = 1/\Gs = \hfagTAUBSMEAN \,,
\end{equation}
and when combined with the flavor-specific 
lifetime
measurements:
\begin{equation}
\bar{\tau}(\Bs) = 1/\Gs = \hfagTAUBSMEANCON \,.
\end{equation}

\mysubsubsection{\Bc lifetime}
\labs{taubc}

There are currently three measurements of the lifetime of the \Bc meson
from CDF~\cite{Abe:1998wi,CDFnote9294:2008,*Abulencia:2006zu_mod_cont} and \dzero~\cite{Abazov:2008rba} using the semileptonic decay
mode \particle{\Bc \to J/\psi \ell} and fitting
simultaneously to the mass and lifetime using the vertex formed
with the leptons from the decay of the \particle{J/\psi} and
the third lepton. Correction factors
to estimate the boost due to the missing neutrino are used.
In the analysis of the CDF Run~I data~\cite{Abe:1998wi},
a mass value of 
$6.40 \pm 0.39 \pm 0.13$~GeV/$c^2$ 
is found by fitting
to the tri-lepton invariant mass spectrum. 
In the CDF and \dzero Run~II results~\cite{CDFnote9294:2008,*Abulencia:2006zu_mod_cont,Abazov:2008rba}, 
the \Bc mass is assumed to be 
$6285.7 \pm 5.3 \pm 1.2$~MeV/$c^2$, taken from a 
CDF result~\cite{Abulencia:2005usa}. 
These mass measurements
are consistent within uncertainties, and also consistent with the
most recent precision determination from CDF of 
$6275.6 \pm 2.9 \pm 2.5$~MeV/$c^2$~\cite{Aaltonen:2007gv}.
Correlated systematic errors include the impact
of the uncertainty of the \Bc $p_T$ spectrum on the correction
factors, the level of feed-down from $\psi(2S)$, 
MC modeling of the decay model varying from phase space
to the ISGW model, and mass variations.
Values of the \particle{\Bc} lifetime are given
in \Table{lifebc} and the world average is
determined to be:
\begin{equation}
\tau(\Bc) = \hfagTAUBC \,.
\end{equation}

\begin{table}[tb]
\caption{Measurements of the \Bc lifetime.}
\labt{lifebc}
\begin{center}
\begin{tabular}{lcccl} \hline
Experiment & Method                    & Data set  & $\tau(\Bc)$ (ps)
      & Ref.\\   \hline
CDF1       & \particle{J/\psi \ell} & 92--95  & $0.46^{+0.18}_{-0.16} \pm
 0.03$   & \cite{Abe:1998wi}  \\ 
CDF2       & \particle{J/\psi \ell} & 02--06  & $0.475^{+0.053}_{-0.049} \pm 0.018$   & \cite{CDFnote9294:2008,*Abulencia:2006zu_mod_cont}$^p$ \\
 \dzero & \particle{J/\psi \mu} & 02--06  & $0.448^{+0.038}_{-0.036} \pm 0.032$
   & \cite{Abazov:2008rba}  \\ \hline
  \multicolumn{2}{l}{Average} &   &  \hfagTAUBCnounit
                 &    \\   \hline
\multicolumn{5}{l}{$^p$ \footnotesize Preliminary.}
\end{tabular}
\end{center}
\end{table}

\mysubsubsection{\Lb and \b-baryon lifetimes}
\labs{taulb}

The first measurements of \b-baryon lifetimes
originate from two classes of partially reconstructed decays.
In the first class, decays with an exclusively 
reconstructed \Lc baryon
and a lepton of opposite charge are used. These products are
more likely to occur in the decay of \Lb baryons.
In the second class, more inclusive final states with a baryon
(\particle{p}, \particle{\bar{p}}, $\Lambda$, or $\bar{\Lambda}$) 
and a lepton have been used, and these final states can generally
arise from any \b baryon.  With the large \b-hadron samples available
at the Tevatron, the most precise measurements of \b-baryons now
come from fully reconstructed exclusive decays.

The following sources of correlated systematic uncertainties have 
been considered:
experimental time resolution within a given experiment, \b-quark
fragmentation distribution into weakly decaying \b baryons,
\Lb polarization, decay model,
and evaluation of the \b-baryon purity in the selected event samples.
In computing the averages
the central values of the masses are scaled to 
$M(\Lb) = 5620 \pm 2\MeVcc$~\cite{Acosta:2005mq} and
$M(\mbox{\b-baryon}) = 5670 \pm 100\MeVcc$.

For the semi-inclusive lifetime measurements, 
the meaning of decay model
systematic uncertainties
and the correlation of these uncertainties between measurements
are not always clear.
Uncertainties related to the decay model are dominated by
assumptions on the fraction of $n$-body semileptonic decays.
To be conservative it is assumed
that these are 100\%  correlated whenever given as an error.
DELPHI varies the fraction of 4-body decays from 0.0 to 0.3. 
In computing the average, the DELPHI
result is corrected to a value of  $0.2 \pm 0.2$ for this fraction.

Furthermore, in computing the average,
the semileptonic decay results from LEP are corrected for a polarization of 
$-0.45^{+0.19}_{-0.17}$~\cite{Abbaneo:2000ej_mod,*Abbaneo:2001bv_mod_cont} and  a 
\Lb fragmentation parameter
$\langle X_E \rangle =0.70\pm 0.03$~\cite{Buskulic:1995mf}.




\begin{table}[t]
\caption{Measurements of the \b-baryon lifetimes.
}
\labt{lifelb}
\begin{center}
\begin{tabular}{lcccl} 
\hline
Experiment&Method                &Data set& Lifetime (ps) & Ref. \\\hline
ALEPH  &$\Lc\ell$             & 91--95 &$1.18^{+0.13}_{-0.12} \pm 0.03$ & \cite{Barate:1997if}$^a$\\
ALEPH  &$\Lambda\ell^-\ell^+$ & 91--95 &$1.30^{+0.26}_{-0.21} \pm 0.04$ & \cite{Barate:1997if}$^a$\\
CDF1   &$\Lc\ell$             & 91--95 &$1.32 \pm 0.15        \pm 0.07$ & \cite{Abe:1996df}\\
CDF2   &$\Lc\pi$              & 02--06 &$1.401 \pm 0.046 \pm 0.035$ & \cite{Aaltonen:2009zn} \\
CDF2   &$J/\psi \Lambda$      & 02--09 &$1.537 \pm 0.045 \pm 0.014$ & \cite{CDFnote10071:2010,*Abulencia:2006dr_mod_cont,*Acosta:2004gt_mod_cont}$^p$ \\
\dzero &$J/\psi \Lambda$      & 02--06 &$1.218^{+0.130}_{-0.115} \pm 0.042$ & \cite{Abazov:2007sf,*Abazov:2004bn_mod_cont}$^b$ \\
\dzero &$\Lc\mu$              & 02--06 &$1.290^{+0.119+0.087}_{-0.110-0.091}$ & \cite{Abazov_mod:2007tha}$^b$ \\
\end{tabular}
\end{center}
\end{table}

Inputs to the averages are given in \Table{lifelb}.
Note that the CDF $\Lambda_b \to J/\psi \Lambda$
lifetime result~\cite{CDFnote10071:2010,*Abulencia:2006dr_mod_cont,*Acosta:2004gt_mod_cont} is
$\hfagNSIGMATAULBCDFTWO\sigma$
larger than the world average computed excluding this result. 
It is nonetheless combined with the rest 
without adjustment of input errors.
The world average lifetime of \b baryons is then:
\begin{equation}
\langle\tau(\mbox{\b-baryon})\rangle = \hfagTAUBB \,.
\end{equation}
Keeping only \particle{\Lambda^{\pm}_c \ell^{\mp}}, 
$\Lambda \ell^- \ell^+$, and fully exclusive
final states, as representative of
the \Lb baryon, the following lifetime is obtained:
\begin{equation}
\tau(\Lb) = \hfagTAULB \,. 
\end{equation}

Averaging the measurements based on the
$\Xi^{\mp} \ell^{\mp}$~\cite{Buskulic:1996sm,Abdallah:2005cw,Abreu:1995kt} 
and $J/\psi\Xi^{\mp}$~\cite{Aaltonen:2009ny}
final states gives
a lifetime value for a sample of events
containing $\Xib^0$ and $\Xib^-$ baryons:
\begin{equation}
\langle\tau(\Xib)\rangle = \hfagTAUXB \,.
\end{equation}
Recent (and first) measurements of fully reconstructed 
$\Xibd \to J/\psi\Xi^-$ and $\Omegab \to J/\psi\Omega^-$
baryons yield~\cite{Aaltonen:2009ny}
\begin{eqnarray}
\tau(\Xibd) &=& \hfagTAUXBD \,, \\
\tau(\Omegab) &=& \hfagTAUOB \,. 
\end{eqnarray}

\mysubsubsection{Summary and comparison with theoretical predictions}
\labs{lifesummary}

Averages of lifetimes of specific \b-hadron species are collected
in \Table{sumlife}.
\begin{table}[t]
\caption{Summary of lifetimes of different \b-hadron species.}
\labt{sumlife}
\begin{center}
\begin{tabular}{lc} \hline
\b-hadron species & Measured lifetime \\ \hline
\Bu                         & \hfagTAUBU   \\
\Bd                         & \hfagTAUBD   \\
\Bs ($\to$ flavor specific) & \hfagTAUBSSL \\
\Bs ($\to J/\psi\phi$)      & \hfagTAUBSJF \\
\Bs ($1/\Gs$)               & \hfagTAUBSMEANCON \\
\Bc                         & \hfagTAUBC   \\ 
\Lb                         & \hfagTAULB   \\
\Xib mixture                & \hfagTAUXB   \\
\b-baryon mixture           & \hfagTAUBB   \\
\b-hadron mixture           & \hfagTAUB    \\
\hline
\end{tabular}
\end{center}
\caption{Measured ratios of \b-hadron lifetimes relative to
the \Bd lifetime and ranges predicted
by theory~\cite{Tarantino:2003qw,*Gabbiani:2003pq,Gabbiani:2004tp}.}
\labt{liferatio}
%
%
%
\begin{center}
\begin{tabular}{lcc} \hline
Lifetime ratio & Measured value & Predicted range \\ \hline
$\tau(\Bu)/\tau(\Bd)$ & \hfagRTAUBU & 1.04 -- 1.08 \\
$\bar{\tau}(\Bs)/\tau(\Bd)^a$ & \hfagRTAUBSMEANCON & 0.99 -- 1.01 \\
$\tau(\Lb)/\tau(\Bd)$ & \hfagRTAULB & 0.86 -- 0.95    \\
$\tau(\mbox{\b-baryon})/\tau(\Bd)$  & \hfagRTAUBB & 0.86 -- 0.95 \\
\hline
\multicolumn{3}{l}{$^a$ \footnotesize 
Using $\bar{\tau}(\Bs) = 1/\Gs = 2/(\Gamma_{\rm L} + \Gamma_{\rm H})$.
}
\end{tabular}
\end{center}
\end{table}
As described in \Sec{lifetimes},
Heavy Quark Effective Theory
can be employed to explain the hierarchy of
$\tau(\Bc) \ll \tau(\Lb) < \bar{\tau}(\Bs) \approx \tau(\Bd) < \tau(\Bu)$,
and used to predict the ratios between lifetimes.
Typical predictions are compared to the measured 
lifetime ratios in \Table{liferatio}.
A recent prediction of the ratio between the \Bu and \Bd lifetimes,
is $1.06 \pm 0.02$~\cite{Tarantino:2003qw,*Gabbiani:2003pq}, in good agreement with experiment. 


The total widths of the \Bs and \Bd mesons
are expected to be very close and differ by at most 
1\%~\cite{Beneke:1996gn,*Keum:1998fd,Gabbiani:2004tp}.
However, the experimental ratio $\bar{\tau}(\Bs)/\tau(\Bd)$,
where $\bar{\tau}(\Bs)=1/\Gs$ is obtained from \DGs and 
flavour-specific lifetime measurements, appears to be 
smaller than 1 by 
\hfagONEMINUSRTAUBSMEANCONpercent, 
at deviation with respect to the prediction. 

The ratio $\tau(\Lb)/\tau(\Bd)$ has particularly
been the source of theoretical
scrutiny since earlier calculations~\cite{Shifman:1986mx,*Chay:1990da,*Bigi:1992su,*Bigi:1992su_erratum,Voloshin:1999pz,*Guberina:1999bw,*Neubert:1996we}
predicted a value greater than 0.90, almost two sigma higher
than the world average at the time. 
Many predictions cluster around a most likely central value
of 0.94~\cite{Uraltsev:1996ta,*Pirjol:1998ur,*Colangelo:1996ta,*DiPierro:1999tb}.
More recent calculations
of this ratio that include higher-order effects predict a
lower ratio between the
\Lb and \Bd lifetimes~\cite{Tarantino:2003qw,*Gabbiani:2003pq,Gabbiani:2004tp}
and reduce this difference.
References~\cite{Tarantino:2003qw,*Gabbiani:2003pq,Gabbiani:2004tp} present probability density functions
of their predictions with variation of theoretical inputs, and the
indicated ranges in \Table{liferatio}
are the RMS of the distributions from the most probable values.
Note that in contrast to the $B$ mesons, complete NLO QCD
corrections and
fully reliable lattice
determinations of the matrix elements for $\Lb$ are not
yet available.
Again, the CDF measurement of the $\Lambda_b$ lifetime
in the exclusive decay mode $J/\psi \Lambda$~\cite{CDFnote10071:2010,*Abulencia:2006dr_mod_cont,*Acosta:2004gt_mod_cont} is significantly 
higher than the world average before inclusion, with a ratio
to the $\tau(\Bd)$ world average of 
$\tau(\Lb)/\tau(\Bd) = 1.012 \pm 0.031$, 
%
resulting in continued interest in lifetimes of $b$ baryons.



\mysubsection{Neutral \B-meson mixing}
\labs{mixing}

The $\Bd-\Bdbar$ and $\Bs-\Bsbar$ systems
both exhibit the phenomenon of particle-antiparticle mixing. For each of them, 
there are two mass eigenstates which are linear combinations of the two flavour states,
\B and $\bar{\B}$. 
The heaviest (lightest) of the these mass states is denoted
$\B_{\rm H}$ ($\B_{\rm L}$),
with mass $m_{\rm H}$ ($m_{\rm L}$)
and total decay width $\Gamma_{\rm H}$ ($\Gamma_{\rm L}$). We define
\begin{eqnarray}
\Delta m = m_{\rm H} - m_{\rm L} \,, &~~~~&  x = \Delta m/\Gamma \,, \labe{dm} \\
\Delta \Gamma \, = \Gamma_{\rm L} - \Gamma_{\rm H} \,, ~ &~~~~&  y= \Delta\Gamma/(2\Gamma) \,, \labe{dg}
\end{eqnarray}
where 
$\Gamma = (\Gamma_{\rm H} + \Gamma_{\rm L})/2 =1/\bar{\tau}(\B)$ 
is the average decay width.
$\Delta m$ is positive by definition, and 
$\Delta \Gamma$ is expected to be positive within
the Standard Model.\footnote{For reason of symmetry in 
\Eqss{dm}{dg}, $\Delta \Gamma$ is sometimes defined with 
the opposite sign. The definition adopted here, \ie\
\Eq{dg}, is the one used by most experimentalists and many
phenomenologists in \B physics.}

There are four different time-dependent probabilities describing the 
case of a neutral \B meson produced 
as a flavour state and decaying to a flavour-specific final state.
If \CPT is conserved (which  
will be assumed throughout), they can be written as 
\begin{equation}
\left\{
\begin{array}{rcl}
{\cal P}(\B\to\B) & = &  \frac{e^{-\Gamma t}}{2} 
\left[ \cosh\!\left(\frac{\Delta\Gamma}{2}t\right) + \cos\!\left(\Delta m t\right)\right]  \\
{\cal P}(\B\to\bar{\B}) & = &  \frac{e^{-\Gamma t}}{2} 
\left[ \cosh\!\left(\frac{\Delta\Gamma}{2}t\right) - \cos\!\left(\Delta m t\right)\right] 
\left|\frac{q}{p}\right|^2 \\
{\cal P}(\bar{\B}\to\B) & = &  \frac{e^{-\Gamma t}}{2} 
\left[ \cosh\!\left(\frac{\Delta\Gamma}{2}t\right) - \cos\!\left(\Delta m t\right)\right] 
\left|\frac{p}{q}\right|^2 \\
{\cal P}(\bar{\B}\to\bar{\B}) & = &  \frac{e^{-\Gamma t}}{2} 
\left[ \cosh\!\left(\frac{\Delta\Gamma}{2}t\right) + \cos\!\left(\Delta m t\right)\right] 
\end{array} \right. \,,
\labe{oscillations}
\end{equation}
where $t$ is the proper time of the system (\ie\ the time interval between the production 
and the decay in the rest frame of the \B meson). 
At the \B factories, only the proper-time difference $\Delta t$ between the decays
of the two neutral \B mesons from the \Ups can be determined, but, 
because the two \B mesons evolve coherently (keeping opposite flavours as long as
none of them has decayed), the 
above formulae remain valid 
if $t$ is replaced with $\Delta t$ and the production flavour is replaced by the flavour 
at the time of the decay of the accompanying \B meson in a flavour-specific state.
As can be seen in the above expressions,
the mixing probabilities 
depend on three mixing observables:
$\Delta m$, $\Delta\Gamma$,
and $|q/p|^2$ which signals \CP violation in the mixing if $|q/p|^2 \ne 1$.

In the next sections we review in turn the experimental knowledge
on these three parameters, separately 
for the \Bd meson (\dmd, \DGd, $|q/p|_{\particle{d}}$) 
and the \Bs meson (\dms, \DGs, $|q/p|_{\particle{s}}$). 

\mysubsubsection{\Bd mixing parameters}
\labs{qpd}
\labs{dmd}
\labs{DGd}

\subsubsubsection{\boldmath \CP violation parameter $|q/p|_{\particle{d}}$}

Evidence for \CP violation in \Bd mixing
has been searched for,
both with flavor-specific and inclusive \Bd decays, 
in samples where the initial 
flavor state is tagged. In the case of semileptonic 
(or other flavor-specific) decays, 
where the final state tag is 
also available, the following asymmetry
\begin{equation} 
\ASLd = \frac{
N(\hbox{\Bdbar}(t) \to \ell^+      \nu_{\ell} X) -
N(\hbox{\Bd}(t)    \to \ell^- \bar{\nu}_{\ell} X) }{
N(\hbox{\Bdbar}(t) \to \ell^+      \nu_{\ell} X) +
N(\hbox{\Bd}(t)    \to \ell^- \bar{\nu}_{\ell} X) } 
= \frac{|p/q|_{\particle{d}}^2 - |q/p|_{\particle{d}}^2}%
{|p/q|_{\particle{d}}^2 + |q/p|_{\particle{d}}^2}
\labe{ASL}
\end{equation} 
has been measured, either in time-integrated analyses at 
CLEO~\cite{Bartelt:1993cf,Behrens:2000qu,Jaffe:2001hz},
CDF~\cite{Abe:1996zt,CDFnote9015:2007} and \dzero~\cite{Abazov:2006qw},
or in time-dependent analyses at 
OPAL~\cite{Ackerstaff:1997vd}, ALEPH~\cite{Barate:2000uk}, 
\babar~\cite{Aubert:2003hd,*Aubert:2004xga_mod_cont,Aubert:2002mn,Aubert:2006nf,Aubert:2006sa}
and \belle~\cite{Nakano:2005jb}.
In the inclusive case, also investigated and published
at ALEPH~\cite{Barate:2000uk} and OPAL~\cite{Abbiendi:1998av},
no final state tag is used, and the asymmetry~\cite{Beneke:1996hv,*Dunietz:1998av}
\begin{equation} 
\frac{
N(\hbox{\Bd}(t) \to {\rm all}) -
N(\hbox{\Bdbar}(t) \to {\rm all}) }{
N(\hbox{\Bd}(t) \to {\rm all}) +
N(\hbox{\Bdbar}(t) \to {\rm all}) } 
\simeq
\ASLd \left[ \frac{\dmd}{2\Gd} \sin(\dmd \,t) - 
\sin^2\left(\frac{\dmd \,t}{2}\right)\right] 
\labe{ASLincl}
\end{equation} 
must be measured as a function of the proper time to extract information 
on \CP violation.
In all cases asymmetries compatible with zero have been found,  
with a precision limited by the available statistics. 

A simple average of all measurements performed at 
\B factories~\cite{Behrens:2000qu,Jaffe:2001hz,Aubert:2003hd,*Aubert:2004xga_mod_cont,Aubert:2006nf,Aubert:2006sa,Nakano:2005jb}
yields 
\begin{equation}
\ASLd = \hfagASLDB 
\labe{ASLDB}
\end{equation}
or, equivalently through \Eq{ASL},
\begin{equation}
|q/p|_{\particle{d}} = \hfagQPDB \,.
\end{equation}
Analyses performed at higher energy, either at LEP or at the Tevatron,
can't separate the contributions from the 
\Bd and \Bs mesons. Under the assumption of no \CP violation in \Bs mixing, a number of 
these analyses~\cite{Abazov:2006qw,Ackerstaff:1997vd,Barate:2000uk,Abbiendi:1998av}
quote a measurement of $\ASLd$ or $|q/p|_{\particle{d}}$ for the \Bd meson. Combining 
these results, as well as that of a 
preliminary CDF analysis~\cite{CDFnote9015:2007}%
\footnote{A low-statistics analysis published by CDF using the Run I data\cite{Abe:1996zt} 
has not been included.},
with the above \B factory averages leads to 
\begin{equation}
\left.
\begin{array}{l}
\ASLd = \hfagASLDA \\
|q/p|_{\particle{d}} = \hfagQPDA 
\end{array} \right\} ~~ \mbox{if $\ASLs =0$, $|q/p|_{\particle{s}}=1$.}
\end{equation}
These results\footnote{Early analyses and (perhaps hence) the PDG use the complex
parameter $\epsilon_{\B} = (p-q)/(p+q)$; if \CP violation in the mixing in small, 
$\ASLd \cong 4 {\rm Re}(\epsilon_{\B})/(1+|\epsilon_{\B}|^2)$ and our 
current averages  
are ${\rm Re}(\epsilon_{\B})/(1+|\epsilon_{\B}|^2)=\hfagREBDB$ (\B factory
measurements only) and $\hfagREBDA$ (all measurements).}, 
summarized in \Table{qoverp},
are compatible 
with no \CP violation in the \Bd mixing, an assumption we make for the rest 
of this section.
Note that as described in \Sec{qps},
a recent update~\cite{Abazov:2010hv,*Abazov:2010hj} of the \dzero dimuon analysis gives a
measurement of the semileptonic charge asymmetry 
at the Tevatron that deviates from
the Standard Model by more than 3$\sigma$, but without a separation
of the asymmetry due to \Bd or \Bs mesons; however, 
the world average value of \ASLd measured at the $B$ factories
is used to extract \ASLs.

\begin{table}
\caption{Measurements of \CP violation in \Bd mixing and their average
in terms of both \ASLd and $|q/p|_{\particle{d}}$.
The individual results are listed as quoted in the original publications, 
or converted\addtocounter{footnote}{-1}\protect\footnotemark\
to an \ASLd value.
When two errors are quoted, the first one is statistical and the 
second one systematic. The second group of measurements, performed 
at high-energy colliders, assume no \CP violation in \Bs mixing, 
\ie\ $|q/p|_{\particle{s}}=1$.}
\labt{qoverp}
\begin{center}
\begin{tabular}{@{}rcl@{$\,\pm$}l@{$\pm$}ll@{$\,\pm$}l@{$\pm$}l@{}}
\hline
Exp.\ \& Ref. & Method & \multicolumn{3}{c}{Measured \ASLd} 
                       & \multicolumn{3}{c}{Measured $|q/p|_{\particle{d}}$} \\
\hline
CLEO   \cite{Behrens:2000qu} & partial hadronic rec. 
                             & $+0.017$ & 0.070 & 0.014 
                             & \multicolumn{3}{c}{} \\
CLEO   \cite{Jaffe:2001hz}   & dileptons 
                             & $+0.013$ & 0.050 & 0.005 
                             & \multicolumn{3}{c}{} \\
CLEO   \cite{Jaffe:2001hz}   & average of above two 
                             & $+0.014$ & 0.041 & 0.006 
                             & \multicolumn{3}{c}{} \\
\babar \cite{Aubert:2003hd,*Aubert:2004xga_mod_cont}   & full hadronic rec. 
                             & \multicolumn{3}{c}{}  
                             & 1.029 & 0.013 & 0.011  \\
\babar \cite{Aubert:2006nf}  & dileptons
                             & \multicolumn{3}{c}{}
                             & 0.9992 & 0.0027 & 0.0019 \\ 
\babar \cite{Aubert:2006sa}$^p$  & part.\ rec.\ $D^{*}\ell\nu$ 
                             & $-0.0130$ & 0.0068 & 0.0040 
                             & 1.0065 & 0.0034 & 0.0020 \\
\end{tabular}
\end{center}
\end{table}

\subsubsubsection{\boldmath Mass and decay width differences \dmd and \DGd}

\begin{table}
\caption{Time-dependent measurements included in the \dmd average.
The results obtained from multi-dimensional fits involving also 
the \Bd (and \Bu) lifetimes
as free parameter(s)~\cite{Aubert:2002sh,Aubert:2005kf,Abe:2004mz} 
have been converted into one-dimensional measurements of \dmd.
All the measurements have then been adjusted to a common set of physics
parameters before being combined. 
The CDF results from Run~II are preliminary.}
\labt{dmd}
\begin{center}
\begin{tabular}{@{}rc@{}cc@{}c@{}cc@{}c@{}c@{}}
\hline
Experiment & \multicolumn{2}{c}{Method} & \multicolumn{3}{l}{\dmd in\invps}   
                                        & \multicolumn{3}{l}{\dmd in\invps}     \\
and Ref.   &  rec. & tag                & \multicolumn{3}{l}{before adjustment} 
                                        & \multicolumn{3}{l}{after adjustment} \\
\hline
 ALEPH~\cite{Buskulic:1996qt}  & \particle{ \ell  } & \particle{ \Qjet  } & $  0.404 $ & $ \pm  0.045 $ & $ \pm  0.027 $ & & & \\
 ALEPH~\cite{Buskulic:1996qt}  & \particle{ \ell  } & \particle{ \ell  } & $  0.452 $ & $ \pm  0.039 $ & $ \pm  0.044 $ & & & \\
 ALEPH~\cite{Buskulic:1996qt}  & \multicolumn{2}{c}{above two combined} & $  0.422 $ & $ \pm  0.032 $ & $ \pm  0.026 $ & $  0.442 $ & $ \pm  0.032 $ & $ ^{+  0.020 }_{-  0.019 } $ \\
 ALEPH~\cite{Buskulic:1996qt}  & \particle{ D^*  } & \particle{ \ell,\Qjet  } & $  0.482 $ & $ \pm  0.044 $ & $ \pm  0.024 $ & $  0.482 $ & $ \pm  0.044 $ & $ \pm  0.024 $ \\
 DELPHI~\cite{Abreu:1997xq}  & \particle{ \ell  } & \particle{ \Qjet  } & $  0.493 $ & $ \pm  0.042 $ & $ \pm  0.027 $ & $  0.503 $ & $ \pm  0.042 $ & $ \pm  0.024 $ \\
 DELPHI~\cite{Abreu:1997xq}  & \particle{ \pi^*\ell  } & \particle{ \Qjet  } & $  0.499 $ & $ \pm  0.053 $ & $ \pm  0.015 $ & $  0.501 $ & $ \pm  0.053 $ & $ \pm  0.015 $ \\
\hline
\end{tabular}
\end{center}
\end{table}

Many time-dependent \Bd--\Bdbar oscillation analyses have been performed by the 
ALEPH, \babar, \belle, CDF, \dzero, DELPHI, L3 and OPAL collaborations. 
The corresponding measurements of \dmd are summarized in 
\Table{dmd},
where only the most recent results
are listed (\ie\ measurements superseded by more recent ones have been omitted). 
Although a variety of different techniques have been used, the 
individual \dmd
results obtained at high-energy colliders have remarkably similar precision.
Their average is compatible with the recent and more precise measurements 
from the asymmetric \B factories.
The systematic uncertainties are not negligible; 
they are often dominated by sample composition, mistag probability,
or \b-hadron lifetime contributions.
Before being combined, the measurements are adjusted on the basis of a 
common set of input values, including the averages of the 
\b-hadron fractions and lifetimes given in this report 
(see \Secss{fractions}{lifetimes}).
Some measurements are statistically correlated. 
Systematic correlations arise both from common physics sources 
(fractions, lifetimes, branching ratios of \b hadrons), and from purely 
experimental or algorithmic effects (efficiency, resolution, flavour tagging, 
background description). Combining all published measurements
listed in \Table{dmd}
and accounting for all identified correlations
as described in Ref.~\cite{Abbaneo:2000ej_mod,*Abbaneo:2001bv_mod_cont} yields $\dmd = \hfagDMDWfull$.

On the other hand, ARGUS and CLEO have published 
measurements of the time-integrated mixing probability 
\chid~\cite{Albrecht:1992yd,*Albrecht:1993gr_cont,Bartelt:1993cf,Behrens:2000qu}, 
which average to $\chid =\hfagCHIDU$.
Following \Ref{Behrens:2000qu}, 
the width difference \DGd could 
in principle be extracted from the
measured value of $\Gd=1/\tau(\Bd)$ and the above averages for 
\dmd and \chid 
(provided that \DGd has a negligible impact on 
the \dmd $\tau(\Bd)$ analyses that have assumed $\DGd=0$), 
using the relation
\begin{equation}
\chid = \frac{\xd^2+\yd^2}{2(\xd^2+1)} ~~~ \mbox{with} ~~ \xd=\frac{\dmd}{\Gd} 
~~~ \mbox{and} ~~ \yd=\frac{\DGd}{2\Gd} \,.
\labe{chid_definition}
\end{equation}
However, direct time-dependent studies provide much stronger constraints: 
$|\DGd|/\Gd < 18\%$ at \CL{95} from DELPHI~\cite{Abdallah:2002mr},
and $-6.8\% < {\rm sign}({\rm Re} \lambda_{\CP}) \DGGd < 8.4\%$
at \CL{90} from \babar~\cite{Aubert:2003hd,*Aubert:2004xga_mod_cont},
where $\lambda_{\CP} = (q/p)_{\particle{d}} (\bar{A}_{\CP}/A_{\CP})$
is defined for a \CP-even final state 
(the sensitivity to the overall sign of 
${\rm sign}({\rm Re} \lambda_{\CP}) \DGGd$ comes
from the use of \Bd decays to \CP final states).
Combining these two results after adjustment to 
$1/\Gd=\tau(\Bd)=\hfagTAUBD$ yields
\begin{equation}
{\rm sign}({\rm Re} \lambda_{\CP}) \DGGd  = \hfagSDGDGD \,.
\end{equation}
The sign of ${\rm Re} \lambda_{\CP}$ is not measured,
but expected to be positive from the global fits
of the Unitarity Triangle within the Standard Model.

Assuming $\DGd=0$ 
and using $1/\Gd=\tau(\Bd)=\hfagTAUBD$,
the \dmd and \chid results are combined through \Eq{chid_definition} 
to yield the 
world average
\begin{equation} 
\dmd = \hfagDMDWU \,,
\labe{dmd}
\end{equation} 
or, equivalently,
\begin{equation} 
\xd= \hfagXDWU ~~~ \mbox{and} ~~~ \chid=\hfagCHIDWU \,.  
\labe{chid}
\end{equation}
\Figure{dmd} compares the \dmd values obtained by the different experiments.

\begin{figure}
\begin{center}
\epsfig{figure=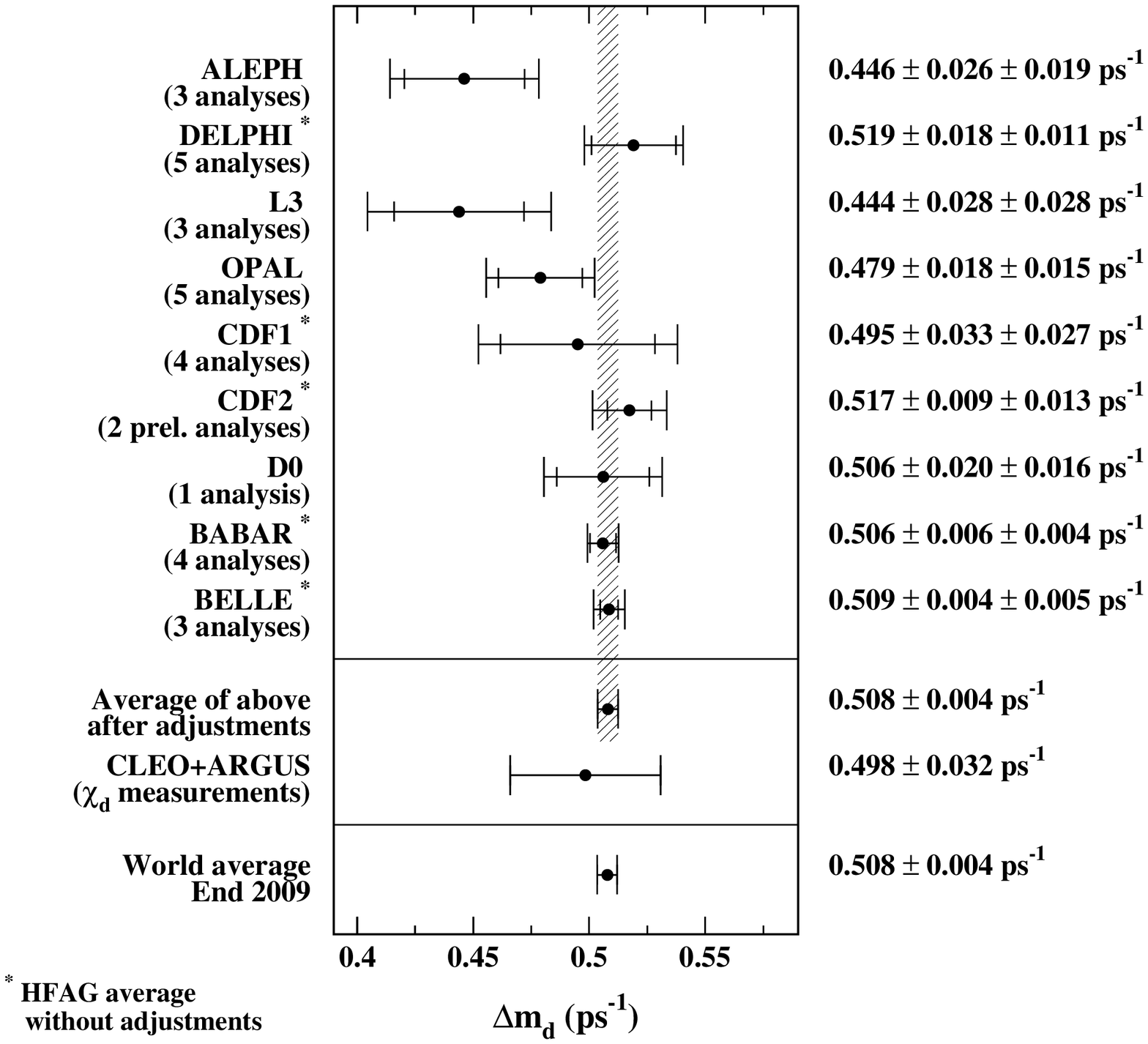,width=\textwidth}
\caption{The \Bd--\Bdbar oscillation frequency \dmd as measured by the different experiments. 
The averages quoted for ALEPH, L3 and OPAL are taken from the original publications, while the 
ones for DELPHI, CDF, \babar, and \belle have been computed from the individual results 
listed in \Table{dmd} without performing any adjustments. The time-integrated measurements 
of \chid from the symmetric \B factory experiments ARGUS and CLEO have been converted 
to a \dmd value using $\tau(\Bd)=\hfagTAUBD$. The two global averages have been obtained 
after adjustments of all the individual \dmd results of \Table{dmd} (see text).}
\labf{dmd}
\end{center}
\end{figure}

The \Bd mixing averages given in \Eqss{dmd}{chid}
and the \b-hadron fractions of \Table{fractions} have been obtained in a fully 
consistent way, taking into account the fact that the fractions are computed using 
the \chid value of \Eq{chid} and that many individual measurements of \dmd
at high energy depend on the assumed values for the \b-hadron fractions.
Furthermore, this set of averages is consistent with the lifetime averages 
of \Sec{lifetimes}.

\begin{table}
\caption{Simultaneous measurements of \dmd and $\tau(\Bd)$, and their average.
The \belle analysis also 
measures $\tau(\Bu)$ at the same time, but it is converted here into a two-dimensional measurement 
of \dmd and $\tau(\Bd)$, for an assumed value of $\tau(\Bu)$. 
The first quoted error on the measurements is statistical
and the second one systematic; in the case of adjusted measurements, the 
latter includes a contribution obtained from the variation of $\tau(\Bu)$ or 
$\tau(\Bu)/\tau(\Bd)$ in the indicated range. Units are\invps\ for \dmd
and\unit{ps} for lifetimes. 
The three different values of $\rho(\dmd,\tau(\Bd))$ correspond 
to the statistical, systematic and total correlation coefficients
between the adjusted measurements of \dmd and $\tau(\Bd)$.}
\labt{dmd2D}
\begin{center}
\begin{tabular}{@{}r@{~}c@{}c@{}c@{~}c@{}c@{}c@{~}c@{}c@{}c@{\hspace{0ex}}c@{}}
\hline
Exp.\ \& Ref.
& \multicolumn{3}{c}{Measured \dmd}   
& \multicolumn{3}{c}{Measured $\tau(\Bd)$}   
& \multicolumn{3}{c}{Measured $\tau(\Bu)$}   
&  Assumed $\tau(\Bu)$ \\
\hline
\babar \cite{Aubert:2002sh}  
      & $0.492$ & $\pm 0.018$ & $\pm 0.013$ 
      & $1.523$ & $\pm 0.024$ & $\pm 0.022$ 
      & \multicolumn{3}{c}{---}
      & $(1.083\pm 0.017)\tau(\Bd)$ \\  
\babar \cite{Aubert:2005kf}  
      & $0.511$ & $\pm 0.007$ & $^{+0.007}_{-0.006}$ 
      & $1.504$ & $\pm 0.013$ & $^{+0.018}_{-0.013}$
      & \multicolumn{3}{c}{---}
      & $1.671\pm 0.018$ \\  
\belle \cite{Abe:2004mz}  
      & $0.511$ & $\pm 0.005$ & $\pm 0.006$
      & $1.534$ & $\pm 0.008$ & $\pm 0.010$
      & $1.635$ & $\pm 0.011$ & $\pm 0.011$
      & --- \\  
\cline{2-10}
& \multicolumn{3}{c}{Adjusted \dmd}   
& \multicolumn{3}{c}{Adjusted $\tau(\Bd)$}   
& \multicolumn{3}{c}{$\rho(\dmd,\Bd)$} 
&  Assumed $\tau(\Bu)$ \\
\cline{2-10}
\babar \cite{Aubert:2002sh}  
      & $0.492$ & $\pm 0.018$ & $\pm 0.013$  
      & $1.523$ & $\pm 0.024$ & $\pm 0.022$  
      & $-0.22$ & $+0.71$ & $+0.16$ 
      & $(\hfagRTAUBUval$$\hfagRTAUBUerr)\tau(\Bd)$ \\  
\babar \cite{Aubert:2005kf} 
      & $0.512$ & $\pm 0.007$ & $\pm 0.007$  
      & $1.506$ & $\pm 0.013$ & $\pm 0.018$ 
      & $+0.01$ & $-0.85$ & $-0.48$ 
      & $\hfagTAUBUval$$\hfagTAUBUerr$ \\
\hline 
\end{tabular}
\end{center}
\end{table}
\begin{figure}
\begin{center}
\vspace{-0.5cm}
\epsfig{figure=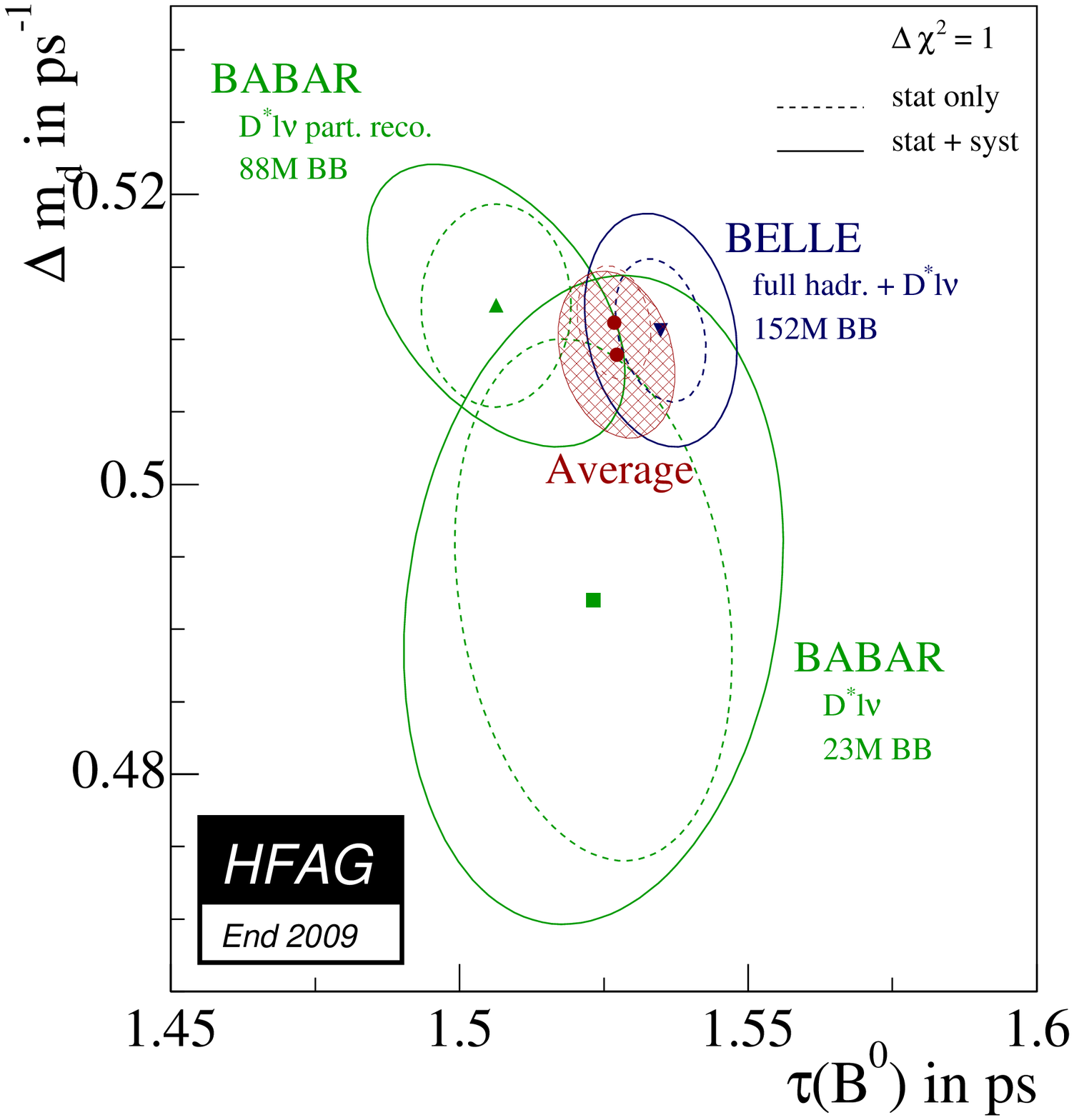,width=0.6\textwidth}
\vspace{-0.5cm}
\caption{Simultaneous measurements of
\dmd and $\tau(\Bd)$~\cite{Aubert:2002sh,Aubert:2005kf,Abe:2004mz}, 
after adjustment to a common set of parameters (see text). 
Statistical and total uncertainties are represented as dashed and
solid contours respectively.
The average of the three measurements
is indicated by a hatched ellipse.}
\labf{dmd2D}
\end{center}
\end{figure}

It should be noted that the most recent (and precise) analyses at the 
asymmetric \B factories measure \dmd
as a result of a multi-dimensional fit. 
Two \babar analyses~\cite{Aubert:2002sh,Aubert:2005kf},  
based on fully and partially reconstructed $\Bd \to D^*\ell\nu$ decays
respectively, 
extract simultaneously \dmd and $\tau(\Bd)$
while the latest \belle analysis~\cite{Abe:2004mz},  
based on fully reconstructed hadronic \Bd decays and $\Bd \to D^*\ell\nu$ decays, 
extracts simultaneously \dmd, $\tau(\Bd)$ and $\tau(\Bu)$.
The measurements of \dmd and $\tau(\Bd)$ of these three analyses 
are displayed in \Table{dmd2D} and in \Fig{dmd2D}. Their two-dimensional average, 
taking into account all statistical and systematic correlations, and expressed
at $\tau(\Bu)=\hfagTAUBU$, is
\begin{equation}
\left.
\begin{array}{r@{}l}
\dmd = \hfagDMDTWODnounit & \invps \\
\tau(\Bd) = \hfagTAUBDTWODnounit & \ps
\end{array}
\right\}
~\mbox{with a total correlation of \hfagRHODMDTAUBD.}
\end{equation}

\mysubsubsection{\Bs mixing parameters}
\labs{qps}
\labs{DGs}
\labs{dms}

\subsubsubsection{\boldmath \CP violation parameter $|q/p|_{\particle{s}}$}

Constraints on a combination of $|q/p|_{\particle{d}}$ and 
$|q/p|_{\particle{s}}$ (or equivalently \ASLd and \ASLs) 
have been explicitly quoted by the Tevatron 
experiments, using inclusive semileptonic decays of \b hadrons:
\begin{eqnarray}
\frac{1}{4}\left(f'_{\particle{d}} \,\chid \ASLd +
                 f'_{\particle{s}} \,\chis \ASLs \right) = 
+0.0015 \pm 0.0038 \mbox{(stat)} \pm 0.0020 \mbox{(syst)}
&~~& \mbox{CDF1~\cite{Abe:1996zt}} \,, 
\labe{CDF_ASLDS} \\
\ASLb = \frac{f'_{\particle{d}}Z_{\particle{d}} \ASLd + f'_{\particle{s}}Z_{\particle{s}} \ASLs}%
{f'_{\particle{d}}Z_{\particle{d}} + f'_{\particle{s}}Z_{\particle{s}}} =
+0.0080 \pm 0.0090 \mbox{(stat)} \pm 0.0068 \mbox{(syst)}
&~~& \mbox{CDF2~\cite{CDFnote9015:2007}} \,,
\labe{CDF2_ASLDS} \\
\ASLb = 
 -0.00957 \pm 0.00251 \mbox{(stat)} \pm 0.00146 \mbox{(syst)}
&~~& \mbox{\dzero~\cite{Abazov:2010hv,*Abazov:2010hj}} \,,
\labe{CDF_Dzero_ASLDS}
\end{eqnarray}
where\footnote{In Ref.~\cite{Abazov:2007zj}, the \dzero result~\cite{Abazov:2006qw}
was reinterpreted by replacing $\chi_{\particle{s}}/\chi_{\particle{d}}$
with $Z_{\particle{s}}/Z_{\particle{d}}$.
For simplicity, and since this has anyway a negligible numerical effect on our
combined result of \Eq{ASLs}, we 
follow the same interpretation and set $\chi_{\particle{q}}=Z_{\particle{q}}/2$
in \Eqss{CDF_ASLDS}{CDF_Dzero_ASLDS}. We also set $f'_{\particle{q}}=f_{\particle{q}}$.}
$Z_{\particle{q}} = 1/(1-y_{\particle{q}}^2)-1/(1+x_{\particle{q}}^2)
= 2 \chi_{\particle{q}}/(1-y_{\particle{q}}^2)$, $q=d,s$.
The \dzero result of \Eq{CDF_Dzero_ASLDS}, obtained by measuring
the charge asymmetry of like-sign dimuons, differs by 3.2 standard
deviations from the Standard Model prediction of
\begin{eqnarray}
\mbox{\hspace{3cm}}
\ASLb({\mathrm{SM}}) = (-2.3^{+0.5}_{-0.6}) \times 10^{-4}
&~~& \mbox{\cite{Lenz:2006hd}}\,.
\end{eqnarray}
In addition a first direct determination of \ASLs
and hence $|q/p|_{\particle{s}}$ has
been obtained by \dzero by measuring the charge asymmetry of
tagged $\Bs \rightarrow D_s \mu X$ decays:
\begin{eqnarray}
\mbox{\hspace{3cm}}
\ASLs = -0.0017 \pm 0.0091 \mbox{(stat)} ^{+0.0014}_{-0.0015} \mbox{(syst)}
&~~& \mbox{\dzero~\cite{Abazov:2009wg,*Abazov:2007nw_mod_cont}}\,.
\end{eqnarray}

Given the average $\ASLd = \hfagASLDB$ of \Eq{ASLDB},
obtained from results at \B factories,
as well as other averages presented in this chapter
for the quantities appearing in 
\Eqsss{CDF_ASLDS}{CDF2_ASLDS}{CDF_Dzero_ASLDS}, 
these four results are turned into measurements of \ASLs 
(displayed in \Fig{ASLs}).
The \dzero result for \ASLb yields
\begin{eqnarray}
\mbox{\hspace{3cm}}
\ASLs = -0.00146 \pm 0.0075 
&~~& \mbox{\dzero~\cite{Abazov:2010hv,*Abazov:2010hj}}\,,
\end{eqnarray}
with an increased uncertainty due to uncertainties
in $f'_{\particle{d}}, f'_{\particle{s}}, Z_{\particle{d}},$ and
$Z_{\particle{s}}$, and does not represent evidence of \CP violation
exclusively in the \Bs system.
The four results of \Fig{ASLs} are
combined to yield
\begin{figure}
\begin{center}
\epsfig{figure=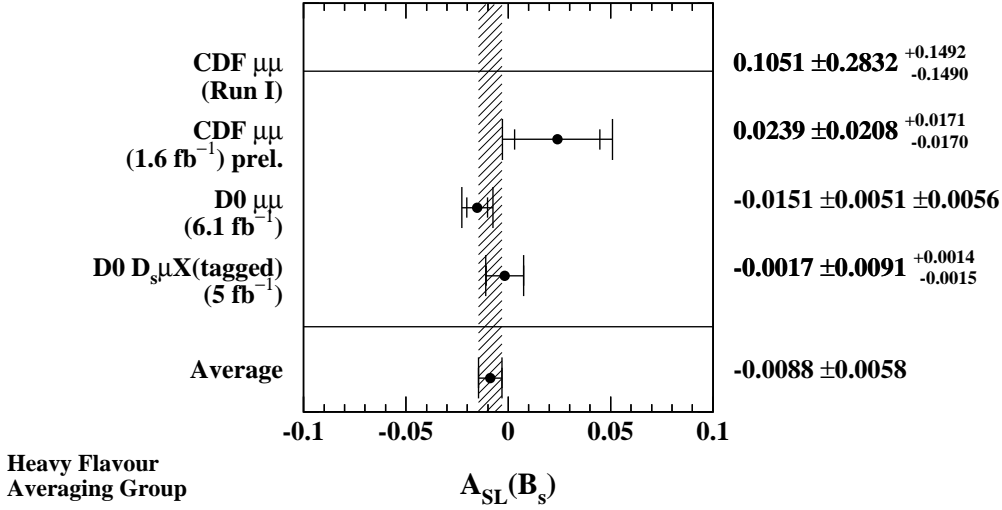,width=0.8\textwidth}
\caption{Measurements of \ASLs, derived from CDF~\cite{Abe:1996zt,CDFnote9015:2007}
and \dzero~\cite{Abazov:2006qw,Abazov:2009wg,*Abazov:2007nw_mod_cont} analyses and adjusted to the latest 
averages of \ASLd, \b-hadron fractions and mixing parameters. 
The combined value of \ASLs is also shown.}
\labf{ASLs}
\end{center}
\end{figure}
\begin{equation}
\ASLs = \hfagASLSval\hfagASLSsta\mbox{(stat)}\hfagASLSsys\mbox{(syst)} = \hfagASLS 
\labe{ASLs}
\end{equation}
or, equivalently through \Eq{ASL},
\begin{equation}
|q/p|_{\particle{s}} = \hfagQPSval\hfagQPSsta\mbox{(stat)}\hfagQPSsys\mbox{(syst)} = \hfagQPS \,.
\end{equation}
The quoted systematic errors include experimental systematics as well as the correlated dependence on external 
parameters. These results are compatible with no \CP violation in \Bs
mixing, an assumption made in almost all of the results described below.


%
%
%

\subsubsubsection{Decay width difference \DGs}



Definitions and an introduction to \DGs can also 
be found in \Sec{taubs}.
Neglecting \CP violation, the mass eigenstates are
also \CP eigenstates, with the short-lived state being
\CP-even and the long-lived one being \CP-odd.
Information on \DGs can be obtained by studying the proper time 
distribution of untagged data samples enriched in 
\Bs mesons~\cite{Hartkorn:1999ga}.
In the case of an inclusive \Bs selection~\cite{Acciarri:1998uv} or a semileptonic 
\Bs decay selection~\cite{Abreu:2000sh,Abe:1998cj,Abazov:2006cb}, 
both the short- and long-lived
components are present, and the proper time distribution is a superposition 
of two exponentials with decay constants 
$\Gs\pm\DGs/2$.
In principle, this provides sensitivity to both \Gs and 
$(\DGGs)^2$. Ignoring \DGs and fitting for 
a single exponential leads to an estimate of \Gs with a 
relative bias proportional to $(\DGGs)^2$. 
An alternative approach, which is directly sensitive to first order in \DGGs, 
is to determine the lifetime of \Bs candidates decaying to \CP
eigenstates; measurements exist for 
\particle{\Bs\to J/\psi\phi}~\cite{Abe:1997bd,CDFnote8524:2007,*CDFnote8524:2007_cont,Abazov:2004ce} and
\particle{\Bs\to D_s^{(*)+} D_s^{(*)-}}, discussed later, which are 
mostly \CP-even states~\cite{Aleksan:1993qp}.
However, later, more sophisticated,
time-dependent angular analyses of \particle{\Bs\to J/\psi\phi} 
allow the simultaneous extraction of \DGs and the \CP-even and \CP-odd 
amplitudes~\cite{CDFnote9458:2008,*Aaltonen:2007gf_mod,*Aaltonen:2007he_mod,Abazov_mod:2008fj,*Abazov:2007tx_mod_cont}.
Flavor tagging the \Bs (or $\bar{B}^0_s$)
that subsequently decays to \particle{J/\psi\phi}
allows for a more effective
extraction of the weak mixing phase as discussed later.
Both the CDF and  \dzero flavor-tagged 
\particle{\Bs\to J/\psi\phi} analyses~\cite{CDFnote9458:2008,*Aaltonen:2007gf_mod,*Aaltonen:2007he_mod,Abazov_mod:2008fj,*Abazov:2007tx_mod_cont} present results first
assuming the very small SM value of mixing-induced 
CP violation in the \Bs system
(effectively zero compared to current experimental resolution)
used in the averaging of \DGs, and then 
also allowing for large \CP violation, used for determining
an average weak mixing phase in the next subsection.

\begin{table}
\caption{Experimental constraints on \DGGs from lifetime 
and $B_s \rightarrow J/\psi \phi$ analyses,
assuming no (or very small SM) \CP violation.
The upper limits,
which have been obtained by the working group, are quoted at the \CL{95}.}
\labt{dgammat}
\begin{center}
\begin{tabular}{l|c|c|c}
\hline
Experiment & Method            & $\Delta \Gs/\Gs$ & Ref.  \\
\hline
L3         & lifetime of inclusive \b-sample              
           & $<0.67$   & \cite{Acciarri:1998uv}      \\
DELPHI     & $\Bsb\to D_s^+\ell^- \overline{\nu_{\ell}} X$, lifetime
	   & $<0.46$   & \cite{Abreu:2000sh} \\
DELPHI     & $\Bsb \to D_s^+$ hadron, lifetime
           & $<0.69$ & \cite{Abreu:2000ev}   \\
CDF1       & $\Bs \to J/\psi\phi$, lifetime
	   & $0.33^{+0.45}_{-0.42}$ & \cite{Abe:1997bd} \\ \hline
	   &                   & $ \DGs$  \\ \hline
CDF2       & $\Bs \to J/\psi\phi$, time-dependent angular analysis
           & $0.02{\pm 0.05}{\pm 0.01\invps}$ & \cite{CDFnote9458:2008,*Aaltonen:2007gf_mod,*Aaltonen:2007he_mod} \\ 
\dzero     & $\Bs \to J/\psi\phi$, time-dependent angular analysis
           & $0.14{\pm 0.07\invps}$ & \cite{Abazov_mod:2008fj,*Abazov:2007tx_mod_cont} \\
	 \hline
	 \end{tabular}
	 \end{center}
	 \end{table}

Measurements quoting \DGGs results from lifetime analyses
and \DGs results from $B^0_s \rightarrow J/\psi \phi$ analyses
under the hypothesis of no (or very small SM) \CP violation
are listed in \Table{dgammat}.  
There is significant correlation
between \DGs and $1/\Gamma_s$. In order to combine these measurements,
the two-dimensional log-likelihood for each measurement
in the $(1/\Gs,\,\DGs)$ plane is summed and the total
normalized with respect to its minimum.  The one-sigma contour (corresponding
to 0.5 units of log-likelihood greater than the minimum) and
95\% CL contour are found. 
Only the \DGs inputs 
from CDF2 and \dzero as indicated in 
\Table{dgammat} were used in the combinations below
(adding the other ones would not change the results).
CDF has very recently made a preliminary
update~\cite{CDFnote10206:2010}
to their
\particle{\Bs \rightarrow J/\psi\phi} analysis to an
integrated luminosity of 5.2~fb$^{-1}$, and assuming no \CP violation, 
find
\begin{eqnarray}
\DGs &=& 0.075 \pm 0.035 \pm 0.01 \thinspace {\mathrm{ps}}^{-1} \,, \\
\bar{\tau}(\Bs) = 1/\Gs &=& 1.530 \pm 0.025 \pm 0.012 \thinspace
{\mathrm{ps}} \,. 
\end{eqnarray}
However, this new update has yet to be included in the following
combinations.

Results of the combination are shown as the one-sigma contour
labeled ``Direct" in both plots of \Fig{DGs}.  Transformation
of variables from $(1/\Gs,\,\DGs)$ space to other pairs
of variables such as $(1/\Gs,\,\DGGs)$ and 
$(\tau_{\rm L} = 1/\Gamma_{\rm L},\,\tau_{\rm H} = 1/\Gamma_{\rm H})$
are also made.
The resulting one-sigma contour for the latter is shown in
\Fig{DGs}(b). 

\begin{figure}
\begin{center}
\epsfig{figure=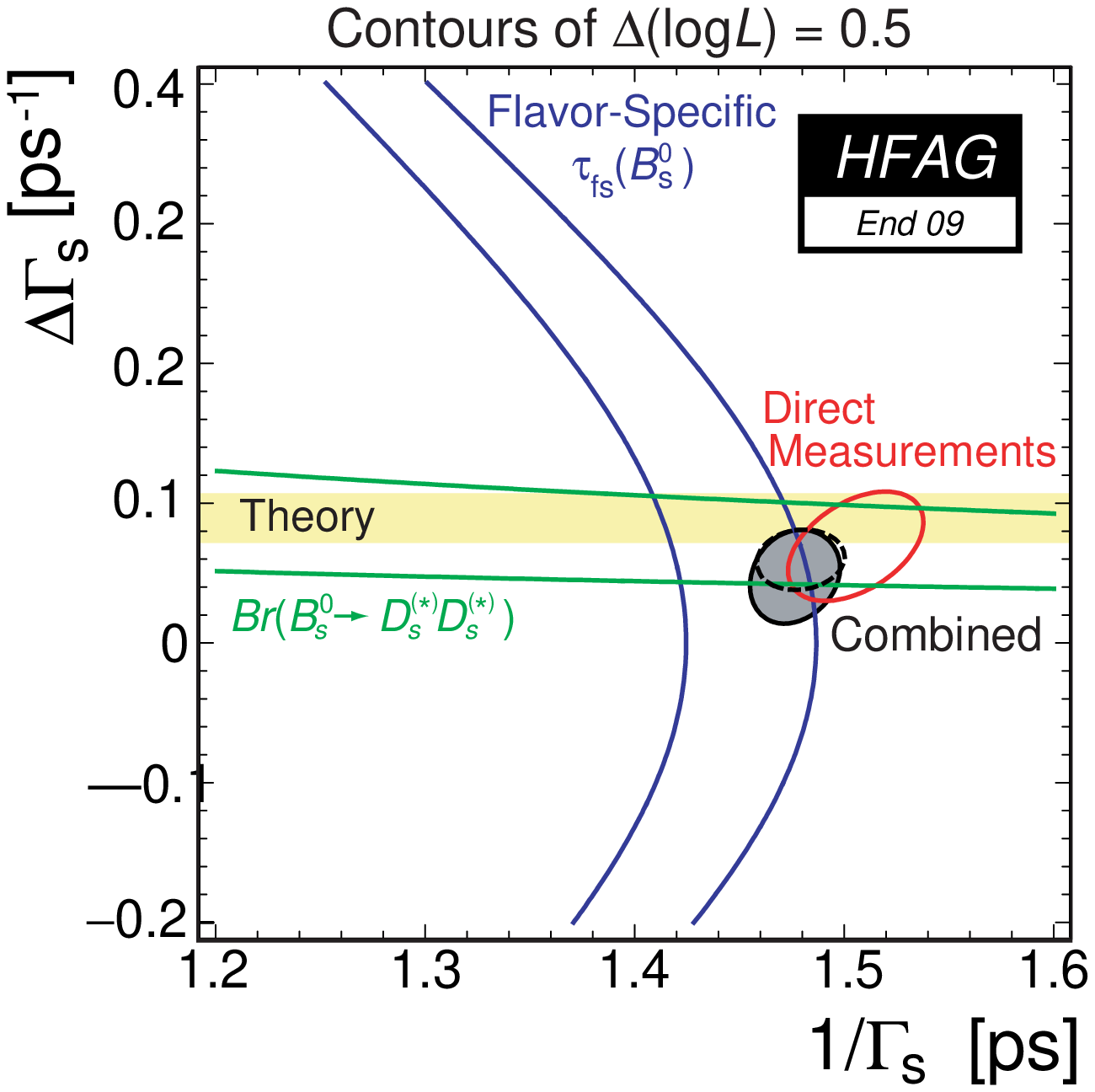,width=0.45\textwidth}
\hfill
\epsfig{figure=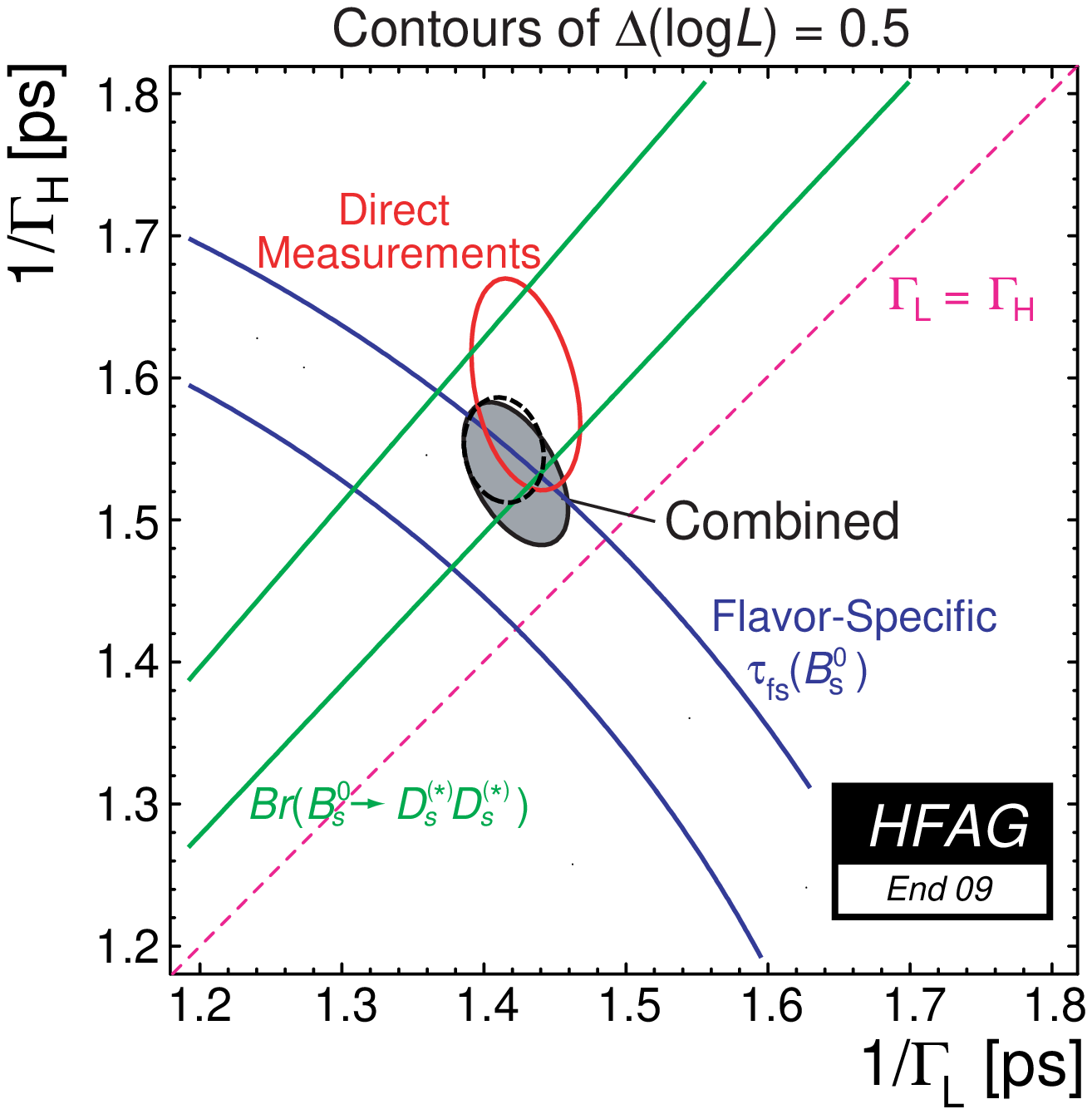,width=0.45\textwidth}
\caption{\DGs combination results with one-sigma contours
($\Delta\log\mathcal{L} = 0.5$) shown for (a) \DGs versus
$\bar{\tau}(\Bs) = 1/\Gs$  and (b)
$\tau_{\rm H} = 1/\Gamma_{\rm H}$ versus $\tau_{\rm L} = 1/\Gamma_{\rm L}$.
The red contours labeled ``Direct" are the result of the combination of
last two measurements of \Table{dgammat}, the blue bands are the one-sigma
contours due to the world average of flavor-specific 
\Bs lifetime measurements,
and the solid and dashed-outlined shaded regions result using
the combination constraints described in the
text.
In (b), the diagonal dashed line indicates 
$\Gamma_{\rm L} = \Gamma_{\rm H}$, \ie, where $\DGs = 0$.}
\labf{DGs}
\end{center}
\end{figure}




\comment{

{\em This text below taken by Donatella from previous \DGs notes \ldots}

\newcommand{\Bh}{B^{\rm heavy}_{d,s}}
\newcommand{\Bl}{B^{\rm light}_{d,s}}
\newcommand{\Mh}{m^{\rm heavy}_{d,s}}
\newcommand{\Ml}{m^{\rm light}_{d,s}}
\newcommand{\Gh}{\Gamma^{\rm heavy}_{d,s}}
\newcommand{\Gl}{\Gamma^{\rm light}_{d,s}}
\newcommand{\Gsho}{\Gamma^{\rm short}_{d,s}}
\newcommand{\Glon}{\Gamma^{\rm long}_{d,s}}
\newcommand{\G}{\Gamma_{d,s}}
\newcommand{\tb}{\tau(B^0_{d,s})}
\newcommand{\dg}{\Delta\Gamma_{d,s}}
\newcommand{\tbssemi}{\tau(B_s)_{\rm semi}}
\newcommand{\Bssh}{B^{\rm short}_s}
\newcommand{\tbsshort}{\tau(\Bssh)}

The Standard Model predicts that the \Bs and \Bd can mix before decay. 
The phenomenology of this interaction can be 
described in terms of a $2\times 2$ effective Hamiltonian matrix, 
$M - i\Gamma /2$.
This results in new states called heavy and light, 
$\Bh$  and $\Bl$, for \Bs and \Bd with masses $\Mh$ ,
$\Ml$. Also the  widths  $\Gh$ and $\Gl$ could be different.

Neglecting \CP violation, the mass eigenstates are also \CP eigenstates, the ``long''  state being 
\CP even and the short one being \CP odd.  For convenience of notation, in the following
we therefore substitute 
$\Gl \equiv \Gsho$ and $\Gh \equiv \Glon$, and 
define $\G=1/\tb=(\Glon+\Gsho)/2$ and 
$\Delta \G = \Gsho-\Glon$ which is positive.

$\dg$ is related to the off-diagonal matrix elements, which have been recently 
calculated at the NLO including NLO QCD correction~\cite{Ciuchini2}. 
The theoretical values are:
\begin{equation}
\DGGs = (7.4\pm 2.4 )\times 10^{-2} \,, \hspace{1truecm} \DGGd = (2.42 \pm 0.59 ) \times 10^{-3} \,.
\end{equation}

In the same work the ratio $\DGd/\DGs$ is evaluated since the uncertainties 
coming from higher orders of QCD and $\Lambda_{\rm QCD}/m_{\b}$ corrections cancel out:
\begin{equation}
\DGd/\DGs = (3.2\pm 0.8)\times10^{-2}
\end{equation}

Experimentally \DGs can be measured fitting the lifetime of the 
light and heavy component of the \Bs.
An alternative method is based on the measurement of the  
branching fraction \particle{\Bs\to D_s^{(*)+}D_s^{(*)-}}.
Methods based on lifetime measurements have two different approaches.
Double exponential lifetime fits to samples containing a mixture of \CP eigenstates like
inclusive or semileptonic \Bs decays or $\Bs\to D_s$-hadron have a quadratic sensitivity 
to \DGs.
Whereas the isolation of a single \CP eigenstate as $\Bs\to\phi\phi$ or 
\particle{\Bs\to J/\psi\phi} to extract the lifetime of the \CP-even or odd state have 
a linear dependence on \DGs and it is more sensitive to \DGs but tend 
to suffer from reduced statistics.
The branching fraction method, exploited by ALEPH~\cite{ALEPH-phiphi}, 
is based on several theoretical assumptions~\cite{theoBR}, and allows to have 
information on \DGs only through the branching fraction measurement:
\begin{equation}
\BR{B_s\to D_s^{(*)+}
D_s^{(*)-}} = \frac{\DGs}{\Gs\left(1+\frac{\DGs}{2\Gs}\right)} \,.
\label{eq:dg_ratio}
\end{equation}

The available results are summarized in \Table{dgammat}. 
The values of the limit on \DGGs quoted in the last column of this 
table have been obtained by the working group.

Details on how these measurements are included in the average can be found  
in the previous summaries~\cite{Pcomb}.

\begin{table}
\caption{Experimental constraints on \DGGs. The upper limits,
which have been obtained by the working group, are quoted at the \CL{95}.}
\labt{dgammat}
\begin{center}
\begin{tabular}{|l|c|c|c|} 
\hline
Experiment & Selection        & Measurement            & $\Delta \Gs/\Gs$ \\ 
\hline
L3~\cite{Acciarri:1998uv}         & inclusive \b-sample              &                               & $<0.67$         \\
DELPHI~\cite{Abreu:2000sh}     & $\Bsb\to D_s^+\ell^- \overline{\nu_{\ell}} X$ & $\tbssemi=(1.42^{+0.14}_{-0.13}\pm0.03)$~ps  & $<0.46$ \\
others~\cite{ref:others}& $\Bsb\to D_s^+\ell^-  \overline{\nu_{\ell}} X$  & $\tbssemi=(1.46\pm{0.07})$~ps & $<0.30$ \\
ALEPH~\cite{ALEPH-phiphi}      & $\Bs\to\phi\phi X$      & 
$\BR{\Bssh \to D_s^{(*)+} D_s^{(*)-}} =(23\pm10^{+19}_{-~9})\%$       & $0.26^{+0.30}_{-0.15}$ \\
ALEPH~\cite{ALEPH-phiphi}      & $\Bs\to\phi\phi X$      & $\tbsshort=(1.27\pm0.33\pm0.07)$~ps           & 
$0.45^{+0.80}_{-0.49}$ \\ 
DELPHI~\cite{Abreu:2000sh}$^a$    & $\Bsb \to D_s^+$ hadron    
&  $\tau_{\rm B^{D_s-had.}_s}=(1.53^{+0.16}_{-0.15}\pm0.07)$~ps                          & $<0.69$         \\
CDF~\cite{CDFB01} & $\Bs \to {\rm J}/\psi\phi$        
& $\tau_{\rm B^{{\rm J}/\psi \phi}_s}=(1.34^{+0.23}_{-0.19}\pm0.05)$~ps & $0.33^{+0.45}_{-0.42}$ \\ 
\hline
\multicolumn{4}{l}{$^a$ \footnotesize 
The value quoted for the measured lifetime differs
slightly from the one quoted in \Table{bs} because it} \\[-1ex]
\multicolumn{4}{l}{~~ \footnotesize 
corresponds to the present status of the analysis in which the information
on \DGs has been obtained.}
\end{tabular}
\end{center}
\end{table}

Here only a short description will be given.

L3 and DELPHI use inclusively reconstructed
\Bs and $\Bs\to \particle{D_s} \ell\nu X$ events respectively.
If those sample are fitted assuming a single exponential lifetime then,
assuming \DGGs is small, the measured lifetime is given by:
\begin{equation}
\tau(\Bs)_{\rm incl.} = \frac{1}{\Gs} \frac{1}{1-\left(\frac{\DGs}{2\Gs}\right)^2}
\quad \quad ; \quad \quad
\tau(\Bs)_{\rm semi.} = \frac{1}{\Gs} 
\frac{{1+\left(\frac{\DGs}{2\Gs}\right)^2}}{{1-\left(\frac{\DGs}{2\Gs}\right)^2}}.     
\end{equation}

The single lifetime fit is thus more sensitive to the effects of 
\DGs in the semileptonic case than in the fully inclusive case.

The same method is used for the \Bs world average lifetime
(recomputed without the DELPHI measurement) obtained
by using only the semileptonic decays and referenced in \Table{dgammat} as {\it others}.

The technique of reconstructing only decays at defined \CP
has been exploited by ALEPH, DELPHI and CDF.

ALEPH reconstructs the decay
$\Bs\to \particle{D_s^{(*)+}D_s^{(*)-}} \to \phi\phi X$
which is predominantly \CP even.
The proper time dependence of the \Bs component is a simple
exponential and the lifetime is related to \DGs via
\begin{equation}
\frac{\Delta \Gs}{\Gs}=2(\frac{1}{\Gs~\tbsshort}-1).  
\end{equation} 
 The same data have been used by ALEPH to exploit the branching fraction method.

DELPHI uses a sample of $\Bs\to D_s$-hadron,
which is expected to have an increased \CP-even component as the contribution
due to \particle{D_s^{(*)+}_s^{(*)-}} events is enhanced by
selection criteria.

CDF reconstructs \particle{\Bs\to J/\psi\phi} with
\particle{J/\psi\to\mu^+\mu^-} and \particle{\phi\to K^+K^-}
where the \CP-even component is equal to $0.84\pm 0.16$ obtained by
combining CLEO~\cite{cleo} measurement of \CP-even fraction in
\particle{\Bd\to J/\psi K^{*0}} and possible SU(3) symmetry
correction.

In order to combine all the measurements~\footnote{L3 is not 
included since the likelihood for the results
was not available} the two-dimensional log-likelihood in the ($1/\Gs$, \DGGs) 
plane is summed and normalized with respect to its minimum.
The 68\%, 95\% and \CL{99} contours of the combined negative 
log-likelihood are shown in \Fig{dgplot} (left)
The corresponding limit on \DGGs is:
\begin{eqnarray}
\DGGs & = & 0.16^{+0.15}_{-0.16}  \,, \\
\DGGs & < & 0.54~\mbox{at \CL{95}} \,. 
\end{eqnarray}

\begin{figure}
\begin{center}
\epsfig{figure=figures/osc/dg_w_notaubd_bw_2d.eps,width=\textwidth}
\epsfig{figure=figures/osc/dg_w_taubd_bw_2d.eps,width=\textwidth}
\end{center}
\caption{Top: 68\%, 95\% and \CL{99} contours of the negative log-likelihood 
distribution in the plane ($1/\Gs$, \DGGs).
Bottom: Same, but with the constraint $1/\Gs \equiv\tau_{\Bd}$} 
\labf{dgplot}
\end{figure}

\begin{figure}
\begin{center}
\epsfig{figure=figures/osc/dg_w_taubd_col_1d,width=\textwidth}
\end{center}
\caption{Probability density distribution for \DGGs after applying the constraint; 
the three shaded regions show the limits at the 68\%, 95\% and \CL{99} respectively.} 
\labf{dgprobplot}
\end{figure}

An improved limit on \DGGs can be obtained by applying the $\tau_{\Bd}=\HFAGtauBd$ constraint.
The world average \Bs lifetime is not used, as its meaning 
is not clear if $\Delta \Gs$ is non-zero.
This is well motivated theoretically, as 
the total widths of the \Bs and \Bd mesons
are expected to be 
equal within less than one percent~\cite{bigilife}, \cite{Beneke}
and \DGd is expected to be small. 
 
The two-dimensional log-likelihood obtained, after including the constraint is shown in 
\Fig{dgplot} (right). The resulting probability density distribution for \DGGs is 
shown in \Fig{dgprobplot}. The corresponding limit on \DGGs is:
\begin{eqnarray}
\DGGs & = & 0.07^{+0.09}_{-0.07} \,, \\
\DGGs & < & 0.29~~\mbox{at \CL{95}} \,.
\end{eqnarray}

} 




Numerical results of the combination of the CDF2 and \dzero inputs
of \Table{dgammat} are:
\begin{eqnarray}
\DGGs &\in& [\hfagDGSGSlow,\hfagDGSGSupp] ~ \mbox{at \CL{95}} \,, \\
\DGGs &=& \hfagDGSGS \,, \\
\DGs &\in& [\hfagDGSlow,\hfagDGSupp]\invps ~ \mbox{at \CL{95}} \,, \\
\DGs &=& \hfagDGS \,, \\
\bar{\tau}(\Bs) = 1/\Gs &=& \hfagTAUBSMEAN \,, \\
1/\Gamma_{\rm L} = \tau_{\rm short} &=& \hfagTAUBSL \,, \\
1/\Gamma_{\rm H} = \tau_{\rm long}  &=& \hfagTAUBSH \,. 
\end{eqnarray}

Flavor-specific lifetime measurements are of an equal mix
of \CP-even and \CP-odd states at time zero, and  
if a single exponential function is used in the likelihood
lifetime fit of such a sample~\cite{Hartkorn:1999ga}, 
\begin{equation}
\tau(\Bs)_{\rm fs} = \frac{1}{\Gs}
\frac{{1+\left(\frac{\DGs}{2\Gs}\right)^2}}{{1-\left(\frac{\DGs}{2\Gs}\right)^2}
} \,.
\labe{fslife_const}
\end{equation}
Using the world average flavor-specific 
lifetime of \Eq{fslife_const2} in \Sec{taubs}
the one-sigma blue bands shown in \Fig{DGs} are obtained. 
Higher-order corrections were checked to be negligible in the
combination.

When the flavor-specific lifetime measurements 
are combined with the 
CDF2 and \dzero measurements of \Table{dgammat}, the solid-outline
shaded
regions of \Fig{DGs} are obtained, with numerical results:
\begin{eqnarray}
\DGGs &\in& [\hfagDGSGSCONlow,\hfagDGSGSCONupp] ~ \mbox{at \CL{95}} \,, \\
\DGGs &=& \hfagDGSGSCON \,, \labe{DGGs_ave} \\
\DGs &\in& [\hfagDGSCONlow,\hfagDGSCONupp]\invps ~ \mbox{at \CL{95}} \,, \\
\DGs &=& \hfagDGSCON \,, \\
\bar{\tau}(\Bs) = 1/\Gs &=& \hfagTAUBSMEANCON \,, \labe{oneoverGs} \\
1/\Gamma_{\rm L} = \tau_{\rm short} &=& \hfagTAUBSLCON \,, \\
1/\Gamma_{\rm H} = \tau_{\rm long}  &=& \hfagTAUBSHCON \,. 
\end{eqnarray}
These results can
be compared with the theoretical prediction of 
$\DGs = 0.096 \pm 0.039\invps$
(or $\DGs = 0.088 \pm 0.017\invps$ if there is no new physics in
\dms)~\cite{Lenz:2006hd,Beneke:1998sy}.

Measurements of $\BR{B^0_s \rightarrow D_s^{(*)+} D_s^{(*)-}}$ can 
also be sensitive to \DGs.
The decay $\Bs \rightarrow D_s^{+} D_s^{-}$ is into
a final state that is purely \CP even. 
Under various theoretical assumptions~\cite{Aleksan:1993qp,Dunietz:2000cr}, the
inclusive decay into this plus the excited states
$\Bs \rightarrow D_s^{(*)+} D_s^{(*)-}$ is also \CP even
to within 5\%, and 
$\Bs \rightarrow D_s^{(*)+} D_s^{(*)-}$ saturates
$\Gamma_s^{\CP \thinspace {\mathrm{even}}}$.
Under these assumptions, for no \CP violation, we have: 
\begin{equation}
\DGGs \approx
\frac{2 \BR{\Bs \rightarrow D_s^{(*)+} D_s^{(*)-}}}
{1 - \BR{\Bs \rightarrow D_s^{(*)+} D_s^{(*)-}}} \,.
\labe{dGsBr}
\end{equation}
However, there are concerns~\cite{Nierste_private:2006} 
that the assumptions needed
for the above are overly restrictive and that the inclusive branching
ratio may be \CP even to only 30\%.
In the application of the constraint as a Gaussian penalty
function, the theoretical uncertainty is dealt with in two ways:
the fraction of the \CP-odd component of the decay~\cite{Dunietz:2000cr} 
is taken
to be a uniform distribution ranging from 0 to 0.05 and
convoluted in the Gaussian, and the fractional uncertainty on the
average measured value is increased in quadrature by 
30\%.

\begin{table}
\caption{Measurements of $\BR{\Bs \rightarrow D_s^{(*)+} D_s^{(*)-}}$.}
\labt{dGsBr}
\begin{center}
\begin{tabular}{l|c|c|c}
\hline
Experiment & Method & Value & Ref.  \\
\hline
ALEPH         & $\phi$-$\phi$ correlations              
           & $0.115 \pm 0.050^{+0.095}_{-0.045}$  & \cite{Barate:2000kd}$^a$     \\
\dzero        & $D_s \rightarrow \phi \pi$, $D_s \rightarrow \phi \mu \nu$            
           & $0.035 \pm 0.010 \pm 0.011$  & \cite{Abazov:2008ig,*Abazov:2007rb_mod_cont}$^{~}$   \\
\belle      & full reco.\ in 6 excl.\ $D_s$ modes 
            & $0.069 ^{+0.015}_{-0.013} \pm 0.019$ & \cite{Esen:2010jq_mod} \\
	 \hline
\multicolumn{2}{l}{Average of above 3} &   \hfagBRDSDS  &   \\
      \hline
\multicolumn{4}{l}{
$^a$ \footnotesize The value quoted in this table is half of 
$\BR{\Bs{\rm(short)} \rightarrow D_s^{(*)+} D_s^{(*)-}}$
given in Ref.~\cite{Barate:2000kd}.} \\[-0.5ex]
\multicolumn{4}{l}{$^{~}$ \footnotesize Before averaging, it has been adjusted the latest values
of \fBs at LEP and \BR{\Ds \to \phi X}.} 
\end{tabular}
\end{center}
\end{table}

Measurements for the branching fraction for this
decay channel are shown in \Table{dGsBr}.
Using their average value of \hfagBRDSDS with \Eq{dGsBr} yields
\begin{equation}
\DGGs = \hfagDGSGSBRDSDS \,,
\end{equation}
consistent with the value given in \Eq{DGGs_ave}. 

As described in \Sec{taubs}
and \Eq{tau_CPeven}, the average of the lifetime
measurements with \Bs $\to K^+ K^-$ and
$\Bs \to D_s^{(*)} D_s^{(*)}$ decays
can be used to measure the lifetime
of the \CP-even (or ``light" mass) eigenstate
$\tau(\Bs \to CP\mbox{-even}) = \tau_L = 1/\Gamma_L =$
\hfagTAUBSSHORT. These decays are assumed to be 100\% \CP even, with
a 5\% theoretical uncertainty on this assumption added in quadrature
for the combination.

When the constraint due this \CP-even lifetime and
the $\BR{B^0_s \rightarrow D_s^{(*)+} D_s^{(*)-}}$
branching fraction are
added to the previous ones,
the dashed-outline
shaded
regions of \Fig{DGs} are obtained, with numerical results:
\begin{eqnarray}
\DGGs &\in& [\hfagDGSGSCONXlow,\hfagDGSGSCONXupp] ~ \mbox{at \CL{95}} \,, \\
\DGGs &=& \hfagDGSGSCONX \,, \labe{DGGs_ave_const} \\
\DGs &\in& [\hfagDGSCONXlow,\hfagDGSCONXupp]\invps ~ \mbox{at \CL{95}} \,, \\
\DGs &=& \hfagDGSCONX \,, \\
\bar{\tau}(\Bs) = 1/\Gs &=& \hfagTAUBSMEANCONX \,, \labe{oneoverGs_const} \\
1/\Gamma_{\rm L} = \tau_{\rm short} &=& \hfagTAUBSLCONX \,, \\
1/\Gamma_{\rm H} = \tau_{\rm long}  &=& \hfagTAUBSHCONX \,. 
\end{eqnarray}

CDF has also measured the exclusive branching fraction 
$\BR{B^0_s \rightarrow D^+_s D^-_s} = 
(9.4^{+4.4}_{-4.2}) \times 10^{-3}$~\cite{Abulencia:2007zz}, and
they use this to set a lower bound of
$\Delta\Gamma_s^{CP}/\Gamma_s \geq 0.012$ at \CL{95} (since
on its own it does not saturate the \CP-even states).

\subsubsubsection{Weak phase in \Bs mixing}
\label{phasebs}
In general there will be a \CP-violating weak phase difference:
\begin{equation}
\phi_s = \arg \left[ -{M_{12}}/{\Gamma_{12}} \right], 
\end{equation}
where $M_{12}$ and $\Gamma_{12}$ are the off-diagonal
elements of the mass and decay matrices of the 
\Bs-\Bsbar system.
This is related to the observed \DGs through the relation:
\begin{equation}
\DGs = 2|\Gamma_{12}|\cos\phi_s.
\labe{new_phys_phase}
\end{equation}
The SM prediction for this phase is tiny,
$\phi_s^{\mathrm{SM}} = 0.004$~\cite{Lenz:2006hd}; however,
new physics in \Bs mixing could change this observed phase to
\begin{equation}
\phi_s = \phi_s^{\mathrm{SM}} + \phi_s^{\mathrm{NP}}.
\end{equation}
The relative phase between the \Bs mixing amplitude and that of
specific $b \rightarrow c\bar{c}s$ quark transitions such as 
for \Bs or \Bsbar $\rightarrow J/\psi \phi$ in the SM 
is~\cite{Lenz:2006hd,Charles:2004jd_mod,*Bona:2006ah}: 
\begin{equation}
2\beta_s^{SM} = 2\arg\left[-\left(V_{ts}V^*_{tb}\right)/\left(V_{cs}V^*_{cb}\right)\right] 
= 0.037 \pm 0.002 \approx 0.04.
\end{equation}
This angle is analogous to the $\beta$ angle in the usual CKM
unitarity triangle aside from the negative sign (resulting in a
positive angle in the SM).
The same additional contribution due to new physics would show up in this
observed phase~\cite{Lenz:2006hd}, i.e.:
\begin{equation}
2\beta_s = 2\beta_s^{\mathrm{SM}} - \phi_s^{\mathrm{NP}}.
\end{equation}
The current experimental precision does not allow these small
\CP-violating phases $\phi_s^{\mathrm{SM}}$ and
$\beta_s^{\mathrm{SM}}$ to be resolved, 
and for large new physics effect, we can approximate
$\phi_s \approx -2\beta_s \approx \phi_s^{\mathrm{NP}}$, i.e., a significantly 
large observed phase would indicate new physics.

For non-zero $|\Gamma_{12}|$, analysis of the time-dependent
decay \particle{\Bs \rightarrow J/\psi\phi} can measure
the weak phase.  Including information on the \Bs flavor at production
time via flavor tagging improves precision and also resolves the 
sign ambiguity on the weak phase angle for a given \DGs.
Both CDF~\cite{CDFnote9458:2008,*Aaltonen:2007gf_mod,*Aaltonen:2007he_mod} 
and \dzero~\cite{Abazov_mod:2008fj,*Abazov:2007tx_mod_cont} have performed 
such analyses and measure the same observed phase that we denote
$\phi_s^{J/\psi \phi} = -2\beta_s^{J/\psi \phi}$ to reflect
the different conventions of the experiments.

Under the assumption of non-zero $\phi_s^{J/\psi\phi}$, 
in addition to the result listed
in \Table{dgammat}, 
the \dzero collaboration~\cite{Abazov_mod:2008fj,*Abazov:2007tx_mod_cont}  has also made simultaneous
fits allowing $\phi_s^{J/\psi\phi}$ to float while weakly 
constraining the strong phases, $\delta_i$ to find: 
\begin{eqnarray}
\DGs &=& +0.19 \pm 0.07 ^{+0.02}_{-0.01}~{\mathrm{ps}}^{-1}\,,  \\ 
\bar{\tau}(\Bs) &= &1/\Gs = 1.52 \pm 0.06~{\mathrm{ps}}\,,  \\
\phi_s^{J/\psi\phi} &=& -0.57 ^{+0.24+0.07}_{-0.30-0.02} \,. 
\end{eqnarray}
If the SM value of $\phi_s^{J/\psi\phi} = -0.04$ is assumed, a probability of 
6.6\% to obtain a value of $\phi_s^{J/\psi\phi}$ lower than $-0.57$ is
found.

The CDF 
analysis~\cite{CDFnote9458:2008,*Aaltonen:2007gf_mod,*Aaltonen:2007he_mod} 
reports confidence regions
in the two-dimensional space of $2\beta_s^{J/\psi\phi}$ and \DGs.
They present a Feldman-Cousins confidence interval of $2\beta_s^{J/\psi\phi}$
where \DGs is treated as a nuisance parameter:
\begin{equation}
2\beta_s^{J/\psi\phi} = -\phi_s^{J/\psi\phi} \in [0.56,2.58]~{\mathrm{at~68\%~CL}}.
\end{equation}
Only a confidence range is quoted and a  point 
estimate is not given since biases were observed in the analysis.
Assuming the SM predictions for $2\beta_s$ and \DGs, they find
that the probability of a deviation as large as the level of the 
observed data is 7\%.
Note that CDF has very recently made a preliminary
update~\cite{CDFnote10206:2010}
to their
\particle{\Bs \rightarrow J/\psi\phi} analysis to an
integrated luminosity of 5.2~fb$^{-1}$ indicating a best-fit
confidence interval of:
\begin{equation}
2\beta_s^{J/\psi\phi} = -\phi_s^{J/\psi\phi} 
\in [0.04,1.04] \cup [2.16,3.10]~{\mathrm{at~68\%~CL}},
\end{equation}
where the probability
of a larger deviation from the SM prediction is 44\% or $0.8\sigma$.
However, this new result has not yet been used in the combinations
below.

Given the consistency of these two measurements of the weak phase,
as well as their
deviations from the SM, there is interest in combining the results and
using in global fits, e.g., see Ref.~\cite{Bona:2008jn}.
To allow a combination on equal footing, the \dzero collaboration
has redone their fits~\cite{D0web:2009} 
allowing  strong phase values, $\delta_i$, to float
as in the CDF analysis.
Ensemble studies to test confidence level coverage were performed 
by both collaborations and used to adjust likelihood
values to correspond to the usual Gaussian confidence levels. 
Two-dimensional likelihoods were 
combined~\cite{CDFnote9787:2009,D0Note5928:2009}
with the result shown in 
\Fig{DGs_phase}(a).  
After the combination, consistency  
of the best fit values for $\phi_s^{J/\psi\phi} = -2\beta_s^{J/\psi\phi}$ with
SM predictions is at the level of $\hfagNSIGMASM\sigma$, with numerical results
for the two solutions given below.
Despite possible biases in the CDF input, point estimates are still
presented and the confidence level regions are straight projections
onto the \DGs or phase angle axes.
\begin{eqnarray}
\DGs &=& \hfagDGSCOMBA 
~ \mbox{or} ~ \hfagDGSCOMBB \,, \\
     &\in& [\hfagDGSCOMBAlow,\hfagDGSCOMBAupp]
     \cup  [\hfagDGSCOMBBlow,\hfagDGSCOMBBupp]\invps ~ \mbox{at \CL{90}} \,, \\
\phi_s^{J/\psi\phi} = -2\beta_s^{J/\psi\phi} &=& \hfagPHISCOMBA 
~ \mbox{or} ~ \hfagPHISCOMBB \,, \\
     &\in& [\hfagPHISCOMBAlow,\hfagPHISCOMBAupp]
     \cup  [\hfagPHISCOMBBlow,\hfagPHISCOMBBupp]~ \mbox{at \CL{90}} \,.
\end{eqnarray}

A comparison between
the above sum of the
CDF and \dzero likelihoods 
and the world average \Bs semileptonic asymmetry of
\Eq{ASLs} through~\cite{Beneke:2003az}:
\begin{equation}
\ASLs = 
\frac{|\Gamma^{12}_s|}{|M^{12}_s|}\sin\phi_s = \frac{\DGs}{\dms}\tan\phi_s
\end{equation}
is also made and shown in 
\Fig{DGs_phase}(a).
Consistency between the two is observed, and the value
of \ASLs is applied as a constraint
resulting in the
confidence level regions 
shown in \Fig{DGs_phase}(b)
including the region delineated by new physics traced by 
the relation of \Eq{new_phys_phase}. Numerical results for the 
two solutions are:
\begin{eqnarray}
\DGs &=& \hfagDGSCOMBACON ~ \mbox{or} ~
         \hfagDGSCOMBBCON \,, \\
     &\in& [\hfagDGSCOMBACONlow,\hfagDGSCOMBACONupp]
     \cup  [\hfagDGSCOMBBCONlow,\hfagDGSCOMBBCONupp]\invps ~ \mbox{at \CL{90}} \,, \\
\phi_s^{J/\psi\phi} = -2\beta_s^{J/\psi\phi} &=& \hfagPHISCOMBACON 
    ~ \mbox{or} \hfagPHISCOMBBCON \,, \\
    &\in& [\hfagPHISCOMBACONlow,\hfagPHISCOMBACONupp]
    \cup  [\hfagPHISCOMBBCONlow,\hfagPHISCOMBBCONupp]~ \mbox{at \CL{90}} \,.
\end{eqnarray}
with a consistency
of the best fit values with
SM predictions of $2\beta_s$ at the level of $\hfagNSIGMASMCON\sigma$.

\begin{figure}
\begin{center}
\epsfig{figure=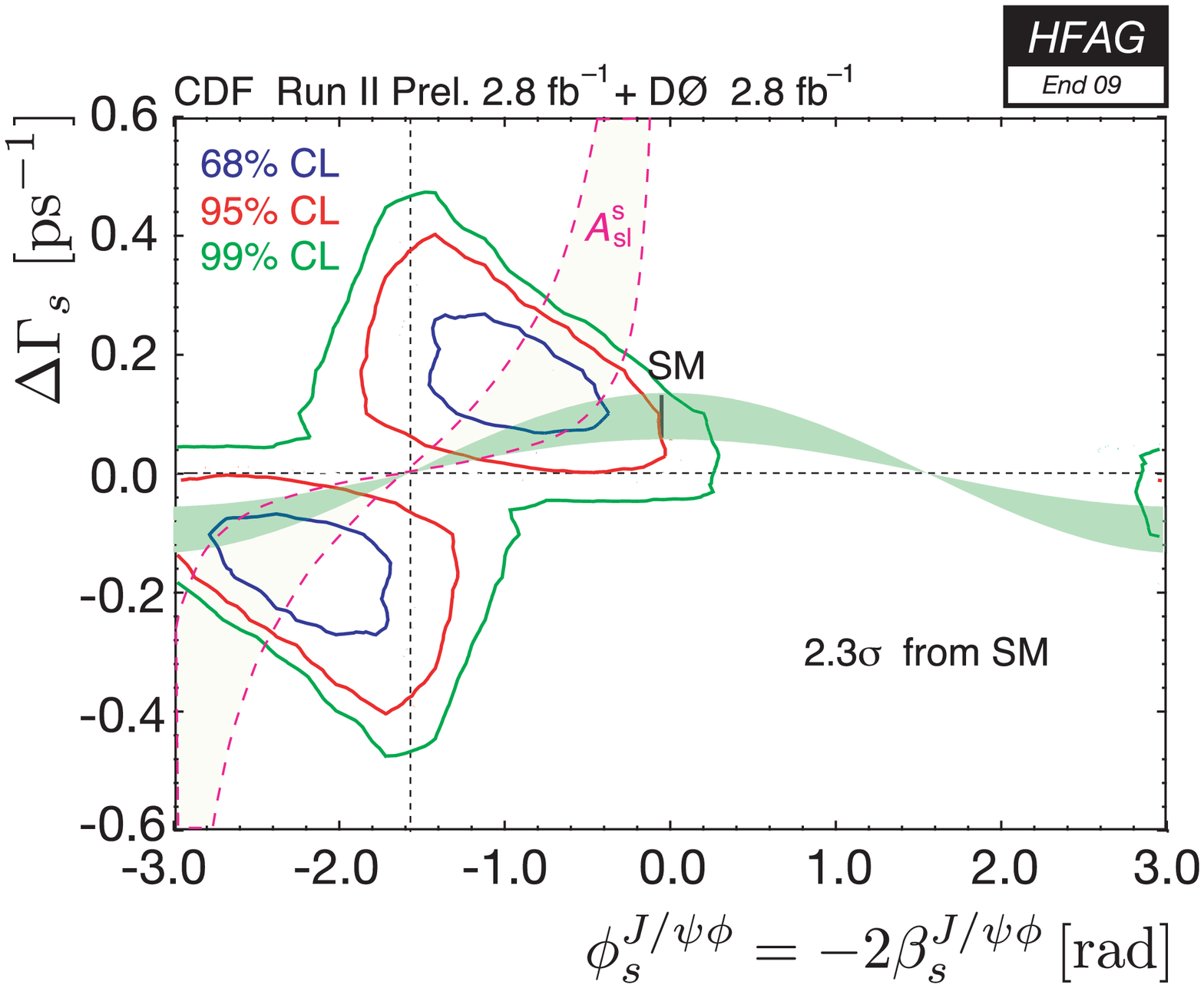,width=0.45\textwidth} 
\hfill    
\epsfig{figure=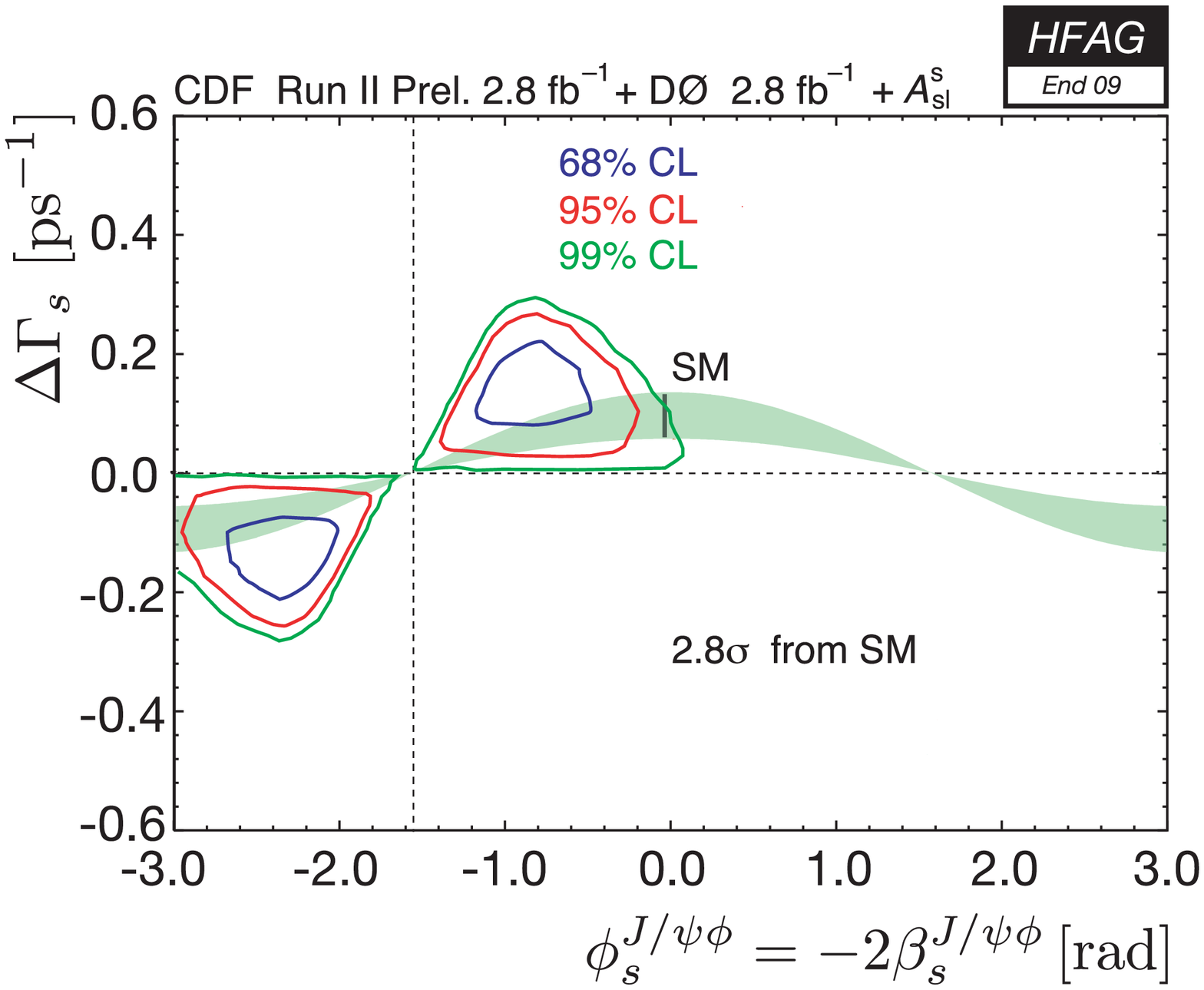,width=0.45\textwidth}
\caption{(a) Confidence regions
in \Bs width difference \DGs and
weak phase angle $\phi_s^{J/\psi\phi} = -2\beta_s^{J/\psi\phi}$
from combined CDF and \dzero
likelihoods determined in  flavor-tagged
\particle{\Bs \rightarrow J/\psi \phi} time-dependent
angular 
analyses~\cite{CDFnote9458:2008,*Aaltonen:2007gf_mod,*Aaltonen:2007he_mod,D0web:2009} compared
to the SM value of $-2\beta_s^{SM}$
and to the world-average value of
the \Bs semileptonic asymmetry, \ASLs (overlaid); 
(b) after adding
the constraint due to the world-average value of
\ASLs.
The region allowed in new physics models given by
$\DGs = 2|\Gamma_{12}|\cos\phi_s$ is also shown (light green band).}
\labf{DGs_phase}
\end{center}
\end{figure}

Finally, additional constraints are added due to
the flavor-specific
\Bs lifetime world average of \Eq{fslife_const2} 
through \Eq{fslife_const}, the \CP-event lifetime
of \Eq{tau_CPeven},
and the world average of the branching fraction
$\BR{B^0_s \rightarrow D_s^{(*)+} D_s^{(*)-}}$
through~\cite{Dunietz:2000cr}:
\begin{eqnarray}
2{\cal{B}}(B^0_s \rightarrow D_s^{(*)+}D_s^{(*)-}) \simeq
\Delta \Gamma_s^{\mathrm{CP}}
\left[
\frac{\frac{1}{1 - 2x_f} + \cos\phi_s}{2 \Gamma_L} +
\frac{\frac{1}{1 - 2x_f} - \cos\phi_s}{2 \Gamma_H}
\right]\,.
\end{eqnarray}
Here $x_f$ is the fraction of the $CP$-odd component
of the decay.
To apply this as a constraint, we expand the above expression 
to second order,
\begin{eqnarray}
2{\cal{B}}(B^0_s \rightarrow D_s^{(*)+}D_s^{(*)-}) \simeq
\frac{\Delta\Gamma_s }{\Gamma_s \cos \phi_s}
\left[
\frac{1}{1 - 2x_f} - \frac{\Delta \Gamma_s \cos \phi_s}{2 \Gamma_s}
\right]\,,
\end{eqnarray}
and use the world average of the branching ratio of \Table{dGsBr}.
Numerical results following these final constraints are:
\begin{eqnarray}
\DGs &=& \hfagDGSCOMBACONX ~ \mbox{or} ~
\hfagDGSCOMBBCONX \,, \\
    &\in& [\hfagDGSCOMBACONXlow,\hfagDGSCOMBACONXupp]
   \cup  [\hfagDGSCOMBBCONXlow,\hfagDGSCOMBBCONXupp]\invps ~ \mbox{at \CL{90}} \
	   ,, \\
 \phi_s^{J/\psi\phi} = -2\beta_s^{J/\psi\phi} &=& \hfagPHISCOMBACONX
       ~ \mbox{or} \hfagPHISCOMBBCONX \,, \\
           &\in& [\hfagPHISCOMBACONXlow,\hfagPHISCOMBACONXupp]
     \cup  [\hfagPHISCOMBBCONXlow,\hfagPHISCOMBBCONXupp]~ \mbox{at \CL{90}} \,.
\end{eqnarray}
with a consistency
of the best fit values with
SM predictions of $2\beta_s$ at the level of $\hfagNSIGMASMCONX\sigma$.

\subsubsubsection{\boldmath Mass difference \dms}

\Bs oscillations have been observed for the first time in 2006
by the CDF collaboration~\cite{Abulencia:2006ze,*Abulencia:2006mq_mod_cont},
based on samples of flavour-tagged hadronic and semileptonic \Bs decays
(in flavour-specific final states), partially or fully reconstructed in 
$1\invfb$ of data collected during Tevatron's Run~II. 
From the proper-time dependence of these \Bs candidates, CDF
observe \Bs oscillations with a significance of at least $5\sigma$ 
and measure $\dms = 17.77 \pm 0.10 \pm 0.07\invps$~\cite{Abulencia:2006ze,*Abulencia:2006mq_mod_cont}.
More recently, the \dzero collaboration has obtained with $2.4\invfb$ an independent  
$\sim 3\sigma$ preliminary evidence for \Bs oscillations;
combining all their results~\cite{D0note5474:2007,*D0note5254:2006,*Abazov:2006dm_mod_cont}
they obtain $\dms = 18.53 \pm 0.93 \pm 0.30\invps$~\cite{D0note5618:2008}.
To a good approximation, both the CDF and \dzero results have Gaussian 
errors, and the world average value of \dms can be obtained as a simple weighted 
average:
\begin{equation}
\dms = \hfagDMS \,.  \labe{dms}
\end{equation}

Multiplying this result with the 
mean \Bs lifetime of \Eq{oneoverGs}, $1/\Gs=\hfagTAUBSMEANCON$,
yields
\begin{equation}
\xs = \frac{\dms}{\Gs} = \hfagXS \,. \labe{xs}
\end{equation}
With $2\ys = \DGGs=\hfagDGSGSCON$ 
(see \Eq{DGGs_ave})
and under the assumption of no \CP violation in \Bs mixing,
this corresponds to
\begin{equation}
\chis = \frac{\xs^2+\ys^2}{2(\xs^2+1)} = \hfagCHIS \,. \labe{chis}
\end{equation}
The ratio of the \Bd and \Bs oscillation frequencies, 
obtained from \Eqss{dmd}{dms}, 
\begin{equation}
\frac{\dmd}{\dms} = \hfagRATIODMDDMS \,, \labe{dmd_over_dms}
\end{equation}
can be used to extract the following ratio of CKM matrix elements, 
\begin{equation}
\left|\frac{V_{td}}{V_{ts}}\right| =
\xi \sqrt{\frac{\dmd}{\dms}\frac{m(\Bs)}{m(\Bd)}} = 
\hfagVTDVTSfull \,, \labe{Vtd_over_Vts}
\end{equation}
where the first quoted error is from experimental uncertainties 
(with the masses $m(\Bs)$ and $m(\Bd)$ taken from~\cite{PDG_2010}),
and where the second quoted error is from theoretical uncertainties 
in the estimation of the SU(3) flavor-symmetry breaking factor
$\xi 
= \hfagXI$
obtained from lattice QCD calculations~\cite{Okamoto:2005zg,*Gray:2005ad_mod_cont,*Aoki:2003xb_mod_cont}.

\Bs mesons were known to mix since many years. Indeed 
the time-integrated measurements of \chibar (see \Sec{chibar}),
when compared to our knowledge
of \chid and the \b-hadron fractions, indicated that \Bs mixing was large,
with a value of \chis close to its maximal possible value of $1/2$.
However, the time dependence of this mixing 
could not be observed until recently, 
mainly because of lack of proper-time resolution to resolve 
the small period of the \Bs\ oscillations.

The statistical significance ${\cal S}$ of a \Bs oscillation signal can be
approximated as~\cite{Moser:1996xf}
\begin{equation}
{\cal S} \approx \sqrt{\frac{N}{2}} \,f_{\rm sig}\, (1-2w)\,
\exp{\left(-\left(\dms\sigma_t\right)^2/2\right)}\,,
\labe{significance}
\end{equation}
where $N$ is 
the number of selected and tagged \Bs candidates, 
$f_{\rm sig}$ is the fraction of \Bs signal
in the selected and tagged sample, $w$ is the total mistag probability, 
and $\sigma_t$ is the resolution on proper time.
As can be seen, the quantity ${\cal S}$ decreases very quickly as 
\dms increases: this dependence is controlled by $\sigma_t$, 
which is therefore the most critical parameter for \dms analyses. 
The method widely used for \Bs oscillation searches
consists of measuring a \Bs oscillation amplitude ${\cal A}$
at several different test values of \dms, 
using a maximum likelihood fit based on the functions 
of \Eq{oscillations} where the cosine terms have been multiplied 
by ${\cal A}$.
One expects ${\cal A}=1$ at the true 
value of \dms and ${\cal A}=0$ at a test value of \dms 
(far) below the true value.
To a good approximation, the statistical uncertainty on ${\cal A}$
is Gaussian and equal to $1/{\cal S}$~\cite{Moser:1996xf}.
In any analysis, a particular value of \dms
can be excluded at \CL{95} if ${\cal A}+ 1.645\,\sigma_{\cal A} < 1$, 
where $\sigma_{\cal A}$ is the total uncertainty on ${\cal A}$.
Because of the proper time resolution, the quantity $\sigma_{\cal A}(\dms)$
is an increasing function of \dms (see \Eq{significance} which merely models  
$1/\sigma_{\cal A}(\dms)$ in an analysis limited 
by the available statistics). Therefore, 
if the true value of \dms were infinitely large, one 
expects to be able to exclude all values of \dms up to $\dms^{\rm sens}$, 
where $\dms^{\rm sens}$, called here the
sensitivity of the analysis, is defined by
$1.645\,\sigma_{\cal A}(\dms^{\rm sens}) = 1$. 

\begin{figure}
\begin{center}
\epsfig{figure=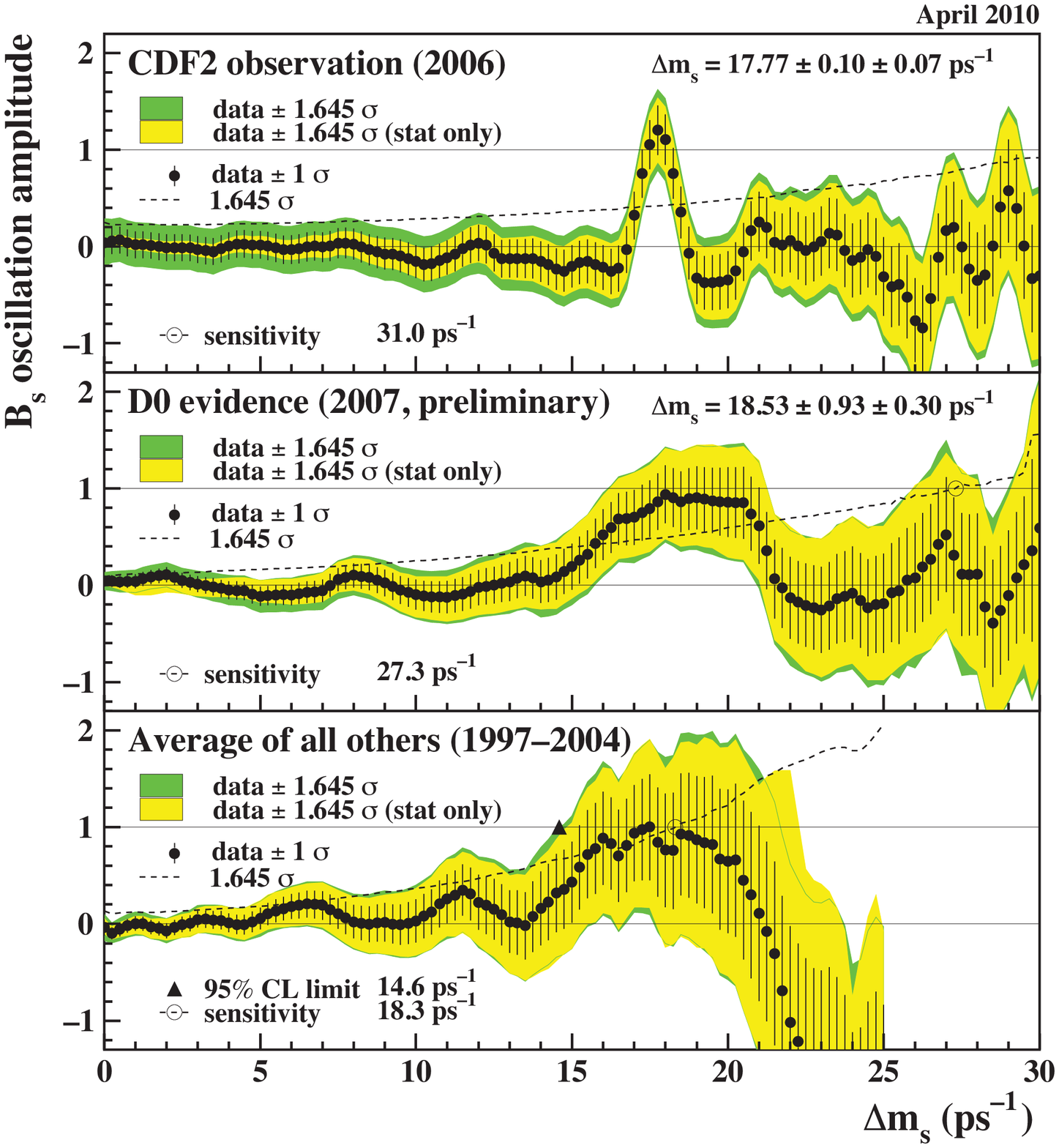,width=\textwidth}
\caption{\Bs oscillation amplitude as a function of \dms,
and measured value of \dms (when available).
\underline{Top:} CDF result based on Run II data, published in 2006~\cite{Abulencia:2006ze,*Abulencia:2006mq_mod_cont}.
\underline{Middle:} Average of the latest \dzero amplitude measurements
released as preliminary results
in 2007~\cite{D0note5474:2007,*D0note5254:2006,*Abazov:2006dm_mod_cont}, 
and corresponding measurement of \dms~\cite{D0note5618:2008}.
\underline{Bottom:} Average of all
ALEPH~\cite{Heister:2002gk},
DELPHI~\cite{Abreu:2000sh,Abreu:2000ev,Abdallah:2002mr,Abdallah:2003we},
OPAL~\cite{Abbiendi:1999gm,Abbiendi:2000bh},
SLD~\cite{Abe:2002ua,Abe:2002wfa},
and CDF Run~I~\cite{Abe:1998qj} results 
published between 1997 and 2004.
Statistical uncertainties dominate. 
Neighboring points are statistically correlated.}
\labf{amplitude}
\end{center}
\end{figure}

\Figure{amplitude} shows the measured \Bs amplitude as a function of \dms, 
as obtained by CDF (top) and \dzero (middle) using Run II data. 
The recent \dzero evidence of a \Bs oscillation signal is consistent
with the 2006 observation by CDF. 
A large number of \Bs oscillation searches,
already based on the amplitude method, 
had been performed previously by ALEPH~\cite{Heister:2002gk},
CDF (Run~I)~\cite{Abe:1998qj},
DELPHI~\cite{Abreu:2000sh,Abreu:2000ev,Abdallah:2002mr,Abdallah:2003we},
OPAL~\cite{Abbiendi:1999gm,Abbiendi:2000bh} and 
SLD~\cite{Abe:2002ua,Abe:2002wfa,Abe:2000gp}
(we omit references to searches that have been superseded
by more recent ones). All the results published by these experiments 
(between 1997 and 2004) have been combined 
by averaging the measured amplitudes ${\cal A}$ at each test value 
of \dms.
The individual results have been adjusted to common physics inputs, 
and all known correlations have been accounted for; 
in the case of the inclusive (lepton) analyses, performed at LEP and SLC, 
the sensitivities (\ie\ 
the statistical uncertainties on ${\cal A}$), which depend directly 
through \Eq{significance} on the assumed fraction $f_{\rm sig}\sim\fBs$
of \Bs mesons in an unbiased sample of weakly-decaying \b hadrons, 
have also been rescaled to the LEP average $\fBs = \hfagLFBS$.
The resulting average amplitude spectrum, completely dominated 
by the $e^+e^-\to\particle{Z}$ experiments, is displayed 
as the bottom plot of \Fig{amplitude}. Although no significant signal 
is seen, it is interesting to note the hint in the region 15--20\invps, 
consistent with the recent results from the Tevatron.

\clearpage
\mysection{Measurements related to Unitarity Triangle angles
}
\label{sec:cp_uta}

The charge of the ``$\CP(t)$ and Unitarity Triangle angles'' group
is to provide averages of measurements 
from time-dependent asymmetry analyses,
and other quantities that are related 
to the angles of the Unitarity Triangle (UT).
In cases where considerable theoretical input is required to 
extract the fundamental quantities, no attempt is made to do so at 
this stage. However, straightforward interpretations of the averages 
are given, where possible.

In Sec.~\ref{sec:cp_uta:introduction} 
a brief introduction to the relevant phenomenology is given.
In Sec.~\ref{sec:cp_uta:notations}
an attempt is made to clarify the various different notations in use.
In Sec.~\ref{sec:cp_uta:common_inputs}
the common inputs to which experimental results are rescaled in the
averaging procedure are listed. 
We also briefly introduce the treatment of experimental errors. 
In the remainder of this section,
the experimental results and their averages are given,
divided into subsections based on the underlying quark-level decays.

\mysubsection{Introduction
}
\label{sec:cp_uta:introduction}

The Standard Model Cabibbo-Kobayashi-Maskawa (CKM) quark mixing matrix $\VCKM$ 
must be unitary. A $3 \times 3$ unitary matrix has four free parameters,\footnote{
  In the general case there are nine free parameters,
  but five of these are absorbed into unobservable quark phases.}
and these are conventionally written by the product
of three (complex) rotation matrices~\cite{Chau:1984fp}, 
where the rotations are characterized by the Euler angles 
$\theta_{12}$, $\theta_{13}$ and $\theta_{23}$, which are the mixing angles
between the generations, and one overall phase $\delta$,
\begin{equation}
\label{eq:ckmPdg}
\VCKM =
        \left(
          \begin{array}{ccc}
            V_{ud} & V_{us} & V_{ub} \\
            V_{cd} & V_{cs} & V_{cb} \\
            V_{td} & V_{ts} & V_{tb} \\
          \end{array}
        \right)
        =
        \left(
        \begin{array}{ccc}
        c_{12}c_{13}    
                &    s_{12}c_{13}   
                        &   s_{13}e^{-i\delta}  \\
        -s_{12}c_{23}-c_{12}s_{23}s_{13}e^{i\delta} 
                &  c_{12}c_{23}-s_{12}s_{23}s_{13}e^{i\delta} 
                        & s_{23}c_{13} \\
        s_{12}s_{23}-c_{12}c_{23}s_{13}e^{i\delta}  
                &  -c_{12}s_{23}-s_{12}c_{23}s_{13}e^{i\delta} 
                        & c_{23}c_{13} 
        \end{array}
        \right)
\end{equation}
where $c_{ij}=\cos\theta_{ij}$, $s_{ij}=\sin\theta_{ij}$ for 
$i<j=1,2,3$. 

Following the observation of a hierarchy between the different
matrix elements, the Wolfenstein parameterization~\cite{Wolfenstein:1983yz}
is an expansion of $\VCKM$ in terms of the four real parameters $\lambda$
(the expansion parameter), $A$, $\rho$ and $\eta$. Defining to 
all orders in $\lambda$~\cite{Buras:1994ec}
\begin{eqnarray}
  \label{eq:burasdef}
  s_{12}             &\equiv& \lambda,\nonumber \\ 
  s_{23}             &\equiv& A\lambda^2, \\
  s_{13}e^{-i\delta} &\equiv& A\lambda^3(\rho -i\eta),\nonumber
\end{eqnarray}
and inserting these into the representation of Eq.~(\ref{eq:ckmPdg}), 
unitarity of the CKM matrix is achieved to all orders.
A Taylor expansion of $\VCKM$ leads to the familiar approximation
\begin{equation}
  \label{eq:cp_uta:ckm}
  \VCKM
  = 
  \left(
    \begin{array}{ccc}
      1 - \lambda^2/2 & \lambda & A \lambda^3 ( \rho - i \eta ) \\
      - \lambda & 1 - \lambda^2/2 & A \lambda^2 \\
      A \lambda^3 ( 1 - \rho - i \eta ) & - A \lambda^2 & 1 \\
    \end{array}
  \right) + {\cal O}\left( \lambda^4 \right) \, .
\end{equation}
At order $\lambda^{5}$, the obtained CKM matrix in this extended
Wolfenstein parametrization is:
{\small
  \begin{equation}
    \label{eq:cp_uta:ckm_lambda5}
    \VCKM
    =
    \left(
      \begin{array}{ccc}
        1 - \frac{1}{2}\lambda^{2} - \frac{1}{8}\lambda^4 &
        \lambda &
        A \lambda^{3} (\rho - i \eta) \\
        - \lambda + \frac{1}{2} A^2 \lambda^5 \left[ 1 - 2 (\rho + i \eta) \right] &
        1 - \frac{1}{2}\lambda^{2} - \frac{1}{8}\lambda^4 (1+4A^2) &
        A \lambda^{2} \\
        A \lambda^{3} \left[ 1 - (1-\frac{1}{2}\lambda^2)(\rho + i \eta) \right] &
        -A \lambda^{2} + \frac{1}{2}A\lambda^4 \left[ 1 - 2(\rho + i \eta) \right] &
        1 - \frac{1}{2}A^2 \lambda^4
      \end{array} 
    \right) + {\cal O}\left( \lambda^{6} \right).
  \end{equation}
}
The non-zero imaginary part of the CKM matrix,
which is the origin of $\CP$ violation in the Standard Model,
is encapsulated in a non-zero value of $\eta$.



The unitarity relation $\VCKM^\dagger\VCKM = {\mathit 1}$
results in a total of nine expressions,
that can be written as
$\sum_{i=u,c,t} V^*_{ij}V_{ik} = \delta_{jk}$,
where $\delta_{jk}$ is the Kronecker symbol.
Of the off-diagonal expressions ($j \neq k$),
three can be transformed into the other three 
leaving six relations, in which three complex numbers sum to zero,
which therefore can be expressed as triangles in the complex plane.
More details about unitarity triangles can be found in~\cite{Jarlskog:1985ht,Jarlskog:2005uq,Bjorken:2005rm,Harrison:2009bz}.

\begin{figure}[t]
  \begin{center}
    \resizebox{0.55\textwidth}{!}{\includegraphics{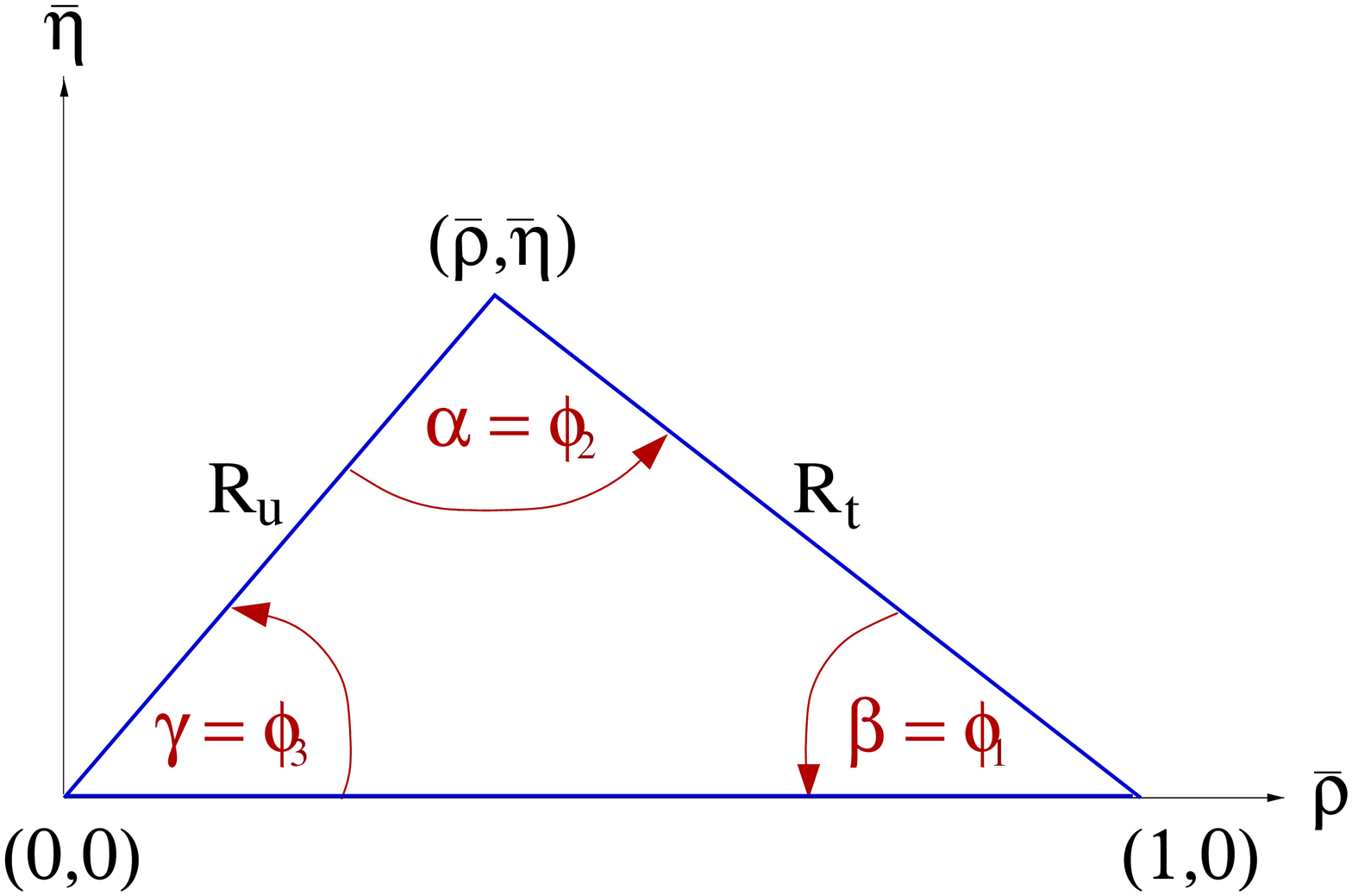}}
    \caption{The Unitarity Triangle.}
    \label{fig:cp_uta:ut}
  \end{center}
\end{figure}

One of these relations,
\begin{equation}
  \label{eq:cp_uta:ut}
  V_{ud}V^*_{ub} + V_{cd}V^*_{cb} + V_{td}V^*_{tb} = 0,
\end{equation}
is of particular importance to the $\B$ system, 
being specifically related to flavour changing 
neutral current $b \to d$ transitions.
The three terms in Eq.~(\ref{eq:cp_uta:ut}) are of the same order 
(${\cal O}\left( \lambda^3 \right)$),
and this relation is commonly known as the Unitarity Triangle.
For presentational purposes,
it is convenient to rescale the triangle by $(V_{cd}V^*_{cb})^{-1}$,
as shown in Fig.~\ref{fig:cp_uta:ut}.

Two popular naming conventions for the UT angles exist in the literature:
\begin{equation}
  \label{eq:cp_uta:abc}
  \alpha  \equiv  \phi_2  = 
  \arg\left[ - \frac{V_{td}V_{tb}^*}{V_{ud}V_{ub}^*} \right],
  \hspace{0.5cm}
  \beta   \equiv   \phi_1 =  
  \arg\left[ - \frac{V_{cd}V_{cb}^*}{V_{td}V_{tb}^*} \right],
  \hspace{0.5cm}
  \gamma  \equiv   \phi_3  =  
  \arg\left[ - \frac{V_{ud}V_{ub}^*}{V_{cd}V_{cb}^*} \right].
\end{equation}
In this document the $\left( \alpha, \beta, \gamma \right)$ set is used.\footnote{
  The relevant unitarity triangle for the $\Bs$ system is obtained 
  by replacing $d \leftrightarrow s$ in Eq.~\ref{eq:cp_uta:ut}.
  Definitions of the set of angles $( \alpha_s, \beta_s, \gamma_s )$ 
  can be obtained using equivalant relations to those of Eq.~\ref{eq:cp_uta:abc},
  for example $\beta_s = \arg\left[ - (V_{cs}V_{cb}^*) / (V_{ts}V_{tb}^*) \right]$.
  This definition gives a value of $\beta_s$ that is negative in the Standard Model,
  so that the sign is often flipped in the literature.
}
The sides $R_u$ and $R_t$ of the Unitarity Triangle 
(the third side being normalized to unity) 
are given by
\begin{equation}
  \label{eq:ru_rt}
  R_u =
  \left|\frac{V_{ud}V_{ub}^*}{V_{cd}V_{cb}^*} \right|
  = \sqrt{\rhobar^2+\etabar^2},
  \hspace{0.5cm}
  R_t = 
  \left|\frac{V_{td}V_{tb}^*}{V_{cd}V_{cb}^*}\right| 
  = \sqrt{(1-\rhobar)^2+\etabar^2}.
\end{equation} 
where $\rhobar$ and $\etabar$ define 
the apex of the Unitarity Triangle~\cite{Buras:1994ec} 
\begin{equation}
  \label{eq:rhoetabar}
  \rhobar + i\etabar
  \equiv -\frac{V_{ud}V_{ub}^*}{V_{cd}V_{cb}^*}
  \equiv 1 + \frac{V_{td}V_{tb}^*}{V_{cd}V_{cb}^*}
  = \frac{\sqrt{1-\lambda^2}\,(\rho + i \eta)}{\sqrt{1-A^2\lambda^4}+\sqrt{1-\lambda^2}A^2\lambda^4(\rho+i\eta)}
\end{equation}
The exact relation between $\left( \rho, \eta \right)$ and 
$\left( \rhobar, \etabar \right)$ is
\begin{equation}
  \label{eq:rhoetabarinv}
  \rho + i\eta \;=\; 
  \frac{ 
    \sqrt{ 1-A^2\lambda^4 }(\rhobar+i\etabar) 
  }{
    \sqrt{ 1-\lambda^2 } \left[ 1-A^2\lambda^4(\rhobar+i\etabar) \right]
  } \, .
\end{equation}

By expanding in powers of $\lambda$, several useful approximate expressions
can be obtained, including
\begin{equation}
  \label{eq:rhoeta_approx}
  \rhobar = \rho (1 - \frac{1}{2}\lambda^{2}) + {\cal O}(\lambda^4) \, ,
  \hspace{0.5cm}
  \etabar = \eta (1 - \frac{1}{2}\lambda^{2}) + {\cal O}(\lambda^4) \, ,
  \hspace{0.5cm}
  V_{td} = A \lambda^{3} (1-\rhobar -i\etabar) + {\cal O}(\lambda^6) \, .
\end{equation}

\mysubsection{Notations
}
\label{sec:cp_uta:notations}

Several different notations for $\CP$ violation parameters
are commonly used.
This section reviews those found in the experimental literature,
in the hope of reducing the potential for confusion, 
and to define the frame that is used for the averages.

In some cases, when $\B$ mesons decay into 
multibody final states via broad resonances ($\rho$, $\Kstar$, \etc),
the experimental analyses ignore the effects of interference 
between the overlapping structures.
This is referred to as the quasi-two-body (Q2B) approximation
in the following.

\mysubsubsection{$\CP$ asymmetries
}
\label{sec:cp_uta:notations:pra}

The $\CP$ asymmetry is defined as the difference between the rate 
involving a $b$ quark and that involving a $\bar b$ quark, divided 
by the sum. For example, the partial rate (or charge) asymmetry for 
a charged $\B$ decay would be given as 
\begin{equation}
  \label{eq:cp_uta:pra}
  \Acp_{f} \;\equiv\; 
  \frac{\Gamma(\Bm \to f)-\Gamma(\Bp \to \bar{f})}{\Gamma(\Bm \to f)+\Gamma(\Bp \to \bar{f})}.
\end{equation}

\mysubsubsection{Time-dependent \CP asymmetries in decays to $\CP$ eigenstates
}
\label{sec:cp_uta:notations:cp_eigenstate}

If the amplitudes for $\Bz$ and $\Bzb$ to decay to a final state $f$, 
which is a $\CP$ eigenstate with eigenvalue $\etacpf$,
are given by $\Af$ and $\Abarf$, respectively, 
then the decay distributions for neutral $\B$ mesons, 
with known flavour at time $\Delta t =0$,
are given by
\begin{eqnarray}
  \label{eq:cp_uta:td_cp_asp1}
  \Gamma_{\Bzb \to f} (\Delta t) & = &
  \frac{e^{-| \Delta t | / \tau(\Bz)}}{4\tau(\Bz)}
  \left[ 
    1 +
    \frac{2\, \Im(\lambda_f)}{1 + |\lambda_f|^2} \sin(\Delta m \Delta t) -
    \frac{1 - |\lambda_f|^2}{1 + |\lambda_f|^2} \cos(\Delta m \Delta t)
  \right], \\
  \label{eq:cp_uta:td_cp_asp2}
  \Gamma_{\Bz \to f} (\Delta t) & = &
  \frac{e^{-| \Delta t | / \tau(\Bz)}}{4\tau(\Bz)}
  \left[ 
    1 -
    \frac{2\, \Im(\lambda_f)}{1 + |\lambda_f|^2} \sin(\Delta m \Delta t) +
    \frac{1 - |\lambda_f|^2}{1 + |\lambda_f|^2} \cos(\Delta m \Delta t)
  \right].
\end{eqnarray}
Here $\lambda_f = \frac{q}{p} \frac{\Abarf}{\Af}$ 
contains terms related to $\Bz$\textendash$\Bzb$ mixing and to the decay amplitude
(the eigenstates of the effective Hamiltonian in the $\BzBzb$ system 
are $\left| B_\pm \right> = p \left| \Bz \right> \pm q \left| \Bzb \right>$).
This formulation assumes $\CPT$ invariance, 
and neglects possible lifetime differences 
(between the eigenstates of the effective Hamiltonian;
see Section~\ref{sec:mixing} where the mass difference $\Delta m$ is also defined)
in the neutral $\B$ meson system.
The case where non-zero lifetime differences are taken into account is 
discussed in Section~\ref{sec:cp_uta:notations:Bs}.
The time-dependent $\CP$ asymmetry,
again defined as the difference between the rate 
involving a $b$ quark and that involving a $\bar b$ quark,
is then given by
\begin{equation}
  \label{eq:cp_uta:td_cp_asp}
  \Acp_{f} \left(\Delta t\right) \; \equiv \;
  \frac{
    \Gamma_{\Bzb \to f} (\Delta t) - \Gamma_{\Bz \to f} (\Delta t)
  }{
    \Gamma_{\Bzb \to f} (\Delta t) + \Gamma_{\Bz \to f} (\Delta t)
  } \; = \;
  \frac{2\, \Im(\lambda_f)}{1 + |\lambda_f|^2} \sin(\Delta m \Delta t) -
  \frac{1 - |\lambda_f|^2}{1 + |\lambda_f|^2} \cos(\Delta m \Delta t).
\end{equation}

While the coefficient of the $\sin(\Delta m \Delta t)$ term in 
Eq.~(\ref{eq:cp_uta:td_cp_asp}) is everywhere\footnote
{
  Occasionally one also finds Eq.~(\ref{eq:cp_uta:td_cp_asp}) written as
  $\Acp_{f} \left(\Delta t\right) = 
  {\cal A}^{\rm mix}_f \sin(\Delta m \Delta t) + {\cal A}^{\rm dir}_f \cos(\Delta m \Delta t)$,
  or similar.
} denoted $S_f$:
\begin{equation}
  \label{eq:cp_uta:s_def}
  S_f \;\equiv\; \frac{2\, \Im(\lambda_f)}{1 + \left|\lambda_f\right|^2},
\end{equation}
different notations are in use for the
coefficient of the $\cos(\Delta m \Delta t)$ term:
\begin{equation}
  \label{eq:cp_uta:c_def}
  C_f \;\equiv\; - A_f \;\equiv\; \frac{1 - \left|\lambda_f\right|^2}{1 + \left|\lambda_f\right|^2}.
\end{equation}
The $C$ notation is used by the \babar\  collaboration 
(see \eg~\cite{Aubert:2001sp}), 
and also in this document.
The $A$ notation is used by the \belle\ collaboration
(see \eg~\cite{Abe:2001xe}).

Neglecting effects due to $\CP$ violation in mixing 
(by taking $|q/p| = 1$),
if the decay amplitude contains terms with 
a single weak (\ie, $\CP$ violating) phase
then $\left|\lambda_f\right| = 1$ and one finds
$S_f = -\etacpf \sin(\phi_{\rm mix} + \phi_{\rm dec})$, $C_f = 0$,
where $\phi_{\rm mix}=\arg(q/p)$ and $\phi_{\rm dec}=\arg(\Abarf/\Af)$.
Note that the $\Bz$--$\Bzb$ mixing phase $\phi_{\rm mix}\approx2\beta$
in the Standard Model (in the usual phase convention)~\cite{Carter:1980tk,Bigi:1981qs}. 

If amplitudes with different weak phases contribute to the decay, 
no clean interpretation of $S_f$ is possible. If the decay amplitudes
have in addition different $\CP$ conserving strong phases,
then $\left| \lambda_f \right| \neq 1$ and no clean interpretation is possible.
The coefficient of the cosine term becomes non-zero,
indicating direct $\CP$ violation.
The sign of $A_f$ as defined above is consistent with that of $\Acp_{f}$ in 
Eq.~(\ref{eq:cp_uta:pra}).

Frequently, we are interested in combining measurements 
governed by similar or identical short-distance physics,
but with different final states
(\eg, $\Bz \to \jpsi \KS$ and $\Bz \to \jpsi \KL$).
In this case, we remove the dependence on the $\CP$ eigenvalue 
of the final state by quoting $-\etacp S_f$.
In cases where the final state is not a $\CP$ eigenstate but has
an effective $\CP$ content (see below),
the reported $-\etacp S$ is corrected by the effective $\CP$.

\mysubsubsection{Time-dependent \CP asymmetries in decays to vector-vector final states
}
\label{sec:cp_uta:notations:vv}

Consider \B decays to states consisting of two spin-1 particles,
such as $\jpsi K^{*0}(\to\KS\piz)$, $D^{*+}D^{*-}$ and $\rho^+\rho^-$,
which are eigenstates of charge conjugation but not of parity.\footnote{
  \noindent
  This is not true of all vector-vector final states,
  \eg, $D^{*\pm}\rho^{\mp}$ is clearly not an eigenstate of 
  charge conjugation.
}
In fact, for such a system, there are three possible final states;
in the helicity basis these can be written $h_{-1}, h_0, h_{+1}$.
The $h_0$ state is an eigenstate of parity, and hence of $\CP$;
however, $\CP$ transforms $h_{+1} \leftrightarrow h_{-1}$ (up to 
an unobservable phase). In the transversity basis, these states 
are transformed into  $h_\parallel =  (h_{+1} + h_{-1})/2$ and 
$h_\perp = (h_{+1} - h_{-1})/2$.
In this basis all three states are $\CP$ eigenstates, 
and $h_\perp$ has the opposite $\CP$ to the others.

The amplitudes to these states are usually given by $A_{0,\perp,\parallel}$
(here we use a normalization such that 
$| A_0 |^2 + | A_\perp |^2 + | A_\parallel |^2 = 1$).
Then the effective $\CP$ of the vector-vector state is known if 
$| A_\perp |^2$ is measured.
An alternative strategy is to measure just the longitudinally polarized 
component,  $| A_0 |^2$
(sometimes denoted by $f_{\rm long}$), 
which allows a limit to be set on the effective $\CP$ since
$| A_\perp |^2 \leq | A_\perp |^2 + | A_\parallel |^2 = 1 - | A_0 |^2$.
The most complete treatment for 
neutral $\B$ decays to vector-vector final states
is time-dependent angular analysis 
(also known as time-dependent transversity analysis).
In such an analysis, 
the interference between the $\CP$-even and $\CP$-odd states 
provides additional sensitivity to the weak and strong phases involved.

In most analyses of time-dependent \CP asymmetries in decays to 
vector-vector final states carried out to date,
an assumption has been made that each helicity (or transversity) amplitude
has the same weak phase.
This is a good approximation for decays that are dominated by 
amplitudes with a single weak phase, such $\Bz \to \jpsi K^{*0}$,
and is a reasonable approximation in any mode for which only 
very limited statistics are available.
However, for modes that have contributions from amplitudes with different 
weak phases, the relative size of these contributions can be different 
for each helicity (or transversity) amplitude,
and therefore the time-dependent \CP asymmetry parameters can also differ.
The most generic analysis, suitable for modes with sufficient statistics,
would allow for this effect;
an intermediate analysis can allow different parameters for the 
$\CP$-even and $\CP$-odd components.
Such an analysis has been carried out by \babar\ for the decay
$\Bz \to D^{*+}D^{*-}$~\cite{:2008aw}.

\mysubsubsection{Time-dependent asymmetries: self-conjugate multiparticle final states
}
\label{sec:cp_uta:notations:dalitz}

Amplitudes for neutral \B decays into 
self-conjugate multiparticle final states
such as $\pi^+\pi^-\pi^0$, $K^+K^-\KS$, $\pi^+\pi^-\KS$,
$\jpsi \pi^+\pi^-$ or $D\pi^0$ with $D \to \KS\pi^+\pi^-$
may be written in terms of \CP-even and \CP-odd amplitudes.
As above, the interference between these terms 
provides additional sensitivity to the weak and strong phases
involved in the decay,
and the time-dependence depends on both the sine and cosine
of the weak phase difference.
In order to perform unbinned maximum likelihood fits,
and thereby extract as much information as possible from the distributions,
it is necessary to select a model for the multiparticle decay,
and therefore the results acquire some model dependence
(binned, model independent methods are also possible,
though are not as statistically powerful).
The number of observables depends on the final state (and on the model used);
the key feature is that as long as there are regions where both
\CP-even and \CP-odd amplitudes contribute,
the interference terms will be sensitive to the cosine 
of the weak phase difference.
Therefore, these measurements allow distinction between multiple solutions
for, \eg, the four values of $\beta$ from the measurement of $\sin(2\beta)$.

We now consider the various notations which have been used 
in experimental studies of
time-dependent asymmetries in decays to self-conjugate multiparticle final states.

\mysubsubsubsection{$\Bz \to D^{(*)}h^0$ with $D \to \KS\pi^+\pi^-$
}
\label{sec:cp_uta:notations:dalitz:dh0}

The states $D\pi^0$, $D^*\pi^0$, $D\eta$, $D^*\eta$, $D\omega$
are collectively denoted $D^{(*)}h^0$.
When the $D$ decay model is fixed,
fits to the time-dependent decay distributions can be performed
to extract the weak phase difference.
However, it is experimentally advantageous to use the sine and cosine of 
this phase as fit parameters, since these behave as essentially 
independent parameters, with low correlations and (potentially)
rather different uncertainties.
A parameter representing direct $\CP$ violation in the $B$ decay 
can also be floated.  
For consistency with other analyses, this could be chosen to be $C_f$,
but could equally well be $\left| \lambda_f \right|$, or other possibilities.

\belle\ performed an analysis of these channels
with $\sin(2\phi_1)$ and $\cos(2\phi_1)$ as free parameters~\cite{Krokovny:2006sv}.
\babar\ have performed an analysis floating also $\left| \lambda_f \right|$~\cite{Aubert:2007rp}
(and, of course, replacing $\phi_1 \Leftrightarrow \beta$).

\mysubsubsubsection{$\Bz \to D^{*+}D^{*-}\KS$
}
\label{sec:cp_uta:notations:dalitz:dstardstarks}

The hadronic structure of the $\Bz \to D^{*+}D^{*-}\KS$ decay
is not sufficiently well understood to perform a full 
time-dependent Dalitz plot analysis.
Instead, following Browder {\it et al.}~\cite{Browder:1999ng},
\babar~\cite{Aubert:2006fh} divide the Dalitz plane in two:
$m(D^{*+}\KS)^2 > m(D^{*-}\KS)^2$ $(\eta_y = +1)$ and 
$m(D^{*+}\KS)^2 < m(D^{*-}\KS)^2$ $(\eta_y = -1)$;
and then fit to a decay time distribution with asymmetry given by
\begin{equation}
  \Acp_{f} \left(\Delta t\right) =
  \eta_y \frac{J_c}{J_0} \cos(\Delta m \Delta t) -  
  \left[ 
    \frac{2J_{s1}}{J_0} \sin(2\beta) + \eta_y \frac{2J_{s2}}{J_0} \cos(2\beta) 
  \right] \sin(\Delta m \Delta t) \, .
\end{equation}
A similar analysis has also been carried out by \belle~\cite{Dalseno:2007hx}.
The measured values are $\frac{J_c}{J_0}$, $\frac{2J_{s1}}{J_0} \sin(2\beta)$
and $\frac{2J_{s2}}{J_0} \cos(2\beta)$, 
where the parameters $J_0$, $J_c$, $J_{s1}$ and $J_{s2}$ are the integrals 
over the half Dalitz plane $m(D^{*+}\KS)^2 < m(D^{*-}\KS)^2$ 
of the functions $|a|^2 + |\bar{a}|^2$, $|a|^2 - |\bar{a}|^2$, 
$\Re(\bar{a}a^*)$ and $\Im(\bar{a}a^*)$ respectively, 
where $a$ and $\bar{a}$ are the decay amplitudes of 
$\Bz \to D^{*+}D^{*-}\KS$ and $\Bzb \to D^{*+}D^{*-}\KS$ respectively. 
The parameter $J_{s2}$ (and hence $J_{s2}/J_0$) is predicted to be positive;
with this assumption is it possible to determine the sign of $\cos(2\beta)$.

\mysubsubsubsection{$\Bz \to K^+K^-\Kz$
}
\label{sec:cp_uta:notations:dalitz:kkk0}

Studies of $\Bz \to K^+K^-\Kz$~\cite{Aubert:2007sd,:2008gv,belle:kkk0:preliminary} 
and of the related decay 
$\Bp \to K^+K^-K^+$~\cite{Garmash:2004wa,Aubert:2006nu},
show that the decay is dominated by components from the 
intermediate $K^+K^-$ resonances $\phi(1020)$, $f_0(980)$,
a poorly understood scalar structure that peaks near 
$m(K^+K^-) \sim 1550 \ {\rm MeV}/c^2$ and is denoted $X_0(1550)$,
as well as a large nonresonant contribution.
There is also a contribution from $\chi_{c0}$.

The full time-dependent Dalitz plot analysis allows 
the complex amplitudes of each contributing term to be determined from data,
including $\CP$ violation effects
(\ie\ allowing the complex amplitude for the $\Bz$ decay to be independent
from that for $\Bzb$ decay), although one amplitude must be fixed 
to give a reference point.
There are several choices for parametrization of the complex amplitudes 
(\eg\ real and imaginary part, or magnitude and phase).
Similarly, there are various approaches to include $\CP$ violation effects.
Note that positive definite parameters such as magnitudes are
disfavoured in certain circumstances 
(they inevitably lead to biases for small values).
In order to compare results between analyses,
it is useful for each experiment to present results in terms of the 
parameters that can be measured in a Q2B analysis
(such as $\Acp_{f}$, $S_f$, $C_f$, 
$\sin(2\beta^{\rm eff})$, $\cos(2\beta^{\rm eff})$, \etc)

In the \babar\ analysis of $\Bz \to K^+K^-\Kz$~\cite{Aubert:2007sd,:2008gv},
the complex amplitude for each resonant contribution is written as
\begin{equation}
  A_f = c_f ( 1 + b_f ) e^{i ( \phi_f + \delta_f )} 
  \ , \ \ \ \ 
  \bar{A}_f = c_f ( 1 - b_f ) e^{i ( \phi_f - \delta_f )} \, ,
\end{equation}
where $b_f$ and $\delta_f$ introduce $\CP$ violation in the magnitude 
and phase respectively.
[The weak phase in $B^0$--$\bar{B}^0$ mixing ($2\beta$) also appears 
in the full formula for the time-dependent decay distribution.]
The Q2B direct $\CP$ violation parameter is directly related to $b_f$
\begin{equation}
  \Acp_{f} = \frac{-2b_f}{1+b_f^2} \approx C_f \, ,
\end{equation}
and the mixing-induced $\CP$ violation parameter can be used to obtain
$\sin(2\beta^{\rm eff})$
\begin{equation}
  -\eta_f S_f \approx \frac{1-b_f^2}{1+b_f^2}\sin(2\beta^{\rm eff}_f) \, ,
\end{equation}
where the approximations are exact in the case that $\left| q/p \right| = 1$.

\babar~\cite{Aubert:2007sd,:2008gv} present results for $c_f$, $\phi_f$,
$\Acp_{f}$ and $\beta^{\rm eff}$ for each resonant contribution,
as well as averaged values of $\Acp_{f}$ and $\beta^{\rm eff}$
for the entire $K^+K^-\Kz$ Dalitz plot.
\belle~\cite{belle:kkk0:preliminary} present results for the resonant
contributions only.

\mysubsubsubsection{$\Bz \to \pi^+\pi^-\KS$
}
\label{sec:cp_uta:notations:dalitz:pipik0}

Studies of $\Bz \to \pi^+\pi^-\KS$~\cite{Aubert:2009me,:2008wwa}
and of the related decay
$\Bp \to \pi^+\pi^-K^+$~\cite{Garmash:2004wa,Garmash:2005rv,Aubert:2005ce,Aubert:2008bj}
show that the decay is dominated by components from intermediate resonances 
in the $K\pi$ ($K^*(892)$, $K^*_0(1430)$) 
and $\pi\pi$ ($\rho(770)$, $f_0(980)$, $f_2(1270)$) spectra,
together with a poorly understood scalar structure that peaks near 
$m(\pi\pi) \sim 1300 \ {\rm MeV}/c^2$ and is denoted $f_X(1300)$
(that could be identified as either the $f_0(1370)$ or $f_0(1500)$),
and a large nonresonant component.
There is also a contribution from the $\chi_{c0}$ state.

The full time-dependent Dalitz plot analysis allows 
the complex amplitudes of each contributing term to be determined from data,
including $\CP$ violation effects.
In the \babar\ analysis~\cite{Aubert:2009me}, 
the magnitude and phase of each component (for both $\Bz$ and $\Bzb$ decays) 
are measured relative to $\Bz \to f_0(980)\KS$, using the following
parameterisation
\begin{equation}
  A_f = \left| A_f \right| e^{i\,{\rm arg}(A_f)}
  \ , \ \ \ \ 
  \bar{A}_f = \left| \bar{A}_f \right| e^{i\,{\rm arg}(\bar{A}_f)} \, .
\end{equation}
In the \belle\ analysis~\cite{:2008wwa}, the $\Bz \to K^{*+}\pi^-$ amplitude
is chosen as the reference, and the amplitudes are parameterised as 
\begin{equation}
  A_f = a_f ( 1 + c_f ) e^{i ( b_f + d_f )} 
  \ , \ \ \ \ 
  \bar{A}_f = a_f ( 1 - c_f ) e^{i ( b_f - d_f )} \, .
\end{equation}
In both cases, the results are translated into quasi-two-body parameters 
such as $2\beta^{\rm eff}_f$, $S_f$, $C_f$ for each \CP\ eigenstate $f$,
and direct \CP\ asymmetries for each flavour-specific state.
Relative phase differences between resonant terms are also extracted.

\mysubsubsubsection{$\Bz \to \pi^+\pi^-\pi^0$
}
\label{sec:cp_uta:notations:dalitz:pipipi0}

The $\Bz \to \pi^+\pi^-\pi^0$ decay is dominated by 
intermediate $\rho$ resonances.
Though it is possible, as above, 
to determine directly the complex amplitudes for each component,
an alternative approach~\cite{Snyder:1993mx,Quinn:2000by},
has been used by both \babar~\cite{Aubert:2007jn}
and \belle~\cite{Kusaka:2007dv,:2007mj}.
The amplitudes for $\Bz$ and $\Bzb$ to $\pi^+\pi^-\pi^0$ are written
\begin{equation}
  A_{3\pi} = f_+ A_+ + f_- A_- + f_0 A_0
  \ , \ \ \ 
  \bar{A}_{3\pi} = f_+ \bar{A}_+ + f_- \bar{A}_- + f_0 \bar{A}_0
\end{equation}
respectively.
$A_+$, $A_-$ and $A_0$
represent the complex decay amplitudes for 
$\Bz \to \rho^+\pi^-$, $\Bz \to \rho^-\pi^+$ and $\Bz \to \rho^0\pi^0$
while 
$\bar{A}_+$, $\bar{A}_-$ and $\bar{A}_0$
represent those for 
$\Bzb \to \rho^+\pi^-$, $\Bzb \to \rho^-\pi^+$ and $\Bzb \to \rho^0\pi^0$
respectively.
$f_+$, $f_-$ and $f_0$ incorporate kinematical and dynamical factors
and depend on the Dalitz plot coordinates.
The full time-dependent decay distribution can then be written 
in terms of 27 free parameters,
one for each coefficient of the form factor bilinears,
as listed in Table~\ref{tab:cp_uta:pipipi0:uandi}.
These parameters are often referred to as ``the $U$s and $I$s'',
and can be expressed in terms of 
$A_+$, $A_-$, $A_0$, $\bar{A}_+$, $\bar{A}_-$ and $\bar{A}_0$.
If the full set of parameters is determined,
together with their correlations,
other parameters, such as weak and strong phases,
direct $\CP$ violation parameters, \etc, 
can be subsequently extracted.
Note that one of the parameters (typically $U_+^+$)
is often fixed to unity to provide a reference point;
this does not affect the analysis.


\begin{table}[htb]
  \begin{center}
    \setlength{\tabcolsep}{0.3pc}
\begin{tabular}{l@{\extracolsep{5mm}}l}
      \hline
      Parameter   & Description \\
      \hline
      $U_+^+$          & Coefficient of $|f_+|^2$ \\
      $U_0^+$          & Coefficient of $|f_0|^2$ \\
      $U_-^+$          & Coefficient of $|f_-|^2$ \\
      [0.15cm]
      $U_0^-$          & Coefficient of $|f_0|^2\cos(\Delta m\Delta t)$ \\
      $U_-^-$          & Coefficient of $|f_-|^2\cos(\Delta m\Delta t)$ \\
      $U_+^-$          & Coefficient of $|f_+|^2\cos(\Delta m\Delta t)$ \\
      [0.15cm]
      $I_0$            & Coefficient of $|f_0|^2\sin(\Delta m\Delta t)$ \\
\hline
    \end{tabular}
    \caption{
      Definitions of the $U$ and $I$ coefficients.
      Modified from~\cite{Aubert:2007jn}.
    }
    \label{tab:cp_uta:pipipi0:uandi}
  \end{center}
\end{table}

\mysubsubsection{Time-dependent \CP asymmetries in decays to non-$\CP$ eigenstates
}
\label{sec:cp_uta:notations:non_cp}

Consider a non-$\CP$ eigenstate $f$, and its conjugate $\bar{f}$. 
For neutral $\B$ decays to these final states,
there are four amplitudes to consider:
those for $\Bz$ to decay to $f$ and $\bar{f}$
($\Af$ and $\Afbar$, respectively),
and the equivalents for $\Bzb$
($\Abarf$ and $\Abarfbar$).
If $\CP$ is conserved in the decay, then
$\Af = \Abarfbar$ and $\Afbar = \Abarf$.


The time-dependent decay distributions can be written in many different ways.
Here, we follow Sec.~\ref{sec:cp_uta:notations:cp_eigenstate}
and define $\lambda_f = \frac{q}{p}\frac{\Abarf}{\Af}$ and
$\lambda_{\bar f} = \frac{q}{p}\frac{\Abarfbar}{\Afbar}$.
The time-dependent \CP asymmetries then follow Eq.~(\ref{eq:cp_uta:td_cp_asp}):
\begin{eqnarray}
\label{eq:cp_uta:non-cp-obs}
  {\cal A}_f (\Delta t) \; \equiv \;
  \frac{
    \Gamma_{\Bzb \to f} (\Delta t) - \Gamma_{\Bz \to f} (\Delta t)
  }{
    \Gamma_{\Bzb \to f} (\Delta t) + \Gamma_{\Bz \to f} (\Delta t)
  } & = & S_f \sin(\Delta m \Delta t) - C_f \cos(\Delta m \Delta t), \\
  {\cal A}_{\bar{f}} (\Delta t) \; \equiv \;
  \frac{
    \Gamma_{\Bzb \to \bar{f}} (\Delta t) - \Gamma_{\Bz \to \bar{f}} (\Delta t)
  }{
    \Gamma_{\Bzb \to \bar{f}} (\Delta t) + \Gamma_{\Bz \to \bar{f}} (\Delta t)
  } & = & S_{\bar{f}} \sin(\Delta m \Delta t) - C_{\bar{f}} \cos(\Delta m \Delta t),
\end{eqnarray}
with the definitions of the parameters 
$C_f$, $S_f$, $C_{\bar{f}}$ and $S_{\bar{f}}$,
following Eqs.~(\ref{eq:cp_uta:s_def}) and~(\ref{eq:cp_uta:c_def}).

The time-dependent decay rates are given by
\begin{eqnarray}
  \Gamma_{\Bzb \to f} (\Delta t) & = &
  \frac{e^{-\left| \Delta t \right| / \tau(\Bz)}}{8\tau(\Bz)} 
  ( 1 + \Adirnoncp ) 
  \left\{ 
    1 + S_f \sin(\Delta m \Delta t) - C_f \cos(\Delta m \Delta t) 
  \right\},
  \\
  \Gamma_{\Bz \to f} (\Delta t) & = &
  \frac{e^{-\left| \Delta t \right| / \tau(\Bz)}}{8\tau(\Bz)} 
  ( 1 + \Adirnoncp ) 
  \left\{ 
    1 - S_f \sin(\Delta m \Delta t) + C_f \cos(\Delta m \Delta t) 
  \right\},
  \\
  \Gamma_{\Bzb \to \bar{f}} (\Delta t) & = &
  \frac{e^{-\left| \Delta t \right| / \tau(\Bz)}}{8\tau(\Bz)} 
  ( 1 - \Adirnoncp ) 
  \left\{ 
    1 + S_{\bar{f}} \sin(\Delta m \Delta t) - C_{\bar{f}} \cos(\Delta m \Delta t) 
  \right\},
  \\
  \Gamma_{\Bz \to \bar{f}} (\Delta t) & = &
    \frac{e^{-\left| \Delta t \right| / \tau(\Bz)}}{8\tau(\Bz)} 
  ( 1 - \Adirnoncp ) 
  \left\{ 
    1 - S_{\bar{f}} \sin(\Delta m \Delta t) + C_{\bar{f}} \cos(\Delta m \Delta t) 
  \right\},
\end{eqnarray}
where the time-independent parameter \Adirnoncp
represents an overall asymmetry in the production of the 
$f$ and $\bar{f}$ final states,\footnote{
  This parameter is often denoted ${\cal A}_f$ (or ${\cal A}_{\CP}$),
  but here we avoid this notation to prevent confusion with the
  time-dependent $\CP$ asymmetry.
}
\begin{equation}
  \Adirnoncp = 
  \frac{
    \left( 
      \left| \Af \right|^2 + \left| \Abarf \right|^2
    \right) - 
    \left( 
      \left| \Afbar \right|^2 + \left| \Abarfbar \right|^2
    \right)
  }{
    \left( 
      \left| \Af \right|^2 + \left| \Abarf \right|^2
    \right) +
    \left( 
      \left| \Afbar \right|^2 + \left| \Abarfbar \right|^2
    \right)
  }.
\end{equation}
Assuming $|q/p| = 1$,
the parameters $C_f$ and $C_{\bar{f}}$
can also be written in terms of the decay amplitudes as follows:
\begin{equation}
  C_f = 
  \frac{
    \left| \Af \right|^2 - \left| \Abarf \right|^2 
  }{
    \left| \Af \right|^2 + \left| \Abarf \right|^2
  }
  \hspace{5mm}
  {\rm and}
  \hspace{5mm}
  C_{\bar{f}} = 
  \frac{
    \left| \Afbar \right|^2 - \left| \Abarfbar \right|^2
  }{
    \left| \Afbar \right|^2 + \left| \Abarfbar \right|^2
  },
\end{equation}
giving asymmetries in the decay amplitudes of $\Bz$ and $\Bzb$
to the final states $f$ and $\bar{f}$ respectively.
In this notation, the direct $\CP$ invariance conditions are
$\Adirnoncp = 0$ and $C_f = - C_{\bar{f}}$.
Note that $C_f$ and $C_{\bar{f}}$ are typically non-zero;
\eg, for a flavour-specific final state, 
$\Abarf = \Afbar = 0$ ($\Af = \Abarfbar = 0$), they take the values
$C_f = - C_{\bar{f}} = 1$ ($C_f = - C_{\bar{f}} = -1$).

The coefficients of the sine terms
contain information about the weak phase. 
In the case that each decay amplitude contains only a single weak phase
(\ie, no direct $\CP$ violation),
these terms can be written
\begin{equation}
  S_f = 
  \frac{ 
    - 2 \left| \Af \right| \left| \Abarf \right| 
    \sin( \phi_{\rm mix} + \phi_{\rm dec} - \delta_f )
  }{
    \left| \Af \right|^2 + \left| \Abarf \right|^2
  } 
  \hspace{5mm}
  {\rm and}
  \hspace{5mm}
  S_{\bar{f}} = 
  \frac{
    - 2 \left| \Afbar \right| \left| \Abarfbar \right| 
    \sin( \phi_{\rm mix} + \phi_{\rm dec} + \delta_f )
  }{
    \left| \Afbar \right|^2 + \left| \Abarfbar \right|^2
  },
\end{equation}
where $\delta_f$ is the strong phase difference between the decay amplitudes.
If there is no $\CP$ violation, the condition $S_f = - S_{\bar{f}}$ holds.
If decay amplitudes with different weak and strong phases contribute,
no clean interpretation of $S_f$ and $S_{\bar{f}}$ is possible.

Since two of the $\CP$ invariance conditions are 
$C_f = - C_{\bar{f}}$ and $S_f = - S_{\bar{f}}$,
there is motivation for a rotation of the parameters:
\begin{equation}
\label{eq:cp_uta:non-cp-s_and_deltas}
  S_{f\bar{f}} = \frac{S_{f} + S_{\bar{f}}}{2},
  \hspace{4mm}
  \Delta S_{f\bar{f}} = \frac{S_{f} - S_{\bar{f}}}{2},
  \hspace{4mm}
  C_{f\bar{f}} = \frac{C_{f} + C_{\bar{f}}}{2},
  \hspace{4mm}
  \Delta C_{f\bar{f}} = \frac{C_{f} - C_{\bar{f}}}{2}.
\end{equation}
With these parameters, the $\CP$ invariance conditions become
$S_{f\bar{f}} = 0$ and $C_{f\bar{f}} = 0$. 
The parameter $\Delta C_{f\bar{f}}$ gives a measure of the ``flavour-specificity''
of the decay:
$\Delta C_{f\bar{f}}=\pm1$ corresponds to a completely flavour-specific decay,
in which no interference between decays with and without mixing can occur,
while $\Delta C_{f\bar{f}} = 0$ results in 
maximum sensitivity to mixing-induced $\CP$ violation.
The parameter $\Delta S_{f\bar{f}}$ is related to the strong phase difference 
between the decay amplitudes of $\Bz$ to $f$ and to $\bar f$. 
We note that the observables of Eq.~(\ref{eq:cp_uta:non-cp-s_and_deltas})
exhibit experimental correlations 
(typically of $\sim 20\%$, depending on the tagging purity, and other effects)
between $S_{f\bar{f}}$ and  $\Delta S_{f\bar{f}}$, 
and between $C_{f\bar{f}}$ and $\Delta C_{f\bar{f}}$. 
On the other hand, 
the final state specific observables of Eq.~(\ref{eq:cp_uta:non-cp-obs})
tend to have low correlations.

Alternatively, if we recall that the $\CP$ invariance
conditions at the decay amplitude level are
$\Af = \Abarfbar$ and $\Afbar = \Abarf$,
we are led to consider the parameters~\cite{Charles:2004jd}
\begin{equation}
  \label{eq:cp_uta:non-cp-directcp}
  {\cal A}_{f\bar{f}} = 
  \frac{
    \left| \Abarfbar \right|^2 - \left| \Af \right|^2 
  }{
    \left| \Abarfbar \right|^2 + \left| \Af \right|^2
  }
  \hspace{5mm}
  {\rm and}
  \hspace{5mm}
  {\cal A}_{\bar{f}f} = 
  \frac{
    \left| \Abarf \right|^2 - \left| \Afbar \right|^2
  }{
    \left| \Abarf \right|^2 + \left| \Afbar \right|^2
  }.
\end{equation}
These are sometimes considered more physically intuitive parameters
since they characterize direct $\CP$ violation 
in decays with particular topologies.
For example, in the case of $\Bz \to \rho^\pm\pi^\mp$
(choosing $f =  \rho^+\pi^-$ and $\bar{f} = \rho^-\pi^+$),
${\cal A}_{f\bar{f}}$ (also denoted ${\cal A}^{+-}_{\rho\pi}$)
parameterizes direct $\CP$ violation
in decays in which the produced $\rho$ meson does not contain the 
spectator quark,
while ${\cal A}_{\bar{f}f}$ (also denoted ${\cal A}^{-+}_{\rho\pi}$)
parameterizes direct $\CP$ violation 
in decays in which it does.
Note that we have again followed the sign convention that the asymmetry 
is the difference between the rate involving a $b$ quark and that
involving a $\bar{b}$ quark, \cf\ Eq.~(\ref{eq:cp_uta:pra}). 
Of course, these parameters are not independent of the 
other sets of parameters given above, and can be written
\begin{equation}
  {\cal A}_{f\bar{f}} =
  - \frac{
    \Adirnoncp + C_{f\bar{f}} + \Adirnoncp \Delta C_{f\bar{f}} 
  }{
    1 + \Delta C_{f\bar{f}} + \Adirnoncp C_{f\bar{f}} 
  }
  \hspace{5mm}
  {\rm and}
  \hspace{5mm}
  {\cal A}_{\bar{f}f} =
  \frac{
    - \Adirnoncp + C_{f\bar{f}} + \Adirnoncp \Delta C_{f\bar{f}} 
  }{
    - 1 + \Delta C_{f\bar{f}} + \Adirnoncp C_{f\bar{f}}  
  }.
\end{equation}
They usually exhibit strong correlations.

We now consider the various notations which have been used 
in experimental studies of
time-dependent $\CP$ asymmetries in decays to non-$\CP$ eigenstates.

\mysubsubsubsection{$\Bz \to D^{*\pm}D^\mp$
}
\label{sec:cp_uta:notations:non_cp:dstard}

The above set of parameters 
($\Adirnoncp$, $C_f$, $S_f$, $C_{\bar{f}}$, $S_{\bar{f}}$),
has been used by both \babar~\cite{:2008aw} 
and \belle~\cite{Aushev:2004uc} in the $D^{*\pm}D^{\mp}$ system
($f = D^{*+}D^-$, $\bar{f} = D^{*-}D^+$).
However, slightly different names for the parameters are used:
\babar\ uses 
(${\cal A}$, $C_{+-}$, $S_{+-}$, $C_{-+}$, $S_{-+}$);
\belle\ uses
(${\cal A}$, $C_{+}$,  $S_{+}$,  $C_{-}$,  $S_{-}$).
In this document, we follow the notation used by \babar.

\mysubsubsubsection{$\Bz \to \rho^{\pm}\pi^\mp$
}
\label{sec:cp_uta:notations:non_cp:rhopi}

In the $\rho^\pm\pi^\mp$ system, the 
($\Adirnoncp$, $C_{f\bar{f}}$, $S_{f\bar{f}}$, $\Delta C_{f\bar{f}}$, 
$\Delta S_{f\bar{f}}$)
set of parameters has been used 
originally by \babar~\cite{Aubert:2003wr} and \belle~\cite{Wang:2004va}, 
in the Q2B approximation; 
the exact names\footnote{
  \babar\ has used the notations
  $A_{\CP}^{\rho\pi}$~\cite{Aubert:2003wr} and 
  ${\cal A}_{\rho\pi}$~\cite{Aubert:2007jn}
  in place of ${\cal A}_{\CP}^{\rho\pi}$.
}
used in this case are
$\left( 
  {\cal A}_{\CP}^{\rho\pi}, C_{\rho\pi}, S_{\rho\pi}, \Delta C_{\rho\pi}, \Delta S_{\rho\pi}
\right)$,
and these names are also used in this document.

Since $\rho^\pm\pi^\mp$ is reconstructed in the final state $\pi^+\pi^-\pi^0$,
the interference between the $\rho$ resonances
can provide additional information about the phases 
(see Sec.~\ref{sec:cp_uta:notations:dalitz}).
Both \babar~\cite{Aubert:2007jn} 
and \belle~\cite{Kusaka:2007dv,:2007mj}
have performed time-dependent Dalitz plot analyses, 
from which the weak phase $\alpha$ is directly extracted.
In such an analysis, the measured Q2B parameters are 
also naturally corrected for interference effects.
See Sec.~\ref{sec:cp_uta:notations:dalitz:pipipi0}.

\mysubsubsubsection{$\Bz \to D^{\pm}\pi^{\mp}, D^{*\pm}\pi^{\mp}, D^{\pm}\rho^{\mp}$
}
\label{sec:cp_uta:notations:non_cp:dstarpi}

Time-dependent $\CP$ analyses have also been performed for the
final states $D^{\pm}\pi^{\mp}$, $D^{*\pm}\pi^{\mp}$ and $D^{\pm}\rho^{\mp}$.
In these theoretically clean cases, no penguin contributions are possible,
so there is no direct $\CP$ violation.
Furthermore, due to the smallness of the ratio of the magnitudes of the 
suppressed ($b \to u$) and favoured ($b \to c$) amplitudes (denoted $R_f$),
to a very good approximation, $C_f = - C_{\bar{f}} = 1$
(using $f = D^{(*)-}h^+$, $\bar{f} = D^{(*)+}h^-$ $h = \pi,\rho$),
and the coefficients of the sine terms are given by
\begin{equation}
  S_f = - 2 R_f \sin( \phi_{\rm mix} + \phi_{\rm dec} - \delta_f )
  \hspace{5mm}
  {\rm and}
  \hspace{5mm}
  S_{\bar{f}} = - 2 R_f \sin( \phi_{\rm mix} + \phi_{\rm dec} + \delta_f ).
\end{equation}
Thus weak phase information can be cleanly obtained from measurements
of $S_f$ and $S_{\bar{f}}$, 
although external information on at least one of $R_f$ or $\delta_f$ is necessary.
(Note that $\phi_{\rm mix} + \phi_{\rm dec} = 2\beta + \gamma$ for all the decay modes 
in question, while $R_f$ and $\delta_f$ depend on the decay mode.)

Again, different notations have been used in the literature.
\babar~\cite{Aubert:2006tw,Aubert:2005yf}
defines the time-dependent probability function by
\begin{equation}
  f^\pm (\eta, \Delta t) = \frac{e^{-|\Delta t|/\tau}}{4\tau} 
  \left[  
    1 \mp S_\zeta \sin (\Delta m \Delta t) \mp \eta C_\zeta \cos(\Delta m \Delta t) 
  \right],
\end{equation} 
where the upper (lower) sign corresponds to 
the tagging meson being a $\Bz$ ($\Bzb$). 
[Note here that a tagging $\Bz$ ($\Bzb$) corresponds to $-S_\zeta$ ($+S_\zeta$).]
The parameters $\eta$ and $\zeta$ take the values $+1$ and $+$ ($-1$ and $-$) 
when the final state is, \eg, $D^-\pi^+$ ($D^+\pi^-$). 
However, in the fit, the substitutions $C_\zeta = 1$ and 
$S_\zeta = a \mp \eta b_i - \eta c_i$ are made.\footnote{
  The subscript $i$ denotes tagging category.
}
[Note that, neglecting $b$ terms, $S_+ = a - c$ and $S_- = a + c$, 
so that $a = (S_+ + S_-)/2$, $c = (S_- - S_+)/2$, in analogy to 
the parameters of Eq.~(\ref{eq:cp_uta:non-cp-s_and_deltas}).] 
The subscript $i$ denotes the tagging category. 
These are motivated by the possibility of 
$\CP$ violation on the tag side~\cite{Long:2003wq}, 
which is absent for semileptonic $\B$ decays (mostly lepton tags). 
The parameter $a$ is not affected by tag side $\CP$ violation. 
The parameter $b$ only depends on tag side $\CP$ violation parameters 
and is not directly useful for determining UT angles.
A clean interpretation of the $c$ parameter is only possible for 
lepton-tagged events,
so the \babar\ measurements report $c$ measured with those events only.

The parameters used by \belle\ in the analysis using 
partially reconstructed $\B$ decays~\cite{:2008kr}, 
are similar to the $S_\zeta$ parameters defined above. 
However, in the \belle\ convention, 
a tagging $\Bz$ corresponds to a $+$ sign in front of the sine coefficient; 
furthermore the correspondence between the super/subscript 
and the final state is opposite, so that $S_\pm$ (\babar) = $- S^\mp$ (\belle). 
In this analysis, only lepton tags are used, 
so there is no effect from tag side $\CP$ violation. 
In the \belle\ analysis using 
fully reconstructed $\B$ decays~\cite{Ronga:2006hv}, 
this effect is measured and taken into account using $\Dstar l \nu$ decays; 
in neither \belle\ analysis are the $a$, $b$ and $c$ parameters used. 
In the latter case, the measured parameters are 
$2 R_{D^{(*)}\pi} \sin( 2\phi_1 + \phi_3 \pm \delta_{D^{(*)}\pi} )$; 
the definition is such that 
$S^\pm$ (\belle) = $- 2 R_{\Dstar \pi} \sin( 2\phi_1 + \phi_3 \pm \delta_{\Dstar \pi} )$. 
However, the definition includes an 
angular momentum factor $(-1)^L$~\cite{Fleischer:2003yb}, 
and so for the results in the $D\pi$ system, 
there is an additional factor of $-1$ in the conversion.

Explicitly, the conversion then reads as given in 
Table~\ref{tab:cp_uta:notations:non_cp:dstarpi}, 
where we have neglected the $b_i$ terms used by \babar
(which are zero in the absence of tag side $\CP$ violation).
For the averages in this document,
we use the $a$ and $c$ parameters,
and give the explicit translations used in 
Table~\ref{tab:cp_uta:notations:non_cp:dstarpi2}.
It is to be fervently hoped that the experiments will
converge on a common notation in future.

\begin{table}
  \begin{center} 
    \caption{
      Conversion between the various notations used for 
      $\CP$ violation parameters in the 
      $D^{\pm}\pi^{\mp}$, $D^{*\pm}\pi^{\mp}$ and $D^{\pm}\rho^{\mp}$ systems.
      The $b_i$ terms used by \babar\ have been neglected.
      Recall that $\left( \alpha, \beta, \gamma \right) = \left( \phi_2, \phi_1, \phi_3 \right)$.
    }
    \vspace{0.2cm}
    \setlength{\tabcolsep}{0.0pc}
\begin{tabular*}{\textwidth}{@{\extracolsep{\fill}}cccc} \hline 
      & \babar\ & \belle\ partial rec. & \belle\ full rec. \\
      \hline
      $S_{D^+\pi^-}$    & $- S_- = - (a + c_i)$ &  N/A  &
      $2 R_{D\pi} \sin( 2\phi_1 + \phi_3 + \delta_{D\pi} )$ \\
      $S_{D^-\pi^+}$    & $- S_+ = - (a - c_i)$ &  N/A  &
      $2 R_{D\pi} \sin( 2\phi_1 + \phi_3 - \delta_{D\pi} )$ \\
      $S_{D^{*+}\pi^-}$ & $- S_- = - (a + c_i)$ & $S^+$ &   
      $- 2 R_{\Dstar \pi} \sin( 2\phi_1 + \phi_3 + \delta_{\Dstar \pi} )$ \\
      $S_{D^{*-}\pi^+}$ & $- S_+ = - (a - c_i)$ & $S^-$ &
      $- 2 R_{\Dstar \pi} \sin( 2\phi_1 + \phi_3 - \delta_{\Dstar \pi} )$ \\
      $S_{D^+\rho^-}$    & $- S_- = - (a + c_i)$ &  N/A  &  N/A  \\
      $S_{D^-\rho^+}$    & $- S_+ = - (a - c_i)$ &  N/A  &  N/A  \\
      \hline 
    \end{tabular*}
    \label{tab:cp_uta:notations:non_cp:dstarpi}
  \end{center}
\end{table}
   
\begin{table}
  \begin{center} 
    \caption{
      Translations used to convert the parameters measured by \belle
      to the parameters used for averaging in this document.
      The angular momentum factor $L$ is $-1$ for $\Dstar\pi$ and $+1$ for $D\pi$.
      Recall that $\left( \alpha, \beta, \gamma \right) = \left( \phi_2, \phi_1, \phi_3 \right)$.
    }
    \vspace{0.2cm}
    \setlength{\tabcolsep}{0.0pc}
    \begin{tabular*}{\textwidth}{@{\extracolsep{\fill}}ccc} \hline 
        & $\Dstar\pi$ partial rec. & $D^{(*)}\pi$ full rec. \\
        \hline
        $a$ & $- (S^+ + S^-)$ &
        $\frac{1}{2} (-1)^{L+1}
        \left(
          2 R_{D^{(*)}\pi} \sin( 2\phi_1 + \phi_3 + \delta_{D^{(*)}\pi} ) + 
          2 R_{D^{(*)}\pi} \sin( 2\phi_1 + \phi_3 - \delta_{D^{(*)}\pi} )
        \right)$ \\
        $c$ & $- (S^+ - S^-)$ & 
        $\frac{1}{2} (-1)^{L+1}
        \left(
          2 R_{D^{(*)}\pi} \sin( 2\phi_1 + \phi_3 + \delta_{D^{(*)}\pi} ) -
          2 R_{D^{(*)}\pi} \sin( 2\phi_1 + \phi_3 - \delta_{D^{(*)}\pi} )
        \right)$ \\
        \hline 
      \end{tabular*}
    \label{tab:cp_uta:notations:non_cp:dstarpi2}
  \end{center}
\end{table}

\mysubsubsubsection{Time-dependent asymmetries in radiative $\B$ decays
}
\label{sec:cp_uta:notations:non_cp:radiative}

As a special case of decays to non-$\CP$ eigenstates,
let us consider radiative $\B$ decays.
Here, the emitted photon has a distinct helicity,
which is in principle observable, but in practice is not usually measured.
Thus the measured time-dependent decay rates 
are given by~\cite{Atwood:1997zr,Atwood:2004jj}
\begin{eqnarray}
  \Gamma_{\Bzb \to X \gamma} (\Delta t) & = &
  \Gamma_{\Bzb \to X \gamma_L} (\Delta t) + \Gamma_{\Bzb \to X \gamma_R} (\Delta t) \\ \nonumber
  & = &
  \frac{e^{-\left| \Delta t \right| / \tau(\Bz)}}{4\tau(\Bz)} 
  \left\{ 
    1 + 
    \left( S_L + S_R \right) \sin(\Delta m \Delta t) - 
    \left( C_L + C_R \right) \cos(\Delta m \Delta t) 
  \right\},
  \\
  \Gamma_{\Bz \to X \gamma} (\Delta t) & = & 
  \Gamma_{\Bz \to X \gamma_L} (\Delta t) + \Gamma_{\Bz \to X \gamma_R} (\Delta t) \\ \nonumber 
  & = &
  \frac{e^{-\left| \Delta t \right| / \tau(\Bz)}}{4\tau(\Bz)} 
  \left\{ 
    1 - 
    \left( S_L + S_R \right) \sin(\Delta m \Delta t) + 
    \left( C_L + C_R \right) \cos(\Delta m \Delta t) 
  \right\},
\end{eqnarray}
where in place of the subscripts $f$ and $\bar{f}$ we have used $L$ and $R$
to indicate the photon helicity.
In order for interference between decays with and without $\Bz$-$\Bzb$ mixing
to occur, the $X$ system must not be flavour-specific,
\eg, in case of $\Bz \to K^{*0}\gamma$, the final state must be $\KS \pi^0 \gamma$.
The sign of the sine term depends on the $C$ eigenvalue of the $X$ system.
At leading order, the photons from 
$b \to q \gamma$ ($\bar{b} \to \bar{q} \gamma$) are predominantly
left (right) polarized, with corrections of order of $m_q/m_b$,
thus interference effects are suppressed.
Higher order effects can lead to corrections of order 
$\Lambda_{\rm QCD}/m_b$~\cite{Grinstein:2004uu,Grinstein:2005nu},
though explicit calculations indicate such corrections are small
for exclusive final states~\cite{Matsumori:2005ax,Ball:2006cva}.
The predicted smallness of the $S$ terms in the Standard Model
results in sensitivity to new physics contributions.

\mysubsubsection{Time-dependent \CP asymmetries in the $B_s$ System}
\label{sec:cp_uta:notations:Bs}

A complete analysis of the time-dependent decay rates of 
neutral $B$ mesons must also take into account the lifetime difference
between the eigenstates of the effective Hamiltonian, 
denoted by $\Delta \Gamma$.
This is particularly important in the $B_s$ system,
since non-negligible values of  $\Delta \Gamma_s$ are expected 
(see Section~\ref{sec:mixing} for the latest experimental constraints).
Neglecting $\CP$ violation in mixing,
the relevant replacements for 
Eqs.~\ref{eq:cp_uta:td_cp_asp1}~\&~\ref{eq:cp_uta:td_cp_asp2} 
are~\cite{Dunietz:2000cr}
\begin{equation}
  \label{eq:cp_uta:td_cp_bs_asp1}
  \begin{array}{lcr}
    \mc{2}{l}{
      \Gamma_{\Bsb \to f} (\Delta t) = 
      {\cal N} 
      \frac{e^{-| \Delta t | / \tau(\Bs)}}{4\tau(\Bs)}
      \Big[ 
      \cosh(\frac{\Delta \Gamma \Delta t}{2}) +
    } & \hspace{40mm} \\
    \hspace{40mm} &
    \mc{2}{r}{
      \frac{2\, \Im(\lambda_f)}{1 + |\lambda_f|^2} \sin(\Delta m \Delta t) -
      \frac{1 - |\lambda_f|^2}{1 + |\lambda_f|^2} \cos(\Delta m \Delta t) -
      \frac{2\, \Re(\lambda_f)}{1 + |\lambda_f|^2} \sinh(\frac{\Delta \Gamma \Delta t}{2})
      \Big],
    } \\
  \end{array}
\end{equation}
and
\begin{equation}
  \label{eq:cp_uta:td_cp_bs_asp2}
  \begin{array}{lcr}
    \mc{2}{l}{
      \Gamma_{\Bs \to f} (\Delta t) =
      {\cal N} 
      \frac{e^{-| \Delta t | / \tau(\Bs)}}{4\tau(\Bs)}
      \Big[ 
      \cosh(\frac{\Delta \Gamma \Delta t}{2}) -
    } & \hspace{40mm} \\
    \hspace{40mm} & 
    \mc{2}{r}{
      \frac{2\, \Im(\lambda_f)}{1 + |\lambda_f|^2} \sin(\Delta m \Delta t) +
      \frac{1 - |\lambda_f|^2}{1 + |\lambda_f|^2} \cos(\Delta m \Delta t) -
      \frac{2\, \Re(\lambda_f)}{1 + |\lambda_f|^2} \sinh(\frac{\Delta \Gamma \Delta t}{2})
      \Big]. 
    } \\
  \end{array}
\end{equation}

To be consistent with our earlier notation,\footnote{
  As ever, alternative and conflicting notations appear in the literature.
  One popular alternative notation for this parameter is 
  ${\cal A}_{\Delta \Gamma}$.
  Particular care must be taken over the signs.
}
we write here the coefficient of the $\sinh$ term as
\begin{equation}
  A^{\Delta \Gamma}_f = - \frac{2\, \Re(\lambda_f)}{1 + |\lambda_f|^2} \, .
\end{equation}
A complete, tagged, time-dependent analysis of \CP asymmetries in 
$B_s$ decays to a \CP eigenstate $f$ can thus obtain the parameters 
$S_f$, $C_f$ and $A^{\Delta \Gamma}_f$.
Note that 
\begin{equation}
  \left( S_f \right)^2 + \left( C_f \right)^2 + \left( A^{\Delta \Gamma}_f \right)^2 = 1 \, .
\end{equation}
Since these parameters have sensitivity to both
$\Im(\lambda_f)$ and $\Re(\lambda_f)$,
alternative choices of parametrization, 
including those directly involving \CP violating phases (such as $\beta_s$), 
are possible.
These can also be adopted for vector-vector final states.

The {\it untagged} time-dependent decay rate is given by
\begin{equation}
  \Gamma_{\Bsb \to f} (\Delta t) + \Gamma_{\Bs \to f} (\Delta t)
  = 
  {\cal N} 
  \frac{e^{-| \Delta t | / \tau(\Bs)}}{2\tau(\Bs)}
  \Big[ 
  \cosh\left(\frac{\Delta \Gamma \Delta t}{2}\right) -
  \frac{2\, \Re(\lambda_f)}{1 + |\lambda_f|^2} \sinh\left(\frac{\Delta \Gamma \Delta t}{2}\right)
  \Big] \, .
\end{equation}
With the requirement
$\int_{-\infty}^{+\infty} \Gamma_{\Bsb \to f} (\Delta t) + \Gamma_{\Bs \to f} (\Delta t) d(\Delta t) = 1$,
the normalization factor ${\cal N}$ 
is fixed to $1 - (\frac{\Delta \Gamma}{2\Gamma})^2$.
Note that an untagged time-dependent analysis can probe
$\lambda_f$, through $\Re(\lambda_f)$, when $\Delta \Gamma \neq 0$.
The tagged analysis is, of course, more sensitive.

Other expressions can be similarly modified to take into account 
non-zero lifetime differences.
Note that when the final state contains 
a mixture of $\CP$-even and $\CP$-odd states
(as, for example, for vector-vector or multibody self-conjugate states),
that $\Re(\lambda_f)$ contains terms proportional to 
both the sine and cosine of the weak phase difference, 
albeit with rather different sensitivities.

\mysubsubsection{Asymmetries in $\B \to \DorDstar K^{(*)}$ decays
}
\label{sec:cp_uta:notations:cus}

$\CP$ asymmetries in $\B \to \DorDstar K^{(*)}$ decays are sensitive to $\gamma$.
The neutral $D^{(*)}$ meson produced 
is an admixture of $\DorDstarz$ (produced by a $b \to c$ transition) and 
$\DorDstarzb$ (produced by a colour-suppressed $b \to u$ transition) states.
If the final state is chosen so that both $\DorDstarz$ and $\DorDstarzb$ 
can contribute, the two amplitudes interfere,
and the resulting observables are sensitive to $\gamma$, 
the relative weak phase between 
the two $\B$ decay amplitudes~\cite{Bigi:1988ym}.
Various methods have been proposed to exploit this interference,
including those where the neutral $D$ meson is reconstructed 
as a $\CP$ eigenstate (GLW)~\cite{Gronau:1990ra,Gronau:1991dp},
in a suppressed final state (ADS)~\cite{Atwood:1996ci,Atwood:2000ck},
or in a self-conjugate three-body final state, 
such as $\KS \pi^+\pi^-$ (Dalitz)~\cite{Giri:2003ty,Poluektov:2004mf}.
It should be emphasised that while each method 
differs in the choice of $D$ decay,
they are all sensitive to the same parameters of the $B$ decay,
and can be considered as variations of the same technique.

Consider the case of $\Bmp \to D \Kmp$,
with $D$ decaying to a final state $f$,
which is accessible to both $\Dz$ and $\Dzb$.
We can write the decay rates for $\Bm$ and $\Bp$ ($\Gamma_\mp$), 
the charge averaged rate ($\Gamma = (\Gamma_- + \Gamma_+)/2$)
and the charge asymmetry 
(${\cal A} = (\Gamma_- - \Gamma_+)/(\Gamma_- + \Gamma_+)$, see Eq.~(\ref{eq:cp_uta:pra})) as 
\begin{eqnarray}
  \label{eq:cp_uta:dk:rate_def}
  \Gamma_\mp  & \propto & 
  r_B^2 + r_D^2 + 2 r_B r_D \cos \left( \delta_B + \delta_D \mp \gamma \right), \\
  \label{eq:cp_uta:dk:av_rate_def}
  \Gamma & \propto &  
  r_B^2 + r_D^2 + 2 r_B r_D \cos \left( \delta_B + \delta_D \right) \cos \left( \gamma \right), \\
  \label{eq:cp_uta:dk:acp_def}
  {\cal A} & = & 
  \frac{
    2 r_B r_D \sin \left( \delta_B + \delta_D \right) \sin \left( \gamma \right)
  }{
    r_B^2 + r_D^2 + 2 r_B r_D \cos \left( \delta_B + \delta_D \right) \cos \left( \gamma \right),  
  }
\end{eqnarray}
where the ratio of $\B$ decay amplitudes\footnote{
  Note that here we use the notation $r_B$ to denote the ratio
  of $\B$ decay amplitudes, 
  whereas in Sec.~\ref{sec:cp_uta:notations:non_cp:dstarpi} 
  we used, \eg, $R_{D\pi}$, for a rather similar quantity.
  The reason is that here we need to be concerned also with 
  $D$ decay amplitudes,
  and so it is convenient to use the subscript to denote the decaying particle.
  Hopefully, using $r$ in place of $R$ will help reduce potential confusion.
} 
is usually defined to be less than one,
\begin{equation}
  \label{eq:cp_uta:dk:rb_def}
  r_B = 
  \frac{
    \left| A\left( \Bm \to \Dzb K^- \right) \right|
  }{
    \left| A\left( \Bm \to \Dz  K^- \right) \right|
  },
\end{equation}
and the ratio of $D$ decay amplitudes is correspondingly defined by
\begin{equation}
  \label{eq:cp_uta:dk:rd_def}
  r_D = 
  \frac{
    \left| A\left( \Dz  \to f \right) \right|
  }{
    \left| A\left( \Dzb \to f \right) \right|
  }.
\end{equation}
The strong phase differences between the $\B$ and $D$ decay amplitudes 
are given by $\delta_B$ and $\delta_D$, respectively.
The values of $r_D$ and $\delta_D$ depend on the final state $f$:
for the GLW analysis, $r_D = 1$ and $\delta_D$ is trivial (either zero or $\pi$),
in the Dalitz plot analysis $r_D$ and $\delta_D$ vary across the Dalitz plot,
and depend on the $D$ decay model used,
for the ADS analysis, the values of $r_D$ and $\delta_D$ are not trivial.

Note that, for given values of $r_B$ and $r_D$, 
the maximum size of ${\cal A}$ (at $\sin \left( \delta_B + \delta_D \right) = 1$)
is $2 r_B r_D \sin \left( \gamma \right) / \left( r_B^2 + r_D^2 \right)$.
Thus even for $D$ decay modes with small $r_D$, 
large asymmetries, and hence sensitivity to $\gamma$, 
may occur for $B$ decay modes with similar values of $r_B$.
For this reason, the ADS analysis of the decay $B^\mp \to D \pi^\mp$ 
is also of interest.

In the GLW analysis, the measured quantities are the 
partial rate asymmetry, and the charge averaged rate,
which are measured both for $\CP$-even and $\CP$-odd $D$ decays.
The former is defined as 
\begin{equation}
  \label{eq:cp_uta:dk:glw-rdef}
  R_{\CP} = 
  \frac{2 \, \Gamma \left( \Bm \to D_{\CP} \Km  \right)}
  {\Gamma\left( \Bm \to \Dz \Km \right)} \, .
\end{equation}
It is experimentally convenient to measure $R_{\CP}$ using a double ratio,
\begin{equation}
  \label{eq:cp_uta:dk:double_ratio}
  R_{\CP} = 
  \frac{
    \Gamma\left( \Bm \to D_{\CP} \Km  \right) \, / \, \Gamma\left( \Bm \to \Dz \Km \right)
  }{
    \Gamma\left( \Bm \to D_{\CP} \pim \right) \, / \, \Gamma\left( \Bm \to \Dz \pim \right)
  }
\end{equation}
that is normalized both to the rate for the favoured $\Dz \to \Km\pip$ decay, 
and to the equivalent quantities for $\Bm \to D\pim$ decays
(charge conjugate modes are implicitly included in 
Eq.~(\ref{eq:cp_uta:dk:glw-rdef}) and~(\ref{eq:cp_uta:dk:double_ratio})).
In this way the constant of proportionality drops out of 
Eq.~(\ref{eq:cp_uta:dk:av_rate_def}).
Eq.~(\ref{eq:cp_uta:dk:double_ratio}) is exact in the limit that the
contribution of the $b \to u$ decay amplitude to $\Bm \to D \pim$ vanishes and
when the flavour-specific rates $\Gamma\left( \Bm \to \Dz h^- \right)$ ($h =
\pi,K$) are determined using appropriately flavour-specific $D$ decays.
The direct \CP\ asymmetry is defined as
\begin{equation}
  \label{eq:cp_uta:dk:glw-adef}
  A_{\CP} = \frac{
    \Gamma\left(\Bm\to D_{\CP}\Km\right) - \Gamma\left(\Bp\to D_{\CP}\Kp\right)
  }{
    \Gamma\left(\Bm\to D_{\CP}\Km\right) + \Gamma\left(\Bp\to D_{\CP}\Kp\right)
  } \, .
\end{equation}

For the ADS analysis, using a suppressed $D \to f$ decay,
the measured quantities are again the partial rate asymmetry, 
and the charge averaged rate.
In this case it is sufficient to measure the rate in a single ratio
(normalized to the favoured $D \to \bar{f}$ decay)
since detection systematics cancel naturally;
the observed quantity is then
\begin{equation}
  \label{eq:cp_uta:dk:r_ads}
  R_{\rm ADS} = 
  \frac{
    \Gamma \left( \Bm \to \left[ f \right]_D \Km \right)
  }{
    \Gamma \left( \Bm \to \left[ \bar{f} \right]_D \Km \right)
  } \, ,
\end{equation}
where inclusion of charge conjugate modes is implied.
The direct \CP\ asymmetry is defined as
\begin{equation}
  \label{eq:cp_uta:dk:a_ads}
  R_{\rm ADS} = 
  \frac{
    \Gamma\left(\Bm\to\left[f\right]_D\Km\right)-
    \Gamma\left(\Bp\to\left[f\right]_D\Kp\right)
  }{
    \Gamma\left(\Bm\to\left[f\right]_D\Km\right)+
    \Gamma\left(\Bp\to\left[f\right]_D\Kp\right)
  }.
\end{equation}
In the ADS analysis, there are an additional two unknowns ($r_D$ and $\delta_D$)
compared to the GLW case.  
However, the value of $r_D$ can be measured using 
decays of $D$ mesons of known flavour.

In the Dalitz plot analysis,
once a model is assumed for the $D$ decay, 
which gives the values of $r_D$ and $\delta_D$ across the Dalitz plot,
it is possible to perform a simultaneous fit to the $B^+$ and $B^-$ samples 
and directly extract $\gamma$, $r_B$ and $\delta_B$.
However, the uncertainties on the phases depend approximately inversely on $r_B$.
Furthermore, $r_B$ is positive definite (and small), 
and therefore tends to be overestimated,
which can lead to an underestimation of the uncertainty.
Some statistical treatment is necessary to correct for this bias.
An alternative approach is to extract from the data the ``Cartesian''
variables
\begin{equation}
  \left( x_\pm, y_\pm \right) = 
  \left( \Re(r_B e^{i(\delta_B\pm\gamma)}), \Im(r_B e^{i(\delta_B\pm\gamma)}) \right) = 
  \left( r_B \cos(\delta_B\pm\gamma), r_B \sin(\delta_B\pm\gamma) \right).
\end{equation}
These are (a) approximately statistically uncorrelated 
and (b) almost Gaussian.
The pairs of variables $\left( x_\pm, y_\pm \right)$ can be extracted
from independent fits of the $B^\pm$ data samples.
Use of these variables makes the combination of results much simpler.

However, if the Dalitz plot is effectively dominated by one $CP$ state,
there will be additional sensitivity to $\gamma$ in the numbers of events
in the $B^\pm$ data samples.
This can be taken into account in various ways.
One possibility is to extract GLW-like variables 
in addition to the $\left( x_\pm, y_\pm \right)$ parameters.
An alternative proceeds by defining $z_\pm = x_\pm + i y_\pm$
and $x_0 = - \int \Re \left[ f(s_1,s_2)f^*(s_2,s_1) \right] ds_1ds_2$,
where $s_1, s_2$ are the coordinates of invariant mass squared that
define the Dalitz plot and $f$ is the complex amplitude for $D$ decay
as a function of the Dalitz plot coordinates.\footnote{
  The $x_0$ parameter is closely related to the $c_i$ parameters of 
  the model dependent Dalitz plot analysis~\cite{Giri:2003ty,Bondar:2005ki,Bondar:2008hh},
  and the coherence factor of inclusive ADS-type analyses~\cite{Atwood:2003mj},
  integrated over the entire Dalitz plot.
}
The fitted parameters ($\rho^\pm, \theta^\pm$) are then defined by
\begin{equation}
  \rho^\pm e^{i \theta^\pm} = z_\pm - x_0 \, .
\end{equation}
Note that the yields of $B^\pm$ decays are proportional 
to $1 + (\rho^\pm)^2 - (x_0)^2$. 
This choice of variables has been used by \babar\ in the analysis of
$\Bmp \to D\Kmp$ with $D \to \pi^+\pi^-\pi^0$~\cite{Aubert:2007ii};
for this $D$ decay, $x_0 = 0.850$.

The relations between the measured quantities and the
underlying parameters are summarized in Table~\ref{tab:cp_uta:notations:dk}.
Note carefully that the hadronic factors $r_B$ and $\delta_B$ 
are different, in general, for each $\B$ decay mode.

\begin{table}[htb]
  \begin{center} 
    \caption{
      Summary of relations between measured and physical parameters 
      in GLW, ADS and Dalitz analyses of $\B \to \DorDstar K^{(*)}$.
    }
    \vspace{0.2cm}
    \setlength{\tabcolsep}{1.0pc}
    \begin{tabular}{cc} \hline 
      \mc{2}{c}{GLW analysis} \\
      $R_{\CP\pm}$ & $1 + r_B^2 \pm 2 r_B \cos \left( \delta_B \right) \cos \left( \gamma \right)$ \\
      $A_{\CP\pm}$ & $\pm 2 r_B \sin \left( \delta_B \right) \sin \left( \gamma \right) / R_{\CP\pm}$ \\
      \hline
      \mc{2}{c}{ADS analysis} \\
      $R_{\rm ADS}$ & $r_B^2 + r_D^2 + 2 r_B r_D \cos \left( \delta_B + \delta_D \right) \cos \left( \gamma \right)$ \\
      $A_{\rm ADS}$ & $2 r_B r_D \sin \left( \delta_B + \delta_D \right) \sin \left( \gamma \right) / R_{\rm ADS}$ \\
      \hline
      \mc{2}{c}{Dalitz analysis ($D \to \KS \pi^+\pi^-$)} \\
      $x_\pm$ & $r_B \cos(\delta_B\pm\gamma)$ \\
      $y_\pm$ & $r_B \sin(\delta_B\pm\gamma)$ \\
      \hline
      \mc{2}{c}{Dalitz analysis ($D \to \pi^+\pi^-\pi^0$)} \\
      $\rho^\pm$ & $|z_\pm - x_0|$ \\
      $\theta^\pm$ & $\tan^{-1}(\Im(z_\pm)/(\Re(z_\pm) - x_0))$ \\
      \hline
    \end{tabular}
    \label{tab:cp_uta:notations:dk}
  \end{center}
\end{table}

\mysubsection{Common inputs and error treatment
}
\label{sec:cp_uta:common_inputs}

The common inputs used for rescaling are listed in 
Table~\ref{tab:cp_uta:common_inputs}.
The $\Bz$ lifetime ($\tau(\Bz)$) and mixing parameter ($\Delta m_d$)
averages are provided by the HFAG Lifetimes and Oscillations 
subgroup (Sec.~\ref{sec:life_mix}).
The fraction of the perpendicularly polarized component 
($\left| A_{\perp} \right|^2$) in $\B \to \jpsi \Kstar(892)$ decays,
which determines the $\CP$ composition, 
is averaged from results by 
\babar~\cite{Aubert:2007hz} and \belle~\cite{Itoh:2005ks}.
See also HFAG $B$ to Charm Decay Parameters subgroup 
(Sec.~\ref{sec:BtoCharm}).

At present, we only rescale to a common set of input parameters
for modes with reasonably small statistical errors
($b \to c\bar{c}s$ transitions).
Correlated systematic errors are taken into account
in these modes as well.
For all other modes, the effect of such a procedure is 
currently negligible.

\begin{table}[htb]
  \begin{center}
    \caption{
      Common inputs used in calculating the averages.
    }
    \vspace{0.2cm}
    \setlength{\tabcolsep}{1.0pc}
    \begin{tabular}{cc} \hline 
      $\tau(\Bz)$ $({\rm ps})$  & $1.530 \pm 0.008$  \\
      $\Delta m_d$ $({\rm ps}^{-1})$ & $0.507 \pm 0.004$ \\
      $\left| A_{\perp} \right|^2 (\jpsi \Kstar)$ & $0.219 \pm 0.009$ \\
      \hline
    \end{tabular}
    \label{tab:cp_uta:common_inputs}
  \end{center}
\end{table}

As explained in Sec.~\ref{sec:intro},
we do not apply a rescaling factor on the error of an average
that has $\chi^2/\dof > 1$ 
(unlike the procedure currently used by the PDG~\cite{PDG_2010}).
We provide a confidence level of the fit so that
one can know the consistency of the measurements included in the average,
and attach comments in case some care needs to be taken in the interpretation.
Note that, in general, results obtained from data samples with low statistics
will exhibit some non-Gaussian behaviour.
We average measurements with asymmetric errors 
using the PDG~\cite{PDG_2010} prescription.
In cases where several measurements are correlated
(\eg\ $S_f$ and $C_f$ in measurements of time-dependent $\CP$ violation
in $B$ decays to a particular $\CP$ eigenstate)
we take these into account in the averaging procedure
if the uncertainties are sufficiently Gaussian.
For measurements where one error is given, 
it represents the total error, 
where statistical and systematic uncertainties have been added in quadrature.
If two errors are given, the first is statistical and the second systematic.
If more than two errors are given,
the origin of the additional uncertainty will be explained in the text.


\clearpage
\mysubsection{Time-dependent asymmetries in $b \to c\bar{c}s$ transitions
}
\label{sec:cp_uta:ccs}

\mysubsubsection{Time-dependent $\CP$ asymmetries in $b \to c\bar{c}s$ decays to $\CP$ eigenstates
}
\label{sec:cp_uta:ccs:cp_eigen}

\begin{figure}[htb]
  \begin{center}
    \resizebox{0.51\textwidth}{!}{
      \includegraphics{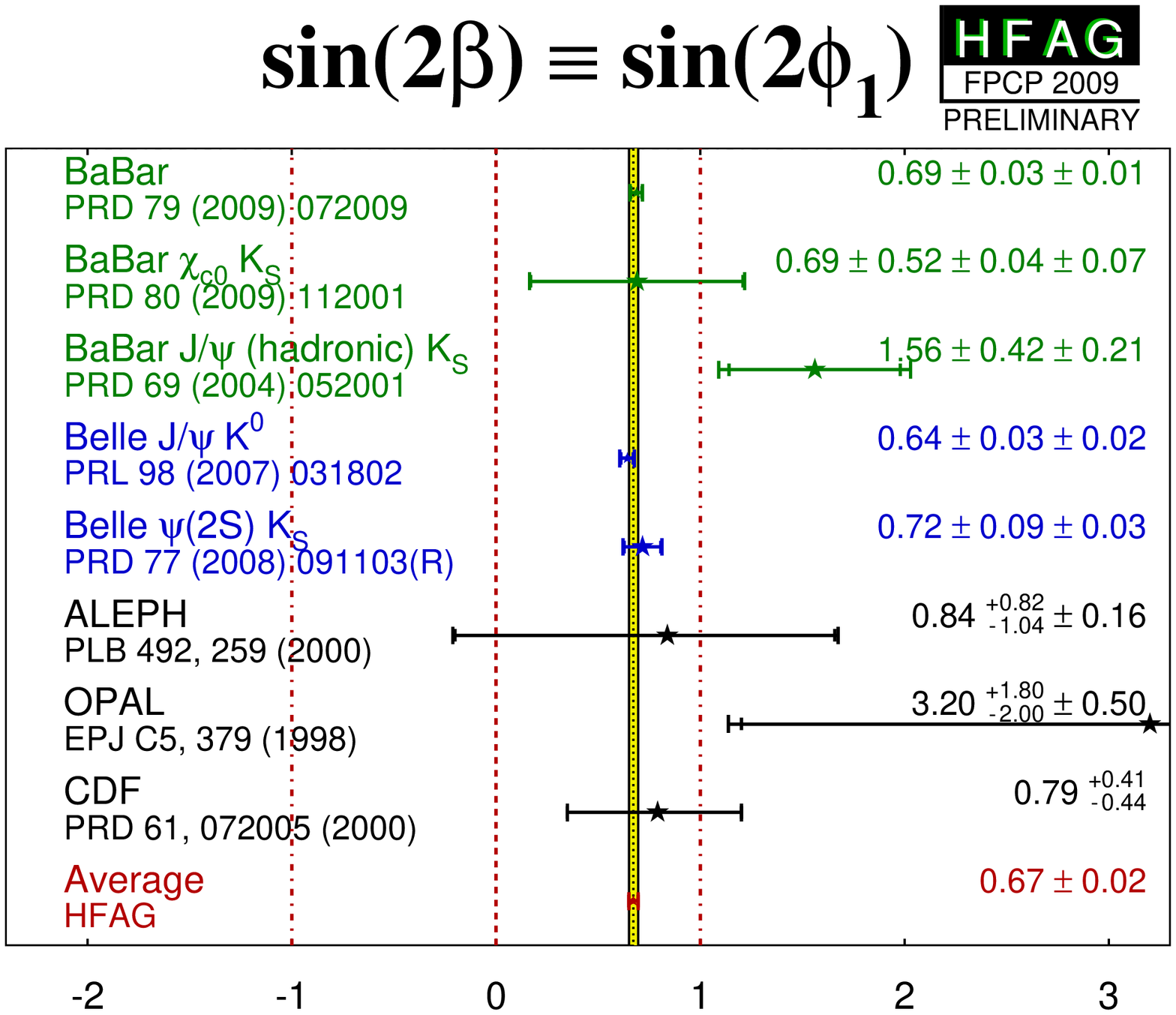}
    }
    \hfill
    \resizebox{0.48\textwidth}{!}{
      \includegraphics{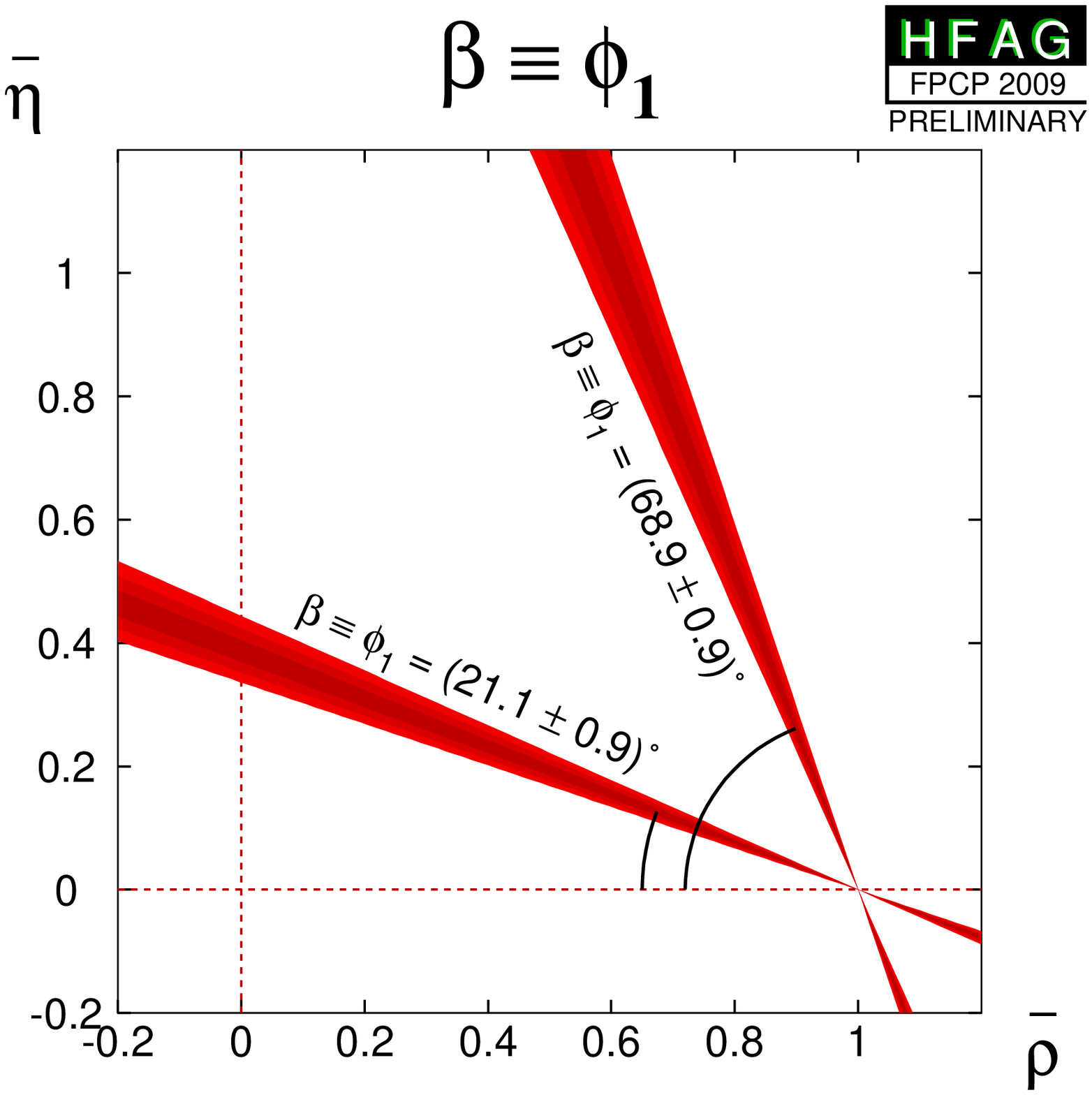}
    }
  \end{center}
  \vspace{-0.5cm}
  \caption{
    (Left) Average of measurements of $S_{b \to c\bar c s}$.
    (Right) Constraints on the $(\rhobar,\etabar)$ plane,
    obtained from the average of $-\etacp S_{b \to c\bar c s}$ 
    and Eq.~\ref{eq:cp_uta:sin2beta}.
  }
  \label{fig:cp_uta:ccs}
\end{figure}

In the Standard Model, the time-dependent parameters for
$b \to c\bar c s$ transitions are predicted to be: 
$S_{b \to c\bar c s} = - \etacp \sin(2\beta)$,
$C_{b \to c\bar c s} = 0$ to very good accuracy.
The averages for $-\etacp S_{b \to c\bar c s}$ and $C_{b \to c\bar c s}$
are provided in Table~\ref{tab:cp_uta:ccs}.
The averages for $-\etacp S_{b \to c\bar c s}$ 
are shown in Fig.~\ref{fig:cp_uta:ccs}.

Both \babar\  and \belle\ have used the $\etacp = -1$ modes
$\jpsi \KS$, $\psi(2S) \KS$, $\chi_{c1} \KS$ and $\eta_c \KS$, 
as well as $\jpsi \KL$, which has $\etacp = +1$
and $\jpsi K^{*0}(892)$, which is found to have $\etacp$ close to $+1$
based on the measurement of $\left| A_\perp \right|$ 
(see Sec.~\ref{sec:cp_uta:common_inputs}).
ALEPH, OPAL and CDF used only the $\jpsi \KS$ final state.
In the latest result from \belle~\cite{Chen:2006nk}, 
only $\jpsi \KS$ and $\jpsi \KL$ are used,
while results from $\psi(2{\rm S}) \KS$ have been presented
separately~\cite{Abe:2007gj}.
\babar\ have also determined the \CP-violation parameters of the
$\Bz\to\chi_{c0} \KS$ decay from the time-dependent Dalitz plot analysis of
$\Bz \to \pi^+\pi^-\KS$ (see subsection~\ref{sec:cp_uta:qqs:dp})~\cite{Aubert:2009me}.
A breakdown of results in each charmonium-kaon final state is given in 
Table~\ref{tab:cp_uta:ccs-BF}.

\begin{table}[htb]
	\begin{center}
		\caption{
                        $S_{b \to c\bar c s}$ and $C_{b \to c\bar c s}$.
                }
		\vspace{0.2cm}
		\setlength{\tabcolsep}{0.0pc}
\begin{tabular*}{\textwidth}{@{\extracolsep{\fill}}lrccc} \hline
      \mc{2}{l}{Experiment} & $N(B\bar{B})$ & $- \etacp S_{b \to c\bar c s}$ & $C_{b \to c\bar c s}$ \\
      \hline
	\babar & \cite{:2009yr} & 465M & $0.687 \pm 0.028 \pm 0.012$ & $0.024 \pm 0.020 \pm 0.016$ \\
	\babar\ $\chi_{c0} \KS$ & \cite{Aubert:2009me} & 383M & $0.690 \pm 0.520 \pm 0.040 \pm 0.070$ & $-0.290 \,^{+0.530}_{-0.440} \pm 0.030 \pm 0.050$ \\
	\babar\ $J/\psi \KS$ ($^{*}$) & \cite{Aubert:2003xn} & 88M & $1.560 \pm 0.420 \pm 0.210$ &  \textendash{} \\
	\belle $J/\psi K^0$ & \cite{Chen:2006nk} & 535M & $0.642 \pm 0.031 \pm 0.017$ & $-0.018 \pm 0.021 \pm 0.014$ \\
	\belle $\psi(2{\rm S}) \KS$ & \cite{Abe:2007gj} & 657M & $0.718 \pm 0.090 \pm 0.031$ & $-0.039 \pm 0.069 \pm 0.0$ \\
	\mc{3}{l}{\bf \boldmath $\B$ factory average} & $0.672 \pm 0.023$ & $0.004 \pm 0.019$ \\
	\mc{3}{l}{\small Confidence level} & {\small $0.30$} & {\small $0.51$} \\
        \hline
        ALEPH & \cite{Barate:2000tf} & \textendash{} & $0.84 \, ^{+0.82}_{-1.04} \pm 0.16$ &  \textendash{} \\
        OPAL  & \cite{Ackerstaff:1998xz} & \textendash{} & $3.2 \, ^{+1.8}_{-2.0} \pm 0.5$ &  \textendash{} \\
        CDF   & \cite{Affolder:1999gg} & \textendash{} & $0.79 \, ^{+0.41}_{-0.44}$ &  \textendash{} \\
        \mc{3}{l}{\bf Average} & $0.673 \pm 0.023$ & $0.004 \pm 0.019 $ \\
		\hline
		\end{tabular*}
                \label{tab:cp_uta:ccs}
        \end{center}
$^{*}$ {\small This result uses "{\it hadronic and previously unused muonic decays of the $J/\psi$}". We neglect a small possible correlation of this result with the main \babar\ result~\cite{:2009yr} that could be caused by reprocessing of the data.}
\end{table}

\begin{table}[htb]
	\begin{center}
		\caption{
                        Breakdown of $B$ factory results on $S_{b \to c\bar c s}$ and $C_{b \to c\bar c s}$.
                }
		\vspace{0.2cm}
		\setlength{\tabcolsep}{0.0pc}
		\begin{tabular*}{\textwidth}{@{\extracolsep{\fill}}lrccc} \hline
        \mc{2}{l}{Mode} & $N(B\bar{B})$ & $- \etacp S_{b \to c\bar c s}$ & $C_{b \to c\bar c s}$ \\
        \hline
        \mc{5}{c}{\babar} \\
        $J/\psi \KS$ & \cite{:2009yr} & 465M & $0.657 \pm 0.036 \pm 0.012$ & $0.026 \pm 0.025 \pm 0.016$ \\
        $J/\psi \KL$ & \cite{:2009yr} & 465M & $0.694 \pm 0.061 \pm 0.031$ & $-0.033 \pm 0.050 \pm 0.027$ \\
        {\bf \boldmath $J/\psi K^0$} & \cite{:2009yr} & 465M & $0.666 \pm 0.031 \pm 0.013$ & $0.016 \pm 0.023 \pm 0.018$ \\
        $\psi(2S) \KS$ & \cite{:2009yr} & 465M & $0.897 \pm 0.100 \pm 0.036$ & $0.089 \pm 0.076 \pm 0.020$ \\
        $\chi_{c1} \KS$ & \cite{:2009yr} & 465M & $0.614 \pm 0.160 \pm 0.040$ & $0.129 \pm 0.109 \pm 0.025$ \\
        $\eta_c \KS$ & \cite{:2009yr} & 465M & $0.925 \pm 0.160 \pm 0.057$ & $0.080 \pm 0.124 \pm 0.029$ \\
        $\jpsi K^{*0}(892)$ & \cite{:2009yr} & 465M & $0.601 \pm 0.239 \pm 0.087$ & $0.025 \pm 0.083 \pm 0.054$ \\
        {\bf All} & \cite{:2009yr} & 465M & $0.687 \pm 0.028 \pm 0.012$ & $0.024 \pm 0.020 \pm 0.016$ \\
	\hline
	\mc{5}{c}{\bf \belle} \\
        $J/\psi \KS$ & \cite{Chen:2006nk} & 535M & $0.643 \pm 0.038_{\rm stat}$ & $0.001 \pm 0.028_{\rm stat}$ \\
        $J/\psi \KL$ & \cite{Chen:2006nk} & 535M & $0.641 \pm 0.057_{\rm stat}$ &$-0.045 \pm 0.033_{\rm stat}$ \\
        {\bf \boldmath $J/\psi K^0$} & \cite{Chen:2006nk} & 535M & $0.642 \pm 0.031 \pm 0.017$ & $-0.018 \pm 0.021 \pm 0.014$ \\
	$\psi(2{\rm S}) \KS$ & \cite{Abe:2007gj} & 657M & $0.718 \pm 0.090 \pm 0.033$ & $-0.039 \pm 0.069 \pm 0.049$ \\
        {\bf All} & \cite{Abe:2007gj} & \textendash{} & $0.650 \pm 0.029 \pm 0.015$ & \textendash{} \\
	\hline
        \mc{3}{l}{\bf \boldmath $J/\psi K^0$ average} & $0.655 \pm 0.0244$ & $-0.003 \pm 0.020$ \\ 
        \mc{3}{l}{\bf \boldmath $\psi(2{\rm S}) K^0$ average} & $0.798 \pm 0.071$ & $0.032 \pm 0.060$ \\ 
		\hline
		\end{tabular*}
                \label{tab:cp_uta:ccs-BF}
        \end{center}
\end{table}

It should be noted that, while the uncertainty in the average for 
$-\etacp S_{b \to c\bar c s}$ is still limited by statistics,
that for $C_{b \to c\bar c s}$ is close to being dominated by systematics.
This occurs due to the possible effect of tag side interference on the
$C_{b \to c\bar c s}$ measurement, an effect which is correlated between
the different experiments.
Understanding of this effect may continue to improve in future,
allowing the uncertainty to reduce.

From the average for $-\etacp S_{b \to c\bar c s}$ above, 
we obtain the following solutions for $\beta$
(in $\left[ 0, \pi \right]$):
\begin{equation}
  \beta = \left( 21.1 \pm 0.9 \right)^\circ
  \hspace{5mm}
  {\rm or}
  \hspace{5mm}
  \beta = \left( 68.9 \pm 0.9 \right)^\circ
  \label{eq:cp_uta:sin2beta}
\end{equation}
In radians, these values are 
$\beta = \left( 0.368 \pm 0.016 \right)$, 
$\beta = \left( 1.203 \pm 0.016 \right)$.

This result gives a precise constraint on the $(\rhobar,\etabar)$ plane,
as shown in Fig.~\ref{fig:cp_uta:ccs}.
The measurement is in remarkable agreement with other constraints from 
$\CP$ conserving quantities, 
and with $\CP$ violation in the kaon system, in the form of the parameter $\epsilon_K$.
Such comparisons have been performed by various phenomenological groups,
such as CKMfitter~\cite{Charles:2004jd} 
and UTFit~\cite{Bona:2005vz}.

\mysubsubsection{Time-dependent transversity analysis of $\Bz \to J/\psi K^{*0}$
}
\label{sec:cp_uta:ccs:vv}

$\B$ meson decays to the vector-vector final state $J/\psi K^{*0}$
are also mediated by the $b \to c \bar c s$ transition.
When a final state which is not flavour-specific ($K^{*0} \to \KS \pi^0$) is used,
a time-dependent transversity analysis can be performed 
allowing sensitivity to both 
$\sin(2\beta)$ and $\cos(2\beta)$~\cite{Dunietz:1990cj}.
Such analyses have been performed by both $\B$ factory experiments.
In principle, the strong phases between the transversity amplitudes
are not uniquely determined by such an analysis, 
leading to a discrete ambiguity in the sign of $\cos(2\beta)$.
The \babar\ collaboration resolves 
this ambiguity using the known variation~\cite{Aston:1987ir}
of the P-wave phase (fast) relative to the S-wave phase (slow) 
with the invariant mass of the $K\pi$ system 
in the vicinity of the $K^*(892)$ resonance. 
The result is in agreement with the prediction from 
$s$ quark helicity conservation,
and corresponds to Solution II defined by Suzuki~\cite{Suzuki:2001za}.
We use this phase convention for the averages given in 
Table~\ref{tab:cp_uta:ccs:psi_kstar}.

\begin{table}[htb]
	\begin{center}
		\caption{
			Averages from $\Bz \to J/\psi K^{*0}$ transversity analyses.
		}
		\vspace{0.2cm}
		\setlength{\tabcolsep}{0.0pc}
		\begin{tabular*}{\textwidth}{@{\extracolsep{\fill}}lrcccc} \hline
		\mc{2}{l}{Experiment} & $N(B\bar{B})$ & $\sin 2\beta$ & $\cos 2\beta$ & Correlation \\
		\hline
	\babar & \cite{Aubert:2004cp} & 88M & $-0.10 \pm 0.57 \pm 0.14$ & $3.32 ^{+0.76}_{-0.96} \pm 0.27$ & $-0.37$ \\
	\belle & \cite{Itoh:2005ks} & 275M & $0.24 \pm 0.31 \pm 0.05$ & $0.56 \pm 0.79 \pm 0.11$ & $0.22$ \\
	\mc{3}{l}{\bf Average} & $0.16 \pm 0.28$ & $1.64 \pm 0.62$ &  \hspace{-8mm} {\small uncorrelated averages}  \\
        \mc{3}{l}{\small Confidence level} & {\small $0.61~(0.5\sigma)$} & {\small $0.03~(2.2\sigma)$} & \\
		\hline
		\end{tabular*}
		\label{tab:cp_uta:ccs:psi_kstar}
	\end{center}
\end{table}

At present the results are dominated by 
large and non-Gaussian statistical errors,
and exhibit significant correlations.
We perform uncorrelated averages, 
the interpretation of which has to be done with the greatest care. 
Nonetheless, it is clear that $\cos(2\beta)>0$ is preferred 
by the experimental data in $J/\psi \Kstar$.
[\babar~\cite{Aubert:2004cp} 
find a confidence level for $\cos(2\beta)>0$ of $89\%$.]

\mysubsubsection{Time-dependent $\CP$ asymmetries in $\Bz \to \Dstarp \Dstarm \KS$ decays
}
\label{sec:cp_uta:ccs:DstarDstarKs}

\begin{figure}[htb]
  \begin{center}
    \begin{tabular}{c@{\hspace{-1mm}}c@{\hspace{-1mm}}c}
      \resizebox{0.32\textwidth}{!}{
        \includegraphics{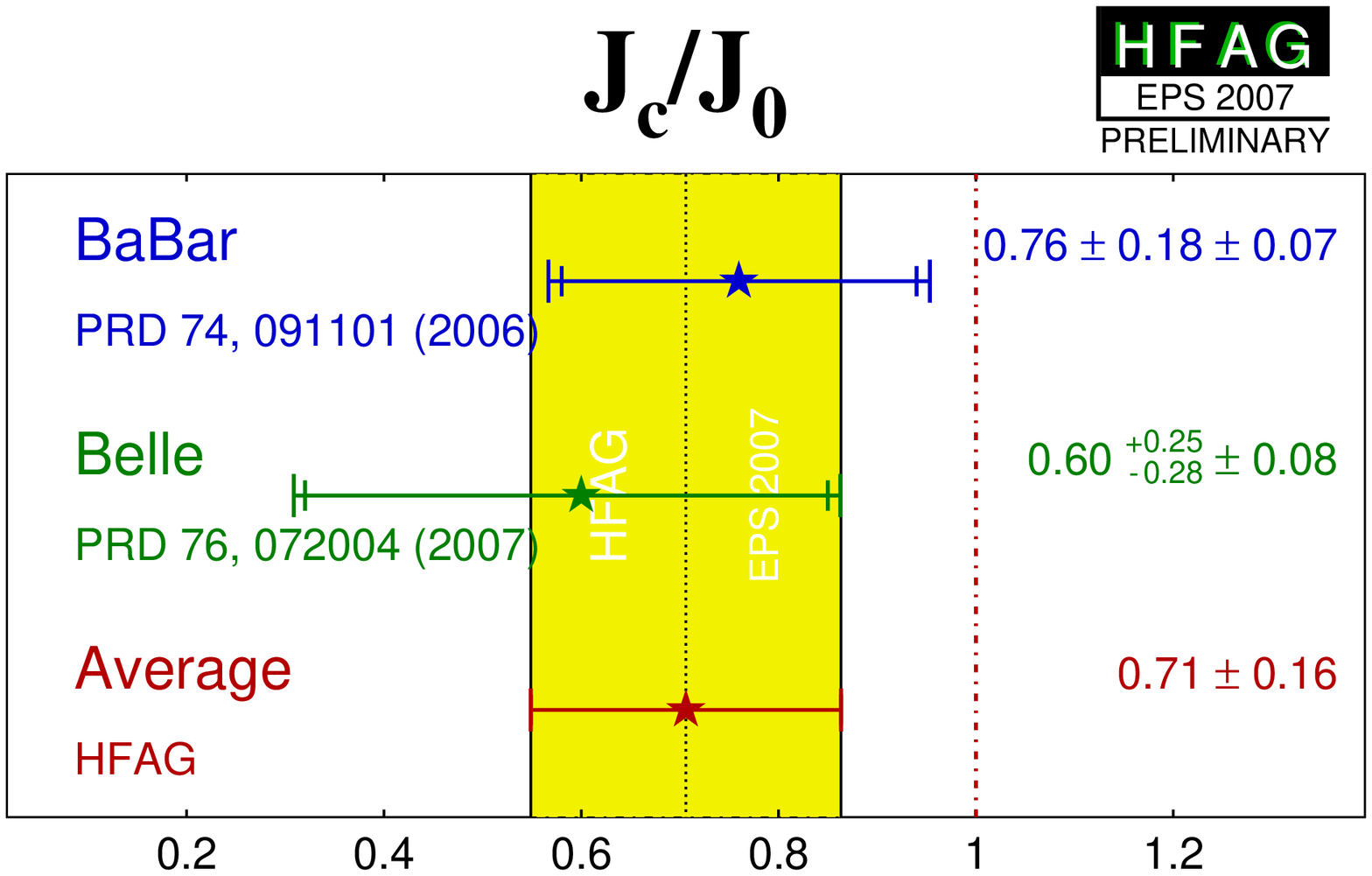}
      }
      &
      \resizebox{0.32\textwidth}{!}{
        \includegraphics{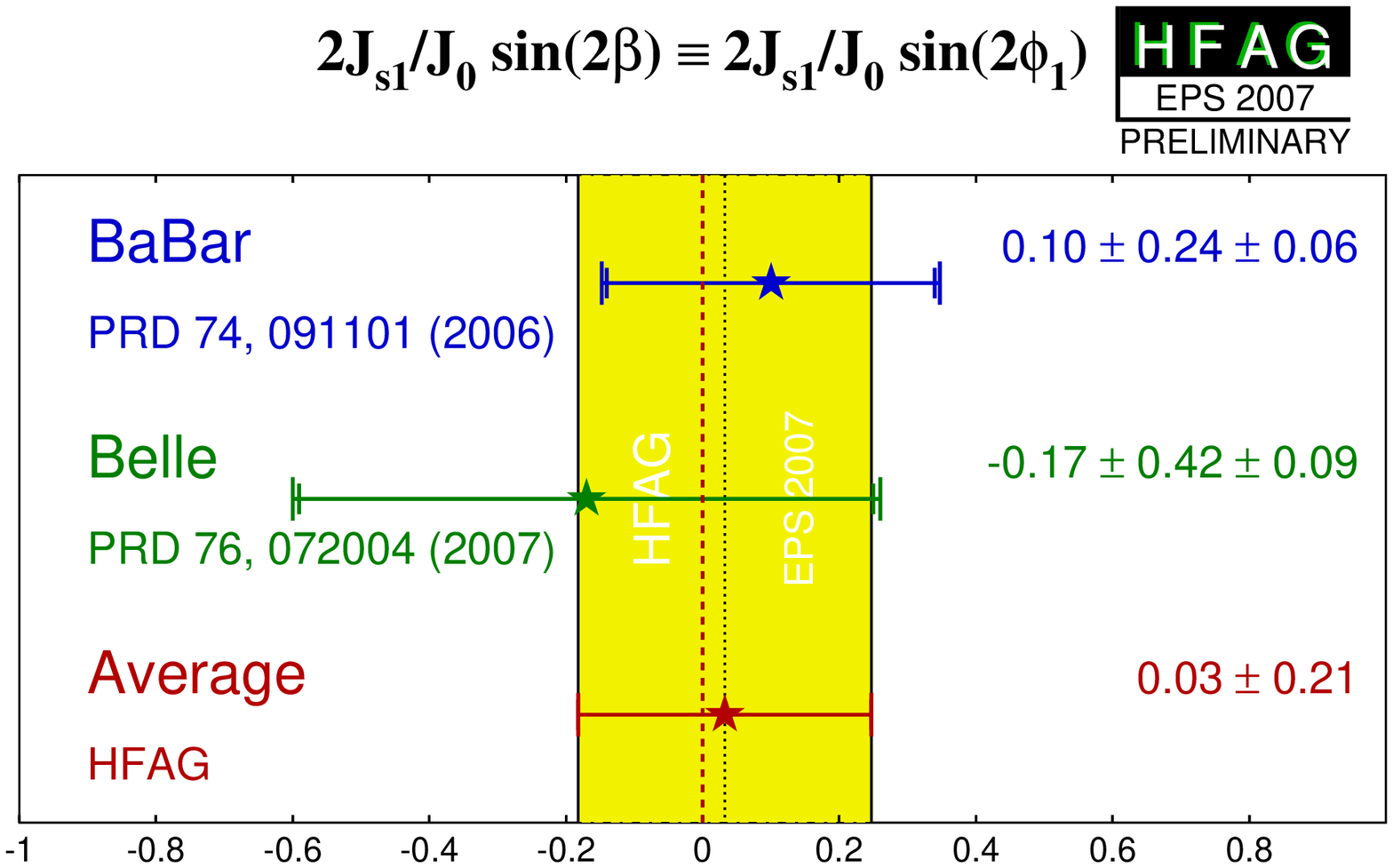}
      }
      &
      \resizebox{0.32\textwidth}{!}{
        \includegraphics{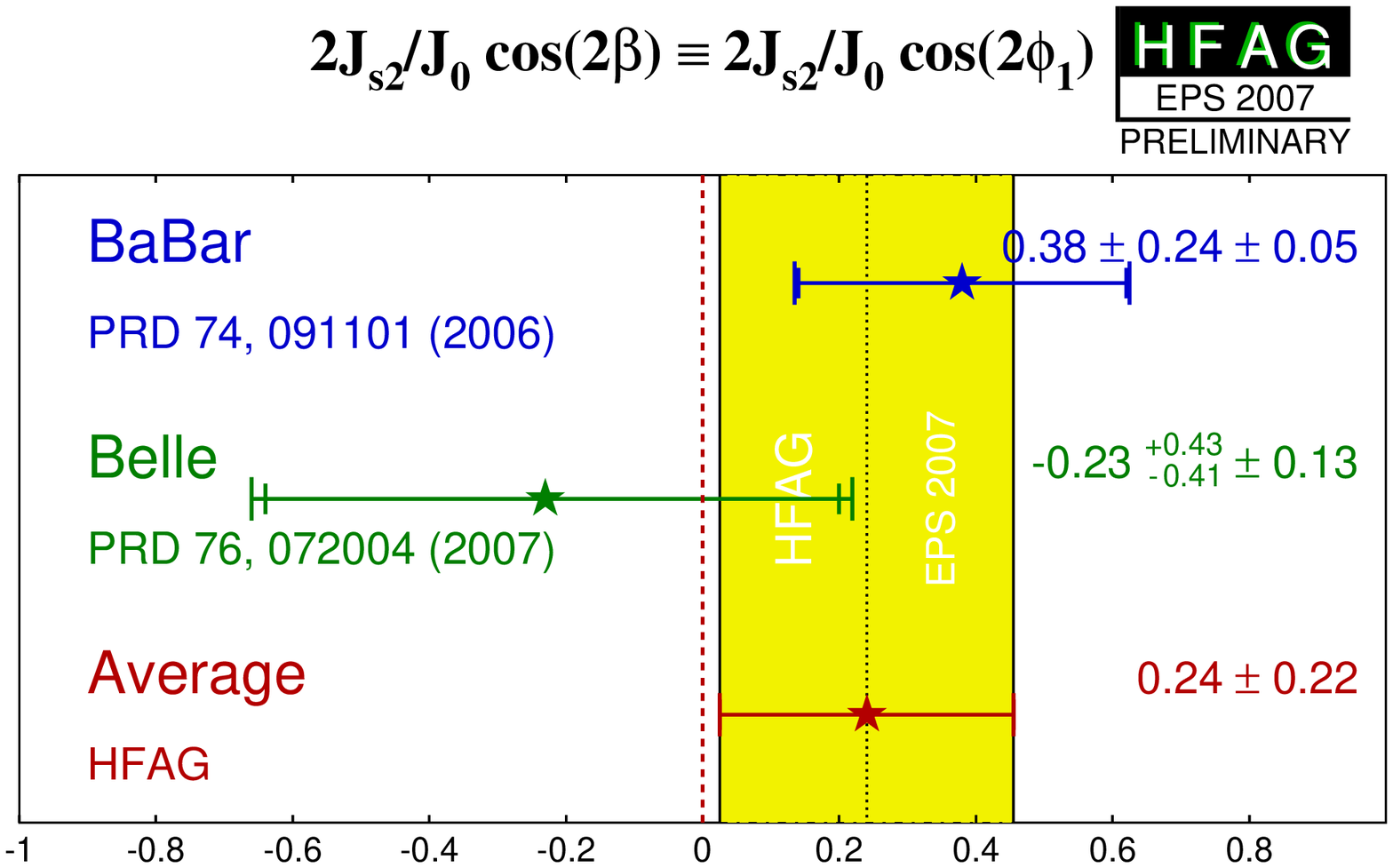}
      }
    \end{tabular}
  \end{center}
  \vspace{-0.8cm}
  \caption{
    Averages of 
    (left) $(J_c/J_0)$, (middle) $(2J_{s1}/J_0) \sin(2\beta)$ and 
    (right) $(2J_{s2}/J_0) \cos(2\beta)$
    from time-dependent analyses of $\Bz \to \Dstarp \Dstarm \KS$ decays.
  }
  \label{fig:cp_uta:ccs:dstardstarks}
\end{figure}

Both \babar~\cite{Aubert:2006fh} and \belle~\cite{Dalseno:2007hx} have performed
time-dependent analyses of the $\Bz \to \Dstarp \Dstarm \KS$ decay,
to obtain information on the sign of $\cos(2\beta)$.
More information can be found in 
Sec.~\ref{sec:cp_uta:notations:dalitz:dstardstarks}.
The results are shown in Table~\ref{tab:cp_uta:ccs:dstardstarks}, 
and Fig.~\ref{fig:cp_uta:ccs:dstardstarks}.

\begin{table}[htb]
	\begin{center}
		\caption{
                        Results from time-dependent analysis of $\Bz \to \Dstarp \Dstarm \KS$.
		}
		\vspace{0.2cm}
		\setlength{\tabcolsep}{0.0pc}
		\begin{tabular*}{\textwidth}{@{\extracolsep{\fill}}lrcccc} \hline
                \mc{2}{l}{Experiment} & $N(B\bar{B})$ & $\frac{J_c}{J_0}$ & $\frac{2J_{s1}}{J_0} \sin(2\beta)$ &  $\frac{2J_{s2}}{J_0} \cos(2\beta)$ \\
		\hline
	\babar & \cite{Aubert:2006fh} & 230M & $0.76 \pm 0.18 \pm 0.07$ & $0.10 \pm 0.24 \pm 0.06$ & $0.38 \pm 0.24 \pm 0.05$ \\
	\belle & \cite{Dalseno:2007hx} & 449M & $0.60 \,^{+0.25}_{-0.28} \pm 0.08$ & $-0.17 \pm 0.42 \pm 0.09$ & $-0.23 \,^{+0.43}_{-0.41} \pm 0.13$ \\
	\mc{3}{l}{\bf Average} & $0.71 \pm 0.16$ & $0.03 \pm 0.21$ & $0.24 \pm 0.22$ \\
	\mc{3}{l}{\small Confidence level} & {\small $0.63~(0.5\sigma)$} & {\small $0.59~(0.5\sigma)$} & {\small $0.23~(1.2\sigma)$} \\
		\hline
		\end{tabular*}
		\label{tab:cp_uta:ccs:dstardstarks}
	\end{center}
\end{table}

From the above result and the assumption that $J_{s2}>0$, 
\babar\ infer that $\cos(2\beta)>0$ at the $94\%$ confidence level~\cite{Aubert:2006fh}.

\mysubsubsection{Time-dependent analysis of $\Bs \to J/\psi \phi$
}
\label{sec:cp_uta:ccs:jpsiphi}

As described in Sec.~\ref{sec:cp_uta:notations:Bs},
time-dependent analysis of $\Bs \to J/\psi \phi$ probes the 
$\CP$ violating phase of $\Bs$--$\Bsb$ oscillations, $\phi_s$.
Within the Standard Model, this parameter is predicted to be small.\footnote{
   We make the approximation $\phi_s = œôø²2 \beta_s$, 
   where $\phi_s \equiv \arg\left[ -M_{12}/\Gamma_{12} \right]$ 
   and $2\beta_s \equiv 2 \arg\left[ -(V_{ts}V_{tb}^*)/(V_{cs}V_{cb}^*) \right]$
   (see Section~\ref{sec:cp_uta:introduction}). 
   This is a reasonable approximation since, 
   although the equality does not hold in the Standard Model~\cite{Lenz:2006hd}, 
   both are much smaller than the current experimental resolution, 
   whereas new physics contributions add a phase $\phi_{\rm NP}$ to $\phi_s$
   and subtract the same phase from $2\beta_s$, 
   so that the approximation remains valid.
}

Both \cdf~\cite{Aaltonen:2007he,cdfandd0:psiphi} and \dzero~\cite{:2008fj,cdfandd0:psiphi} 
have performed full tagged, time-dependent angular analyses of 
$\Bs \to J/\psi \phi$ decays.
Both experiments perform analyses that take into account the correlations
between the average $\Bs$ lifetime $\tau(\Bs)$, 
$\Delta \Gamma_s$, $\phi_s$, 
the magnitude of the perpendicularly polarized component $A_\perp$, 
the difference in the fractions of the two $CP$-even components $|A_0|^2 - |A|_\parallel^2$, 
and the strong phases associated with the two $CP$-even components 
$\delta_0$ and $\delta_\parallel$.
Both experiments find that the likelihood function has a highly non-Gaussian shape, so that central values and uncertainties are not presented.
The combination of results is performed by the experiments themselves~\cite{cdfandd0:psiphi},
and the results are summarised by the HFAG Lifetimes and Oscillations group, 
see Sec.~\ref{sec:life_mix}.








\clearpage
\mysubsection{Time-dependent $\CP$ asymmetries in colour-suppressed $b \to c\bar{u}d$ transitions
}
\label{sec:cp_uta:cud_beta}

Decays of $\B$ mesons to final states such as $D\pi^0$ are 
governed by $b \to c\bar{u}d$ transitions. 
If the final state is a $\CP$ eigenstate, \eg\ $D_{\CP}\pi^0$, 
the usual time-dependence formulae are recovered, 
with the sine coefficient sensitive to $\sin(2\beta)$. 
Since there is no penguin contribution to these decays, 
there is even less associated theoretical uncertainty 
than for $b \to c\bar{c}s$ decays like $\B \to \jpsi \KS$.
Such measurements therefore allow to test the Standard Model prediction
that the $\CP$ violation parameters in $b \to c\bar{u}d$ transitions
are the same as those in $b \to c\bar{c}s$~\cite{Grossman:1996ke}.

Note that there is an additional contribution from CKM suppressed
$b \to u \bar{c} d$ decays.
The effect of this contribution is small, and can be taken into 
account in the analysis~\cite{Fleischer:2003ai,Fleischer:2003aj}.

Results of such an analysis are available from \babar~\cite{Aubert:2007mn}.
The decays $\Bz \to D\pi^0$, $\Bz \to D\eta$, $\Bz \to D\omega$,
$\Bz \to D^*\pi^0$ and $\Bz \to D^*\eta$ are used.
The daughter decay $D^* \to D\pi^0$ is used.
The $\CP$-even $D$ decay to $K^+K^-$ is used for all decay modes,
with the $\CP$-odd $D$ decay to $\KS\omega$ also used in $\Bz \to D^{(*)}\pi^0$
and the additional $\CP$-odd $D$ decay to $\KS\pi^0$ 
also used in $\Bz \to D\omega$.
Results are presented separately for $\CP$-even and $\CP$-odd 
$D^{(*)}$ decays (denoted $D^{(*)}_+ h^0$ and $D^{(*)}_- h^0$ respectively),
and for both combined, with the different $\CP$ factors accounted for
(denoted $D^{(*)}_{CP} h^0$).
The results are summarized in Table~\ref{tab:cp_uta:cud_cp_beta}.

\begin{table}[htb]
	\begin{center}
		\caption{
			Results from analyses of $\Bz \to D^{(*)}h^0$, $D \to CP$ eigenstates decays.
		}
		\vspace{0.2cm}
		\setlength{\tabcolsep}{0.0pc}
		\begin{tabular*}{\textwidth}{@{\extracolsep{\fill}}lrcccc} \hline
	\mc{2}{l}{Experiment} & $N(B\bar{B})$ & $S_{CP}$ & $C_{CP}$ & Correlation \\
	\hline
        \mc{6}{c}{$D^{(*)}_+ h^0$}  \\
	\babar & \cite{Aubert:2007mn} & 383M & $-0.65 \pm 0.26 \pm 0.06$ & $-0.33 \pm 0.19 \pm 0.04$ & $0.04$ \\
	\hline

        \mc{6}{c}{$D^{(*)}_- h^0$} \\
	\babar & \cite{Aubert:2007mn} & 383M & $-0.46 \pm 0.46 \pm 0.13$ & $-0.03 \pm 0.28 \pm 0.07$ & $-0.14$ \\
	\hline

        \mc{6}{c}{$D^{(*)}_{CP} h^0$} \\
	\babar & \cite{Aubert:2007mn} & 383M & $-0.56 \pm 0.23 \pm 0.05$ & $-0.23 \pm 0.16 \pm 0.04$ & $-0.02$ \\
	\hline
		\end{tabular*}
		\label{tab:cp_uta:cud_cp_beta}
	\end{center}
\end{table}

When multibody $D$ decays, such as $D \to \KS\pi^+\pi^-$ are used, 
a time-dependent analysis of the Dalitz plot of the neutral $D$ decay 
allows a direct determination of the weak phase: $2\beta$. 
(Equivalently, both $\sin(2\beta)$ and $\cos(2\beta)$ can be measured.)
This information allows to resolve the ambiguity in the 
measurement of $2\beta$ from $\sin(2\beta)$~\cite{Bondar:2005gk}.

\begin{figure}[htb]
  \begin{center}
    \begin{tabular}{cc}
      \resizebox{0.46\textwidth}{!}{
        \includegraphics{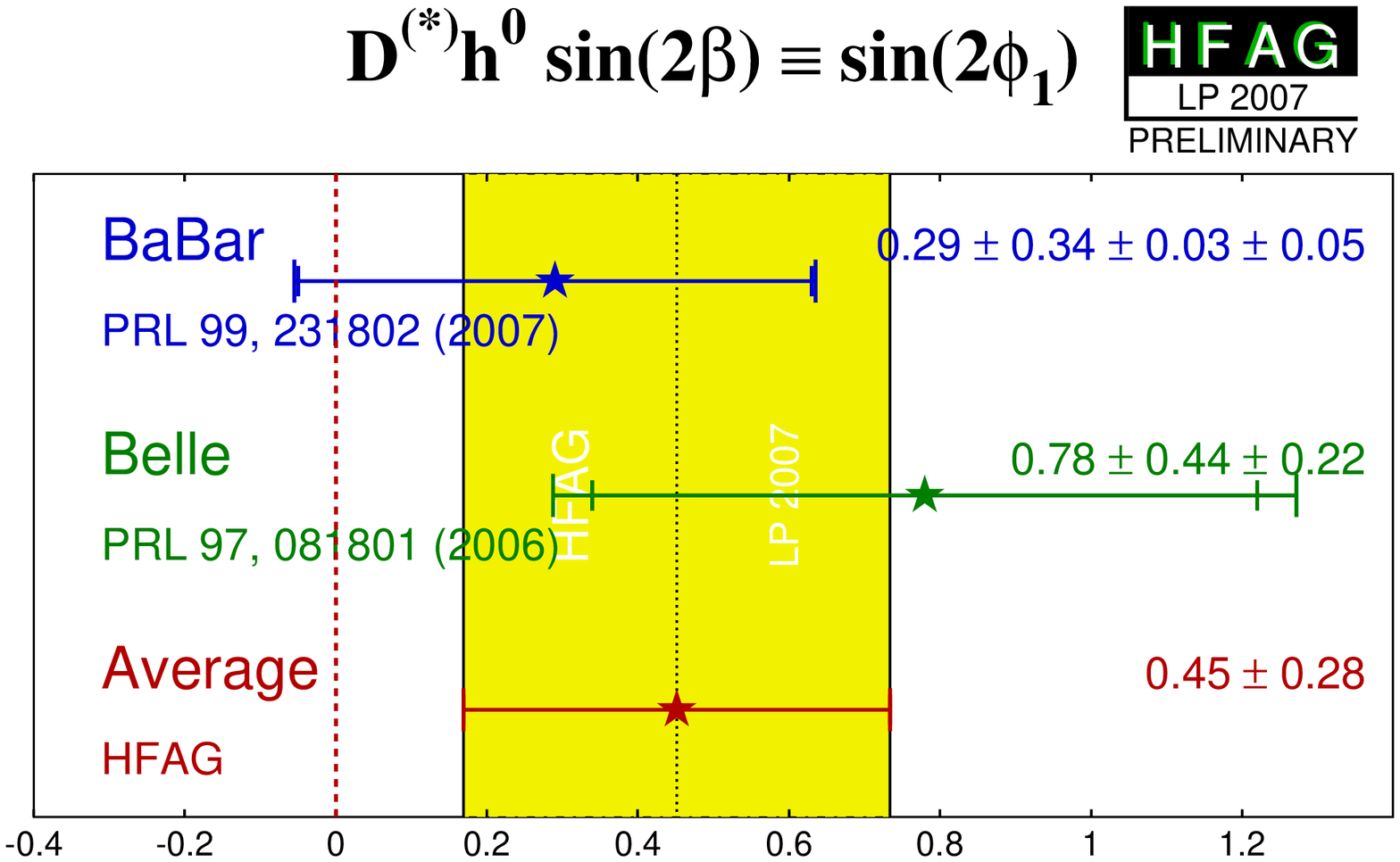}
      }
      &
      \resizebox{0.46\textwidth}{!}{
        \includegraphics{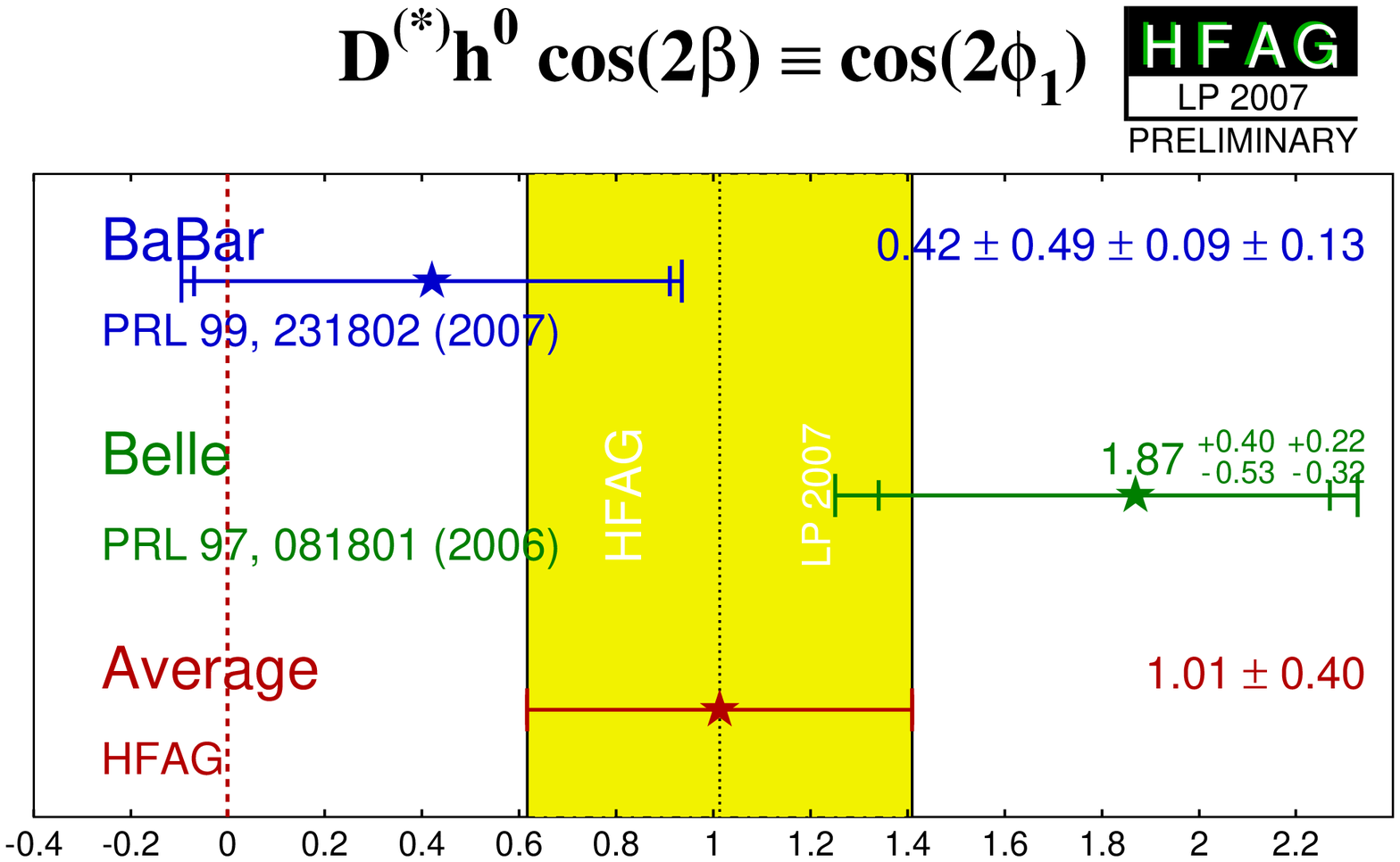}
      }
    \end{tabular}
  \end{center}
  \vspace{-0.8cm}
  \caption{
    Averages of 
    (left) $\sin(2\beta)$ and (right) $\cos(2\beta)$
    measured in colour-suppressed $b \to c\bar{u}d$ transitions.
  }
  \label{fig:cp_uta:cud_beta}
\end{figure}

Results of such analyses are available from both 
\belle~\cite{Krokovny:2006sv} and \babar~\cite{Aubert:2007rp}.
The decays $\B \to D\pi^0$, $\B \to D\eta$, $\B \to D\omega$, 
$\B \to D^*\pi^0$ and $\B \to D^*\eta$ are used. 
[This collection of states is denoted by $D^{(*)}h^0$.]
The daughter decays are $D^* \to D\pi^0$ and $D \to \KS\pi^+\pi^-$.
The results are shown in Table~\ref{tab:cp_uta:cud_beta},
and Fig.~\ref{fig:cp_uta:cud_beta}.
Note that \babar\ quote uncertainties due to the $D$ decay model 
separately from other systematic errors, while \belle\ do not.

\begin{table}[htb]
	\begin{center}
		\caption{
			Averages from $\Bz \to D^{(*)}h^0$, $D \to K_S\pi^+\pi^-$ analyses.
		}
		\vspace{0.2cm}
		\setlength{\tabcolsep}{0.0pc}
    \resizebox{\textwidth}{!}{
      		\begin{tabular*}{\textwidth}{@{\extracolsep{\fill}}lrcccc} \hline
	\mc{2}{l}{Experiment} & $N(B\bar{B})$ & $\sin 2\beta$ & $\cos 2\beta$ & $|\lambda|$ \\
		\hline
	\babar & \cite{Aubert:2007rp} & 383M & $0.29 \pm 0.34 \pm 0.03 \pm 0.05$ & $0.42 \pm 0.49 \pm 0.09 \pm 0.13$ & $1.01 \pm 0.08 \pm 0.02$ \\
	\belle & \cite{Krokovny:2006sv} & 386M & $0.78 \pm 0.44 \pm 0.22$ & $1.87 \,^{+0.40}_{-0.53} \,^{+0.22}_{-0.32}$ & \textendash{} \\
	\mc{3}{l}{\bf Average} & $0.45 \pm 0.28$ & $1.01 \pm 0.40$ & $1.01 \pm 0.08$ \\
	\mc{3}{l}{\small Confidence level} & {\small $0.59~(0.5\sigma)$} & {\small $0.12~(1.6\sigma)$} & \textendash{} \\
		\hline
		\end{tabular*}
    }
		\label{tab:cp_uta:cud_beta}
	\end{center}
\end{table}

Again, it is clear that the data prefer $\cos(2\beta)>0$.
Indeed, \belle~\cite{Krokovny:2006sv} 
determine the sign of $\cos(2\phi_1)$ to be positive at $98.3\%$ confidence level,
while \babar~\cite{Aubert:2007rp} 
favour the solution of $\beta$ with $\cos(2\beta)>0$ at $87\%$ confidence level.
Note, however, that the Belle measurement has strongly non-Gaussian behaviour. 
Therefore, we perform uncorrelated averages, 
from which any interpretation has to be done with the greatest care. 

\clearpage
\mysubsection{Time-dependent $\CP$ asymmetries in charmless $b \to q\bar{q}s$ transitions
}
\label{sec:cp_uta:qqs}

The flavour changing neutral current $b \to s$ penguin
can be mediated by any up-type quark in the loop, 
and hence the amplitude can be written as
\begin{equation}
  \label{eq:cp_uta:b_to_s}
  \begin{array}{ccccc}
    A_{b \to s} & = & 
    \mc{3}{l}{F_u V_{ub}V^*_{us} + F_c V_{cb}V^*_{cs} + F_t V_{tb}V^*_{ts}} \\
    & = & (F_u - F_c) V_{ub}V^*_{us} & + & (F_t - F_c) V_{tb}V^*_{ts} \\
    & = & {\cal O}(\lambda^4) & + & {\cal O}(\lambda^2) \\
  \end{array}
\end{equation}
using the unitarity of the CKM matrix.
Therefore, in the Standard Model, 
this amplitude is dominated by $V_{tb}V^*_{ts}$, 
and to within a few degrees 
($\delta\beta \lesssim 2^\circ$ for $\beta \approx 20^\circ$) 
the time-dependent parameters can be written\footnote
{
  The presence of a small (${\cal O}(\lambda^2)$) weak phase in 
  the dominant amplitude of the $s$ penguin decays introduces 
  a phase shift given by
  $S_{b \to q\bar q s} = -\eta\sin(2\beta)\cdot(1 + \Delta)$. 
  Using the CKMfitter results for the Wolfenstein 
  parameters~\cite{Charles:2004jd}, one finds: 
  $\Delta \simeq 0.033$, which corresponds to a shift of 
  $2\beta$ of $+2.1$ degrees. 
  Nonperturbative contributions can alter this result.
}
$S_{b \to q\bar q s} \approx - \etacp \sin(2\beta)$,
$C_{b \to q\bar q s} \approx 0$,
assuming $b \to s$ penguin contributions only ($q = u,d,s$).

Due to the large virtual mass scales occurring in the penguin loops,
additional diagrams from physics beyond the Standard Model,
with heavy particles in the loops, may contribute.
In general, these contributions will affect the values of 
$S_{b \to q\bar q s}$ and $C_{b \to q\bar q s}$.
A discrepancy between the values of 
$S_{b \to c\bar c s}$ and $S_{b \to q\bar q s}$
can therefore provide a clean indication of new physics~\cite{Grossman:1996ke,Fleischer:1996bv,London:1997zk,Ciuchini:1997zp}.

However, there is an additional consideration to take into account.
The above argument assumes only the $b \to s$ penguin contributes
to the $b \to q\bar q s$ transition.
For $q = s$ this is a good assumption, which neglects only rescattering effects.
However, for $q = u$ there is a colour-suppressed $b \to u$ tree diagram
(of order ${\cal O}(\lambda^4)$), 
which has a different weak (and possibly strong) phase.
In the case $q = d$, any light neutral meson that is formed from $d \bar{d}$ 
also has a $u \bar{u}$ component, and so again there is ``tree pollution''. 
The \Bz decays to $\piz\KS$, $\rhoz\KS$ and $\omega\KS$ belong to this category.
The mesons $\phi$, $f_0$ and $\etapr$ are expected to have predominant
$s\bar{s}$ parts, which reduces the relative size of the possible tree
pollution. 
If the inclusive decay $\Bz\to\Kp\Km\Kz$ (excluding $\phi\Kz$) is dominated by
a nonresonant three-body transition, 
an OZI-rule suppressed tree-level diagram can occur 
through insertion of an $s\sbar$ pair. 
The corresponding penguin-type transition 
proceeds via insertion of a $u\ubar$ pair, which is expected
to be favored over the $s\sbar$ insertion by fragmentation models.
Neglecting rescattering, the final state $\Kz\Kzb\Kz$ 
(reconstructed as $\KS\KS\KS$) has no tree pollution~\cite{Gershon:2004tk}.
Various estimates, using different theoretical approaches,
of the values of $\Delta S = S_{b \to q\bar q s} - S_{b \to c\bar c s}$
exist in the literature~\cite{Grossman:2003qp,Gronau:2003ep,Gronau:2003kx,Gronau:2004hp,Cheng:2005bg,Gronau:2005gz,Buchalla:2005us,Beneke:2005pu,Engelhard:2005hu,Cheng:2005ug,Engelhard:2005ky,Gronau:2006qh,Silvestrini:2007yf,Dutta:2008xw}.
In general, there is agreement that the modes
$\phi\Kz$, $\etapr\Kz$ and $\Kz\Kzb\Kz$ are the cleanest,
with values of $\left| \Delta S \right|$ at or below the few percent level 
($\Delta S$ is usually positive).

\mysubsubsection{Time-dependent $\CP$ asymmetries: $b \to q\bar{q}s$ decays to $\CP$ eigenstates
}
\label{sec:cp_uta:qqs:cp_eigen}

\begin{figure}[htb]
  \begin{center}
    \resizebox{0.45\textwidth}{!}{
      \includegraphics{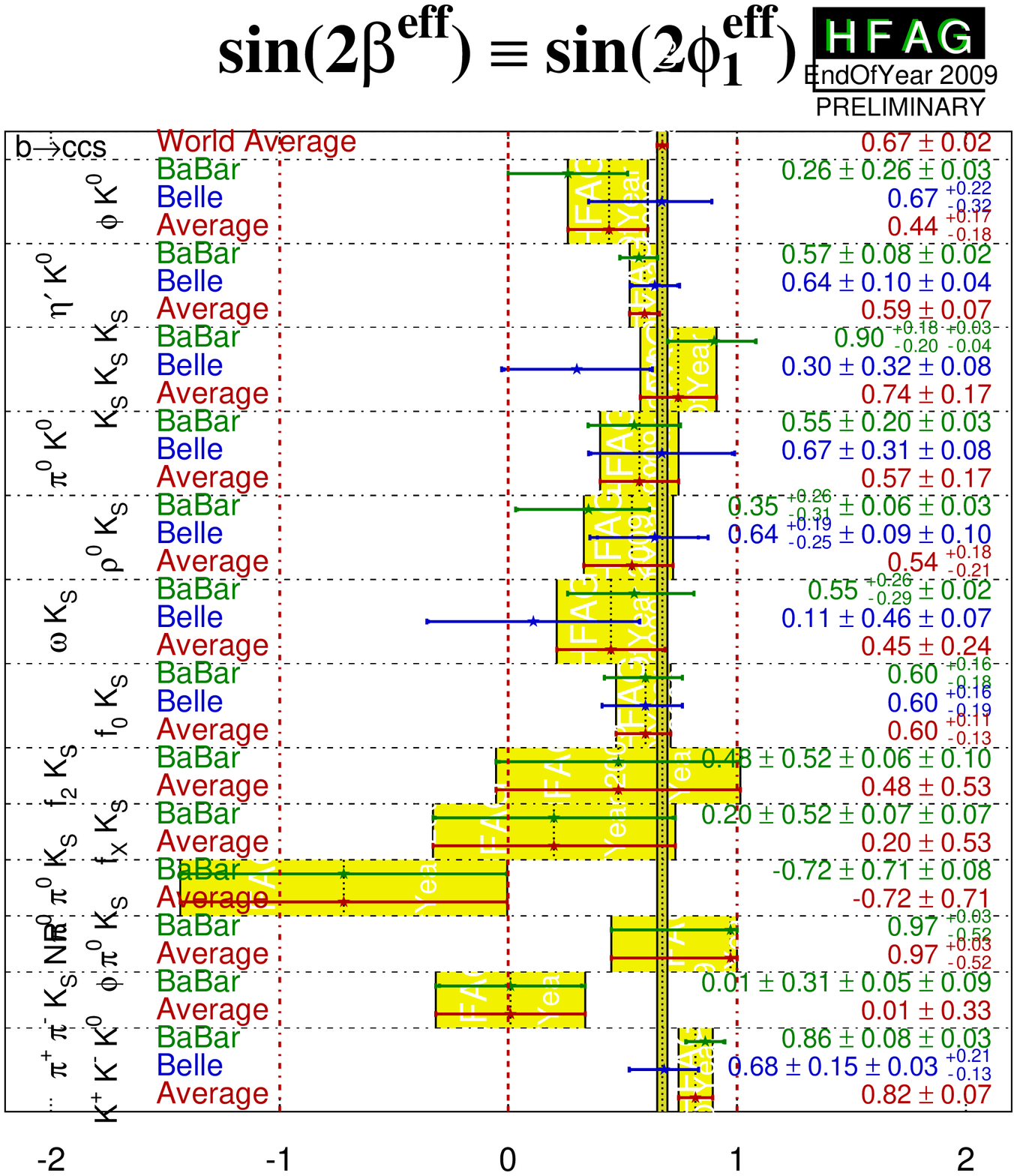}
    }
    \hfill
    \resizebox{0.45\textwidth}{!}{
      \includegraphics{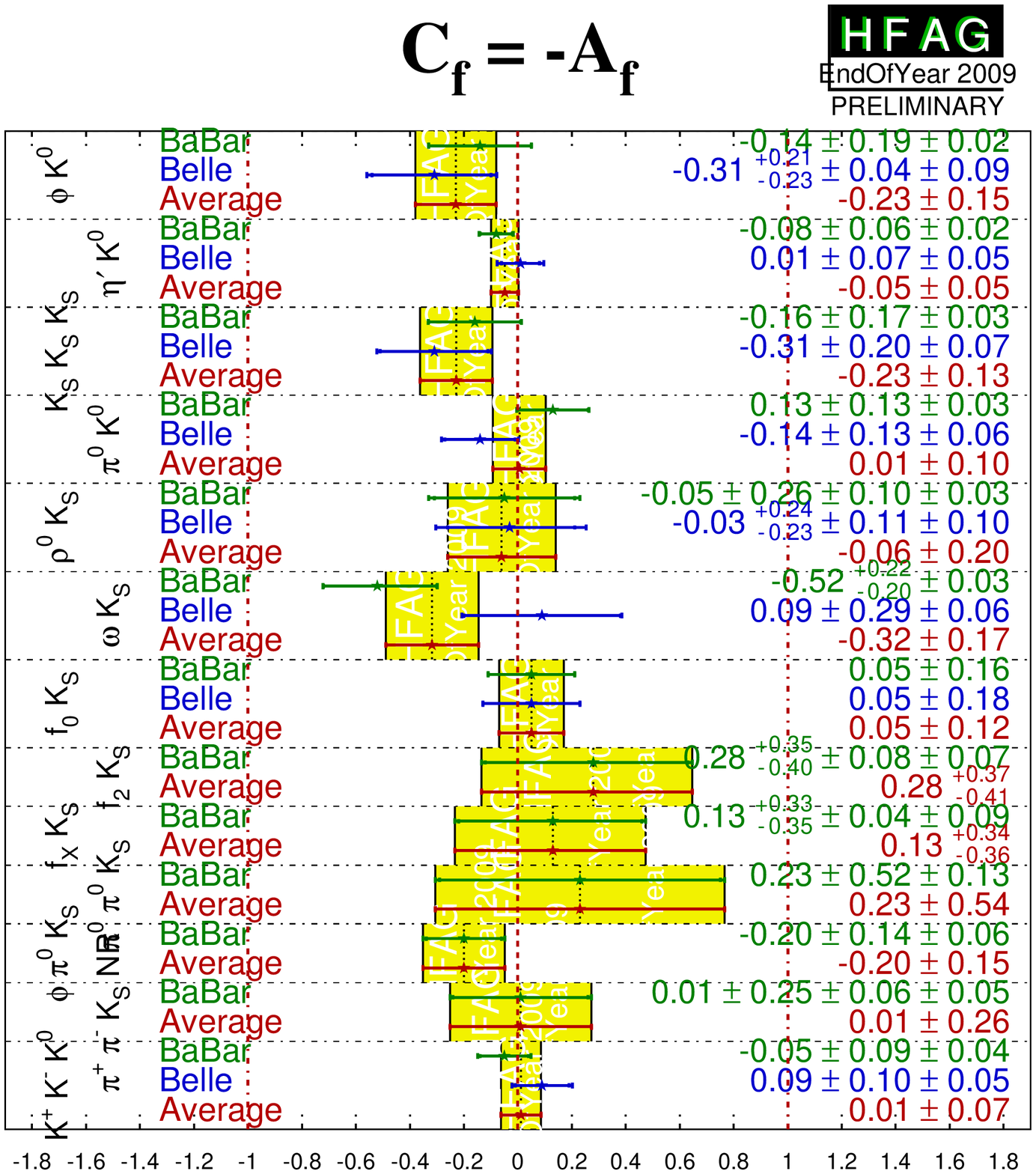}
    }
    \\
    \resizebox{0.45\textwidth}{!}{
      \includegraphics{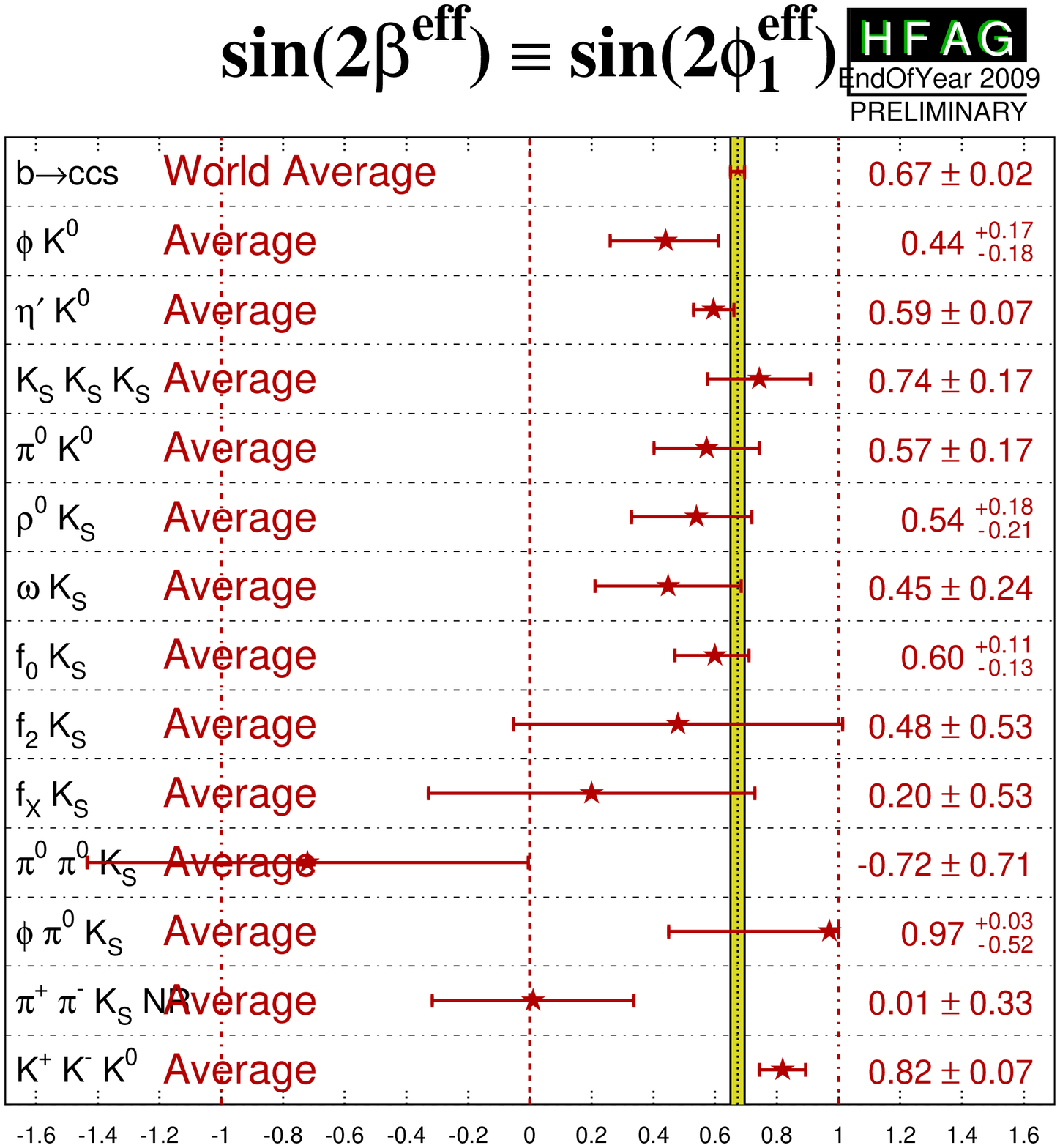}
    }
    \hfill
    \resizebox{0.45\textwidth}{!}{
      \includegraphics{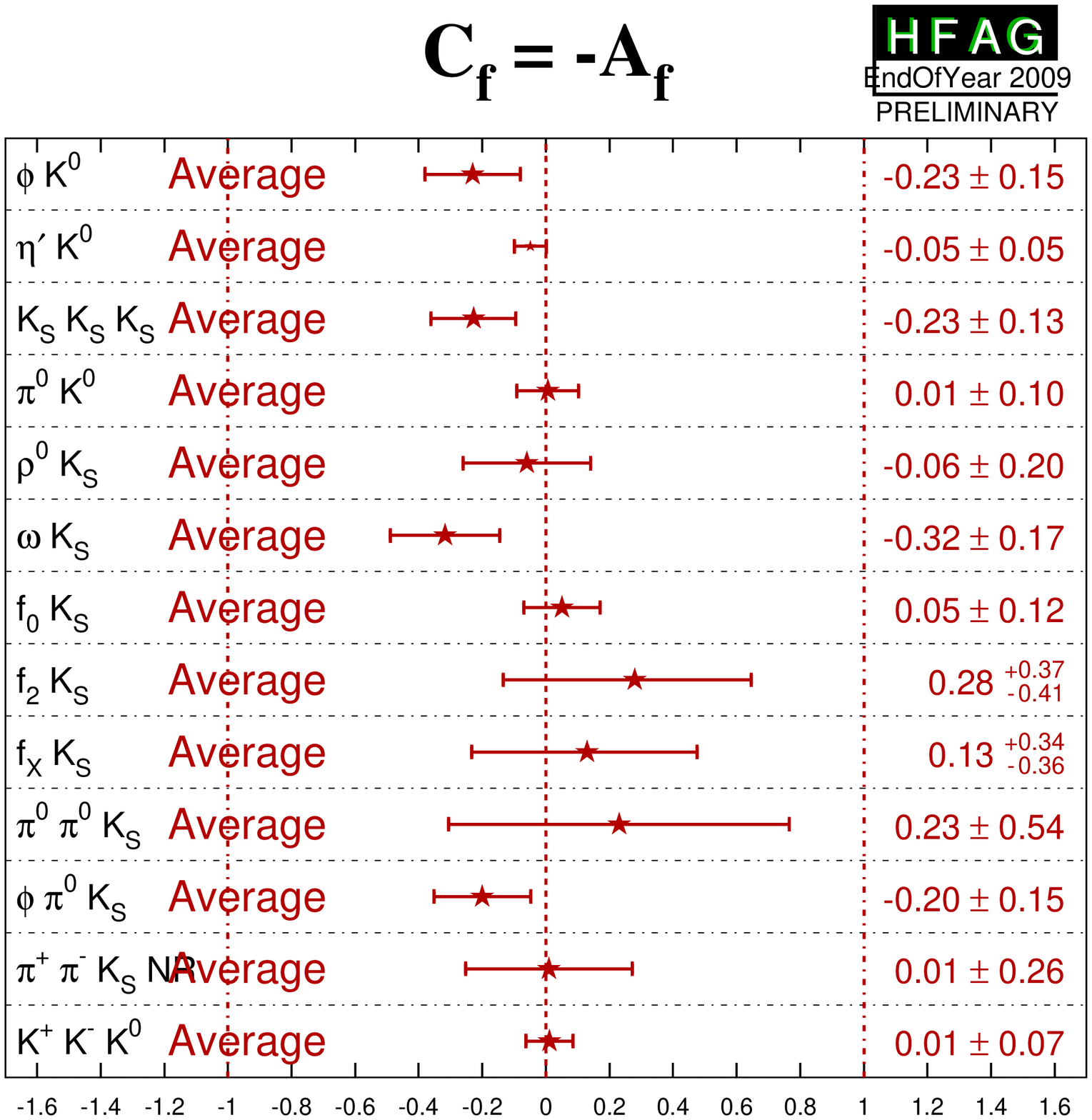}
    }
  \end{center}
  \vspace{-0.8cm}
  \caption{
    (Top)
    Averages of 
    (left) $-\etacp S_{b \to q\bar q s}$ and (right) $C_{b \to q\bar q s}$.
    The $-\etacp S_{b \to q\bar q s}$ figure compares the results to 
    the world average 
    for $-\etacp S_{b \to c\bar c s}$ (see Section~\ref{sec:cp_uta:ccs:cp_eigen}).
    (Bottom) Same, but only averages for each mode are shown.
    More figures are available from the HFAG web pages.
  }
  \label{fig:cp_uta:qqs}
\end{figure}

\begin{figure}[htb]
  \begin{center}
    \resizebox{0.33\textwidth}{!}{
      \includegraphics{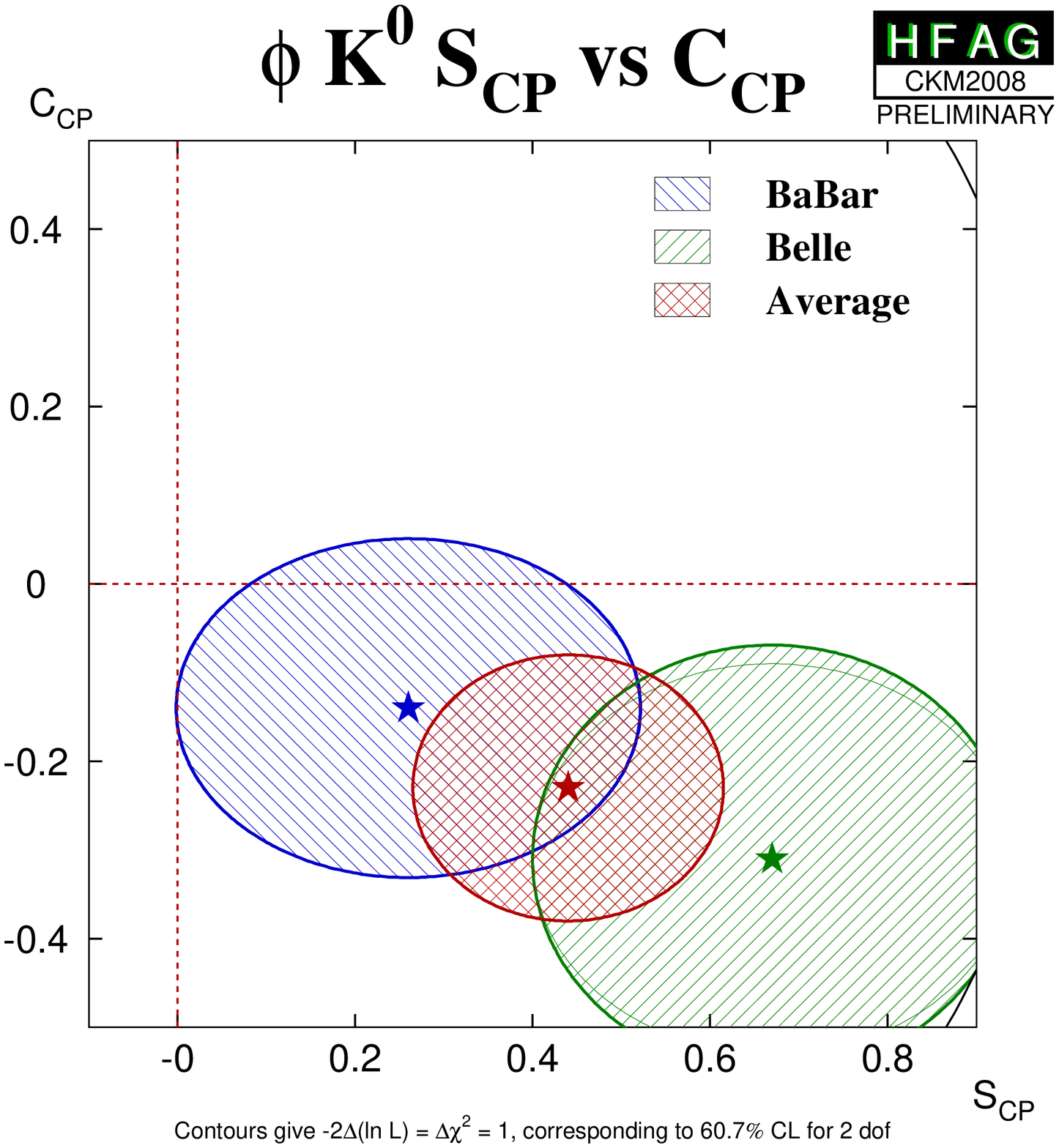}
    }
    \hspace{0.08\textwidth}
    \resizebox{0.33\textwidth}{!}{
      \includegraphics{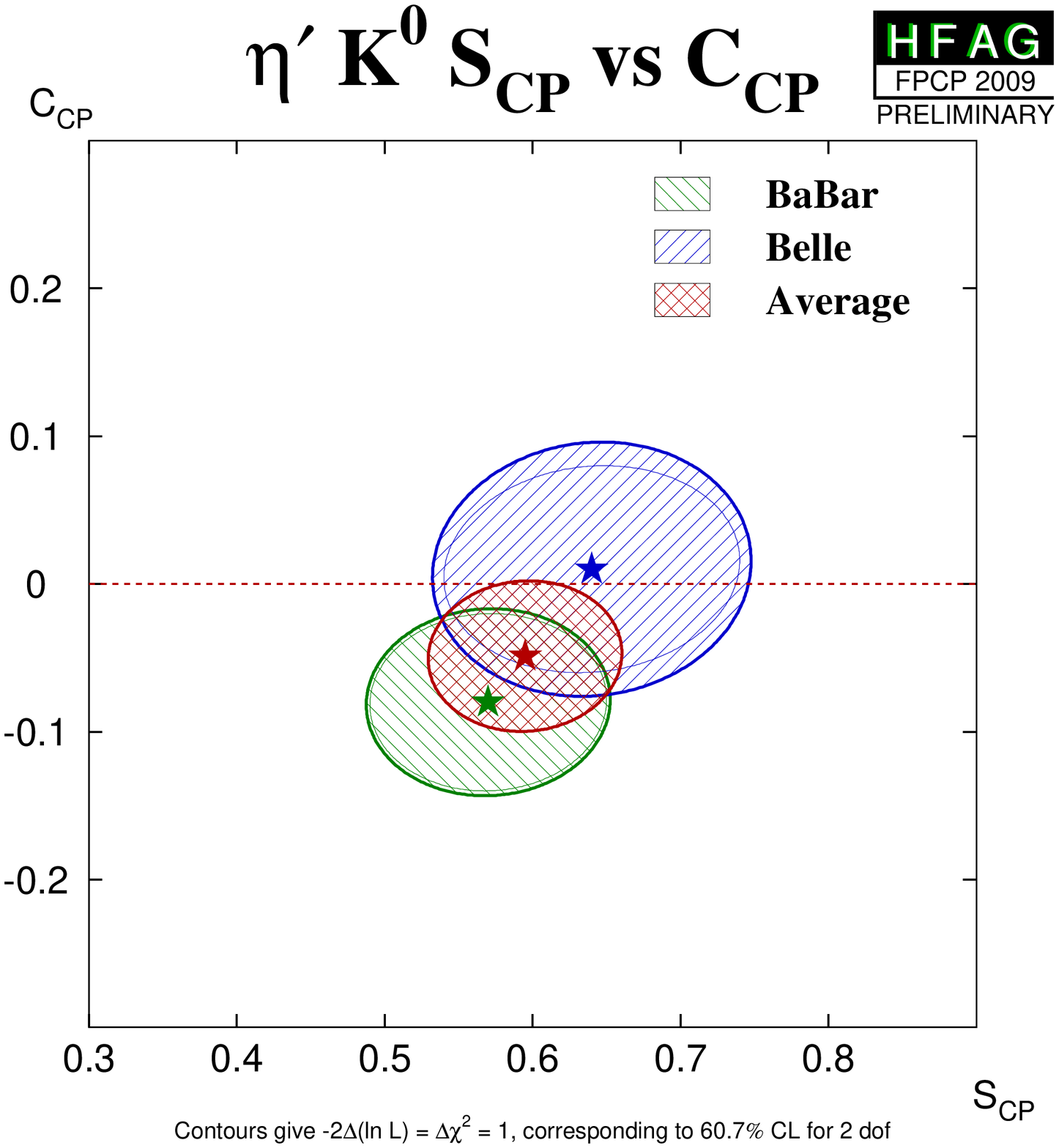}
    }
    \\
    \resizebox{0.33\textwidth}{!}{
      \includegraphics{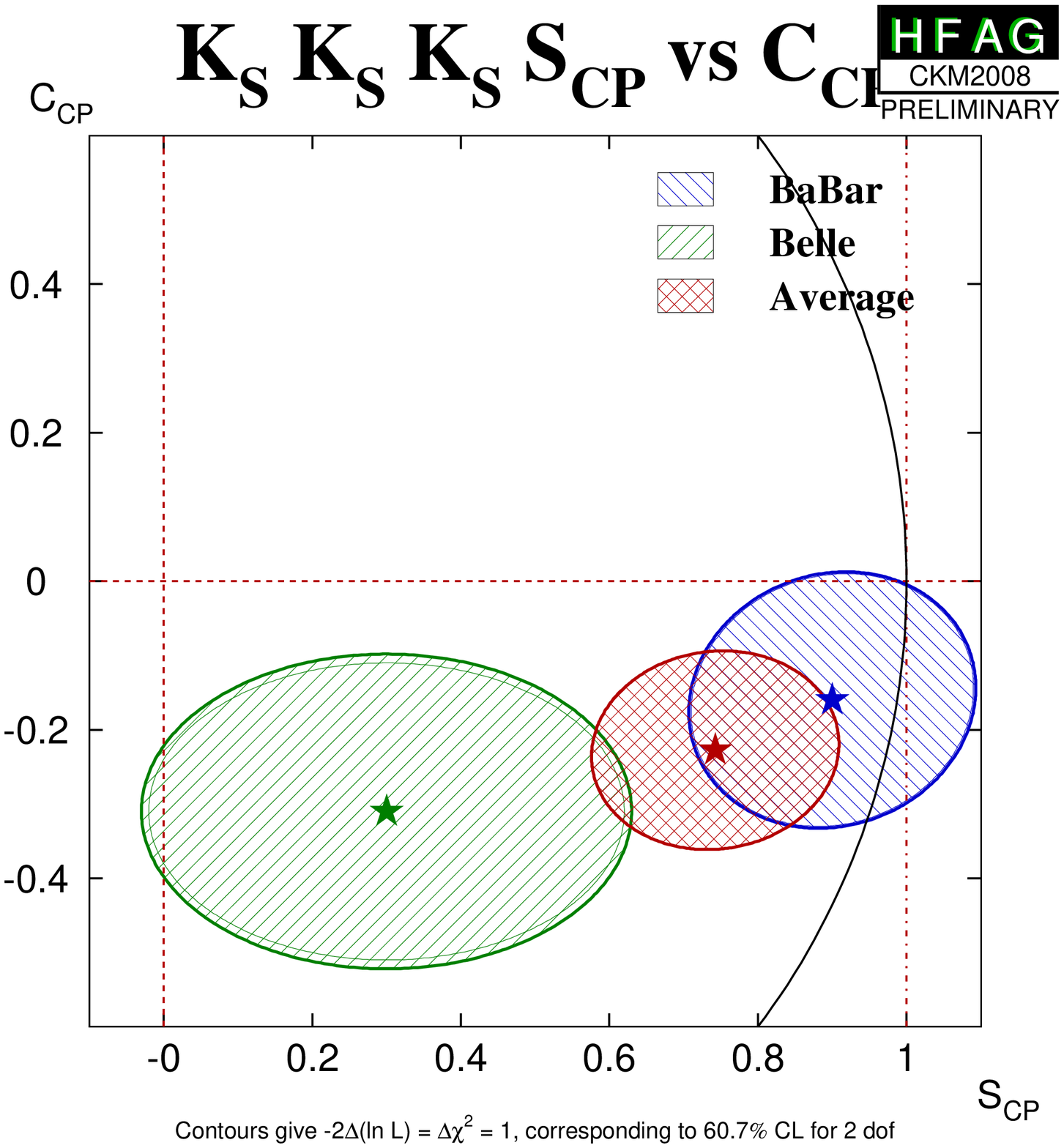}
    }
    \hspace{0.08\textwidth}
    \resizebox{0.33\textwidth}{!}{
      \includegraphics{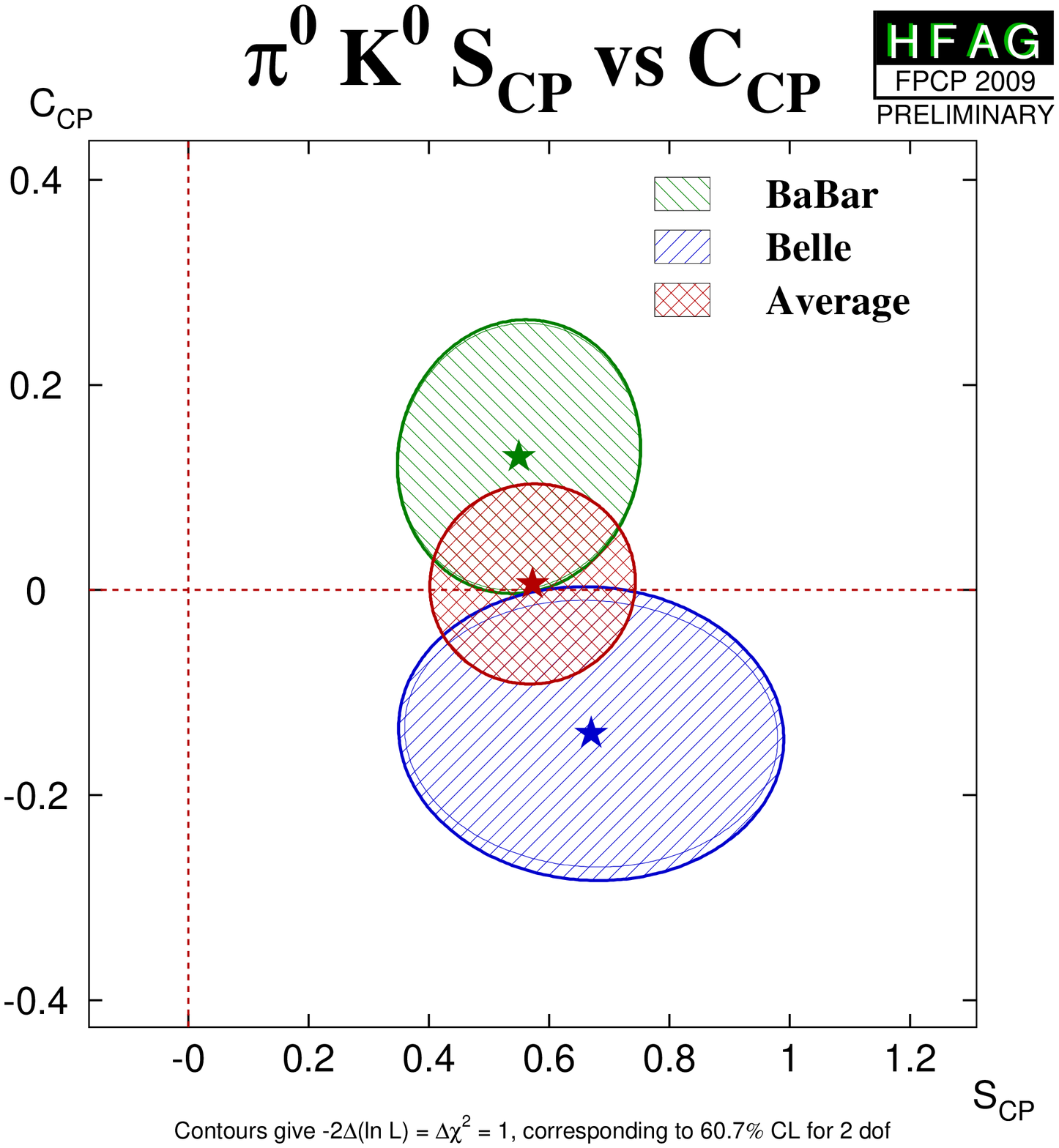}
    }
  \end{center}
  \vspace{-0.8cm}
  \caption{
    Averages of four $b \to q\bar q s$ dominated channels,
    for which correlated averages are performed,
    in the $S_{\CP}$ \vs\ $C_{\CP}$ plane,
    where $S_{\CP}$ has been corrected by the $\CP$ eigenvalue to give
    $\sin(2\beta^{\rm eff})$.
    (Top left) $\Bz \to \phi\Kz$,
    (top right) $\Bz \to \eta^\prime\Kz$,
    (bottom left) $\Bz \to \KS\KS\KS$,
    (bottom right) $\Bz \to \pi^0\KS$.
    More figures are available from the HFAG web pages.
  }
  \label{fig:cp_uta:qqs_SvsC}
\end{figure}

\begin{figure}[htb]
  \begin{center}
    \resizebox{0.66\textwidth}{!}{
      \includegraphics{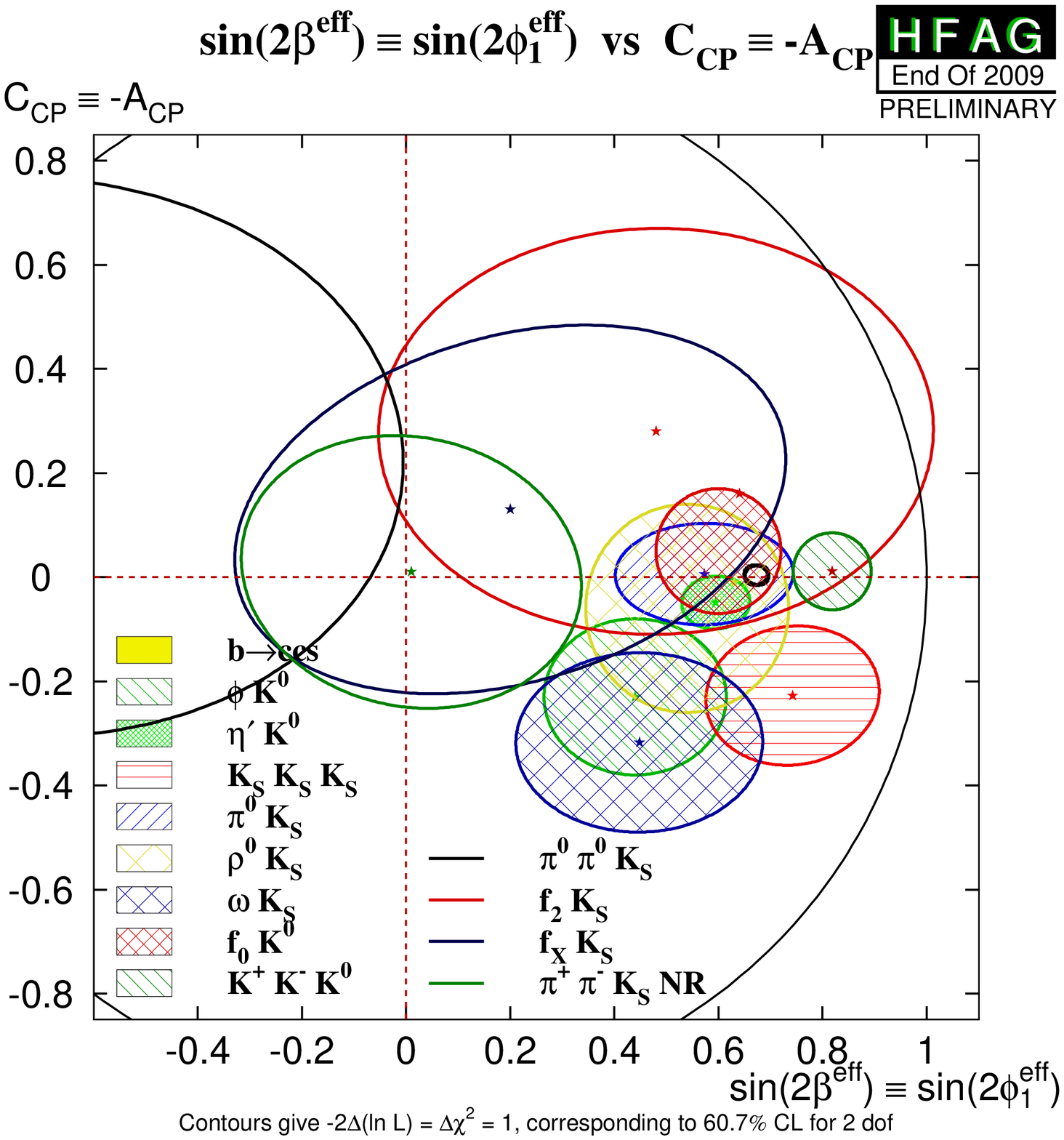}
    }
  \end{center}
  \vspace{-0.8cm}
  \caption{
    Compilation of constraints in the 
    $-\etacp S_{b \to q\bar q s}$ \vs\ $C_{b \to q\bar q s}$ plane.
  }
  \label{fig:cp_uta:qqs_SvsC-all}
\end{figure}

The averages for $-\etacp S_{b \to q\bar q s}$ and $C_{b \to q\bar q s}$
can be found in Table~\ref{tab:cp_uta:qqs},
and are shown in Figs.~\ref{fig:cp_uta:qqs},~\ref{fig:cp_uta:qqs_SvsC} 
and~\ref{fig:cp_uta:qqs_SvsC-all}.
Results from both \babar\  and \belle\ are averaged for the modes
$\phi\Kz$, $\etapr\Kz$, $f_0\Kz$ and $K^+K^-\Kz$
($\Kz$ indicates that both $\KS$ and $\KL$ are used, 
although \belle\ use neither $f_0\KL$ nor $K^+K^-\KL$), 
$\KS\KS\KS$, $\pi^0 \KS$,\footnote{
  \belle~\cite{Fujikawa:2008pk} include the $\pi^0\KL$ final state in order to
  improve the constraint on the direct \CP\ violation parameter; these events
  cannot be used for time-dependent analysis.
} $\rho^0\KS$ and $\omega\KS$.
\babar\ also has presented results with the final states
$\pi^0\pi^0\KS$,\footnote{
  We do not include a preliminary result from \belle~\cite{:2007xd}, which
  remains unpublished after more than two years.
}
$f_2 \KS$, $f_{\rm X} \KS$, $\pi^+ \pi^- \KS$ nonresonant and $\phi \KS \pi^0$. 
Results for $f_0\Kz$, $K^+K^-\Kz$, $\rho^0\KS$, $f_2 \KS$, $f_{\rm X} \KS$ and
$\pi^+ \pi^- \KS$ nonresonant are taken from time-dependent Dalitz plot
analyses of $\Bz \to K^+K^-\Kz$ and $\Bz \to \pi^+\pi^-\KS$ (see
subsection~\ref{sec:cp_uta:qqs:dp}). 
The results presented in Table~\ref{tab:cp_uta:qqs} for $f_0\Kz$ are for both
experiments combinations of the results determined in the $K^+K^-\Kz$ and
$\pi^+\pi^-\KS$ final states.

Of these final states,
$\phi\KS$, $\etapr\KS$, $\pi^0 \KS$, $\rho^0\KS$, $\omega\KS$ and $f_0\KL$
have $\CP$ eigenvalue $\etacp = -1$, 
while $\phi\KL$, $\etapr\KL$, $\KS\KS\KS$, $f_0 \KS$, $f_2 \KS$, 
$f_{\rm X} \KS$,\footnote{ 
  The $f_{\rm X}$ is assumed to be spin even.
} $\pi^0\pi^0\KS$ and $\pi^+ \pi^- \KS$ nonresonant have $\etacp = +1$.

\begin{table}[!htb]
	\begin{center}
		\caption{
      Averages of $-\etacp S_{b \to q\bar q s}$ and $C_{b \to q\bar q s}$.
		}
		\vspace{0.2cm}
		\setlength{\tabcolsep}{0.0pc}

		\label{tab:cp_uta:qqs}
	\end{center}
\end{table}

\begin{table}[!htb]
	\begin{center}
		\caption{
      Averages of $-\etacp S_{b \to q\bar q s}$ and $C_{b \to q\bar q s}$ (continued).
		}
		\vspace{0.2cm}
		\setlength{\tabcolsep}{0.0pc}
		%
		\label{tab:cp_uta:qqs2}
	\end{center}
\end{table}

The final state $K^+K^-\Kz$ 
(contributions from $\phi\Kz$ are implicitly excluded) 
is not a $\CP$ eigenstate.
However, it can be treated as a quasi-two-body decay, 
with the $\CP$ composition determined using either an 
isospin argument (used by \belle\ to determine a $\CP$-even fraction of 
$0.93 \pm 0.09 \pm 0.05$~\cite{Abe:2006gy})
or a moments analysis 
(previously used by \babar\ to find a $\CP$-even fractions of 
$0.89 \pm 0.08 \pm 0.06$ in $K^+K^-\KS$~\cite{Aubert:2005ja}).
Note that uncertainty in the $\CP$ composition of the final state leads to 
a third source of uncertainty on the \belle\ results for $-\etacp S_{K^+K^-\Kz}$.
\babar\ results for $K^+K^-\Kz$ are determined from the inclusive ``high-mass'' 
($m_{K^+K^-} > 1.1 \ {\rm GeV}/c^2$) region in their
$\Bz \to K^+K^-\Kz$ time-dependent Dalitz plot analysis~\cite{:2008gv}
(this approach automatically corrects for the 
$\CP$ composition of the final state).
\belle\ have also performed a time-dependent Dalitz plot analysis of $\Bz \to
K^+K^-\Kz$~\cite{belle:kkk0:preliminary}, but the results presented in
Table~\ref{tab:cp_uta:qqs} are from their previous analysis~\cite{Abe:2006gy}.

The final state $\phi \KS \pi^0$ is also not a \CP-eigenstate but its
\CP-composition can be determined from an angular analysis.
Since the angular parameters are common to the $\Bz\to\phi \KS \pi^0$ and
$\Bz\to \phi \Kp\pim$ decays (because only $K\pi$ resonance contribute),
\babar\ perform a simultaneous analysis of the two final
states~\cite{Aubert:2008zza} (see subsection~\ref{sec:cp_uta:qqs:vv}).

It must be noted that Q2B parameters extracted from Dalitz plot analyses 
are constrained to lie within the physical boundary ($S_{\CP}^2 + C_{\CP}^2 < 1$)
and consequenty the obtained errors are highly non-Gaussian when
the central value is close to the boundary.  
This is particularly evident in the \babar\ results for 
$\Bz \to f_0\Kz$ with $f_0 \to \pi^+\pi^-$~\cite{Aubert:2009me}.
These results must be treated with extreme caution.

As explained above,
each of the modes listed in Table~\ref{tab:cp_uta:qqs} has
different uncertainties within the Standard Model,
and so each may have a different value of $-\etacp S_{b \to q\bar q s}$.
Therefore, there is no strong motivation to make a combined average
over the different modes.
We refer to such an average as a ``na\"\i ve $s$-penguin average.''
It is na\"\i ve not only because of the neglect of the theoretical uncertainty,
but also since possible correlations of systematic effects 
between different modes are neglected.
In spite of these caveats, there remains substantial interest 
in the value of this quantity,
and therefore it is given here:
$\langle -\etacp S_{b \to q\bar q s} \rangle = 0.62 \pm 0.04$,
with confidence level $0.18~(1.3\sigma)$.
This value is in agreement with the average 
$-\etacp S_{b \to c\bar c s}$ given in Sec.~\ref{sec:cp_uta:ccs:cp_eigen}.
%
(The average for $C_{b \to q\bar q s}$ is 
$\langle C_{b \to q\bar q s} \rangle = -0.05 \pm 0.03$
with confidence level $0.78~(0.3\sigma)$.)
%
We emphasise again that we do not advocate the use of these averages,
and that the values should be treated with {\it extreme caution}, if at all.

From Table~\ref{tab:cp_uta:qqs} it may be noted 
that the average for $-\etacp S_{b \to q\bar q s}$ in $\etapr \Kz$ 
($0.59 \pm 0.07$),
is now more than $5\sigma$ away from zero, 
so that $\CP$ violation in this mode is well established.
Amongst other modes,
$\CP$ violation effects in both $f_0 \Kz$ and $K^+K^- \Kz$ appear 
to be established --
\babar\ have claimed $5.1\sigma$ observation of $\CP$ violation in 
$\Bz \to K^+K^- \Kz$~\cite{Aubert:2007sd} and 
$4.3\sigma$ evidence of $\CP$ violation in 
$\Bz \to f_0\KS$ with $f_0 \to \pi^+\pi^-$~\cite{Aubert:2009me}.
Due to possible non-Gaussian errors in these results
it may be prudent to defer any strong conclusions 
on the numerical significance of the averages.
The average for $-\etacp S_{b \to q\bar q s}$ in $\KS\KS\KS$
also appears to have significance greater than $4\sigma$.
There is no evidence (above $2\sigma$) for direct $\CP$ violation 
in any $b \to q \bar q s$ mode.

\mysubsubsection{Time-dependent Dalitz plot analyses: $\Bz \to K^+K^-\Kz$ and $\Bz \to \pi^+\pi^-\KS$}
\label{sec:cp_uta:qqs:dp}

\begin{figure}[htb]
  \begin{center}
    \resizebox{0.45\textwidth}{!}{
      \includegraphics{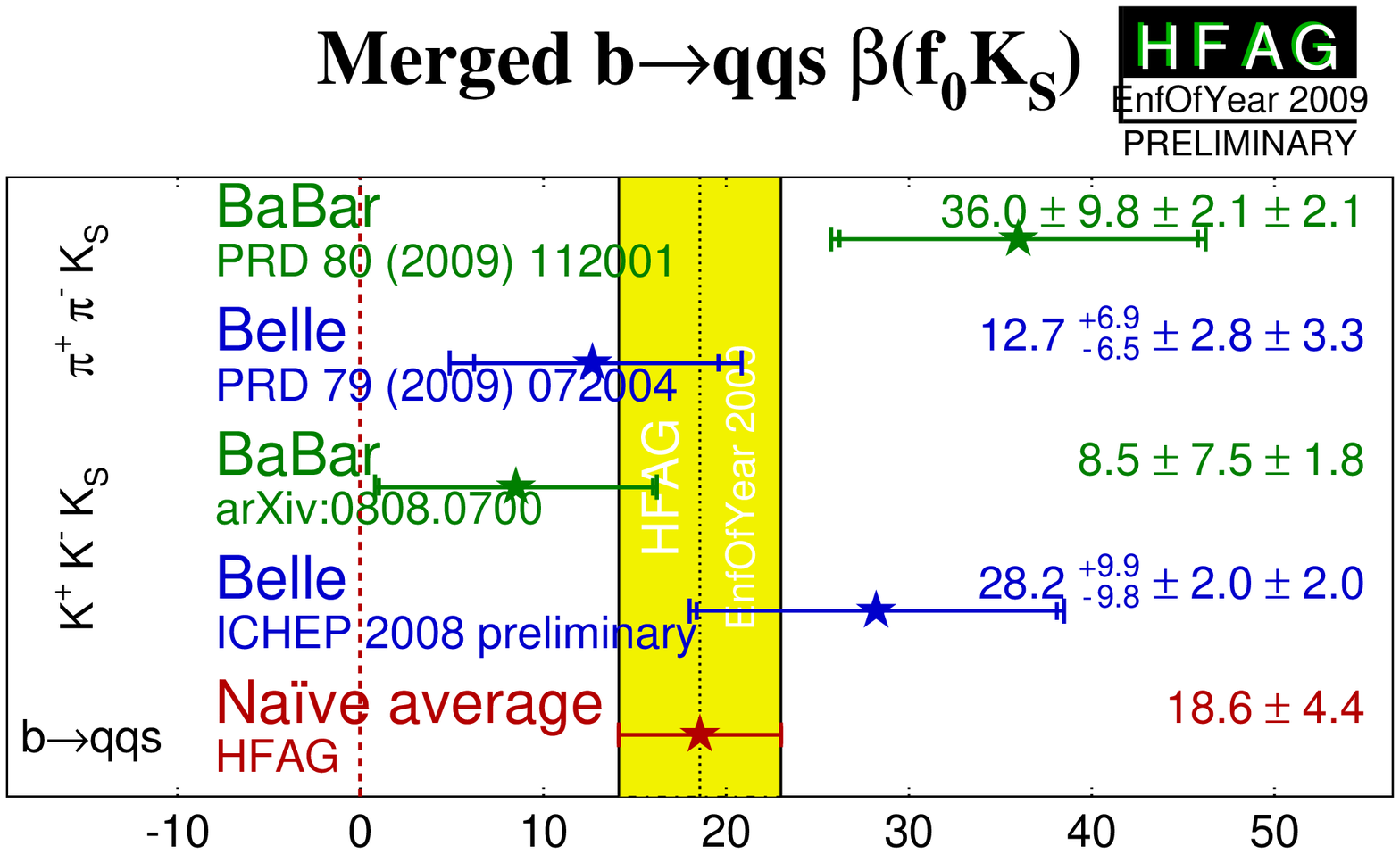}
    }
    \hfill
    \resizebox{0.45\textwidth}{!}{
      \includegraphics{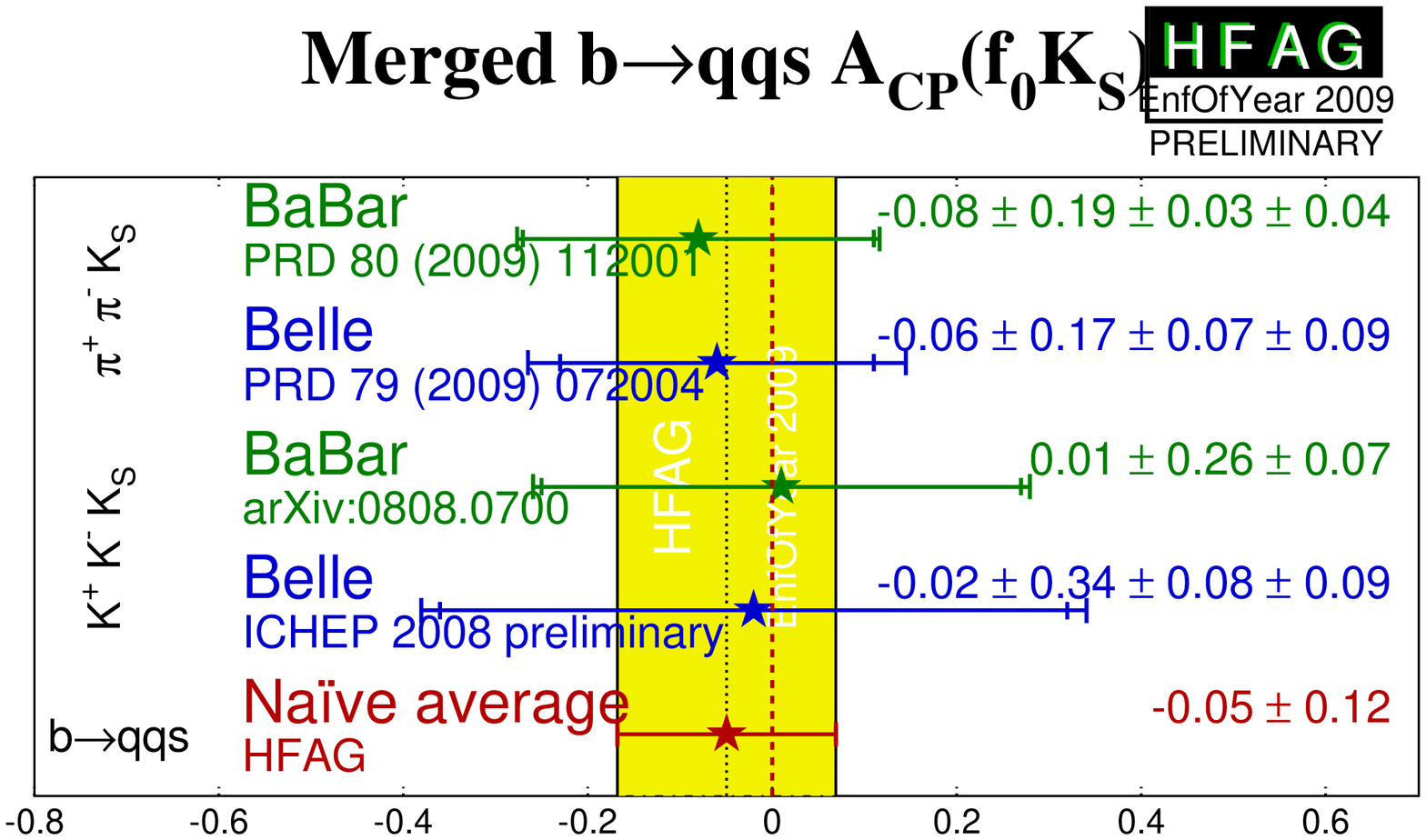}
    }
  \end{center}
  \vspace{-0.8cm}
  \caption{
    (Top)
    Averages of 
    (left) $\beta^{\rm eff} \equiv \phi_1^{\rm eff}$ and (right) $A_{CP}$
    for the $\Bz\to f_0\KS$ decay including measurements from Dalitz plot analyses of both $\Bz\to K^+K^-\KS$ and $\Bz\to \pi^+\pi^-\KS$.
  }
  \label{fig:cp_uta:qqs:f0KS}
\end{figure}

As mentioned in Sec.~\ref{sec:cp_uta:notations:dalitz:kkk0} and above,
both \babar\ and \belle\ have performed time-dependent Dalitz plot analysis of
$\Bz \to K^+K^-\Kz$ and $\Bz \to \pi^+\pi^-\KS$ decays.
The results are summarized in Tabs.~\ref{tab:cp_uta:kkk0_tddp} 
and~\ref{tab:cp_uta:pipik0_tddp}.
Averages for the $\Bz\to f_0 \KS$ decay, which contributes to both Dalitz
plots, are shown in Fig.~\ref{fig:cp_uta:qqs:f0KS}.
Results are presented in terms of the effective weak phase (from mixing and
decay) difference $\beta^{\rm eff}$ and the direct $\CP$ violation parameter
$\Acp$ ($\Acp = -C$) for each of the resonant contributions.
Note that Dalitz plot analyses, including all those included in these
averages, often suffer from ambiguous solutions -- we quote the results
corresponding to those presented as solution 1 in all cases.
Results on flavour specific amplitudes that may contribute to these Dalitz
plots (such as $K^{*+}\pi^-$) are averaged by the HFAG Rare Decays subgroup 
(Sec.~\ref{sec:rare}).

 \begin{table}[htb]
  \begin{center}
    \caption{
      Results from time-dependent Dalitz plot analysis of 
      the $\Bz \to K^+K^-\Kz$ decay.
    }
    \vspace{0.2cm}
    \setlength{\tabcolsep}{0.0pc}
\begin{tabular}{l@{\hspace{2mm}}r@{\hspace{2mm}}c@{\hspace{2mm}}|@{\hspace{2mm}}c@{\hspace{2mm}}c} 
      \hline 
      \mc{2}{l}{Experiment} & $N(B\bar{B})$ &
      \mc{2}{c}{$K^+K^-\Kz$ (whole Dalitz plot)} \\
      & & & $\beta^{\rm eff}$ & $\Acp$ \\
      \babar & \cite{:2008gv} & 465M &
      $(25.3 \pm 3.9 \pm 0.9)^\circ$ & $0.03 \pm 0.07 \pm 0.02$ \\
      \hline
    \end{tabular}

    \vspace{2ex}

    \resizebox{\textwidth}{!}{
      \begin{tabular}{l@{\hspace{2mm}}r@{\hspace{2mm}}c@{\hspace{2mm}}|@{\hspace{2mm}}c@{\hspace{2mm}}c@{\hspace{2mm}}|@{\hspace{2mm}}c@{\hspace{2mm}}c|@{\hspace{2mm}}c@{\hspace{2mm}}c} 
        \hline 
        \mc{2}{l}{Experiment} & $N(B\bar{B})$ &
        \mc{2}{c}{$\phi\Kz$} & \mc{2}{c}{$f_0\Kz$} & \mc{2}{c}{$K^+K^-\Kz~(m_{K^+K^-}>1.1\gevcc)$} \\
        & & & $\beta^{\rm eff}$ & $\Acp$ & $\beta^{\rm eff}$ & $\Acp$ & $\beta^{\rm eff}$ & $\Acp$ \\
        \babar & \cite{:2008gv} & 465M & $(7.7 \pm 7.7 \pm 0.9)^\circ$ & $0.14 \pm 0.19 \pm 0.02$ & $(8.5 \pm 7.5 \pm 1.8)^\circ$ & $0.01 \pm 0.26 \pm 0.07$ & $(29.5 \pm 4.5 \pm 1.5)^\circ$ & $0.05 \pm 0.09 \pm 0.04$ \\
        \belle & \cite{belle:kkk0:preliminary} & 657M & $(21.2 \,^{+9.8}_{-10.4} \pm 2.0 \pm 2.0)^\circ$ & $0.31 \,^{+0.21}_{-0.23} \pm 0.04 \pm 0.09$ & $(28.2 \,^{+9.9}_{-9.8} \pm 2.0 \pm 2.0)^\circ$ & $-0.02 \pm 0.34 \pm 0.08 \pm 0.09$ \\ 
        \mc{3}{l}{\bf Average} & $(12.9 \pm 5.6)^\circ$ & $0.23 \pm 0.15$ & $(16.3 \pm 6.0)^\circ$ & $0.06 \pm 0.19$ \\
	\mc{3}{l}{\small Confidence level} & \mc{4}{c}{\small $0.58~(0.6\sigma)$} \\
        \hline
      \end{tabular}
    }
    
    \label{tab:cp_uta:kkk0_tddp}
  \end{center}
\end{table}

\begin{table}[htb]
  \begin{center}
    \caption{
      Results from time-dependent Dalitz plot analysis of 
      the $\Bz \to \pi^+\pi^-\KS$ decay.
    }
    \vspace{0.2cm}
    \setlength{\tabcolsep}{0.0pc}
    \resizebox{\textwidth}{!}{
      \begin{tabular}{l@{\hspace{2mm}}r@{\hspace{2mm}}c@{\hspace{2mm}}|@{\hspace{2mm}}c@{\hspace{2mm}}c|@{\hspace{2mm}}c@{\hspace{2mm}}c} 
        \hline 
        \mc{2}{l}{Experiment} & $N(B\bar{B})$ & 
        \mc{2}{c}{$\rho^0\KS$} & \mc{2}{c}{$f_0\KS$} \\
        & & & $\beta^{\rm eff}$ & $\Acp$ & $\beta^{\rm eff}$ & $\Acp$ \\
        \hline
        \babar & \cite{Aubert:2009me} & 383M & $(10.2 \pm 8.9 \pm 3.0 \pm 1.9)^\circ$ & $0.05 \pm 0.26 \pm 0.10 \pm 0.03$ & $(36.0 \pm 9.8 \pm 2.1 \pm 2.1)^\circ$ & $-0.08 \pm 0.19 \pm 0.03 \pm 0.04$ \\
        \belle & \cite{:2008wwa} & 657M & $(20.0 \,^{+8.6}_{-8.5} \pm 3.2 \pm 3.5)^\circ$ & $0.03 \,^{+0.23}_{-0.24} \pm 0.11 \pm 0.10$ & $(12.7 \,^{+6.9}_{-6.5} \pm 2.8 \pm 3.3)^\circ$ & $-0.06 \pm 0.17 \pm 0.07 \pm 0.09$ \\
        \hline
        \mc{3}{l}{\bf Average} & $16.4 \pm 6.8$ & $0.06 \pm 0.20$ & $20.6 \pm 6.2$ & $-0.07 \pm 0.14$  \\
        \mc{3}{l}{\small Confidence level} & \mc{4}{c}{\small $0.39~(0.9\sigma)$} \\
        \hline
      \end{tabular}
    }

    \vspace{2ex}

    \setlength{\tabcolsep}{0.0pc}
    \resizebox{\textwidth}{!}{
      \begin{tabular}{l@{\hspace{2mm}}r@{\hspace{2mm}}c@{\hspace{2mm}}|@{\hspace{2mm}}c@{\hspace{2mm}}c|@{\hspace{2mm}}c@{\hspace{2mm}}c} 
        \hline 
        \mc{2}{l}{Experiment} & $N(B\bar{B})$ & 
        \mc{2}{c}{$f_2\KS$} & \mc{2}{c}{$f_{\rm X}\KS$} \\
        & & & $\beta^{\rm eff}$ & $\Acp$ & $\beta^{\rm eff}$ & $\Acp$ \\
        \babar & \cite{Aubert:2009me} & 383M & $(14.9 \pm 17.9 \pm 3.1 \pm 5.2)^\circ$ & $-0.28 \,^{+0.40}_{-0.35} \pm 0.08 \pm 0.07$ & $(5.8 \pm 15.2 \pm 2.2 \pm 2.3)^\circ$ & $-0.13 \,^{+0.35}_{-0.33} \pm 0.04 \pm 0.09$ \\
        \hline
      \end{tabular}
    }

    \vspace{2ex}

    \setlength{\tabcolsep}{0.0pc}
    \resizebox{\textwidth}{!}{
      \begin{tabular}{l@{\hspace{2mm}}r@{\hspace{2mm}}c@{\hspace{2mm}}|@{\hspace{2mm}}c@{\hspace{2mm}}c|@{\hspace{2mm}}c@{\hspace{2mm}}c} 
        \hline 
        \mc{2}{l}{Experiment} & $N(B\bar{B})$ & 
        \mc{2}{c}{$\Bz \to \pi^+\pi^-\KS$ nonresonant} & \mc{2}{c}{$\chi_{c0}\KS$} \\
        & & & $\beta^{\rm eff}$ & $\Acp$ & $\beta^{\rm eff}$ & $\Acp$ \\
        \babar & \cite{Aubert:2009me} & 383M & $(0.4 \pm 8.8 \pm 1.9 \pm 3.8)^\circ$ & $-0.01 \pm 0.25 \pm 0.06 \pm 0.05$ & $(23.2 \pm 22.4 \pm 2.3 \pm 4.2)^\circ$ & $0.29 \,^{+0.44}_{-0.53} \pm 0.03 \pm 0.05$ \\
        \hline
      \end{tabular}
    }

    \label{tab:cp_uta:pipik0_tddp}
  \end{center}
\end{table}

From the results in Tab.~\ref{tab:cp_uta:pipik0_tddp},
\babar\ infer that the trigonometric reflection 
at $\pi/2 - \beta^{\rm eff}$ in $\Bz \to K^+K^-\Kz$,
which is inconsistent with the Standard Model expectation,
is disfavoured at $4.8\sigma$.

\mysubsubsection{Time-dependent analyses of $\Bz \to \phi \KS \pi^0$}
\label{sec:cp_uta:qqs:vv}

The final state in the decay $\Bz \to \phi \KS \pi^0$ is a mixture of \CP-even
and \CP-odd amplitudes. However, since only $\phi K^{*0}$ resonant states
contribute (in particular, $\phi K^{*0}(892)$, $\phi K^{*0}_0(1430)$ and $\phi
K^{*0}_2(1430)$ are seen), the composition can be determined from the analysis
of $B \to \phi K^+ \pi^-$, assuming only that the ratio of branching fractions
${\cal B}(K^{*0} \to \KS \pi^0)/{\cal B}(K^{*0} \to K^+ \pi^-)$ is the same
for each exited kaon state. 

\babar~\cite{Aubert:2008zza} have performed a simultaneous analysis of 
$\Bz \to \phi \KS \pi^0$ and $\Bz \to \phi K^+ \pi^-$ that is time-dependent
for the former mode and time-integrated for the latter. Such an analysis
allows, in principle, all parameters of the $\Bz \to \phi K^{*0}$ system to be
determined, including mixing-induced \CP violation effects. The latter is
determined to be $\Delta\phi_{00} = 0.28 \pm 0.42 \pm 0.04$, where
$\Delta\phi_{00}$ is half the weak phase difference between $\Bz$ and $\Bzb$
decays to $\phi K^{*0}_0(1430)$. As discussed above, this can also be
presented in terms of the quasi-two-body parameter $\sin(2\beta^{\rm eff}_{00}) =
\sin(2\beta+2\Delta\phi_{00}) = 0.97 \,^{+0.03}_{-0.52}$. The highly asymmetric
uncertainty arises due to the conversion from the phase to the sine of the
phase, and the proximity of the physical boundary. 

Similar $\sin(2\beta^{\rm eff})$ parameters can be defined for each of the
helicity amplitudes for both $\phi K^{*0}(892)$ and $\phi
K^{*0}_2(1430)$. However, the relative phases between these decays are
constrained due to the nature of the simultaneous analysis of $\Bz \to \phi
\KS \pi^0$ and $\Bz \to \phi K^+ \pi^-$, and therefore these measurements are
highly correlated. Instead of quoting all these results, BaBar provide an
illustration of their measurements with the following differences: 
\begin{eqnarray}
  \sin(2\beta - 2\Delta\delta_{01}) - \sin(2\beta) & = & -0.42\,^{+0.26}_{-0.34} \, \\
  \sin(2\beta - 2\Delta\phi_{\parallel1}) - \sin(2\beta) & = & -0.32\,^{+0.22}_{-0.30} \, \\
  \sin(2\beta - 2\Delta\phi_{\perp1}) - \sin(2\beta) & = & -0.30\,^{+0.23}_{-0.32} \, \\
  \sin(2\beta - 2\Delta\phi_{\perp1}) - \sin(2\beta - 2\Delta\phi_{\parallel1})
  & = & 0.02 \pm 0.23 \, \\
  \sin(2\beta - 2\Delta\delta_{02}) - \sin(2\beta) & = & -0.10\,^{+0.18}_{-0.29} \,
\end{eqnarray}
where the first subscript indicates the helicity amplitude and the second
indicates the spin of the kaon resonance. For the complete definitions of the
$\Delta\delta$ and $\Delta\phi$ parameters, please refer to the \babar\ paper~\cite{Aubert:2008zza}.

Direct \CP violation parameters for each of the contributing helicity
amplitudes can also be measured. Again, these are determined from a
simultaneous fit of $\Bz \to \phi \KS \pi^0$ and $\Bz \to \phi K^+ \pi^-$,
with the precision being dominated by the statistics of the latter
mode. Direct \CP violation measurements are tabulated by HFAG - Rare Decays 
(Sec.~\ref{sec:rare}). 

\clearpage
\mysubsection{Time-dependent $\CP$ asymmetries in $b \to c\bar{c}d$ transitions
}
\label{sec:cp_uta:ccd}

The transition $b \to c\bar c d$ can occur via either a $b \to c$ tree
or a $b \to d$ penguin amplitude.  
Similarly to Eq.~(\ref{eq:cp_uta:b_to_s}), the amplitude for 
the $b \to d$ penguin can be written
\begin{equation}
  \label{eq:cp_uta:b_to_d}
  \begin{array}{ccccc}
    A_{b \to d} & = & 
    \mc{3}{l}{F_u V_{ub}V^*_{ud} + F_c V_{cb}V^*_{cd} + F_t V_{tb}V^*_{td}} \\
    & = & (F_u - F_c) V_{ub}V^*_{ud} & + & (F_t - F_c) V_{tb}V^*_{td} \\
    & = & {\cal O}(\lambda^3) & + & {\cal O}(\lambda^3). \\
  \end{array}
\end{equation}
From this it can be seen that the $b \to d$ penguin amplitude 
contains terms with different weak phases at the same order of
CKM suppression.

In the above, we have followed Eq.~(\ref{eq:cp_uta:b_to_s}) 
by eliminating the $F_c$ term using unitarity.
However, we could equally well write
\begin{equation}
  \label{eq:cp_uta:b_to_d_alt}
  \begin{array}{ccccc}
    A_{b \to d} 
    & = & (F_u - F_t) V_{ub}V^*_{ud} & + & (F_c - F_t) V_{cb}V^*_{cd}, \\
    & = & (F_c - F_u) V_{cb}V^*_{cd} & + & (F_t - F_u) V_{tb}V^*_{td}. \\
  \end{array}
\end{equation}
Since the $b \to c\bar{c}d$ tree amplitude 
has the weak phase of $V_{cb}V^*_{cd}$,
either of the above expressions allow the penguin to be decomposed into 
parts with weak phases the same and different to the tree amplitude
(the relative weak phase can be chosen to be either $\beta$ or $\gamma$).
However, if the tree amplitude dominates,
there is little sensitivity to any phase 
other than that from $\Bz$\textendash$\Bzb$ mixing.

\begin{figure}[htb]
  \begin{center}
\begin{tabular}{cc}
      \resizebox{0.46\textwidth}{!}{
        \includegraphics{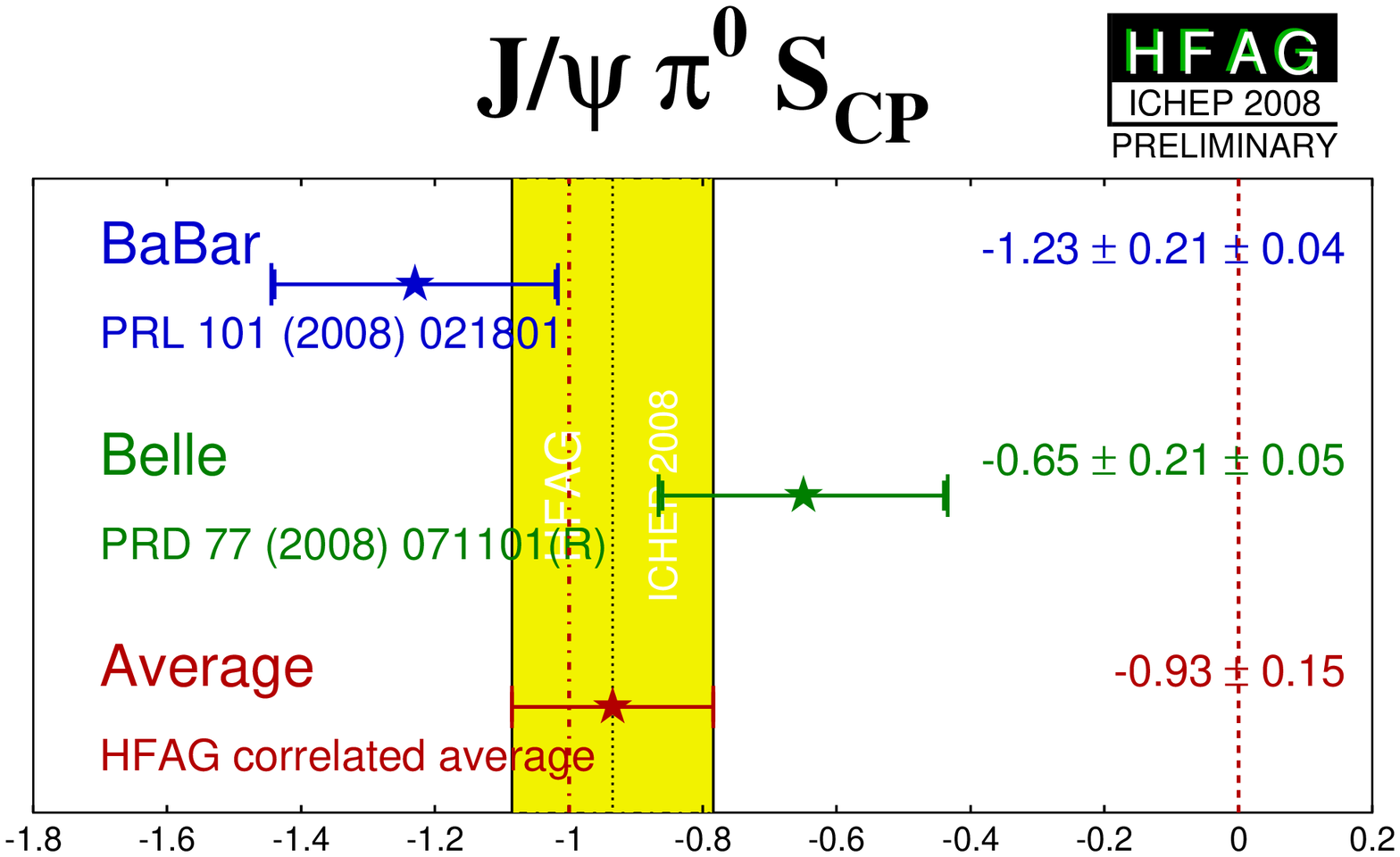}
      }
      &
      \resizebox{0.46\textwidth}{!}{
        \includegraphics{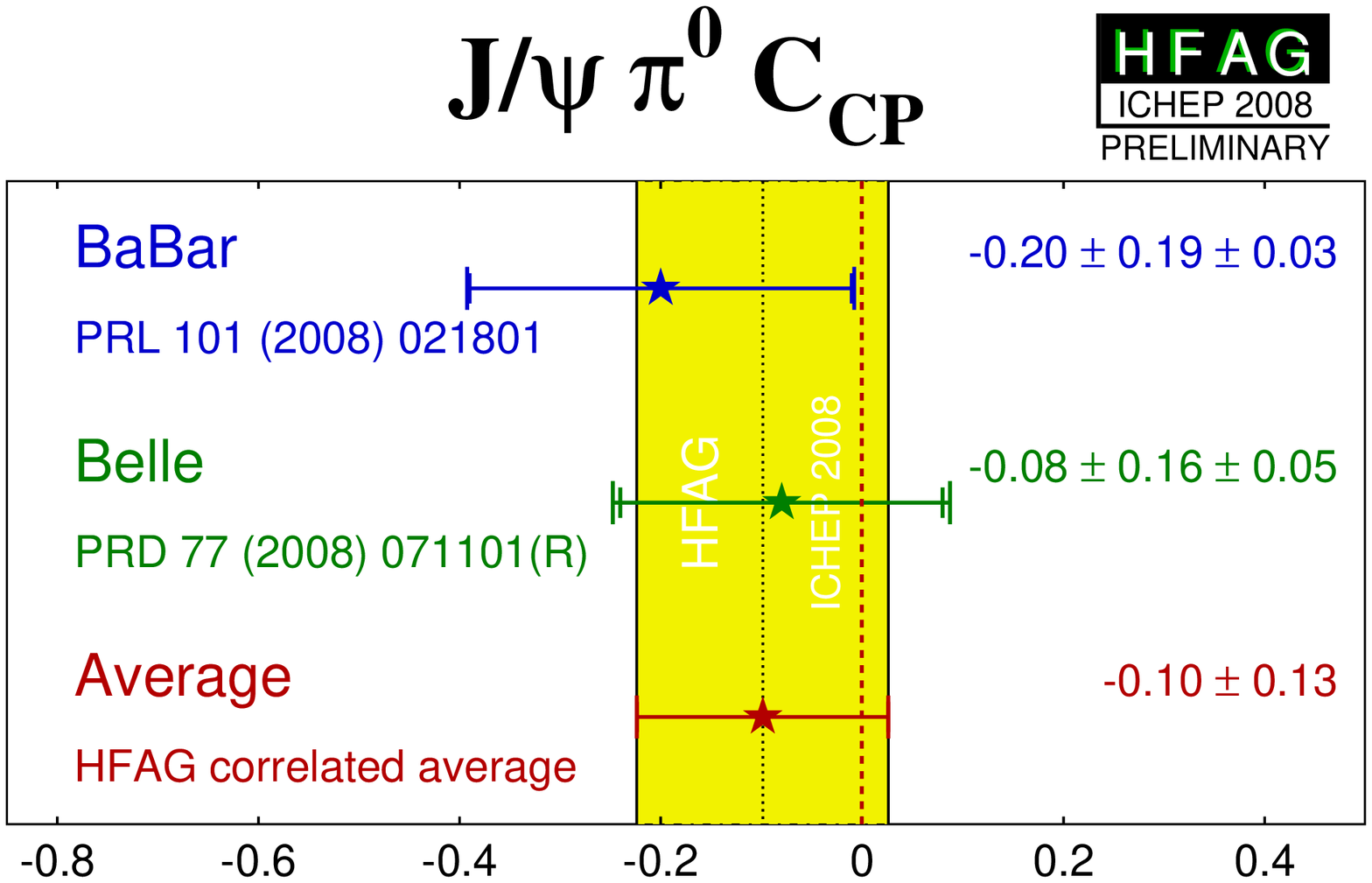}
      }
    \end{tabular}
  \end{center}
  \vspace{-0.8cm}
  \caption{
    Averages of 
    (left) $S_{b \to c\bar c d}$ and (right) $C_{b \to c\bar c d}$ 
    for the mode $\Bz \to J/ \psi \pi^0$.
  }
  \label{fig:cp_uta:ccd:psipi0}
\end{figure}

\begin{figure}[htb]
  \begin{center}
    \begin{tabular}{cc}
      \resizebox{0.46\textwidth}{!}{
        \includegraphics{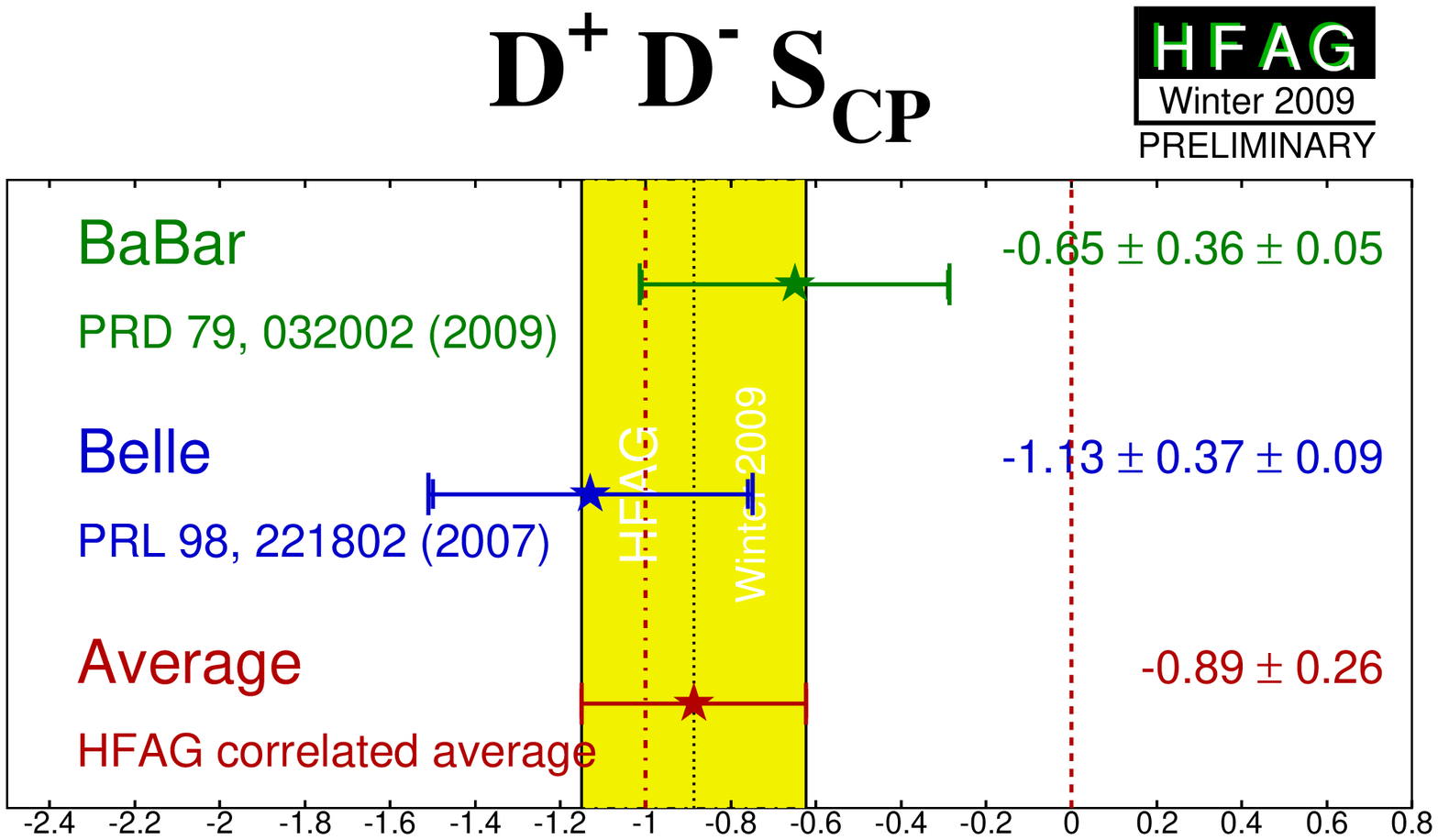}
      }
      &
      \resizebox{0.46\textwidth}{!}{
        \includegraphics{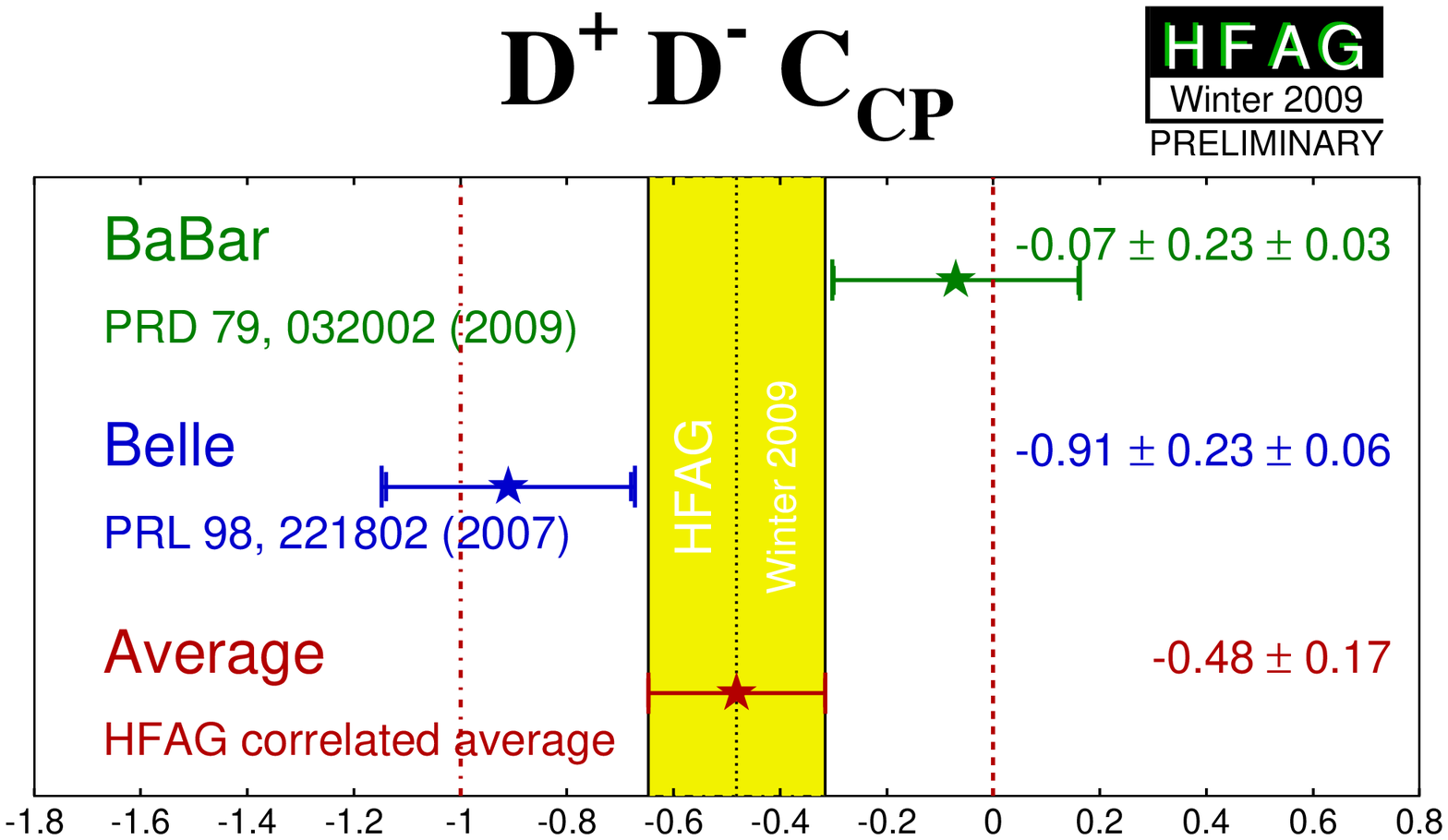}
      }
    \end{tabular}
  \end{center}
  \vspace{-0.8cm}
  \caption{
    Averages of 
    (left) $S_{b \to c\bar c d}$ and (right) $C_{b \to c\bar c d}$ 
    for the mode $\Bz \to D^+D^-$.
  }
  \label{fig:cp_uta:ccd:dd}
\end{figure}

\begin{figure}[htb]
  \begin{center}
    \begin{tabular}{cc}
      \resizebox{0.46\textwidth}{!}{
        \includegraphics{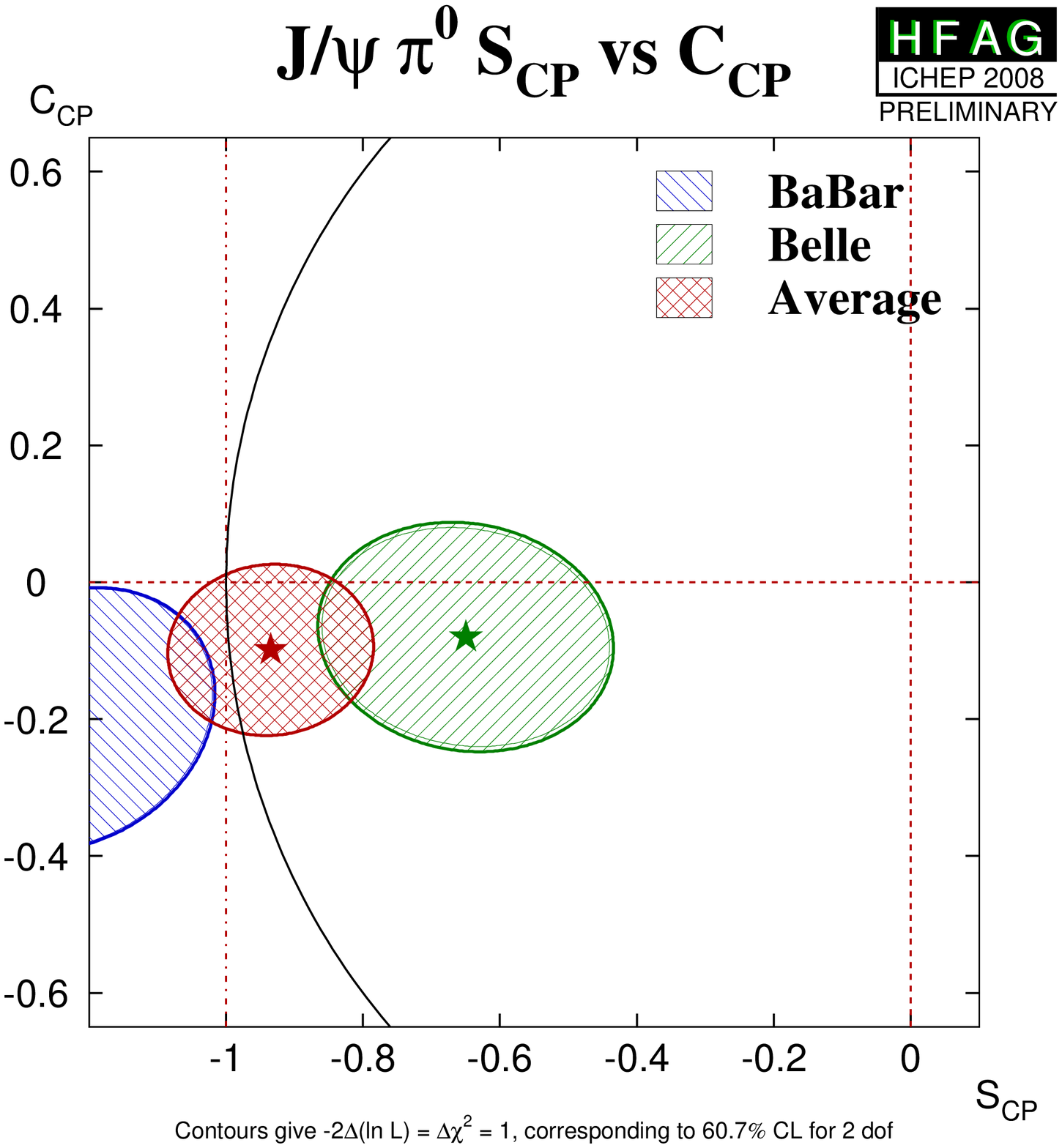}
      }
      &
      \resizebox{0.46\textwidth}{!}{
        \includegraphics{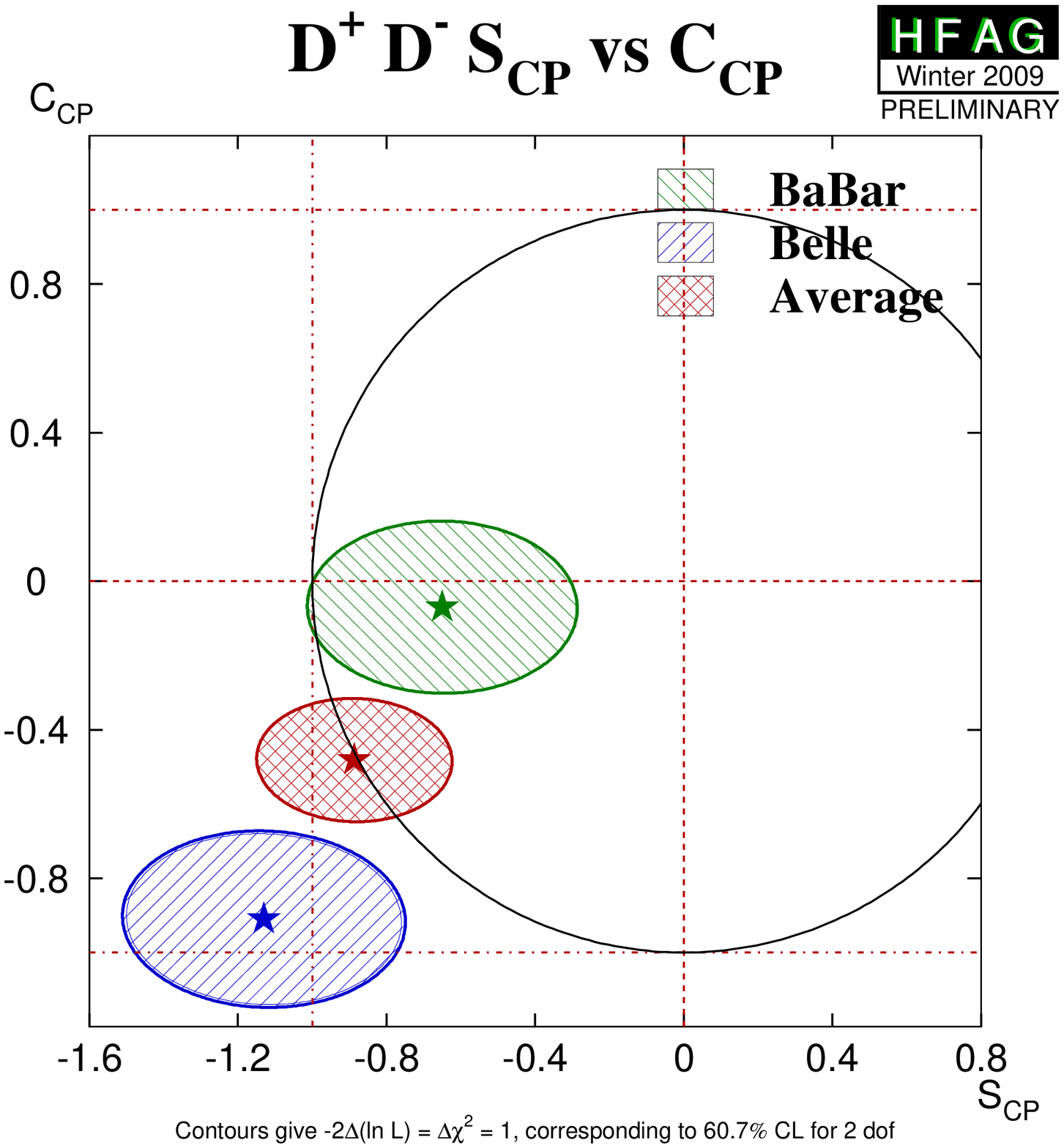}
      }
    \end{tabular}
  \end{center}
  \vspace{-0.8cm}
  \caption{
    Averages of two $b \to c\bar c d$ dominated channels,
    for which correlated averages are performed,
    in the $S_{\CP}$ \vs\ $C_{\CP}$ plane.
    (Left) $\Bz \to J/ \psi \pi^0$ and (right) $\Bz \to D^+D^-$.
  }
  \label{fig:cp_uta:ccd_SvsC}
\end{figure}

The $b \to c\bar{c}d$ transitions can be investigated with studies 
of various different final states. 
Results are available from both \babar\  and \belle\ 
using the final states $\jpsi \pi^0$, $D^+D^-$, 
$D^{*+}D^{*-}$ and $D^{*\pm}D^{\mp}$,
the averages of these results are given in Table~\ref{tab:cp_uta:ccd}.
The results using the $\CP$ eigenstate ($\etacp = +1$) modes
$\jpsi \pi^0$ and $D^+D^-$
are shown in Fig.~\ref{fig:cp_uta:ccd:psipi0} and 
Fig.~\ref{fig:cp_uta:ccd:dd} respectively,
with two-dimensional constraints shown in Fig.~\ref{fig:cp_uta:ccd_SvsC}.

\begin{figure}[htb]
  \begin{center}
    \begin{tabular}{cc}
      \resizebox{0.46\textwidth}{!}{
        \includegraphics{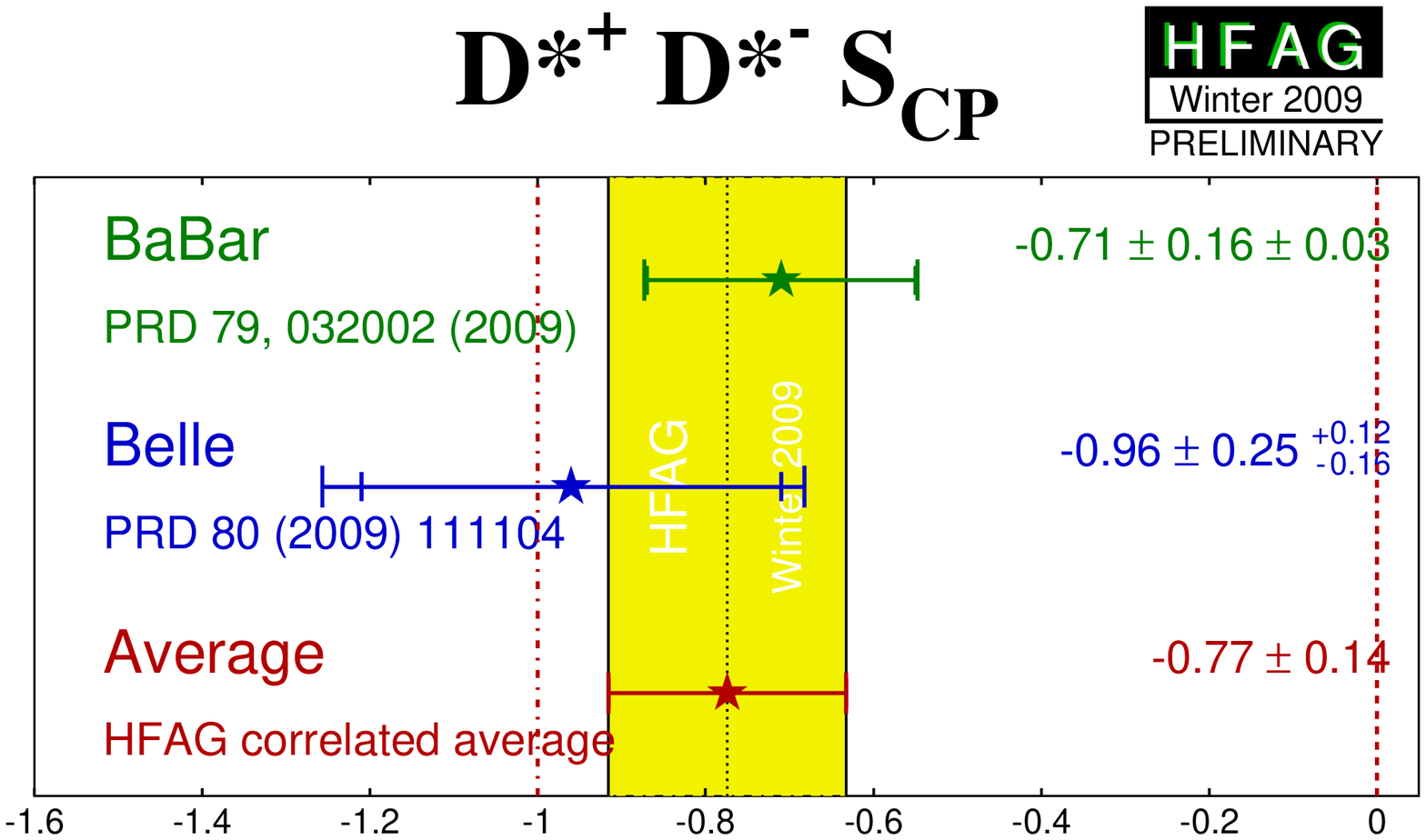}
      }
      &
      \resizebox{0.46\textwidth}{!}{
        \includegraphics{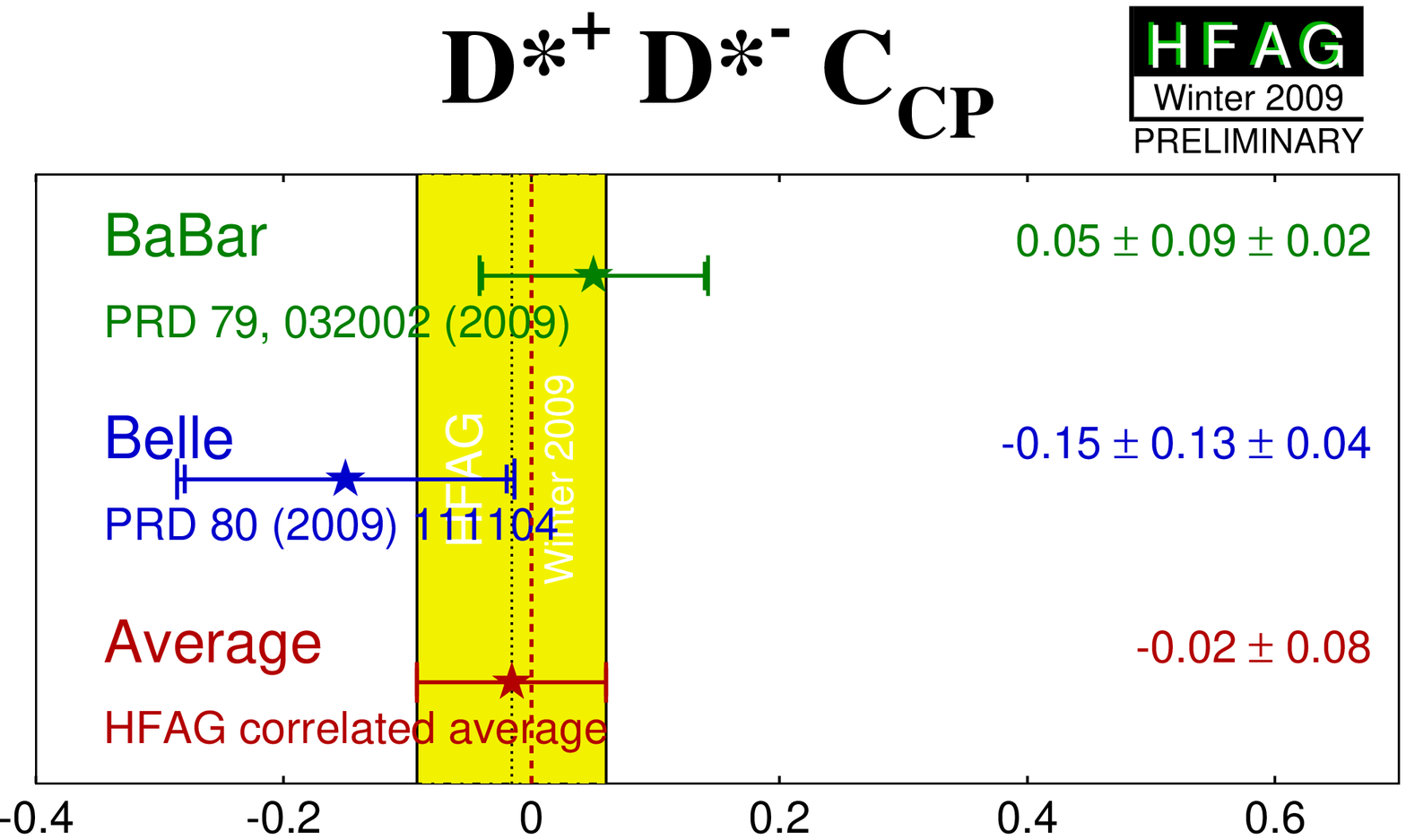}
      }
    \end{tabular}
  \end{center}
  \vspace{-0.8cm}
  \caption{
    Averages of 
    (left) $S_{b \to c\bar c d}$ and (right) $C_{b \to c\bar c d}$ 
    for the mode $\Bz \to D^{*+}D^{*-}$.
  }
  \label{fig:cp_uta:ccd:dstardstar}
\end{figure}

The vector-vector mode $D^{*+}D^{*-}$ 
is found to be dominated by the $\CP$-even longitudinally polarized component;
\babar\ measures a $\CP$-odd fraction of 
$0.158 \pm 0.028 \pm 0.006$~\cite{:2008aw} while
\belle\ measures a $\CP$-odd fraction of 
$0.125 \pm 0.043 \pm 0.023$~\cite{:2009za}.
These values, listed as $R_\perp$, are included in the averages which ensures
the correlations to be taken into account.\footnote{
  Note that the \babar\ value given in Table~\ref{tab:cp_uta:ccd} differs from
  that given above, since that in the table is not corrected for efficiency.
}
\babar\ have also performed an additional fit in which the 
$\CP$-even and $\CP$-odd components are allowed to have different 
$\CP$ violation parameters $S$ and $C$.  
These results are included in Table~\ref{tab:cp_uta:ccd}.
Results using $D^{*+}D^{*-}$ are shown in Fig.~\ref{fig:cp_uta:ccd:dstardstar}.

For the non-$\CP$ eigenstate mode $D^{*\pm}D^{\mp}$
\babar\ uses fully reconstructed events while 
\belle\ combines both fully and partially reconstructed samples.
At present we perform uncorrelated averages of the parameters in the 
$D^{*\pm}D^{\mp}$ system.

\begin{table}[htb]
	\begin{center}
		\caption{
     Averages for the $b \to c\bar{c}d$ modes,
     $\Bz \to J/\psi \pi^{0}$, $D^+D^-$, $D^{*+}D^{*-}$ and $D^{*\pm}D^\mp$.
		}
		\vspace{0.2cm}
		\setlength{\tabcolsep}{0.0pc}
\begin{tabular*}{\textwidth}{@{\extracolsep{\fill}}lrcccc} \hline
	\mc{2}{l}{Experiment} & $N(B\bar{B})$ & $S_{CP}$ & $C_{CP}$ & Correlation \\
	\hline
        \mc{6}{c}{$J/\psi \pi^{0}$} \\
	\babar & \cite{Aubert:2008bs} & 466M & $-1.23 \pm 0.21 \pm 0.04$ & $-0.20 \pm 0.19 \pm 0.03$ & $0.20$ \\
	\belle & \cite{:2007wd} & 535M & $-0.65 \pm 0.21 \pm 0.05$ & $-0.08 \pm 0.16 \pm 0.05$ & $-0.10$ \\
	\mc{3}{l}{\bf Average} & $-0.93 \pm 0.15$ & $-0.10 \pm 0.13$ & $0.04$ \\
	\mc{3}{l}{\small Confidence level} & \mc{2}{c}{\small $0.15~(1.4\sigma)$} & \\
		\hline

        \mc{6}{c}{$D^{+} D^{-}$} \\
	\babar & \cite{:2008aw} & 467M & $-0.65 \pm 0.36 \pm 0.05$ & $-0.07 \pm 0.23 \pm 0.03$ & $-0.01$ \\
	\belle & \cite{Fratina:2007zk} & 535M & $-1.13 \pm 0.37 \pm 0.09$ & $-0.91 \pm 0.23 \pm 0.06$ & $-0.04$ \\
	\mc{3}{l}{\bf Average} & $-0.89 \pm 0.26$ & $-0.48 \pm 0.17$ & $-0.02$ \\
	\mc{3}{l}{\small Confidence level} & \mc{2}{c}{\small $0.025~(2.2\sigma)$} & \\
		\hline
 		\end{tabular*}

                \vspace{2ex}


 		\begin{tabular*}{\textwidth}{@{\extracolsep{\fill}}lrcccc} \hline
 		\mc{2}{l}{Experiment} & $N(B\bar{B})$ & $S_{CP}$ & $C_{CP}$ & $R_\perp$ \\
 		\hline
        \mc{6}{c}{$D^{*+} D^{*-}$} \\
	\babar & \cite{:2008aw} & 467M & $-0.71 \pm 0.16 \pm 0.03$ & $0.05 \pm 0.09 \pm 0.02$ & $0.17 \pm 0.03$ \\
	\belle & \cite{:2009za} & 657M & $-0.96 \pm 0.25 \,^{+0.12}_{-0.16}$ & $-0.15 \pm 0.13 \pm 0.04$ & $0.12 \pm 0.04 \pm 0.02$ \\
	\mc{3}{l}{\bf Average} & $-0.77 \pm 0.14$ & $-0.02 \pm 0.08$ & $0.16 \pm 0.02$ \\
	\mc{3}{l}{\small Confidence level} & \mc{3}{c}{\small $0.41~(0.8\sigma)$} \\
		\hline
		\end{tabular*}

                \vspace{2ex}

    \resizebox{\textwidth}{!}{
		\begin{tabular}{@{\extracolsep{2mm}}lrcccccc} \hline
	\mc{2}{l}{Experiment} & $N(B\bar{B})$ & $S_{CP+}$ & $C_{CP+}$ & $S_{CP-}$ & $C_{CP-}$ & $R_\perp$ \\
	\hline
        \mc{7}{c}{$D^{*+} D^{*-}$} \\
	\babar & \cite{:2008aw} & 467M & $-0.76 \pm 0.16 \pm 0.04$ & $0.02 \pm 0.12 \pm 0.02$ & $-1.81 \pm 0.71 \pm 0.16$ & $0.41 \pm 0.50 \pm 0.08$ & $0.15 \pm 0.03$ \\
		\hline
		\end{tabular}
    }

                \vspace{2ex}

    \resizebox{\textwidth}{!}{
		\begin{tabular}{@{\extracolsep{2mm}}lrcccccc} \hline
	\mc{2}{l}{Experiment} & $N(B\bar{B})$ & $S_{+-}$ & $C_{+-}$ & $S_{-+}$ & $C_{-+}$ & ${\cal A}$ \\
        \hline
        \mc{8}{c}{$D^{*\pm} D^{\mp}$} \\
	\babar & \cite{:2008aw} & 467M & $-0.63 \pm 0.21 \pm 0.03$ & $0.08 \pm 0.17 \pm 0.04$ & $-0.74 \pm 0.23 \pm 0.05$ & $0.00 \pm 0.17 \pm 0.03$ & $0.01 \pm 0.05 \pm 0.01$ \\
	\belle & \cite{Aushev:2004uc} & 152M & $-0.55 \pm 0.39 \pm 0.12$ & $-0.37 \pm 0.22 \pm 0.06$ & $-0.96 \pm 0.43 \pm 0.12$ & $0.23 \pm 0.25 \pm 0.06$ & $0.07 \pm 0.08 \pm 0.04$ \\
	\mc{3}{l}{\bf Average} & $-0.61 \pm 0.19$ & $-0.09 \pm 0.14$ & $-0.79 \pm 0.21$ & $0.07 \pm 0.14$ & $0.02 \pm 0.04$ \\
	\mc{3}{l}{\small Confidence level} & {\small $0.86~(0.2\sigma)$} & {\small $0.12~(1.6\sigma)$} & {\small $0.66~(0.4\sigma)$} & {\small $0.46~(0.7\sigma)$} & {\small $0.54~(0.6\sigma)$} \\
        \hline
                \end{tabular}
    }
		\label{tab:cp_uta:ccd}
	\end{center}
\end{table}

In the absence of the penguin contribution (tree dominance),
the time-dependent parameters would be given by
$S_{b \to c\bar c d} = - \etacp \sin(2\beta)$,
$C_{b \to c\bar c d} = 0$,
$S_{+-} = \sin(2\beta + \delta)$,
$S_{-+} = \sin(2\beta - \delta)$,
$C_{+-} = - C_{-+}$ and 
${\cal A} = 0$,
where $\delta$ is the strong phase difference between the 
$D^{*+}D^-$ and $D^{*-}D^+$ decay amplitudes.
In the presence of the penguin contribution,
there is no clean interpretation in terms of CKM parameters,
however
direct $\CP$ violation may be observed as any of
$C_{b \to c\bar c d} \neq 0$, $C_{+-} \neq - C_{-+}$ or $A_{+-} \neq 0$.

\begin{figure}[htb]
  \begin{center}
    \begin{tabular}{cc}
      \resizebox{0.46\textwidth}{!}{
        \includegraphics{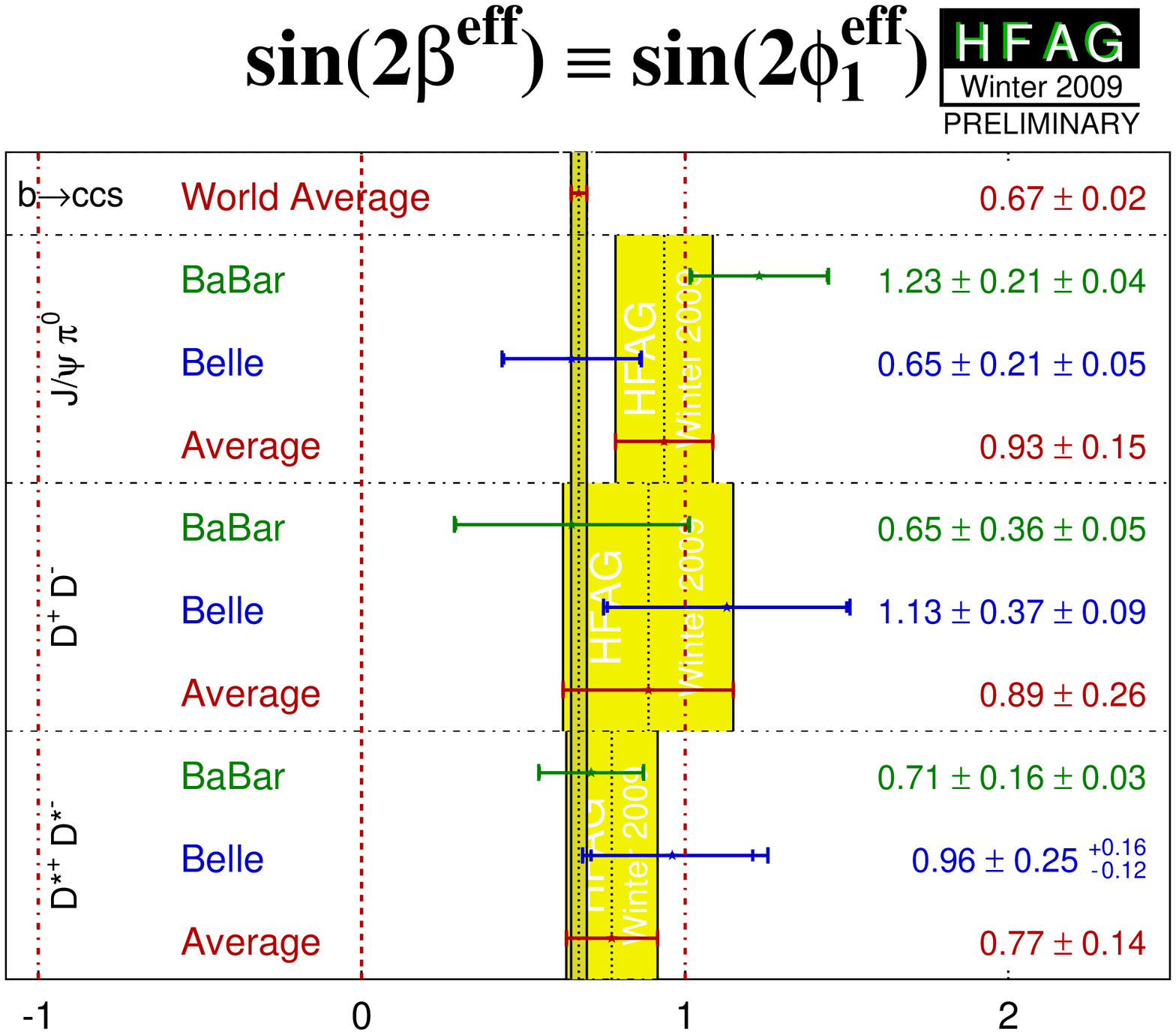}
      }
      &
      \resizebox{0.46\textwidth}{!}{
        \includegraphics{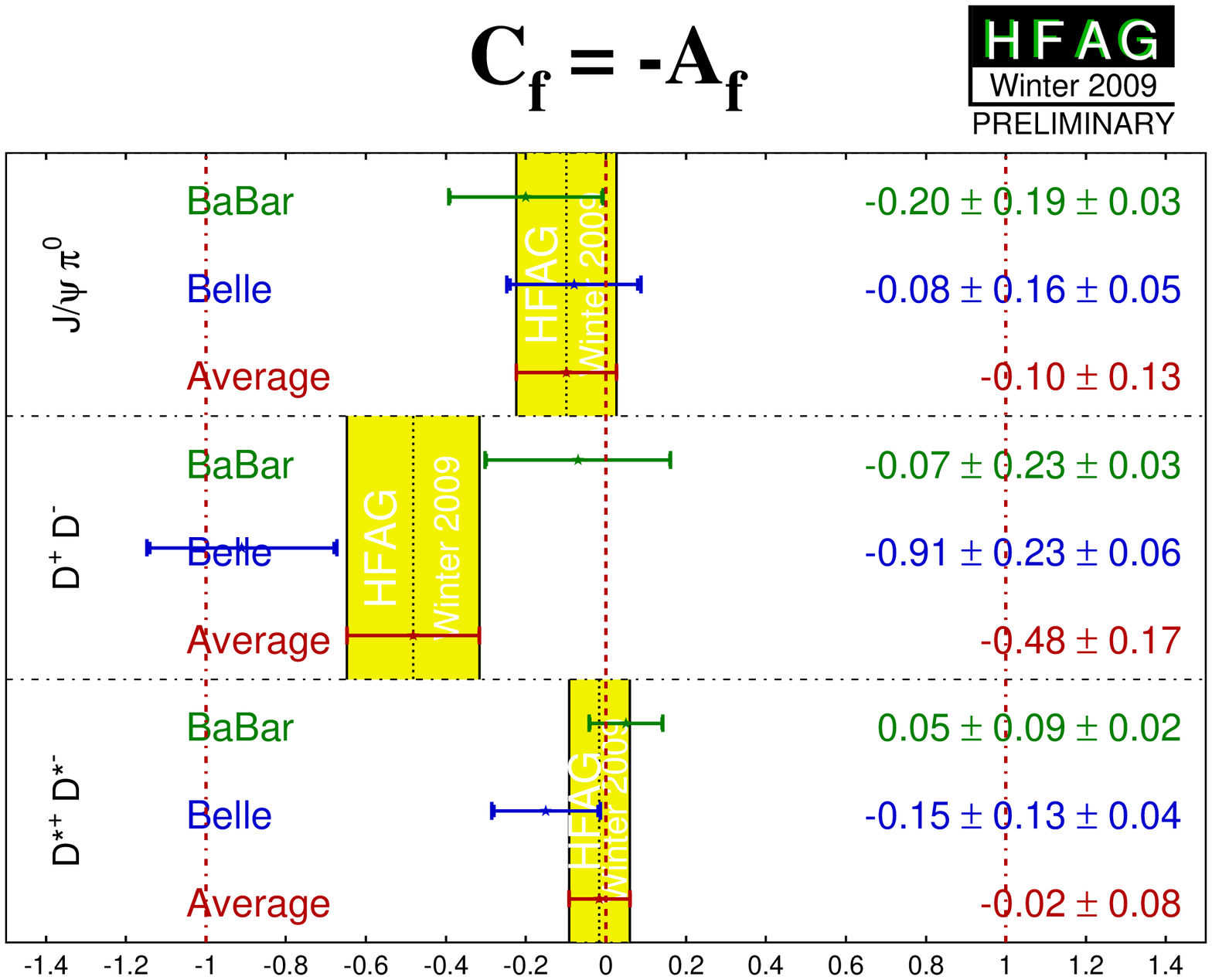}
      }
    \end{tabular}
  \end{center}
  \vspace{-0.8cm}
  \caption{
    Averages of 
    (left) $-\etacp S_{b \to c\bar c d}$ and (right) $C_{b \to c\bar c d}$.
    The $-\etacp S_{b \to q\bar q s}$ figure compares the results to 
    the world average 
    for $-\etacp S_{b \to c\bar c s}$ (see Section~\ref{sec:cp_uta:ccs:cp_eigen}).
  }
  \label{fig:cp_uta:ccd}
\end{figure}

\begin{figure}[htb]
  \begin{center}
    \resizebox{0.66\textwidth}{!}{
      \includegraphics{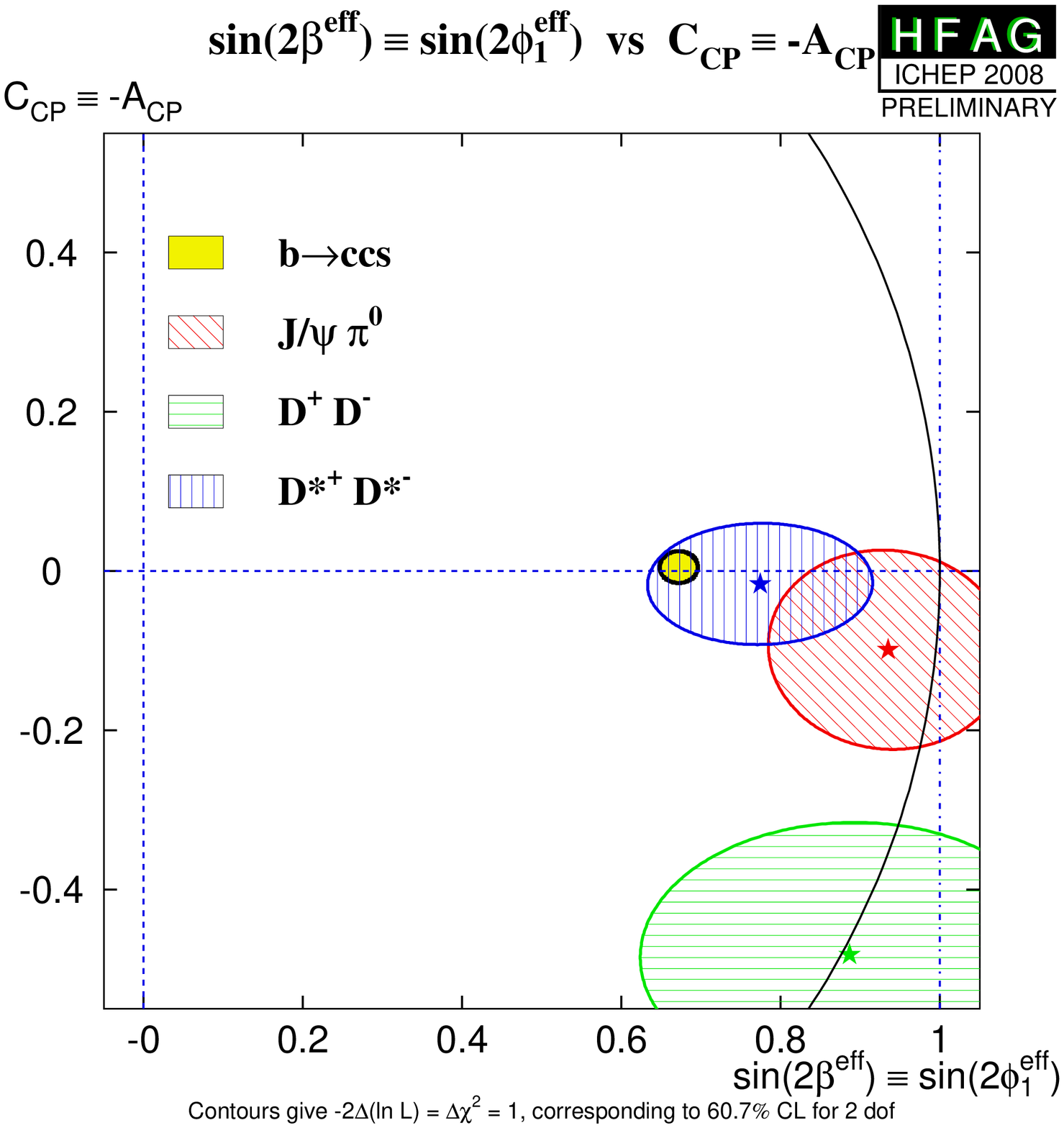}
    }
  \end{center}
  \vspace{-0.8cm}
  \caption{
    Compilation of constraints in the 
    $-\etacp S_{b \to c\bar c d}$ \vs\ $C_{b \to c\bar c d}$ plane.
  }
  \label{fig:cp_uta:ccd_SvsC-all}
\end{figure}

The averages for the $b \to c\bar c d$ modes 
are shown in Figs.~\ref{fig:cp_uta:ccd} and~\ref{fig:cp_uta:ccd_SvsC-all}.
Results are consistent with tree dominance,
and with the Standard Model,
though the \belle\ results in $\Bz \to D^+D^-$~\cite{Fratina:2007zk}
show an indication of direct $\CP$ violation,
and hence a non-zero penguin contribution.
The average of $S_{b \to c\bar c d}$ in both $J/\psi \pi^{0}$ and
$D^{*+}D^{*-}$ final states is more than $5\sigma$ from zero, corresponding to
observations of \CP violation in these decay channels.,
That in the $D^+D^-$ final state is more than $3\sigma$ from zero;
however, due to the large uncertainty and possible non-Gaussian effects,
any strong conclusion should be deferred.


\clearpage
\mysubsection{Time-dependent $\CP$ asymmetries in $b \to q\bar{q}d$ transitions
}
\label{sec:cp_uta:qqd}

Decays such as $\Bz\to\KS\KS$ are pure $b \to q\bar{q}d$ penguin transitions.
As shown in Eq.~\ref{eq:cp_uta:b_to_d},
this diagram has different contributing weak phases,
and therefore the observables are sensitive to the difference 
(which can be chosen to be either $\beta$ or $\gamma$).
Note that if the contribution with the top quark in the loop dominates,
the weak phase from the decay amplitudes should cancel that from mixing,
so that no $\CP$ violation (neither mixing-induced nor direct) occurs.
Non-zero contributions from loops with intermediate up and charm quarks
can result in both types of effect 
(as usual, a strong phase difference is required for direct $\CP$ violation
to occur).

\begin{figure}[htb]
  \begin{center}
    \begin{tabular}{cc}
      \resizebox{0.46\textwidth}{!}{
        \includegraphics{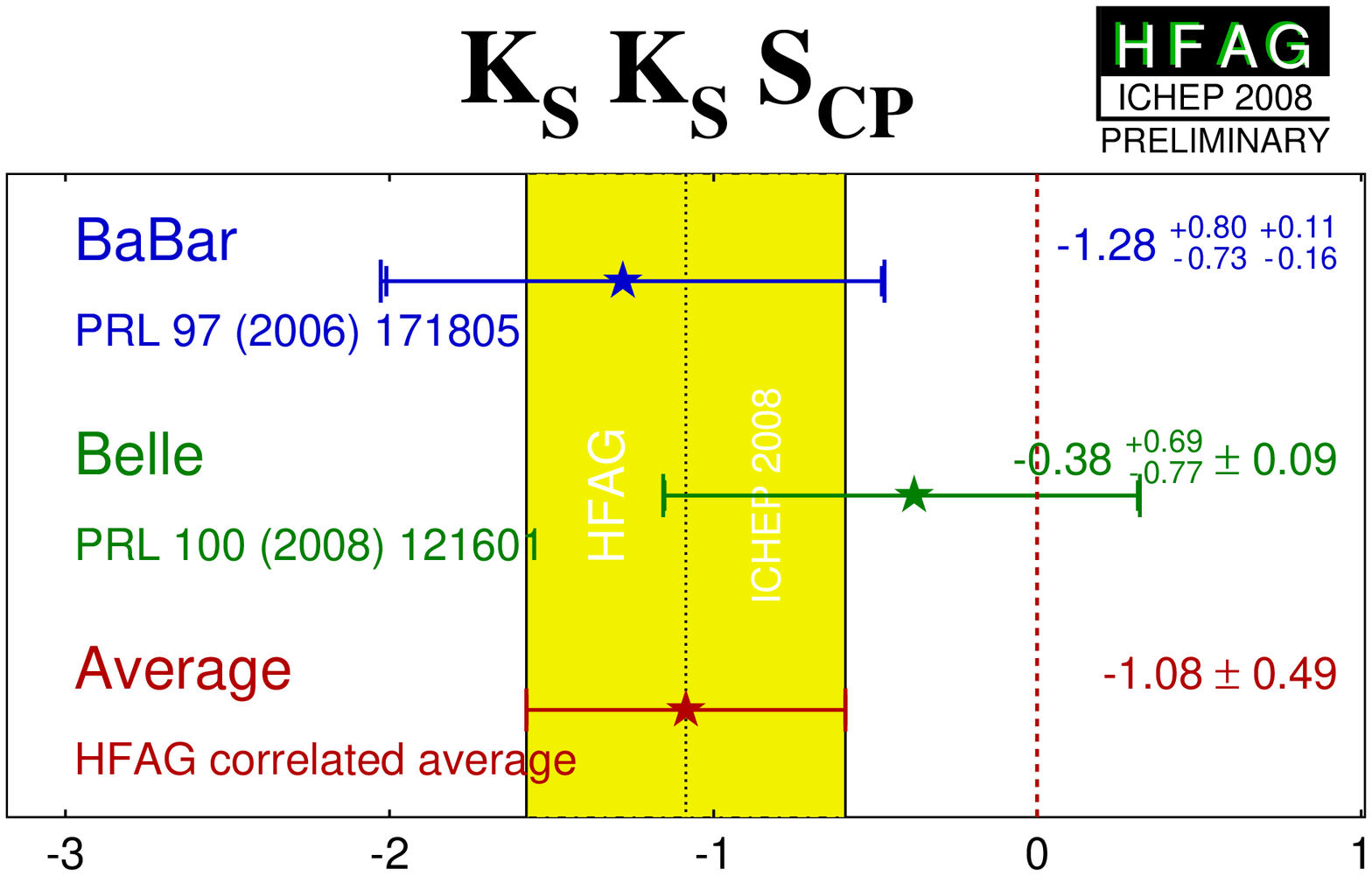}
      }
      &
      \resizebox{0.46\textwidth}{!}{
        \includegraphics{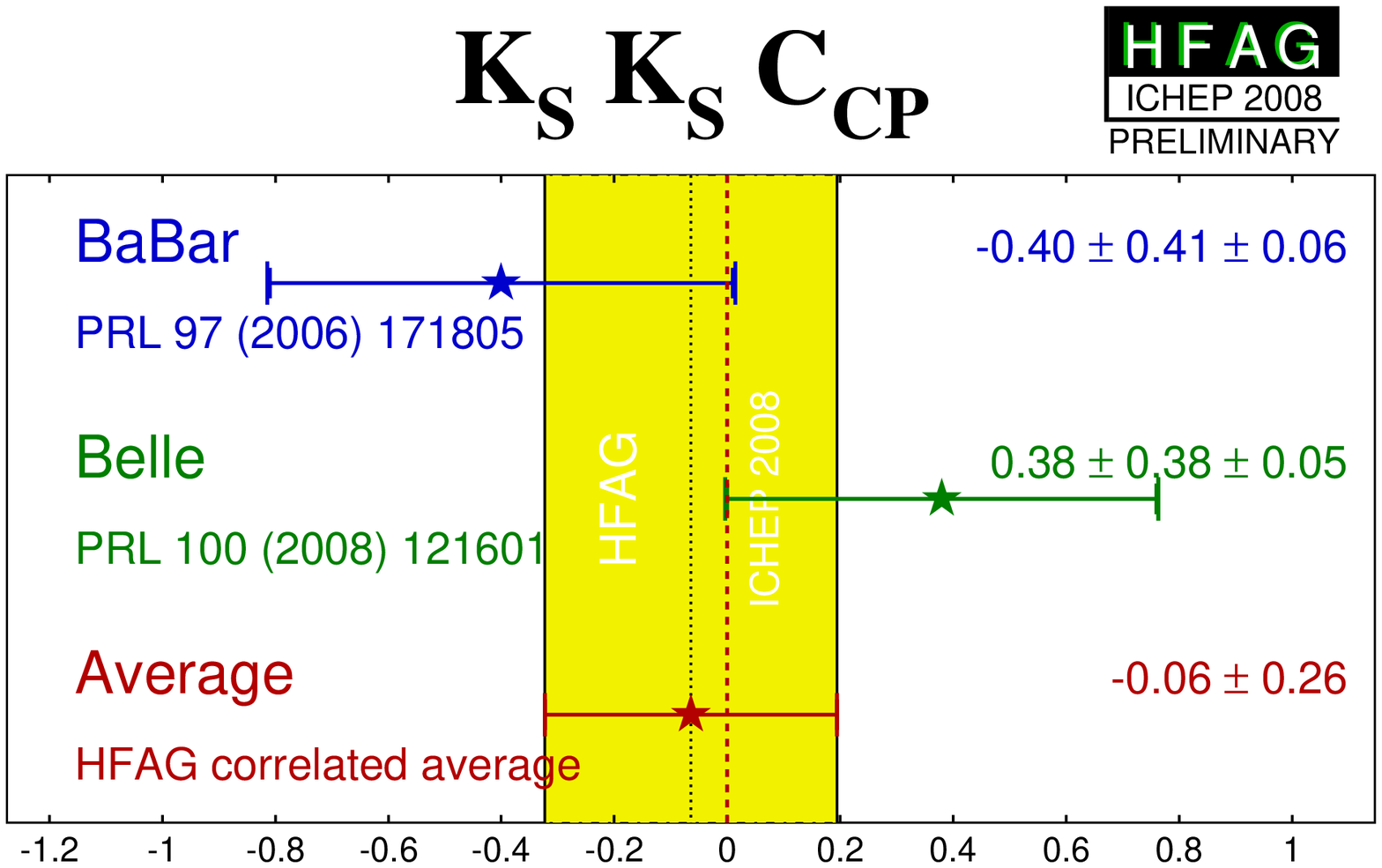}
      }
    \end{tabular}
  \end{center}
  \vspace{-0.8cm}
  \caption{
    Averages of 
    (left) $S_{b \to q\bar q d}$ and (right) $C_{b \to q\bar q d}$ 
    for the mode $\Bz \to \KS\KS$.
  }
  \label{fig:cp_uta:qqd:ksks}
\end{figure}

Both \babar~\cite{Aubert:2006gm} and \belle~\cite{Nakahama:2007dg}
have performed time-dependent analyses of $\Bz\to\KS\KS$.
The results are shown in Table~\ref{tab:cp_uta:qqd}
and Fig.~\ref{fig:cp_uta:qqd:ksks}.

\begin{table}[htb]
	\begin{center}
		\caption{
			Results for $\Bz \to \KS\KS$.
		}
		\vspace{0.2cm}
		\setlength{\tabcolsep}{0.0pc}
		\begin{tabular*}{\textwidth}{@{\extracolsep{\fill}}lrcccc} \hline
	\mc{2}{l}{Experiment} & $N(B\bar{B})$ & $S_{CP}$ & $C_{CP}$ & Correlation \\
	\hline
	\babar & \cite{Aubert:2006gm} & 350M & $-1.28 \,^{+0.80}_{-0.73} \,^{+0.11}_{-0.16}$ & $-0.40 \pm 0.41 \pm 0.06$ & $-0.32$ \\
	\belle & \cite{Nakahama:2007dg} & 657M & $-0.38 \,^{+0.69}_{-0.77} \pm 0.09$ & $0.38 \pm 0.38 \pm 0.05$ & $0.48$ \\
	\hline
	\mc{3}{l}{\bf Average} & $-1.08 \pm 0.49$ & $-0.06 \pm 0.26$ & $0.14$ \\
	\mc{3}{l}{\small Confidence level} & \mc{2}{c}{\small $0.29~(1.1\sigma)$} & \\
		\hline
		\end{tabular*}
		\label{tab:cp_uta:qqd}
	\end{center}
\end{table}

\clearpage
\mysubsection{Time-dependent asymmetries in $b \to s\gamma$ transitions
}
\label{sec:cp_uta:bsg}

The radiative decays $b \to s\gamma$ produce photons 
which are highly polarized in the Standard Model.
The decays $\Bz \to F \gamma$ and $\Bzb \to F \gamma$ 
produce photons with opposite helicities, 
and since the polarization is, in principle, observable,
these final states cannot interfere.
The finite mass of the $s$ quark introduces small corrections
to the limit of maximum polarization,
but any large mixing induced $\CP$ violation would be a signal for new physics.
Since a single weak phase dominates the $b \to s \gamma$ transition in the 
Standard Model, the cosine term is also expected to be small.

Atwood {\it et al.}~\cite{Atwood:2004jj} have shown that 
an inclusive analysis with respect to $\KS\pi^0\gamma$ can be performed,
since the properties of the decay amplitudes 
are independent of the angular momentum of the $\KS\pi^0$ system. 
However, if non-dipole operators contribute significantly to the amplitudes, 
then the Standard Model mixing-induced $\CP$ violation could be larger 
than the na\"\i ve expectation 
$S \simeq -2 (m_s/m_b) \sin \left(2\beta\right)$~\cite{Grinstein:2004uu,Grinstein:2005nu}.
In this case, 
the $\CP$ parameters may vary over the $\KS\pi^0\gamma$ Dalitz plot, 
for example as a function of the $\KS\pi^0$ invariant mass.
Explicit calculations indicate such corrections are small
for exclusive final states~\cite{Matsumori:2005ax,Ball:2006cva}.

With the above in mind, 
we quote two averages: one for $K^*(892)$ candidates only, 
and the other one for the inclusive $\KS\pi^0\gamma$ decay (including the $K^*(892)$).
If the Standard Model dipole operator is dominant, 
both should give the same quantities 
(the latter naturally with smaller statistical error). 
If not, care needs to be taken in interpretation of the inclusive parameters, 
while the results on the $K^*(892)$ resonance remain relatively clean.
Results from \babar~\cite{Aubert:2008gy} and \belle~\cite{Ushiroda:2006fi} are
used for both averages; both experiments use the invariant mass range 
$0.60 \ {\rm GeV}/c^2 < M_{\KS\pi^0} < 1.80 \ {\rm GeV}/c^2$
in the inclusive analysis.
In addition to the $\KS\pi^0\gamma$ decay, \babar\ have presented results
using $\KS\eta\gamma$~\cite{Aubert:2008js}, and \belle\ have presented results
using $\KS\rho\gamma$~\cite{Li:2008qma}.

\begin{table}[htb]
	\begin{center}
		\caption{
      Averages for $b \to s \gamma$ modes.
		}
		\vspace{0.2cm}
		\setlength{\tabcolsep}{0.0pc}
\begin{tabular*}{\textwidth}{@{\extracolsep{\fill}}lrcccc} \hline
	\mc{2}{l}{Experiment} & $N(B\bar{B})$ & $S_{CP} (b \to s \gamma)$ & $C_{CP} (b \to s \gamma)$ & Correlation \\
        \hline
        \mc{6}{c}{$\Kstar(892)\gamma$} \\
	\babar & \cite{Aubert:2008gy} & 467M & $-0.03 \pm 0.29 \pm 0.03$ & $-0.14 \pm 0.16 \pm 0.03$ & $0.05$ \\
	\belle & \cite{Ushiroda:2006fi} & 535M & $-0.32 \,^{+0.36}_{-0.33} \pm 0.05$ & $0.20 \pm 0.24 \pm 0.05$ & $0.08$ \\
	\mc{3}{l}{\bf Average} & $-0.16 \pm 0.22$ & $-0.04 \pm 0.14$ & $0.06$ \\
	\mc{3}{l}{\small Confidence level} & \mc{2}{c}{\small $0.40~(0.9\sigma)$} & \\
		\hline

        \mc{6}{c}{$\KS \pi^0 \gamma$ (including $\Kstar(892)\gamma$)} \\
	\babar & \cite{Aubert:2008gy} & 467M & $-0.17 \pm 0.26 \pm 0.03$ & $-0.19 \pm 0.14 \pm 0.03$ & $0.04$ \\
	\belle & \cite{Ushiroda:2006fi} & 535M & $-0.10 \pm 0.31 \pm 0.07$ & $0.20 \pm 0.20 \pm 0.06$ & $0.08$ \\
	\mc{3}{l}{\bf Average} & $-0.15 \pm 0.20$ & $-0.07 \pm 0.12$ & $0.05$ \\
        \mc{3}{l}{\small Confidence level} & \mc{2}{c}{\small $0.30~(1.0\sigma)$} & \\

		\hline

        \mc{6}{c}{$\KS \eta \gamma$} \\
	\babar & \cite{Aubert:2008js} & 465M & $-0.18 \,^{+0.49}_{-0.46} \pm 0.12$ & $-0.32 \,^{+0.40}_{-0.39} \pm 0.07$ & $-0.17$ \\
	\hline

        \mc{6}{c}{$\KS \rho^0 \gamma$} \\
	\belle & \cite{Li:2008qma} & 657M & $0.11 \pm 0.33 \,^{+0.05}_{-0.09}$ & $-0.05 \pm 0.18 \pm 0.06$ & $0.04$ \\
	\hline
		\end{tabular*}
		\label{tab:cp_uta:bsg}
	\end{center}
\end{table}

\begin{figure}[htb]
  \begin{center}
    \begin{tabular}{cc}
      \resizebox{0.46\textwidth}{!}{
        \includegraphics{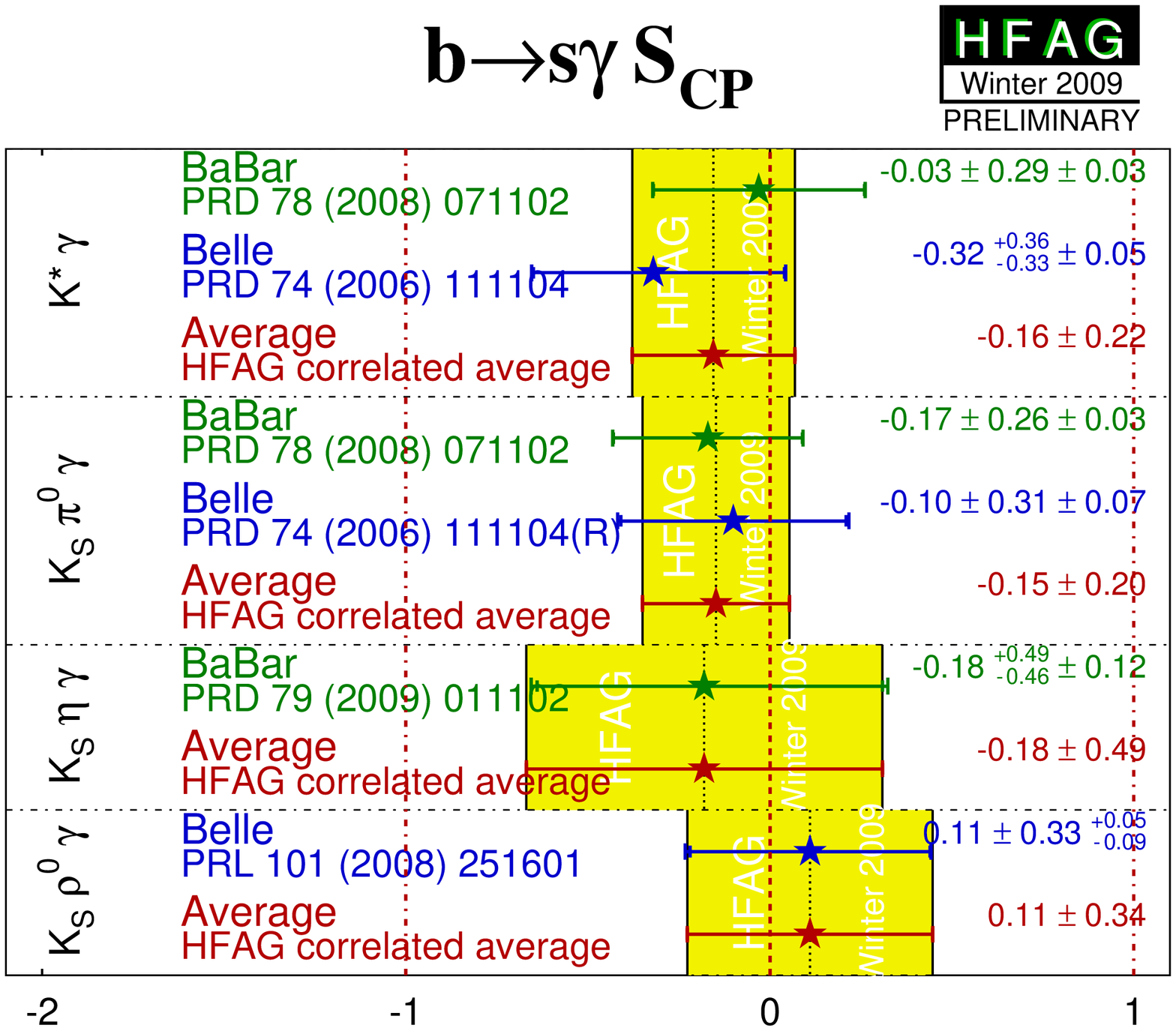}
      }
      &
      \resizebox{0.46\textwidth}{!}{
        \includegraphics{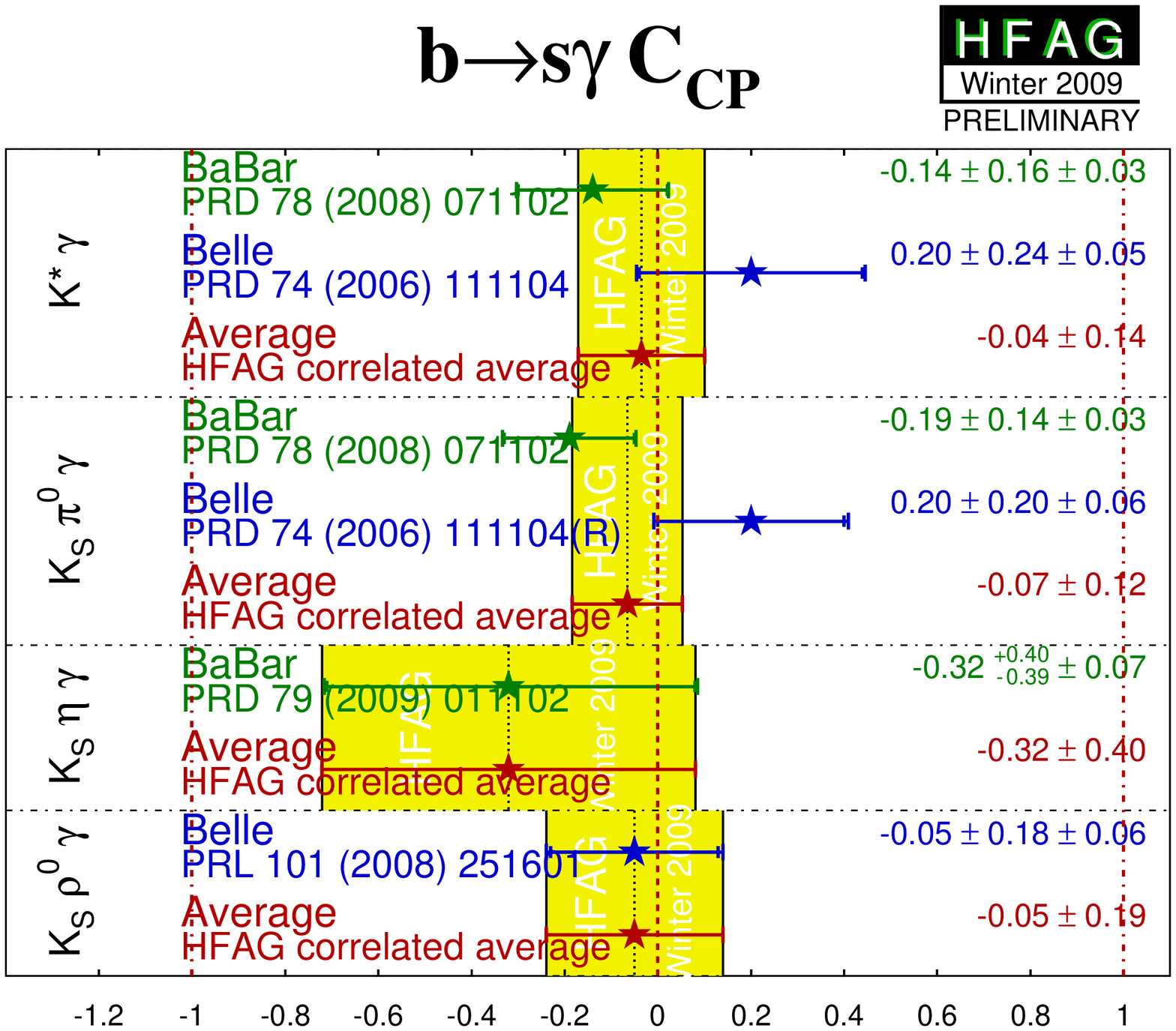}
      }
    \end{tabular}
  \end{center}
  \vspace{-0.8cm}
  \caption{
    Averages of (left) $S_{b \to s \gamma}$ and (right) $C_{b \to s \gamma}$.
    Recall that the data for $K^*\gamma$ is a subset of that for $\KS\pi^0\gamma$.
  }
  \label{fig:cp_uta:bsg}
\end{figure}

\begin{figure}[htb]
  \begin{center}
    \begin{tabular}{cc}
      \resizebox{0.46\textwidth}{!}{
        \includegraphics{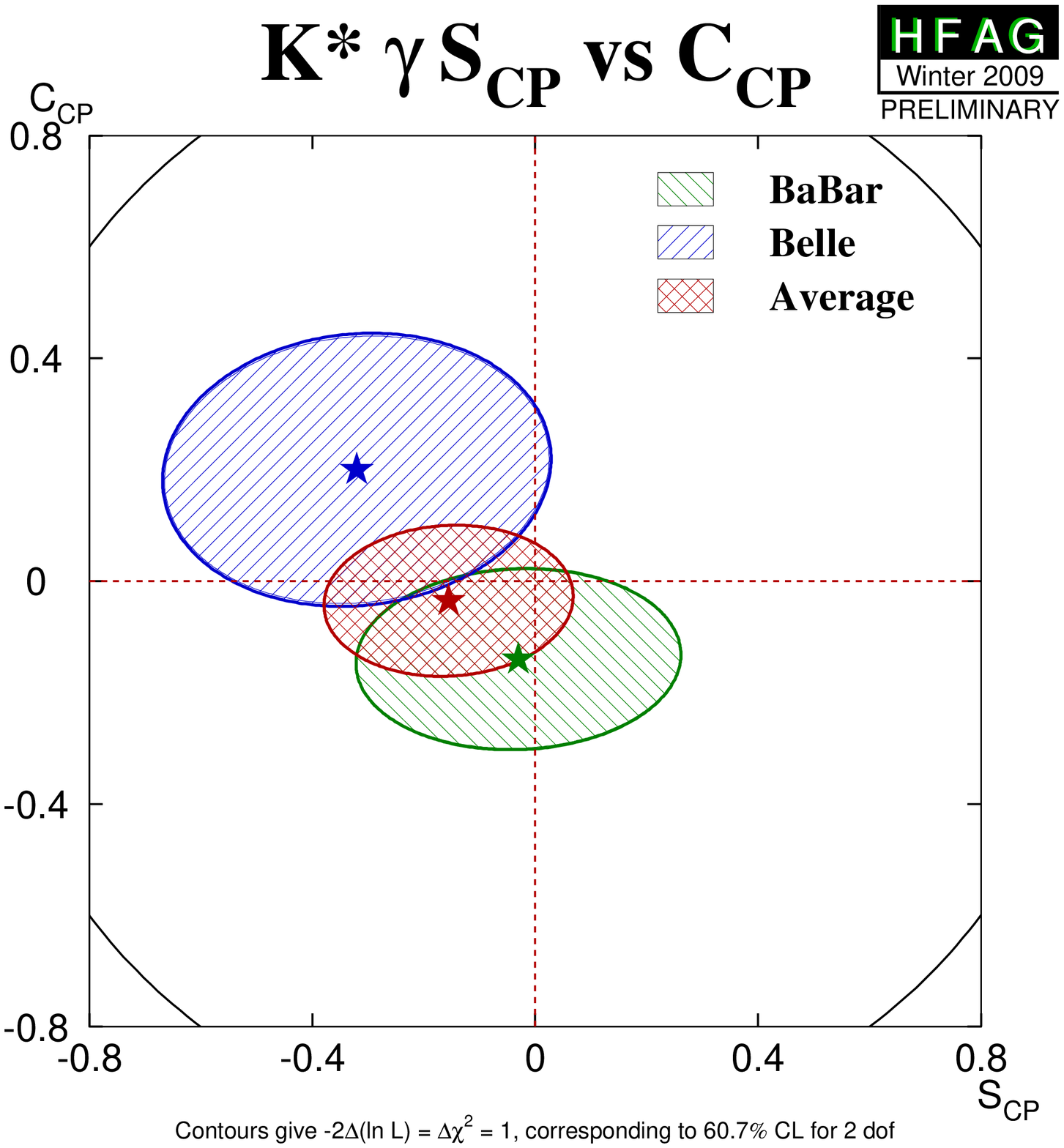}
      }
      &
      \resizebox{0.46\textwidth}{!}{
        \includegraphics{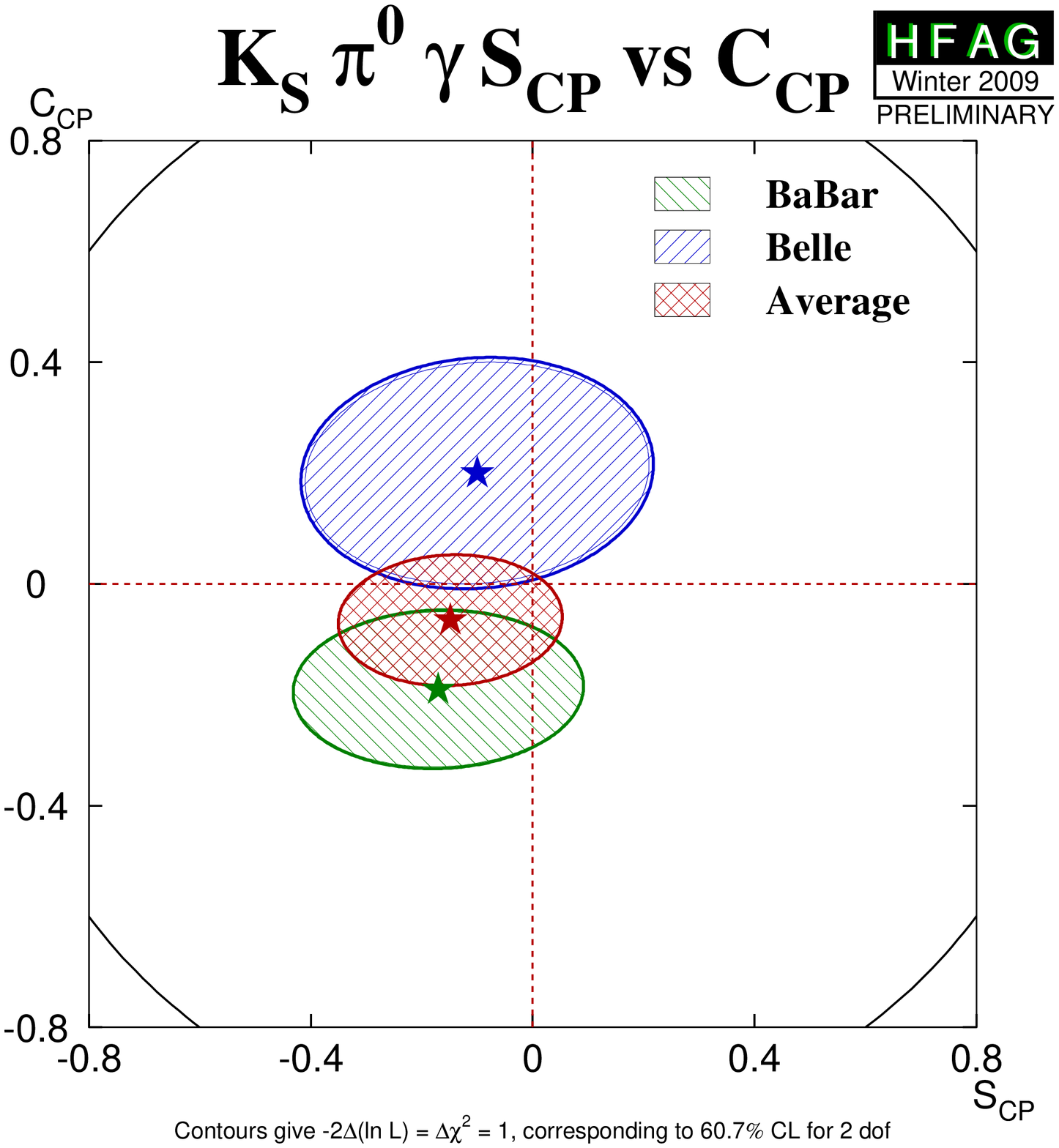}
      }
    \end{tabular}
  \end{center}
  \vspace{-0.8cm}
  \caption{
    Averages of $b \to s\gamma$ dominated channels,
    for which correlated averages are performed,
    in the $S_{\CP}$ \vs\ $C_{\CP}$ plane.
    (Left) $\Bz \to K^*\gamma$ and 
    (right) $\Bz \to \KS\pi^0\gamma$ (including $K^*\gamma$).
  }
  \label{fig:cp_uta:bsg_SvsC}
\end{figure}

The results are shown in Table~\ref{tab:cp_uta:bsg},
and in Figs.~\ref{fig:cp_uta:bsg} and~~\ref{fig:cp_uta:bsg_SvsC}.
No significant $\CP$ violation results are seen;
the results are consistent with the Standard Model
and with other measurements in the $b \to s\gamma$ system (see Sec.~\ref{sec:rare}).

\mysubsection{Time-dependent asymmetries in $b \to d\gamma$ transitions
}
\label{sec:cp_uta:bdg}

The formalism for the radiative decays $b \to d\gamma$ is much the same
as that for $b \to s\gamma$ discussed above.
Assuming dominance of the top quark in the loop,
the weak phase in decay should cancel with that from mixing,
so that the mixing-induced \CP\ violation parameter $S_{\CP}$ 
should be very small.
Corrections due to the finite light quark mass are smaller
compared to $b \to s\gamma$, since $m_d < m_s$,
and although QCD corrections may still play a role,
they cannot significantly affect the prediction $S_{b \to d \gamma} \simeq 0$.
Large direction \CP\ violation effects could, however, be seen through
a non-zero value of $C_{b \to d \gamma}$, 
since the top loop is not the only contribution.

Results using the mode $\Bz \to \rho^0\gamma$ are available from 
\belle\ and are shown in Table~\ref{tab:cp_uta:bdg}.

\begin{table}[htb]
	\begin{center}
		\caption{
			Averages for $\Bz \to \rho^{0} \gamma$.
		}
		\vspace{0.2cm}
		\setlength{\tabcolsep}{0.0pc}
		\begin{tabular*}{\textwidth}{@{\extracolsep{\fill}}lrcccc} \hline
	\mc{2}{l}{Experiment} & $N(B\bar{B})$ & $S_{CP}$ & $C_{CP}$ & Correlation \\
	\hline
	\belle & \cite{:2007jf} & 657M & $-0.83 \pm 0.65 \pm 0.18$ & $0.44 \pm 0.49 \pm 0.14$ & $-0.08$ \\
		\hline
		\end{tabular*}
		\label{tab:cp_uta:bdg}
	\end{center}
\end{table}

\clearpage
\mysubsection{Time-dependent $\CP$ asymmetries in $b \to u\bar{u}d$ transitions
}
\label{sec:cp_uta:uud}

The $b \to u \bar u d$ transition can be mediated by either 
a $b \to u$ tree amplitude or a $b \to d$ penguin amplitude.
These transitions can be investigated using 
the time dependence of $\Bz$ decays to final states containing light mesons.
Results are available from both \babar\ and \belle\ for the 
$\CP$ eigenstate ($\etacp = +1$) $\pi^+\pi^-$ final state
and for the vector-vector final state $\rho^+\rho^-$,
which is found to be dominated by the $\CP$-even
longitudinally polarized component
(\babar\ measure $f_{\rm long} = 
0.992 \pm 0.024 \, ^{+0.026}_{-0.013}$~\cite{Aubert:2007nua}
while \belle\ measure $f_{\rm long} = 
0.941 \, ^{+0.034}_{-0.040} \pm 0.030$~\cite{Somov:2006sg}).
\babar\ have also performed a time-dependent analysis of the 
vector-vector final state $\rho^0\rho^0$~\cite{:2008iha},
in which they measure  $f_{\rm long} = 0.70 \pm 0.14 \pm 0.05$;
\belle\ measures a smaller branching fraction than \babar\ for
$\Bz\to\rho^0\rho^0$~\cite{:2008et} with corresponding signal yields too small
to perform time-dependent or angular analyses.
\babar\ have furthermore performed a time-dependent analysis of the 
$\Bz \to a_1^\pm \pi^\mp$ decay~\cite{Aubert:2006gb}; further experimental
input for the extraction of $\alpha$ from this channel is reported in a later
publication~\cite{:2009ii}.

\begin{figure}[htb]
  \begin{center}
    \begin{tabular}{cc}
      \resizebox{0.46\textwidth}{!}{
        \includegraphics{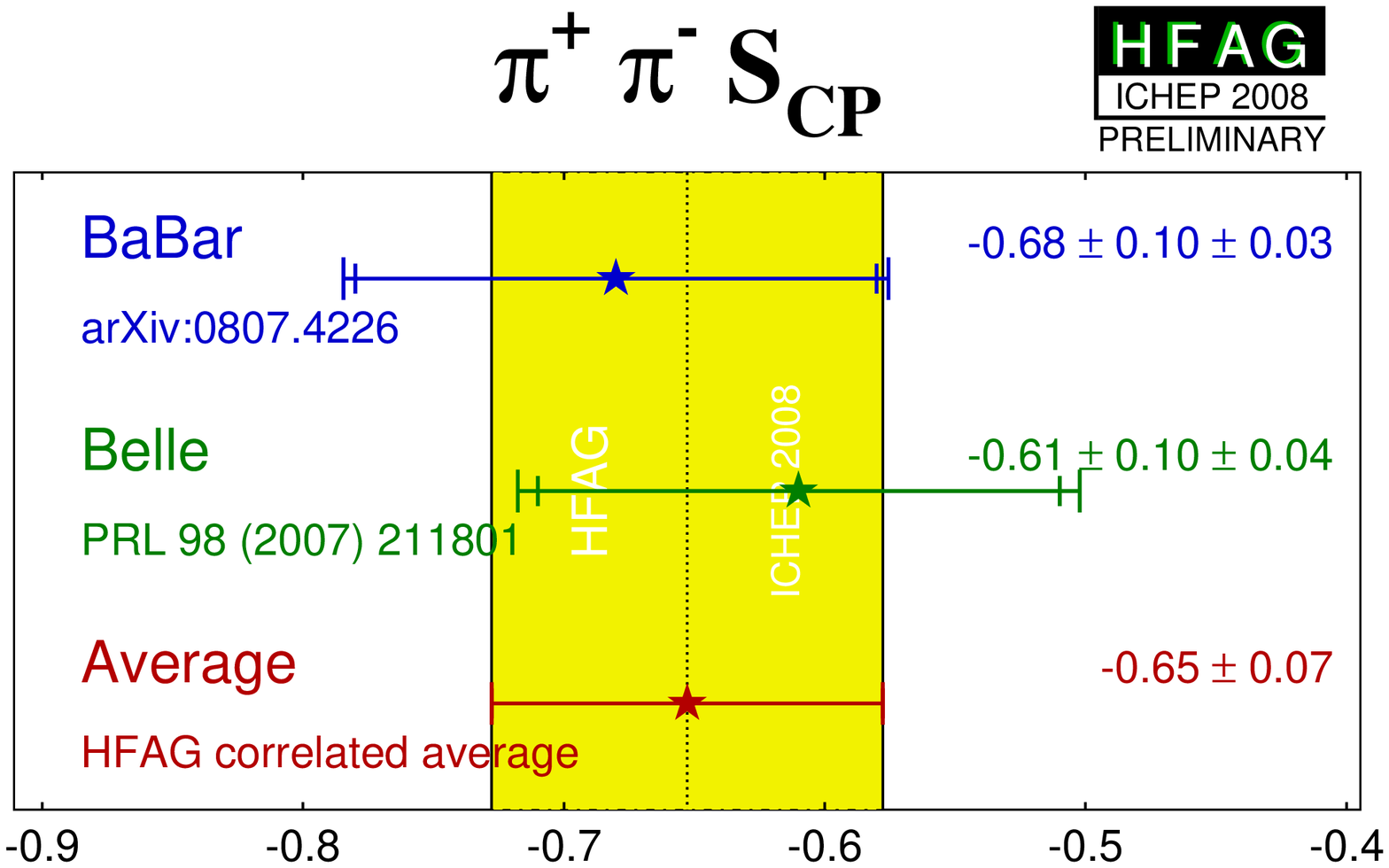}
      }
      &
      \resizebox{0.46\textwidth}{!}{
        \includegraphics{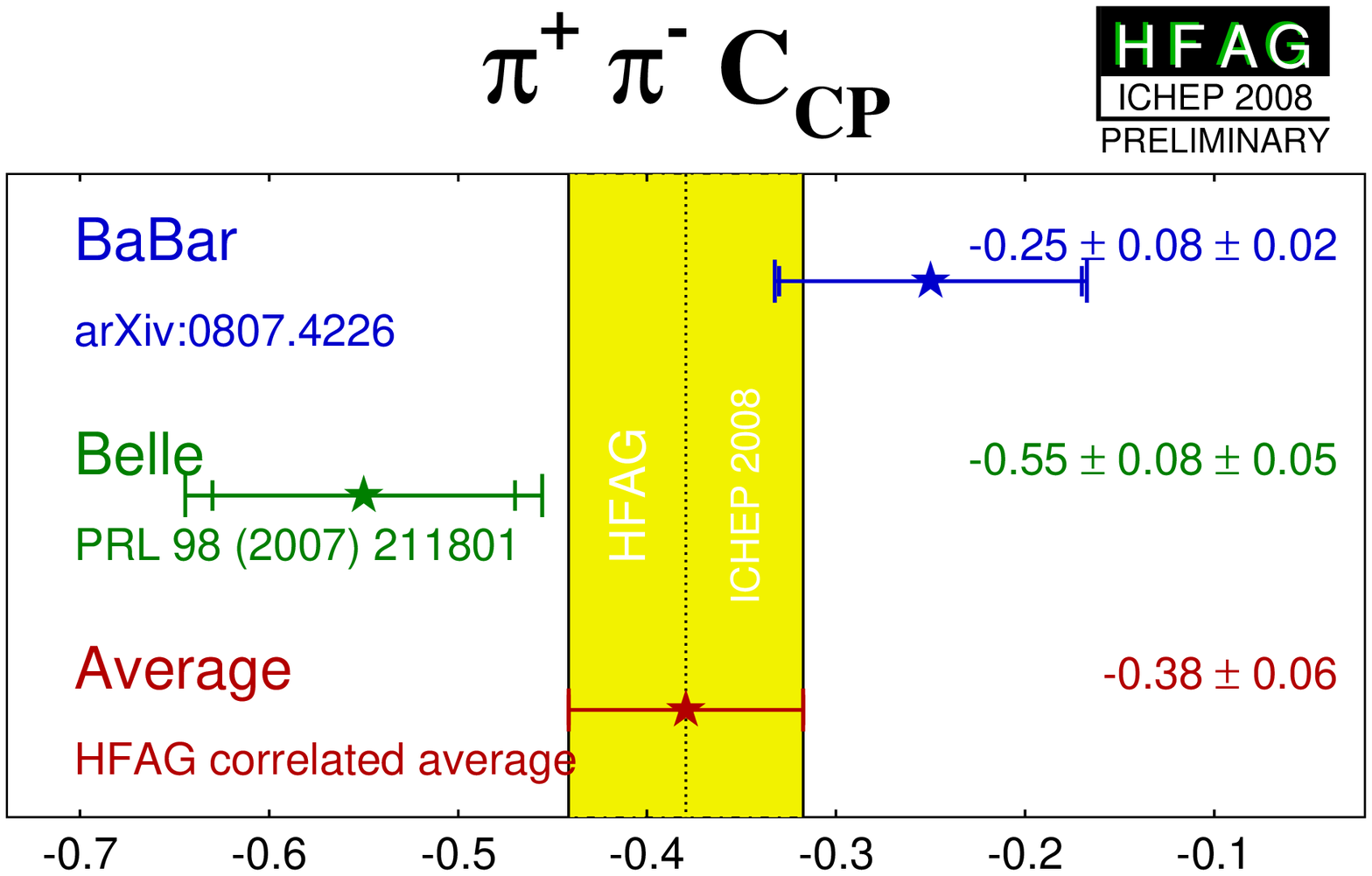}
      }
    \end{tabular}
  \end{center}
  \vspace{-0.8cm}
  \caption{
    Averages of (left) $S_{b \to u\bar u d}$ and (right) $C_{b \to u\bar u d}$
    for the mode $\Bz \to \pi^+\pi^-$.
  }
  \label{fig:cp_uta:uud:pipi}
\end{figure}

\begin{figure}[htb]
  \begin{center}
    \begin{tabular}{cc}
      \resizebox{0.46\textwidth}{!}{
        \includegraphics{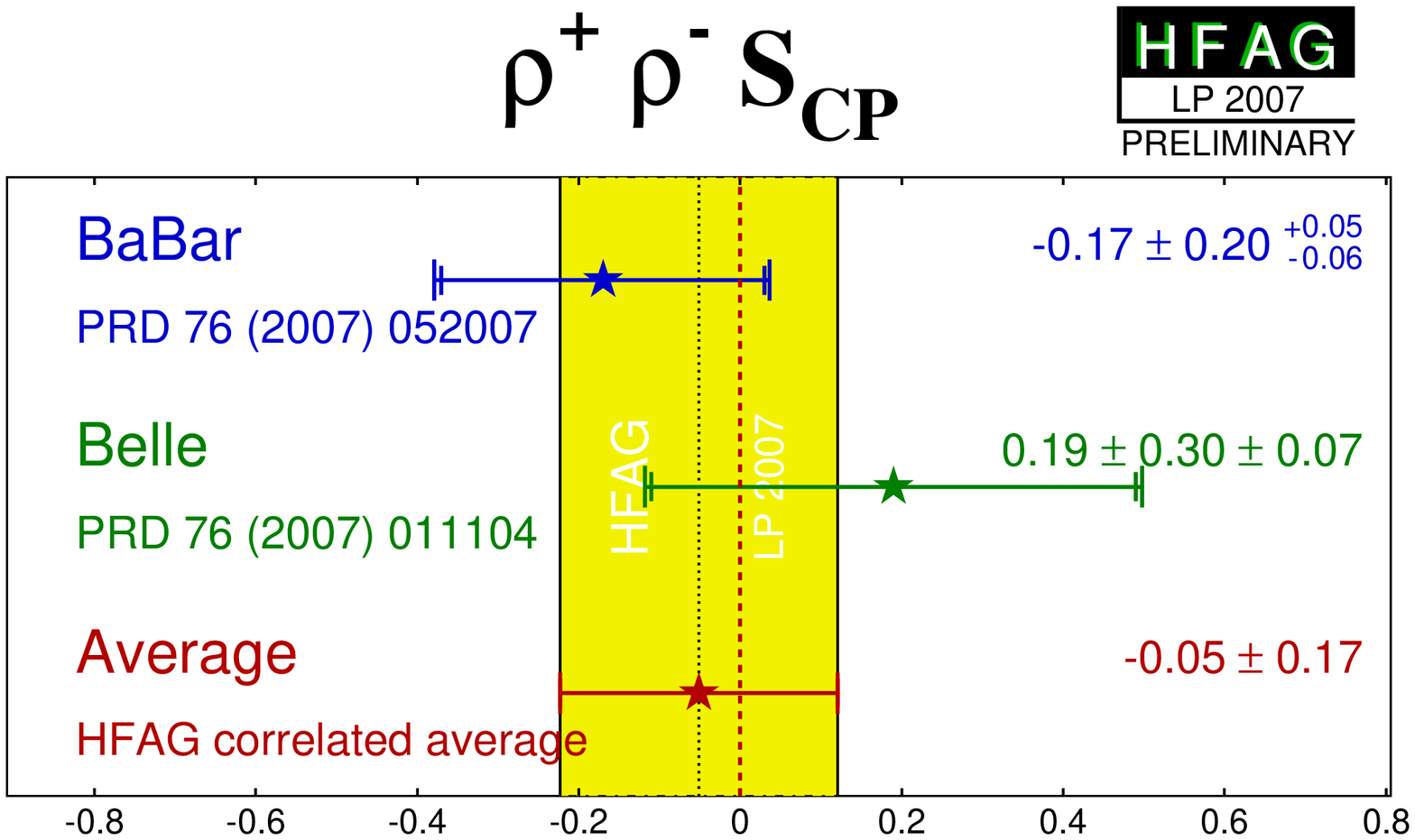}
      }
      &
      \resizebox{0.46\textwidth}{!}{
        \includegraphics{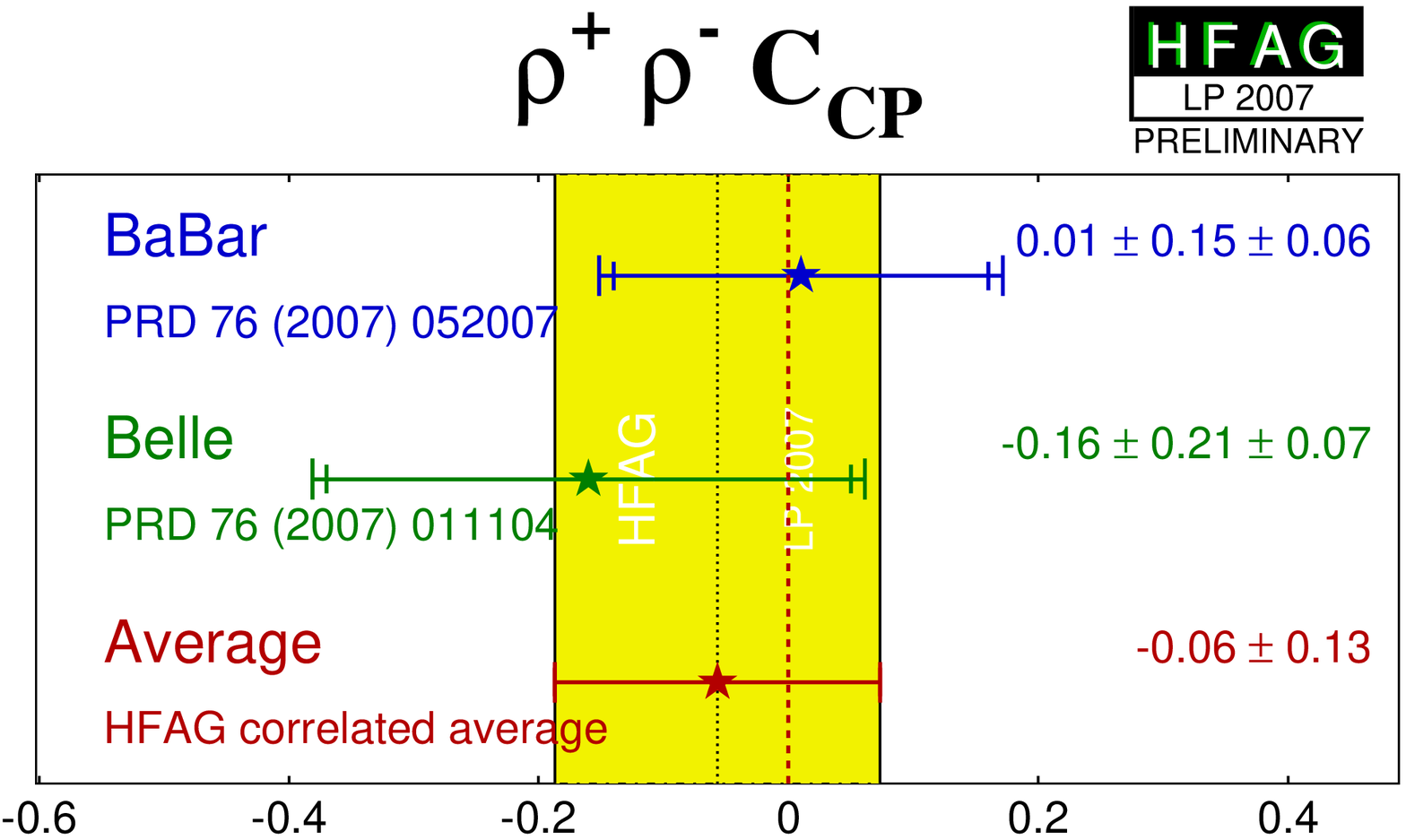}
      }
    \end{tabular}
  \end{center}
  \vspace{-0.8cm}
  \caption{
    Averages of (left) $S_{b \to u\bar u d}$ and (right) $C_{b \to u\bar u d}$
    for the mode $\Bz \to \rho^+\rho^-$.
  }
  \label{fig:cp_uta:uud:rhorho}
\end{figure}

\begin{figure}[htb]
  \begin{center}
    \begin{tabular}{cc}
      \resizebox{0.46\textwidth}{!}{
        \includegraphics{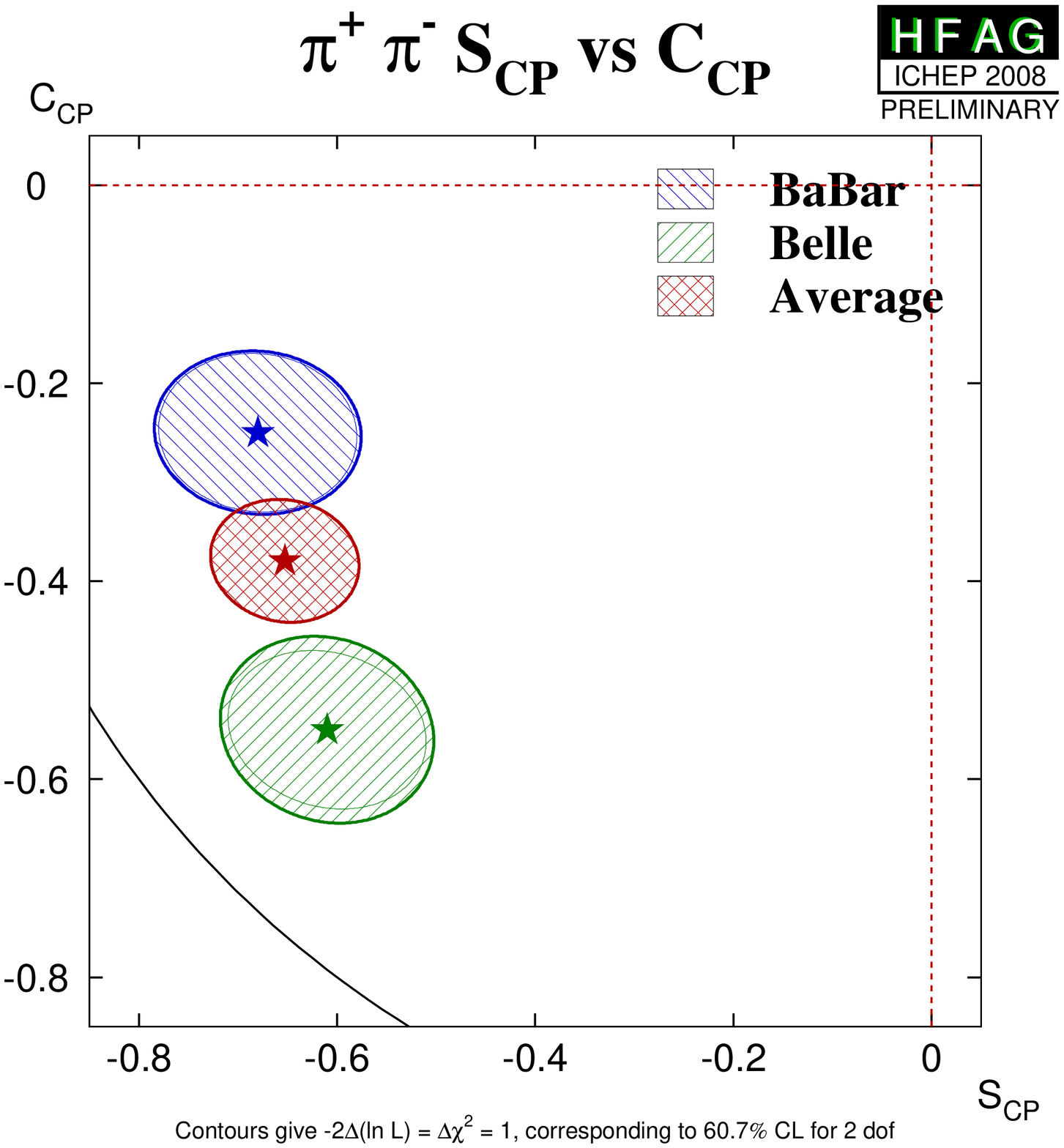}
      }      
      &
      \resizebox{0.46\textwidth}{!}{
        \includegraphics{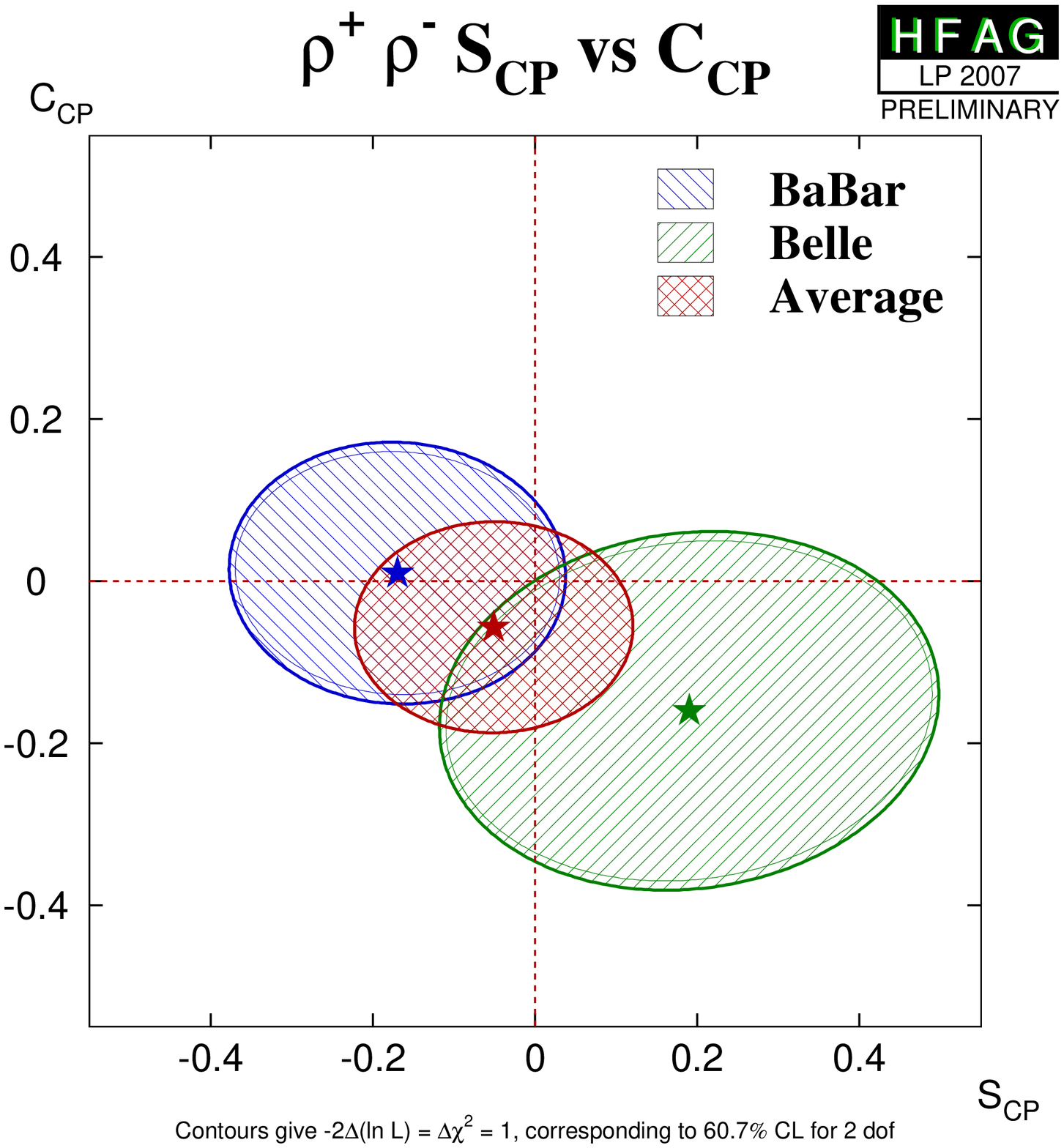}
      }
    \end{tabular}
  \end{center}
  \vspace{-0.8cm}
  \caption{
    Averages of $b \to u\bar u d$ dominated channels,
    for which correlated averages are performed,
    in the $S_{\CP}$ \vs\ $C_{\CP}$ plane.
    (Left) $\Bz \to \pi^+\pi^-$ and (right) $\Bz \to \rho^+\rho^-$.
  }
  \label{fig:cp_uta:uud_SvsC}
\end{figure}

Results, and averages, of time-dependent \CP-violation parameters in 
$b \to u \bar u d$ transitions are listed in Table~\ref{tab:cp_uta:uud}.
The averages for $\pi^+\pi^-$ are shown in Fig.~\ref{fig:cp_uta:uud:pipi},
and those for $\rho^+\rho^-$ are shown in Fig.~\ref{fig:cp_uta:uud:rhorho},
with the averages in the $S_{\CP}$ \vs\ $C_{\CP}$ plane 
shown in Fig.~\ref{fig:cp_uta:uud_SvsC}.

\begin{table}[htb]
	\begin{center}
		\caption{
      Averages for $b \to u \bar u d$ modes.
		}
		\vspace{0.2cm}
		\setlength{\tabcolsep}{0.0pc}
\begin{tabular*}{\textwidth}{@{\extracolsep{\fill}}lrcccc} \hline
	\mc{2}{l}{Experiment} & $N(B\bar{B})$ & $S_{CP}$ & $C_{CP}$ & Correlation \\
	\hline
      \mc{6}{c}{$\pi^{+} \pi^{-}$} \\
	\babar & \cite{Aubert:2008sb} & 467M & $-0.68 \pm 0.10 \pm 0.03$ & $-0.25 \pm 0.08 \pm 0.02$ & $-0.06$ \\
	\belle & \cite{Ishino:2006if} & 535M & $-0.61 \pm 0.10 \pm 0.04$ & $-0.55 \pm 0.08 \pm 0.05$ & $-0.15$ \\
	\mc{3}{l}{\bf Average} & $-0.65 \pm 0.07$ & $-0.38 \pm 0.06$ & $-0.08$ \\
	\mc{3}{l}{\small Confidence level} & \mc{2}{c}{\small $0.055~(1.9\sigma)$} & \\
		\hline

      \mc{6}{c}{$\rho^{+} \rho^{-}$} \\
	\babar & \cite{Aubert:2007nua} & 387M & $-0.17 \pm 0.20 \,^{+0.05}_{-0.06}$ & $0.01 \pm 0.15 \pm 0.06$ & $-0.04$ \\
	\belle & \cite{Abe:2007ez} & 535M & $0.19 \pm 0.30 \pm 0.07$ & $-0.16 \pm 0.21 \pm 0.07$ & $0.10$ \\
	\mc{3}{l}{\bf Average} & $-0.05 \pm 0.17$ & $-0.06 \pm 0.13$ & $0.01$ \\
	\mc{3}{l}{\small Confidence level} & \mc{2}{c}{\small $0.50~(0.7\sigma)$} & \\
		\hline

      \mc{6}{c}{$\rho^{0} \rho^{0}$} \\
	\babar & \cite{:2008iha} & 465M & $0.30 \pm 0.70 \pm 0.20$ & $0.20 \pm 0.80 \pm 0.30$ & $-0.04$ \\
 		\hline
 		\end{tabular*}

                \vspace{2ex}

    \resizebox{\textwidth}{!}{
 		\begin{tabular}{@{\extracolsep{2mm}}lrcccccc} \hline
 		\mc{2}{l}{Experiment} & $N(B\bar{B})$ & $A_{CP}$ & $C$ & $S$ & $\Delta C$ & $\Delta S$ \\
 		\hline
      \mc{8}{c}{$a_1^{\pm} \pi^{\mp}$} \\
	\babar & \cite{Aubert:2006gb} & 384M & $-0.07 \pm 0.07 \pm 0.02$ & $-0.10 \pm 0.15 \pm 0.09$ & $0.37 \pm 0.21 \pm 0.07$ & $0.26 \pm 0.15 \pm 0.07$ & $-0.14 \pm 0.21 \pm 0.06$ \\
 	\hline
		\end{tabular}
              }
		\label{tab:cp_uta:uud}
	\end{center}
\end{table}

If the penguin contribution is negligible, 
the time-dependent parameters for $\Bz \to \pi^+\pi^-$ 
and $\Bz \to \rho^+\rho^-$ are given by
$S_{b \to u\bar u d} = \etacp \sin(2\alpha)$ and
$C_{b \to u\bar u d} = 0$.
In the presence of the penguin contribution, 
direct $\CP$ violation may arise, 
and there is no straightforward interpretation 
of $S_{b \to u\bar u d}$ and $C_{b \to u\bar u d}$.
An isospin analysis~\cite{Gronau:1990ka} 
can be used to disentangle the contributions and extract $\alpha$.

\begin{figure}[htb]
  \begin{center}
    \begin{tabular}{cc}
      \resizebox{0.46\textwidth}{!}{
        \includegraphics{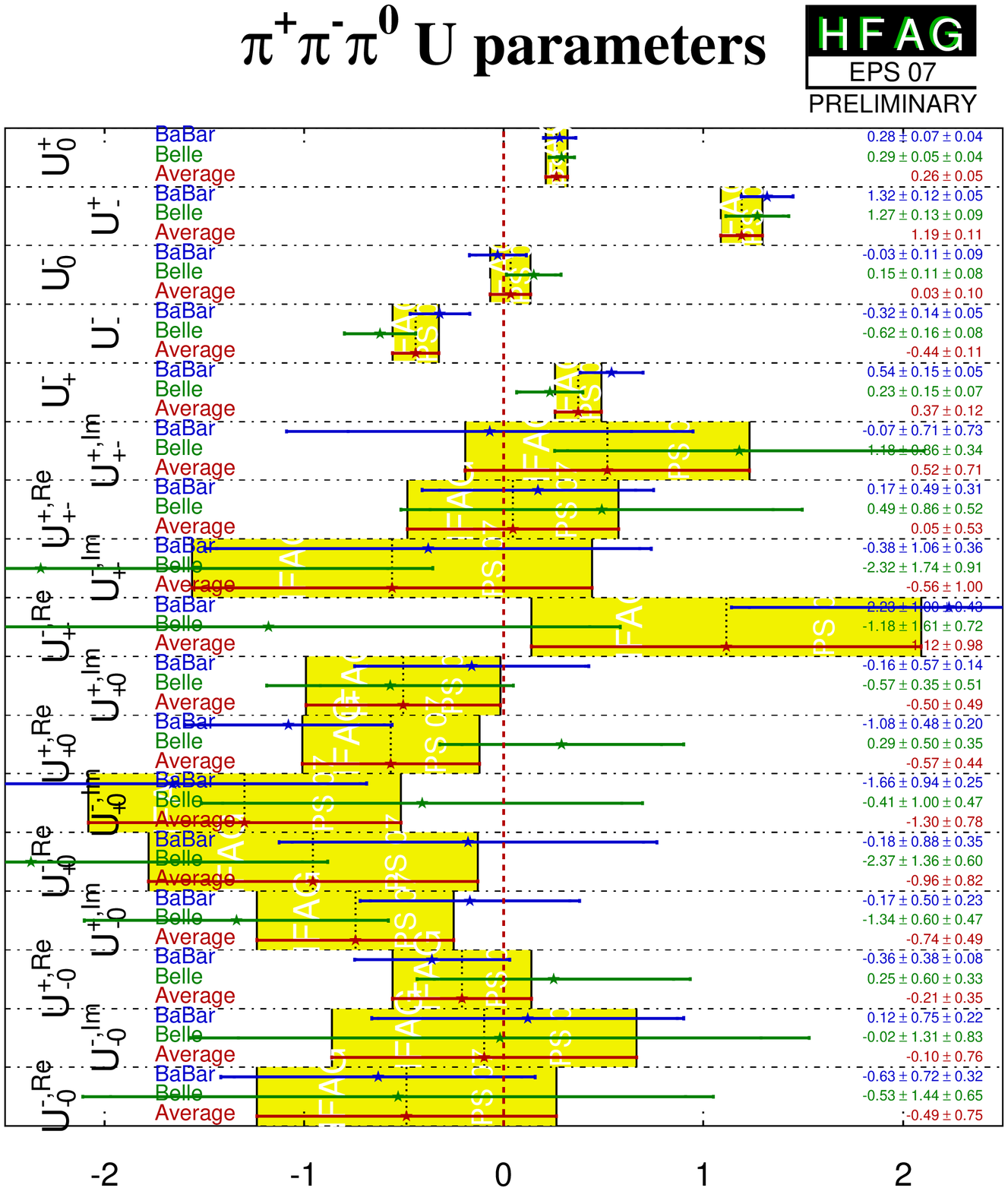}
      }
      &
      \resizebox{0.46\textwidth}{!}{
        \includegraphics{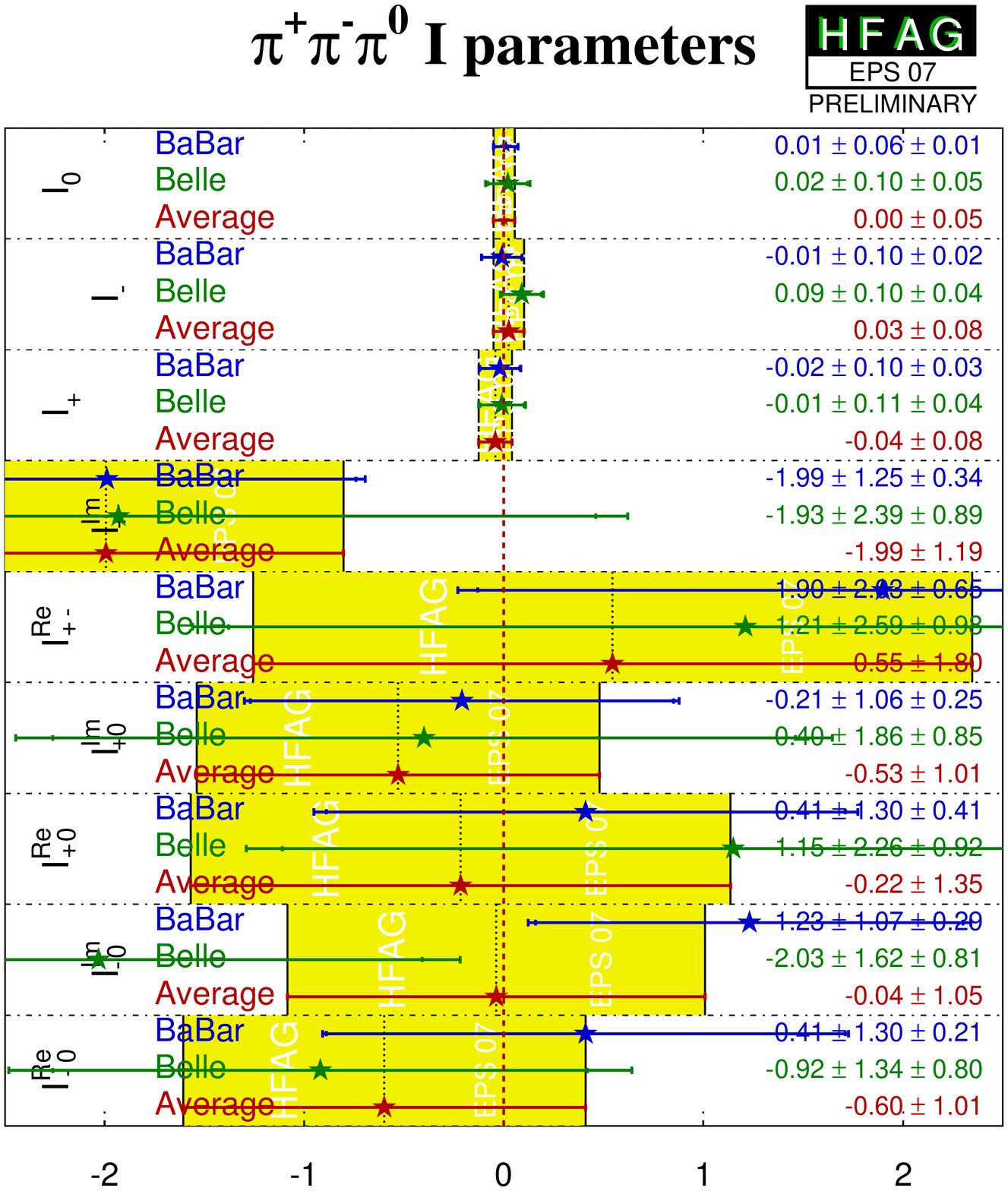}
      }
    \end{tabular}
  \end{center}
  \vspace{-0.8cm}
  \caption{
    Summary of the $U$ and $I$ parameters measured in the 
    time-dependent $\Bz \to \pi^+\pi^-\pi^0$ Dalitz plot analysis.
  }
  \label{fig:cp_uta:uud:uandi}
\end{figure}

For the non-$\CP$ eigenstate $\rho^{\pm}\pi^{\mp}$, 
both \babar~\cite{Aubert:2007jn} 
and \belle~\cite{Kusaka:2007dv,:2007mj} have performed 
time-dependent Dalitz plot (DP) analyses
of the $\pi^+\pi^-\pi^0$ final state~\cite{Snyder:1993mx};
such analyses allow direct measurements of the phases.
Both experiments have measured the $U$ and $I$ parameters discussed in 
Sec.~\ref{sec:cp_uta:notations:dalitz:pipipi0} and defined in 
Table~\ref{tab:cp_uta:pipipi0:uandi}.
We have performed a full correlated average of these parameters,
the results of which are summarized in Fig.~\ref{fig:cp_uta:uud:uandi}.

\begin{figure}[htb]
  \begin{center}
    \begin{tabular}{cc}
      \resizebox{0.46\textwidth}{!}{
        \includegraphics{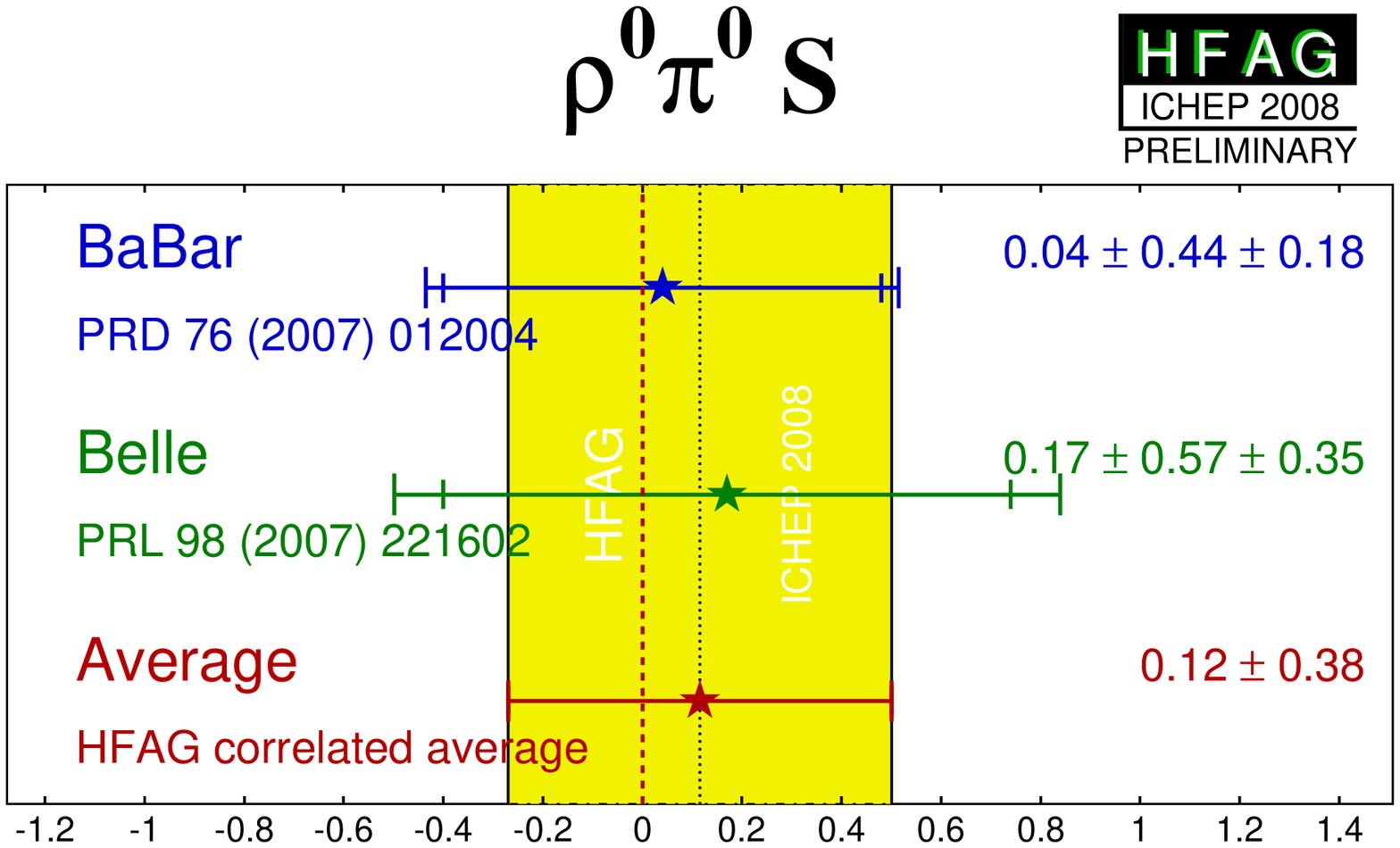}
      }
      &
      \resizebox{0.46\textwidth}{!}{
        \includegraphics{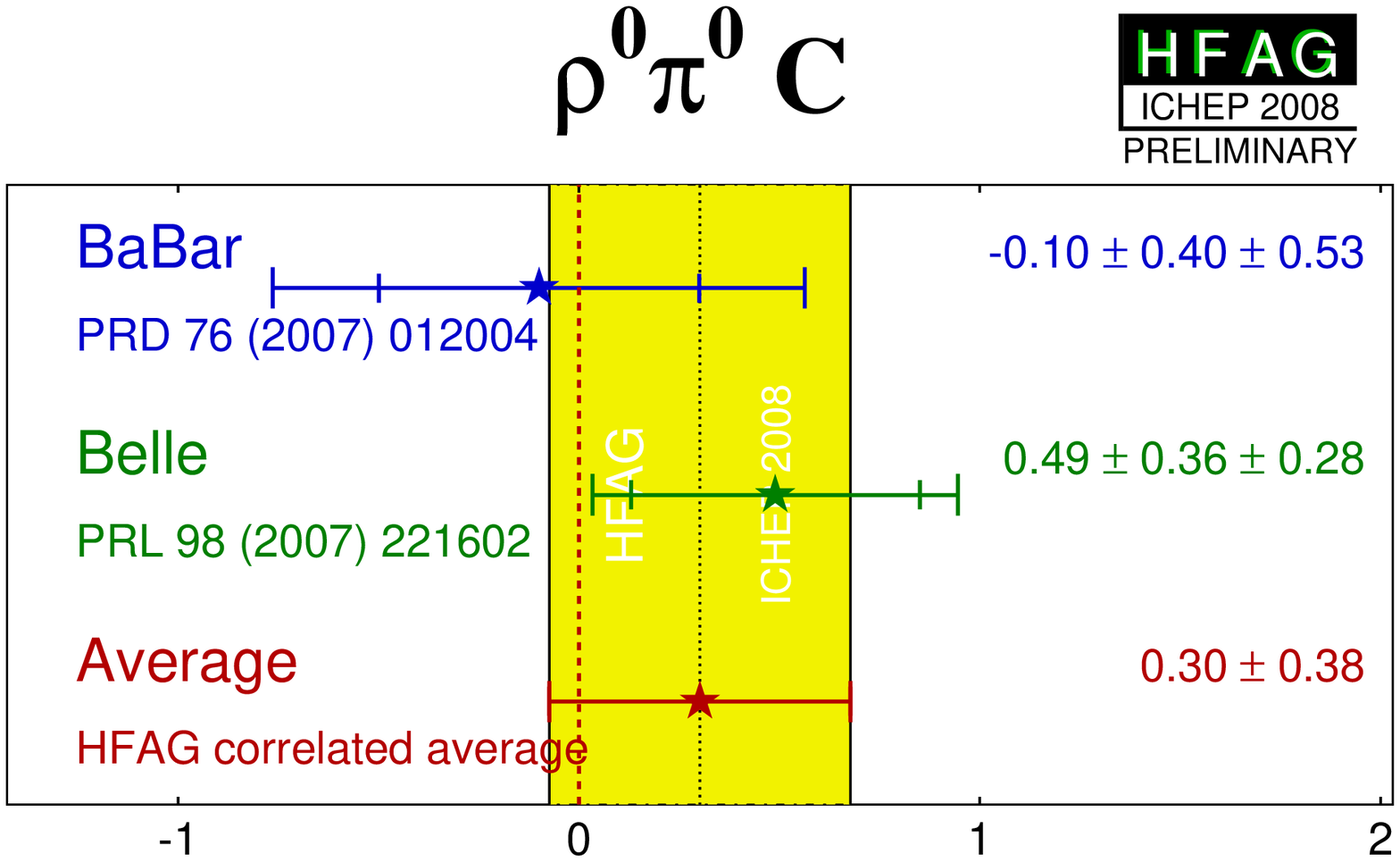}
      }
    \end{tabular}
  \end{center}
  \vspace{-0.8cm}
  \caption{
    Averages of (left) $S_{b \to u\bar u d}$ and (right) $C_{b \to u\bar u d}$
    for the mode $\Bz \to \rho^0\pi^0$.
  }
  \label{fig:cp_uta:uud:rho0pi0}
\end{figure}

\begin{figure}[htb]
  \begin{center}
    \resizebox{0.46\textwidth}{!}{
      \includegraphics{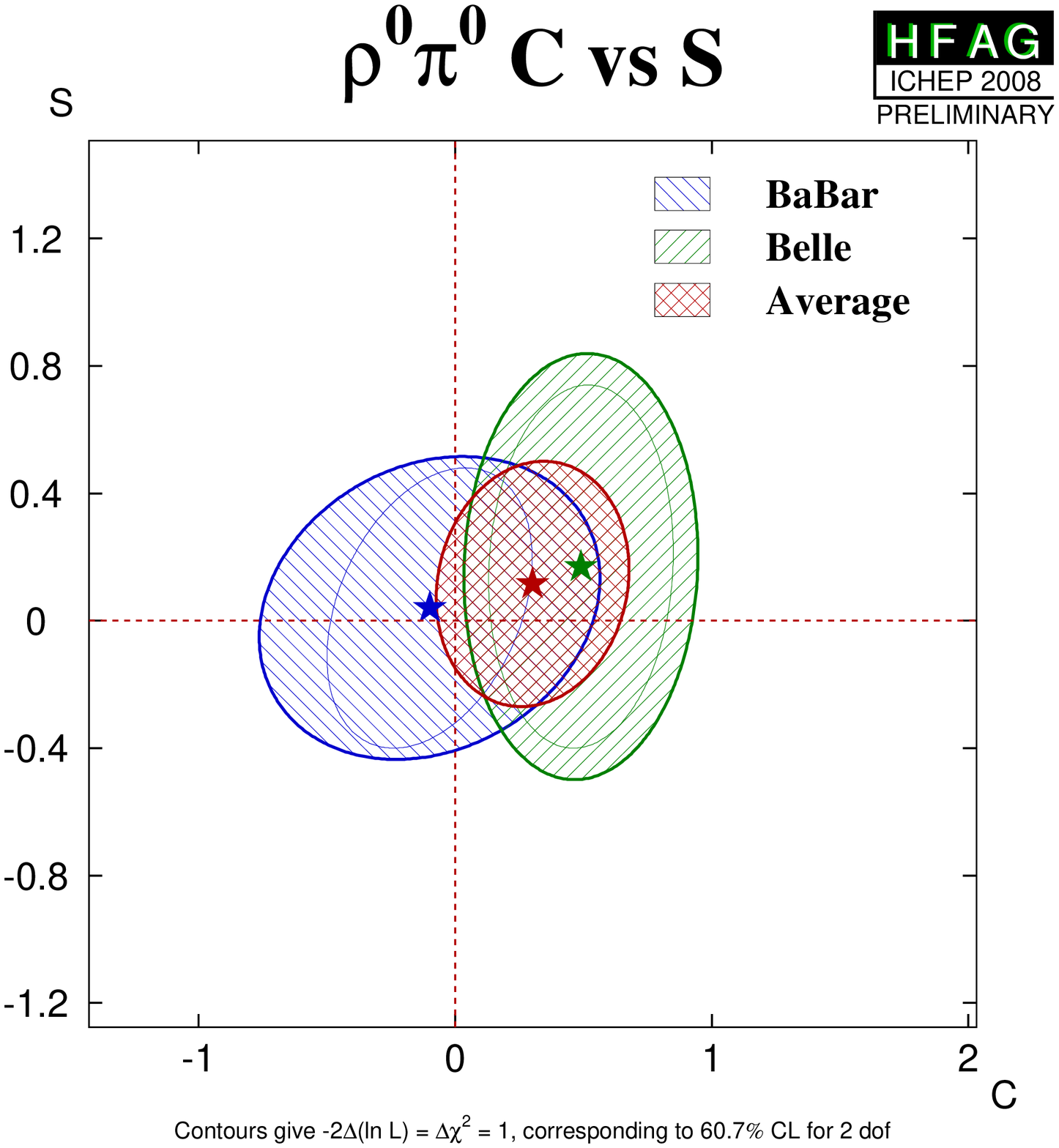}
    }      
  \end{center}
  \vspace{-0.8cm}
  \caption{
    Averages of $b \to u\bar u d$ dominated channels,
    for the mode $\Bz \to \rho^0\pi^0$
    in the $S_{\CP}$ \vs\ $C_{\CP}$ plane.
  }
  \label{fig:cp_uta:uud:rho0pi0_SvsC}
\end{figure}

Both experiments have also extracted the Q2B parameters.
We have performed a full correlated average of these parameters,
which is equivalent to determining the values from the 
averaged $U$ and $I$ parameters.
The results are shown in Table.~\ref{tab:cp_uta:uud:rhopi_q2b}.
Averages of the $\Bz \to \rho^0\pi^0$ Q2B parameters are shown in 
Figs.~\ref{fig:cp_uta:uud:rho0pi0} and~\ref{fig:cp_uta:uud:rho0pi0_SvsC}.

\begin{table}[htb]
	\begin{center}
		\caption{
                  Averages of quasi-two-body parameters extracted
                  from time-dependent Dalitz plot analysis of 
                  $\Bz \to \pi^+\pi^-\pi^0$.
		}
		\vspace{0.2cm}
		\setlength{\tabcolsep}{0.0pc}
    \resizebox{\textwidth}{!}{
\begin{tabular}{@{\extracolsep{2mm}}lrcccccc} \hline
		\mc{2}{l}{Experiment} & $N(B\bar{B})$ & ${\cal A}_{CP}^{\rho\pi}$ & $C_{\rho\pi}$ & $S_{\rho\pi}$ & $\Delta C_{\rho\pi}$ & $\Delta S_{\rho\pi}$ \\
	\hline
	\babar & \cite{Aubert:2007jn} & 375M & $-0.14 \pm 0.05 \pm 0.02$ & $0.15 \pm 0.09 \pm 0.05$ & $-0.03 \pm 0.11 \pm 0.04$ & $0.39 \pm 0.09 \pm 0.09$ & $-0.01 \pm 0.14 \pm 0.06$ \\
	\belle & \cite{Kusaka:2007dv,:2007mj} & 449M & $-0.12 \pm 0.05 \pm 0.04$ & $-0.13 \pm 0.09 \pm 0.05$ & $0.06 \pm 0.13 \pm 0.05$ & $0.36 \pm 0.10 \pm 0.05$ & $-0.08 \pm 0.13 \pm 0.05$ \\
	\mc{3}{l}{\bf Average} & $-0.13 \pm 0.04$ & $0.01 \pm 0.07$ & $0.01 \pm 0.09$ & $0.37 \pm 0.08$ & $-0.04 \pm 0.10$ \\
	\mc{3}{l}{\small Confidence level} & \mc{5}{c}{\small $0.52~(0.6\sigma)$} \\
        \hline
		\end{tabular}
              }

                \vspace{2ex}

		\begin{tabular*}{\textwidth}{@{\extracolsep{\fill}}lrcccc} \hline
		\mc{2}{l}{Experiment} & $N(B\bar{B})$ & ${\cal A}^{-+}_{\rho\pi}$ & ${\cal A}^{+-}_{\rho\pi}$ & Correlation \\
		\hline
	\babar & \cite{Aubert:2007jn} & 375M & $-0.37 \,^{+0.16}_{-0.10} \pm 0.09$ & $0.03 \pm 0.07 \pm 0.04$ & $0.62$ \\
	\belle & \cite{Kusaka:2007dv,:2007mj} & 449M & $0.08 \pm 0.16 \pm 0.11$ & $0.21 \pm 0.08 \pm 0.04$ & $0.47$ \\
	\mc{3}{l}{\bf Average} & $-0.18 \pm 0.12$ & $0.11 \pm 0.06$ & $0.40$ \\
        \mc{3}{l}{\small Confidence level} & \mc{2}{c}{\small $0.14~(1.5\sigma)$} \\
		\hline
		\end{tabular*}

                \vspace{2ex}

		\begin{tabular*}{\textwidth}{@{\extracolsep{\fill}}lrcccc} \hline
		\mc{2}{l}{Experiment} & $N(B\bar{B})$ & $C_{\rho^0\pi^0}$ & $S_{\rho^0\pi^0}$ & Correlation \\
		\hline
	\babar & \cite{Aubert:2007jn} & 375M & $-0.10 \pm 0.40 \pm 0.53$ & $0.04 \pm 0.44 \pm 0.18$ & $0.35$ \\
	\belle & \cite{Kusaka:2007dv,:2007mj} & 449M & $0.49 \pm 0.36 \pm 0.28$ & $0.17 \pm 0.57 \pm 0.35$ & $0.08$ \\
	\mc{3}{l}{\bf Average} & $0.30 \pm 0.38$ & $0.12 \pm 0.38$ & $0.12$ \\
	\mc{3}{l}{\small Confidence level} & \mc{2}{c}{\small $0.76~(0.3\sigma)$} \\
		\hline
		\end{tabular*}
		\label{tab:cp_uta:uud:rhopi_q2b}
	\end{center}
\end{table}

\begin{figure}[htb]
  \begin{center}
    \resizebox{0.46\textwidth}{!}{
      \includegraphics{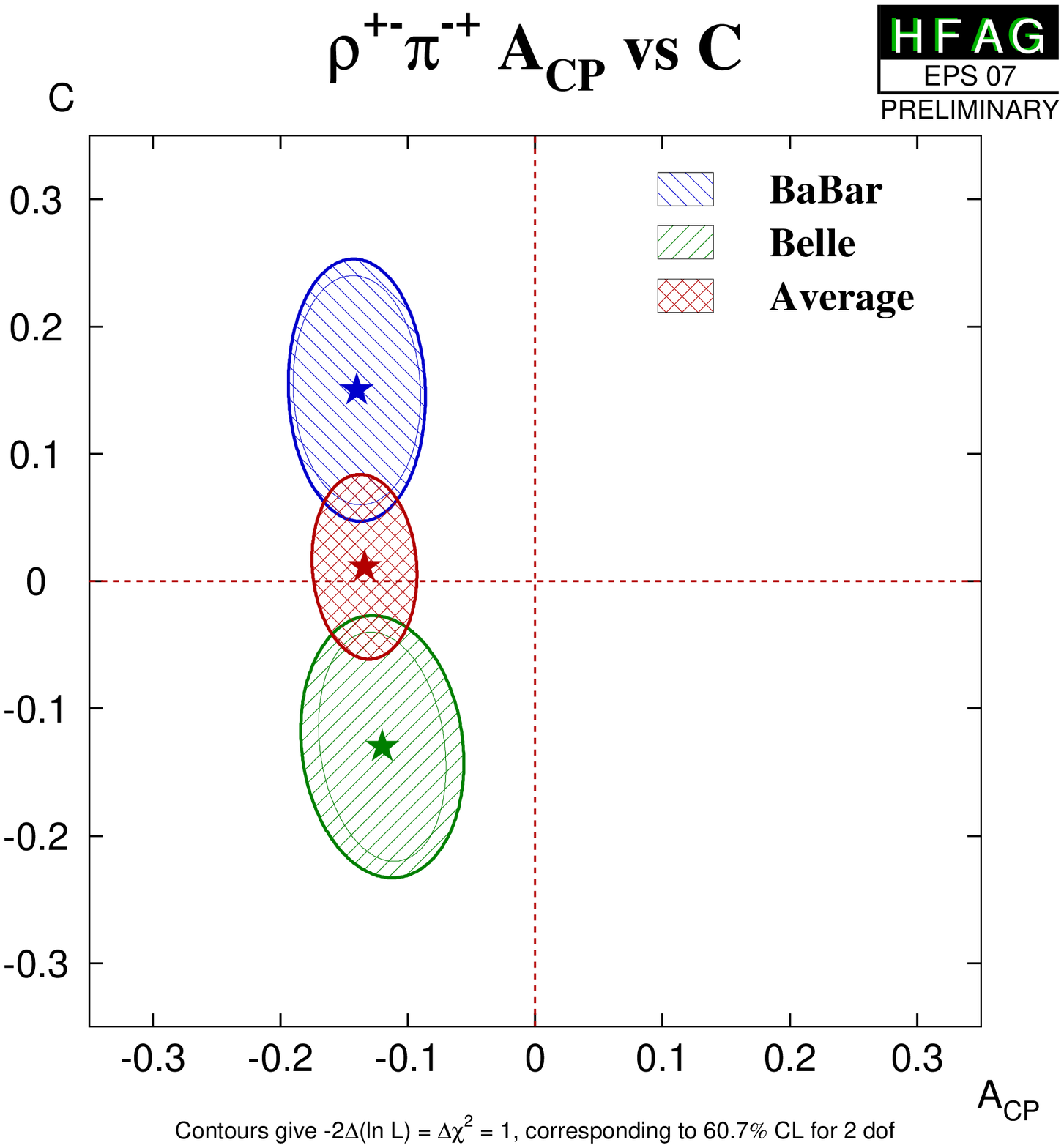}
    }
    \hfill
    \resizebox{0.46\textwidth}{!}{
      \includegraphics{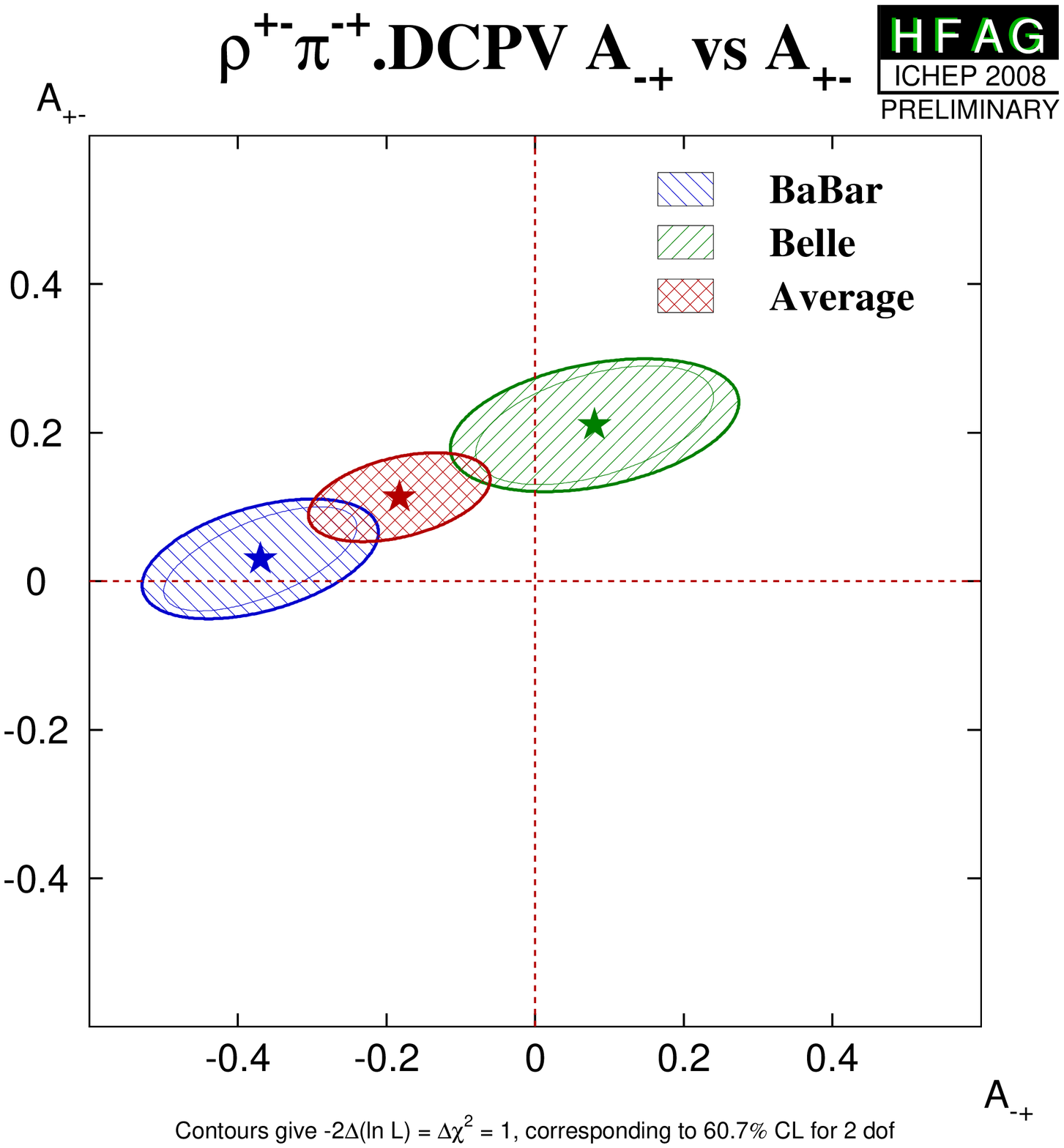}
    }
  \end{center}
  \vspace{-0.8cm}
  \caption{
    Direct $\CP$ violation in $\Bz\to\rho^\pm\pi^\mp$.
    (Left) ${\cal A}^{\rho\pi}_{\CP}$ \vs\ $C_{\rho\pi}$ space,
    (right) ${\cal A}^{-+}_{\rho\pi}$ \vs\ ${\cal A}^{+-}_{\rho\pi}$ space.
  }
  \label{fig:cp_uta:uud:rhopi-dircp}
\end{figure}

With the notation described in Sec.~\ref{sec:cp_uta:notations}
(Eq.~(\ref{eq:cp_uta:non-cp-s_and_deltas})), 
the time-dependent parameters for the Q2B $\Bz \to \rho^\pm\pi^\mp$ analysis are,
neglecting penguin contributions, given by
\begin{equation}
  S_{\rho\pi} = 
  \sqrt{1 - \left(\frac{\Delta C}{2}\right)^2}\sin(2\alpha)\cos(\delta)
  \ , \ \ \ 
  \Delta S_{\rho\pi} = 
  \sqrt{1 - \left(\frac{\Delta C}{2}\right)^2}\cos(2\alpha)\sin(\delta)
\end{equation} 
and $C_{\rho\pi} = {\cal A}_{\CP}^{\rho\pi} = 0$,
where $\delta=\arg(A_{-+}A^*_{+-})$ is the strong phase difference 
between the $\rho^-\pi^+$ and $\rho^+\pi^-$ decay amplitudes.
In the presence of the penguin contribution, there is no straightforward 
interpretation of the Q2B observables in the $\Bz \to \rho^\pm\pi^\mp$ system
in terms of CKM parameters.
However direct $\CP$ violation may arise,
resulting in either or both of $C_{\rho\pi} \neq 0$ and ${\cal A}_{\CP}^{\rho\pi} \neq 0$.
Equivalently,
direct $\CP$ violation may be seen by either of
the decay-type-specific observables ${\cal A}^{+-}_{\rho\pi}$ 
and ${\cal A}^{-+}_{\rho\pi}$, defined in Eq.~(\ref{eq:cp_uta:non-cp-directcp}), 
deviating from zero.
Results and averages for these parameters
are also given in Table~\ref{tab:cp_uta:uud:rhopi_q2b}.
Averages of the direct $\CP$ violation effect in $\Bz \to \rho^\pm\pi^\mp$
are shown in Fig.~\ref{fig:cp_uta:uud:rhopi-dircp},
both in 
${\cal A}^{\rho\pi}_{\CP}$ \vs\ $C_{\rho\pi}$ space and in 
${\cal A}^{-+}_{\rho\pi}$ \vs\ ${\cal A}^{+-}_{\rho\pi}$ space.

Some difference is seen between the 
\babar\ and \belle\ measurements in the $\pi^+\pi^-$ system.
The confidence level of the average is $0.034$,
which corresponds to a $2.1\sigma$ discrepancy.  Since there is no
evidence of systematic problems in either analysis,
we do not rescale the errors of the averages.
The averages for $S_{b \to u\bar u d}$ and $C_{b \to u\bar u d}$ 
in $\Bz \to \pi^+\pi^-$ are both more than $5\sigma$ away from zero,
suggesting that both mixing-induced and direct $\CP$ violation 
are well-established in this channel.
Nonetheless, due to the possible discrepancy mentioned above,
a slightly cautious interpretation should be made 
with regard to the significance of direct $\CP$ violation.

In $\Bz \to \rho^\pm\pi^\mp$, however,
both experiments see an indication of direct $\CP$ violation in the 
${\cal A}^{\rho\pi}_{\CP}$ parameter 
(as seen in Fig.~\ref{fig:cp_uta:uud:rhopi-dircp}).
The average is more than $3\sigma$ from zero,
providing evidence of direct $\CP$ violation in this channel.

\vspace{3ex}

\noindent
\underline{\large Constraints on $\alpha$}

The precision of the measured $\CP$ violation parameters in
$b \to u\bar{u}d$ transitions allows 
constraints to be set on the UT angle $\alpha$. 
Constraints have been obtained with various methods:
\begin{itemize}\setlength{\itemsep}{0.5ex}
\item 
  Both \babar~\cite{Aubert:2007hh}
  and  \belle~\cite{Ishino:2006if} have performed 
  isospin analyses in the $\pi\pi$ system.
  \belle\ exclude $9^\circ < \phi_2 < 81^\circ$ at the $95.4\%$  C.L. while
  \babar\ give a confidence level interpretation for $\alpha$,
  exclude the range $23^\circ < \alpha < 67^\circ$ at the $90\%$  C.L.
  In both cases, only solutions in $0^\circ$--$180^\circ$ are considered.

\item
  Both experiments have also performed isospin analyses in the $\rho\rho$
  system. 
  The most recent result from \babar\ is given in an update of the
  measurements of the $B^+\to\rho^+\rho^0$ decay~\cite{Aubert:2009it}, and
  sets the constraint $\alpha = \left( 92.4 \,^{+6.0}_{-6.5}\right)^\circ$.
  The most recent result from \belle\ is given in an update of the
  search for the $\Bz \to \rho^0\rho^0$ decay and sets the constraint
  $\phi_2 = \left( 91.7 \pm 14.9 \right)^\circ$~\cite{:2008et}.

\item
  The time-dependent Dalitz plot analysis of the $\Bz \to \pi^+\pi^-\pi^0$
  decay allows a determination of $\alpha$ without input from any other 
  channels.
  \babar~\cite{Aubert:2007jn} obtain the constraint 
  $75^\circ < \alpha < 152^\circ$ at $68\%$ C.L.
  \belle~\cite{Kusaka:2007dv,:2007mj} have performed a similar analysis,
  and in addition have included information from the SU(2) partners of 
  $B \to \rho\pi$, which can be used to constrain $\alpha$
  via an isospin pentagon relation~\cite{Lipkin:1991st}. 
  With this analysis,
  \belle\ obtain the tighter constraint $\phi_2 = (83 \, ^{+12}_{-23})^\circ$
  (where the errors correspond to $1\sigma$, \ie\ $68.3\%$ confidence level).

\item 
  The results from \babar\ on $\Bz \to a_1^\pm \pi^\mp$~\cite{Aubert:2006gb} can be
  combined with results from modes related by isospin~\cite{Gronau:2005kw}
  leading to the following constraint: 
  $\alpha = \left( 79 \pm 7 \pm 11 \right)^\circ$~\cite{:2009ii}.

\item 
  Each experiment has obtained a value of $\alpha$ from combining its 
  results in the different $b \to u \bar{u} d$ modes 
  (with some input also from HFAG).
  These values have appeared in talks, but not in publications,
  and are not listed here.

\item 
  The CKMfitter~\cite{Charles:2004jd} and 
  UTFit~\cite{Bona:2005vz} groups use the measurements 
  from \belle\ and \babar\ given above
  with other branching fractions and \CP asymmetries in 
  $\B\to\pi\pi$, $\rho\pi$ and $\rho\rho$ modes, 
  to perform isospin analyses for each system, 
  and to make combined constraints on $\alpha$.
\end{itemize}

Note that methods based on isospin symmetry make extensive use of 
measurements of branching fractions and direct $\CP$ asymmetries,
as averaged by the HFAG Rare Decays subgroup (Sec.~\ref{sec:rare}).
Note also that each method suffers from discrete ambiguities in the solutions.
The model assumption in the $\Bz \to \pi^+\pi^-\pi^0$ analysis 
allows to resolve some of the multiple solutions, 
and results in a single preferred value for $\alpha$ in $\left[ 0, \pi \right]$.
All the above measurements correspond to the choice
that is in agreement with the global CKM fit.

At present we make no attempt to provide an HFAG average for $\alpha$.
More details on procedures to calculate a best fit value for $\alpha$ 
can be found in Refs.~\cite{Charles:2004jd,Bona:2005vz}.

\clearpage
\mysubsection{Time-dependent $\CP$ asymmetries in $b \to c\bar{u}d / u\bar{c}d$ transitions
}
\label{sec:cp_uta:cud}

Non-$\CP$ eigenstates such as $D^\pm\pi^\mp$, $D^{*\pm}\pi^\mp$ and $D^\pm\rho^\mp$ can be produced 
in decays of $\Bz$ mesons either via Cabibbo favoured ($b \to c$) or
doubly Cabibbo suppressed ($b \to u$) tree amplitudes. 
Since no penguin contribution is possible,
these modes are theoretically clean.
The ratio of the magnitudes of the suppressed and favoured amplitudes, $R$,
is sufficiently small (predicted to be about $0.02$),
that terms of ${\cal O}(R^2)$ can be neglected, 
and the sine terms give sensitivity to the combination of UT angles $2\beta+\gamma$.

\begin{figure}[htb]
  \begin{center}
    \begin{tabular}{cc}
      \resizebox{0.46\textwidth}{!}{
        \includegraphics{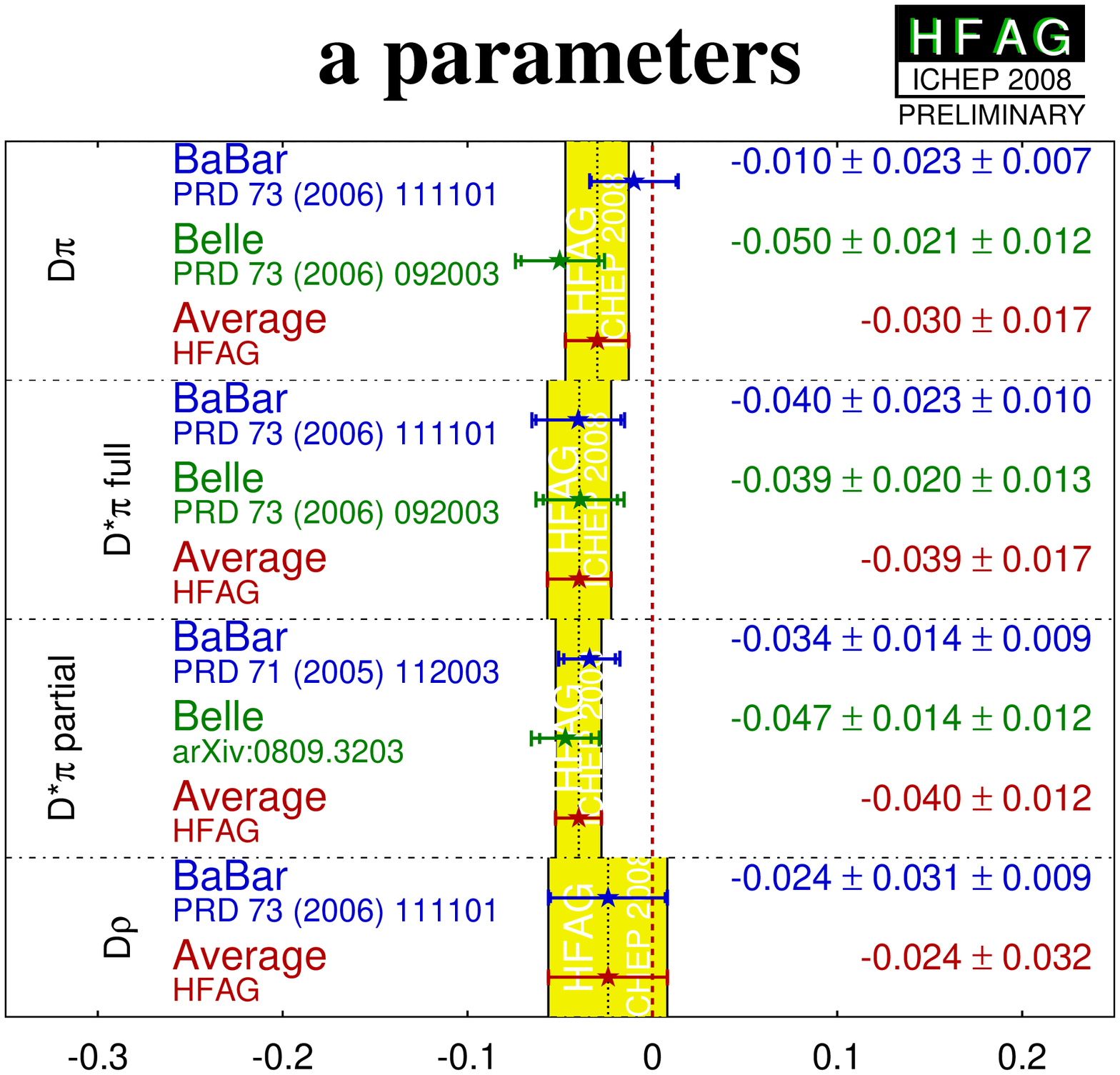}
      }
      &
      \resizebox{0.46\textwidth}{!}{
        \includegraphics{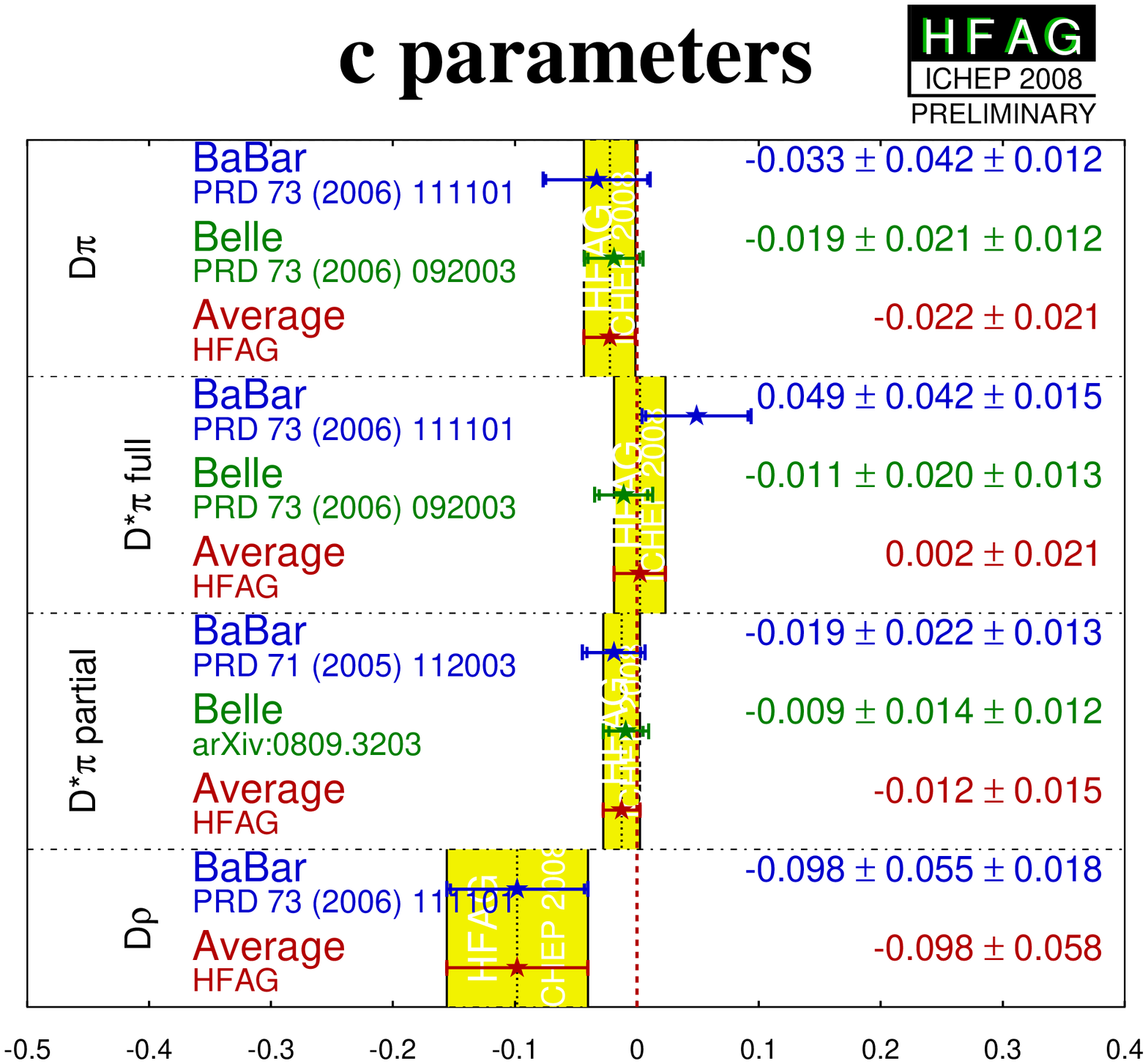}
      }
    \end{tabular}
  \end{center}
  \vspace{-0.8cm}
  \caption{
    Averages for $b \to c\bar{u}d / u\bar{c}d$ modes.
  }
  \label{fig:cp_uta:cud}
\end{figure}

\begin{figure}[htb]
  \begin{center}
    \begin{tabular}{cc}
      \resizebox{0.46\textwidth}{!}{
        \includegraphics{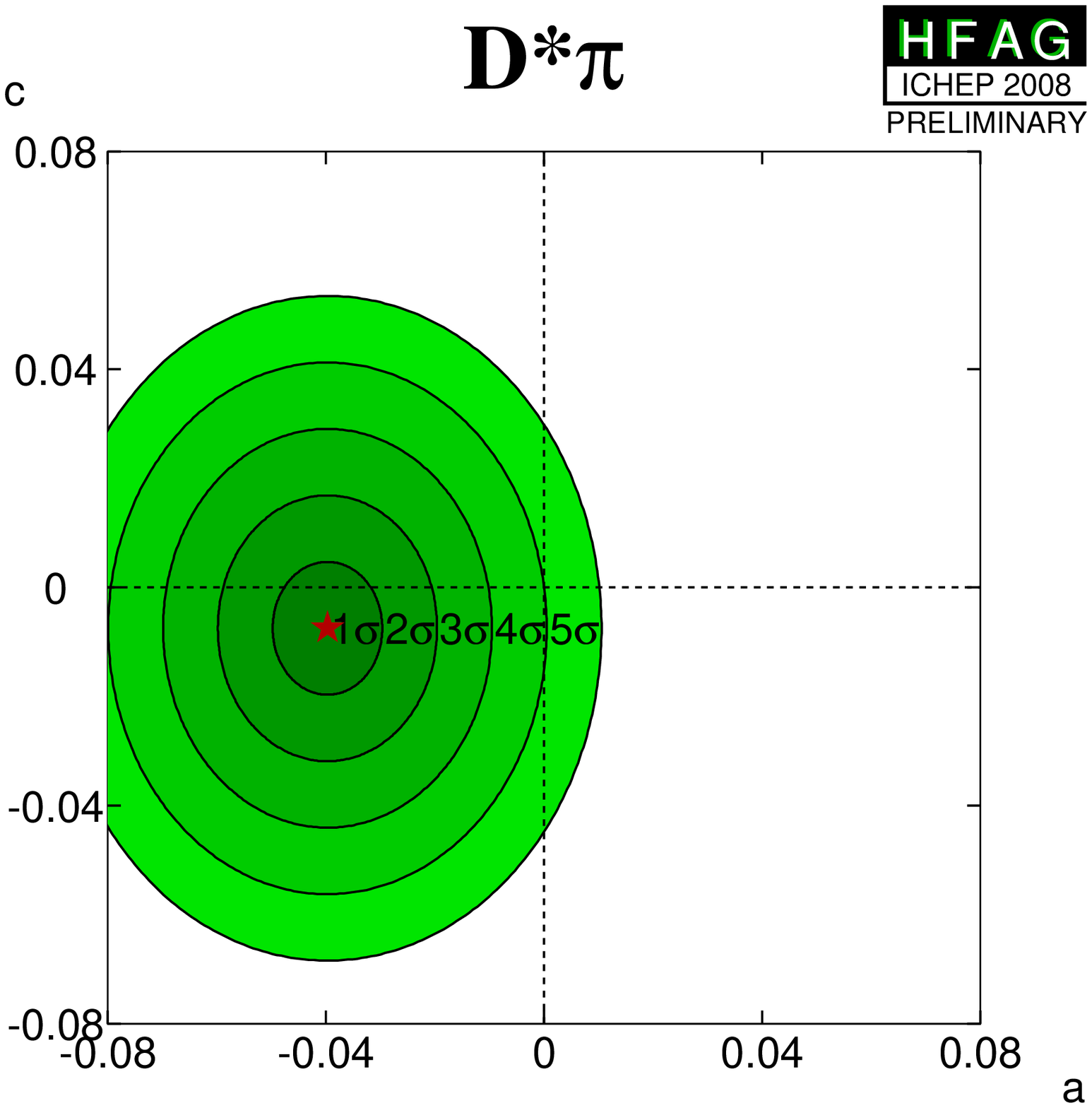}
      }
      &
      \resizebox{0.46\textwidth}{!}{
        \includegraphics{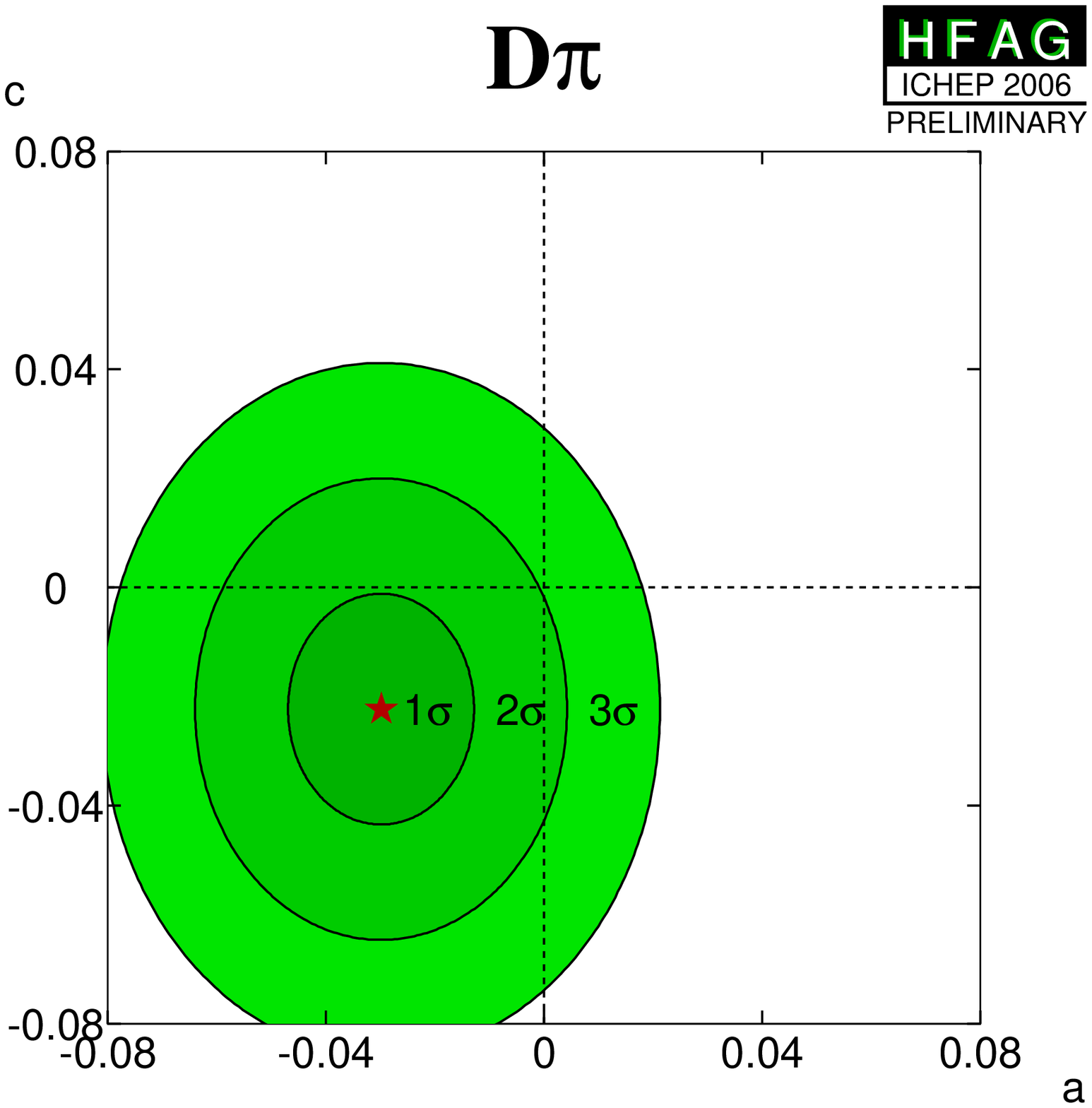}
      } \\
      \resizebox{0.46\textwidth}{!}{
        \includegraphics{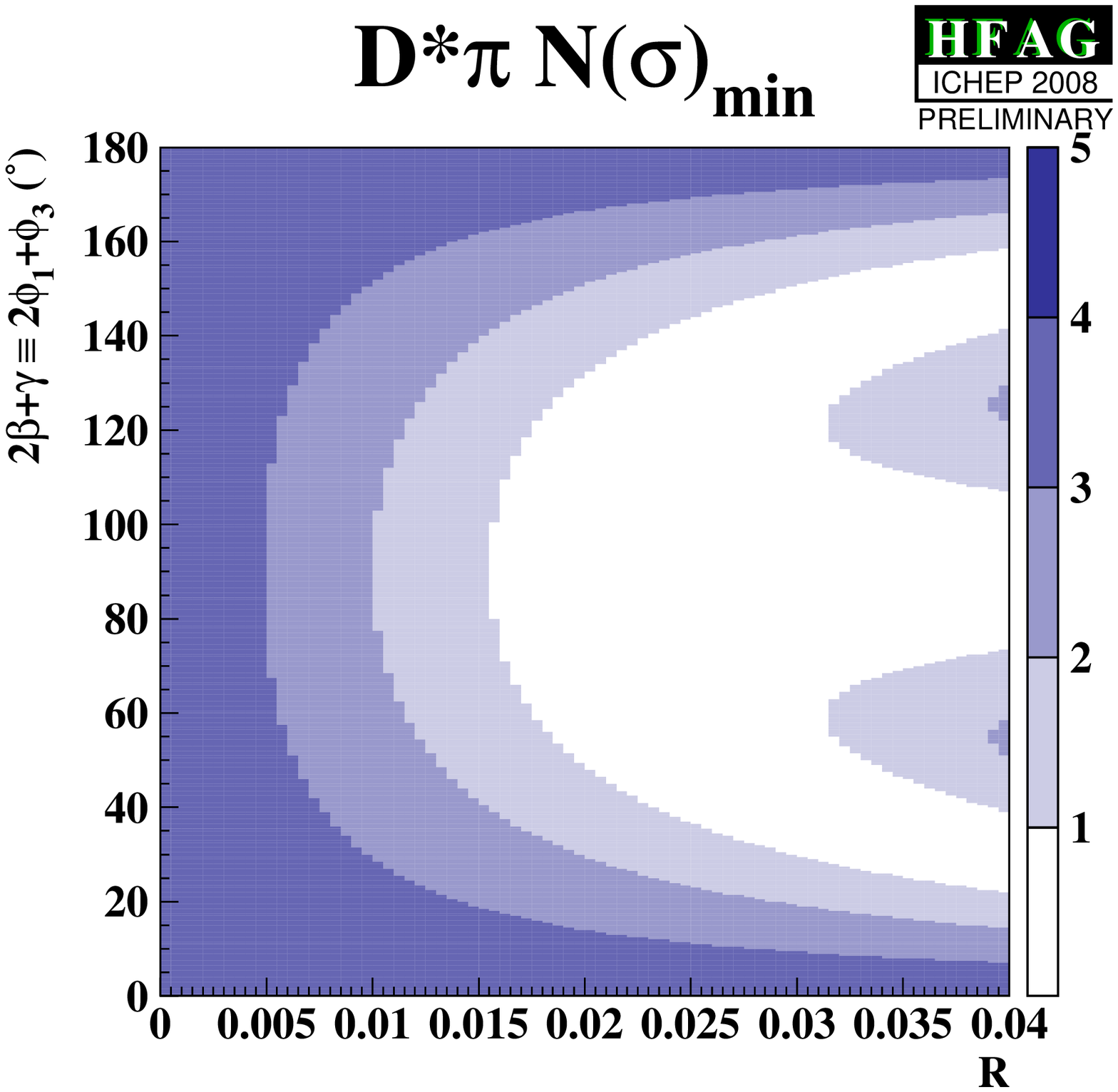}
      }
      &
      \resizebox{0.46\textwidth}{!}{
        \includegraphics{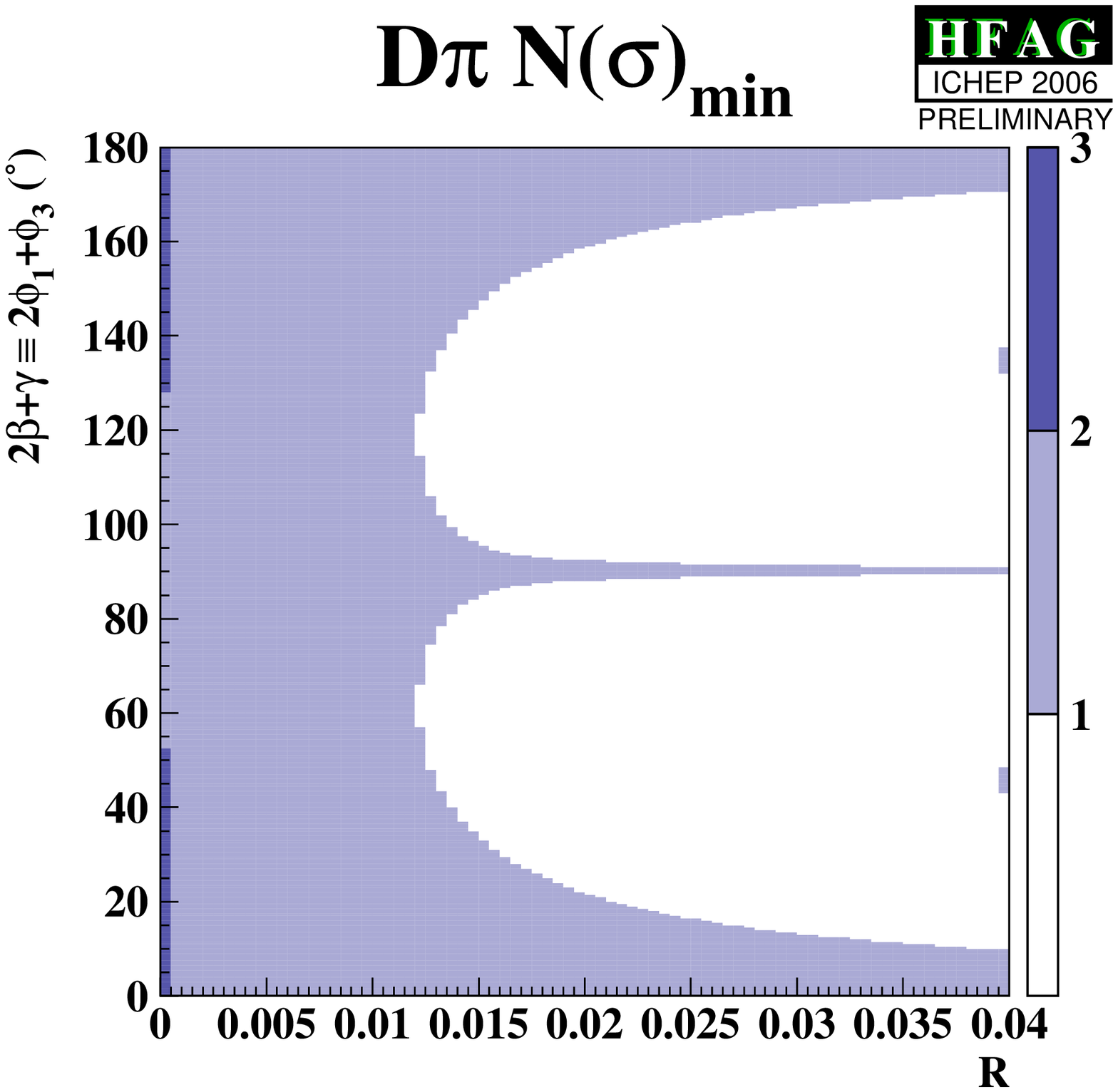}
      }          
    \end{tabular}
  \end{center}
  \vspace{-0.8cm}
  \caption{
    Results from $b \to c\bar{u}d / u\bar{c}d$ modes.
    (Top) Constraints in $c$ {\it vs.} $a$ space.
    (Bottom) Constraints in $2\beta+\gamma$ {\it vs.} $R$ space.
    (Left) $D^*\pi$ and (right) $D\pi$ modes.
  }
  \label{fig:cp_uta:cud_constraints}
\end{figure}

As described in Sec.~\ref{sec:cp_uta:notations:non_cp:dstarpi},
the averages are given in terms of parameters $a$ and $c$.
$\CP$ violation would appear as $a \neq 0$.
Results are available from both \babar\ and \belle\ in the modes
$D^\pm\pi^\mp$ and $D^{*\pm}\pi^\mp$; for the latter mode both experiments 
have used both full and partial reconstruction techniques.
Results are also available from \babar\ using $D^\pm\rho^\mp$.
These results, and their averages, are listed in Table~\ref{tab:cp_uta:cud},
and are shown in Fig.~\ref{fig:cp_uta:cud}.
The constraints in $c$ \vs\ $a$ space for the $D\pi$ and $D^*\pi$ modes
are shown in Fig.~\ref{fig:cp_uta:cud_constraints}.
It is notable that the average value of $a$ from $D^*\pi$ is more than
$3\sigma$ from zero, providing evidence of $\CP$ violation in this channel.

\begin{table}[htb]
	\begin{center}
		\caption{
      Averages for $b \to c\bar{u}d / u\bar{c}d$ modes.
                }
                \vspace{0.2cm}
                \setlength{\tabcolsep}{0.0pc}
\begin{tabular*}{\textwidth}{@{\extracolsep{\fill}}lrccc} \hline 
	\mc{2}{l}{Experiment} & $N(B\bar{B})$ & $a$ & $c$ \\
	\hline
      \mc{5}{c}{$D^{\pm}\pi^{\mp}$} \\
	\babar (full rec.) & \cite{Aubert:2006tw} & 232M & $-0.010 \pm 0.023 \pm 0.007$ & $-0.033 \pm 0.042 \pm 0.012$ \\
	\belle (full rec.) & \cite{Ronga:2006hv} & 386M & $-0.050 \pm 0.021 \pm 0.012$ & $-0.019 \pm 0.021 \pm 0.012$ \\
        \mc{3}{l}{\bf Average} & $ -0.030 \pm 0.017$ & $ -0.022 \pm 0.021 $ \\
        \mc{3}{l}{\small Confidence level} & {\small $0.24~(1.2\sigma)$} & {\small $0.78~(0.3\sigma)$} \\
        \hline
      \mc{5}{c}{$D^{*\pm}\pi^{\mp}$} \\
      \babar (full rec.) & \cite{Aubert:2006tw} & 232M & $-0.040 \pm 0.023 \pm 0.010$ & $0.049 \pm 0.042 \pm 0.015$ \\
      \babar (partial rec.)  & \cite{Aubert:2005yf} & 232M & $-0.034 \pm 0.014 \pm 0.009$ & $-0.019 \pm 0.022 \pm 0.013$ \\
      \belle (full rec.) & \cite{Ronga:2006hv} & 386M & $-0.039 \pm 0.020 \pm 0.013$ & $-0.011 \pm 0.020 \pm 0.013$ \\
      \belle (partial rec.) & \cite{:2008kr} & 657M & $-0.047 \pm 0.014 \pm 0.012$ & $-0.009 \pm 0.014 \pm 0.012$ \\
      \mc{3}{l}{\bf Average} & $-0.040 \pm 0.010$ & $-0.007 \pm 0.012$ \\
      \mc{3}{l}{\small Confidence level} & {\small $0.96~(0.0\sigma)$} & {\small $0.61~(0.5\sigma)$} \\
      \hline
      \mc{5}{c}{$D^{\pm}\rho^{\mp}$} \\
      \babar (full rec.) & \cite{Aubert:2006tw} & 232M & $-0.024 \pm 0.031 \pm 0.009$ & $-0.098 \pm 0.055 \pm 0.018$ \\
      \hline 
    \end{tabular*}
    \label{tab:cp_uta:cud}
  \end{center}
\end{table}

For each of $D\pi$, $D^*\pi$ and $D\rho$, 
there are two measurements ($a$ and $c$, or $S^+$ and $S^-$) 
which depend on three unknowns ($R$, $\delta$ and $2\beta+\gamma$), 
of which two are different for each decay mode. 
Therefore, there is not enough information to solve directly for $2\beta+\gamma$. 
However, for each choice of $R$ and $2\beta+\gamma$, 
one can find the value of $\delta$ that allows $a$ and $c$ to be closest 
to their measured values, 
and calculate the distance in terms of numbers of standard deviations.
(We currently neglect experimental correlations in this analysis.) 
These values of $N(\sigma)_{\rm min}$ can then be plotted 
as a function of $R$ and $2\beta+\gamma$
(and can trivially be converted to confidence levels). 
These plots are given for the $D\pi$ and $D^*\pi$ modes 
in Figure~\ref{fig:cp_uta:cud_constraints}; 
the uncertainties in the $D\rho$ mode are currently too large 
to give any meaningful constraint.

The constraints can be tightened if one is willing 
to use theoretical input on the values of $R$ and/or $\delta$. 
One popular choice is the use of SU(3) symmetry to obtain 
$R$ by relating the suppressed decay mode to $\B$ decays 
involving $D_s$ mesons. 
More details can be found 
in Refs.~\cite{Charles:2004jd,Bona:2005vz}.

\mysubsection{Time-dependent $\CP$ asymmetries in $b \to c\bar{u}s / u\bar{c}s$ transitions
}
\label{sec:cp_uta:cus-td}

Time-dependent analyses of transitions such as $\Bz \to D^\pm \KS \pi^\mp$ can
be used to probe $\sin(2\beta+\gamma)$ in a similar way to that discussed
above (Sec.~\ref{sec:cp_uta:cud}). Since the final state contains three
particles, a Dalitz plot analysis is necessary to maximise the
sensitivity. \babar~\cite{Aubert:2007qe} have carried out such an
analysis. They obtain $2\beta+\gamma = \left( 83 \pm 53 \pm 20 \right)^\circ$
(with an ambiguity $2\beta+\gamma \leftrightarrow 2\beta+\gamma+\pi$) assuming
the ratio of the $b \to u$ and $b \to c$ amplitude to be constant across the
Dalitz plot at 0.3.

\clearpage
\mysubsection{Rates and asymmetries in $\Bmp \to \DorDstar K^{(*)\mp}$ decays
}
\label{sec:cp_uta:cus}

As explained in Sec.~\ref{sec:cp_uta:notations:cus},
rates and asymmetries in $\Bmp \to \DorDstar K^{(*)\mp}$ decays
are sensitive to $\gamma$.
Various methods using different $\DorDstar$ final states exist.

\mysubsubsection{$D$ decays to $\CP$ eigenstates
}
\label{sec:cp_uta:cus:glw}

\begin{figure}[htb]
  \begin{center}
    \begin{tabular}{cc}
      \resizebox{0.46\textwidth}{!}{
        \includegraphics{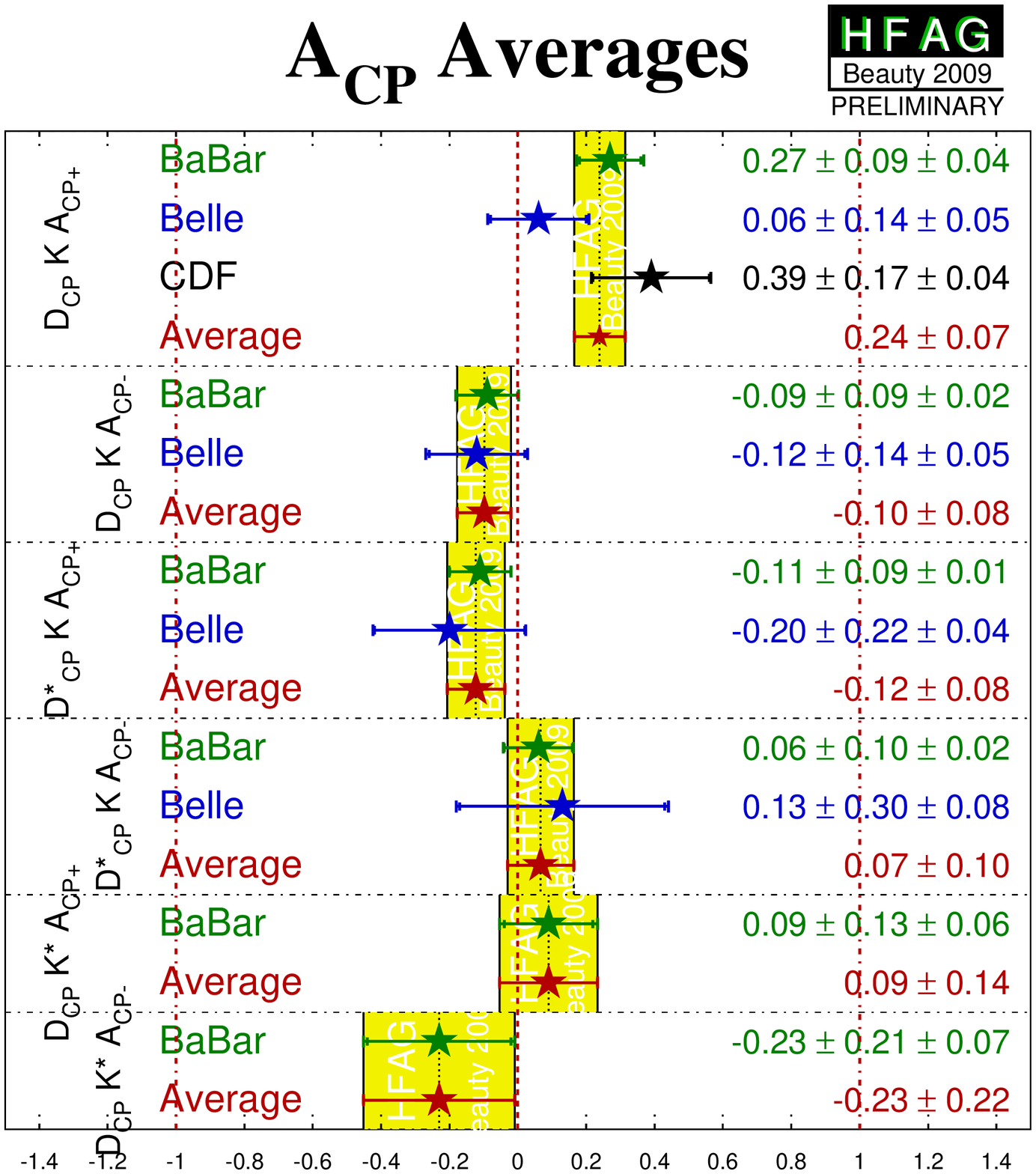}
      }
      &
      \resizebox{0.46\textwidth}{!}{
        \includegraphics{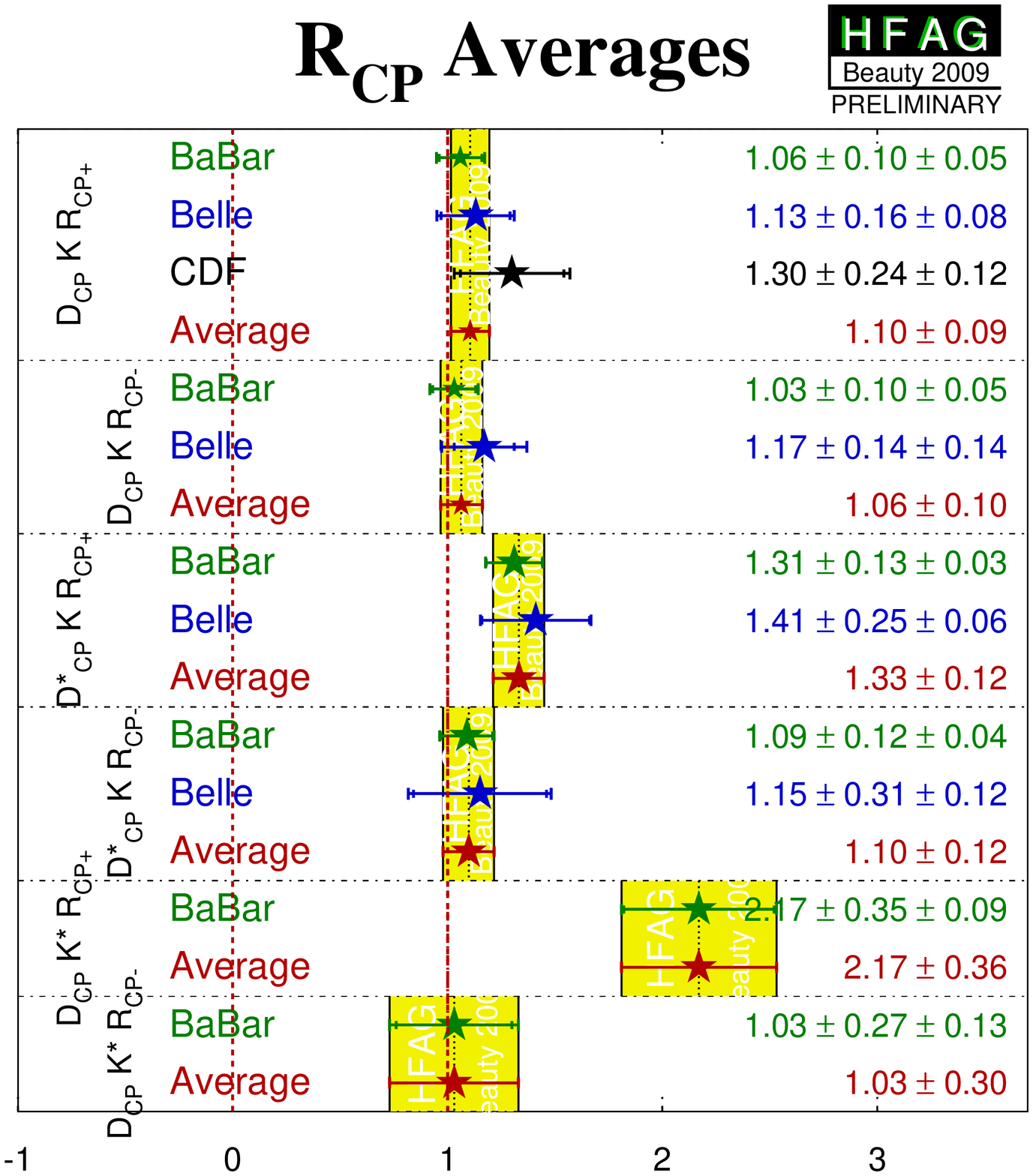}
      }
    \end{tabular}
  \end{center}
  \vspace{-0.8cm}
  \caption{
    Averages of $A_{\CP}$ and $R_{\CP}$ from GLW analyses.
  }
  \label{fig:cp_uta:cus:glw}
\end{figure}

Results are available from both \babar\ and \belle\ on GLW analyses in the
decay modes $\Bmp \to D\Kmp$, $\Bmp \to \Dstar\Kmp$ and 
$\Bmp \to D\Kstarmp$.\footnote{
  We do not include a preliminary result from \belle~\cite{Abe:2003rg}, which
  remains unpublished after more than two years.
}
Both experiments use the $\CP$-even $D$ decay final states $K^+K^-$ and
$\pi^+\pi^-$ in all three modes; both experiments generally use the \CP-odd
decay modes $\KS\pi^0$, $\KS\omega$ and $\KS\phi$, though care is taken to
avoid statistical overlap with the $\KS K^+K^-$ sample used for Dalitz plot
analysis (see Sec.~\ref{sec:cp_uta:cus:dalitz}), 
and asymmetric systematic errors are assigned due to $\CP$-even pollution
under the $\KS\omega$ and $\KS\phi$ signals.
Both experiments also use the $\Dstar \to D\pi^0$ decay, 
which gives $\CP(\Dstar) = \CP(D)$;
\babar\ in addition use the $\Dstar \to D\gamma$ decays, 
which gives $\CP(\Dstar) = -\CP(D)$.
In addition, results from CDF, using $1 \ {\rm fb}^{-1}$, are available in the
decay mode $\Bmp \to D\Kmp$, 
for $\CP$-even final states ($K^+K^-$ and $\pipi$) only.
The results and averages are given in Table~\ref{tab:cp_uta:cus:glw}
and shown in Fig.~\ref{fig:cp_uta:cus:glw}.

\begin{table}[htb]
	\begin{center}
		\caption{
                        Averages from GLW analyses of $b \to c\bar{u}s / u\bar{c}s$ modes.
                }
                \vspace{0.2cm}
    \resizebox{\textwidth}{!}{
      \setlength{\tabcolsep}{0.0pc}
\begin{tabular}{@{\extracolsep{2mm}}lrccccc} \hline 
        \mc{2}{l}{Experiment} & $N(B\bar{B})$ & $A_{\CP+}$ & $A_{\CP-}$ & $R_{\CP+}$ & $R_{\CP-}$ \\
        \hline
        \mc{7}{c}{$D_{\CP} K^-$} \\
	\babar & \cite{:2008yk} & 382M & $0.27 \pm 0.09 \pm 0.04$ & $-0.09 \pm 0.09 \pm 0.02$ & $1.06 \pm 0.10 \pm 0.05$ & $1.03 \pm 0.10 \pm 0.05$ \\
	\belle & \cite{Abe:2006hc} & 275M & $0.06 \pm 0.14 \pm 0.05$ & $-0.12 \pm 0.14 \pm 0.05$ & $1.13 \pm 0.16 \pm 0.08$ & $1.17 \pm 0.14 \pm 0.14$ \\
	CDF & \cite{Aaltonen:2009hz} & \textendash{} & $0.39 \pm 0.17 \pm 0.04$ & \textendash{} & $1.30 \pm 0.24 \pm 0.12$ &  \textendash{} \\
	\mc{3}{l}{\bf Average} & $0.24 \pm 0.07$ & $-0.10 \pm 0.08$ & $1.10 \pm 0.09$ & $1.06 \pm 0.10$ \\
	\mc{3}{l}{\small Confidence level} & {\small $0.32~(1.0\sigma)$} & {\small $0.86~(0.2\sigma)$} & {\small $0.70~(0.4\sigma)$} & {\small $0.54~(0.6\sigma)$} \\
		\hline

        \mc{7}{c}{$\Dstar_{\CP} K^-$} \\
	\babar & \cite{:2008jd} & 383M & $-0.11 \pm 0.09 \pm 0.01$ & $0.06 \pm 0.10 \pm 0.02$ & $1.31 \pm 0.13 \pm 0.03$ & $1.09 \pm 0.12 \pm 0.04$ \\
	\belle & \cite{Abe:2006hc} & 275M & $-0.20 \pm 0.22 \pm 0.04$ & $0.13 \pm 0.30 \pm 0.08$ & $1.41 \pm 0.25 \pm 0.06$ & $1.15 \pm 0.31 \pm 0.12$ \\
	\mc{3}{l}{\bf Average} & $-0.12 \pm 0.08$ & $0.07 \pm 0.10$ & $1.33 \pm 0.12$ & $1.10 \pm 0.12$ \\
	\mc{3}{l}{\small Confidence level} & {\small $0.71~(0.4\sigma)$} & {\small $0.83~(0.2\sigma)$} & {\small $0.73~(0.4\sigma)$} & {\small $0.87~(0.2\sigma)$} \\
		\hline

        \mc{7}{c}{$D_{\CP} K^{*-}$} \\
	\babar & \cite{Aubert:2009yw} & 379M & $0.09 \pm 0.13 \pm 0.06$ & $-0.23 \pm 0.21 \pm 0.07$ & $2.17 \pm 0.35 \pm 0.09$ & $1.03 \pm 0.27 \pm 0.13$ \\
		\hline
      \end{tabular}
    }
    \label{tab:cp_uta:cus:glw}
	\end{center}
\end{table}

\mysubsubsection{$D$ decays to suppressed final states}
\label{sec:cp_uta:cus:ads}

\begin{figure}[htb]
  \begin{center}
    \resizebox{0.46\textwidth}{!}{
      \includegraphics{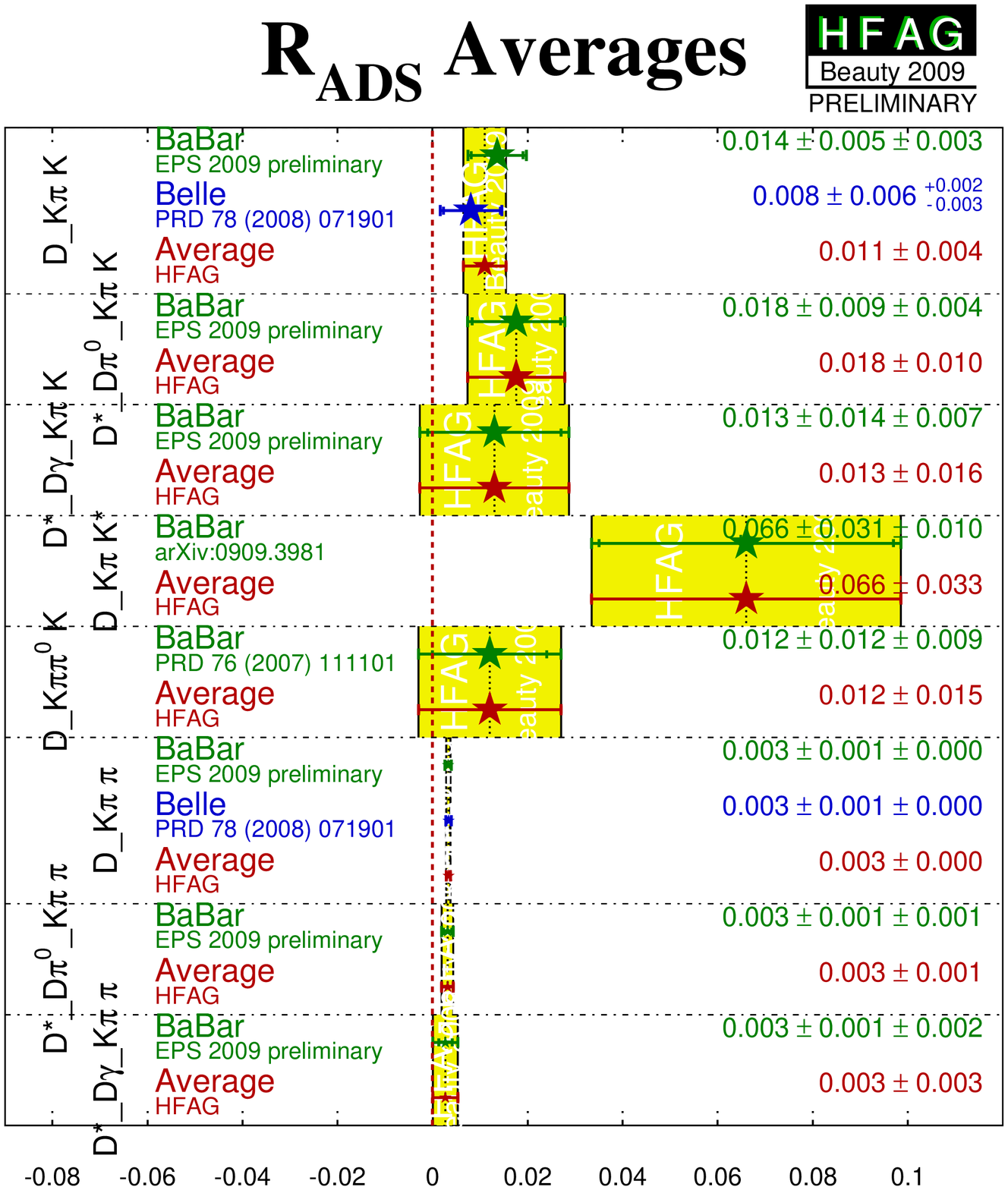}
    }
  \end{center}
  \vspace{-0.8cm}
  \caption{
    Averages of $R_{\rm ADS}$.
  }
  \label{fig:cp_uta:cus:ads}
\end{figure}

For ADS analysis, both \babar\ and \belle\ have studied the modes 
$\Bmp \to D\Kmp$ and $\Bmp \to D\pi^\mp$. \babar\ has also analyzed the 
$\Bmp \to \Dstar\Kmp$ and $\Bmp \to D\Kstarmp$ modes.
There is an effective shift of $\pi$ in the strong phase difference between
the cases that the $\Dstar$ is reconstructed as $D\pi^0$ and
$D\gamma$~\cite{Bondar:2004bi}, therefore these modes are studied separately.
$\Kstarmp$ is reconstructed as $\KS\pi^\mp$.
In all cases the suppressed decay $D \to K^+\pi^-$ has been used.
\babar\ also has results using $\Bmp \to D\Kmp$ with $D \to K^+\pi^-\pi^0$.
The results and averages are given in Table~\ref{tab:cp_uta:cus:ads}
and shown in Fig.~\ref{fig:cp_uta:cus:ads}.
Note that although no clear signals for these modes have yet been seen,
the central values are given.

\begin{table}[htb]
	\begin{center}
		\caption{
      Averages from ADS analyses of $b \to c\bar{u}s / u\bar{c}s$ and 
      $b \to c\bar{u}d / u\bar{c}d$ modes.
                }
                \vspace{0.2cm}
                \setlength{\tabcolsep}{0.0pc}
\begin{tabular*}{\textwidth}{@{\extracolsep{\fill}}lrccc} \hline 
        \mc{2}{l}{Experiment} & $N(B\bar{B})$ & $A_{\rm ADS}$ & $R_{\rm ADS}$ \\
	\hline
        \mc{5}{c}{$D K^-$, $D \to K^+\pi^-$} \\
	\babar & \cite{babar:ads:preliminary} & 426M & $-0.70 \pm 0.35 \,^{+0.09}_{-0.14}$ & $0.0136 \pm 0.0055 \pm 0.0027$ \\
	\belle & \cite{:2008as} & 657M & $-0.13 \,^{+0.97}_{-0.88} \pm 0.26$ & $0.0080 \,^{+0.0063}_{-0.0057} \,^{+0.0020}_{-0.0028}$ \\
        \mc{3}{l}{\bf Average} & $-0.62 \pm 0.34$ & $0.0110 \pm 0.0045$ \\
	\mc{3}{l}{\small Confidence level} & {\small $0.57~(0.6\sigma)$} & {\small $0.53~(0.6\sigma)$} \\
		\hline

        \mc{5}{c}{$\Dstar K^-$, $\Dstar \to D\pi^0$, $D \to K^+\pi^-$} \\
	\babar & \cite{babar:ads:preliminary} & 426M & $0.77 \pm 0.35 \pm 0.12$ & $0.018 \pm 0.009 \pm 0.004$ \\
        \hline

        \mc{5}{c}{$\Dstar K^-$, $\Dstar \to D\gamma$, $D \to K^+\pi^-$} \\
	\babar & \cite{babar:ads:preliminary} & 426M & $0.36 \pm 0.94 \,^{+0.25}_{-0.41}$ & $0.013 \pm 0.014 \pm 0.007$ \\
        \hline

        \mc{5}{c}{$D K^{*-}$, $D \to K^+\pi^-$, $K^{*-} \to \KS \pi^-$} \\
	\babar & \cite{Aubert:2009yw} & 379M & $-0.34 \pm 0.43 \pm 0.16$ & $0.066 \pm 0.031 \pm 0.010$ \\
        \hline

        \mc{5}{c}{$D K^{-}$, $D \to K^+\pi^-\pi^0$} \\
        \babar & \cite{Aubert:2007nu} & 226M & \textendash{} & $0.012 \pm 0.012 \pm 0.009$ \\
        \hline 

        \vspace{1ex} \\

        \hline
       \mc{5}{c}{$D \pi^-$, $D \to K^+\pi^-$} \\
       \babar & \cite{babar:ads:preliminary} & 426M & \textendash{} & $0.0033 \pm 0.0006 \pm 0.0003$ \\
       \belle & \cite{:2008as} & 657M & $-0.023 \pm 0.218 \pm 0.071$ & $0.0034 \,^{+0.0006}_{-0.0005} \,^{+0.0001}_{-0.0002}$ \\
	\mc{3}{l}{\bf Average} & \textendash{} & $0.0034 \pm 0.0004$ \\
	\mc{3}{l}{\small Confidence level} & \textendash{} & {\small $0.91~(0.1\sigma)$} \\
       \hline 
       \mc{5}{c}{$\Dstar \pi^-$, $\Dstar \to D\pi^0$, $D \to K^+\pi^-$} \\
       \babar & \cite{babar:ads:preliminary} & 426M & \textendash{} & $0.0032 \pm 0.0009 \pm 0.0009$ \\
       \hline 
       \mc{5}{c}{$\Dstar \pi^-$, $\Dstar \to D\gamma$, $D \to K^+\pi^-$} \\
       \babar & \cite{babar:ads:preliminary} & 426M & \textendash{} & $0.0027 \pm 0.0014 \pm 0.0022$ \\
        \hline

 		\end{tabular*}
                \label{tab:cp_uta:cus:ads}
 	\end{center}
 \end{table}

\babar~\cite{:2009au} have also presented results on a similar analysis with
self-tagging neutral $B$ decays: $\Bz \to DK^{*0}$ with $D \to K^-\pi^+$, 
$D \to K^-\pi^+\pi^0$ and $D \to K^-\pi^+\pi^+\pi^-$ 
(all with $K^{*0} \to K^+\pi^-$). 
Effects due to the natural width of the $K^{*0}$ are
handled using the parametrization suggested by Gronau~\cite{Gronau:2002mu}. 

The following 95\% C.L. limits are set:
\begin{equation}
  R_{\rm ADS}(K\pi) < 0.244 \hspace{5mm}
  R_{\rm ADS}(K\pi\pi^0) < 0.181 \hspace{5mm}
  R_{\rm ADS}(K\pi\pi\pi) < 0.391 \, .
\end{equation}

Combining the results and using additional input from
CLEOc~\cite{Asner:2008ft,Lowery:2009id} a limit on the ratio between the 
$b \to u$ and $b \to c$ amplitudes of $r_s \in \left[ 0.07,0.41 \right]$ 
at 95\% C.L. limit is set.

\belle~\cite{Krokovny:2002ua} have also presented results that set constraints
on $r_s$. 

\mysubsubsection{$D$ decays to multiparticle self-conjugate final states}
\label{sec:cp_uta:cus:dalitz}

For the Dalitz plot analysis, both 
\babar~\cite{Aubert:2008bd} and
\belle~\cite{Poluektov:2010wz,Poluektov:2006ia} have studied the modes 
$\Bmp \to D\Kmp$, $\Bmp \to \Dstar\Kmp$ and $\Bmp \to D\Kstarmp$.
For $\Bmp \to \Dstar\Kmp$,
both experiments have used both $\Dstar$ decay modes, $\Dstar \to D\pi^0$ and
$\Dstar \to D\gamma$, taking the effective shift in the strong phase
difference into account. 
In all cases the decay $D \to \KS\pi^+\pi^-$ has been used.
\babar\ also used the decay $D \to \KS K^+K^-$ .
\babar\ has also performed an analysis of $\Bmp \to D\Kmp$ with 
$D \to \pi^+\pi^-\pi^0$~\cite{Aubert:2007ii}.
Results and averages are given in Table~\ref{tab:cp_uta:cus:dalitz}.
The third error on each measurement is due to $D$ decay model uncertainty.

The parameters measured in the analyses are explained in
Sec.~\ref{sec:cp_uta:notations:cus}.
Both \babar\ and \belle\ have measured the ``Cartesian''
$(x_\pm,y_\pm)$ variables,
and perform frequentist statistical procedures,
to convert these into measurements of $\gamma$, $r_B$ and $\delta_B$.
In the $\Bmp \to D\Kmp$ with $D \to \pi^+\pi^-\pi^0$ analysis,
the parameters $(\rho^{\pm}, \theta^\pm)$ are used instead.

Both experiments reconstruct $\Kstarmp$ as $\KS\pi^\mp$,
but the treatment of possible nonresonant $\KS\pi^\mp$ differs:
\belle\ assign an additional model uncertainty,
while \babar\ use a reparametrization suggested by Gronau~\cite{Gronau:2002mu}.
The parameters $r_B$ and $\delta_B$ are replaced with 
effective parameters $\kappa r_s$ and $\delta_s$;
no attempt is made to extract the true hadronic parameters 
of the $\Bmp \to D\Kstarmp$ decay.

We perform averages using the following procedure, which is based on a set of
(more or less) reasonable, though imperfect, assumptions. 

\begin{itemize}\setlength{\itemsep}{0.5ex}
\item 
  It is assumed that effects due to the different $D$ decay models 
  used by the two experiments are negligible. 
  Therefore, we do not rescale the results to a common model.
\item 
  It is further assumed that the model uncertainty is $100\%$ 
  correlated between experiments, 
  and therefore this source of error is not used in the averaging procedure.
  (This approximation is significantly less valid now that the \babar\ results
  include $D \to \KS K^+K^-$ decays in addition to $D \to \KS\pi^+\pi^-$.)
\item 
  We include in the average the effect of correlations 
  within each experiments set of measurements.
\item 
  At present it is unclear how to assign an average model uncertainty. 
  We have not attempted to do so. 
  Our average includes only statistical and systematic error. 
  An unknown amount of model uncertainty should be added to the final error.
\item 
  We follow the suggestion of Gronau~\cite{Gronau:2002mu} 
  in making the $DK^*$ averages. 
  Explicitly, we assume that the selection of $K^{*\pm} \to \KS\pi^\pm$
  is the same in both experiments 
  (so that $\kappa$, $r_s$ and $\delta_s$ are the same), 
  and drop the additional source of model uncertainty 
  assigned by Belle due to possible nonresonant decays.
\item 
  We do not consider common systematic errors, 
  other than the $D$ decay model. 
\end{itemize}

\begin{table}[htb]
	\begin{center}
		\caption{
      Averages from Dalitz plot analyses of $b \to c\bar{u}s / u\bar{c}s$ modes.
      Note that the uncertainities assigned to the averages do not include model errors.	
		}
		\vspace{0.2cm}
		\setlength{\tabcolsep}{0.0pc}
    \resizebox{\textwidth}{!}{

              }
		\label{tab:cp_uta:cus:dalitz}
	\end{center}
\end{table}

\begin{figure}[htb]
  \begin{center}
    \resizebox{0.30\textwidth}{!}{
      \includegraphics{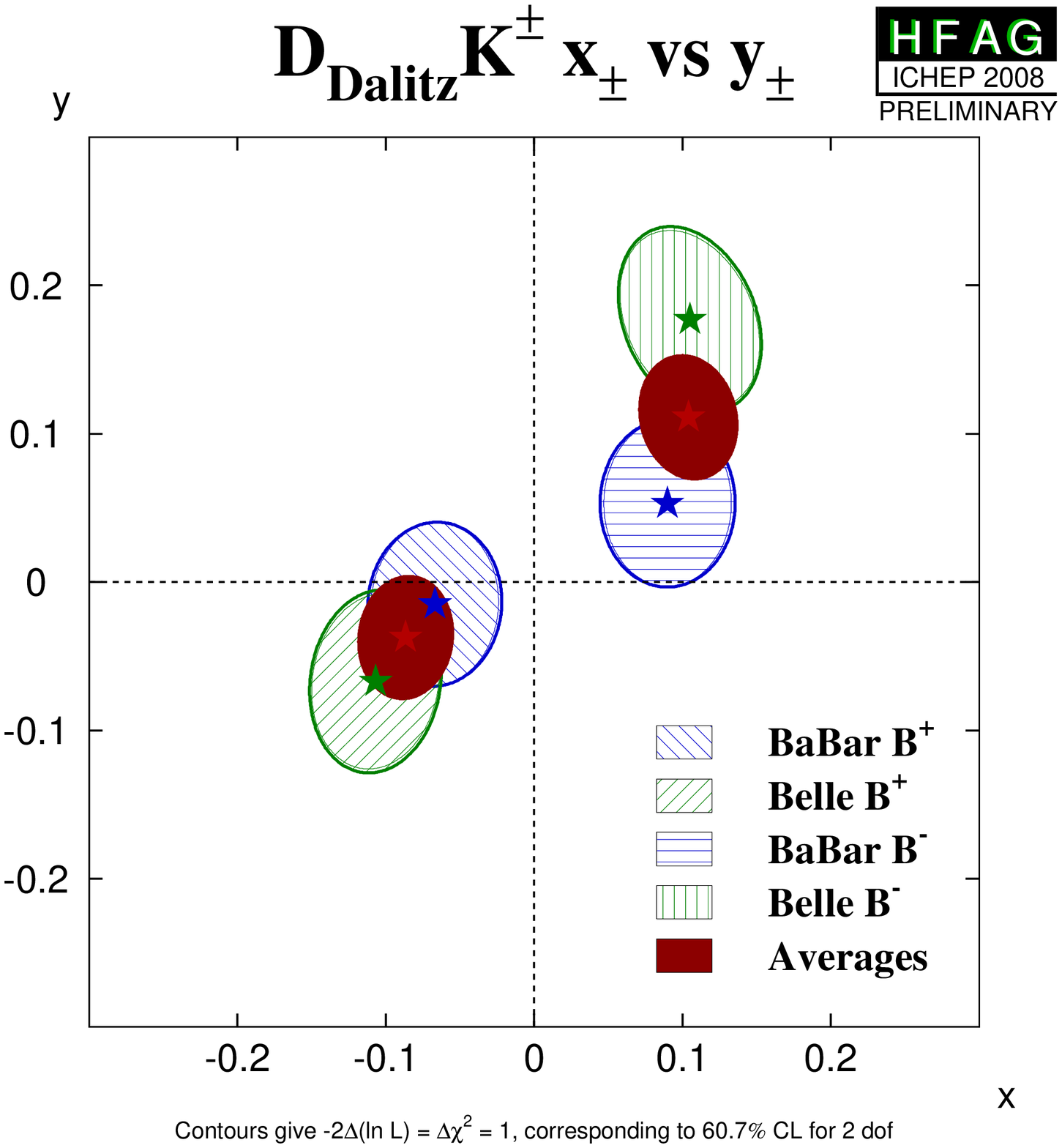}
    }
    \hfill
    \resizebox{0.30\textwidth}{!}{
      \includegraphics{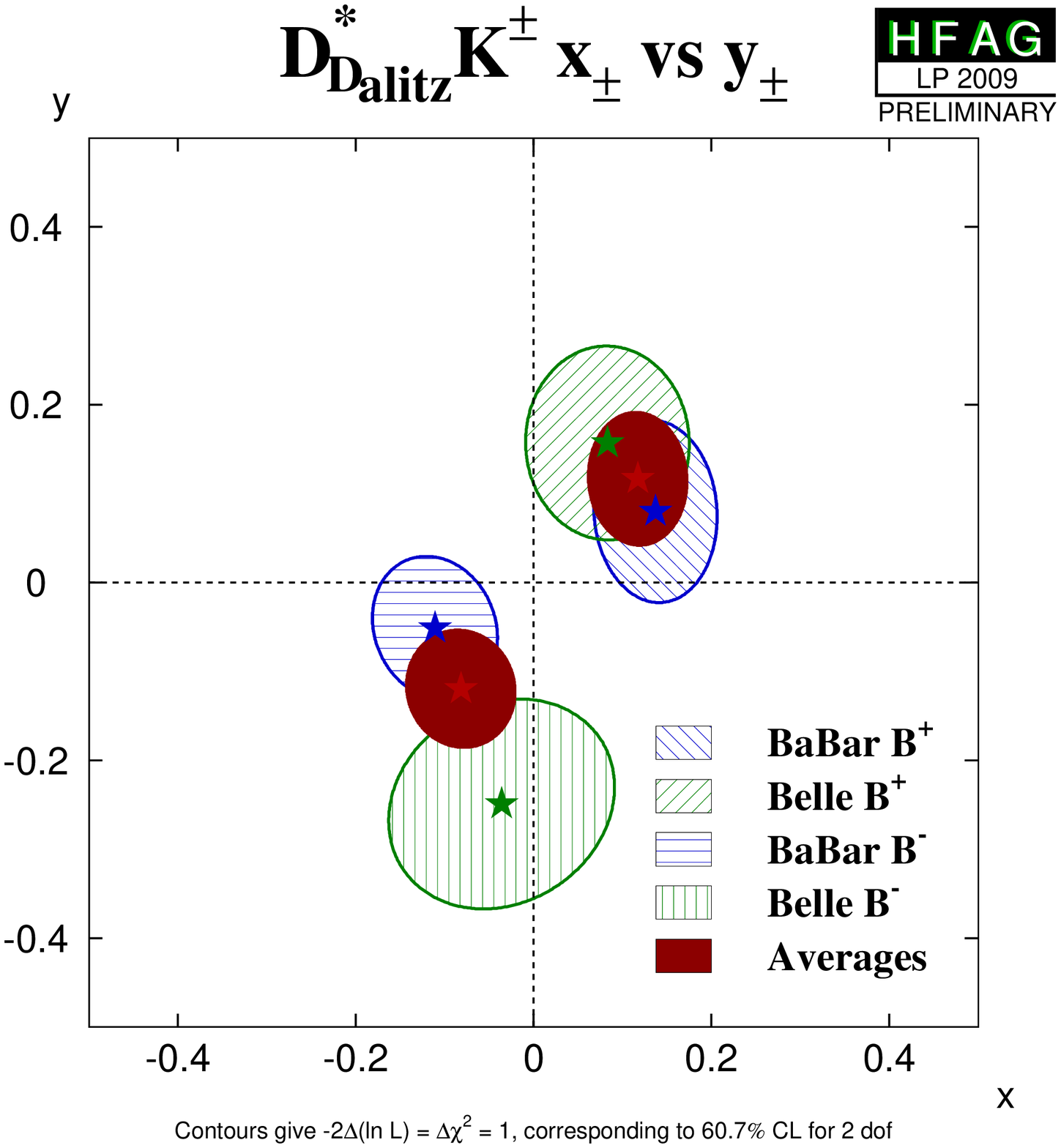}
    }
    \hfill
    \resizebox{0.30\textwidth}{!}{
      \includegraphics{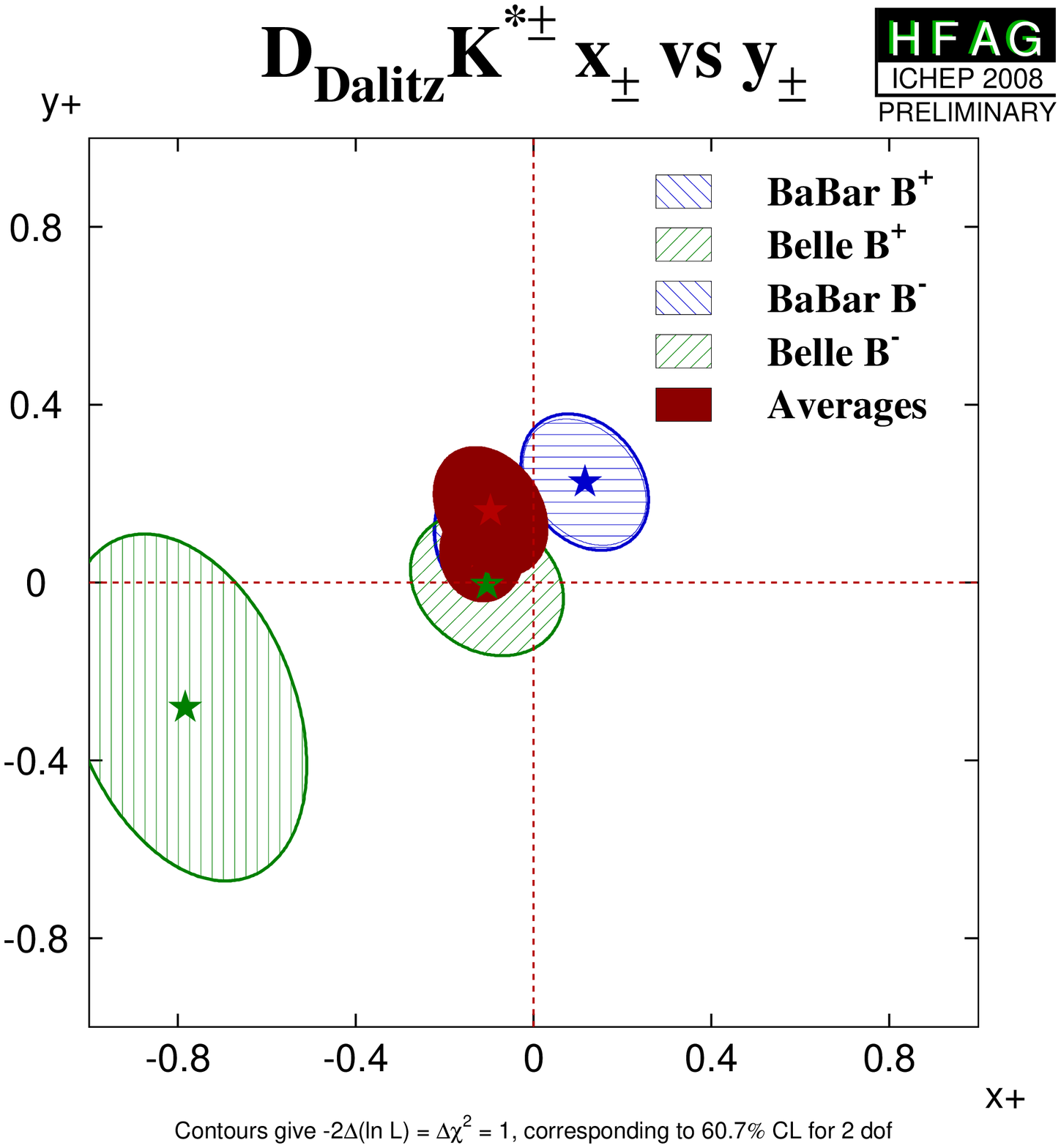}
    }
  \end{center}
  \vspace{-0.8cm}
  \caption{
    Contours in the $(x_\pm, y_\pm)$ from $\Bmp \to D^{(*)}K^{(*)\pm}$.
    (Left) $\Bmp \to D\Kmp$, 
    (middle) $\Bmp \to \Dstar\Kmp$,
    (right) $\Bmp \to D\Kstarmp$.
    Note that the uncertainities assigned to the averages given in these plots
    do not include model errors.        
  }
  \label{fig:cp_uta:cus:dalitz_2d}
\end{figure}

\begin{figure}[htb]
  \begin{center}
    \resizebox{0.40\textwidth}{!}{
      \includegraphics{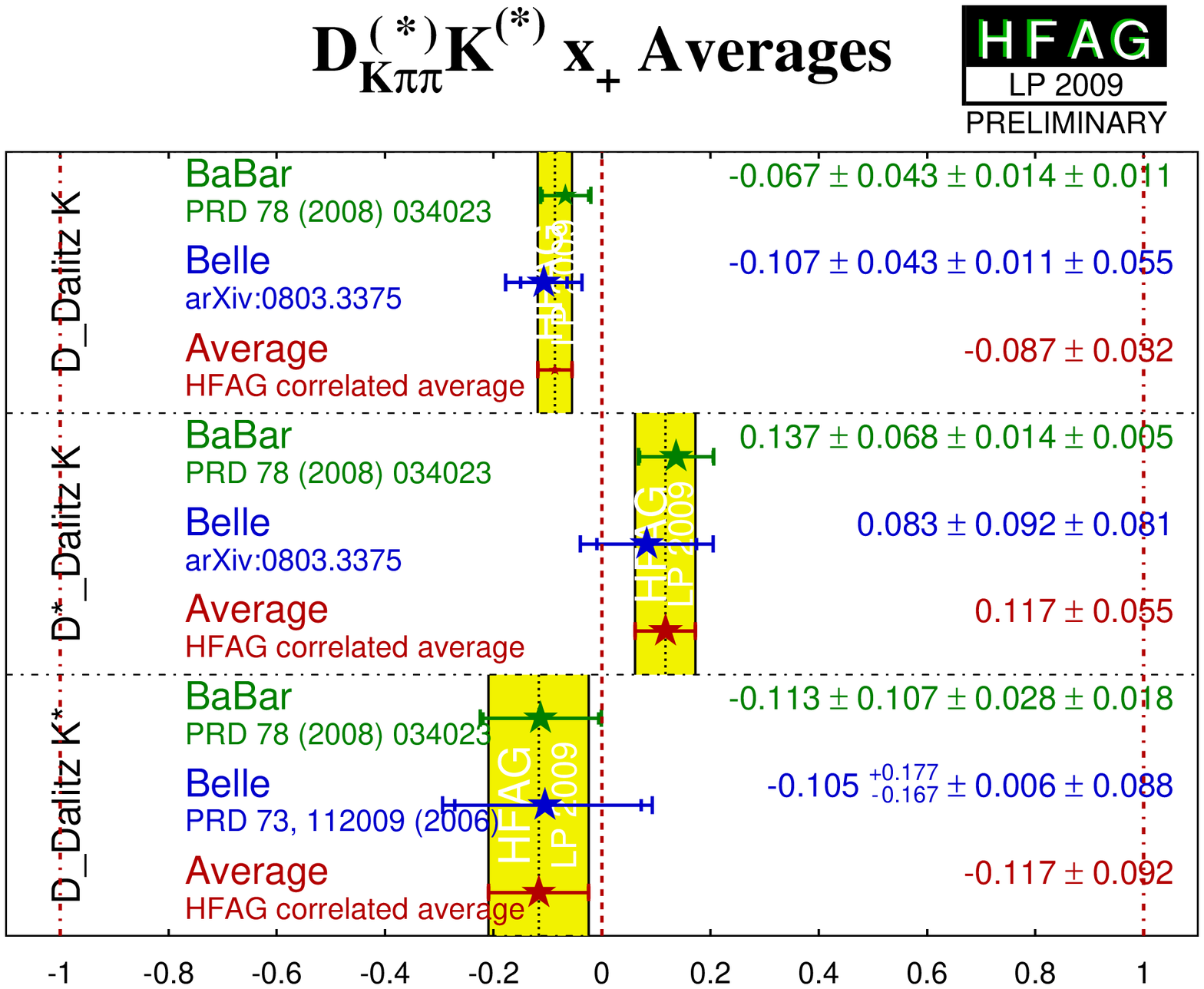}
    }
    \hspace{0.1\textwidth}
    \resizebox{0.40\textwidth}{!}{
      \includegraphics{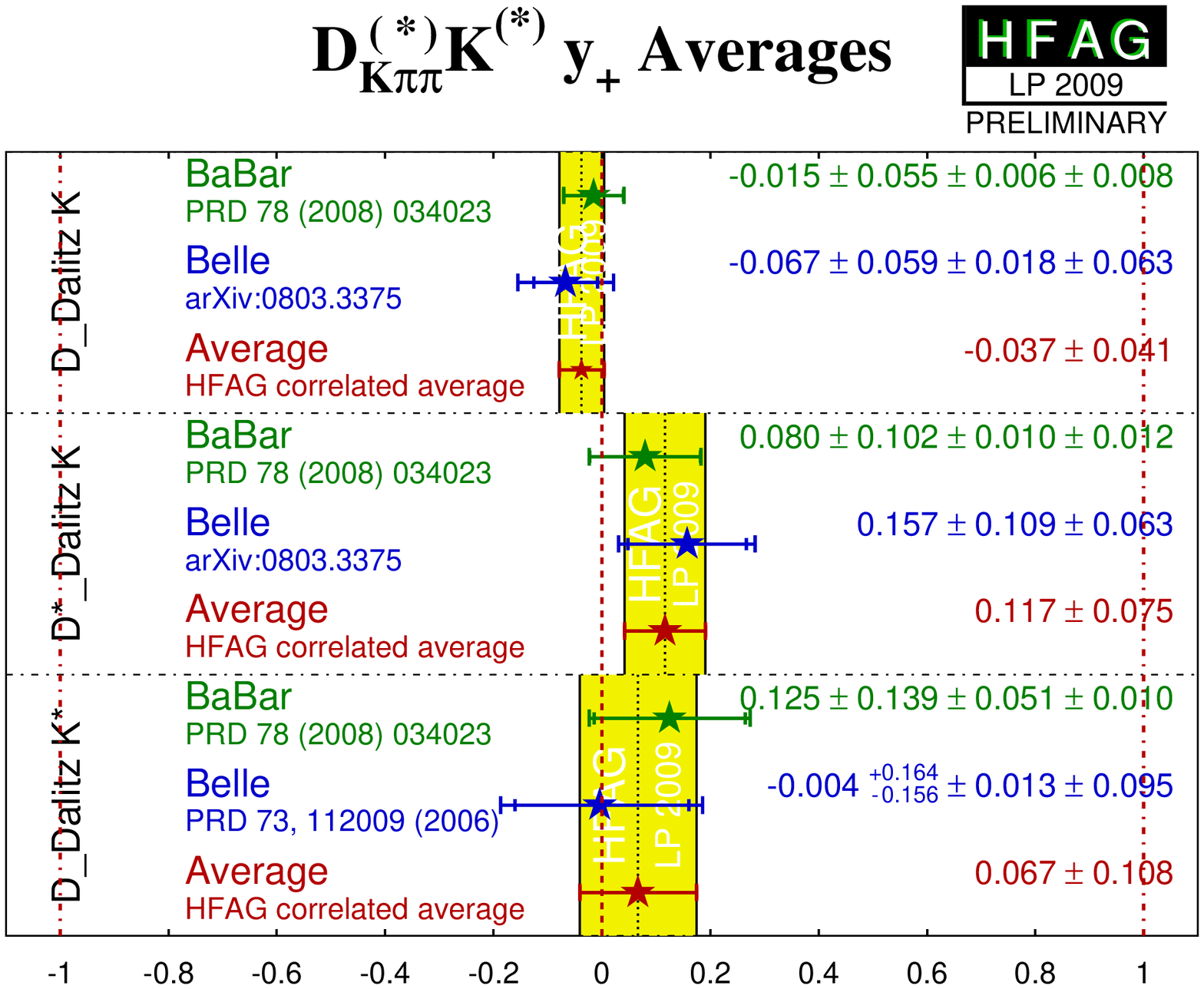}
    }
    \\
    \resizebox{0.40\textwidth}{!}{
      \includegraphics{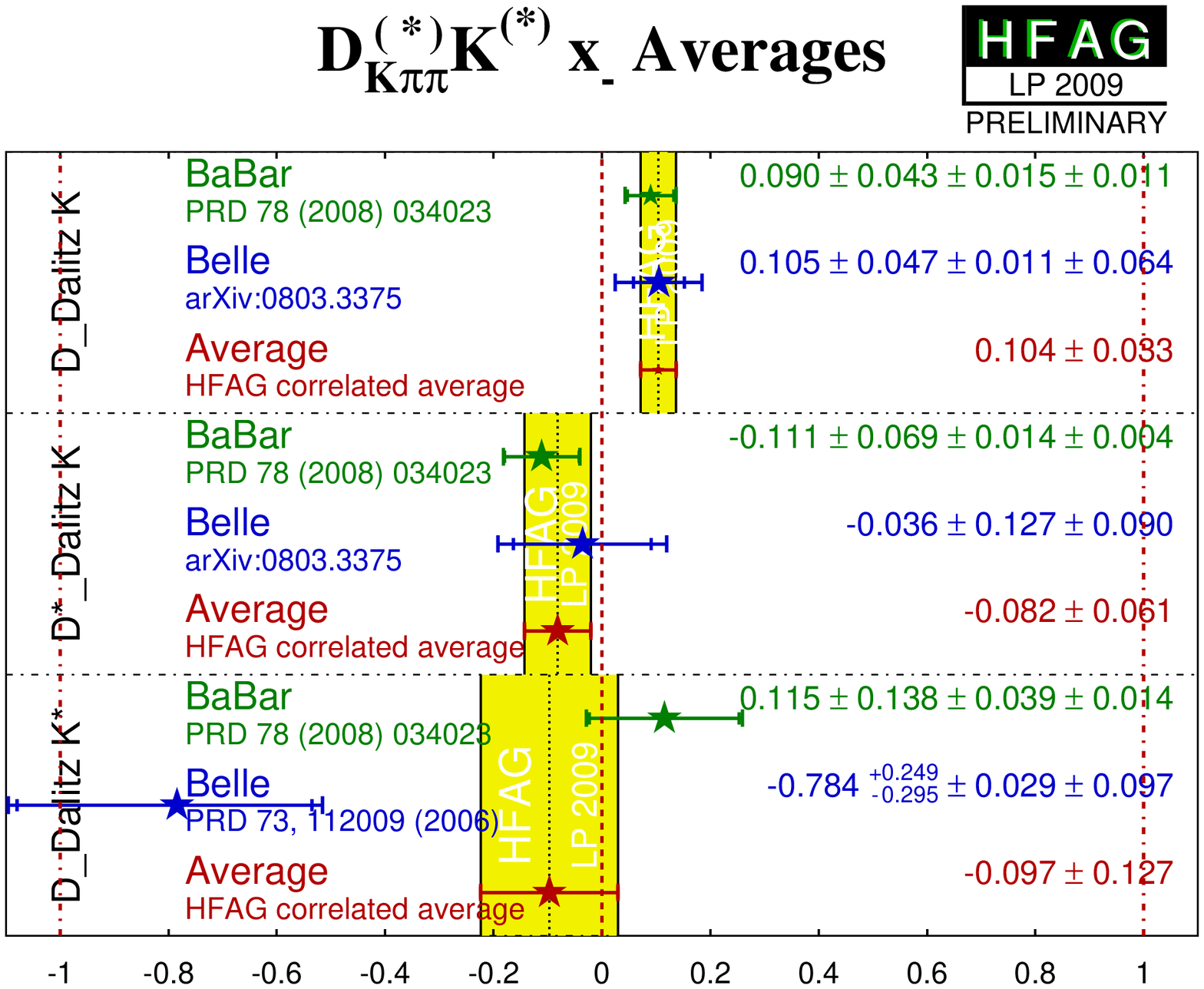}
    }
    \hspace{0.1\textwidth}
    \resizebox{0.40\textwidth}{!}{
      \includegraphics{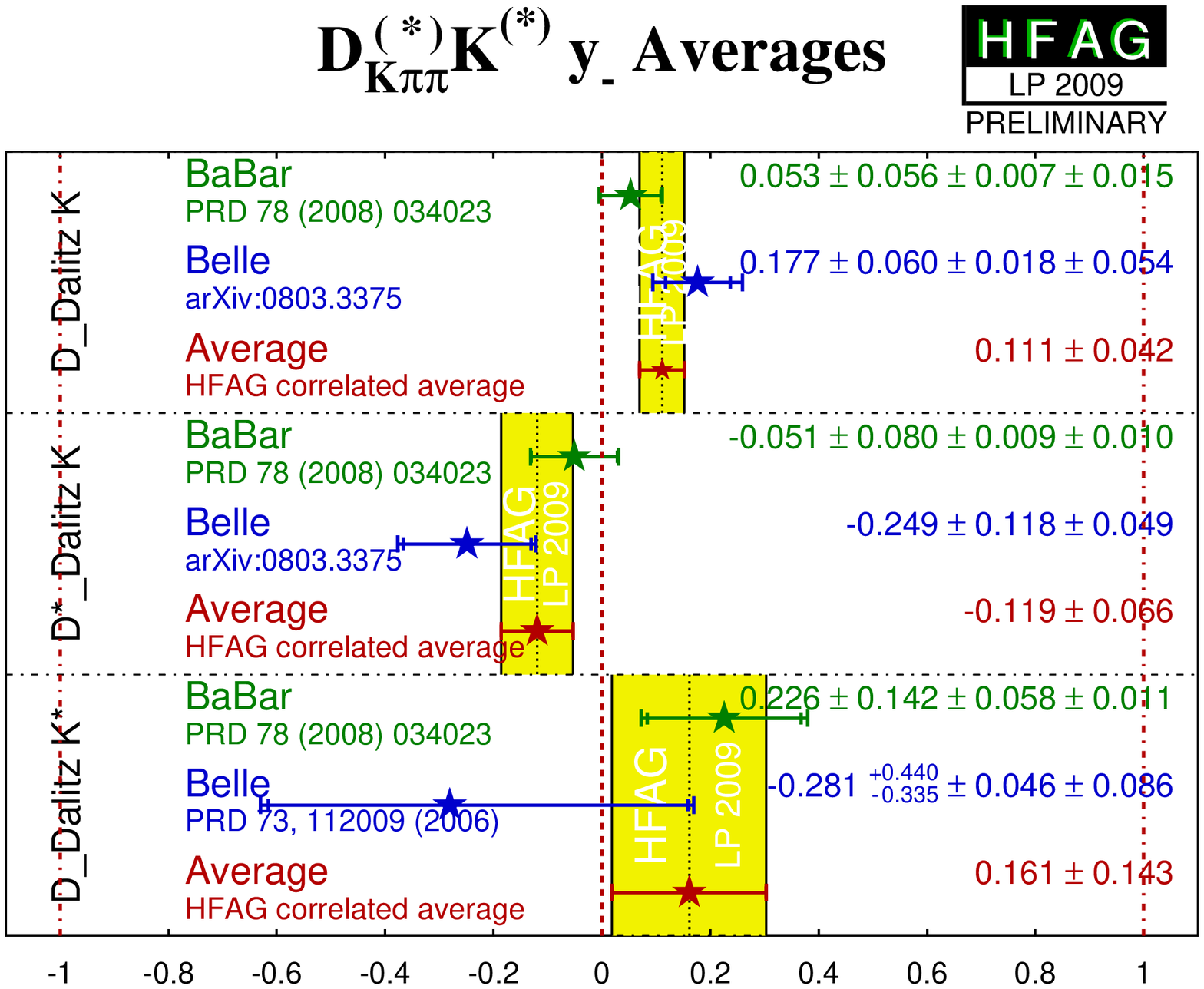}
    }
  \end{center}
  \vspace{-0.8cm}
  \caption{
    Averages of $(x_\pm, y_\pm)$ from $\Bpm \to D^{(*)}K^{(*)\pm}$.
    (Top left) $x_+$, (top right) $y_+$,
    (bottom left) $x_-$, (bottom right) $y_-$.
    Note that the uncertainities assigned to the averages given in these plots
    do not include model errors.        
  }
  \label{fig:cp_uta:cus:dalitz_1d}
\end{figure}

\vspace{3ex}

\noindent
\underline{\large Constraints on $\gamma$}

The measurements of $(x_\pm, y_\pm)$ can be used to obtain constraints on 
$\gamma$, as well as the hadronic parameters $r_B$ and $\delta_B$.
Both
\babar~\cite{Aubert:2008bd} and 
\belle~\cite{Poluektov:2010wz,Poluektov:2006ia} 
have done so using a frequentist procedure 
(there are some differences in the details of the techniques used).

\begin{itemize}\setlength{\itemsep}{0.5ex}

\item 
  \babar\ obtain $\gamma = (76 \pm 22 \pm 5 \pm 5)^\circ$
  from $D\Kpm$, $\Dstar\Kpm$ and $D\Kstarpm$

\item
  \belle\ obtain $\phi_3 = (78 \,^{+11}_{-12} \pm 4 \pm 9)^\circ$
  from $D\Kpm$ and $\Dstar\Kpm$

\item
  The experiments also obtain values for the hadronic parameters as detailed
  in Tab.~\ref{tab:cp_uta:rBdeltaB_summary}.

\item 
  Improved constraints can be achieved combining the information from
  $\Bpm \to D\Kpm$ analysis with different $D$ decay modes.
  The experiments have not yet published such results,
  and none are listed here.

\item 
  The CKMfitter~\cite{Charles:2004jd} and 
  UTFit~\cite{Bona:2005vz} groups use the measurements 
  from \belle\ and \babar\ given above
  to make combined constraints on $\gamma$.

\item 
  In the \babar\ analysis of $\Bmp \to D\Kmp$ with 
  $D \to \pi^+\pi^-\pi^0$~\cite{Aubert:2007ii},
  a constraint of $-30^\circ < \gamma < 76^\circ$ is obtained 
  at the 68\% confidence level.

\end{itemize}

\begin{table}
  \begin{center}
  \begin{tabular}{lcc}
    \hline
    & $r_B$ & $\delta_B$ \\
    \hline
    \multicolumn{3}{c}{In $D\Kpm$} \\
    \babar & $0.086 \pm 0.035$ & $\delta_B (D\Kpm) = (109 \,^{+28}_{-31})^\circ$ \\
    \belle & $0.160 \,^{+0.040}_{-0.038} \pm 0.011 \,^{+0.05}_{-0.010}$ & 
    $(138 \,^{+13}_{-16} \pm 4 \pm 23)^\circ$ \\
    \hline
    \multicolumn{3}{c}{In $\Dstar\Kpm$} \\
    \babar & $0.135 \pm 0.051$ & $(297 \,^{+30}_{-28})^\circ$ \\
    \belle & $0.196 \,^{+0.072}_{-0.069} \pm 0.012 \,^{+0.062}_{-0.012}$ &
    $(342 \,^{+19}_{-21} \pm 3 \pm 23)^\circ$ \\
    \hline
    \multicolumn{3}{c}{In $D\Kstarpm$} \\
    \babar & $\kappa r_S = 0.163 \,^{+0.088}_{-0.105}$ &
    $\delta_S = (104 \,^{+43}_{-41})^\circ$ \\
    \belle & $0.56 \,^{+0.22}_{-0.16} \pm 0.04 \pm 0.08$ & 
    $(243 \,^{+20}_{-23} \pm 3 \pm 50)^\circ$ \\
    \hline
  \end{tabular}
  \caption{
    Summary of constraints on hadronic parameters 
    in $\Bpm \to \DorDstar\KorKstarpm$ decays.
    Note the alternative parametrisation of the hadronic parameters used by
    \babar\ in the $D\Kstarpm$ mode.
  }
  \label{tab:cp_uta:rBdeltaB_summary}
  \end{center}
\end{table}

At present we make no attempt to provide an HFAG average for $\gamma$,
nor indeed for the hadronic parameters.
More details on procedures to calculate a best fit value for $\gamma$ 
can be found in Refs.~\cite{Charles:2004jd,Bona:2005vz}.

\babar~\cite{Aubert:2008yn} have also performed a similar Dalitz plot analysis
to that described above using the self-tagging neutral $B$ decay $\Bz \to
DK^{*0}$ (with $K^{*0} \to K^+\pi^-$). Effects due to the natural width of the
$K^{*0}$ are handled using the parametrization suggested by
Gronau~\cite{Gronau:2002mu}.

\babar\ extract the three-dimensional likelihood for the parameters 
$\left( \gamma, \delta_S, r_S \right)$ and, combining with a separately
measured PDF for $r_S$ (using a Bayesian technique), obtain bounds on each of
the three parameters. 
\begin{equation}
  \gamma = (162 \pm 56)^\circ \hspace{5mm}
  \delta_S = (62 \pm 57)^\circ \hspace{5mm}
  r_S < 0.55  \, ,
\end{equation}
where the limit on $r_S$ is at 95\% probability.
Note that there is an ambiguity in the solutions 
$\left( \gamma, \delta_S \leftrightarrow \gamma+\pi, \delta_S+\pi \right)$.

\clearpage


\section{Semileptonic $B$ decays}
\label{sec:slbdecays}

Measurements of semileptonic $B$-meson decays are an important tool to
study the magnitude of the CKM matrix elements $|V_{cb}|$ and
$|V_{ub}|$, the Heavy Quark parameters (e.g. $b$ and $c$--quark masses),
QCD form factors, QCD dynamics, new physics, etc.

In the following, we provide averages of exclusive and inclusive
 branching fractions, the product of $|V_{cb}|$ and the form factor
 normalization ${\cal F}(1)$ and ${\cal G}(1)$ for $\bar{B} \to D^* \ell^-\bar{\nu}_{\ell}$ and
$\bar{B} \to D \ell^-\bar{\nu}_{\ell}$ decays, respectively, and $|V_{ub}|$ as determined from
inclusive and exclusive measurements of $\B\to X_u \ell \nul$ decays.
We will compute Heavy Quark parameters and extract QCD form factors
for $\bar{B} \to D^* \ell^-\bar{\nu}_{\ell}$ decays.
Throughout this section, charge conjugate states are implicitly included, 
unless otherwise indicated.

Brief descriptions of all parameters
and analyses (published or preliminary) relevant for the
determination of the combined results are given.  The descriptions are
based on the information available on the web page at\\
 \centerline{\tt http://www.slac.stanford.edu/xorg/hfag/semi/EndOfYear09}
A description of the technique employed for calculating averages
was presented in the previous update~\cite{Barberio:2008fa}. 
Asymmetric errors have been introduced in the current averages
for $\Bb\to X_u\ell\nub$ decays to take into account theoretical 
asymmetric errors. 






\subsection{Exclusive CKM-favored decays}
\label{slbdecays_b2cexcl}
Averages are provided for the branching fractions
$\cbf(\bar{B} \to D \ell^-\bar{\nu}_{\ell})$ and $\cbf(\bar{B} \to D^* \ell^-\bar{\nu}_{\ell})$.  
We then provide averages for the inclusive branching
fractions $\cbf(\bar{B} \to D^{(*)}\pi \ell^-\bar{\nu}_{\ell})$ and for $B$ semileptonic 
decays into orbitally-excited $P$-wave charm mesons ($D^{**}$). As the $D^{**}$ branching 
fraction is poorly known, we report the averages for the products 
$\cbf(B^- \to D^{**}(D^{(*)}\pi)\ell^-\bar{\nu}_{\ell})
\times \cbf(D^{**} \to D^{(*)}\pi)$.
In addition, averages are provided for $ {\cal F}(1)\vcb$ vs $\rho^2$, where ${\cal F}(1)$ and $\rho^2$ are the normalization and slope of the form factor at zero recoil in $\bar{B} \to D^* \ell^-\bar{\nu}_{\ell}$ decays, and for the corresponding quantities ${\cal G}(1) \vcb$ vs $\rho^2$ in $\bar{B} \to D \ell^-\bar{\nu}_{\ell}$ decays.

\mysubsubsection{$\bar{B} \to D \ell^-\bar{\nu}_{\ell}$}
\label{slbdecays_dlnu}

The average branching fraction $\cbf(\bar{B} \to D \ell^-\bar{\nu}_{\ell})$ 
is determined by the combination of the results provided in Table~\ref{tab:dplnu} and
~\ref{tab:d0lnu}, for $\bar{B}^0 \to D^+ \ell^-\bar{\nu}_{\ell}$ and $B^- \to D^0
\ell^-\bar{\nu}_{\ell}$, respectively.
The measurements included in the average 
are scaled to a consistent set of input
parameters and their errors~\cite{HFAG_sl:inputparams}.  
Therefore some of the (older) measurements are subject to considerable adjustments.
The branching fractions are obtained from the integral over the measured differential decay rates,
 apart for the \babar\ results, for which the semileptonic $B$ signal yields are
 extracted from a fit to the missing mass squared in a sample of fully
 reconstructed \BB\ events. 
 
Figure~\ref{fig:brdl} illustrates the measurements and the
resulting average.

\begin{table}[!htb]
\caption{
Average of the branching fraction $\cbf(\BzbDplnu)$ and individual
results. }
\begin{center}
\begin{tabular}{|l|c|c|}\hline
Experiment                                 &$\cbf(\BzbDplnu) [\%]$ (rescaled) &$\cbf(\BzbDplnu) [\%]$ (published) \\
\hline\hline 
ALEPH ~\hfill\cite{Buskulic:1996yq}        &$2.25 \pm0.18_{\rm stat} \pm0.36_{\rm syst}$  &$2.35 \pm0.18_{\rm stat} \pm0.44_{\rm syst}$ \\
CLEO  ~\hfill\cite{Bartelt:1998dq}         &$2.12 \pm0.13_{\rm stat} \pm0.15_{\rm syst}$  &$2.20 \pm0.13_{\rm stat} \pm0.18_{\rm syst}$ \\
\belle  ~\hfill\cite{Abe:2001yf}           &$2.10 \pm0.12_{\rm stat} \pm0.39_{\rm syst}$  &$2.13 \pm0.12_{\rm stat} \pm0.41_{\rm syst}$ \\
\babar  ~\hfill\cite{Aubert:vcbExcl}       &$2.20 \pm0.11_{\rm stat} \pm0.12_{\rm syst}$  &$2.22 \pm0.11_{\rm stat} \pm0.12_{\rm syst}$ \\
\hline 
{\bf Average}                              &\mathversion{bold}$2.17 \pm0.12$     &\mathversion{bold}$\chi^2/\dof = 0.2/3$ (CL=$98\%$)  \\
\hline 
\end{tabular}
\end{center}
\label{tab:dplnu}
\end{table}

\begin{table}[!htb]
\caption{Average of the branching fraction $\cbf(B^- \to D^0
\ell^-\bar{\nu}_{\ell})$ and individual
results. }
\begin{center}
%
  \caption{Average branching fraction  of exclusive semileptonic $B$ decays
(a) $\bar{B}^0 \to D^+ \ell^-\bar{\nu}_{\ell}$ and (b) 
$B^- \to D^0 \ell^-\bar{\nu}_{\ell}$ and individual
  results.}
  \label{fig:brdl}
 \end{center}
\end{figure}

Recent measurement~\cite{Aubert:2009_1,Aubert:2009_2} assume isospin conservation for the $\bar{B} \to D \ell^-\bar{\nu}_{\ell}$ decays and are averaged independently from the previous determinations of  the $\cbf(\bar{B} \to D \ell^-\bar{\nu}_{\ell})$ average branching fraction. Figure~\ref{fig:brdlIso} and table~\ref{tab:dlnuIso}  illustrates the measurements and the
resulting average.

\begin{table}[!htb]
\caption{Average of the branching fraction $\cbf(B^- \to D^0
\ell^-\bar{\nu}_{\ell})$ and individual
results. }
\begin{center}
\begin{tabular}{|l|c|c|}\hline
Experiment                                 &$\cbf(B^- \to D^0 \ell^-\bar{\nu}_{\ell}) [\%]$ (rescaled) &$\cbf(B^- \to D^0\ell^-\bar{\nu}_{\ell}) [\%]$ (published) \\
\hline\hline 
\babar  ~\hfill\cite{Aubert:2009_1}       &$2.34 \pm 0.03_{\rm stat} \pm0.12_{\rm syst}$  &$2.34 \pm0.03_{\rm stat} \pm0.13_{\rm syst}$ \\
\babar  ~\hfill\cite{Aubert:2009_2}       &$2.30 \pm 0.06_{\rm stat} \pm0.08_{\rm syst}$  &$2.30 \pm0.06_{\rm stat} \pm0.08_{\rm syst}$ \\
\hline 
{\bf Average}                              &\mathversion{bold}$2.31 \pm0.09$    &\mathversion{bold}$\chi^2/\dof = 0.11$ (CL=$71\%$)  \\
\hline 
\end{tabular}
\end{center}
\label{tab:dlnuIso}
\end{table}

\begin{figure}[!ht]
 \begin{center}
  \unitlength1.0cm 
  \begin{picture}(8.,8.0)  
  \put( -0.5,  0.0){\includegraphics[width=7.55cm]{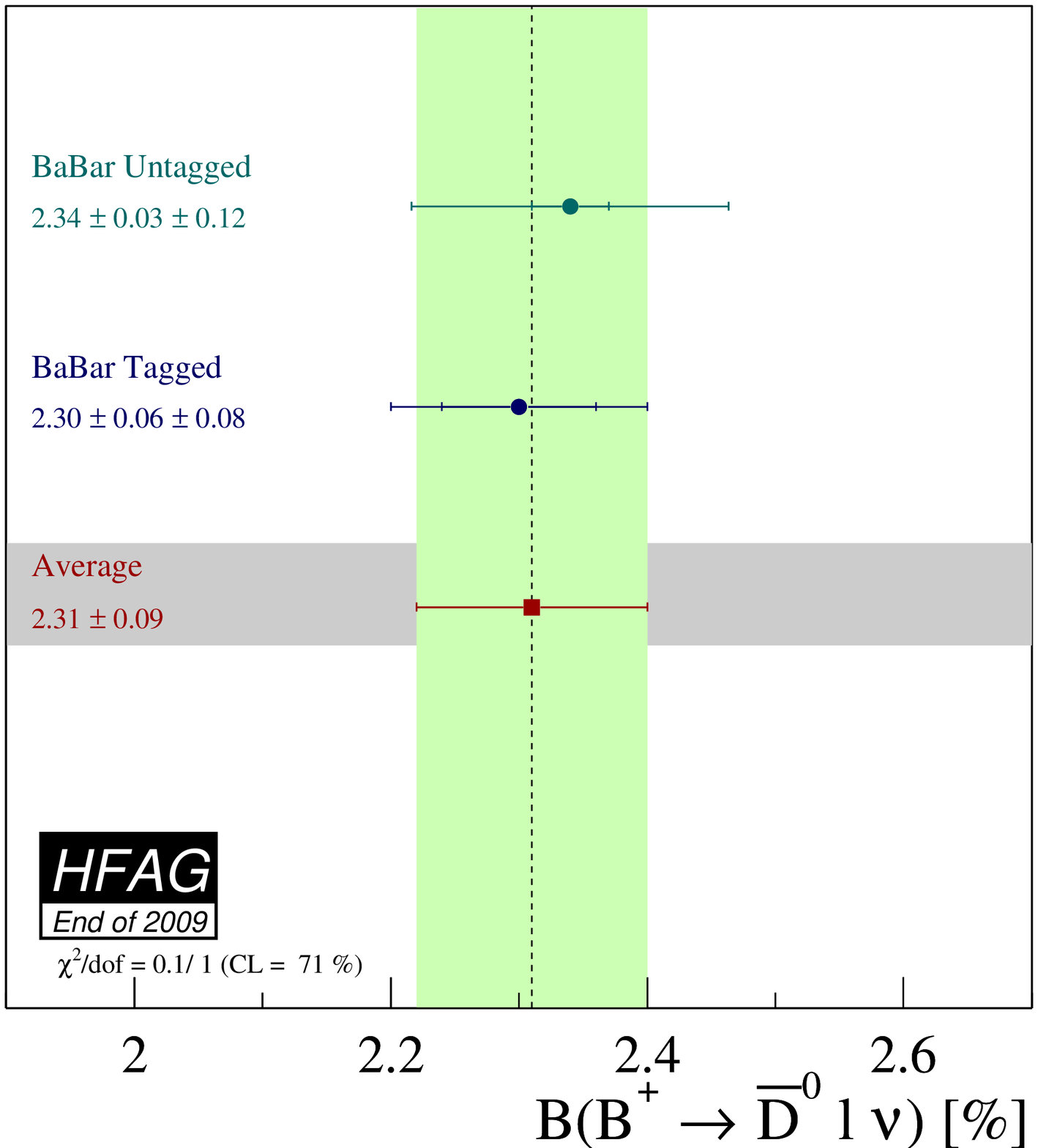}}

    
  \end{picture}
  \caption{Average branching fraction  of exclusive semileptonic $B$ decays
$\bar{B} \to D \ell^-\bar{\nu}_{\ell}$ and individual
  results assuming isospin conservation.}
  \label{fig:brdlIso}
 \end{center}
\end{figure}

The average for ${\cal G}(1)\vcb$ is determined by the two-dimensional
combination of the results provided in Table~\ref{tab:vcbg1}.
Figure~\ref{fig:vcbg1} (a)
provides a one-dimensional projection for illustrative purposes,
(b) illustrates the average ${\cal G}(1)\vcb$ and the measurements included in the average. 

\begin{table}[!htb]
\caption{Average  of $G(1)\vcb$  determined  in the  decay \BzbDplnu\  and
individual  results. The  fit  for the  average  has $\chi^2/\dof  =
0.3/4$.   The total  correlation between  the average  $G(1)\vcb$ and
$\rho^2$ is $0.93$.}
\begin{center}
\begin{tabular}{|l|c|c|}\hline
Experiment &$G(1)\vcb [10^{-3}]$ (rescaled)  &$\rho^2$ (rescaled) \\ 
           &$G(1)\vcb [10^{-3}]$ (published) &$\rho^2$ (published) \\
\hline\hline 
ALEPH~\hfill\cite{Buskulic:1996yq}  &$38.3 \pm11.8_{\rm stat} \pm6.2_{\rm syst}$   &$0.92 \pm0.98_{\rm stat} \pm0.36_{\rm syst}$ \\
                                    &$31.1 \pm9.9_{\rm stat}  \pm8.6_{\rm syst}$&$0.70 \pm0.98_{\rm stat} \pm0.50_{\rm syst}$ \\
\hline
CLEO ~\hfill\cite{Bartelt:1998dq}   &$44.7 \pm5.9_{\rm stat} \pm3.45_{\rm syst}$    &$1.27 \pm0.25_{\rm stat} \pm0.14_{\rm syst}$ \\
                                    &$44.8 \pm6.1_{\rm stat} \pm3.7_{\rm syst}$  &$1.30 \pm0.27_{\rm stat} \pm0.14_{\rm syst}$ \\
\hline
\belle~\hfill\cite{Abe:2001yf}       &$40.85 \pm4.4_{\rm stat} \pm5.14_{\rm syst}$    &$1.12 \pm0.22_{\rm stat} \pm0.14_{\rm syst}$ \\
                                    &$41.1 \pm4.4_{\rm stat} \pm5.1_{\rm syst}$    &$1.12 \pm0.22_{\rm stat} \pm0.14_{\rm syst}$ \\
\hline
\babar~\hfill\cite{Aubert:2009_1}       &$43.1 \pm0.8_{\rm stat} \pm2.1_{\rm syst}$    &$1.20 \pm0.04_{\rm stat} \pm0.06_{\rm syst}$ \\
                                    &$43.1 \pm0.8_{\rm stat} \pm2.3_{\rm syst}$    &$1.20 \pm0.04_{\rm stat} \pm0.07_{\rm syst}$ \\
\hline
\babar~\hfill\cite{Aubert:2009_2}       &$42.3 \pm1.9_{\rm stat} \pm1.0_{\rm syst}$    &$1.20 \pm0.09_{\rm stat} \pm0.04_{\rm syst}$ \\
                                    &$42.3 \pm1.9_{\rm stat} \pm1.0_{\rm syst}$    &$1.20 \pm0.09_{\rm stat} \pm0.04_{\rm syst}$ \\
\hline 
{\bf Average }                      &\mathversion{bold}$42.3 \pm 1.5$       &\mathversion{bold}$1.18 \pm0.06$      \\
\hline 
\end{tabular}
\end{center}
\label{tab:vcbg1}
\end{table}

For a determination of \vcb, the form factor at zero recoil $G(1)$
needs to be computed.  
Using an unquenched lattice calculation \cite{Okamoto:2004xg}, corrected by a factor of 1.007 for QED effects, we obtain 

\begin{displaymath}
\vcb = (39.2 \pm 1.4_{\rm exp} \pm 0.9_{\rm theo}) \times 10^{-3},
\end{displaymath}

\noindent where the third error is due to the theoretical uncertainty in ${\cal G}(1)$. As an alternative, we use a quenched lattice calculation based on the Step Scaling Method (SSM)~\cite{Nazario}, and obtain 

\begin{displaymath}
\vcb = (40.9 \pm 1.5_{\rm exp} \pm 0.7_{\rm theo}) \times 10^{-3}.
\end{displaymath}

\begin{figure}[!ht]
 \begin{center}
  \unitlength1.0cm 
  \begin{picture}(14.,8.) 
   \put(  8.0, -0.2){\includegraphics[width=8.0cm]{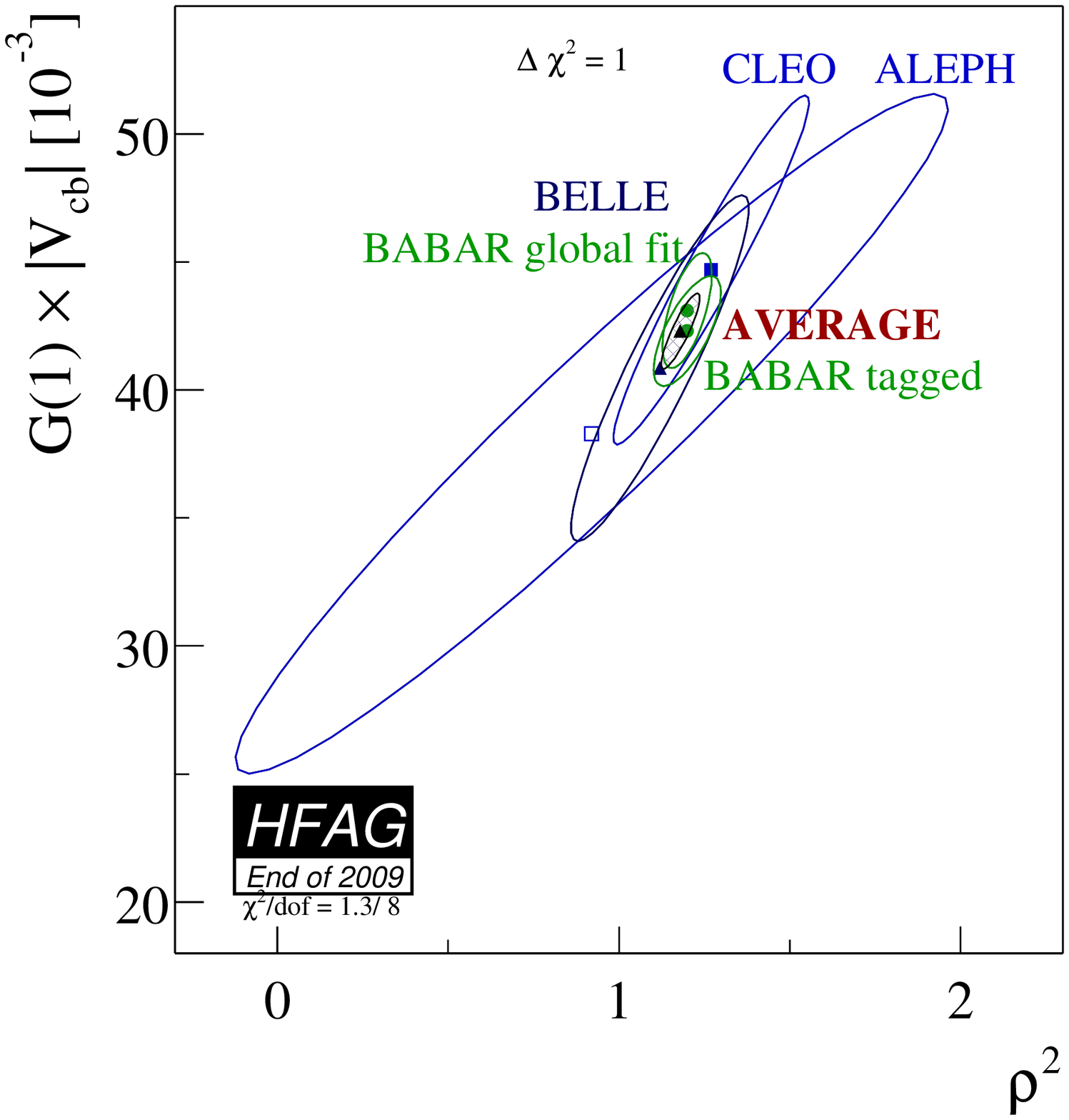}}
   \put( -0.5,  0.0){\includegraphics[width=7.5cm]{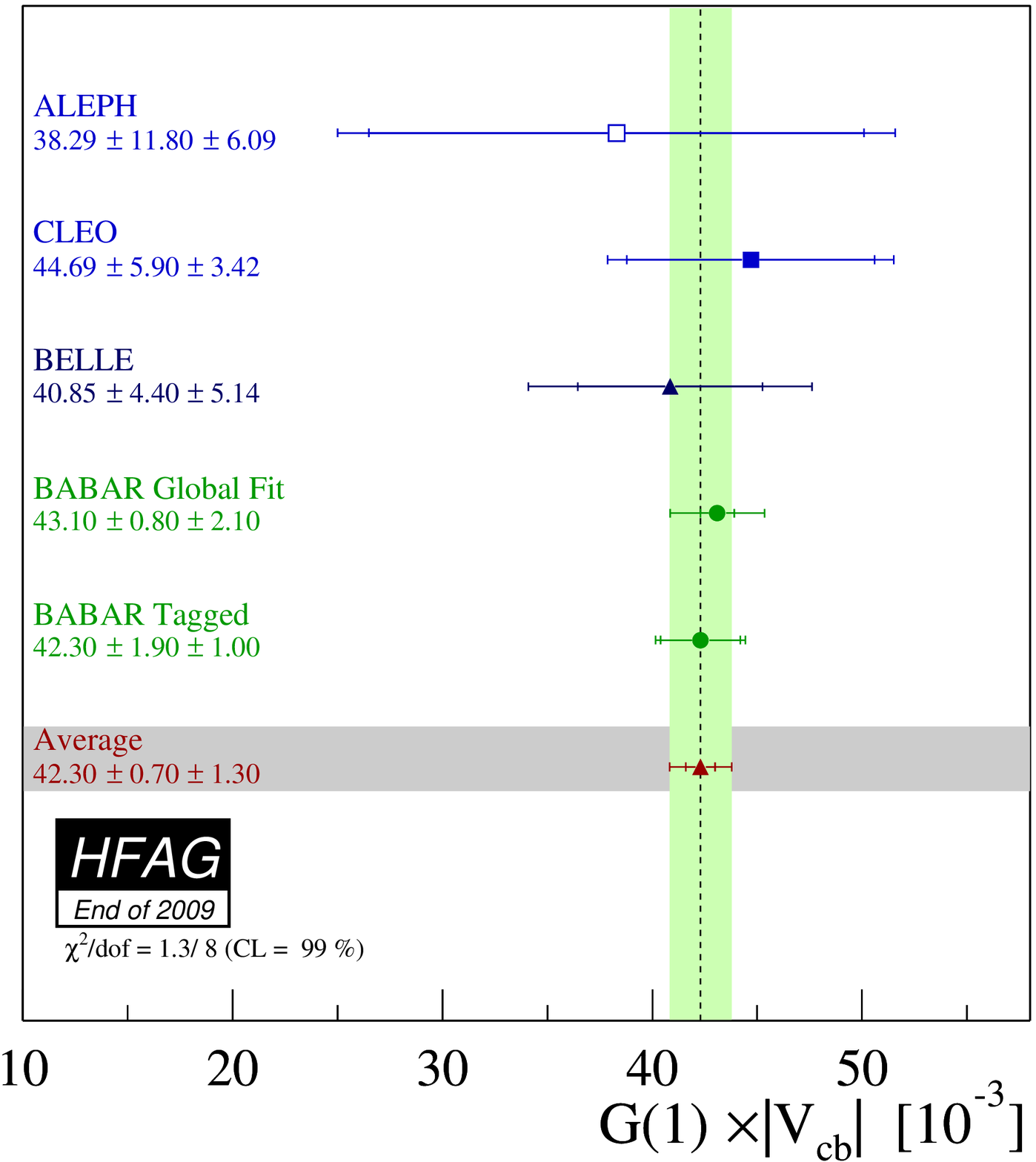}}
   \put(  5.5,  6.8){{\large\bf a)}}
   \put( 9.4,  6.8){{\large\bf b)}}
  \end{picture}
  \caption{(a) Illustration  of
   $G(1)\vcb$   vs.  $\rho^2$.  The   error  ellipses   correspond  to
   $\Delta\chi^2 = 1$. 
  (b) Illustration of the  average $G(1)\vcb$
   and   rescaled  measurements   of   exclusive  $\Bbar \to D\ell\bar{\nu}_{\ell}$   decays
   determined   in  a  two-dimensional   fit.   }
  \label{fig:vcbg1}
 \end{center}
\end{figure}

\mysubsubsection{$\bar{B} \to D^* \ell^-\bar{\nu}_{\ell}$}
\label{slbdecays_dstarlnu}

The average branching fraction $\cbf(\bar{B} \to D^{*} \ell^-\bar{\nu}_{\ell})$ 
is determined by the
combination of the results provided in Table~\ref{tab:dstarlnu} and~\ref{tab:dstar0lnu}, 
for $\bar{B}^0 \to D^{*+} \ell^-\bar{\nu}_{\ell}$ and 
$B^- \to D^{*0}\ell^-\bar{\nu}_{\ell}$, respectively.
Advances have also been made in the determination of $|V_{cb}|$
from exclusive $\bar{B} \to D^* \ell^- \bar{nu}_{\ell}$ 
decays with substantially improved
measurements of the form factor ratios $R_1$ and $R_2$.

\begin{table}[!htb]
\caption{Average branching fraction $\cbf(\BzbDstarlnu)$ and individual
  results, where ``excl'' and ``partial reco'' refer to full and
  partial reconstruction of the \BzbDstarlnu\ decay, respectively.}
\begin{center} 
\begin{tabular}{|l|c|c|}\hline
Experiment    &$\cbf(\BzbDstarlnu) [\%]$    (rescaled)     &$\cbf(\BzbDstarlnu) [\%]$  (published) \\
\hline\hline 
ALEPH (excl)~\hfill\cite{Buskulic:1996yq}
&$5.29\pm 0.24_{\rm stat} \pm 0.19_{\rm syst}$
&$5.53\pm 0.26_{\rm stat} \pm 0.52_{\rm syst}$ \\
OPAL (excl)~\hfill\cite{Abbiendi:2000hk}
&$5.10\pm 0.19_{\rm stat} \pm 0.37_{\rm syst}$
&$5.11\pm 0.20_{\rm stat} \pm 0.49_{\rm syst}$ \\
OPAL (partial reco)~\hfill\cite{Abbiendi:2000hk}
&$5.52\pm 0.26_{\rm stat} \pm 0.42_{\rm syst}$
&$5.92\pm 0.28_{\rm stat} \pm 0.68_{\rm syst}$ \\
DELPHI (partial reco)~\hfill\cite{Abreu:2001ic}
&$4.93\pm 0.15_{\rm stat} \pm 0.17_{\rm syst}$
&$4.70\pm 0.14_{\rm stat} \ {}^{+0.36}_{-0.31}\ {}_{\rm syst}$ \\
\belle  (excl)~\hfill\cite{Abe:2009New}
&$4.40\pm 0.03_{\rm stat} \pm 0.25_{\rm syst}$
&$4.42\pm 0.03_{\rm stat} \pm 0.25_{\rm syst}$ \\
CLEO  (excl)~\hfill\cite{Adam:2002uw}
&$5.65\pm 0.18_{\rm stat} \pm 0.19_{\rm syst}$
&$6.09\pm 0.19_{\rm stat} \pm 0.40_{\rm syst}$ \\
DELPHI (excl)~\hfill\cite{Abdallah:2004rz}
&$5.41\pm 0.18_{\rm stat} \pm 0.33_{\rm syst}$
&$5.90\pm 0.20_{\rm stat} \pm 0.50_{\rm syst}$ \\
\babar\ (excl)~\hfill\cite{Aubert:2006mb}
&$4.59\pm 0.04_{\rm stat}\pm 0.24_{\rm syst}$
&$4.69\pm 0.04_{\rm stat} \pm 0.34_{\rm syst}$ \\
\babar\ (tagged)~\hfill\cite{Aubert:vcbExcl}
&$5.40\pm 0.16_{\rm stat}\pm 0.25_{\rm syst}$
&$5.49\pm 0.16_{\rm stat} \pm 0.25_{\rm syst}$ \\
\hline 
{\bf Average}                              &\mathversion{bold}$5.05\pm0.12$     
     &\mathversion{bold}$\chi^2/\dof = 16.6/9$ (CL=$5.5\%$)  \\
\hline 
\end{tabular}
\end{center}
\label{tab:dstarlnu}
\end{table}

\begin{table}[!htb]
\caption{Average of the branching fraction $\cbf(B^-\to D^{*0}\ell^-\bar{\nu})$ 
and individual results. }
\begin{center}
%
\end{center}
\label{tab:dstar0lnu}
\end{table}

For the $\cbf(B^- \to D^{*0} \ell^-\bar{\nu}_{\ell})$, the average is performed
as for the $\bar{B} \to D \ell^- \bar{\nu}_{\ell}$ modes, by scaling the
different measurements to a common set of input parameters. 
For the $\cbf(\bar{B}^0 \to D^{*+} \ell^-\bar{\nu}_{\ell})$, the average 
is performed with a new method that combines all the information available from
the different experiments regarding the measurements of $|V_{cb}|$, the slope
parameter $\rho^2$ and the other form-factor parameters $R_1$ and $R_2$
A global $\chi^2$ is built incorporating all the inputs provided by each
experiment. 
The dependence of the ${\cal F} (1)|V_{cb}| - \rho^2$ only
measurements on the global values of $R_1$ and $R_2$ is explicitly included
through a Taylor expansion in $\Delta R_{1,2} = R_{1,2} -R^{nom}_{1,2}$ where 
$R^{nom}_{1,2}$ are some nominal values for the form-factor parameters.
Statistical correlations between measurements from the same experiment
are taken into account. The form-factor parametrization derived 
by Caprini, Lellouch and Neubert~\cite{CLN} is used. 

The $\chi^2$ minimization gives values for the form-factor parameters 
equal to $R1=1.410 \pm 0.049$ and $R_2=0.844 \pm 0.027$. 
 The errors contain both the common and the experiment dependent 
 systematic uncertainties. 

The values extracted from the fit for ${\cal F} (1)|V_{cb}|$ and the
form-factor parameters are used to obtain the 
$\cbf(\bar{B}^0 \to D^{*+} \ell^-\bar{\nu}_{\ell})$  
branching fractions by computing 
the integral over the measured differential decay rates. The $\cbf(\bar{B}^0 \to
D^{*+} \ell^-\bar{\nu}_{\ell})$ average is computed form these inputs, 
 apart for the \babar\ results~\cite{Aubert:vcbExcl}, 
 for which the semileptonic $B$ signal yields are
 extracted from a fit to the missing mass squared in a sample of fully
 reconstructed \BB\ events. This measurement is rescaled to the common set of
 input parameters, and then averaged with the other ones, neglecting at this
 stage remaining correlations.
Figure~\ref{fig:brdsl} illustrates the
measurements and the resulting average for the $\cbf(\bar{B} \to D^{*} \ell^-
\bar{\nu}_{\ell})$.  

The average for $F(1)\vcb$ is determined by the two-dimensional
combination of the results provided by the global $\chi^2$ minimization
described above: the corresponding values are reported 
in Table~\ref{tab:vcbf1}.  This
allows the correlation between $F(1)\vcb$ and $\rho^2$ to be
maintained. Figure~\ref{fig:vcbf1}(a) illustrates the average
$F(1)\vcb$ and the measurements included in the
average. Figure~\ref{fig:vcbf1}(b) provides a one-dimensional
projection for illustrative purposes. The largest 
systematic errors correlated between measurements are owing to
uncertainties on: $R_b$,
the ratio of production cross-sections
$\sigma_{b\bar{b}}/\sigma_{\rm{had}}$, the $B^0$ fraction at
$\sqrt{s}=m_{Z^0}$, the branching fractions $\cbf(D^0\to K^-\pi^+)$ and
$\cbf(D^0\to K^-\pi^+\pi^0)$, the correlated background from $D^{**}$,
and the $D^*$ form factor ratios $R_1$ and $R_2$.
Together these uncertainties account for about two thirds of the
systematic error. In all the measurements the total systematic errors are
reduced with respect to the published values because the values and
uncertainties assumed for parameters on which these measurements
depend, for example $R_1$ and $R_2$,  have since been better determined. 
 The $\chi^2/\dof = 38.7/23$
corresponds to a 2.1\% confidence level, suggesting some caution
in interpreting the errors on the average.


\begin{table}[!htb]
\caption{Average of $F(1)\vcb$ determined in the decay \BzbDstarlnu\ and
individual  results, where ``excl'' and ``partial reco'' refer to full and
  partial reconstruction of the \BzbDstarlnu\ decay, respectively.
The  fit  for  the average  has  $\chi^2/\dof  =
38.7/23$ (CL=$2.1\%$).  The total  correlation between  the average  $F(1)\vcb$ and
$\rho^2$ is 0.23.}
\begin{center}

\end{center}
\label{tab:vcbf1}
\end{table}

For a determination of \vcb, the form factor at zero recoil $F(1)$
needs to be computed.  A possible choice is
$F(1) = 0.921^{+0.013}_{-0.020}$~\cite{FNAL2009},
which, taking into account the QED correction($+0.7\%$), gives in 

\begin{displaymath}
\vcb = (38.9 \pm 0.6_{\rm exp} \pm 1.0_{\rm theo}) \times 10^{-3},
\end{displaymath}

\noindent where the errors are from experiment and theory, respectively.

\noindent

\begin{figure}[!ht]
 \begin{center}
  \unitlength1.0cm 
  \begin{picture}(14.,8.0)  
   \put( -0.5,  0.0){\includegraphics[width=7.5cm]{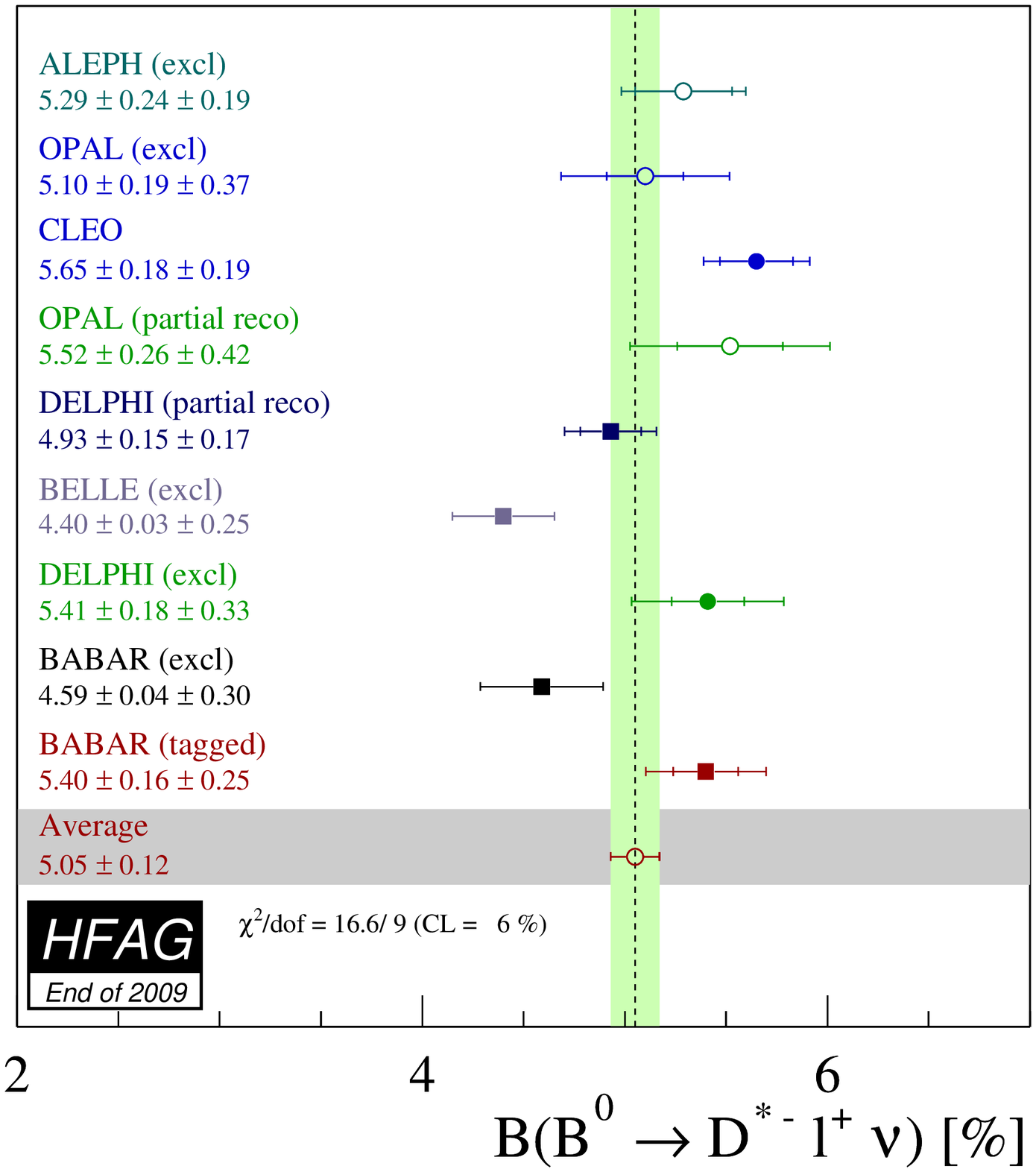}}

   \put(  8.0,  0.0){\includegraphics[width=7.5cm]{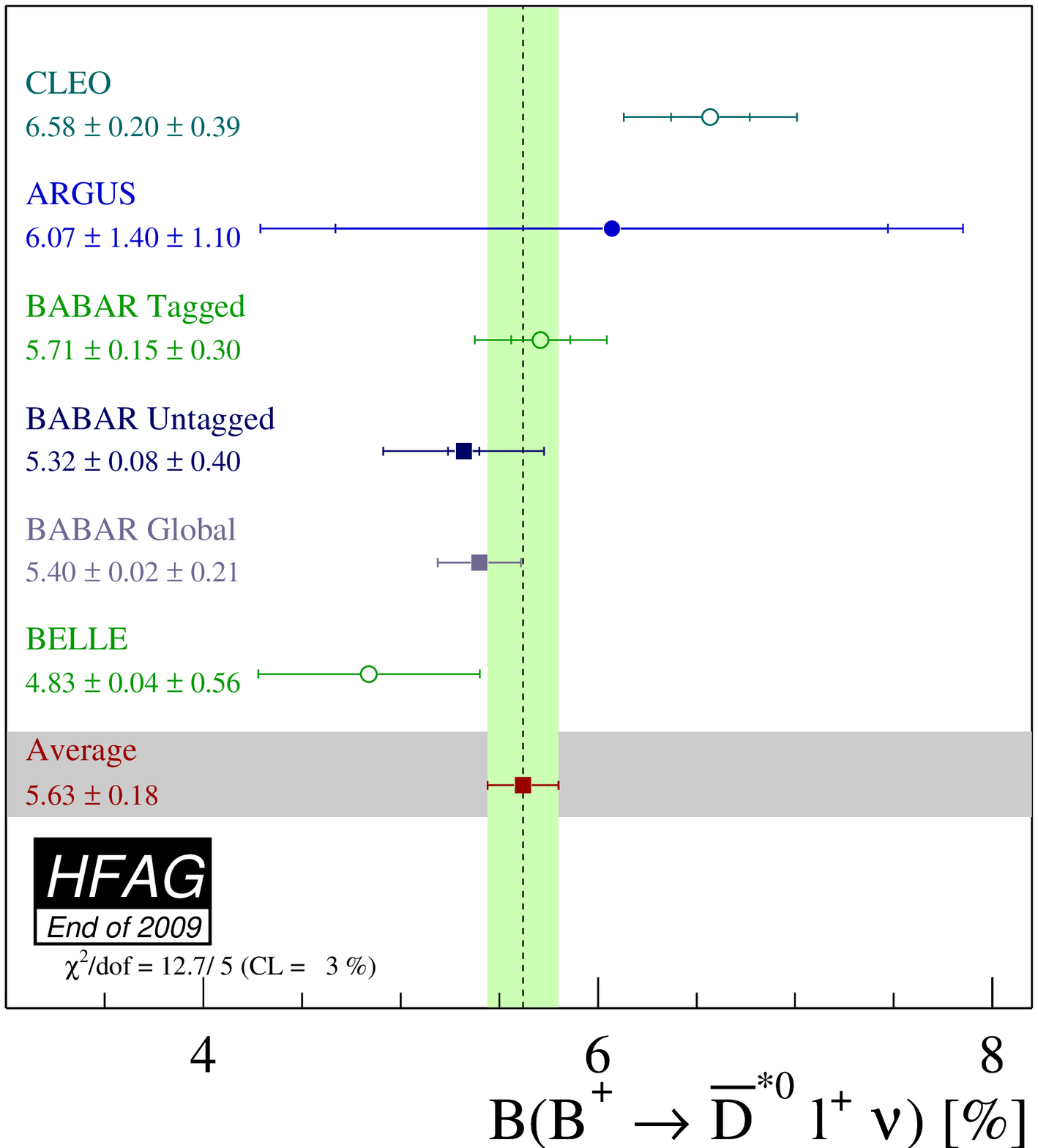}}
    
   \put(  5.5,  6.8){{\large\bf a)}}
   \put( 14.0,  6.8){{\large\bf b)}}
  \end{picture}
  \caption{Average branching fraction  of exclusive semileptonic $B$ decays
(a) $\bar{B}^0 \to D^{*+} \ell^- \bar{\nu}_{\ell}$ and (b) 
$B^- \to D^{*0} \ell^- \bar{\nu}_{\ell}$ and individual
  results.  At LEP, the measurements of \BzbDstarlnu\ decays have been done both
with inclusive (``partial reco'') and exclusive (``excl'') analyses based 
on a partial and full reconstruction of the
\BzbDstarlnu\ decay, respectively.}
  \label{fig:brdsl}
 \end{center}
\end{figure}

\begin{figure}[!ht]
 \begin{center}
  \unitlength1.0cm 
  \begin{picture}(14.,8.0)  
   \put(  8.0, -0.2){\includegraphics[width=8.0cm]{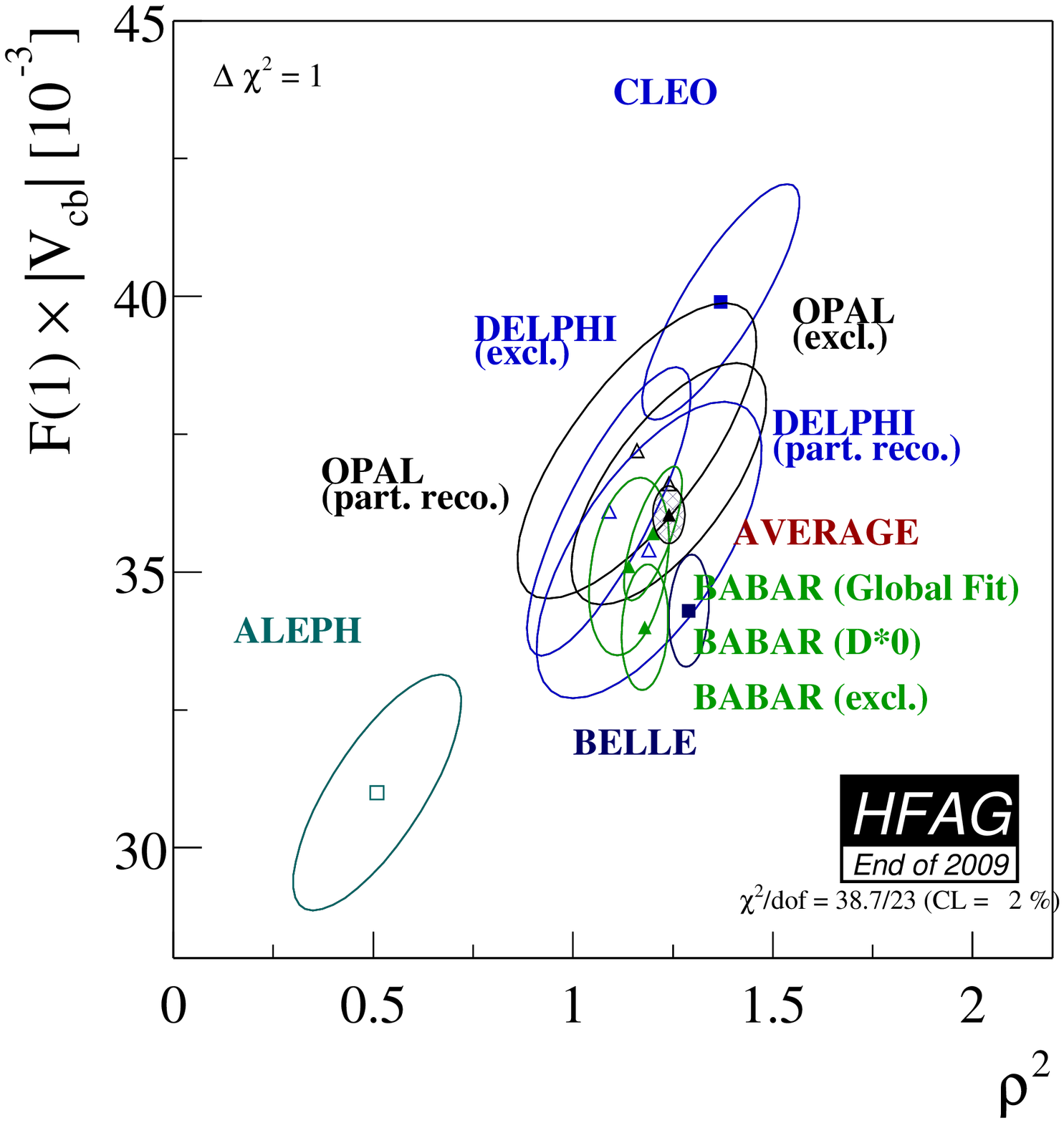}}
   \put( -0.5,  0.0){\includegraphics[width=7.5cm]{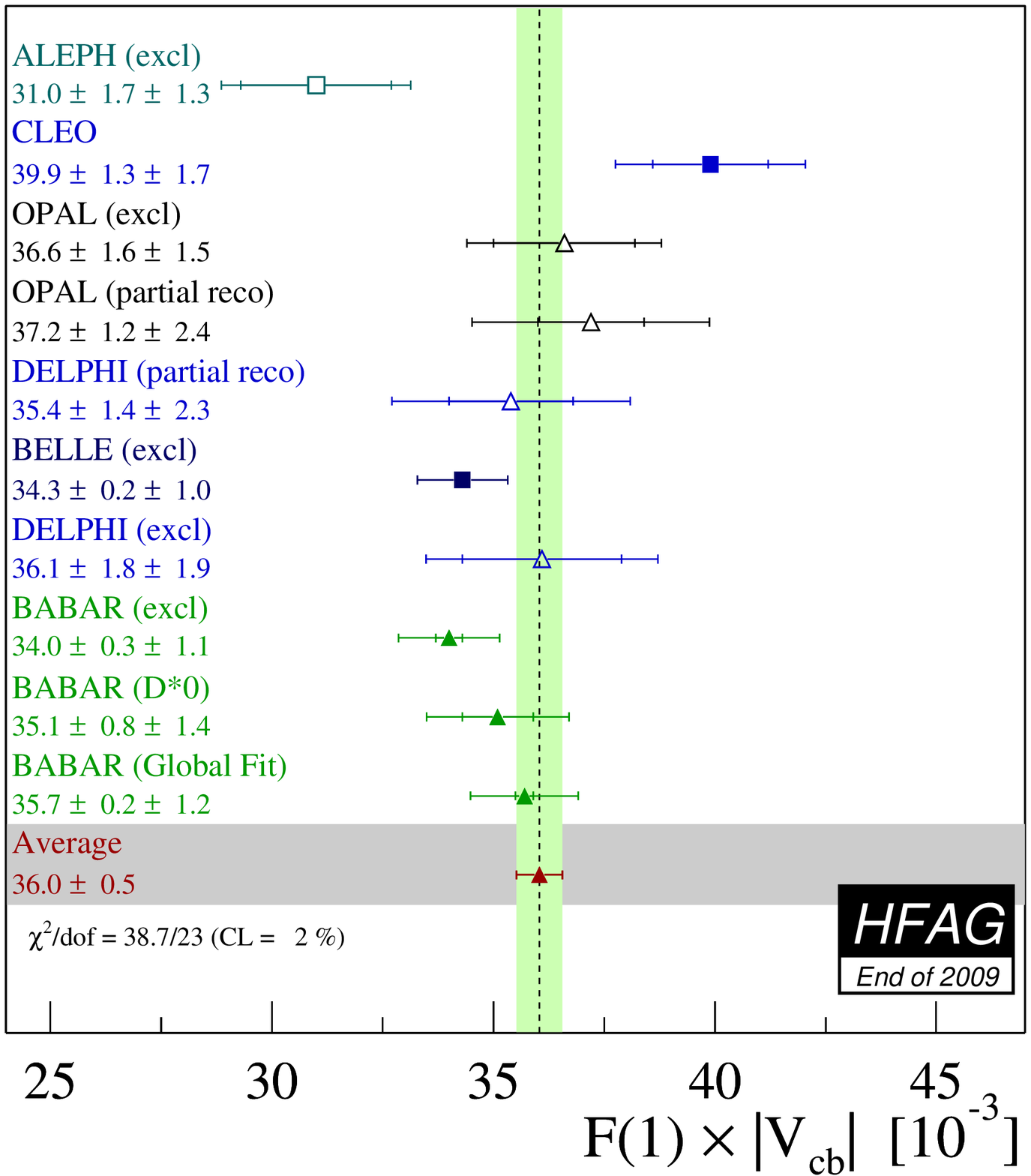}}
   \put(  5.5,  6.8){{\large\bf a)}}  
   \put( 14.4,  6.8){{\large\bf b)}}
   \end{picture} \caption{(a)   Illustration  of
   $F(1)\vcb$   vs.  $\rho^2$.  The   error  ellipses   correspond  to
   $\Delta\chi^2 = 1$ (CL=39\%).
   (b)  Illustration of the  average $F(1)\vcb$
   and   rescaled  measurements   of   exclusive  \BzbDstarlnu\   decays
   determined   in  a  two-dimensional   fit, where ``excl'' 
   and ``partial reco'' refer to full and partial reconstruction.
  }  \label{fig:vcbf1} \end{center}
\end{figure}

\mysubsubsection{$\bar{B} \to D^{(*)}\pi \ell^-\bar{\nu}_{\ell}$}
\label{slbdecays_dpilnu}

The average inclusive branching fractions for $\bar{B} \to D^{*}\pi \ell^-\bar{\nu}_{\ell}$
decays
, where no constrain is applied to the hadronic $D^{(*)}\pi$ system, 
are determined by the
combination of the results provided in Table~\ref{tab:dpi1lnu} -
\ref{tab:dpi4lnu}  for 
$\bar{B}^0 \to D^0 \pi^+ \ell^-\bar{\nu}_{\ell}$, $\bar{B}^0 \to D^{*0} \pi^+
\ell^-\bar{\nu}_{\ell}$, 
$B^- \to D^+ \pi^-
\ell^-\bar{\nu}_{\ell}$, and $B^- \to D^{*+} \pi^- \ell^-\bar{\nu}_{\ell}$.
The measurements included in the average 
are scaled to a consistent set of input
parameters and their errors~\cite{HFAG_sl:inputparams}.

For both the \babar\ and Belle results, the $B$ semileptonic signal yields are
 extracted from a fit to the missing mass squared in a sample of fully
 reconstructed \BB\ events. 
 
Figure~\ref{fig:brdpil} illustrates the measurements and the
resulting average.

\begin{table}[!htb]
\caption{Average of the branching fraction $\bar{B}^0 \to D^0 \pi^+ \ell^-\bar{\nu}_{\ell}$ and individual
results. }
\begin{center}

\end{center}
\label{tab:dpi1lnu}
\end{table}

\begin{table}[!htb]
\caption{Average of the branching fraction $\bar{B}^0 \to D^{*0} \pi^+
\ell^-\bar{\nu}_{\ell}$ and individual
results. }
\begin{center}
%
\end{center}
\label{tab:dpi2lnu}
\end{table}

\begin{table}[!htb]
\caption{Average of the branching fraction $B^- \to D^+ \pi^-
\ell^-\bar{\nu}_{\ell}$ and individual
results. }
\begin{center}
%
\end{center}
\label{tab:dpi3lnu}
\end{table}

\begin{table}[!htb]
\caption{Average of the branching fraction $B^- \to D^{*+} \pi^- \ell^-\bar{\nu}_{\ell}$ and individual
results. }
\begin{center}
%
  \caption{Average branching fraction  of exclusive semileptonic $B$ decays
(a) $\bar{B}^0 \to D^0 \pi^+ \ell^-\bar{\nu}_{\ell}$, (b) $\bar{B}^0 \to D^{*0} \pi^+
\ell^-\bar{\nu}_{\ell}$, (c) 
$B^- \to D^+ \pi^-
\ell^-\bar{\nu}_{\ell}$, and (d) $B^- \to D^{*+} \pi^- \ell^-\bar{\nu}_{\ell}$.
The corresponding individual
  results are also shown.}
  \label{fig:brdpil}
 \end{center}
\end{figure}

\mysubsubsection{$\bar{B} \to D^{**} \ell^-\bar{\nu}_{\ell}$}
\label{slbdecays_dsslnu}

The $D^{**}$ mesons contain one charm quark and one light quark with relative angular momentum $L=1$. According to Heavy Quark Symmetry (HQS)~\cite{IW}, they form one doublet of states with angular momentum $j \equiv s_q + L= 3/2$  $\left[D_1(2420), D_2^*(2460)\right]$ and another doublet with $j=1/2$ $\left[D^*_0(2400), D_1'(2430)\right]$, where $s_q$ is the light quark spin. Parity and angular momentum conservation constrain the decays allowed for each state. The $D_1$ and $D_2^*$ states decay through a D-wave to $D^*\pi$ and $D^{(*)}\pi$, respectively, and have small decay widths, while the $D_0^*$ and $D_1'$ 
states decay through an S-wave to $D\pi$ and $D^*\pi$ and are very broad.
For the narrow states, the average 
are determined by the
combination of the results provided in Table~\ref{tab:dss1lnu} and \ref{tab:dss2lnu} for 
$\cbf(B^- \to D_1^0(D^{*+}\pi^-)\ell^-\bar{\nu}_{\ell})
\times \cbf(D_1^0 \to D^{*+}\pi^-)$ and $\cbf(B^- \to D_2^0(D^{*+}\pi^-)\ell^-\bar{\nu}_{\ell})
\times \cbf(D_2^0 \to D^{*+}\pi^-)$. 
For the broad states, the average 
are determined by the
combination of the results provided in Table~\ref{tab:dss1plnu} and \ref{tab:dss0lnu} for 
$\cbf(B^- \to D_1'^0(D^{*+}\pi^-)\ell^-\bar{\nu}_{\ell})
\times \cbf(D_1'^0 \to D^{*+}\pi^-)$ and $\cbf(B^- \to D_0^{*0}(D^{+}\pi^-)\ell^-\bar{\nu}_{\ell})
\times \cbf(D_0^{*0} \to D^{+}\pi^-)$. 
The measurements included in the average 
are scaled to a consistent set of input
parameters and their errors~\cite{HFAG_sl:inputparams}.  

For both the B-factory and the LEP and Tevatron results, the $B$ semileptonic 
signal yields are
 extracted from a fit to the invariant mass distribution of the $D^{(*)+}\pi^-$ system.
 Apart for the CLEO and BELLE results, the other measurements 
 are for the final state $\bar{B} \to D_2(D^{*+}\pi^-)X \ell^- \bar{\nu}_{\ell}$. 
 We assume that no particle is left in the X system. 
Figure~\ref{fig:brdssl} and ~\ref{fig:brdssl2} illustrate the measurements and the
resulting average.

\begin{table}[!htb]
\caption{Average of the branching fraction $\cbf(B^- \to D_1^0(D^{*+}\pi^-)\ell^-\bar{\nu}_{\ell})
\times \cbf(D_1^0 \to D^{*+}\pi^-))$ and individual
results. }
\begin{center}

\end{center}
\label{tab:dss1lnu}
\end{table}

\begin{table}[!htb]
\caption{Average of the branching fraction $\cbf(B^- \to D_2^0(D^{*+}\pi^-)\ell^-\bar{\nu}_{\ell})
\times \cbf(D_2^0 \to D^{*+}\pi^-))$ and individual
results. }
\begin{center}
%
\end{center}
\label{tab:dss2lnu}
\end{table}

\begin{table}[!htb]
\caption{Average of the branching fraction $\cbf(B^- \to D_1^{'0}(D^{*+}\pi^-)\ell^-\bar{\nu}_{\ell})
\times \cbf(D_1^{'0} \to D^{*+}\pi^-))$ and individual
results. }
\begin{center}
%
\end{center}
\label{tab:dss1plnu}
\end{table}

\begin{table}[!htb]
\caption{Average of the branching fraction $\cbf(B^- \to D_0^{*0}(D^{+}\pi^-)\ell^-\bar{\nu}_{\ell})
\times \cbf(D_0^{*0} \to D^{+}\pi^-))$ and individual
results. }
\begin{center}
%
  \caption{Average of the product of branching fraction (a) 
  $\cbf(B^- \to D_1^0(D^{*+}\pi^-)\ell^-\bar{\nu}_{\ell})
\times \cbf(D_1^0 \to D^{*+}\pi^-)$ and (b) $\cbf(B^- \to D_2^0(D^{*+}\pi^-)\ell^-\bar{\nu}_{\ell})
\times \cbf(D_2^0 \to D^{*+}\pi^-)$
The corresponding individual
  results are also shown.}
  \label{fig:brdssl}
 \end{center}
\end{figure}

\begin{figure}[!ht]
 \begin{center}
  \unitlength1.0cm 
  %
  \caption{Average of the product of branching fraction (a) 
  $\cbf(B^- \to D_1'^0(D^{*+}\pi^-)\ell^-\bar{\nu}_{\ell})
\times \cbf(D_1'^0 \to D^{*+}\pi^-)$ and (b) $\cbf(B^- \to D_0^{*0}(D^{*+}\pi^-)\ell^-\bar{\nu}_{\ell})
\times \cbf(D_0^{*0} \to D^{+}\pi^-)$
The corresponding individual
  results are also shown.}
  \label{fig:brdssl2}
 \end{center}
\end{figure}

%
\subsection{Inclusive CKM-favored decays}
\label{slbdecays_b2cincl}

\subsubsection{Inclusive Semileptonic Branching Fraction}
\label{ref:Xlcnu}

In our previous update~\cite{Barberio:2008fa}, the branching fraction 
of inclusive semileptonic $B$~decays $B\to X\ell\nul$, where $B$ refers 
to both charged and neutral $B$~mesons, was averaged for a lepton momentum
threshold of 0.6~GeV/$c$, as measured in the rest frame of the
$B$~meson. A value of $(10.23\pm 0.15)\%$ was found with a
$\chi^2/$dof of the combination of $4.2/5$. Since no new measurements
have become available, we do not update this average and just refer to our
previous update~\cite{Barberio:2008fa}. This partial branching
fraction corresponds to an inclusive semileptonic branching fraction
of $(10.74\pm 0.16)\%$.

For the same reason, we do not update our averages of the ratio
$\cbf(\Bu\to X^0\ellp\nul)/\cbf(\Bz\to X^-\ellp\nul)$, of $\cbf(\Bu\to
X^0\ellp\nul)$, and of $\cbf(\Bz\to X^-\ellp\nul)$. For these
averages the reader is referred to our previous update~\cite{Barberio:2008fa}.

\subsubsection{Determination of \vcb}

The magnitude of the CKM matrix element \vcb\ can be determined from
inclusive semileptonic $B$~decays $B\to X_c\ell\nul$ using
calculations based on the Heavy Quark Effective Theory and the
Operator Production
Expansion~\cite{Benson:2003kp,Bauer:2004ve}. However, these
expressions depend also on non-perturbative parameters such as the
$b$-quark mass $m_b$ which can be determined from inclusive
observables in $B$~decays. In practice, \vcb\ and these parameters are
determined simultaneously from a global fit to measured moments of the
inclusive lepton and hadronic mass spectrum in semileptonic decays,
and of the inclusive photon spectrum in radiative $B$~meson penguin
decays. The moments are measured as a function of the minimum lepton or
photon energy.

Two independent sets of theoretical expressions, refered to as
kinetic~\cite{Benson:2003kp,Gambino:2004qm,Benson:2004sg} and 1S
schemes~\cite{Bauer:2004ve} are available for this kind of
analysis. The HFAG fit presented here is done in the kinetic scheme
and follows closely the approach of Ref.~\cite{Schwanda:2008kw}. The
fit is based on the experimental data given in
Table~\ref{tab:icln_1}. The only external input is the average
lifetime~$\tau_B$ of neutral and charged $B$~mesons, taken to be
$(1.582\pm 0.007)$~ps (Sect.~\ref{sec:life_mix}).
\begin{table}[!htb]
\caption{Experimental inputs used in the global fit analysis. $n$ is
the order of the moment, $c$ is the threshold value in GeV. In total,
there are 29 measurements from BaBar, 25 measurements from Belle and
12 from other experiments.} \label{tab:icln_1}
\begin{center}
\begin{small}
\begin{tabular}{l|lll}
  \hline
  Experiment & Hadron moments $\langle M^n_X\rangle$ & Lepton moments
  $\langle E^n_\ell\rangle$ & Photons moment $\langle
  E^n_\gamma\rangle$\\
  \hline
  BaBar & $n=2$, $c=0.9,1.1,1.3,1.5$ & $n=0$, $c=0.6,1.2,1.5$ & $n=1$,
  $c=1.9,2.0$\\
  & $n=4$, $c=0.8,1.0,1.2,1.4$ & $n=1$, $c=0.6,0.8,1.0,1.2,1.5$ & $n=2$,
  $c=1.9$~\cite{Aubert:2005cua,Aubert:2006gg}\\
  & $n=6$, $c=0.9,1.3$~\cite{Aubert:2009qda} & $n=2$, $c=0.6,1.0,1.5$
  & \\
  & & $n=3$, $c=0.8,1.2$~\cite{Aubert:2009qda,Aubert:2004td} & \\
  \hline
  Belle & $n=2$, $c=0.7,1.1,1.3,1.5$ & $n=0$, $c=0.6,1.0,1.4$ & $n=1$,
  $c=1.8,1.9$\\
  & $n=4$, $c=0.7,0.9,1.3$~\cite{Schwanda:2006nf} & $n=1$,
  $c=0.6,0.8,1.0,1.2,1.4$ & $n=2$, $c=1.8,2.0$~\cite{:2009qg}\\
  & & $n=2$, $c=0.6,1.0,1.4$ & \\
  & & $n=3$, $c=0.8,1.0, 1.2$~\cite{Urquijo:2006wd} & \\
  \hline
  CDF & $n=2$, $c=0.7$ & & \\
  & $n=4$, $c=0.7$~\cite{Acosta:2005qh} & & \\
  \hline
  CLEO & $n=2$, $c=1.0,1.5$ & & $n=1$, $c=2.0$~\cite{Chen:2001fja}\\
  & $n=4$, $c=1.0,1.5$~\cite{Csorna:2004kp} & & \\
  \hline
  DELPHI & $n=2$, $c=0.0$ & $n=1$, $c=0.0$ & \\
  & $n=4$, $c=0.0$~\cite{Abdallah:2005cx} & $n=2$, $c=0.0$ & \\
  & & $n=3$, $c=0.0$~\cite{Abdallah:2005cx} & \\
  \hline
\end{tabular}
\end{small}
\end{center}
\end{table}

\subsubsection{Global Fit in the Kinetic Scheme}
\label{globalfitsKinetic}

This fit relies on the calculations of the spectral moments in $B\to
X_c\ell\nul$~decays in the kinetic mass
scheme~\cite{Gambino:2004qm}. Compared to the original publication,
the expressions have been updated by the authors. For the moments in
$B\to X_s\gamma$, the (biased) OPE prediction and the bias correction
have been calculated~\cite{Benson:2004sg}. All these expressions
depend on the following set of parameters: the $b$- and $c$-quark
masses $m_b^\mathrm{kin}$ and $m_c^\mathrm{kin}$, $\mu^2_\pi$ and
$\mu^2_G$ at $\mathcal{O}(1/m^2_b)$ and $\rho^3_D$ and $\rho^3_{LS}$
at $\mathcal{O}(1/m^3_b)$\footnote{All non-perturbative parameters in
the kinetic scheme are defined at the scale $\mu=1$~GeV.}. In our
analysis, we determine these six parameters together with the
semileptonic branching fraction (over the full lepton energy range)
${\mathcal B}(B\to X_c\ell\nul)$. The total number of parameters in
the fit is thus seven. The conversion from ${\mathcal B}(B\to
X_c\ell\nul)$ to \vcb\ is done using the expression in
Ref.~\cite{Benson:2003kp}.

The results of the fit in the kinetic scheme to the $X_c\ell\nul$ and
$X_s\gamma$~data (Table~\ref{tab:icln_1}) are given in
Table~\ref{tab:icln_2}. For the semileptonic branching fraction we
obtain ${\mathcal B}(B\to X_c\ell\nul)=(10.55\pm 0.14)\%$. The
$\chi^2$ of the fit is 29.7 for $(66-7)$ degrees of freedom. The
predictions of the $B\to X_s\gamma$~moments are not entirely OPE-based
and involve some amount of modeling. Therefore, we have also
performed a fit to the $X_c\ell\nul$~data only,
Table~\ref{tab:icln_3}. The comparison of the
$\Delta\chi^2=1$~ellipses of these two fits in the
$(m_b^\mathrm{kin},\mu^2_\pi)$ and $(m_b^\mathrm{kin},\vcb)$~planes is
shown in Fig.~\ref{fig:icln_1}.
\begin{table}[!htb]
  \caption{Result of the kinetic scheme fit to all moments in
  Table~\ref{tab:icln_1}. The $\sigma(\mathrm{fit})$~error contains
  the experimental and theoretical uncertainties in the moments. The
  $\sigma(\tau_B)$ and $\sigma$(th)~errors on \vcb\ are due to the
  uncertainty in the average $B$~meson lifetime and the limited
  accuracy of the expression for \vcb~\cite{Benson:2003kp},
  respectively. In the lower part of the table, the correlation matrix
  of the parameters is given.} \label{tab:icln_2}
  \begin{center}
    \begin{tabular}{l|ccccccc}
      \hline
      & \vcb\ (10$^{-3}$) & $m_b^\mathrm{kin}$ (GeV) &
  $m_c^\mathrm{kin}$ (GeV) & $\mu^2_\pi$ (GeV$^2$) & $\rho^3_D$
  (GeV$^3$) & $\mu^2_G$ (GeV$^2$) & $\rho^3_{LS}$ (GeV$^3$)\\
      \hline
      value & 41.85 & \phantom{$-$}4.591 & \phantom{$-$}1.152 &
      \phantom{$-$}0.454 & \phantom{$-$}0.193 & \phantom{$-$}0.262 &
      $-$0.177\\
      $\sigma$(fit) & 0.42 & \phantom{$-$}0.031 &
      \phantom{$-$}0.046 & \phantom{$-$}0.038 & \phantom{$-$}0.020 &
      \phantom{$-$}0.044 & \phantom{$-$}0.085\\
      $\sigma(\tau_B)$ & 0.09 & & & & & & \\
      $\sigma$(th) & 0.59 & & & & & & \\
      \hline
      $|V_{cb}|$ & 1.000 & $-$0.169 & $-$0.024 & \phantom{$-$}0.118 &
      \phantom{$-$}0.297 & $-$0.248 & \phantom{$-$}0.109\\
      $m_b^\mathrm{kin}$ & & \phantom{$-$}1.000 & \phantom{$-$}0.926 &
      $-$0.405 & $-$0.126 & $-$0.021 & $-$0.266\\
      $m_c^\mathrm{kin}$ & & & \phantom{$-$}1.000 & $-$0.464 &
      $-$0.051 & $-$0.307 & $-$0.056\\
      $\mu^2_\pi$ & & & & \phantom{$-$}1.000 & \phantom{$-$}0.392 &
      $-$0.010 & $-$0.077\\
      $\rho^3_D$ & & & & & \phantom{$-$}1.000 & $-$0.236 & $-$0.318\\
      $\mu^2_G$ & & & & & & \phantom{$-$}1.000 & $-$0.203\\
      $\rho^3_{LS}$ & & & & & & & \phantom{$-$}1.000\\
      \hline
    \end{tabular}
  \end{center}
\end{table}
\begin{table}[!htb]
  \caption{Kinetic fit results for $B\to X_c\ell\nul$ and $B\to
    X_s\gamma$, and for $B\to X_c\ell\nul$ only.}
    \label{tab:icln_3}
  \begin{center}
    \begin{tabular}{c|cccc}
      \hline
      Data & $\chi^2/$dof & \vcb\ (10$^{-3}$) & $m_b^\mathrm{kin}$ (GeV) &
    $\mu^2_\pi$ (GeV$^2$)\\
      \hline
      All moments ($X_c\ell\nul$ and $X_s\gamma$) & $29.7/(66-7)$ &
    $41.85\pm 0.73$ & $4.591\pm 0.031$ & $0.454\pm 0.038$\\
      $X_c\ell\nul$ only & $24.2/(55-7)$ & $41.68\pm 0.74$ & $4.646\pm
    0.047$ & $0.439\pm 0.042$\\
      \hline
    \end{tabular}
  \end{center}
\end{table}
\begin{figure}
  \begin{center}
    \includegraphics[width=0.45\columnwidth]{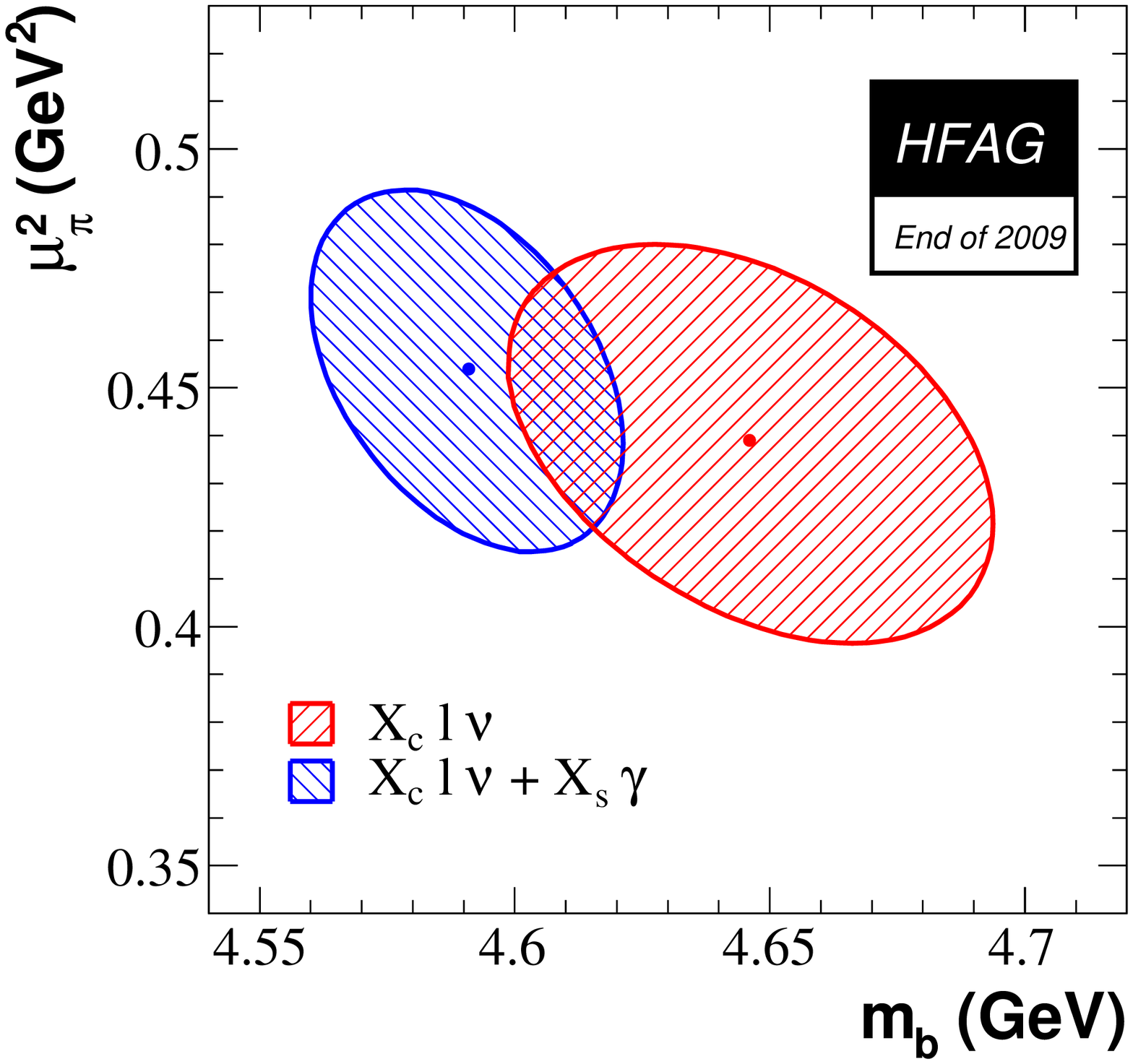}
    \includegraphics[width=0.45\columnwidth]{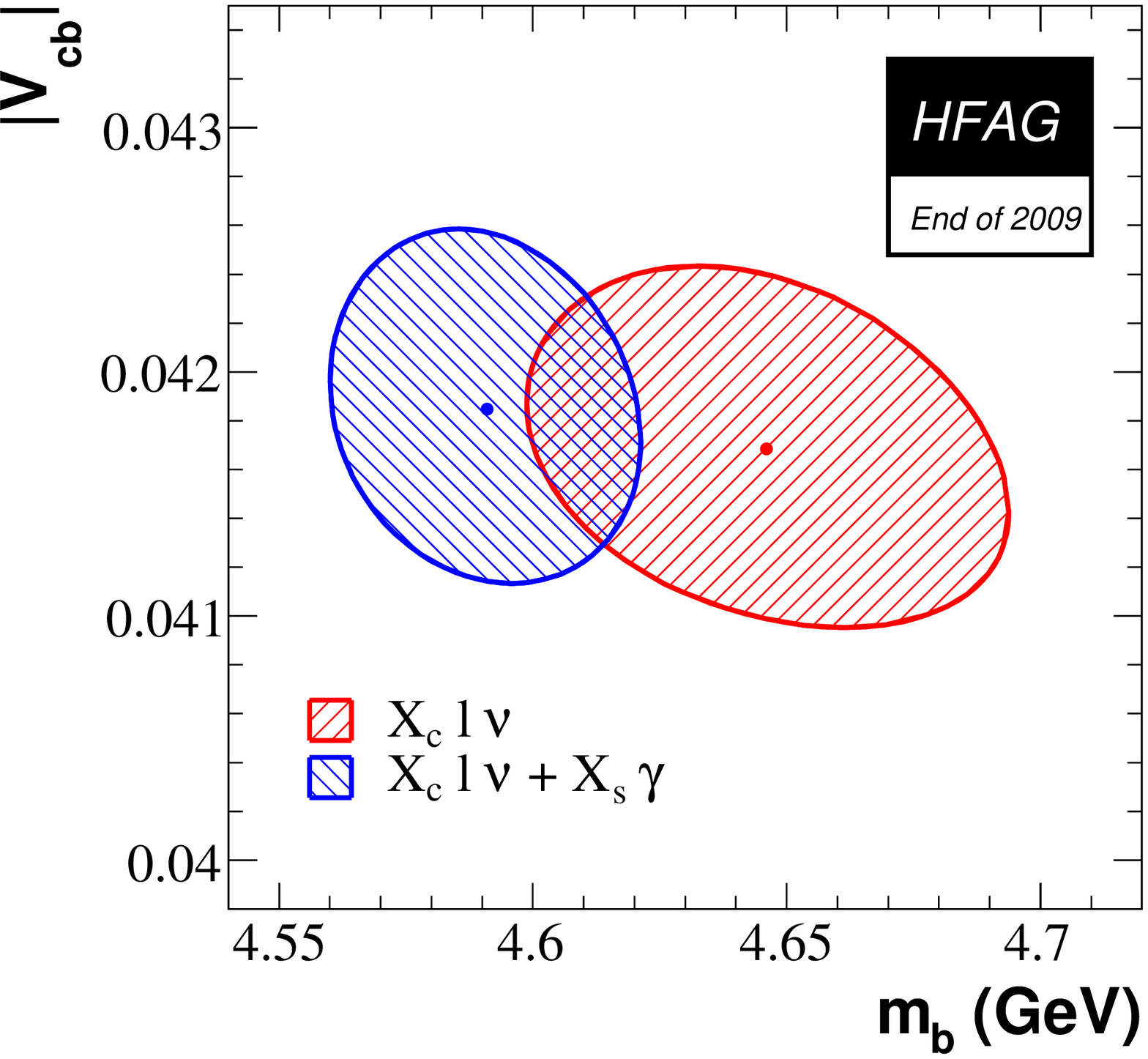}
  \end{center}
  \caption{$\Delta\chi^2=1$~contours for the fit in the kinetic mass
    scheme.} \label{fig:icln_1}
\end{figure}

\subsection{Exclusive CKM-suppressed decays}
\label{slbdecays_b2uexcl}
In this section, we list results on exclusive charmless semileptonic branching fractions
and determinations of $\vub$ based on $\Bb\to\pi\ell\nub$ decays.
The measurements are based on two different event selections: tagged
events, in which case the second $B$ meson in the event is fully
reconstructed in either a hadronic decay (``$B_{reco}$'') or in a 
CKM-favored semileptonic decay (``SL''); and untagged events, in which case the selection infers the momentum
of the undetected neutrino based on measurements of the total 
momentum sum of detected particles and knowledge of the initial state.
We present averages for $\Bb\to\rho\ell\nub$ and $\Bb\to\omega\ell\nub$. Moreover, the average for the  branching fraction $\Bb\to\eta\ell\nub$ is presented for the first time. 

The results for the full and partial branching fraction for $\Bb\to\pi\ell\nub$ are given
in Table~\ref{tab:pilnubf} and shown in Figure~\ref{fig:xlnu} (a).   

When averaging these results, systematic uncertainties due to external
inputs, e.g., form factor shapes and background estimates from the
modeling of $\Bb\to X_c\ell\nub$ and $\Bb\to X_u\ell\nub$ decays, are
treated as fully correlated (in the sense of Eq.~\ref{eq:correlrho}).
Uncertainties due to experimental reconstruction effects are treated
as fully correlated among measurements from a given experiment.  Varying
the assumed dependence of the quoted errors on the measured value
for error sources where the dependence was not obvious had no significant impact.

\begin{table}[!htb]
\begin{center}
\caption{\label{tab:pilnubf}
Summary of exclusive determinations of $\cbf(\Bb\to\pi
\ell\nub)$. The errors quoted
correspond to statistical and systematic uncertainties, respectively.
Measured branching fractions for $B\rightarrow \pi^0 l \nu$ have been
multiplied by $2\times \tau_{B^0}/\tau_{B^+}$ in accordance with
isospin symmetry. The labels ``$B_{reco}$'' and ``SL'' tags refer to
the type of $B$
decay tag used in a measurement, and ``untagged'' refers to an untagged measurement.}
\begin{small}
\begin{tabular}{|lccc|}
\hline
& $\cbf [10^{-4}]$
& $\cbf(q^2>16\,\gev^2/c^2) [10^{-4}]$
& $\cbf(q^2<16\,\gev^2/c^2) [10^{-4}]$
\\
\hline\hline
CLEO $\pi^+,\pi^0$~\cite{Adam:2007pv}
& $1.38\pm 0.15\pm 0.11\ $ 
& $0.41\pm 0.08\pm 0.04$
& $0.97\pm 0.13\pm 0.09$
\\ 
BABAR $\pi^+$~\cite{Aubert:2006px}
& $1.45\pm 0.07\pm 0.11\ $
& $0.38\pm 0.04\pm 0.05$
& $1.08\pm 0.06\pm 0.09$
\\  
BELLE SL $\pi^+$~\cite{Hokuue:2006nr}
& $1.38\pm 0.19\pm 0.15\ $
& $0.36\pm 0.10\pm 0.04$
& $1.02\pm 0.16\pm 0.11$
\\ 
BELLE SL $\pi^0$~\cite{Hokuue:2006nr}
& $1.43\pm 0.26\pm 0.15\ $
& $0.37\pm 0.15\pm 0.04$
& $1.05\pm 0.23\pm 0.11$
\\ 
BABAR SL $\pi^+$~\cite{:2008gka}
& $1.39\pm 0.21\pm 0.08\ $
& $0.46\pm 0.13\pm 0.03$
& $0.92\pm 0.16\pm 0.05$
\\ 
BABAR SL $\pi^0$~\cite{:2008gka}
& $1.80\pm 0.28\pm 0.15\ $
& $0.45\pm 0.17\pm 0.06$
& $1.38\pm 0.23\pm 0.11$
\\ 
BABAR $B_{reco}$ $\pi^+$~\cite{Aubert:2006ry}
& $1.07\pm 0.27\pm 0.19\ $
& $0.65\pm 0.20\pm 0.13$
& $0.42\pm 0.18\pm 0.06$
\\ 
BABAR $B_{reco}$ $\pi^0$~\cite{Aubert:2006ry}
& $1.54\pm 0.41 \pm0.30\ $
& $0.49\pm 0.23\pm 0.12$
& $1.05\pm 0.36\pm 0.19$
\\ 
BELLE $B_{reco}$ $\pi^+$~\cite{:2008kn}
& $1.12\pm 0.18\pm 0.05\ $
& $0.26\pm 0.08\pm 0.01$
& $0.85\pm 0.16\pm 0.04$
\\ 
BELLE $B_{reco}$ $\pi^0$~\cite{:2008kn}
& $1.24\pm 0.23\pm 0.05\ $
& $0.41\pm 0.11\pm 0.02$
& $0.85\pm 0.16\pm 0.04$
\\  \hline
{\bf Average}
& \mathversion{bold}$1.36\pm 0.05\pm 0.05\ $
& \mathversion{bold}$0.37\pm 0.02\pm 0.02$
& \mathversion{bold}$0.94\pm 0.05\pm 0.04$
\\ 
\hline
\end{tabular}\\
\end{small}
\end{center}
\end{table}

\begin{figure}[!ht]
 \begin{center}
  \unitlength1.0cm 
  \begin{picture}(14.,8.0)  
   \put(  8.0,  0.0){\includegraphics[width=8.0cm]{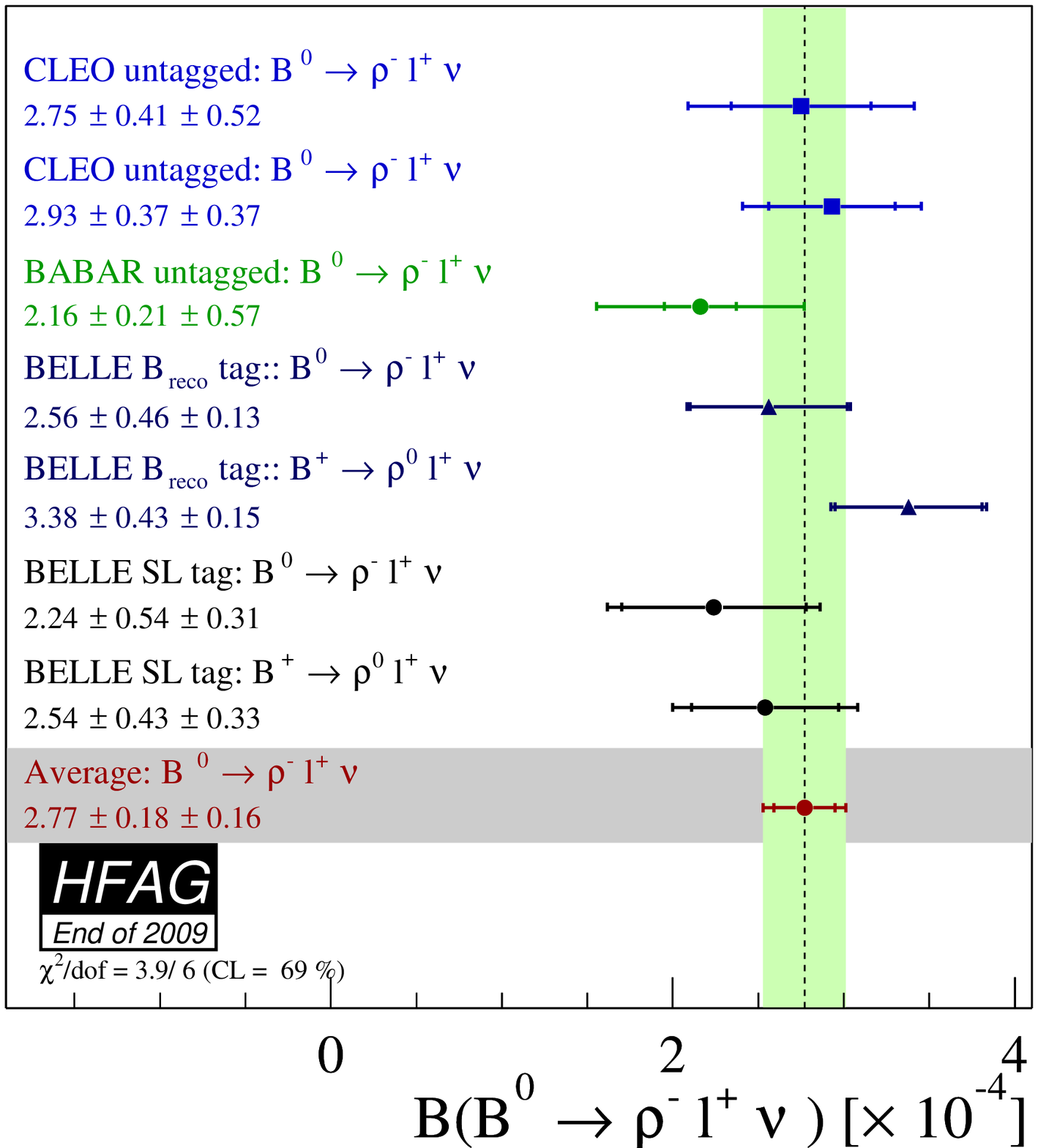}}
   \put( -0.5,  0.0){\includegraphics[width=8.0cm]{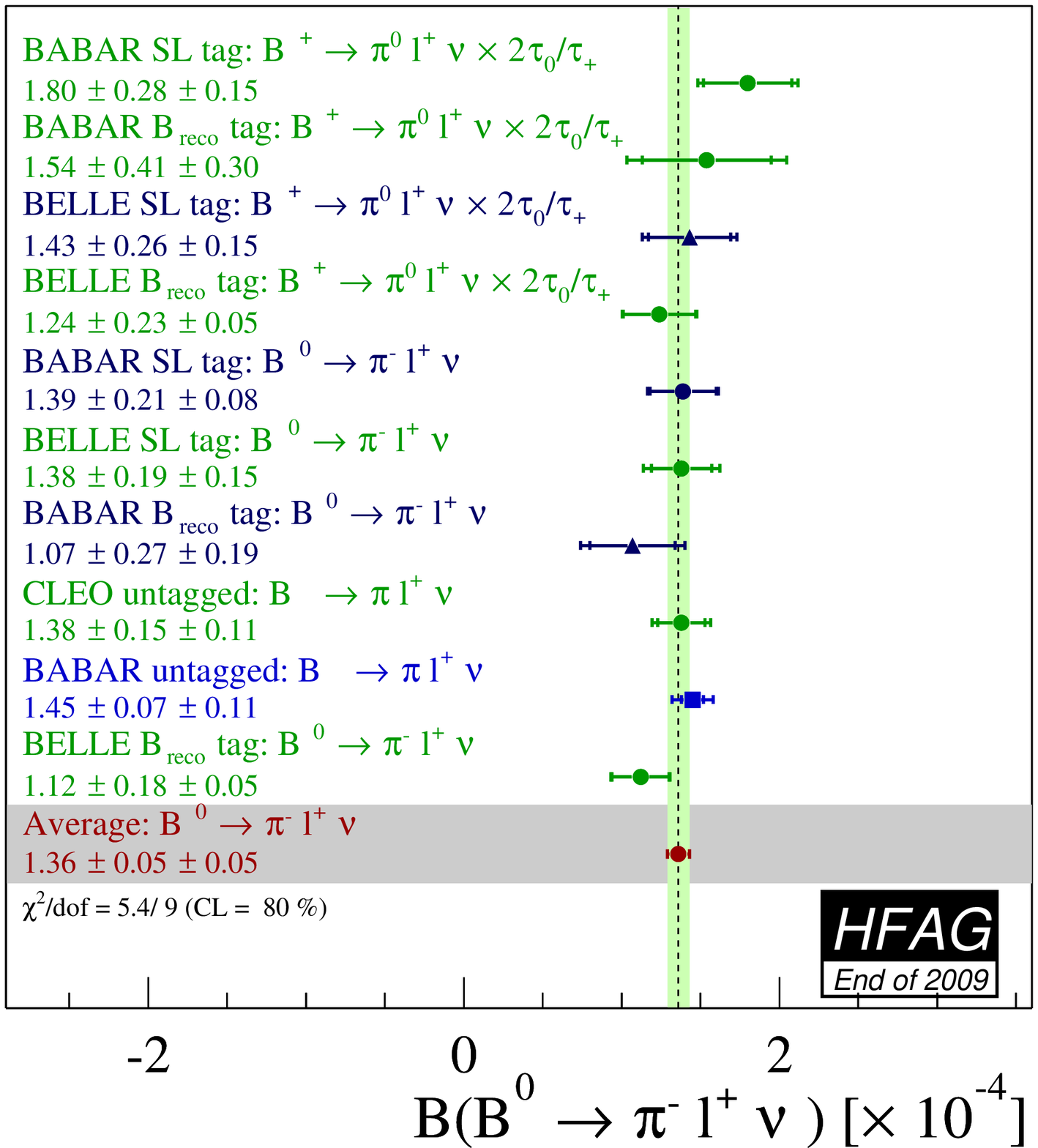}}
   \put(  5.5,  7.3){{\large\bf a)}}  
   \put( 14.4,  7.3){{\large\bf b)}}
   \end{picture} \caption{
(a) Summary of exclusive determinations of $\cbf(\Bb\to\pi
\ell\nub)$ and their average.
Measured branching fractions for $B\rightarrow \pi^0 l \nu$ have been
multiplied by $2\times \tau_{B^0}/\tau_{B^+}$ in accordance with
isospin symmetry. The labels ``$B_{reco}$'' and ``SL''
refer to type of $B$ decay tag used in a measurement. ``untagged'' refers to an untagged measurement.
(b) Summary of exclusive determinations of $\cbf(\Bb\to\rho\ell\nub)$ and their average.
}
\label{fig:xlnu}
\end{center}
\end{figure}

The determination of \vub\ from the $\Bb\to\pi\ell\nub$ decays is
shown in Table~\ref{tab:pilnuvub}, and uses our average for the branching
fraction given in Table~\ref{tab:pilnubf}. Two theoretical approaches are
used: Lattice QCD (quenched and unquenched) and QCD sum rules.
Lattice calculations of the Form Factors (FF) are limited to small hadron momenta, i.e.
large $q^2$, while calculations based on light cone sum rules are restricted
to small $q^2$. 

\begin{table}[hbtf]
\caption{\label{tab:pilnuvub}
Determinations of \vub\ based on the average total and partial
$\Bb\to\pi\ell\nub$ decay branching fraction stated in
Table~\ref{tab:pilnubf}. The first
uncertainty is experimental, and the second is from theory.  The
full or partial branching fractions are used as indicated. 
Acronyms for the calculations refer to either the method (LCSR) or
the collaboration working on it (HPQCD, FNAL, APE).
}
\begin{center}
\begin{tabular}{|lc|}
\hline
Method & $\Vub [10^{-3}]$ \\\hline\hline
LCSR, full $q^2$~\cite{Ball:2004ye} & $3.45 \pm 0.11 {}^{+0.67}_{-0.42}$ \\ 
LCSR, $q^2<16\,\gev^2/c^2$~\cite{Ball:2004ye}   & $3.34\pm 0.12 {}^{+0.55}_{-0.37}$ \\ \hline
HPQCD, full $q^2$~\cite{Dalgic:2006dt}& $3.05 \pm 0.10 {}^{+0.73}_{-0.43}$ \\ 
HPQCD, $q^2>16\,\gev^2/c^2$~\cite{Dalgic:2006dt}  & $3.40\pm 0.20 {}^{+0.59}_{-0.39}$ \\  \hline
FNAL, full $q^2$~\cite{Okamoto:2004xg}  & $3.73 \pm 0.12 {}^{+0.88}_{-0.52}$ \\ 
FNAL, $q^2>16\,\gev^2/c^2$~\cite{Okamoto:2004xg}    & $3.62\pm 0.22 {}^{+0.63}_{-0.41}$ \\  \hline
APE, full $q^2$~\cite{Abada:2000ty}  & $3.59\pm 0.11 {}^{+1.11}_{-0.57}$ \\ 
APE, $q^2>16\,\gev^2/c^2$~\cite{Abada:2000ty}    & $3.72\pm 0.21 {}^{+1.43}_{-0.66}$ \\ 
\hline
\end{tabular}
\end{center}
\end{table}

The branching fractions for 
$\Bb\to \rho\ell\nub$ decays is computed based on the measurements in
Table~\ref{tab:rholnu} and is shown in Figure~\ref{fig:xlnu} (b). The determination of $\Vub$
from these other channels looks less promising than for
$\Bb\to\pi\ell\nub$ and at the moment it is not extracted.

\begin{table}[!htb]
\begin{center}
\caption{Summary of exclusive determinations of $\cbf(\Bb\to\rho
\ell\nub)$. The errors quoted
correspond to statistical and systematic uncertainties, respectively.}
\begin{small}
\begin{tabular}{|lc|}
\hline
& $\cbf [10^{-4}]$
\\
\hline\hline
CLEO $\rho^+$~\cite{Behrens:1999vv}
& $2.75\pm 0.41\pm 0.52\ $ 
\\ 
CLEO $\rho^+$~\cite{Adam:2007pv}
& $2.93\pm 0.37\pm 0.37\ $ 
\\ 
BABAR $\rho^+$~\cite{Aubert:2005cd}
& $2.16\pm 0.21\pm 0.57\ $
\\
BELLE $\rho^+$~\cite{:2008kn}
& $2.56\pm 0.46\pm 0.13\ $
\\
BELLE $\rho^0$~\cite{:2008kn}
& $3.38\pm 0.43\pm 0.15\ $
\\
BELLE $\rho^+$~\cite{Hokuue:2006nr}
& $2.24\pm 0.54\pm 0.31\ $
\\
BELLE $\rho^0$~\cite{Hokuue:2006nr}
& $2.54\pm 0.43\pm 0.33\ $
\\  \hline
{\bf Average}
& \mathversion{bold}$2.77 \pm 0.18\pm 0.16\ $
\\ 
\hline
\end{tabular}\\
\end{small}
\end{center}
\label{tab:rholnu}
\end{table}

We also report the branching fraction average for $\Bb\to\omega\ell\nub$  and $\Bb\to\eta\ell\nub$. 
The measurements for $\Bb\to\omega\ell\nub$ are reported in Table and shown in Figure~\ref{fig:xlnu2}, while the ones for $\Bb\to\eta\ell\nub$ are reported in Table and shown in Figure~\ref{fig:xlnu2}. 

\begin{table}[!htb]
\begin{center}
\caption{Summary of exclusive determinations of $\cbf(\Bb\to\omega
\ell\nub)$. The errors quoted
correspond to statistical and systematic uncertainties, respectively.}
\begin{small}
\begin{tabular}{|lc|}
\hline
& $\cbf [10^{-4}]$
\\
\hline\hline
BELLE $\omega$~\cite{:2008kn}
& $1.19\pm 0.32\pm 0.06\ $
\\
BABAR $\omega$~\cite{Aubert:2008ct}
& $1.14\pm 0.16\pm 0.08\ $
\\  \hline
{\bf Average}
& \mathversion{bold}$1.15 \pm 0.16\ $
\\ 
\hline
\end{tabular}\\
\end{small}
\end{center}
\label{tab:omegalnu}
\end{table}

\begin{table}[!htb]
\begin{center}
\caption{Summary of exclusive determinations of $\cbf(\Bb\to\eta
\ell\nub)$. The errors quoted
correspond to statistical and systematic uncertainties, respectively.}
\begin{small}
\begin{tabular}{|lc|}
\hline
& $\cbf [10^{-4}]$
\\
\hline\hline
CLEO $\eta$~\cite{Gray:2007pw}
& $0.45\pm 0.23\pm 0.11\ $
\\
BABAR $\eta$~\cite{Aubert:2008ct}
& $0.31\pm 0.06\pm 0.08\ $
\\ 
BABAR $\eta$~\cite{:2008gka}
& $0.64\pm 0.20\pm 0.04\ $
\\  
 \hline
{\bf Average}
& \mathversion{bold}$0.38 \pm 0.09\ $
\\ 
\hline
\end{tabular}\\
\end{small}
\end{center}
\label{tab:etalnu}
\end{table}

\begin{figure}[!ht]
 \begin{center}
  \unitlength1.0cm 
  \begin{picture}(14.,8.0)  
   \put(  8.0,  0.0){\includegraphics[width=8.0cm]{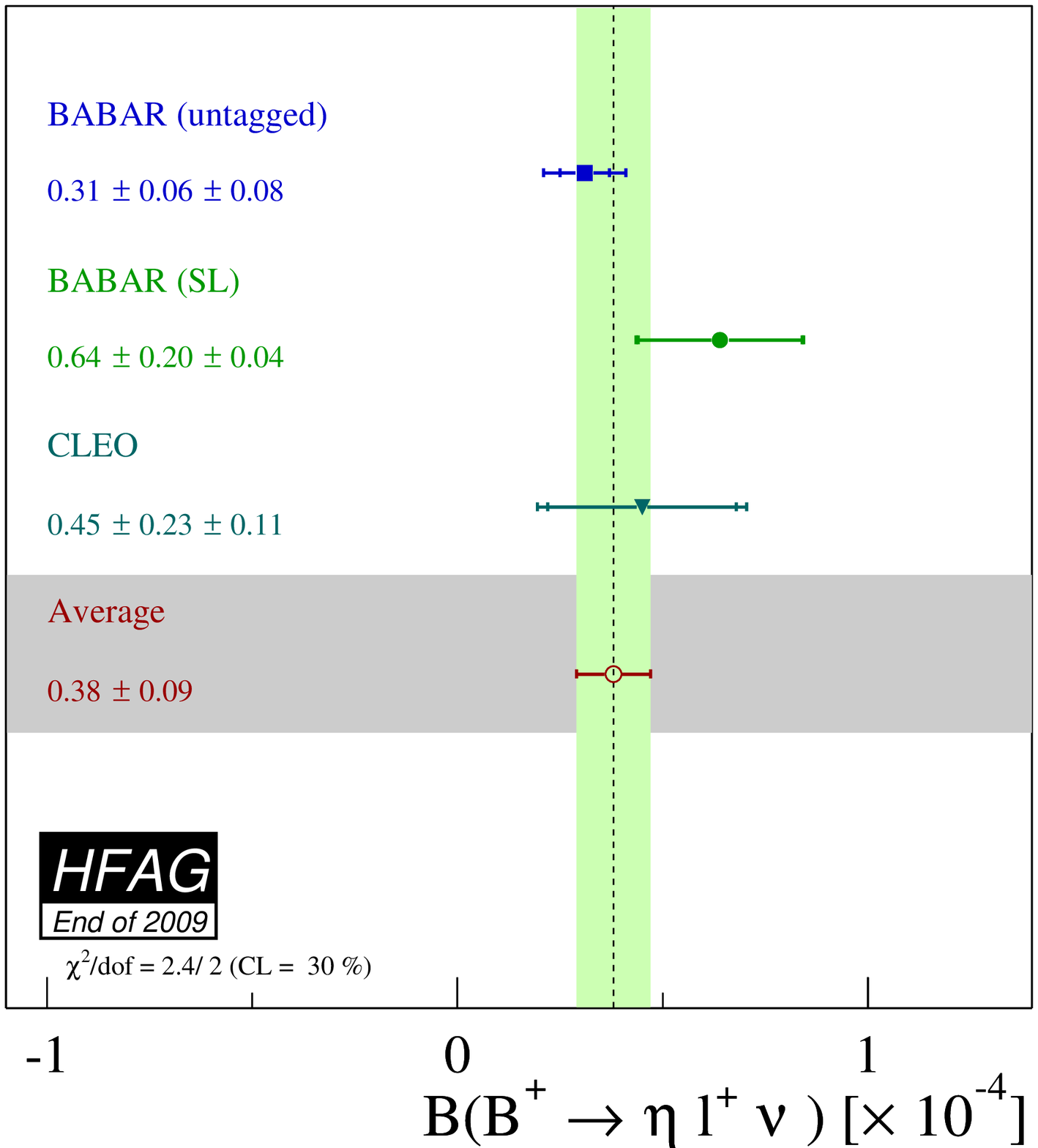}}
   \put( -0.5,  0.0){\includegraphics[width=8.0cm]{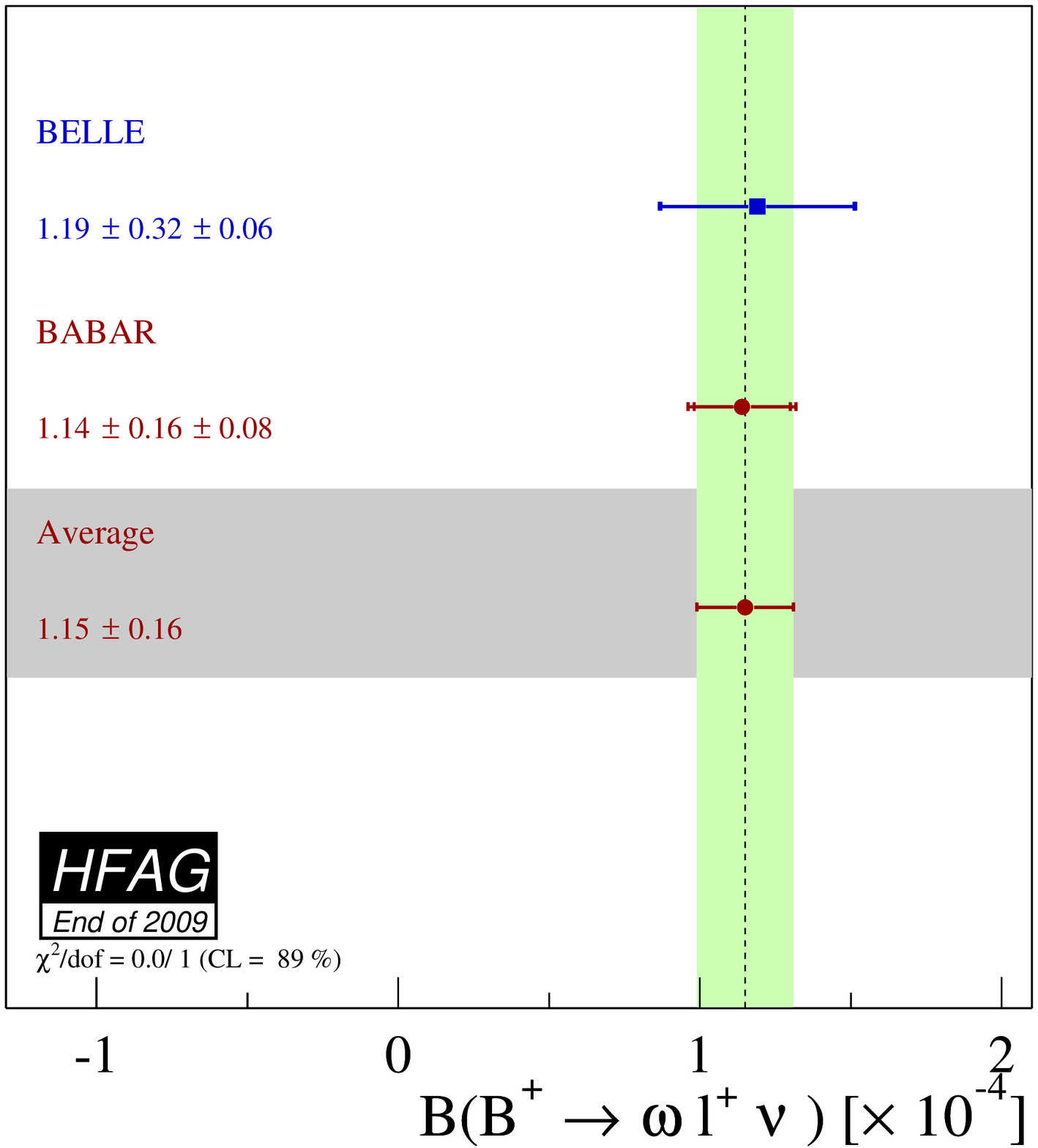}}
   \put(  5.5,  7.3){{\large\bf a)}}  
   \put( 14.4,  7.3){{\large\bf b)}}
   \end{picture} \caption{
(a) Summary of exclusive determinations of $\cbf(\Bb\to\omega\ell\nub)$ and their average.
(b) Summary of exclusive determinations of $\cbf(\Bb\to\eta\ell\nub)$ and their average.
}
\label{fig:xlnu2}
\end{center}
\end{figure}

Branching fractions for other $\Bb\to X_u\ell\nub$ decays are given in
Table~\ref{tab:xslother}. 

\begin{table}[!htb]
\caption{Summary of other branching fractions to $\cbf(\Bb\to X\ell\nub)$ decays
not included in the averages. The errors quoted
correspond to statistical and systematic
uncertainties, respectively.  Where a third uncertainty is quoted, it
corresponds to uncertainties from form factor shapes.}
\begin{center}
\begin{small}
\begin{tabular}{|llc|}
\hline
Experiment & Mode & $\cbf [10^{-4}]$  \\\hline\hline
CLEO~\cite{Athar:2003yg} & $\Bp\to\eta\ell\nub$ & $0.84\pm 0.31\pm 0.16 \pm 0.09$ \\ 
BABAR~\cite{Aubert:2006gba} & $\Bp\to\eta\ell\nub$ & $0.84\pm 0.27\pm 0.21$ \\ 
CLEO~\cite{Gray:2007pw} & $\Bp\to\eta^\prime\ell\nub$ & $2.66\pm 0.80\pm 0.56$ \\
BABAR~\cite{Aubert:2006gba} & $\Bp\to\eta^\prime\ell\nub$ & $0.33\pm 0.60\pm 0.30$ \\ 
BABAR~\cite{:2008gka} & $\Bp\to\eta^\prime\ell\nub$ & $<0.47$ @90 CL \\
\hline
\end{tabular}\\
\end{small}
\end{center}
\label{tab:xslother}
\end{table}


%
\subsection{Inclusive CKM-suppressed decays}
\label{slbdecays_b2uincl}
The large background from $\B\to X_c\ell^+\nul$ decays is the chief
experimental limitation in determinations of $\vub$.  Cuts designed to
reject this background limit the acceptance for $\B\to X_u\ell^+\nul$
decays. The calculation of partial rates for these restricted
acceptances is more complicated and requires substantial theoretical machinery.
In this update, we use several theoretical calculations
to extract \vub. We do not advocate the use of one method over another.
The authors for the different calculations have provided 
codes to compute
the partial rates in limited regions of phase space covered by the measurements.
A recent result by Belle~\cite{ref:belle-multivariate}, superceding the previous three 
measurements of \cite{ref:belle-mx}, selects a big portion of the phase space by using 
a multivariate technique to reject background, with a consequent reduction of the theoretical 
uncertainties. 

For the averages we performed, the systematic errors associated with the
modeling of $\B\to X_c\ell^+\nul$ and $\B\to X_u\ell^+\nul$ decays and the theoretical
uncertainties are taken as fully correlated among all measurements.
Reconstruction-related
uncertainties are taken as fully correlated within a given experiment.
We use all three results published by \babar\ in~\cite{ref:babar-mx}, since the 
statistical correlations are given. 
To make use of the theoretical calculations of Ref.~\cite{ref:BLL}, we restrict the
kinematic range in $M_X$ and $q^2$, thereby reducing the size of the data
sample significantly, but also the theoretical uncertainty, as stated by the
authors~\cite{ref:BLL}.
The dependence of the quoted error on the measured value for each source of error
is taken into account in the calculation of the averages.
Measurements of partial branching fractions for $\B\to X_u\ell^+\nul$
transitions from $\Upsilon(4S)$ decays, together with the corresponding accepted region, 
are given in Table~\ref{tab:BFbulnu}.  
The signal yields for all the measurements shown in Table~\ref{tab:BFbulnu}
are not rescaled to common input values of the $B$ meson 
lifetime~\cite{HFAG_sl:inputparams} and the semileptonic width~\cite{PDG_2010}.

It has been first suggested by Neubert~\cite{Neubert:1993um} and later detailed by Leibovich, 
Low, and Rothstein (LLR)~\cite{Leibovich:1999xf} and Lange, Neubert and Paz (LNP)~\cite{Lange:2005qn}, 
that the uncertainty of
the leading shape functions can be eliminated by comparing inclusive rates for
$\B\to X_u\ell^+\nul$ decays with the inclusive photon spectrum in $\B\to X_s\gamma$,
based on the assumption that the shape functions for transitions to light
quarks, $u$ or $s$, are the same to first order.
However, shape function uncertainties are only eliminated at the leading order
and they still enter via the signal models used for the determination of efficiency. 
For completeness, we provide a comparison of the results using 
calculations with reduced dependence on the shape function, as just
introduced, with our averages using different theoretical approaches.
Results are presented by \babar\ in Ref.\cite{Aubert:2006qi} using the LLR prescription. 
More recently, V.B.Golubev, V.G.Luth and Yu.I.Skovpen (Ref.~\cite{Golubev:2007cs})
extracted \vub\ from the 
endpoint spectrum of $\B\to X_u\ell^+\nul$ from \babar~\cite{ref:babar-endpoint}, 
using several theoretical approaches with reduced dependence on the shape function.
In both cases, the photon energy spectrum in the
rest frame of the $B$-meson by \babar~\cite{BABARSEMI} has been used.

\begin{table}[!htb]
\caption{\label{tab:BFbulnu}
Summary of inclusive determinations of partial branching
fractions for $B\rightarrow X_u \ell^+ \nu_{\ell}$ decays.
The errors quoted on $\Delta\cbf$ correspond to
statistical and systematic uncertainties.
The statistical correlations between the analysis are given where applicable. 
The $s_\mathrm{h}^{\mathrm{max}}$ variable is described in Refs.~\cite{ref:shmax,ref:babar-elq2}. }
\begin{center}
\begin{small}
\begin{tabular}{|llcp{5.5cm}|}
\hline
Measurement & Accepted region &  $\Delta\cbf [10^{-4}]$ & Notes\\
\hline\hline
CLEO~\cite{ref:cleo-endpoint}
& $E_e>2.1\,\gev$ & $3.3\pm 0.2\pm 0.7$ &  \\ 
\babar~\cite{ref:babar-elq2}
& $E_e>2.0\,\gev$, $s_\mathrm{h}^{\mathrm{max}}<3.5\,\mathrm{GeV^2}$ & $4.4\pm 0.4\pm 0.4$ & \\
\babar~\cite{ref:babar-endpoint}
& $E_e>2.0\,\gev$  & $5.7\pm 0.4\pm 0.5$ & \\
BELLE~\cite{ref:belle-endpoint}
& $E_e>1.9\,\gev$  & $8.5\pm 0.4\pm 1.5$ & \\
\babar~\cite{ref:babar-mx}
& $M_X<1.7\,\gev/c^2, q^2>8\,\gev^2/c^2$ & $7.7\pm 0.7\pm 0.7$ & 
\begin{minipage}[t]{\linewidth} 65\% correlation with \babar\ $M_X$ analysis \end{minipage} \\ 
BELLE~\cite{ref:belle-mxq2Anneal}
& $M_X<1.7\,\gev/c^2, q^2>8\,\gev^2/c^2$ & $7.4\pm 0.9\pm 1.3$ & \\
\babar~\cite{ref:babar-mx}
& $P_+<0.66\,\gev$  & $9.4\pm 0.9\pm 0.8 $ & 
\begin{minipage}[t]{\linewidth} 38\% correlation with \babar\ $(M_X-q^2)$ analysis \end{minipage} \\
\babar~\cite{ref:babar-mx}
& $M_X<1.55\,\gev/c^2$ & $11.7\pm 0.9\pm 0.7 $ & 
\begin{minipage}[t]{\linewidth} 67\% correlation with \babar\ $P_+$ analysis \end{minipage} \\
BELLE~\cite{ref:belle-multivariate}
& $p^*_{\ell} > 1 \gev/c$ & $19.6\pm 1.7\pm 1.6$ & \\ \hline
\end{tabular}\\
\end{small}
\end{center}
\end{table}

\subsubsection{BLNP}
Bosch, Lange, Neubert and Paz (BLNP)~\cite{ref:BLNP,
  ref:Neubert-new-1,ref:Neubert-new-2,ref:Neubert-new-3}
provide theoretical expressions for the triple
differential decay rate for $B\to X_u \ell^+ \nul$ events, incorporating all known
contributions, whilst smoothly interpolating between the 
``shape-function region'' of large hadronic
energy and small invariant mass, and the ``OPE region'' in which all
hadronic kinematical variables scale with the $b$-quark mass. BLNP assign
uncertainties to the $b$-quark mass which enters through the leading shape function, 
to sub-leading shape function forms, to possible weak annihilation
contribution, and to matching scales. The extracted values
of \vub\, for each measurement along with their average are given in
Table~\ref{tab:bulnu} and illustrated in Figure~\ref{fig:BLNP}.
The total uncertainty is $^{+6.2}_{-6.6}\%$ and is due to:
statistics ($^{+2.4}_{-2.6}\%$),
detector ($^{+1.9}_{-1.9}\%$),
$B\to X_c \ell^+ \nul$ model ($^{+0.9}_{-0.9}\%$),
$B\to X_u \ell^+ \nul$ model ($^{+1.7}_{-1.6}\%$),
heavy quark parameters ($^{+3.3}_{-3.8}\%$),
SF functional form ($^{+0.5}_{-0.5}\%$),
sub-leading shape functions ($^{+0.8}_{-0.8}\%$),
BLNP theory: matching scales $\mu,\mu_i,\mu_h$ ($^{+3.5}_{-3.5}\%$), and
weak annihilation ($^{+1.3}_{-1.2}\%$).
The error on the HQE parameters ($b$-quark mass and $\mu_\pi^2)$ 
and the uncertainty assigned for the matching scales are the dominant 
contribution to the total uncertainty. 

\begin{table}[!htb]
\caption{\label{tab:bulnu}
Summary of input parameters used by the different theory calculations,
corresponding inclusive determinations of $\vub$ and their average.
The errors quoted on \vub\ correspond to
experimental and theoretical uncertainties, respectively.}
\begin{center}
\begin{small}
\\
\end{small}
\end{center}
\end{table}

\begin{figure}
\begin{center}
\includegraphics[width=0.48\textwidth]{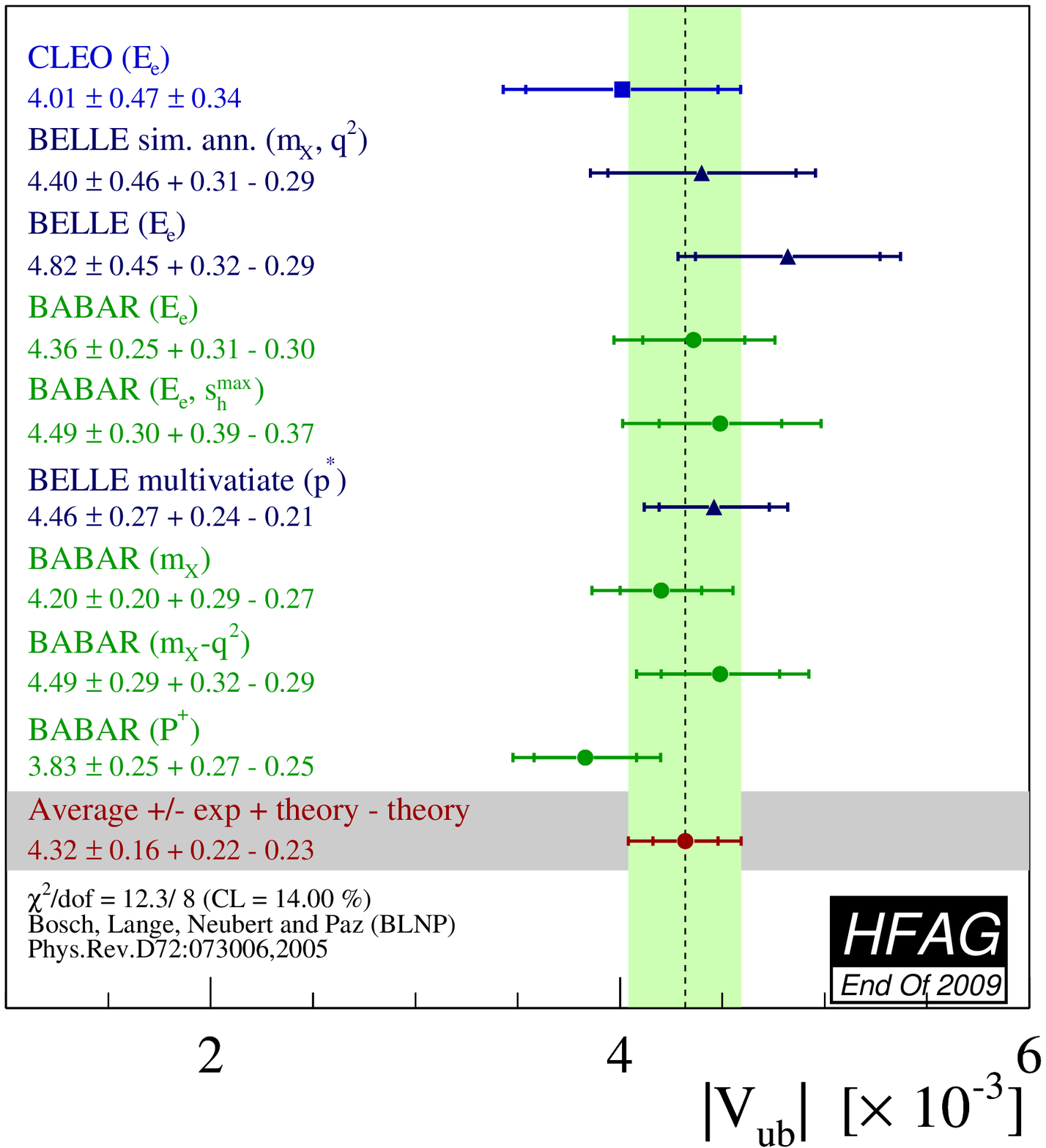}
\end{center}
\caption{Measurements of $\vub$ from inclusive semileptonic decays 
and their average based on the BLNP prescription.
``$E_e$'', ``$M_X$'', ``$(M_X,q^2)$'' and ``($E_e,s^{max}_h$)'' indicate the 
distributions and cuts used for the measurement of the partial decay rates.}
\label{fig:BLNP}
\end{figure}

\subsubsection{DGE}
J.R.~Andersen and E.~Gardi (Dressed Gluon Exponentiation, DGE)~\cite{ref:DGE} provide
a framework where the on-shell $b$-quark calculation, converted into hadronic variables, is
directly used as an approximation to the meson decay spectrum without
the use of a leading-power non-perturbative function (or, in other words,
a shape function). The on-shell mass of the $b$-quark within the $B$-meson ($m_b$) is
required as input. 
The extracted values
of \vub\, for each measurement along with their average are given in
Table~\ref{tab:bulnu} and illustrated in Figure~\ref{fig:DGE}.
The total error is $^{+5.4}_{-5.2}\%$, whose breakdown is:
statistics ($^{+2.4}_{-2.3}\%$),
detector ($^{+1.8}_{-1.8}\%$),
$B\to X_c \ell^+ \nul$ model ($^{+0.8}_{-0.9}\%$),
$B\to X_u \ell^+ \nul$ model ($^{+1.7}_{-1.6}\%$),
strong coupling $\alpha_s$ ($^{+0.5}_{-0.5}\%$),
$m_b$ ($^{+3.8}_{-3.6}\%$),
weak annihilation ($^{+1.3}_{-1.3}\%$),
DGE theory: matching scales ($^{+0.7}_{-0.7}\%$).
The largest contribution to the total error is due to the effect of the uncertainty 
on $m_b$  

\begin{figure}
\begin{center}
\includegraphics[width=0.48\textwidth]{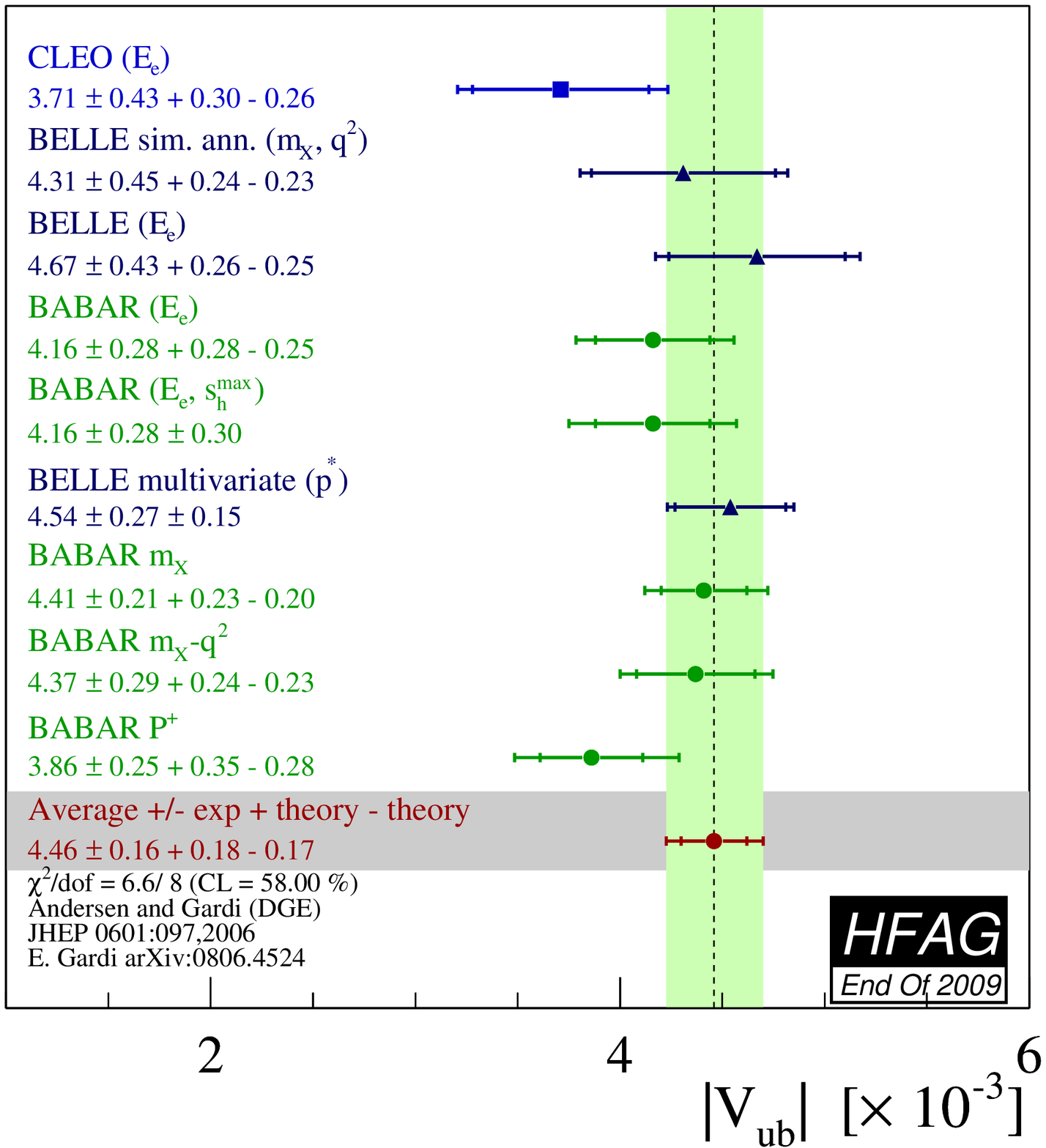}
\end{center}
\caption{Measurements of $\vub$ from inclusive semileptonic decays 
and their average based on the DGE prescription.
``$E_e$'', ``$M_X$'', ``$(M_X,q^2)$'' and ``($E_e,s^{max}_h$)'' indicate the 
analysis type.}
\label{fig:DGE}
\end{figure}

\subsubsection{GGOU}
Gambino, Giordano, Ossola and Uraltsev (GGOU)~\cite{Gambino:2007rp} 
compute the triple differential decay rates of $B \to X_u \ell^+ \nul$, 
including all perturbative and non--perturbative effects through $O(\alphas^2 \beta_0)$ 
and $O(1/m_b^3)$. 
The Fermi motion is parameterized in terms of a single light--cone function 
for each structure function and for any value of $q^2$, accounting for all subleading effects. 
The calculations are performed in the kinetic scheme, a framework characterized by a Wilsonian 
treatment with a hard cutoff $\mu \sim 1 $ GeV.
At present, GGOU have not included calculations for the ``($E_e,s^{max}_h$)'' analysis, but
this addition is planned.
The extracted values
of \vub\, for each measurement along with their average are given in
Table~\ref{tab:bulnu} and illustrated in Figure~\ref{fig:GGOU}.
The total error is $^{+4.9}_{-6.3}\%$ whose breakdown is:
statistics ($^{+2.3}_{-2.3}\%$),
detector ($^{+1.9}_{-1.9}\%$),
$B\to X_c \ell^+ \nul$ model ($^{+1.2}_{-1.2}\%$),
$B\to X_u \ell^+ \nul$ model ($^{+1.6}_{-1.6}\%$),
$\alpha_s$, $m_b$ and other non--perturbative parameters ($^{+2.5}_{-2.5}\%$), 
higher order perturbative and non--perturbative corrections ($^{+1.5}_{-1.5}\%$), 
modelling of the $q^2$ tail and choice of the scale $q^{2*}$ ($^{+1.7}_{-1.7}\%$), 
weak annihilations matrix element ($^{+0}_{-3.9}\%$), 
functional form of the distribution functions ($^{+0.5}_{-0.2}\%$), 
The leading uncertainties
on  \vub\ are both from theory, and are due to perturbative and non--perturbative
parameters and the modelling of the $q^2$ tail and choice of the scale $q^{2*}$. 

\begin{figure}
\begin{center}
\includegraphics[width=0.48\textwidth]{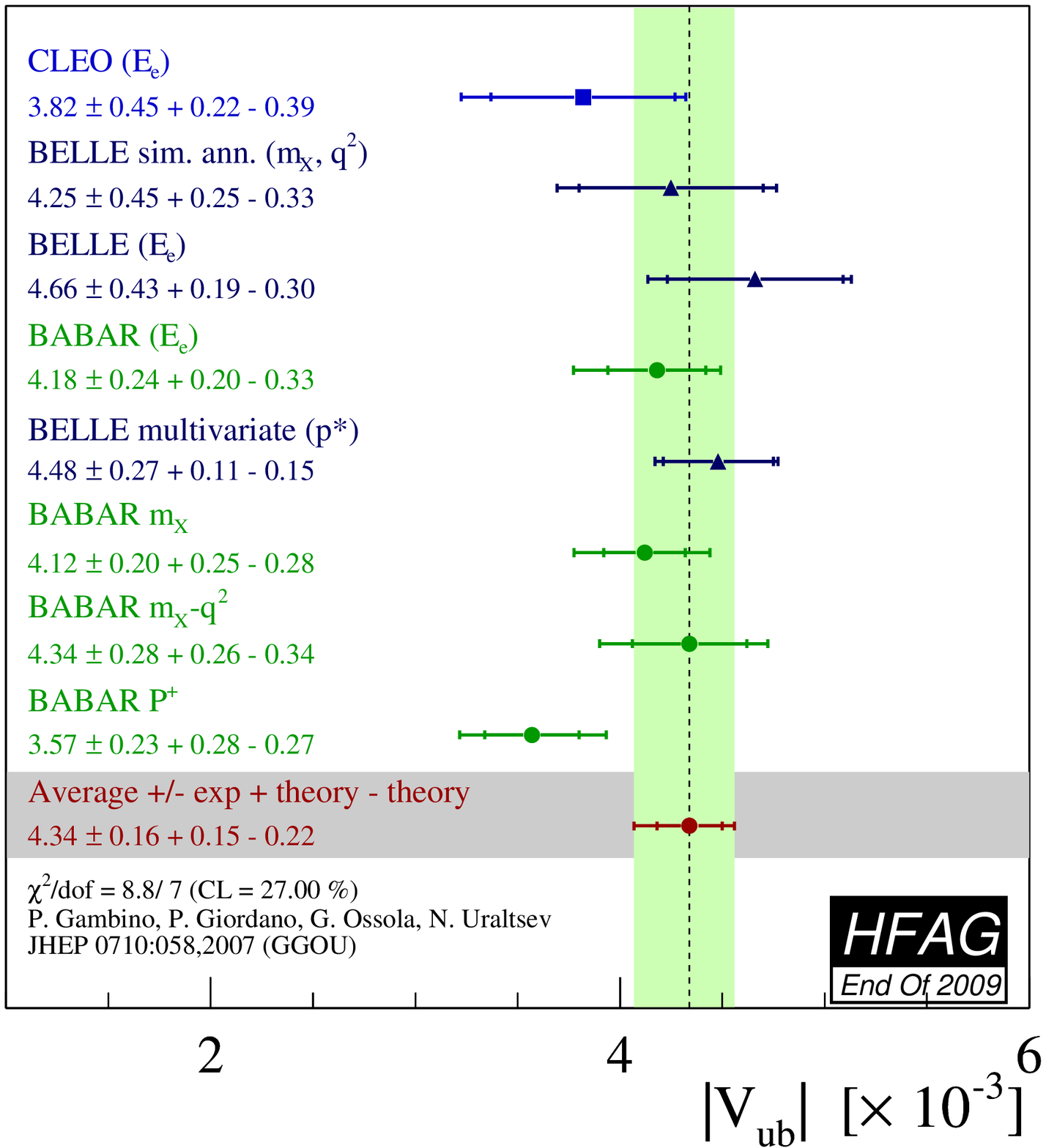}
\end{center}
\caption{Measurements of $\vub$ from inclusive semileptonic decays 
and their average based on the GGOU prescription.
``$E_e$'', ``$M_X$'', ``$(M_X,q^2)$'' and ``($E_e,s^{max}_h$)''  indicate the
analysis type.}
\label{fig:GGOU}
\end{figure}

\subsubsection{ADFR}
Aglietti, Di Lodovico, Ferrera and Ricciardi (ADFR)~\cite{Aglietti:2007ik}
use an approach to extract \vub, which makes use of the ratio
of the  $B \to X_c \ell^+ \nul$ and $B \to X_u \ell^+ \nul$ widths. 
The normalized triple differential decay rate for 
$B \to X_u \ell^+ \nul$~\cite{Aglietti:2006yb,Aglietti:2005mb, Aglietti:2005bm, Aglietti:2005eq}
is calculated with a model based on (i) soft--gluon resummation 
to next--to--next--leading order and (ii) an effective QCD coupling without
Landau pole. This coupling is constructed by means of an extrapolation to low
energy of the high--energy behaviour of the standard coupling. More technically,
an analyticity principle is used.
Following a recommendation by the ADFR authors, we lowered the cut on the electron energy for the endpoint analyses 
from 2.3~GeV to 2.1~GeV and recomputed the \vub\ values accordingly. 

The extracted values
of \vub\, for each measurement along with their average are given in
Table~\ref{tab:bulnu} and illustrated in Figure~\ref{fig:AC}.
The total error is $^{+6.8}_{-6.9}\%$ whose breakdown is:
statistics ($^{+1.9}_{-1.9}\%$),
detector ($^{+2.0}_{-2.0}\%$),
$B\to X_c \ell^+ \nul$ model ($^{+1.2}_{-1.2}\%$),
$B\to X_u \ell^+ \nul$ model ($^{+1.3}_{-1.3}\%$),
$\alpha_s$ ($^{+1.0}_{-1.6}\%$), 
$|V_{cb}|$ ($^{+1.7}_{-1.7}\%$), 
$m_b$ ($^{+0.7}_{-0.8}\%$), 
$m_c$ ($^{+4.5}_{-4.4}\%$), 
semileptonic branching fraction ($^{+0.9}_{-1.0}\%$), 
theory model ($^{+3.2}_{-3.2}\%$).
The leading
uncertainties, both from theory, are due to the $m_c$ mass and the theory model.

\begin{figure}
\begin{center}
\includegraphics[width=0.48\textwidth]{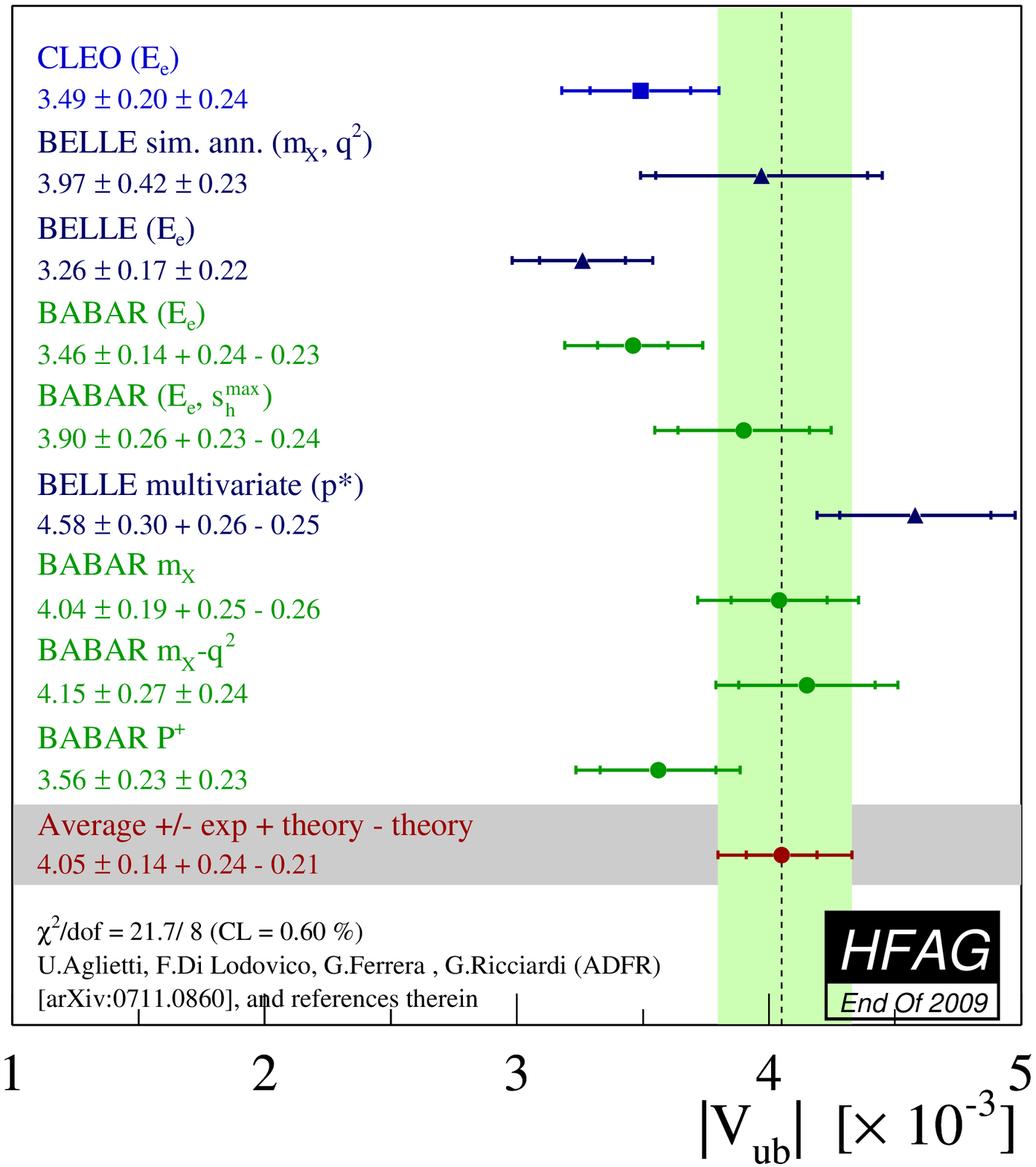}
\end{center}
\caption{Measurements of $\vub$ from inclusive semileptonic decays 
and their average based on the ADFR prescription.
``$E_e$'', ``$M_X$'', ``$(M_X,q^2)$'' and ``($E_e,s^{max}_h$)'' indicate the 
analysis type.}
\label{fig:AC}
\end{figure}

\subsubsection{BLL}
Bauer, Ligeti, and Luke (BLL)~\cite{ref:BLL} give a
HQET-based prescription that advocates combined cuts on the dilepton invariant mass, $q^2$,
and hadronic mass, $m_X$, to minimise the overall uncertainty on \vub.
In their reckoning a cut on $m_X$ only, although most efficient at
preserving phase space ($\sim$80\%), makes the calculation of the partial
rate untenable due to uncalculable corrections
to the $b$-quark distribution function or shape function. These corrections are
suppressed if events in the low $q^2$ region are removed. The cut combination used
in measurements is $M_x<1.7$ GeV/$c^2$ and $q^2 > 8$ GeV$^2$/$c^2$.  
The extracted values
of \vub\, for each measurement along with their average are given in
Table~\ref{tab:bulnu} and illustrated in Figure~\ref{fig:BLL}.
The total error is $^{+9.0}_{-9.0}\%$ whose breakdown is:
statistics ($^{+3.2}_{-3.2}\%$),
detector ($^{+3.8}_{-3.8}\%$),
$B\to X_c \ell^+ \nul$ model ($^{+1.3}_{-1.3}\%$),
$B\to X_u \ell^+ \nul$ model ($^{+2.1}_{-2.1}\%$),
spectral fraction ($m_b$) ($^{+3.0}_{-3.0}\%$),
perturbative : strong coupling $\alpha_s$ ($^{+3.0}_{-3.0}\%$),
residual shape function ($^{+4.5}_{-4.5}\%$),
third order terms in the OPE ($^{+4.0}_{-4.0}\%$),
The leading
uncertainties, both from theory, are due to residual shape function
effects and third order terms in the OPE expansion. The leading
experimental uncertainty is due to statistics. 

\begin{figure}
\begin{center}
\includegraphics[width=0.48\textwidth]{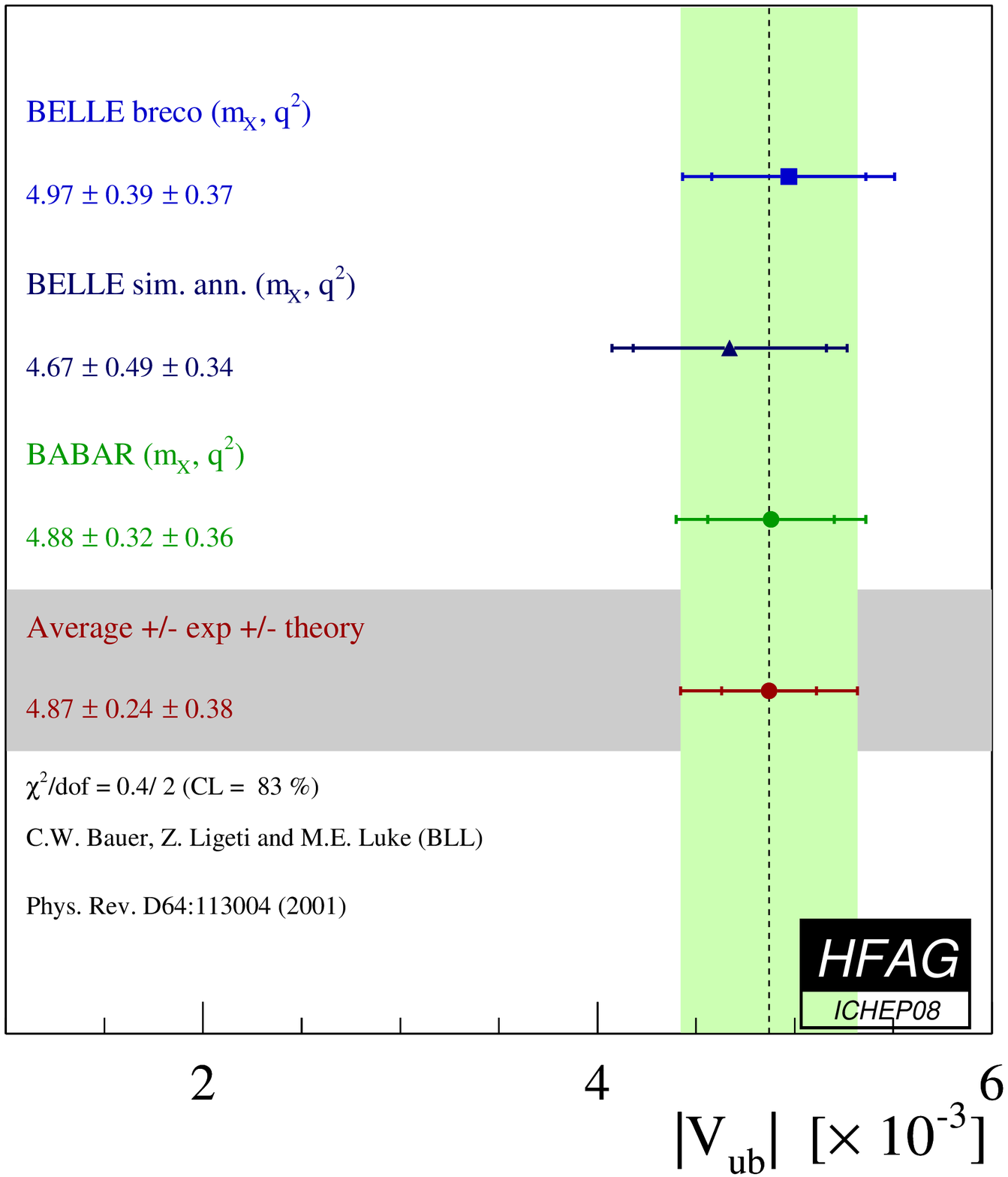}
\end{center}
\caption{Measurements of $\vub$ from inclusive semileptonic decays 
and their average in the BLL prescription.
``$(M_X, q^2)$'' indicates the analysis type.}
\label{fig:BLL}
\end{figure}

\subsubsection{Summary}
A summary of the averages presented in several different
frameworks and results by V.B.Golubev, V.G.Luth and Yu.I.Skovpen~\cite{Golubev:2007cs},
based on prescriptions by LLR~\cite{Leibovich:1999xf} and LNP~\cite{Lange:2005qn} 
to reduce the leading shape function uncertainties are presented in 
Table~\ref{tab:vubcomparison}.
It is difficult to quote a preferred value:
the experimental and theoretical uncertainties play out
differently among the schemes, and the theoretical assumptions 
underlying the theory calculations are different.

\begin{table}[!htb]
\caption{\label{tab:vubcomparison}
Summary of inclusive determinations of $\vub$.
The errors quoted on \vub\ correspond to experimental and theoretical uncertainties, except for the last two 
measurements where the errors are due to the \babar\ endpoint analysis, the \babar $b\to s\gamma$ analysis~\cite{Aubert:2006qi}, 
the theoretical errors and $V_{ts}$ for the last averages. 
}
\begin{center}
\begin{small}
\begin{tabular}{|lc|}
\hline
Framework
&  $\Vub [10^{-3}]$\\
\hline\hline
BLNP
& $4.32 \pm 0.16 ^{+0.22}_{-0.23}$ \\ 
DGE
& $4.46 \pm 0.16 ^{+0.18}_{-0.17}$ \\
GGOU
& $4.34 \pm 0.16 ^{+0.15}_{-0.22}$ \\
ADFR
& $4.16 \pm 0.14 ^{+0.25}_{-0.22}$ \\
BLL ($m_X/q^2$ only)
& $4.87 \pm 0.24 \pm 0.38$ \\ 
LLR (\babar)~\cite{Aubert:2006qi}
& $4.43 \pm 0.45 \pm 0.29$ \\
LLR (\babar)~\cite{Golubev:2007cs}
& $4.28 \pm 0.29 \pm 0.29 \pm 0.26 \pm0.28$ \\
LNP (\babar)~\cite{Golubev:2007cs}
& $4.40 \pm 0.30 \pm 0.41 \pm 0.23$ \\
\hline
\end{tabular}\\
\end{small}
\end{center}
\end{table}

%
%

\clearpage
\mysection{$B$ decays to charmed hadrons}
\label{sec:BtoCharm}


\definecolor{hfviolet}{rgb}{0.5,0,0.5}
\definecolor{hflightcyan}{rgb}{0.1,1.0,1.0}

\def\hflmode{Mode}
\def\hflpdg#1{PDG #1}
\def\hflbabar{\mbox{\slshape B\kern-0.1em{\smaller A}\kern-0.1em B\kern-0.1em{\smaller A\kern-0.2em R}}}
\def\hflbelle{\mbox{Belle}}
\def\hflcdf{\mbox{CDF}}
\def\hflcleo{\mbox{CLEO}}
\def\hfld0{\mbox{D0}}
\def\hflavg{Average}

\def\hfpubhotcolor{red}
\def\hfpubcolor{magenta}
\def\hfpuboldcolor{black}

\def\hfprehotcolor{blue}
\def\hfprecolor{cyan}
\def\hfpreoldcolor{hflightcyan}

\def\hfdefcolor{black}
\def\hferrcolor{yellow}
\def\hfsuperceededcolor{hfviolet}
\def\hfwaitingcolor{green}
\def\hfinactivecolor{hfviolet}
\def\hfnoquocolor{hfviolet}
\def\hfdeftext#1{\textcolor{\hfdefcolor}{#1}}
\def\hflabel#1{\textcolor{\hfdefcolor}{$#1$}}
\def\hfavg#1{\textcolor{\hfdefcolor}{$#1$}}
\def\hfnewavg#1{\textcolor{\hfdefcolor}{\boldmath$#1$}}
\def\hfdefault#1{\textcolor{\hfdefcolor}{$#1$}}
\def\hfpdg#1{\textcolor{\hfdefcolor}{$#1$}}
\def\hfwaiting#1{\textcolor{\hfwaitingcolor}{$#1$}}
\def\hfpubhot#1{\textcolor{\hfpubhotcolor}{$#1$}}
\def\hfprehot#1{\textcolor{\hfprehotcolor}{$#1$}}
\def\hfwaitingtext#1{\textcolor{\hfwaitingcolor}{#1}}
\def\hfpubhottext#1{\textcolor{\hfpubhotcolor}{#1}}
\def\hfprehottext#1{\textcolor{\hfprehotcolor}{#1}}
\def\hfpub#1{\textcolor{\hfpubcolor}{$#1$}}
\def\hfpre#1{\textcolor{\hfprecolor}{$#1$}}
\def\hfpubold#1{\textcolor{\hfpuboldcolor}{$#1$}}
\def\hfpreold#1{\textcolor{\hfpreoldcolor}{$#1$}}
\def\hfpubtext#1{\textcolor{\hfpubcolor}{#1}}
\def\hfpretext#1{\textcolor{\hfprecolor}{#1}}
\def\hfpuboldtext#1{\textcolor{\hfpuboldcolor}{#1}}
\def\hfpreoldtext#1{\textcolor{\hfpreoldcolor}{#1}}
\def\hferror#1{\textcolor{\hferrcolor}{$#1$}}
\def\hfsuperceeded#1{\textcolor{\hfsuperceededcolor}{$#1$}}
\def\hfinactive#1{\textcolor{\hfinactivecolor}{$#1$}}
\def\hfnoquo#1{\textcolor{\hfnoquocolor}{$#1$}}
\def\hferrortext#1{\textcolor{\hferrcolor}{#1}}
\def\hfsuperceededtext#1{\textcolor{\hfsuperceededcolor}{#1}}
\def\hfinactivetext#1{\textcolor{\hfinactivecolor}{#1}}
\def\hfnoquotext#1{\textcolor{\hfnoquocolor}{#1}}
\def\hfbb#1{#1M $B\bar{B}$ pairs}

\def\hffootnotemark#1{\tiny{$^{#1}$}}
\def\hffootnotetext#1{\tiny{#1}}
\def\hffootspacing{-5pt}
\def\hffootitemsep{-0.7}

\def\hfnewp{new particles }
\def\hfdstr{strange D mesons }
\def\hfbary{baryons }
\def\hfjpsi{$J/\psi(1S)$ }
\def\hfochm{charmonium other than $J/\psi(1S)$ }
\def\hfmuld{multiple $D$, $D^{*}$ or $D^{**}$ mesons }
\def\hfsgdx{a single $D^{*}$ or $D^{**}$ meson }
\def\hfsgld{a single D meson }
\def\hfothe{charmed particles } 

\def\hfsitebase{http://hfag.phys.ntu.edu.tw/b2charm/}
\def\hfurl#1{\href{\hfsitebase#1}{\hfsitebase#1}} 
\def\hfhref#1#2{\href{\hfsitebase#1}{#2}} 
\def\hftabletype{sidewaystable}
\def\hftableposn{!htbp}
\def\hfaftercapspace{\vspace*{2mm}}

\def\hfcaption#1#2#3#4#5{\textcolor{\hfdefcolor}{#1 of #2 modes producing #3 #4, #5.}}
\def\hfnewcaption#1#2#3#4#5#6#7{\caption{ #1 of #2 modes producing #3 #4, #5. The latest version is available at: \hfurl{#6} \label{#7}  }\hfaftercapspace}
\newcommand\hftabletlcell{\rule{0pt}{2.6ex}}  
\newcommand\hftableblcell{\rule[-1.2ex]{0pt}{0pt}}

\def\hfmetadata#1{}      

\def\hfcBR{{\cal{B}}}

%
%
%
%
%
%

This section reports the updated contribution to the HFAG report from the ``$B \to$ charm" group\footnote{The HFAG/BtoCharm group was formed in the spring of 2005; it performs its work using an XML database backed web application.}. The mandate of the group is to compile measurements and perform averages of all available quantities related to $B$ decays to charmed particles, excluding CP related quantities. To date the group has analyzed a total of 492 measurements reported in 148 papers, principally branching fractions. The group aims to organize and present the copious information on $B$ decays to charmed particles obtained from a combined sample of about two billion $B$ mesons from the BABAR, Belle and CDF Collaborations. 

This huge sample of $B$ mesons allows to measure decays to states with open or hidden charm content with unprecedented precision. Branching fractions for rare $B$-meson decays or decay chains of a few $10^{-7}$ are being measured with statistical uncertainties typically below $30\%$, and new decay chains can be accessed with branching fractions down to $10^{-8}$. Results for more common decay chains, with branching fractions around $10^{-4}$, are becoming precision measurements, with uncertainties typically at the $3\%$ level. Some decays have been observed for the first time, for example $B^0 \to J/\psi\eta$ or $\bar{B^0}\to \Lambda_c^+\bar{p}K^-\pi^+$, with a branching fraction of $(9.6\pm 1.8)\times 10^{-6}$ and $(4.3\pm 1.4)\times 10^{-5}$, respectively.

The large sample of $B$ mesons allows to greatly improve our understanding of recently discovered new states with either hidden or open charm content, such as the $X(3872)$, the $Y(3940)$,the $Z(4430)^-$, the $D_{sJ}^{*-}(2317)$ and $D_{sJ}^-(2460)$ mesons. Measurements with many different final states for these particles are reported, allowing to shed more light on their nature. The $D^0\bar{D}^{*0}(2007)$ decay of the $X(3872)$ has been observed for the first time, as well as the decay into $\psi(2S) \gamma$. Using the branching fraction products $\hfcBR(B^-\to X(3872) K^-)\times\hfcBR(X(3872)\to f)$, a hierarchy can be established between the decay modes $f$: these branching fraction products are found to be $(1.67\pm0.59)\times 10^{-4}$, $(0.12\pm0.02)\times 10^{-4}$, and $(0.022\pm0.005)\times 10^{-4}$, for $D^0\bar{D}^{*0}(2007)$, $J/\psi\pi^+\pi^-$ and $J/\psi\gamma$, respectively. This is an important piece of information to discriminate between various interpretations for the $X(3872)$ state.

The measurements are classified according to the decaying particle: Charged B, Neutral B or Miscellaneous; the decay products 
and the type of quantity: branching fraction, product of branching fractions, ratio of branching fractions or other quantities. 
For the decay product classification the below precedence order is used to ensure that each measurement appears in only one category. 
\begin{itemize}\addtolength{\itemsep}{-0.4\baselineskip}
\item new particles
\item strange $D$ mesons
\item baryons
\item $J/\psi$
\item charmonium other than $J/\psi$
\item multiple $D$, $D^{*}$ or $D^{**}$ mesons
\item a single $D^{*}$ or $D^{**}$ meson
\item a single $D$ meson
\item other particles
\end{itemize}
  
Within each table the measurements are color coded according to the
publication status and age. Table~\ref{tab:hfc99999} provides a key to the
color scheme and categories used. When viewing the tables with most pdf
viewers every number, label and average provides hyperlinks to the corresponding 
reference and individual quantity web pages on the HFAG/BtoCharm group website
\hfhref{}{http://hfag.phys.ntu.edu.tw}.
The links provided in the captions of the table lead to the corresponding compilation
pages.  Both the individual and compilation webpages provide a graphical view
of the results, in a variety of formats.

Tables \ref{tab:hfc01101} to \ref{tab:hfc03300} provide either limits at 90\%
confidence level or measurements with statistical and systematic uncertainties 
and in some cases a third error corresponding to correlated systematics. 
For details on the meanings of the uncertainties and access to the references 
click on the numbers to visit the corresponding web pages.  Where there are
multiple determinations of the same quantity by one experiment the table
footnotes act to distinguish the methods or datasets used; such cases are
visually highlighted in the table by presenting the measurements on the lines
beneath the quantity label.
Where both limits and measured values of a quantity are available the limits 
are presented in the tables but are not used in the determination of the
average. Where only limits are available the most stringent is presented in
the Average column of the tables.
Where available the PDG 2008 result is also presented.

\clearpage 

\begin{\hftabletype}[\hftableposn]

\begin{center}
\caption{Key to the colors used to classify the results presented in tables \ref{tab:hfc01101} to \ref{tab:hfc03300}. When viewing these tables in a pdf viewer each number, label and average provides a hyperlink to the corresponding online version provided by the charm subgroup website \hfurl{}. Where an experiment has multiple determinations of a single quantity they are distinguished by the table footnotes.}
 \label{tab:hfc99999} 
 
\hfmetadata{  }



\end{center}

\end{\hftabletype}
\clearpage  
\begin{\hftabletype}[\hftableposn]

\begin{center}
\hfnewcaption{Branching fractions}{charged B}{\hfnewp}{in units of $10^{-3}$}{upper limits are at 90\% CL}{00101.html}{tab:hfc01101}
\hfmetadata{ [ .tml created 2010-04-27T00:46:54.714+08:00] }

%

\end{center}

\end{\hftabletype}
\clearpage  
\begin{\hftabletype}[\hftableposn]

\begin{center}
\hfnewcaption{Product branching fractions}{charged B}{\hfnewp}{in units of $10^{-4}$}{upper limits are at 90\% CL}{00101.html}{tab:hfc02101}
\hfmetadata{ [ .tml created 2010-04-27T00:47:38.913+08:00] }

%

\begin{description}\addtolength{\itemsep}{\hffootitemsep\baselineskip}

	  \item[ \hffootnotemark{1} ] \hffootnotetext{ Observation of $Y(3940) \rightarrow J/\psi \omega$ in $B \rightarrow J/\psi \omega K$ at BaBar } 
    
	  \item[ \hffootnotemark{2} ] \hffootnotetext{ Observation of $Y(3940)\to J/\psi\omega$ in $B\to J/\psi\omega K$ at BABAR (\hfbb{383}) } 
     ;  \hffootnotetext{ by3940kjpsiomega } 
\end{description}

\end{center}

\end{\hftabletype}
\clearpage  
\begin{\hftabletype}[\hftableposn]

\begin{center}
\hfnewcaption{Branching fractions}{charged B}{\hfdstr}{in units of $10^{-4}$}{upper limits are at 90\% CL}{00102.html}{tab:hfc01102}
\hfmetadata{ [ .tml created 2010-04-27T01:44:47.148+08:00] }



\end{center}

\end{\hftabletype}
\clearpage  
\begin{\hftabletype}[\hftableposn]

\begin{center}
\hfnewcaption{Product branching fractions}{charged B}{\hfdstr}{in units of $10^{-4}$}{upper limits are at 90\% CL}{00102.html}{tab:hfc02102}
\hfmetadata{ [ .tml created 2010-04-27T01:45:07.876+08:00] }

%

\end{center}

\end{\hftabletype}
\clearpage  
\begin{\hftabletype}[\hftableposn]

\begin{center}
\hfnewcaption{Branching fractions}{charged B}{\hfbary}{in units of $10^{-5}$}{upper limits are at 90\% CL}{00103.html}{tab:hfc01103}
\hfmetadata{ [ .tml created 2010-04-27T02:09:55.975+08:00] }

%

\end{center}

\end{\hftabletype}
\clearpage  
\begin{\hftabletype}[\hftableposn]

\begin{center}
\hfnewcaption{Product branching fractions}{charged B}{\hfbary}{in units of $10^{-5}$}{upper limits are at 90\% CL}{00103.html}{tab:hfc02103}
\hfmetadata{ [ .tml created 2010-04-27T02:10:24.178+08:00] }

%

\end{center}


\begin{center}
\hfnewcaption{Ratios of branching fractions}{charged B}{\hfbary}{in units of $10^{-1}$}{upper limits are at 90\% CL}{00103.html}{tab:hfc03103}
\hfmetadata{ [ .tml created 2010-04-27T02:10:43.48+08:00] }

%

\end{center}

\end{\hftabletype}
\clearpage  
\begin{\hftabletype}[\hftableposn]

\begin{center}
\hfnewcaption{Branching fractions}{charged B}{\hfjpsi}{in units of $10^{-4}$}{upper limits are at 90\% CL}{00104.html}{tab:hfc01104}
\hfmetadata{ [ .tml created 2010-04-27T02:30:35.115+08:00] }

%

\begin{description}\addtolength{\itemsep}{\hffootitemsep\baselineskip}

	  \item[ \hffootnotemark{1} ] \hffootnotetext{ Observation of $Y(3940) \rightarrow J/\psi \omega$ in $B \rightarrow J/\psi \omega K$ at BaBar } 
    
	  \item[ \hffootnotemark{2} ] \hffootnotetext{ Observation of $Y(3940)\to J/\psi\omega$ in $B\to J/\psi\omega K$ at BABAR (\hfbb{383}) } 
     ;  \hffootnotetext{ bjpsiomegak } 
	  \item[ \hffootnotemark{3} ] \hffootnotetext{ MEASUREMENT OF BRANCHING FRACTIONS AND CHARGE ASYMMETRIES FOR EXCLUSIVE B DECAYS TO CHARMONIUM (\hfbb{124}) } 
     ;  \hffootnotetext{ $B^- \rightarrow J/\psi K^-$ with $J/\psi$ to leptons } 
	  \item[ \hffootnotemark{4} ] \hffootnotetext{ MEASUREMENT OF THE $B^+ \rightarrow p \overline{p} K^+$ BRANCHING FRACTION AND STUDY OF THE DECAY DYNAMICS (\hfbb{232}) } 
     ;  \hffootnotetext{ $B^- \rightarrow J/\psi K^-$ with $J/\psi \rightarrow p\overline{p}$ } 
	  \item[ \hffootnotemark{5} ] \hffootnotetext{ Measurements of the absolute branching fractions of $B^\pm \rightarrow K^\pm X_{c\overline{c}}$ (\hfbb{231.8}) } 
     ;  \hffootnotetext{ $B^- \rightarrow J/\psi K^-$ (inclusive) } 
\end{description}

\end{center}


\begin{center}
\hfnewcaption{Product branching fractions}{charged B}{\hfjpsi}{in units of $10^{-4}$}{upper limits are at 90\% CL}{00104.html}{tab:hfc02104}
\hfmetadata{ [ .tml created 2010-04-27T02:30:56.683+08:00] }



\end{center}

\end{\hftabletype}
\clearpage  
\begin{\hftabletype}[\hftableposn]

\begin{center}
\hfnewcaption{Ratios of branching fractions}{charged B}{\hfjpsi}{in units of $10^{0}$}{upper limits are at 90\% CL}{00104.html}{tab:hfc03104}
\hfmetadata{ [ .tml created 2010-04-27T02:31:13.838+08:00] }

%

\begin{description}\addtolength{\itemsep}{\hffootitemsep\baselineskip}

	  \item[ \hffootnotemark{1} ] \hffootnotetext{ Measurement of the Branching Fraction $B(B^+ \rightarrow J/\psi \pi^+)$ and Search for $B^{c+} \rightarrow J/\psi \pi^+$ } 
	  \item[ \hffootnotemark{1} ] \hffootnotetext{ Measurement of the Branching Fraction $B(B^+ \rightarrow J/\psi \pi^+)$ and Search for $B^{c+} \rightarrow J/\psi \pi^+$ } 
    
	  \item[ \hffootnotemark{2} ] \hffootnotetext{ Measurement of the Ratio of Branching Fractions B(B -- J/psi Pi)/B(B -- J/psi K) } 
     ;  \hffootnotetext{ Br(B--J/psiPi)/Br(B--J/psi K) } 
	  \item[ \hffootnotemark{3} ] \hffootnotetext{ Branching Fraction Measurements of $B \rightarrow \eta_c K$ Decays (\hfbb{86.1}) } 
     ;  \hffootnotetext{ Ratio $B^- \rightarrow \eta_{c} K^-$ to $B^- \rightarrow J/\psi K^-$ with $\eta_c \rightarrow K\overline{K}\pi$ } 
	  \item[ \hffootnotemark{4} ] \hffootnotetext{ Measurements of the absolute branching fractions of $B^\pm \rightarrow K^\pm X_{c\overline{c}}$ (\hfbb{231.8}) } 
     ;  \hffootnotetext{ Ratio $B^- \rightarrow \eta_c K^-$ to $B^- \rightarrow J/\psi K^-$  (inclusive analysis) } 
\end{description}

\end{center}

\end{\hftabletype}
\clearpage  
\begin{\hftabletype}[\hftableposn] 
\tiny  


\begin{center}
\hfnewcaption{Branching fractions}{charged B}{\hfochm}{in units of $10^{-4}$}{upper limits are at 90\% CL}{00105.html}{tab:hfc01105}
\hfmetadata{ [ .tml created 2010-04-27T02:53:45.519+08:00] }



\begin{description}\addtolength{\itemsep}{\hffootitemsep\baselineskip}

	  \item[ \hffootnotemark{1} ] \hffootnotetext{ Dalitz plot analysis of the decay $B^\pm\rightarrow K^\pm K^\pm K^\mp$ (\hfbb{226}) } 
     ;  \hffootnotetext{ $B^\pm\rightarrow K^\pm \chi_{c0}$, with $chi_c0\rightarrow K^+K^-$ (Dalitz analysis) } 
	  \item[ \hffootnotemark{2} ] \hffootnotetext{ Measurements of the absolute branching fractions of $B^\pm \rightarrow K^\pm X_{c\overline{c}}$ (\hfbb{231.8}) } 
     ;  \hffootnotemark{2a}  \hffootnotetext{ $B^- \rightarrow \chi_{c0} K^-$ (inclusive) }  ;  \hffootnotemark{2b}  \hffootnotetext{ $B^- \rightarrow \psi(2S) K^-$ (inclusive) }  ;  \hffootnotemark{2c}  \hffootnotetext{ $B^- \rightarrow \chi_{c1} K^-$ (inclusive) }  ;  \hffootnotemark{2d}  \hffootnotetext{ $B^- \rightarrow \eta_c K^-$ (inclusive) } 
	  \item[ \hffootnotemark{3} ] \hffootnotetext{ MEASUREMENT OF THE BRANCHING FRACTION FOR $B^\pm \rightarrow \chi_{c0} K^\pm$. (\hfbb{88.9}) } 
     ;  \hffootnotetext{ $B^- \rightarrow \chi_{c0} K^-$ with $\chi_{c0}\rightarrow K^+ K^-, \pi^+ \pi^-$ } 
	  \item[ \hffootnotemark{4} ] \hffootnotetext{ Dalitz-plot analysis of the decays $B^\pm \rightarrow K^\pm \pi^\mp \pi^\pm$ (\hfbb{226}) } 
     ;  \hffootnotetext{ $B^- \rightarrow \chi_{c0} K^-$ with $\chi_{c0} \rightarrow \pi^+ \pi^-$ (Dalitz analysis) } 
	  \item[ \hffootnotemark{5} ] \hffootnotetext{ Search for $X(3872) \rightarrow \psi(2S) \gamma$ in $B^{\pm} \rightarrow X(3872) K^{\pm}$ decays, and a study of $B \rightarrow \overline{c}c\gamma K$ } 
	  \item[ \hffootnotemark{5} ] \hffootnotetext{ Search for $X(3872) \rightarrow \psi(2S) \gamma$ in $B^{\pm} \rightarrow X(3872) K^{\pm}$ decays, and a study of $B \rightarrow \overline{c}c\gamma K$ } 
    
	  \item[ \hffootnotemark{6} ] \hffootnotetext{ MEASUREMENT OF BRANCHING FRACTIONS AND CHARGE ASYMMETRIES FOR EXCLUSIVE B DECAYS TO CHARMONIUM (\hfbb{124}) } 
     ;  \hffootnotetext{ $B^- \rightarrow \psi(2S) K^-$ with $\psi(2S)$ to leptons } 
	  \item[ \hffootnotemark{7} ] \hffootnotetext{ Branching Fraction Measurements of $B \rightarrow \eta_c K$ Decays (\hfbb{86.1}) } 
     ;  \hffootnotetext{ $B^- \rightarrow \eta_{c} K^-$ with $\eta_c \rightarrow K\overline{K}\pi$ } 
	  \item[ \hffootnotemark{8} ] \hffootnotetext{ MEASUREMENT OF THE $B^+ \rightarrow p \overline{p} K^+$ BRANCHING FRACTION AND STUDY OF THE DECAY DYNAMICS (\hfbb{232}) } 
     ;  \hffootnotetext{ $B^- \rightarrow \eta_c K^-$ with $\eta_c \rightarrow p\overline{p}$ } 
\end{description}

\end{center}

\normalsize 

\end{\hftabletype}
\clearpage  
\begin{\hftabletype}[\hftableposn]

\begin{center}
\hfnewcaption{Product branching fractions}{charged B}{\hfochm}{in units of $10^{-4}$}{upper limits are at 90\% CL}{00105.html}{tab:hfc02105}
\hfmetadata{ [ .tml created 2010-04-27T02:54:14.972+08:00] }



\end{center}


\begin{center}
\hfnewcaption{Ratios of branching fractions}{charged B}{\hfochm}{in units of $10^{-1}$}{upper limits are at 90\% CL}{00105.html}{tab:hfc03105}
\hfmetadata{ [ .tml created 2010-04-27T02:54:34.717+08:00] }

%

\end{center}

\end{\hftabletype}
\clearpage  
\begin{\hftabletype}[\hftableposn]

\begin{center}
\hfnewcaption{Branching fractions}{charged B}{\hfmuld}{in units of $10^{-3}$}{upper limits are at 90\% CL}{00106.html}{tab:hfc01106}
\hfmetadata{ [ .tml created 2010-04-27T03:24:33.765+08:00] }

%

\begin{description}\addtolength{\itemsep}{\hffootitemsep\baselineskip}

	  \item[ \hffootnotemark{1} ] \hffootnotetext{ Measurement of B+ - D+ D0bar branching fraction and charge asymmetry and search for B0 - D0 D0bar (\hfbb{656.7}) } 
    
	  \item[ \hffootnotemark{2} ] \hffootnotetext{ Observation of $B^0 \to D^+ D^-$, $B^- \to D^0 D^-$ and
$B^- \to D^0 D^{*-}$ decays (\hfbb{152}) } 
    
\end{description}

\end{center}

\end{\hftabletype}
\clearpage  
\begin{\hftabletype}[\hftableposn]

\begin{center}
\hfnewcaption{Product branching fractions}{charged B}{\hfmuld}{in units of $10^{-4}$}{upper limits are at 90\% CL}{00106.html}{tab:hfc02106}
\hfmetadata{ [ .tml created 2010-04-27T03:25:15.537+08:00] }



\end{center}


\begin{center}
\hfnewcaption{Ratios of branching fractions}{charged B}{\hfmuld}{in units of $10^{0}$}{upper limits are at 90\% CL}{00106.html}{tab:hfc03106}
\hfmetadata{ [ .tml created 2010-04-27T03:25:47.597+08:00] }

%

\end{center}

\end{\hftabletype}
\clearpage  
\begin{\hftabletype}[\hftableposn]

\begin{center}
\hfnewcaption{Branching fractions}{charged B}{\hfsgdx}{in units of $10^{-4}$}{upper limits are at 90\% CL}{00107.html}{tab:hfc01107}
\hfmetadata{ [ .tml created 2010-04-27T03:59:06.433+08:00] }

%

\begin{description}\addtolength{\itemsep}{\hffootitemsep\baselineskip}

	  \item[ \hffootnotemark{1} ] \hffootnotetext{ Branching fraction measurements and isospin analyses for $\bar{B} \rightarrow D^{(*)}\pi^-$
decays (\hfbb{65}) } 
     ;  \hffootnotetext{ $B^- \rightarrow D^{*0}\pi^-$ } 
	  \item[ \hffootnotemark{2} ] \hffootnotetext{ Measurement of the Absolute Branching Fractions $B \rightarrow D^{(*,**)}\pi$ with a Missing Mass method (\hfbb{231}) } 
     ;  \hffootnotetext{ $B^- \rightarrow D^{*0} \pi^-$ } 
\end{description}

\end{center}

\end{\hftabletype}
\clearpage  
\begin{\hftabletype}[\hftableposn]

\begin{center}
\hfnewcaption{Branching fractions}{charged B}{\hfsgld}{in units of $10^{-3}$}{upper limits are at 90\% CL}{00108.html}{tab:hfc01108}
\hfmetadata{ [ .tml created 2010-04-27T04:14:40.634+08:00] }



\begin{description}\addtolength{\itemsep}{\hffootitemsep\baselineskip}

	  \item[ \hffootnotemark{1} ] \hffootnotetext{ Branching fraction measurements and isospin analyses for $\bar{B} \rightarrow D^{(*)}\pi^-$
decays (\hfbb{65}) } 
     ;  \hffootnotetext{ $B^- \rightarrow D^0\pi^-$ } 
	  \item[ \hffootnotemark{2} ] \hffootnotetext{ Measurement of the Absolute Branching Fractions $B \rightarrow D^{(*,**)}\pi$ with a Missing Mass method (\hfbb{231}) } 
     ;  \hffootnotetext{ $B^- \rightarrow D^0\pi^-$ } 
\end{description}

\end{center}


\begin{center}
\hfnewcaption{Branching fractions}{charged B}{\hfothe}{in units of $10^{-6}$}{upper limits are at 90\% CL}{00109.html}{tab:hfc01109}
\hfmetadata{ [ .tml created 2010-04-27T04:25:25.66+08:00] }

\begin{tabular}{lccccc}  
\hflmode
&  \hflpdg2008
&  \hflbelle
&  \hflbabar
&  \hflcdf
&  \hflavg
\\\hline 

 \hfhref{BR_-521_-321+333+30223.html}{ \hflabel{K^- \omega(1650) \phi(1020)} } \hftabletlcell\hftableblcell 
 &  { \,}  
 &  \hfhref{0905014.html}{ \hfprehot{<1.90} }  
 &  { \,}  
 &  { \,}  
 &  \hfhref{BR_-521_-321+333+30223.html}{ \hfnewavg{<1.90} } 
\\\hline 

\end{tabular}

\end{center}

\end{\hftabletype}
\clearpage  
\begin{\hftabletype}[\hftableposn]

\begin{center}
\hfnewcaption{Branching fractions}{neutral B}{\hfnewp}{in units of $10^{-4}$}{upper limits are at 90\% CL}{00201.html}{tab:hfc01201}
\hfmetadata{ [ .tml created 2010-04-27T04:28:20.826+08:00] }



\end{center}

\end{\hftabletype}
\clearpage  
\begin{\hftabletype}[\hftableposn]

\begin{center}
\hfnewcaption{Product branching fractions}{neutral B}{\hfnewp}{in units of $10^{-4}$}{upper limits are at 90\% CL}{00201.html}{tab:hfc02201}
\hfmetadata{ [ .tml created 2010-04-27T04:28:55.146+08:00] }

%

\end{center}

\end{\hftabletype}
\clearpage  
\begin{\hftabletype}[\hftableposn]

\begin{center}
\hfnewcaption{Ratios of branching fractions}{neutral B}{\hfnewp}{in units of $10^{0}$}{upper limits are at 90\% CL}{00201.html}{tab:hfc03201}
\hfmetadata{ [ .tml created 2010-04-27T04:29:17.817+08:00] }

%

\end{center}

\end{\hftabletype}
\clearpage  
\begin{\hftabletype}[\hftableposn] 
\tiny  


\begin{center}
\hfnewcaption{Branching fractions}{neutral B}{\hfdstr}{in units of $10^{-3}$}{upper limits are at 90\% CL}{00202.html}{tab:hfc01202}
\hfmetadata{ [ .tml created 2010-04-27T05:05:11.674+08:00] }

%

\begin{description}\addtolength{\itemsep}{\hffootitemsep\baselineskip}

	  \item[ \hffootnotemark{1} ] \hffootnotetext{ Measurement of the Branching Fractions of the Rare Decays $B^0\rightarrow D_s^{(*)+}\pi^-$, $B^0\rightarrow D_s^{(*)+}\rho^-$, and $B^0\rightarrow D_s^{(*)-}K^{(*)+}$ (\hfbb{381}) } 
    
	  \item[ \hffootnotemark{2} ] \hffootnotetext{ A search for the rare decay $\bar{B}^0 \rightarrow D_s^- \rho^+$ (\hfbb{90}) } 
     ;  \hffootnotetext{ $\bar{B}^0 \rightarrow D_s^- \rho^+$ } 
	  \item[ \hffootnotemark{3} ] \hffootnotetext{ Observation of B0bar to Ds+ Lambda pbar (\hfbb{447}) } 
    
	  \item[ \hffootnotemark{4} ] \hffootnotetext{ Observation of B0bar - Ds+ Lambda pbar decay (\hfbb{449}) } 
    
	  \item[ \hffootnotemark{5} ] \hffootnotetext{ Improved measurement of B0bar - Ds-D+ and search for B0bar - Ds+Ds- (\hfbb{449}) } 
    
	  \item[ \hffootnotemark{6} ] \hffootnotetext{ Improved measurement of $\bar{B}^0\to D_s^-D^+$ and search for $\bar{B}0\to D_s^+D_s^-$ at Belle } 
    
	  \item[ \hffootnotemark{7} ] \hffootnotetext{ Study of $\bar{B} \rightarrow D^{(*)+,-}X^{-}$ and $\bar{B} \rightarrow D_s^{(*)-}X^{+,0}$ decays and measurement of $D_s^-$ and $D_{sJ}^-(2460)$ absolute branching fractions (\hfbb{230}) } 
     ;  \hffootnotemark{7a}  \hffootnotetext{ $\bar{B}^0 \rightarrow D_s^-D^{*+}$ }  ;  \hffootnotemark{7b}  \hffootnotetext{ $\bar{B}^0 \rightarrow D_s^{*-}D^{*+})$ } 
	  \item[ \hffootnotemark{8} ] \hffootnotetext{ Measurement of $\bar{B}^0 \rightarrow D_s^{(*)}D^*$ Branching Fractions and $D_s^*D^*$
 Polarization with a Partial Reconstruction technique (\hfbb{22.7}) } 
     ;  \hffootnotemark{8a}  \hffootnotetext{ $\bar{B}^0 \rightarrow D_s^- D^{*+}$ }  ;  \hffootnotemark{8b}  \hffootnotetext{ $\bar{B}^0 \rightarrow D_s^{*-} D^{*+}$ } 
	  \item[ \hffootnotemark{9} ] \hffootnotetext{ Measurement of the $\bar{B}^0 \rightarrow D_s^{*-} D^+$ and $D_s^+ \rightarrow \phi \pi^+$ branching
fractions (\hfbb{123}) } 
     ;  \hffootnotetext{ $\bar{B}^0 \rightarrow D_s^{*-} D^{*+}$ } 
\end{description}

\end{center}

\normalsize 

\end{\hftabletype}
\clearpage  
\begin{\hftabletype}[\hftableposn]

\begin{center}
\hfnewcaption{Product branching fractions}{neutral B}{\hfdstr}{in units of $10^{-4}$}{upper limits are at 90\% CL}{00202.html}{tab:hfc02202}
\hfmetadata{ [ .tml created 2010-04-27T05:05:43.926+08:00] }



\end{center}


\begin{center}
\hfnewcaption{Ratios of branching fractions}{neutral B}{\hfdstr}{in units of $10^{0}$}{upper limits are at 90\% CL}{00202.html}{tab:hfc03202}
\hfmetadata{ [ .tml created 2010-04-27T05:06:14.756+08:00] }

%

\end{center}

\end{\hftabletype}
\clearpage  
\begin{\hftabletype}[\hftableposn] 
\tiny  


\begin{center}
\hfnewcaption{Branching fractions}{neutral B}{\hfbary}{in units of $10^{-5}$}{upper limits are at 90\% CL}{00203.html}{tab:hfc01203}
\hfmetadata{ [ .tml created 2010-04-27T05:49:21.96+08:00] }

%

\begin{description}\addtolength{\itemsep}{\hffootitemsep\baselineskip}

	  \item[ \hffootnotemark{1} ] \hffootnotetext{ STUDY OF EXCLUSIVE B DECAYS TO CHARMED BARYONS AT BELLE. (\hfbb{31.7}) } 
    
	  \item[ \hffootnotemark{2} ] \hffootnotetext{ Study of the charmed baryonic decays $\bar{B}^0\to\Sigma_c^{++}\bar{p}\pi^-$ and $\bar{B}^0\to\Sigma_c^{0}\bar{p}\pi^+$ (\hfbb{386}) } 
     ;  \hffootnotemark{2a}  \hffootnotetext{ B0bar to Sigmac(2455)++ pbar pi } 
\end{description}

\end{center}

\normalsize 

\end{\hftabletype}
\clearpage  
\begin{\hftabletype}[\hftableposn]

\begin{center}
\hfnewcaption{Product branching fractions}{neutral B}{\hfbary}{in units of $10^{-5}$}{upper limits are at 90\% CL}{00203.html}{tab:hfc02203}
\hfmetadata{ [ .tml created 2010-04-27T05:49:44.834+08:00] }

\begin{tabular}{lccccc}  
\hflmode
&  \hflpdg2008
&  \hflbelle
&  \hflbabar
&  \hflcdf
&  \hflavg
\\\hline 

 \hfhref{BR_-511_-4122+4232xBR_4232_211+211+3312.html}{ \hflabel{\Lambda_c^- \Xi_c^+ [ \Xi^- \pi^+ \pi^+ ]} } \hftabletlcell\hftableblcell 
 &  \hfhref{BR_-511_-4122+4232xBR_4232_211+211+3312.html}{ \hfpdg{9.0\pm5.0} }  
 &  \hfhref{0607006.html}{ \hfpre{9.3\pm^{3.7}_{2.8}\pm1.9\pm2.4} }  
 &  \hfhref{0905001.html}{ \hfpub{<5.6} }  
 &  { \,}  
 &  \hfhref{BR_-511_-4122+4232xBR_4232_211+211+3312.html}{ \hfnewavg{9.3\pm^{4.8}_{4.1}} } 
\\\hline 

\end{tabular}

\end{center}

\end{\hftabletype}
\clearpage  
\begin{\hftabletype}[\hftableposn] 
\tiny  


\begin{center}
\hfnewcaption{Branching fractions}{neutral B}{\hfjpsi}{in units of $10^{-5}$}{upper limits are at 90\% CL}{00204.html}{tab:hfc01204}
\hfmetadata{ [ .tml created 2010-04-27T06:12:20.487+08:00] }



\begin{description}\addtolength{\itemsep}{\hffootitemsep\baselineskip}

	  \item[ \hffootnotemark{1} ] \hffootnotetext{ Study of $B^0 \to J/\psi \pi^+ \pi^-$ decays with 449 million $B\bar{B}$ pairs at Belle (\hfbb{449}) } 
    
	  \item[ \hffootnotemark{2} ] \hffootnotetext{ MEASUREMENT OF BRANCHING FRACTIONS IN B0 --- J/PSI PI+ PI- DECAY. (\hfbb{152}) } 
    
	  \item[ \hffootnotemark{3} ] \hffootnotetext{ Observation of $Y(3940) \rightarrow J/\psi \omega$ in $B \rightarrow J/\psi \omega K$ at BaBar } 
    
	  \item[ \hffootnotemark{4} ] \hffootnotetext{ Observation of $Y(3940)\to J/\psi\omega$ in $B\to J/\psi\omega K$ at BABAR (\hfbb{383}) } 
     ;  \hffootnotetext{ bjpsiomegak0 } 
\end{description}

\end{center}

\normalsize 

\end{\hftabletype}
\clearpage  
\begin{\hftabletype}[\hftableposn]

\begin{center}
\hfnewcaption{Ratios of branching fractions}{neutral B}{\hfjpsi}{in units of $10^{0}$}{upper limits are at 90\% CL}{00204.html}{tab:hfc03204}
\hfmetadata{ [ .tml created 2010-04-27T06:12:45.626+08:00] }



\end{center}


\begin{center}
\hfnewcaption{Miscellaneous quantities}{neutral B}{\hfjpsi}{in units of $10^{0}$}{upper limits are at 90\% CL}{00204.html}{tab:hfc04204}
\hfmetadata{ [ .tml created 2010-04-27T06:13:04.144+08:00] }

%

\end{center}

\end{\hftabletype}
\clearpage  
\begin{\hftabletype}[\hftableposn]

\begin{center}
\hfnewcaption{Branching fractions}{neutral B}{\hfochm}{in units of $10^{-4}$}{upper limits are at 90\% CL}{00205.html}{tab:hfc01205}
\hfmetadata{ [ .tml created 2010-04-27T06:46:03.338+08:00] }

%

\begin{description}\addtolength{\itemsep}{\hffootitemsep\baselineskip}

	  \item[ \hffootnotemark{1} ] \hffootnotetext{ Study of B-meson decays to etac K(*), etac(2S) K(*) and etac gamma K(*) } 
    
	  \item[ \hffootnotemark{2} ] \hffootnotetext{ Branching Fraction Measurements of $B \rightarrow \eta_c K$ Decays (\hfbb{86.1}) } 
    
	  \item[ \hffootnotemark{3} ] \hffootnotetext{ Evidence for the $B^0\to p \bar{p} K^{*0}$ and $B^+\to \eta_cK^{*+}$ decays and Study of the Decay Dynamics of $B$ Meson Decays into $p\bar{p}h$ Final States. (\hfbb{232}) } 
     ;  \hffootnotemark{3a}  \hffootnotetext{ betackzero }  ;  \hffootnotemark{3b}  \hffootnotetext{ betackstarzppbar } 
\end{description}

\end{center}

\end{\hftabletype}
\clearpage  
\begin{\hftabletype}[\hftableposn]

\begin{center}
\hfnewcaption{Product branching fractions}{neutral B}{\hfochm}{in units of $10^{-4}$}{upper limits are at 90\% CL}{00205.html}{tab:hfc02205}
\hfmetadata{ [ .tml created 2010-04-27T06:46:38.558+08:00] }



\end{center}


\begin{center}
\hfnewcaption{Ratios of branching fractions}{neutral B}{\hfochm}{in units of $10^{0}$}{upper limits are at 90\% CL}{00205.html}{tab:hfc03205}
\hfmetadata{ [ .tml created 2010-04-27T06:46:59.578+08:00] }

%

\end{center}

\end{\hftabletype}
\clearpage  
\begin{\hftabletype}[\hftableposn]

\begin{center}
\hfnewcaption{Branching fractions}{neutral B}{\hfmuld}{in units of $10^{-3}$}{upper limits are at 90\% CL}{00206.html}{tab:hfc01206}
\hfmetadata{ [ .tml created 2010-04-27T07:14:02.703+08:00] }

%

\begin{description}\addtolength{\itemsep}{\hffootitemsep\baselineskip}

	  \item[ \hffootnotemark{1} ] \hffootnotetext{ Evidence for CP Violation in B0 - D+D- Decays (\hfbb{535}) } 
    
	  \item[ \hffootnotemark{2} ] \hffootnotetext{ Observation of $B^0 \to D^+ D^-$, $B^- \to D^0 D^-$ and
$B^- \to D^0 D^{*-}$ decays (\hfbb{152}) } 
    
\end{description}

\end{center}

\end{\hftabletype}
\clearpage  
\begin{\hftabletype}[\hftableposn]

\begin{center}
\hfnewcaption{Product branching fractions}{neutral B}{\hfmuld}{in units of $10^{-4}$}{upper limits are at 90\% CL}{00206.html}{tab:hfc02206}
\hfmetadata{ [ .tml created 2010-04-27T07:14:26.628+08:00] }



\end{center}


\begin{center}
\hfnewcaption{Ratios of branching fractions}{neutral B}{\hfmuld}{in units of $10^{0}$}{upper limits are at 90\% CL}{00206.html}{tab:hfc03206}
\hfmetadata{ [ .tml created 2010-04-27T07:15:18.46+08:00] }

%

\end{center}

\end{\hftabletype}
\clearpage  
\begin{\hftabletype}[\hftableposn]

\begin{center}
\hfnewcaption{Branching fractions}{neutral B}{\hfsgdx}{in units of $10^{-4}$}{upper limits are at 90\% CL}{00207.html}{tab:hfc01207}
\hfmetadata{ [ .tml created 2010-04-27T07:55:48.14+08:00] }

%

\begin{description}\addtolength{\itemsep}{\hffootitemsep\baselineskip}

	  \item[ \hffootnotemark{1} ] \hffootnotetext{ Study of $\bar{B^{0}} \to D^{(*) 0} \pi^+ \pi^-$ Decays (\hfbb{31.3}) } 
    
	  \item[ \hffootnotemark{2} ] \hffootnotetext{ Study of $\bar{B}^{0} \to  D^{(*)0} \pi^{+} \pi^{-}$ decays ; Dalitz fit analysis (\hfbb{152}) } 
    
	  \item[ \hffootnotemark{3} ] \hffootnotetext{ Branching fraction measurements and isospin analyses for $\bar{B} \rightarrow D^{(*)}\pi^-$
decays (\hfbb{65}) } 
     ;  \hffootnotetext{ $\bar{B}^0 \rightarrow D^{*+}\pi^-$ } 
	  \item[ \hffootnotemark{4} ] \hffootnotetext{ Measurement of the Absolute Branching Fractions $B \rightarrow D^{(*,**)}\pi$ with a Missing Mass method (\hfbb{231}) } 
     ;  \hffootnotetext{ $\bar{B}^0 \rightarrow D^{*+}\pi^-$ } 
\end{description}

\end{center}

\end{\hftabletype}
\clearpage  

\clearpage 
\begin{\hftabletype}[\hftableposn]

\begin{center}
\hfnewcaption{Branching fractions}{neutral B}{\hfsgld}{in units of $10^{-4}$}{upper limits are at 90\% CL}{00208.html}{tab:hfc01208}
\hfmetadata{ [ .tml created 2010-04-27T08:34:35.84+08:00] }



\begin{description}\addtolength{\itemsep}{\hffootitemsep\baselineskip}

	  \item[ \hffootnotemark{1} ] \hffootnotetext{ Study of $\bar{B^{0}} \to D^{(*) 0} \pi^+ \pi^-$ Decays (\hfbb{31.3}) } 
    
	  \item[ \hffootnotemark{2} ] \hffootnotetext{ Study of $\bar{B}^{0} \to  D^{(*)0} \pi^{+} \pi^{-}$ decays ; Dalitz fit analysis (\hfbb{152}) } 
    
	  \item[ \hffootnotemark{3} ] \hffootnotetext{ Branching fraction measurements and isospin analyses for $\bar{B} \rightarrow D^{(*)}\pi^-$
decays (\hfbb{65}) } 
     ;  \hffootnotetext{ $\bar{B}^0 \rightarrow D^+\pi^-$ } 
	  \item[ \hffootnotemark{4} ] \hffootnotetext{ Measurement of the Absolute Branching Fractions $B \rightarrow D^{(*,**)}\pi$ with a Missing Mass method (\hfbb{231}) } 
     ;  \hffootnotetext{ $\bar{B}^0 \rightarrow D^+\pi^-$ } 
\end{description}

\end{center}

\end{\hftabletype}
\clearpage  
\begin{\hftabletype}[\hftableposn]

\begin{center}
\hfnewcaption{Product branching fractions}{neutral B}{\hfsgld}{in units of $10^{-5}$}{upper limits are at 90\% CL}{00208.html}{tab:hfc02208}
\hfmetadata{ [ .tml created 2010-04-27T08:35:21.098+08:00] }



\end{center}


\begin{center}
\hfnewcaption{Branching fractions}{miscellaneous}{charmed particles }{in units of $10^{-3}$}{upper limits are at 90\% CL}{00300.html}{tab:hfc01300}
\hfmetadata{ [ .tml created 2010-04-27T09:09:36.05+08:00] }

%

\end{center}

\end{\hftabletype}
\clearpage  
\begin{\hftabletype}[\hftableposn]

\begin{center}
\hfnewcaption{Product branching fractions}{miscellaneous}{charmed particles }{in units of $10^{-5}$}{upper limits are at 90\% CL}{00300.html}{tab:hfc02300}
\hfmetadata{ [ .tml created 2010-04-27T09:09:59.957+08:00] }

%

\end{center}


\begin{center}
\hfnewcaption{Ratios of branching fractions}{miscellaneous}{charmed particles }{in units of $10^{0}$}{upper limits are at 90\% CL}{00300.html}{tab:hfc03300}
\hfmetadata{ [ .tml created 2010-04-27T09:11:11.682+08:00] }

%

\end{center}

\end{\hftabletype}
\clearpage  
\begin{\hftabletype}[\hftableposn]

\begin{center}
\hfnewcaption{Miscellaneous quantities}{miscellaneous}{charmed particles }{in units of $10^{0}$}{upper limits are at 90\% CL}{00300.html}{tab:hfc04300}
\hfmetadata{ [ .tml created 2010-04-27T09:11:57.85+08:00] }

%

\end{center}

\end{\hftabletype}
\clearpage  
 


\clearpage 
\mysection{$B$ decays to charmless final states}


\label{sec:rare}

The aim of this section is to provide the branching fractions and
the partial rate asymmetries ($A_{CP}$) of charmless 
$B$ decays. The asymmetry is defined as 
$A_{CP} = \frac{N_{\Bbar} -N_B}{N_{\Bbar} +N_B}$, where $N_{\Bbar}$ 
and $N_B$ are respectively number of $\Bzb/\Bm$ and $\Bz/\Bp$ decaying
into a specific final state. 
Four different $B$ decay categories are considered: 
charmless mesonic, baryonic, radiative and leptonic. Measurements supported 
with  written documents are accepted in  
the averages; written documents could be journal papers, 
conference contributed papers, preprints or conference proceedings.  
Results from  $A_{CP}$ measurements  obtained from time dependent analyses 
are listed and described in Sec.~\ref{sec:cp_uta}. Measurements of  charmful 
baryonic $B$ decays, which were included in our previous averages
\cite{Barberio:2008fa}, are now shown in Section 7, which deals 
with $B$ decays to charm.  

So far all branching fractions assume equal production of charged and
neutral $B$ pairs.  The best measurements to date show that this is
still a good approximation (see Sec.~\ref{sec:life_mix}).
For branching fractions, we provide either averages or the most stringent
90\% confidence level upper limits.  If one or more experiments have
measurements with $>$4$\sigma$ for a decay channel, all available central values
for that channel are used in the averaging.  We also give central values
and errors for cases where the significance of the average value is at
least $3 \sigma$, even if no single measurement is above $4 \sigma$. 
Since a few decay modes are sensitive to the contribution of
 new physics and the current experimental upper limits are not far from the 
Standard Model expectation, we provide the combined upper limits or
averages in these cases.
Their upper limits can be estimated assuming that the errors are 
Gaussian.  For $A_{CP}$ we provide averages in all cases.  

Our averaging is performed by maximizing the likelihood,
   $\displaystyle {\mathcal L} = \prod_i {\mathcal P}_i(x),$  
where ${\mathcal P_i}$ is the probability density function (PDF) of the
$i$th  measurement, and $x$ is the branching fraction or $A_{CP}$.
The PDF is modeled by an asymmetric Gaussian function with the measured
central value as its mean and the quadratic sum of the statistical
and systematic errors as the standard deviations. The experimental
uncertainties are considered to be uncorrelated with each other when the 
averaging is performed. No error scaling is applied when the fit $\chi^2$ is 
greater than 1 since we believe that tends to overestimate the errors
except in cases of extreme disagreement (we have no such cases).
One exception to consider the correlated systematic errors is the inclusive
$B\to X_s\gamma$ mode, which is sensitive to physics beyond the Standard Model.
In this update, we have included new measurements from both Belle and BaBar
to perform the average. The detail is  
described  in Sec. ~\ref{sec:btosg}.

At present, we have measurements of about 400 decay modes, reported in
more than 200 papers. Because the number of references is so large, we do
not include them with the tables shown here but the full set of
references is available quickly from active gifs at the 
``Winter 2010'' link on 
the rare web page: {\tt http://www.slac.stanford.edu/xorg/hfag/rare/index.html}.
Finally many new  measurements involving scalar and tensor mesons are   
included for the first time.  

\mysubsection{Mesonic charmless decays}

\begin{table}
\begin{center}
\caption{Branching fractions (BF) of charmless mesonic  
$B^+$ decays with kaons (in units of $\times 10^6)$). Upper limits are
at 90\% CL. Values in {\red red} ({\blue blue}) are new {\red published}
({\blue preliminary}) results since PDG2008  [as of March 12, 2010].
}
\scriptsize

\end{center}

\vspace{-0.4cm} 
\hspace*{-0.8cm} 
\hspace{0.4cm}
$\dag$Product BF - daughter BF taken to be 100\%
\end{table}

\begin{table}
\begin{center}
\caption{Branching Fractions (BF) of charmless mesonic $B^+$ decays 
 with kaons - part 2 (in units of $10^{-6}$). Upper limits are at 90\% CL.
Values in {\red red} ({\blue blue}) are new {\red published}
({\blue preliminary}) results since PDG2008  [as of March 12, 2010].
}
\scriptsize
%
\end{center}

\vspace{-0.4cm} 
\hspace*{-0.8cm} 
$\dag$Product BF - daughter BF taken to be 100\%; 
~\S $M_{\phi\phi}<2.85$ GeV/$c^2$
\end{table}

\begin{table}
\begin{center}
\caption{Branching Fractions (BF) of charmless mesonic $B^+$ decays 
 without kaons (in units of $10^{-6}$). Upper limits are at 90\% CL.
Values in {\red red} ({\blue blue}) are new {\red published}
({\blue preliminary}) results since PDG2008  [as of March 12, 2010].
}
\scriptsize
%
\end{center}

\vspace{-0.4cm} 
\hspace*{0.8cm} 
$\dag$Product BF - daughter BF taken to be 100\%; 
\end{table}
\clearpage

\begin{table}
\begin{center}
\caption{Branching fractions of charmless mesonic $B^0$ decays with kaons 
(in units of $10^{-6}$).
Upper limits are at 90\% CL.
Values in {\red red} ({\blue blue}) are new {\red published}
({\blue preliminary}) results since PDG2008  [as of March 12, 2010].
}

\scriptsize
%
\end{center}
\vspace{-0.4cm}
$\dag$Product BF - daughter BF taken to be 100\%, 
 $\ddag$Relative BF converted to absolute BF
 $^1 0.755<M(K\pi)<1.250$ GeV/$c^2$.
 $^2$Excludes $M(K_SK_S)$ regions [3.400,3.429] and [3.540,3.585] and $M(K_SK_L)<1.049$ GeV/$c^2$
 $^3$Includes $K\pi$ S-wave contribution and uncorrected for K*(1430) BF
\end{table}
\clearpage

\begin{table}
\begin{center}
\caption{
Branching fractions of charmless mesonic $B^0$ decays with kaons - part 2
(in units of $10^{-6}$).
Upper limits are at 90\% CL.
Values in {\red red} ({\blue blue}) are new {\red published}
({\blue preliminary}) results since PDG2008  [as of March 12, 2010].
}

\scriptsize

\end{center}
\vspace{-0.4cm}
$\dag$Product BF - daughter BF taken to be 100\%, 
~\S $M_{\phi\phi}<2.85$ GeV/$c^2$
 $\ddag 0.55<M(\pi\pi)<1.42$ GeV/$c^2$ and $0.75<M(K\pi)<1.20$ GeV/$c^2$;
 $^1 0.55<M(\pi\pi)<1.42$ GeV/$c^2$;
 $^2 0.75<M(K\pi)<1.20$ GeV/$c^2$
\end{table}
\clearpage

\begin{table}
\begin{center}
\caption{
Branching fractions of charmless mesonic $B^0$ decays without kaons
(in units of $10^{-6}$).
Upper limits are at 90\% CL.
Values in {\red red} ({\blue blue}) are new {\red published}
({\blue preliminary}) results since PDG2008  [as of March 12, 2010].
}

\scriptsize

\end{center}
\vspace{-0.4cm}
$\dag$Product BF - daughter BF taken to be 100\%, 
 $\ddag$Relative BF converted to absolute BF
\end{table}

\begin{table}
\begin{center}
\caption{
 Relative branching fractions of $B^0\to K^+K^-, K^+\pi^-, \pi^+\pi^-$.
Values in {\red red} ({\blue blue}) are new {\red published}
({\blue preliminary}) result since PDG2008  [as of March 15, 2007].
}
\small

%
\end{center}
\end{table}
\clearpage

\mysubsection{Radiative and leptonic decays}

\begin{table}[!hb]
\begin{center}
\caption{
Branching fractions of semileptonic and radiative $B^+$ decays
(in units of $10^{-6}$). Upper limits are at 90\% CL.
Values in {\red red} ({\blue blue}) are new {\red published}
({\blue preliminary}) results since PDG2008  [as of March 12, 2010].
}
\scriptsize
%
\vspace{-0.4cm}
\end{center}
~\dag $M_{K\pi\pi} < 1.8$ GeV/$c^2$;~\ddag~$1.0 < M_{K\pi\pi} < 2.0$ GeV/$c^2$;
~\S~$M_{K\pi\pi} < 2.4$ GeV/$c^2$
\end{table}
\clearpage

\begin{table}
\begin{center}
\caption{
Branching fractions of semileptonic and radiative $B^0$ decays
(in units of $10^{-6}$).  Upper limits are at 90\% CL.
Values in {\red red} ({\blue blue}) are new {\red published}
({\blue preliminary}) results since PDG2008  [as of March 12, 2010].
}

\scriptsize
%
\vspace{-0.4cm}
\end{center}
~\dag $M_{K\pi\pi} < 1.8$ GeV/$c^2$;~\ddag~$1.0 < M_{K\pi\pi} < 2.0$ GeV/$c^2$;
~\S~$1.25$ GeV/$c^2 < M_{K\pi} < 1.6$ GeV/$c^2$

\end{table}

\begin{table}
\begin{center}
\caption{
Branching fractions of  semileptonic and radiative $B$ decays
(in units of $10^{-6}$).  Upper limits are at 90\% CL.
Values in {\red red} ({\blue blue}) are new {\red published}
({\blue preliminary}) results since PDG2008  [as of March 12, 2010].
}
\end{center}

\scriptsize
%
\vspace{-0.7cm}
 ~\dag $E_\gamma > 2.0$ GeV; ~\ddag $M(\ell^+\ell^-)>0.2$ GeV/$c^2$
\end{table}

\begin{table}
\begin{center}
\caption{
Branching fractions of  inclusive $B$ decays
(in units of $10^{-6}$). Values in {\red red} ({\blue blue}) 
are new {\red published}
({\blue preliminary}) results since PDG2008  [as of March 12, 2010].
}

\vspace{0.3cm}
%
\end{center}
\vspace{-0.3cm}
~~~~~~~~~~~~~~~~~~\dag~$p^* > 2.34$ GeV;
~\S~$0.4 < M_{X_s} < 2.6$ GeV;
~\ddag~$2.0 < p^* < 2.7$ GeV
\end{table}

\begin{table}[!htb]
\begin{center}

\caption{
 Branching fractions of leptonic  $B$ decays
(in units of $10^{-6}$). Upper limits are at 90\% CL.
Values in {\red red} ({\blue blue}) are new {\red published}
({\blue preliminary}) results since PDG2008  [as of March 12, 2010].
}

%
\end{center}
\end{table}

\mysubsection{$B\to X_s\gamma$}
\label{sec:btosg}
\newcommand{\BFcnv}{\mathcal{B}^{\mathrm{cnv}}}
\newcommand{\Egamma}{E_{\gamma}}
\newcommand{\Emin}{E_{\mathrm{min}}}

The decay $b \to s\gamma$ proceeds through a process of 
flavor changing neutral current. Since the charged Higgs or SUSY particles may
contribute in the penguin loop, the branching fraction is sensitive to physics
beyond the Standard Model. Experimentally, the branching fraction is measured
using either a semi-inclusive or an inclusive approach. A minimum 
photon energy requirement is applied in the analysis and the branching fraction
is corrected based on the theoretical model for the photon energy spectrum 
(shape function). Where there are multiple experimental results from an 
experiment, we use only the ones that are independent for \babar\ and Belle
to avoid dealing with correlated errors. Furthermore, the 
model uncertainties from the shape function should be highly 
correlated but no proper action was made in our older averages. 
To perform the average with better precision and good accuracy, it is 
important to use as many experimental 
results as possible and to handle the shape function issue in a proper
way. In this note, we report the updated average of $b\to s\gamma$ branching
fraction by implementing a common  shape function.  

Several shape function schemes are commonly used.
Usually one is chosen to obtain the extrapolation factor, 
defined as the ratio of the $b\to s\gamma$ branching fractions 
with minimum photon energies above and at 1.6 GeV,
and the difference between various schemes are treated as the 
model uncertainty. O. Buchm\"uller and H. Fl\"acher have calculated the
extrapolation factors \cite{Buchmuller:2005zv}.
Table \ref{tab:factor} lists the extrapolation factors with various photon
energy cuts for three different schemes and the average. The appropriate
approach to average the experimental results is to first convert them 
according to the average extrapolation factors and then perform the average,
assuming that the errors of the extrapolation factors are 100\% correlated. 

\begin{table}[h]
\caption{Extrapolation factor in various scheme with various minimum
   photon energy requirement (in GeV).}
\begin{tabular}{lccccc} \hline\hline
Scheme & $E_\gamma < 1.7$ & $E_\gamma < 1.8$ & $E_\gamma < 1.9$ & 
$E_\gamma < 2.0$ & $E_\gamma < 2.242$ \\ \hline
Kinetic & $0.986\pm 0.001$ & $0.968\pm 0.002$ & $0.939\pm 0.005$ & $0.903\pm0.009$ & $0.656\pm 0.031$ \\
Neubert SF & $0.982\pm 0.002$ & $0.962\pm 0.004$ & $0.930\pm 0.008$ & $0.888\pm 0.014$ & $0.665\pm 0.035$ \\
Kagan-Neubert & $0.988\pm 0.002$ & $0.970\pm 0.005$ & $0.940\pm 0.009$ & $0.892\pm 0.014$ & $0.643\pm 0.033$\\ \hline
Average & $0.985\pm 0.004$ & $0.967\pm 0.006$ & $0.936\pm 0.010$ & $0.894\pm 0.016$ &  $0.655\pm 0.037$ \\ \hline
\end{tabular}
\label{tab:factor}
\end{table}        
                  
After surveying all available experimental results, the six shown in 
Table \ref{tab:measurement} are selected for the average.  They
have provided in their papers either the $b\to s\gamma$ branching fraction at 
a certain photon energy cut or the extrapolation factor used.
Therefore we are able to convert them to the values at $E_{\rm min}= 1.6$ 
GeV using the information in Table \ref{tab:factor}.  
In the inclusive and full hadronic tag analysis, a possible $B\to X_d\gamma$
contamination has been considered according to the expectation
of $(4.5\pm 0.3)$\%.
The central value is slightly higher than $4.0\%$ used in our 2006 average,
and the uncertainty shrinks by a factor of five, due to better understanding of
$|V_{td}/V_{ts}|$ from the $B_s$-$\overline{B}_s$ mixing and $B\to \rho/\omega
\;\gamma$ measurements. Compared to the other systematic uncertainties,
the error that arises from the $B\to X_d\gamma$ fraction is too small to be
considered.
We perform the average assuming that the systematic errors of the shape 
function and the $d\gamma$ fraction are correlated, and the other systematic
errors and the statistical errors are Gaussian and uncorrelated.     
The obtained average is 
${\cal B}(B\to X_s\gamma) = (355 \pm 24 \pm 9)\times 10^{-6}$ with
a $\chi^2$/DOF$= 0.85/5$, where the
errors are combined statistical and systematic, systematic due to the shape 
function. The second error is estimated to
be the difference of the average after simultaneously varying the central
value of each experimental result by $\pm 1\sigma$. Although  a small fraction
of events was used in multiple analyses in the same experiment, 
we neglect their statistical correlations. Some
other correlated systematic errors, such as photon detection and the background
suppression, are not considered in our new average. 

\begin{table}[h]
\caption{Reported branching fraction, minimum photon energy, branching fraction
at minimum photon energy  and converted branching fraction $\BFcnv$ for the 
decay $b\to s\gamma$. All the branching fractions are in units of $10^{-6}$.
The errors are, in order, statistical, systematic and theoretical (if exists)
for $\cal{B}$, and statistical, systematic and shape-function systematic
for $\BFcnv$.
Theoretical errors in $\cal{B}(\Egamma>\Emin)$ are merged
into the systematic error of $\BFcnv$ during conversion.
The CLEO measurement on the branching fraction at $\Emin$ includes
$B\to X_d \gamma$ events.\newline
} 

\label{tab:measurement}
\end{table}

\mysubsection{Baryonic decays}

\begin{table}
\begin{center}
\caption{
 Branching fractions of  baryonic $B^+$ decays (in units of $10^{-6}$).
Upper limits are at 90\% CL.
values in {\red red} ({\blue blue}) are new {\red published}
({\blue preliminary}) results since PDG2008  [as of March 12, 2010].
}
\end{center} 
\footnotesize
\begin{center}
\vspace{-0.3cm}
%
\end{center}
\vspace{-0.3cm}
\S Di-baryon mass is less than 2.85 GeV/$c^2$; 
$\dag$ Charmonium decays to $p\bar p$ have been statistically subtracted;\\
$\ddag$ The charmonium mass region has been vetoed;
$^1~\Theta(1540)^{++}\to K^+p$ (pentaquark candidate); \\
$^2$ Product BF --- daughter BF taken to be 100\% \\
\end{table}

\begin{table}
\begin{center}
\caption{
Branching fractions of  baryonic $B^0$ decays
(in units of $10^{-6}$). Upper limits are at 90\% CL.
values in {\red red} ({\blue blue}) are new {\red published}
({\blue preliminary}) results since PDG2008  [as of March 12, 2010].
}
\end{center}
\footnotesize
\begin{center}
\vspace{-0.3cm}

\end{center}
\hspace*{-1.0cm}

$\dag$ Charmonium decays to $p\bar p$ have been statistically subtracted;
$\ddag$ The charmonium mass region has been vetoed;
$^1~\Theta(1540)^+\to p K^0$ (pentaquark candidate);
$^2$ Product BF --- daughter BF taken to be 100\%.
\end{table}

\clearpage
\mysubsection{$B_s$ decays}

\begin{table}[h]
\begin{center}
\caption{
 $B_s$  branching fractions (in units of $10^{-6}$). 
Upper limits are at 90\% CL.
Values in {\red red} ({\blue blue}) are new {\red published}
({\blue preliminary}) results since PDG2008  [as of March 12, 2010].
}
\vskip 0.25cm


\end{center}
\vspace{-0.3cm}
\hspace{1.cm} $\dag$Relative BF converted to absolute BF
\end{table}

\begin{table}[h]
\begin{center}
\caption{
 $B_s$ rare relative branching fractions.
Values in {\red red} ({\blue blue}) are new {\red published}
({\blue preliminary}) results since PDG2008  [as of March 12, 2010].
}
\vskip 0.25cm

\scriptsize
%
\end{center}
\end{table}

\mysubsection{Charge asymmetries}

\begin{sidewaystable}[!htbp]
\parbox{8.2in}{\caption{
 $CP$ asymmetries for charmless hadronic charged $B$ decays (part I).
Values in {\red red} ({\blue blue}) are 
new {\red published}
({\blue preliminary}) results since PDG2008  [as of March 12, 2010].
}}

\scriptsize
%
\end{sidewaystable}

\begin{sidewaystable}[!htbp]
\begin{center}
\parbox{7.5in}{\caption{$CP$ asymmetries for 
charmless hadronic charged  $B$ decays (part II).
Values in {\red red} ({\blue blue}) are new {\red published}
({\blue preliminary}) results since PDG2008  [as of March 12, 2010].
}}

\scriptsize
%
\end{center}
\end{sidewaystable}

\begin{sidewaystable}[!htbp]
\begin{center}
\parbox{7.5in}{\caption{$CP$ asymmetries for 
charmless hadronic neutral $B$ decays.
Values in {\red red} ({\blue blue}) are new {\red published}
({\blue preliminary}) results since PDG2008  [as of March 12, 2010].
}}

\scriptsize
%
\end{center}

\vspace{-0.3cm}
\hspace{1.cm}
\dag~Measurements of time-dependent $CP$ asymmetries are listed in the section
of the Unitarity Triangle.

\end{sidewaystable}

\begin{sidewaystable}[!htbp]
\begin{center}

\parbox{7.5in}{\caption{
Charmless hadronic $CP$ asymmetries for $B^\pm/B^0$ admixtures.
Values in {\red red} ({\blue blue}) are new {\red published}
({\blue preliminary}) results since PDG2008  [as of March 12, 2010].
}}

\vspace*{ 0.4cm}
\scriptsize
\hspace*{-1.4cm}

\end{center}

\vspace{1cm}
\begin{center}
\parbox{7.5in}{\caption{
 $CP$ asymmetries for charmless hadronic neutral $B^0_S$ decays.
Values in {\red red} ({\blue blue}) are new {\red published}
({\blue preliminary}) results since PDG2008  [as of March 12, 2010].
}}

\vspace*{ 0.4cm}
%
\end{center}
\end{sidewaystable}

 \clearpage
\mysubsection{Polarization measurements}
%

%

\begin{table}[!htbp]
\caption{
 Longitudinal polarization fraction $f_L$ for $B^+$ decays.
Values in {\red red} ({\blue blue}) are new {\red published}
({\blue preliminary}) results since PDG2008 [as of March 12, 2010].
\vspace{0.3cm}
}

\begin{center}


\end{center}
\end{table}

\begin{table}[!htbp]
\caption{
Full angular analysis of $B^+\to\phi K^{*+}$.
Values in {\red red} ({\blue blue}) are new {\red published}
({\blue preliminary}) results since PDG2008  [as of March 12, 2010].
\vspace{0.3cm}
}

\begin{center}
%
\end{center}
\vspace{-0.3cm}
\hspace*{1.0cm}BR, $f_L$ and $A_{CP}$ are tabulated separately.
\end{table}
\clearpage


\begin{table}[!htbp]
\caption{
Longitudinal polarization fraction $f_L$ for $B^0$ decays.
Values in {\red red} ({\blue blue}) are new {\red published}
({\blue preliminary}) results since PDG2008  [as of March 12, 2010].
\vspace{0.3cm}
}

\begin{center}

\footnotesize
%
\end{center}
\end{table}

\begin{table}[!htbp]
\caption{
Full angular analysis of $B^0 \to \phi K^{*0}$.
Values in {\red red} ({\blue blue}) are new {\red published}
({\blue preliminary}) results since PDG2008  [as of March 12, 2010].
\vspace{0.3cm}
}

\begin{center}
\footnotesize
%
\end{center}
\vspace{-0.3cm}
\hspace{1.3cm}
BR, $f_L$ and $A_{CP}$ are tabulated separately.
\end{table}

\begin{table}[!htbp]
\caption{
Full angular analysis of $B^0 \to \phi K_2^*(1430)^0$.
Values in {\red red} ({\blue blue}) are new {\red published}
({\blue preliminary}) results since PDG2008  [as of March 12, 2010].
\vspace{0.3cm}
}

\begin{center}
\footnotesize
%
\end{center}
\vspace{-0.3cm}
\hspace{2.3cm}
BR, $f_L$ and $A_{CP}$ are tabulated separately.
\end{table}

\clearpage
\section{$D$ decays}
\label{sec:charm_physics}

\def\kbar{\overline{K}{}^{\,0}}
\def\dbar{\overline{D}{}^{\,0}}
\def\bbar{\overline{B}{}^{\,0}}
\def\cp{$CP$}
\def\cpv{$CPV$}
\def\ra{\!\rightarrow\!}
\def\ddbar{$D^0$-$\dbar$}
\def\ycp{$y^{}_{\rm CP}$}

\def\dklnu{$D^0\ra K^+\ell^-\nu$}
\def\dkpi{$D^0\ra K^+\pi^-$}
\def\dkk{$D^0\ra K^+K^-$}
\def\dpipi{$D^0\ra\pi^+\pi^-$}
\def\dkkpp{$D^0\ra K^+K^-/\pi^+\pi^-$}
\def\dkspp{$D^0\ra K^0_S\,\pi^+\pi^-$}
\def\dkskk{$D^0\ra K^0_S\,K^+ K^-$}

\def\dsphipi{$D^+_s\ra\phi\,\pi^+$}
\def\dsmunu{$D^+_s\ra\mu^+\nu$}
\def\dstaunu{$D^+_s\ra\tau^+\nu$}

\def\gevm{~GeV/$c^2$}
\def\gevp{~GeV/$c$}
\def\geve{~GeV}
\def\mevm{~MeV/$c^2$}
\def\meve{~MeV}

\def\babar{Babar}

\def\simge{\mathrel{%
   \rlap{\raise 0.511ex \hbox{$>$}}{\lower 0.511ex \hbox{$\sim$}}}}
\def\simle{\mathrel{
   \rlap{\raise 0.511ex \hbox{$<$}}{\lower 0.511ex \hbox{$\sim$}}}}

\newcommand{\Dnan}{\ensuremath{D_0^\ast(2400)^0}}
\newcommand{\Dtan}{\ensuremath{D_2^\ast(2460)^0}}
\newcommand{\Don}{\ensuremath{D_1(2420)^{0}}}
\newcommand{\Dopn}{\ensuremath{D_1(2430)^{0}}}
\newcommand{\Dnap}{\ensuremath{D_0^\ast(2400)^\pm}}
\newcommand{\Dtap}{\ensuremath{D_2^\ast(2460)^\pm}}
\newcommand{\Dop}{\ensuremath{D_1(2420)^{\pm}}}
\newcommand{\Dopp}{\ensuremath{D_1(2430)^{\pm}}}

\newcommand{\Dsa}{\ensuremath{D_s^{\ast\pm}}}
\newcommand{\Dsna}{\ensuremath{D_{s0}^\ast(2317)^{\pm}}}
\newcommand{\Dsop}{\ensuremath{D_{s1}(2460)^{\pm}}}
\newcommand{\Dso}{\ensuremath{D_{s1}(2536)^{\pm}}}
\newcommand{\Dst}{\ensuremath{D_{s2}(2573)^{\pm}}}
\newcommand{\Dsts}{\ensuremath{D_{sJ}(2700)^{\pm}}}
\newcommand{\Dste}{\ensuremath{D_{sJ}(2860)^{\pm}}}
\newcommand{\Dstsi}{\ensuremath{D_{sJ}(2632)^{\pm}}}

\newcommand{\citep}{\cite}

\subsection{\emph{$D^0$-$\dbar$} Mixing and \emph{\cp}\ Violation}

\subsubsection{Introduction}

In 2007, Belle~\cite{Staric:2007dt}and Babar~\cite{Aubert:2007wf} 
obtained the first evidence for $D^0$-$\dbar$ mixing, which 
had been searched for for more than two decades without success. 
These results were later confirmed by CDF~\cite{Aaltonen:2007uc}.
There are now numerous measurements of $D^0$-$\dbar$ mixing, 
with various levels of sensitivity. All the results can be 
combined to yield world average (WA) values for the mixing 
parameters $x\equiv(m^{}_1-m^{}_2)/\Gamma$ and 
$y\equiv (\Gamma^{}_1-\Gamma^{}_2)/(2\Gamma)$, where 
$m^{}_1,\,m^{}_2$ and $\Gamma^{}_1,\,\Gamma^{}_2$ are
the masses and decay widths for the mass eigenstates
$D^{}_1\equiv p|D^0\rangle-q|\dbar\rangle$ and
$D^{}_2\equiv p|D^0\rangle+q|\dbar\rangle$,
and $\Gamma=(\Gamma^{}_1+\Gamma^{}_2)/2$. Here we use 
the phase convention $CP|D^0\rangle=-|\dbar\rangle$,
$CP|\dbar\rangle=-|D^0\rangle$; in the absence of \cp\ violation (\cpv), 
$q/p\!=\!1$, $D^{}_1$ is \cp-even, and $D^{}_2$ is \cp-odd.

The WA values are calculated by performing a global fit to
more than two dozen measured observables. Assuming no \cpv, 
the fit parameters are $x$, $y$, $\delta$ 
(the strong phase difference between amplitudes 
${\cal A}(\dbar\ra K^+\pi^-)$ and ${\cal A}(D^0\ra K^+\pi^-)$), 
a strong phase $\delta^{}_{K\pi\pi}$ entering
$D^0\ra K^+\pi^-\pi^0$ decays, and 
$R^{}_D\equiv\left|{\cal A}(D^0\ra K^+\pi^-)/
              {\cal A}(\dbar\ra K^+\pi^-)\right|^2$. 
To account for possible \cpv, three additional parameters 
are added: $|q/p|$, $\phi\equiv {\rm Arg}(q/p)$, and
$A^{}_D\equiv (R^+_D-R^-_D)/(R^+_D+R^-_D)$, where the $+\,(-)$
superscript corresponds to $D^0\,(\dbar)$ decays. 

The observables used are from measurements of \dklnu, 
\dkkpp, \dkpi, $D^0\ra K^+\pi^-\pi^0$, 
\dkspp, and \dkskk\ decays; and from double-tagged branching 
fractions measured at the $\psi(3770)$ resonance. Correlations 
among observables are accounted for by using covariance matrices 
provided by the experimental collaborations. Systematic errors 
among different experiments are assumed uncorrelated as no 
significant correlations have been identified.
We have checked this method with a second method that adds
together three-dimensional log-likelihood functions 
for $x$, $y$, and $\delta$ obtained from several analyses;
this combination accounts for non-Gaussian errors.
When both methods are applied to the same set of 
observables and data, equivalent results are obtained.

Mixing in heavy flavor systems such as those of $B^0$ and $B^0_s$ 
is governed by the short-distance box diagram. In the $D^0$ system,
however, this diagram is doubly-Cabibbo-suppressed relative to 
amplitudes dominating the decay width, and it is also GIM-suppressed.
Thus the short-distance mixing rate is tiny, and $D^0$-$\dbar$ 
mixing is expected to be dominated by long-distance processes. 
These are difficult to calculate reliably, and theoretical estimates 
for $x$ and $y$ range over two-three orders of 
magnitude~\cite{Bigi:2000wn,Petrov:2003un,Petrov:2004rf,Falk:2004wg,
Bobrowski:2010xg}.

With the exception of $\psi(3770)\ra DD$ measurements, all methods 
identify the flavor of the $D^0$ or $\dbar$ when produced by 
reconstructing the decay $D^{*+}\ra D^0\pi^+$ or $D^{*-}\ra\dbar\pi^-$; 
the charge of the pion 
(which has low momentum and is often referred to 
as the ``soft'' pion $\pi^{}_s$)
identifies the $D$ flavor. For signal 
decays, $M^{}_{D^*}-M^{}_{D^0}-M^{}_{\pi^+}\equiv Q\approx 6$\meve, 
which is close to the threshold; thus analyses typically
require that the reconstructed $Q$ be small to suppress backgrounds. 
For time-dependent measurements, the $D^0$ decay time is 
calculated as $(d/p)\times M^{}_{D^0}$, where $d$ is
the distance between the $D^*$ and $D^0$ decay vertices and 
$p$ is the $D^0$ momentum. The $D^*$ vertex position is 
taken to be at the primary vertex for $\bar{p}p$ collider 
experiments~\cite{Aaltonen:2007uc}, and at the intersection 
of the $D^0$ momentum vector with the beamspot profile for 
$e^+e^-$ experiments.

\subsubsection{Input Observables}

The global fit determines central values and errors for
the underlying parameters using a $\chi^2$ statistic
constructed from 30 observables. The
parameters are $x,\,y,\,\delta,\,R^{}_D,
A^{}_D,\,|q/p|,\,\phi$, and $\delta^{}_{K\pi\pi}$.
Parameters $x$ and $y$ govern mixing, while
$A^{}_D$, $|q/p|$, and $\phi$ govern \cpv.
The parameter $\delta$ is the strong 
phase difference between the amplitudes 
${\cal A}(\dbar\ra K^+\pi^-)$ and 
${\cal A}(D^0\ra K^+\pi^-)$, and $\delta^{}_{K\pi\pi}$ 
is the analogous strong phase difference between 
the amplitudes ${\cal A}(\dbar\ra K^+\rho^-)$ and 
${\cal A}(D^0\ra K^+\rho^-)$.\footnote{In the
$D\ra K^+\pi^-\pi^0$ Dalitz plot analysis yielding 
sensitivity to $x$ and $y$, the
$\dbar\ra K^+\pi^-\pi^0$ isobar phases are determined 
relative to that for ${\cal A}(\dbar\ra K^+\rho^-)$, and 
the $D^0\ra K^+\pi^-\pi^0$ isobar phases are determined 
relative to that for ${\cal A}(D^0\ra K^+\rho^-)$. 
As the $\dbar$ and $D^0$ Dalitz plots are fit independently, 
the phase difference $\delta^{}_{K\pi\pi}$ between the
two ``normalizing'' amplitudes cannot be determined
from these fits.}

\begin{figure}
\begin{center}
\includegraphics[width=4.2in]{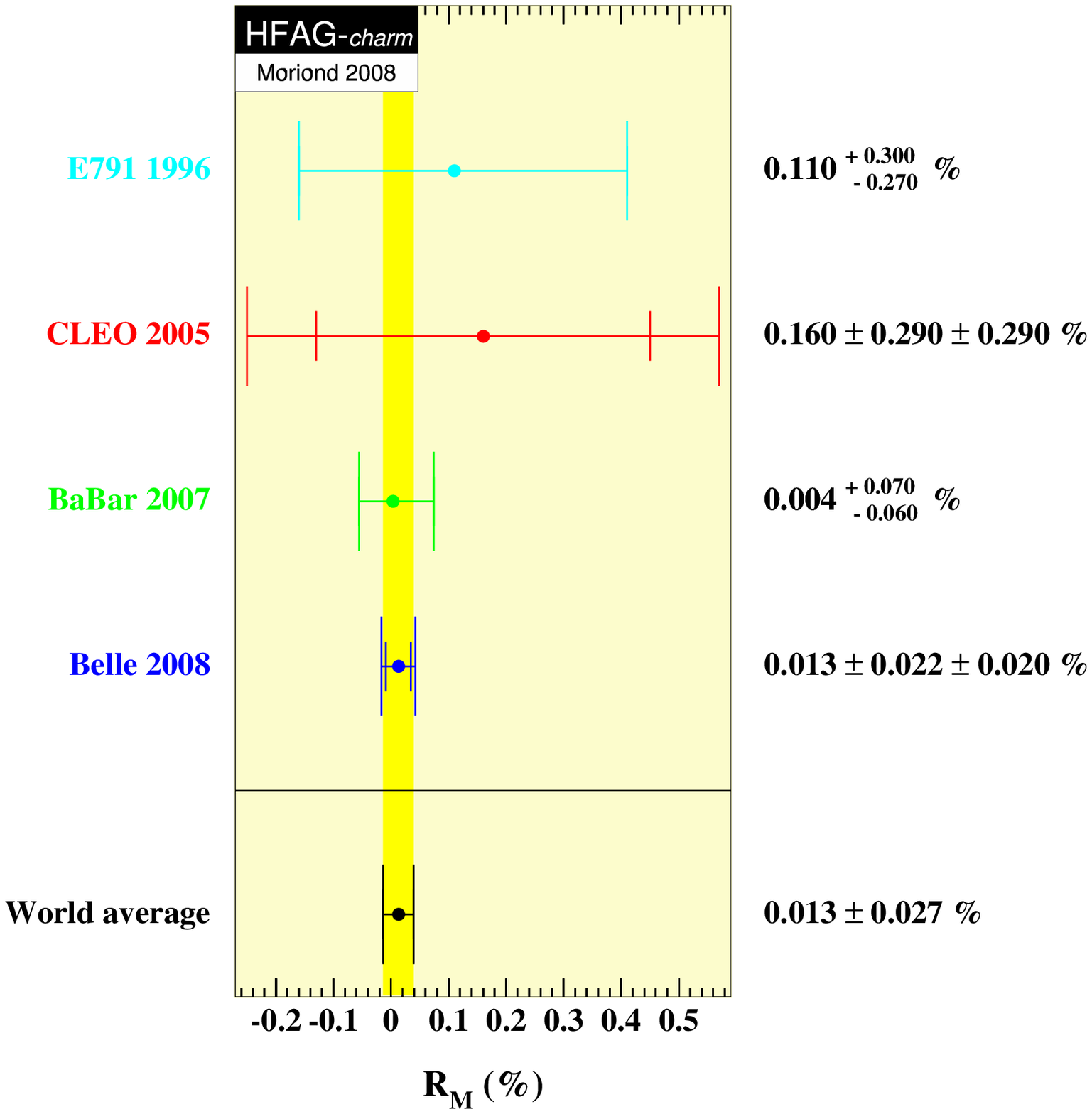}
\end{center}
\vskip-0.20in
\caption{\label{fig:rm_semi}
WA value of $R^{}_M$ from Ref.~\cite{HFAG_charm:webpage},
as calculated from $D^0\ra K^+\ell^-\nu$ 
measurements~\cite{Aitala:1996vz,Cawlfield:2005ze,Aubert:2007aa,Bitenc:2008bk}. }
\end{figure}

All input values are listed in Tables~\ref{tab:observables1} 
and~\ref{tab:observables2}. 
The observable $R^{}_M=(x^2+y^2)/2$ calculated from \dklnu\ 
decays~\cite{Aitala:1996vz,Cawlfield:2005ze,Aubert:2007aa,Bitenc:2008bk}
is the WA value calculated by HFAG~\cite{HFAG_charm:webpage} 
(see Fig.~\ref{fig:rm_semi}). The observables $y^{}_{CP}$ and 
$A^{}_\Gamma$ are also HFAG WA values~\cite{HFAG_charm:webpage}
(see Fig.~\ref{fig:ycp}).
The \dkpi\ observables used are from Belle~\cite{Zhang:2006dp}, 
Babar~\cite{Aubert:2007wf}, and CDF~\cite{Aaltonen:2007uc};
earlier measurements have much less precision and are not used.
The observables from \dkspp\ decays for no-\cpv\ are from 
Belle~\cite{Abe:2007rd} and BaBar~\cite{delAmoSanchez:2010xz}, 
but for the \cpv-allowed case only Belle measurements~\cite{Abe:2007rd} 
are available. The $D^0\ra K^+\pi^-\pi^0$ results are from 
Babar~\cite{Aubert:2008zh}, and the $\psi(3770)\ra\overline{D}D$ 
results are from CLEOc~\cite{Asner:2008ft}.

The relationships between the observables and the fitted
parameters are listed in Table~\ref{tab:relationships}. 
For each set of correlated observables we construct a
difference vector $\vec{V}$; e.g., for 
$D^0\ra K^0_S\,\pi^+\pi^-$ decays
$\vec{V}=(\Delta x,\Delta y,\Delta |q/p|,\Delta \phi)$,
where $\Delta$ represents the difference between the 
measured value and the fitted parameter value. The 
contribution of a set of observables to the $\chi^2$ 
is calculated as $\vec{V}\cdot (M^{-1})\cdot\vec{V}^T$, 
where $M^{-1}$ is the inverse of the covariance matrix 
for the measurement. All covariance matrices 
are listed in Tables~\ref{tab:observables1} 
and~\ref{tab:observables2}.

\begin{figure}
\begin{center}
\vbox{
\includegraphics[width=4.2in]{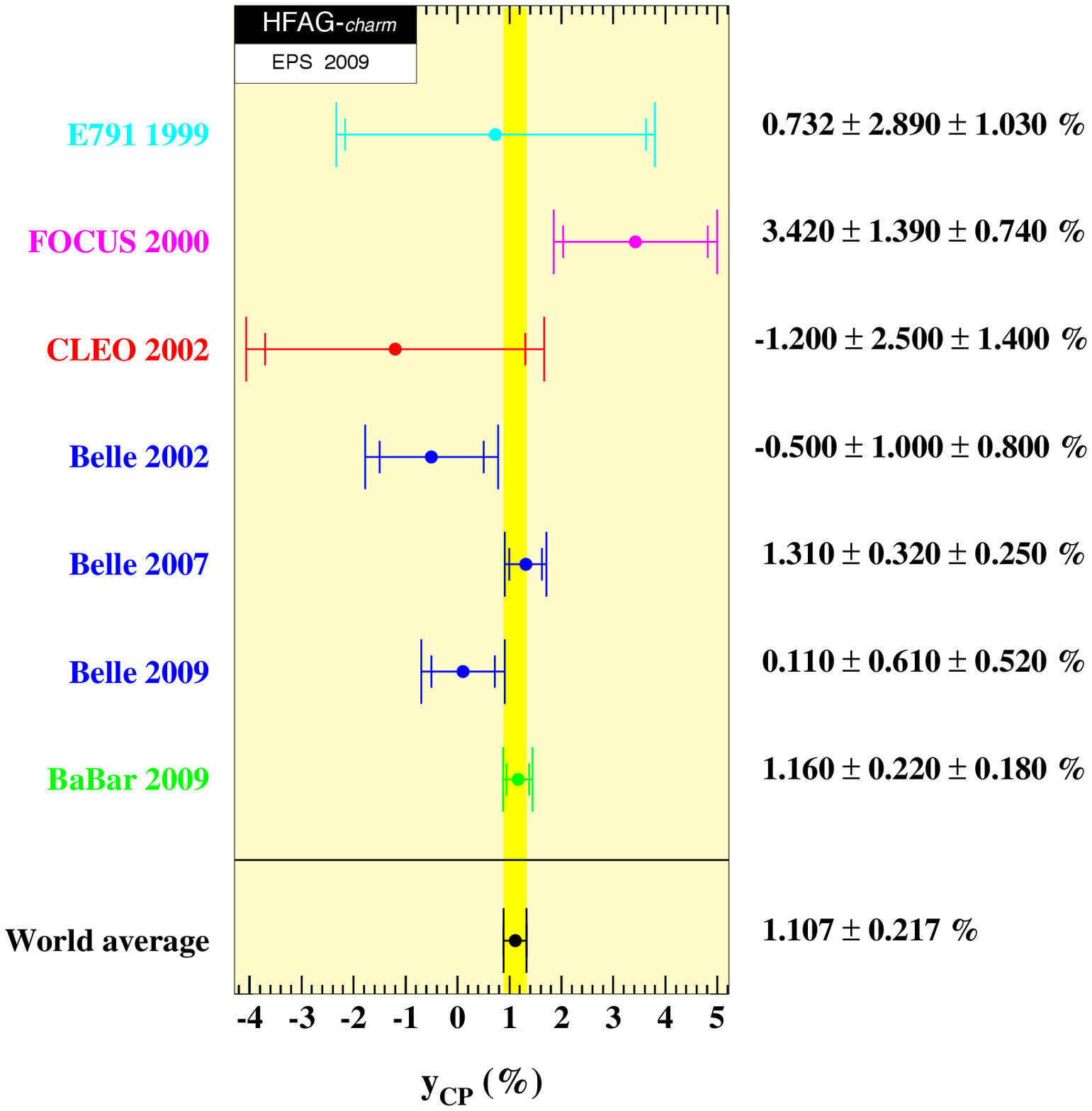}
\vskip0.10in
\includegraphics[width=4.2in]{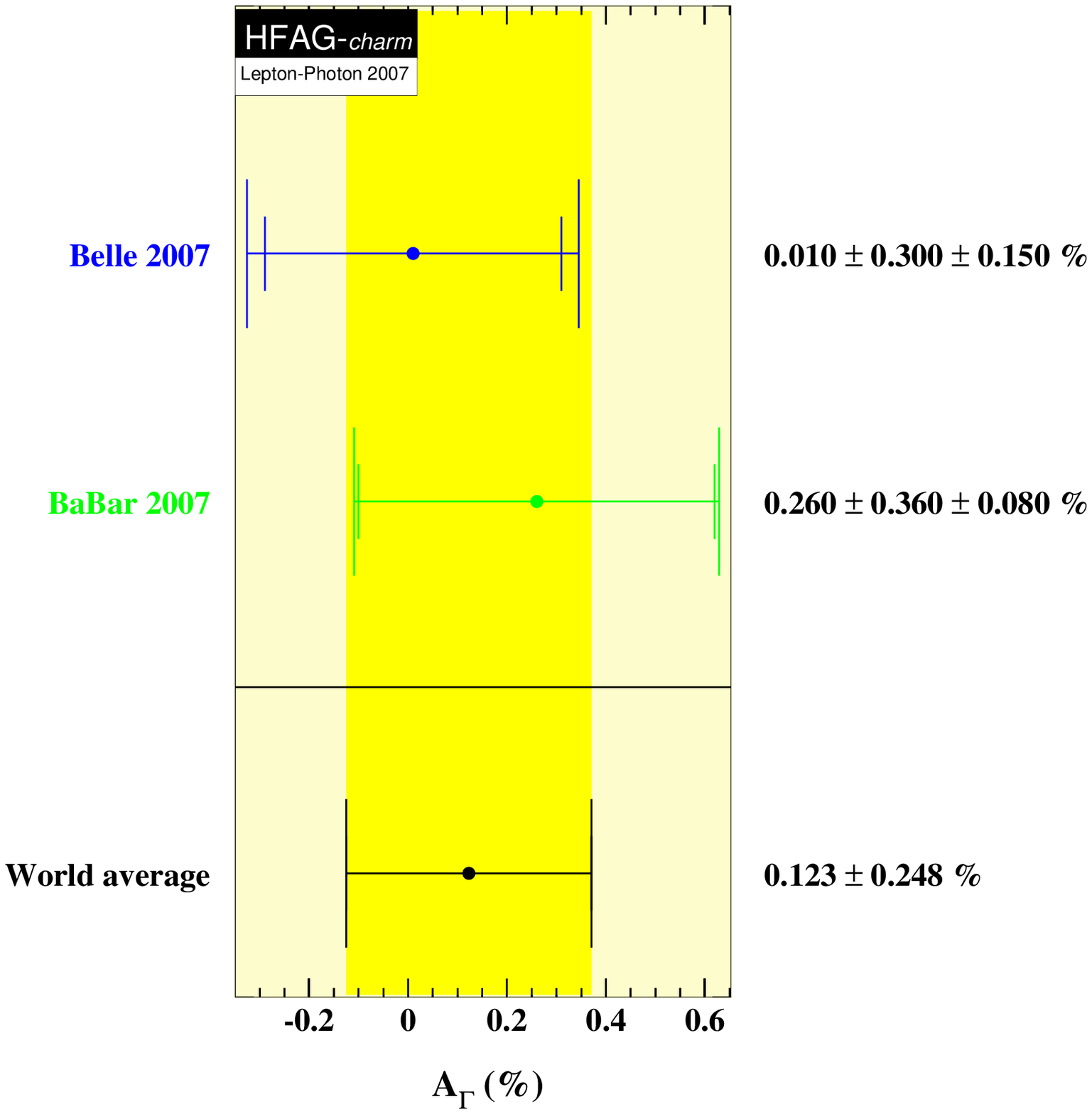}
}
\end{center}
\vskip-0.20in
\caption{\label{fig:ycp}
WA values of $y^{}_{CP}$ (top) and $A^{}_\Gamma$ (bottom)
from Ref.~\cite{HFAG_charm:webpage}, as calculated from 
\dkkpp\ measurements~\cite{Staric:2007dt,Aitala:1999dt,
Link:2000cu,Csorna:2001ww,Aubert:2007en,Zupanc:2009sy}.  }
\end{figure}

\begin{table}
\renewcommand{\arraystretch}{1.3}
\renewcommand{\arraycolsep}{0.02in}
\renewcommand{\tabcolsep}{0.05in}
\caption{\label{tab:observables1}
All observables except those for \dkpi\ 
used for the global fit, from 
Refs.~\cite{Staric:2007dt,
Aitala:1996vz,
Cawlfield:2005ze,
Aubert:2007aa,
Bitenc:2008bk,
Abe:2007rd,
delAmoSanchez:2010xz,
Aubert:2008zh,
Asner:2008ft,
Aitala:1999dt,
Link:2000cu,
Csorna:2001ww,
Aubert:2007en}.}
\vspace*{6pt}
\footnotesize

\end{table}

\begin{table}
\renewcommand{\arraystretch}{1.3}
\renewcommand{\arraycolsep}{0.02in}
\caption{\label{tab:observables2}
\dkpi\ observables used for the global fit, from
Refs.~\cite{Aubert:2007wf,Aaltonen:2007uc,Zhang:2006dp}.}
\vspace*{6pt}
\footnotesize
\begin{center}
%
\end{center}
\end{table}

\begin{table}
\renewcommand{\arraycolsep}{0.02in}
\renewcommand{\arraystretch}{1.3}
\begin{center}
\caption{\label{tab:relationships}
Left: decay modes used to determine fitted parameters 
$x,\,y,\,\delta,\,\delta^{}_{K\pi\pi},\,R^{}_D,\,A^{}_D,\,|q/p|$, and $\phi$.
Middle: the observables measured for each decay mode. Right: the 
relationships between the observables measured and the fitted parameters.}
\vspace*{6pt}
\footnotesize
%
$ \\
\hline
\end{tabular}
\end{center}
\end{table}

\subsubsection{Fit results}

The global fit uses MINUIT with the MIGRAD minimizer, 
and all errors are obtained from MINOS~\cite{MINUIT:webpage}. 
Three separate fits are performed: 
{\it (a)}\ assuming \cp\ conservation ($A^{}_D$ 
and $\phi$ are fixed to zero, $|q/p|$ is fixed to one);
{\it (b)}\ assuming no direct \cpv\ ($A^{}_D$ is 
fixed to zero); and
{\it (c)}\ allowing full \cpv\ (all parameters
floated). The results are listed in 
Table~\ref{tab:results}. For the \cpv-allowed fit,
individual contributions to the $\chi^2$ are listed 
in Table~\ref{tab:results_chi2}. The total $\chi^2$ 
is 31.9 for $30-8=22$ degrees of freedom; this 
corresponds to a confidence level of~0.08, which 
is small but acceptable given the variety of 
measurements and systematic uncertainties.

Confidence contours in the two dimensions $(x,y)$ or 
in $(|q/p|,\phi)$ are obtained by letting, for any point in the
two-dimensional plane, all other fitted parameters take their 
preferred values. The resulting $1\sigma$-$5\sigma$ contours 
are shown in Fig.~\ref{fig:contours_ncpv} for the \cp-conserving
case, and in Fig.~\ref{fig:contours_cpv} for the \cpv-allowed 
case. The contours are determined from the increase of the
$\chi^2$ above the minimum value.
One observes that the $(x,y)$ contours for the no-\cpv\ fit 
are almost identical to those for the \cpv-allowed fit. 
In the latter fit, the
$\chi^2$ at the no-mixing point $(x,y)\!=\!(0,0)$ is 110 units above 
the minimum value; for two degrees of freedom this has a confidence 
level corresponding to $10.2\sigma$. Thus, no mixing is excluded 
at this high level. In the $(|q/p|,\phi)$ plot, the point $(1,0)$ 
is within the $1\sigma$ contour; thus the data is consistent 
with \cp\ conservation.

One-dimensional confidence curves for individual parameters 
are obtained by letting, for any value of the parameter, all other 
fitted parameters take their preferred values. The resulting
functions $\Delta\chi^2=\chi^2-\chi^2_{\rm min}$ ($\chi^2_{\rm min}$
is the minimum value) are shown in Fig.~\ref{fig:1dlikelihood}.
The points where $\Delta\chi^2=3.84$ determine 95\% C.L. intervals 
for the parameters; these intervals are listed in Table~\ref{tab:results}.

\begin{figure}
\begin{center}
\includegraphics[width=4.2in]{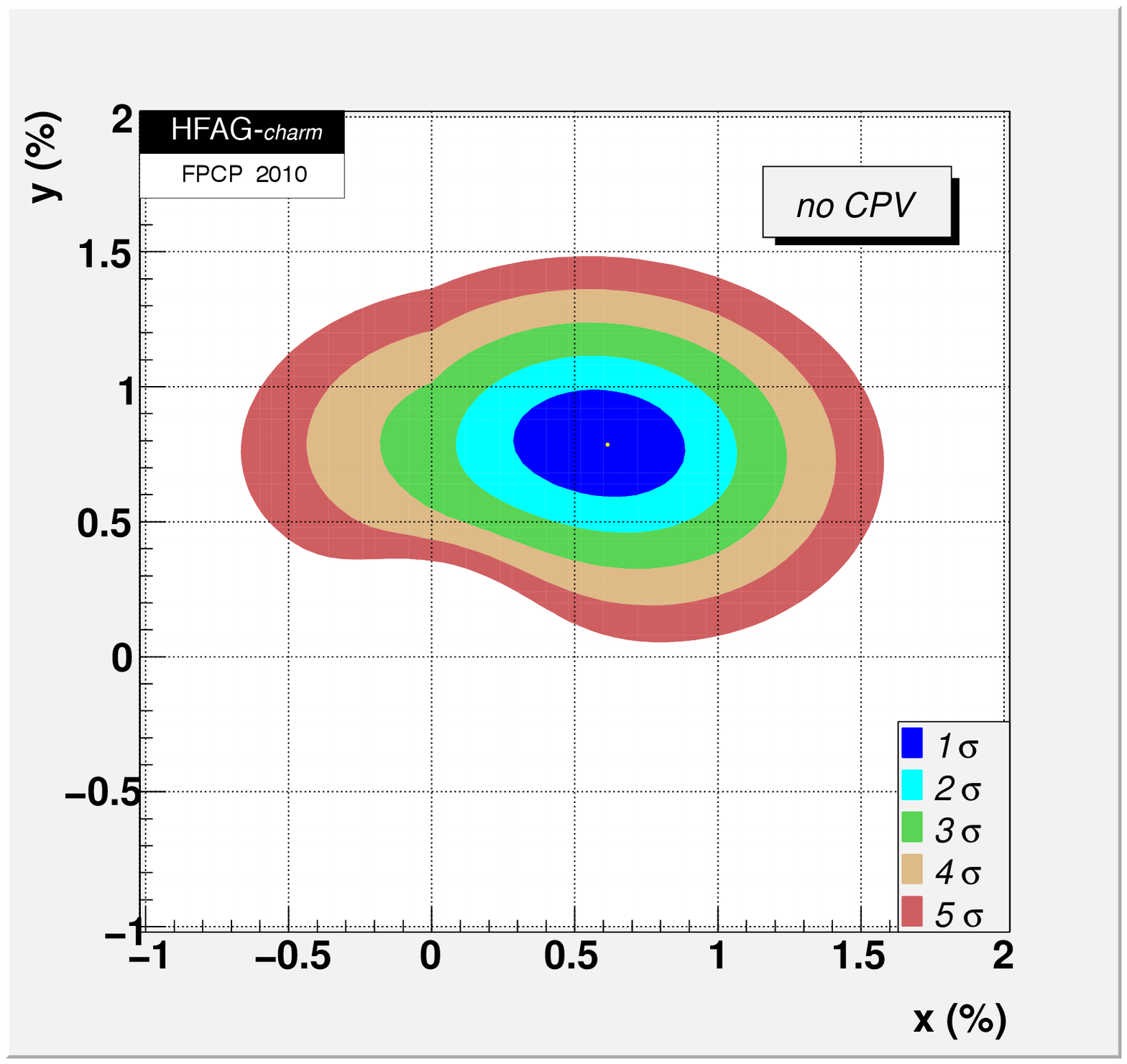}
\end{center}
\vskip-0.20in
\caption{\label{fig:contours_ncpv}
Two-dimensional contours for mixing parameters $(x,y)$, for no \cpv. }
\end{figure}

\begin{figure}
\begin{center}
\vbox{
\includegraphics[width=4.2in]{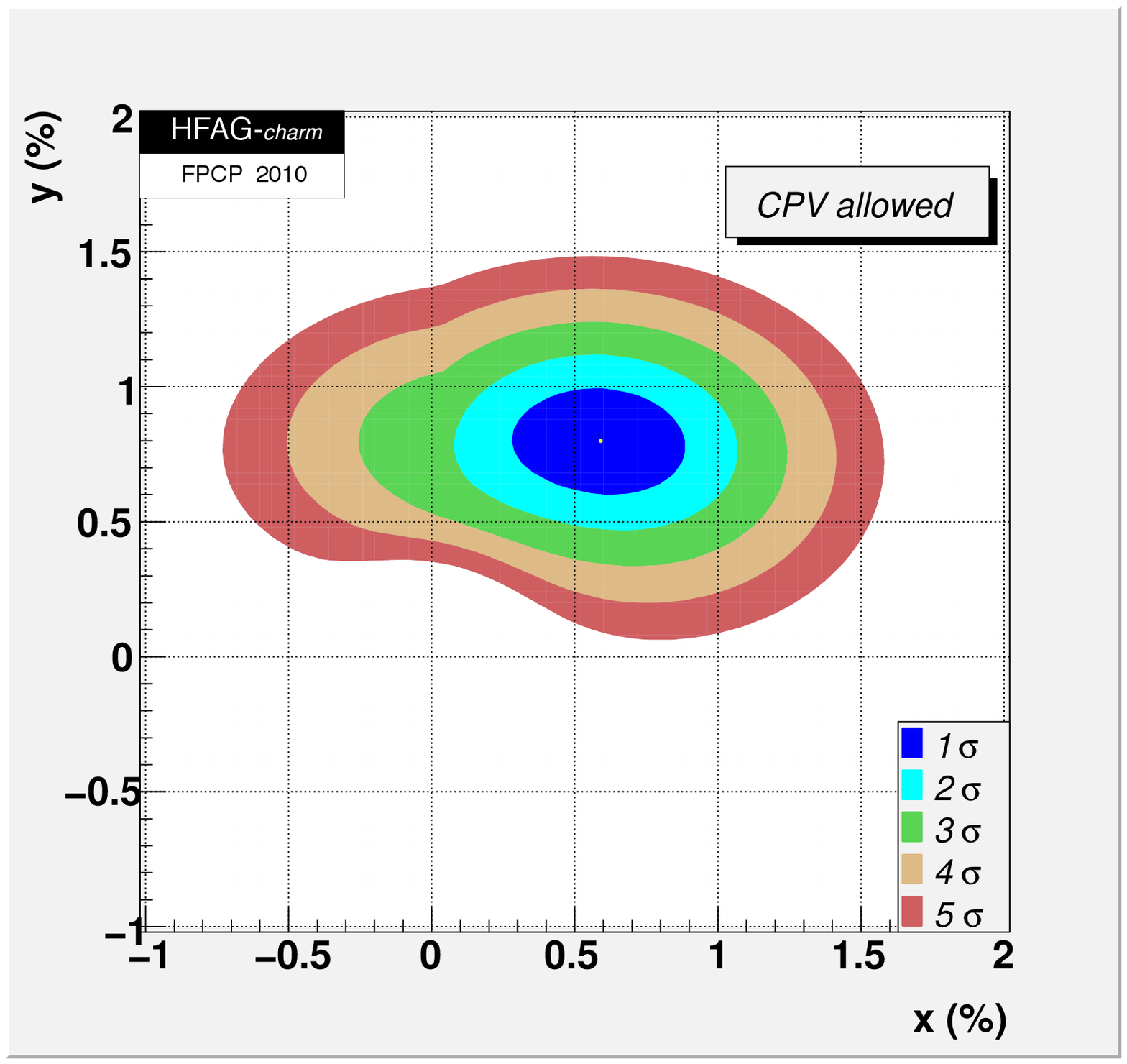}
\vskip0.10in
\includegraphics[width=4.2in]{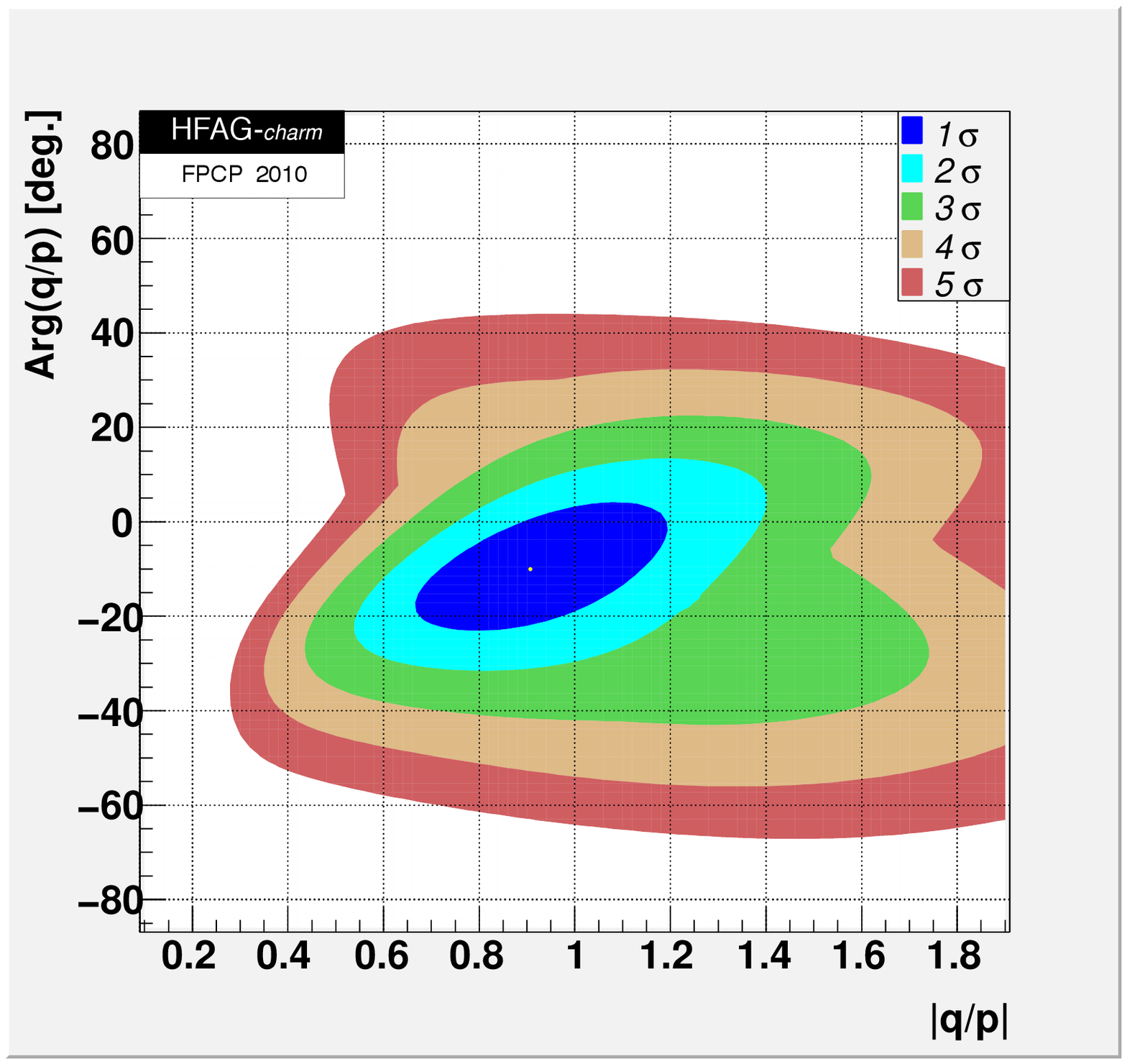}
}
\end{center}
\vskip-0.10in
\caption{\label{fig:contours_cpv}
Two-dimensional contours for parameters $(x,y)$ (top) 
and $(|q/p|,\phi)$ (bottom), allowing for \cpv.}
\end{figure}

\begin{figure}
\begin{center}
\hbox{\hskip0.50in
\includegraphics[width=72mm]{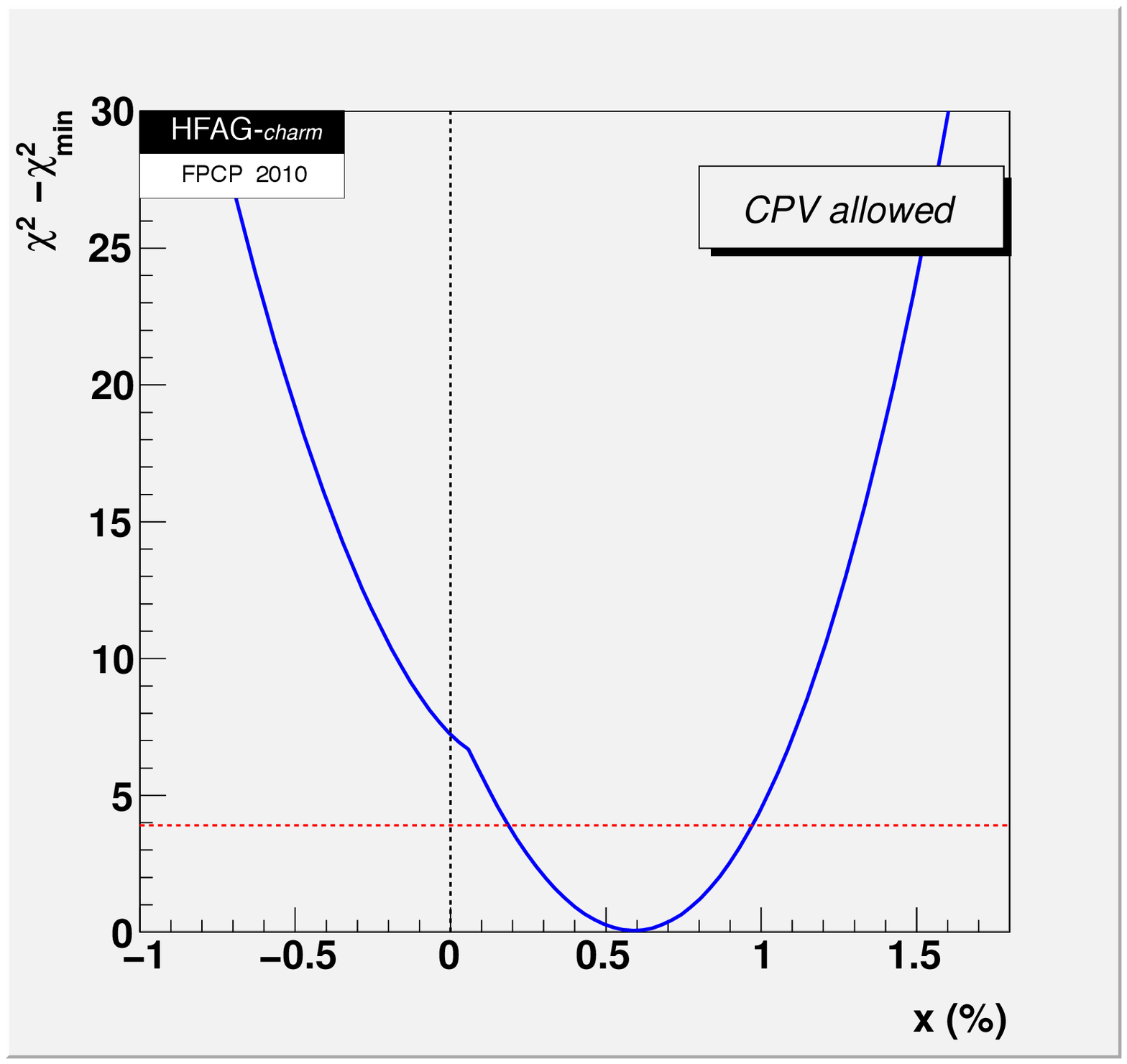}
\hskip0.20in
\includegraphics[width=72mm]{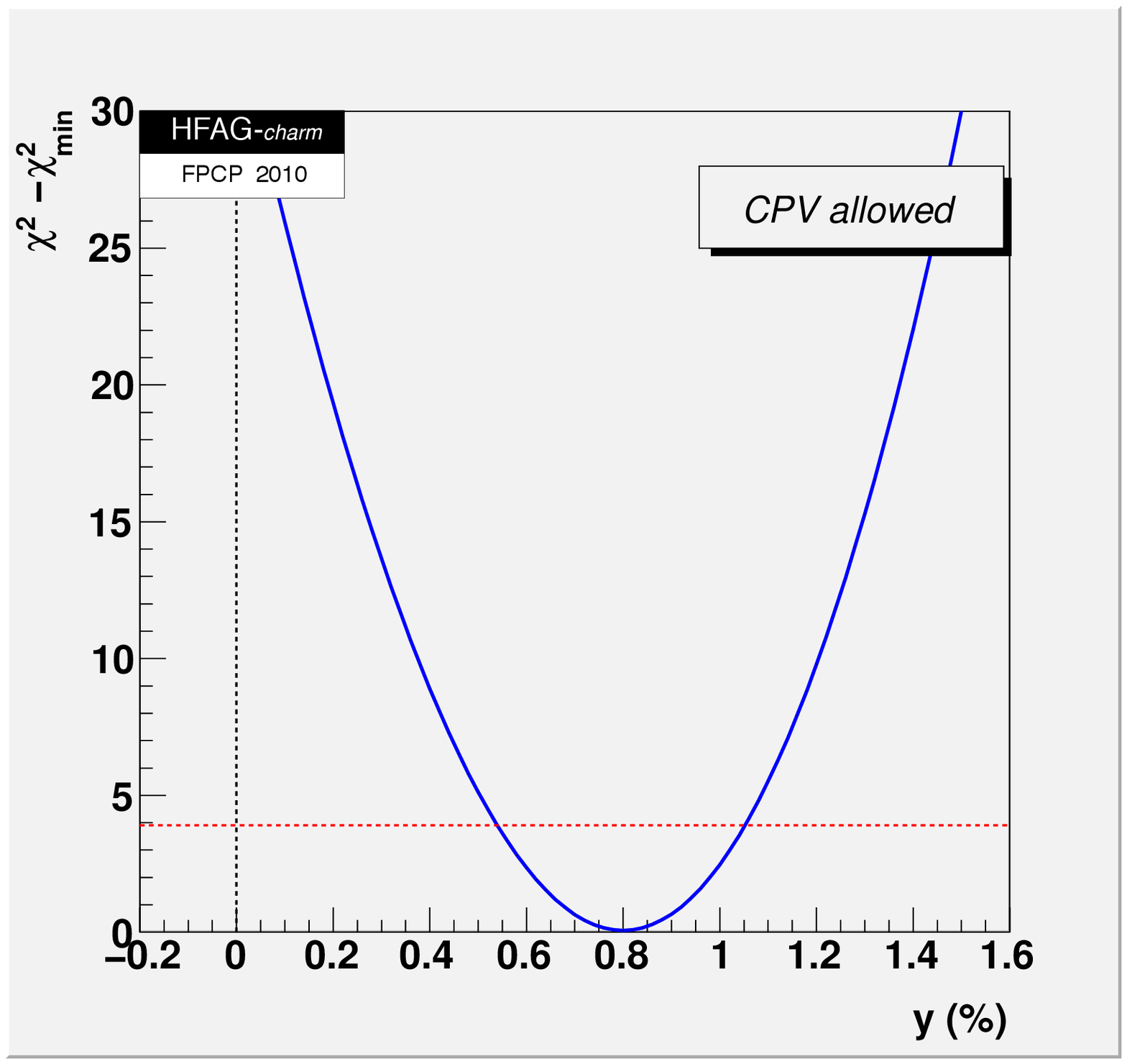}}
\hbox{\hskip0.50in
\includegraphics[width=72mm]{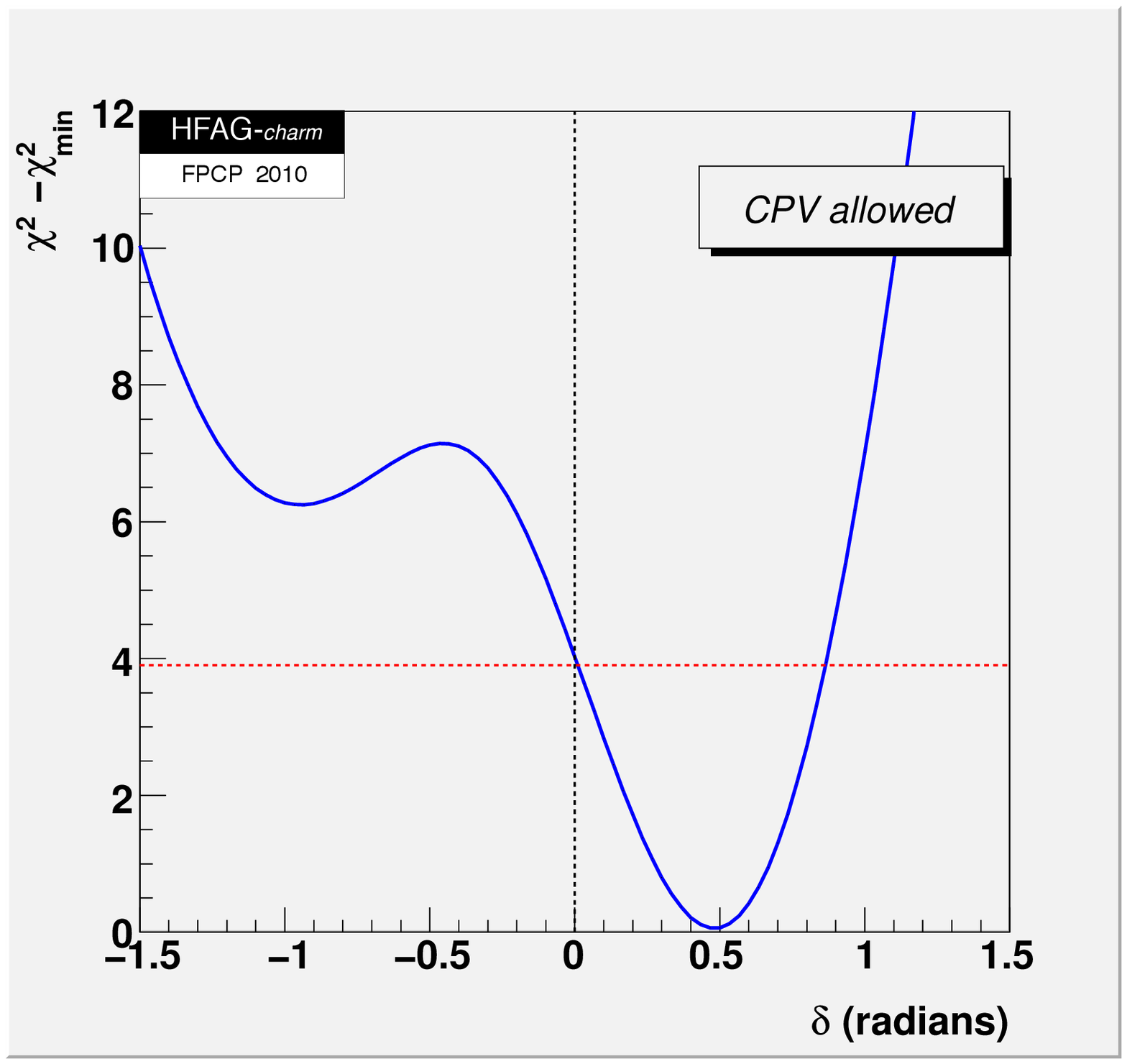}
\hskip0.20in
\includegraphics[width=72mm]{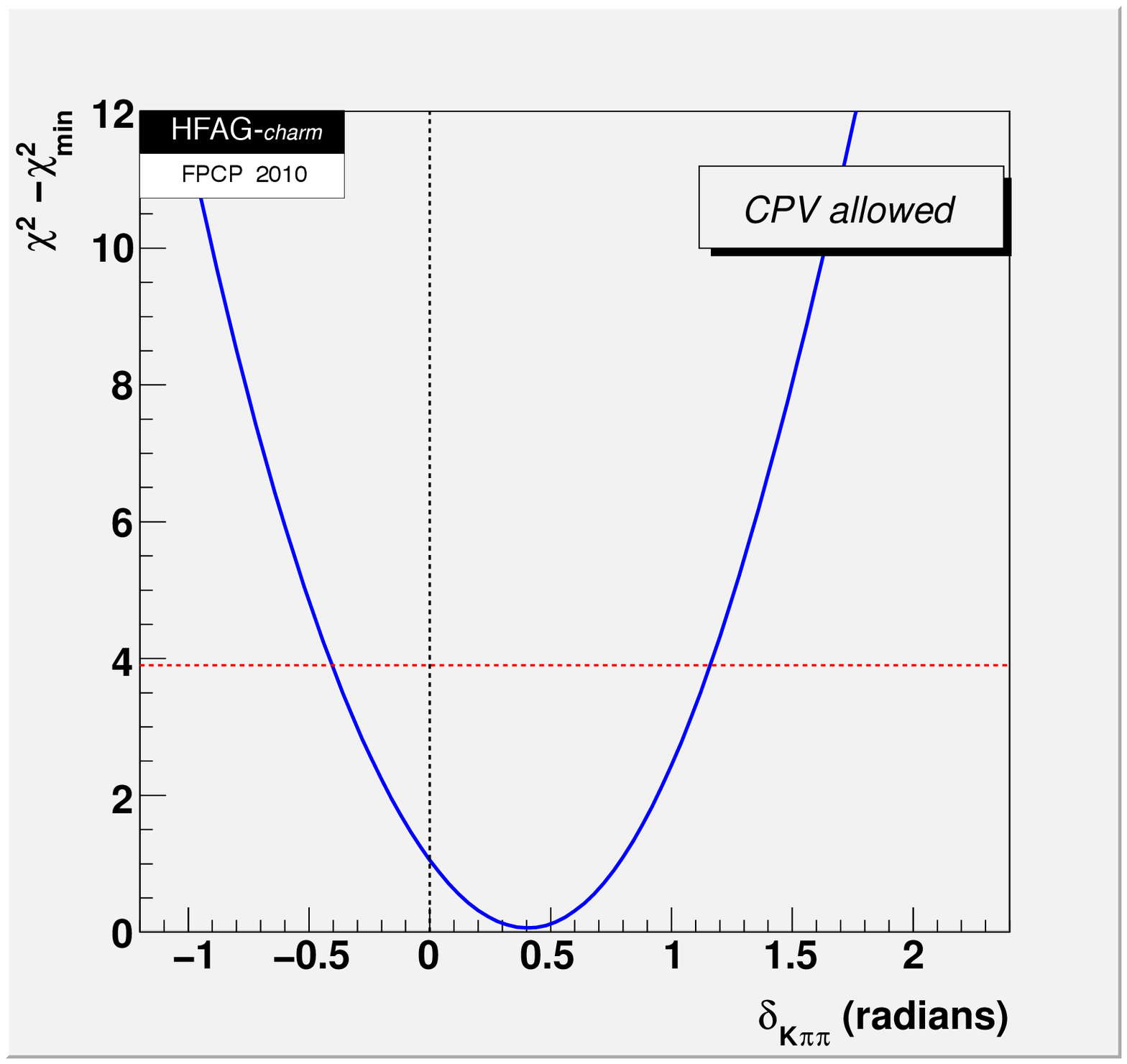}}
\hbox{\hskip0.50in
\includegraphics[width=72mm]{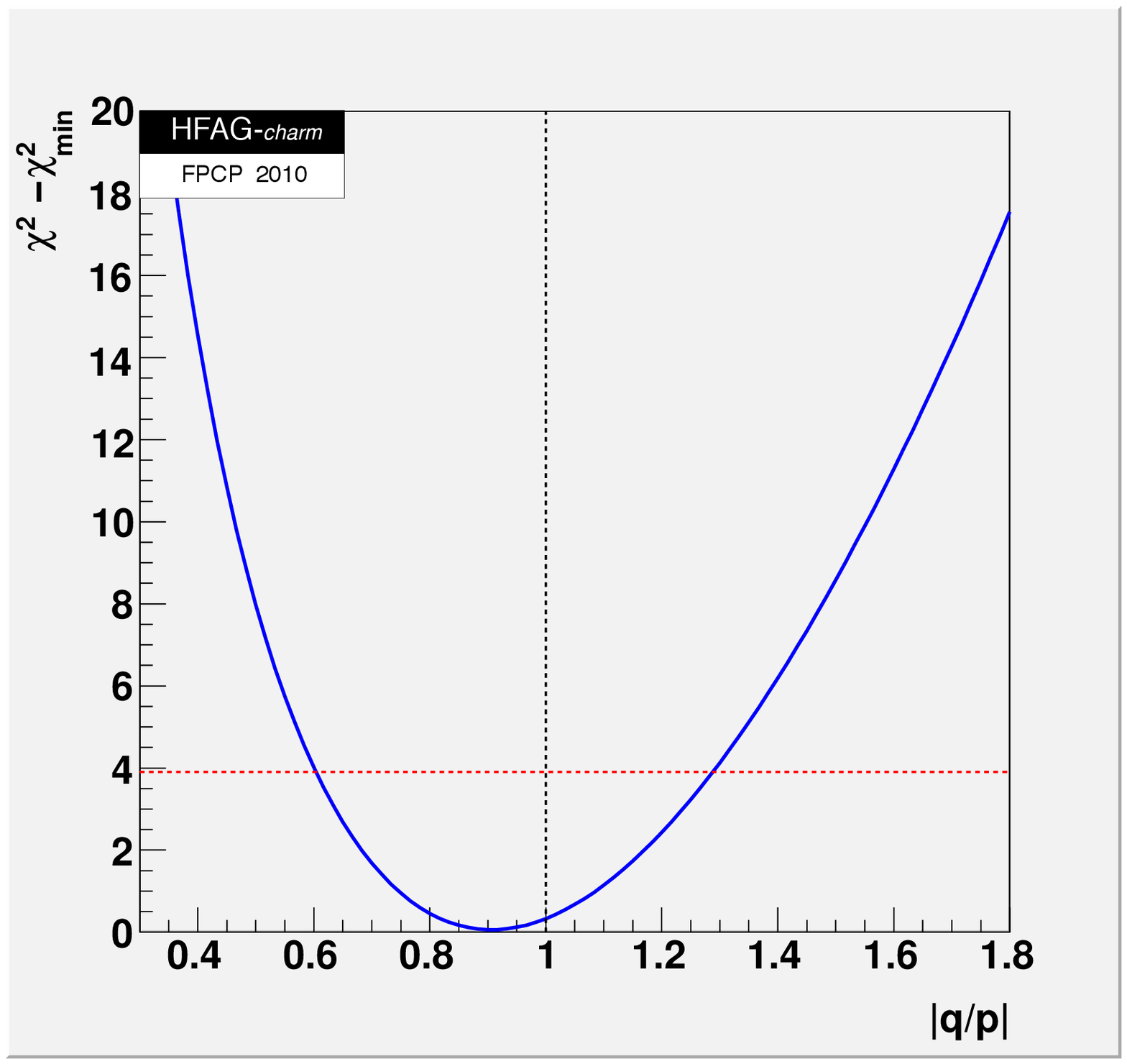}
\hskip0.20in
\includegraphics[width=72mm]{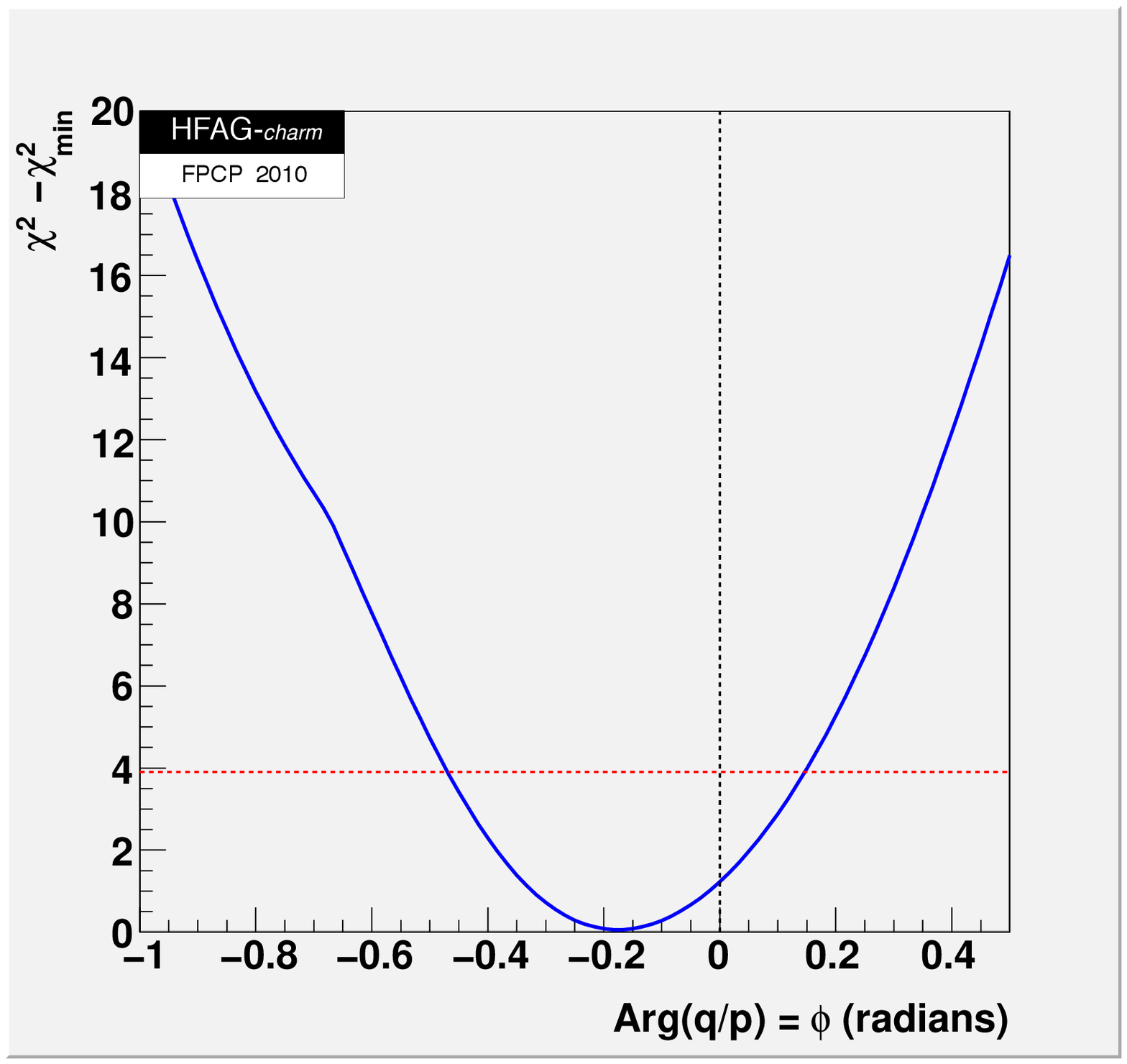}}
\end{center}
\vskip-0.30in
\caption{\label{fig:1dlikelihood}
The function $\Delta\chi^2=\chi^2-\chi^2_{\rm min}$ 
for fitted parameters
$x,\,y,\,\delta,\,\delta^{}_{K\pi\pi},\,|q/p|$, and $\phi$.
The points where $\Delta\chi^2=3.84$ (denoted by the dashed 
horizontal line) determine a 95\% C.L. interval. }
\end{figure}

\begin{table}
\renewcommand{\arraystretch}{1.4}
\begin{center}
\caption{\label{tab:results}
Results of the global fit for different assumptions concerning~\cpv.}
\vspace*{6pt}
\footnotesize

\end{center}
\end{table}

\begin{table}
\renewcommand{\arraystretch}{1.4}
\begin{center}
\caption{\label{tab:results_chi2}
Individual contributions to the $\chi^2$ for the \cpv-allowed fit.}
\vspace*{6pt}
\footnotesize
%
\end{center}
\end{table}


\subsubsection{Conclusions}

From the fit results listed in Table~\ref{tab:results}
and shown in Figs.~\ref{fig:contours_cpv} and \ref{fig:1dlikelihood},
we conclude the following:
\begin{itemize}
\item the experimental data consistently indicate that 
$D^0$ mesons undergo mixing. The no-mixing point $x=y=0$
is excluded at $10.2\sigma$. The parameter $x$ differs from
zero by $2.5\sigma$, and $y$ differs from zero by
$5.7\sigma$. This mixing is presumably dominated 
by long-distance processes, which are difficult to calculate.
Unless it turns out that $|x|\gg |y|$~\cite{Bigi:2000wn},
which is not indicated, it will probably be difficult to 
identify new physics from mixing alone.
\item Since \ycp\ is positive, the \cp-even state is shorter-lived,
as in the $K^0$-$\kbar$ system. However, since $x$ also appears
to be positive, the \cp-even state is heavier, 
unlike in the $K^0$-$\kbar$ system.
\item There is no evidence (yet) for \cpv\ in the $D^0$-$\dbar$ system.
Observing \cpv\ at the current level of sensitivity would indicate 
new physics.
\end{itemize}

\clearpage
\subsection{Excited \emph{$D_{(s)}$} Mesons}

Tables \ref{tab1}--\ref{tab3} represent a summary of recent results. 
For a complete list of related publications, see Ref.~\cite{PDG_2010}.
All upper limits (U.L.) correspond to 90\% confidence (C.L.) unless 
otherwise noted. The significances listed are approximate; they
are calculated as either
$\sqrt{-2\Delta\log{\cal{L}}}$ or $\sqrt{\Delta\chi^2}$, where 
$\Delta$ represents the change in the corresponding minimized 
function between two hypotheses, e.g., those for different spin states.

The broad charged $J^P\!=\!1^+$ $c\bar{d}$ state is denoted 
$D_1(2430)^+$, although it has not yet been observed. The masses and
widths of narrow states $\Dop$, $\Don$, $\Dtan$, $\Dtap$ are
well-measured, and thus only their averages 
are given\cite{PDG_2010}. The same holds for the wide state $\Dnan$. On
the other hand for $\Dnap$ and $\Dopn$ the only dedicated measurements available
are from \cite{Abe:2003zm} and \cite{Link:2003bd}, respectively, and hence these
measurements are quoted separately. New precise measurements of masses
and widths of $\Dtan$ and $\Dnan$ became available recently
\cite{Aubert:2009wg} and are included in the weighted averages\footnote{We
  calculate the weighted average of the PDG \cite{Amsler:2008zzb} and
  Ref.~\cite{Aubert:2009wg} values.} shown in
Fig.~\ref{fig1}. In these averages also the mass of $\Don$ from
\cite{Abulencia:2005ry} is used\footnote{PDG does not use values from
  \cite{Abulencia:2005ry} since they are measured relative to the mass 
of $D^{(\ast)\pm}$ mesons.}. 

The masses and widths of narrow ($\Gamma\sim$ 20--40~MeV) orbitally 
excited $D$ mesons (denoted $D^{\ast\ast}$), both neutral and charged, 
are well established. Measurements of broad states ($\Gamma\sim$ 200--400~MeV)
are less abundant, as identifying the signal is more challenging. 
There is a slight discrepancy between the 
\Dnan\ masses measured by the Belle\cite{Abe:2003zm} and 
FOCUS\cite{Link:2003bd} experiments. No data exists yet for the 
\Dopp\ state. Dalitz plot analyses of $B\to D^{(\ast)}\pi\pi$ decays
strongly favor the assignments $0^+$ and $1^+$ for the spin-parity 
quantum numbers of the \Dnan/\Dnap\ and \Dopn\ states, respectively. 
The measured masses and widths, as well as the $J^P$ values, are 
in agreement with theoretical predictions based on potential 
models\cite{Godfrey:1985xj,Rosner:1985dx,Godfrey:1986wj,Isgur:1991wq,DiPierro:2001uu}. The quantitative information on the values of branching
fractions for all $D^{\ast\ast}$ mesons is scarce. In Fig.~\ref{fig1}
we include the available measurements from \cite{Abe:2003zm,Abe:2004sm} for
$\Don$ and from \cite{Abe:2003zm,Aubert:2009wg} for $\Dtan$. 
While the branching fractions for $B$
mesons decaying to a narrow $D^{\ast\ast}$ state and a pion are similar 
for charged and neutral $B$ initial states, the branching fractions 
to a broad $D^{\ast\ast}$ state and $\pi^+$ are much larger for $B^+$ 
than for $B^0$. This may be due to the fact that color-suppressed 
amplitudes contribute only to the $B^+$ decay and not to the $B^0$ 
decay (for a theoretical discussion, see Ref.~\cite{Colangelo:2004vu,Jugeau:2005yr}).

\begin{figure}[htbp]
  \begin{center}
    \includegraphics[width=14cm]{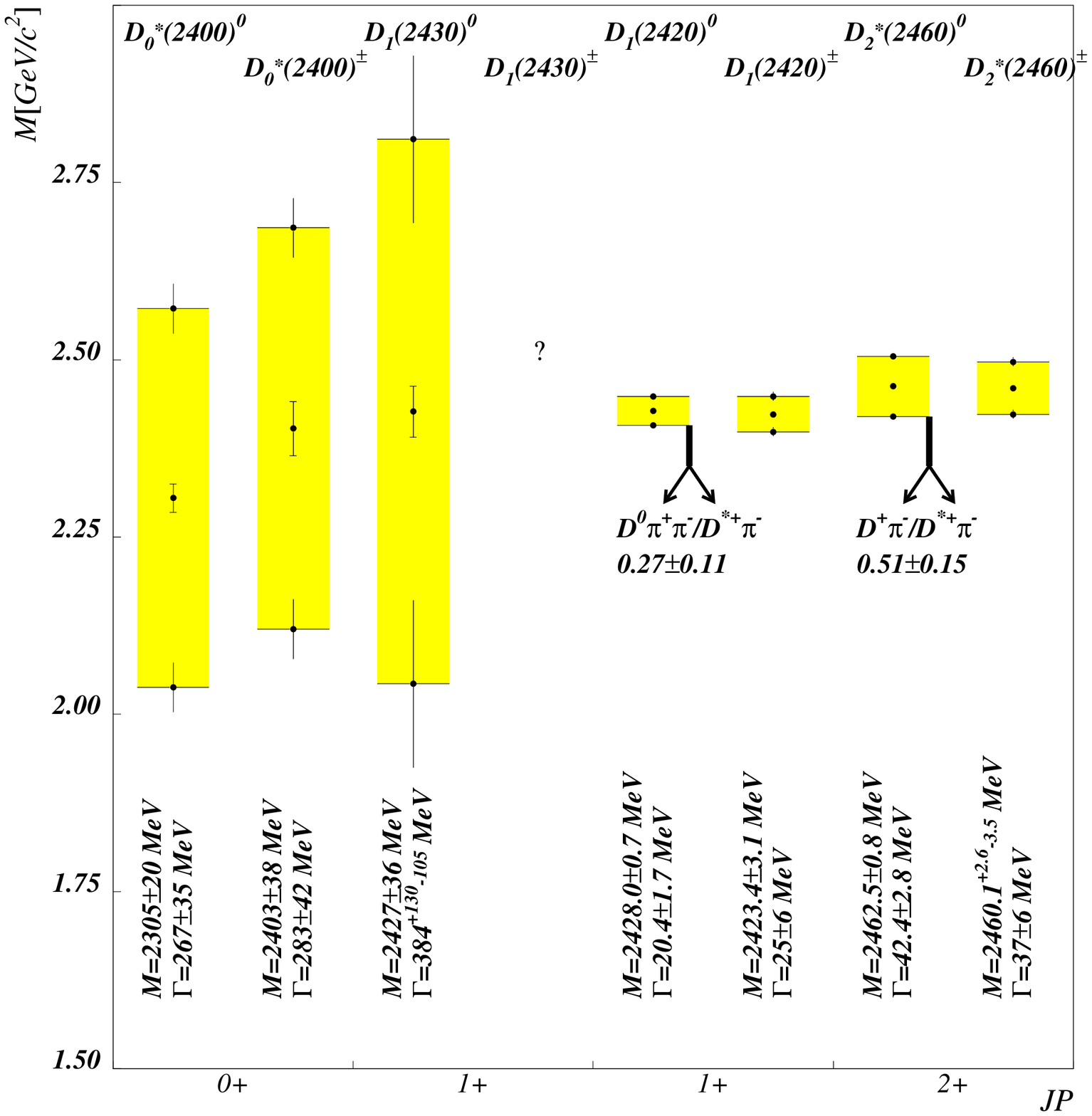}
    \caption{
      Masses, widths and some branching fractions of orbitally excited $D$ mesons. Shaded
      regions show the masses and widths of individual states. The 
      central point with error bars denotes the measured mass of each
      state. Error bars at the edges of the shaded regions denote the
      uncertainties of the width determination. Divided arrows denote
      relative branching ratios for the final states marked.
    }
    \label{fig1}
  \end{center}
\end{figure}

The first observations of $\Dsop$ and $\Dsna$ 
states are described in Refs.~\cite{Aubert:2003fg} and \cite{Besson:2003cp}, 
respectively.
The discoveries of the \Dsna\ and \Dsop\ have triggered increased 
interest in properties of, and searches for, excited $D_s$ mesons 
(here generically denoted $D_s^{\ast\ast}$). While the masses and 
widths of \Dso\ and \Dst\ states are in relatively good agreement 
with potential model predictions, 
the masses of \Dsna\ and \Dsop\ states (and consequently their 
widths, less than around 5~MeV) are significantly lower than expected 
(see Ref.~\cite{Cahn:2003cw} for a discussion of $c\bar{s}$ models). 
Moreover, the mass splitting between these two states greatly exceeds 
that between the \Dso\ and~\Dst. These unexpected properties have led 
to interpretations of the \Dsna\ and \Dsop\ as exotic four-quark
states. Measurements of masses (and the width
of $\Dst$) are averaged by the PDG~\cite{PDG_2010}. In the averages shown
in Fig.~\ref{fig2} we include the mass measurement of $\Dsna$ and
$\Dsop$ from \cite{Abe:2003jk}\footnote{We calculate the weighted average
  of the PDG \cite{Amsler:2008zzb} and Ref. \cite{Abe:2003jk} values. The latter are 
  excluded from the PDG average since they are measured relative to the mass of
  $D_s^{(\ast)}$ mesons.}. Widths of other $D_s^{\ast\ast}$ mesons are
below the current experimental sensitivity and the obtained upper
limits are quoted separately. 

While there are few measurements with respect to the
$J^P$ values of \Dsna~and \Dsop, 
the available data favors $0^+$ and $1^+$, respectively. 
A molecule-like ($DK$) interpretation of the \Dsna\ and 
\Dsop\cite{Barnes:2003dj,Lipkin:2003zk} that can account for their low masses 
and isospin-breaking decay modes is tested by searching for 
charged and neutral isospin partners of these states; thus far 
such searches have yielded negative results. 
Hence the subset of models that predict equal production rates for
different charged states is nominally excluded. The molecular picture
can also be tested by measuring the rates for the radiative processes 
$\Dsna/\Dsop\to D_s^{(\ast)}\gamma$ and comparing to theoretical 
predictions. The predicted rates, however, are below the sensitivity 
of current experiments. 
Another model successful in explaining the total widths and the
\Dsna-\Dsop\ mass splitting is based on the assumption that these 
states are chiral partners of the ground states \Ds\ and~\Dsa\cite{Bardeen:2003kt}. 
While some measured branching fraction ratios agree with predicted
values, further experimental tests with better sensitivity are 
needed to confirm or refute this scenario.
 
In addition to the \Dsna\ and \Dsop\ states, other excited $D_s$ 
states may have been observed. SELEX has reported a \Dstsi\ candidate \cite{Evdokimov:2004iy},
but this has not been confirmed by other experiments. 
Belle and BaBar have observed \Dsts\ and \Dste\ states 
\cite{:2007aa,Aubert:2006mh}, 
which may be radial excitations of the \Dsa\ and \Dsna, 
respectively (see for example \cite{Matsuki:2006rz}). 
However, the \Dste\ has been searched for in
$B$ decays and not observed, which may indicate that this 
state has higher spin. Recently new precise measurements of \Dsts\ and
\Dste\ properties were performed by BaBar \cite{:2009di}. The weighted
average of 
the results from \cite{:2007aa, :2009di} is
$M(D_{s1}(2700)^\pm)=(2709\pm 8)$~MeV/$c^2$ and
$\Gamma(D_{s1}(2700)^\pm)=(126\pm 31)$~MeV. In the same paper BaBar
observes another state, denoted $D_{sJ}(3040)^\pm$, with a
significance of 6 standard deviations. According to calculations of 
\cite{Matsuki:2006rz} this state is a candidate for the radial excitation of 
\Dsop ~or \Dso. 

The existing studies of \Dsop ~provide for sufficient information that 
the individual branching fractions are
calculated by HFAG; they are shown in Fig.~\ref{fig2}. Beside this the relative
branching ratios of \Dso ~are shown \cite{:2007dya,Heister:2001nj}. Measurements
of individual branching fractions of $D_s^{\ast\ast}$ are difficult
due to the unknown fragmentation of $c\bar{c}\to D_s^{\ast\ast}$ (in
the studies where $D_s^{\ast\ast}$ mesons are produced in $c\bar{c}$ fragmentation) or due
to the unknown $B\to D_s^{\ast\ast}X$ branching fractions (in the
studies where $D_s^{\ast\ast}$ are produced in $B$ meson decays).

\begin{figure}[htbp]
  \begin{center}
    \includegraphics[width=14cm]{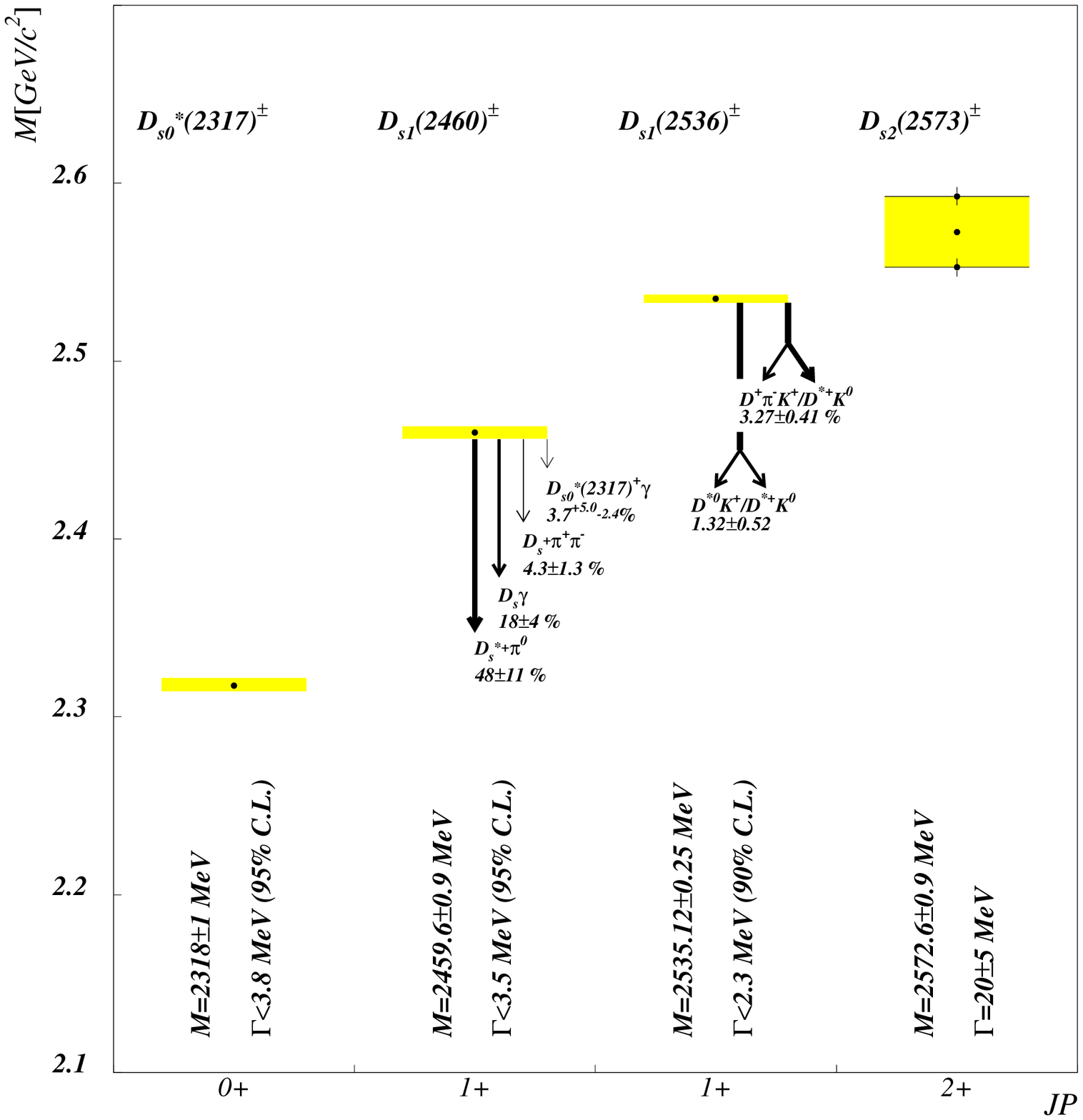}
    \caption{
      Masses, widths and some branching fractions of orbitally excited $D_s$ mesons. Shaded
      regions show the masses and widths of individual states (U.L. on the widths in case
      of $\Dsna,~\Dsop$, and $\Dso$). The 
      central point with error bars denotes the measured mass of each
      state. Error bars at the edges of the shaded regions denote the
      uncertainties of the width determination. Arrows (divided arrows) denote
      branching fractions (relative branching ratios) for the final states marked.
    }
    \label{fig2}
  \end{center}
\end{figure}


\begin{table}
\caption{\label{tab1} Recent results for properties of $D^{\ast\ast}$
  mesons.}
\vspace*{6pt}
\footnotesize
\hskip-0.25in

\end{table}

\begin{table}[h]
\caption{\label{tab2} Recent results for masses and branching
fractions of excited $D_s$ mesons.}
\vspace*{6pt}
\footnotesize
\hskip-0.35in
%
\end{table}

\begin{table}[t]
\caption{\label{tab3} Recent results for quantum numbers of 
excited $D_s$ mesons.}
\footnotesize
\begin{center}
%
\end{center}
\end{table}

\clearpage
\subsection{Semileptonic Decays}

\subsubsection{Introduction}

Semileptonic decays of $D$ mesons involve the interaction of a leptonic
current with a hadronic current. The latter is nonperturbative
and cannot be calculated from first principles; thus it is usually
parameterized in terms of form factors. The transition matrix element 
is written
\begin{eqnarray}
  {\cal M} & = & -i\,{G_F\over\sqrt{2}}\,V^{}_{cq}\,L^\mu H_\mu\,,
  \label{Melem}
\end{eqnarray}
where $G_F$ is the Fermi constant and $V^{}_{cq}$ is a CKM matrix element.
The leptonic current $L_\mu$ is evaluated directly from the lepton spinors 
and has a simple structure; this allows one to extract information about 
the form factors (in $H^{}_\mu$) from data on semileptonic decays~\cite{Becher:2005bg}.  
Conversely, because there are no final-state interactions between the
leptonic and hadronic systems, semileptonic decays for which the form 
factors can be calculated allow one to 
determine~$V^{}_{cq}$~\cite{Kobayashi:1973fv}.

\subsubsection{$D\ra P\ell\nu$ Decays}

When the final state hadron is a pseudoscalar, the hadronic 
current is given by
\begin{eqnarray}
H_\mu & = & \left< P(p) | \bar{q}\gamma^\mu c | D(p') \right> \ =\  
f_+(q^2)\left[ (p' + p)^\mu -\frac{M_D^2-m_P^2}{q^2}q^\mu\right] + 
 f_0(q^2)\frac{M_D^2-m_P^2}{q^2}q^\mu\,,
\label{eq:hadronic}
\end{eqnarray}
where $M_D$ and $p'$ are the mass and four momentum of the 
parent $D$ meson, $m_P$ and $p$ are those of the daughter meson, 
$f_+(q^2)$ and $f_0(q^2)$ are form factors, and $q = p' - p$.  
Kinematics require that $f_+(0) = f_0(0)$.
The contraction $q^\mu L_\mu$ results in terms proportional 
to $m^{}_\ell$\cite{Gilman:1989uy}, and thus for $\ell=e,\mu$
the last two terms in Eq.~(\ref{eq:hadronic}) are negligible. 
Thus, only the $f_+(q^2)$ form factor 
is relevant. The differential partial width is
\begin{eqnarray}
\frac{d\Gamma(D \to P\ell\bar\nu_\ell)}{dq^2\, d\cos\theta_\ell} & = & 
   \frac{G_F^2|V_{cq}|^2}{32\pi^3} p^{*\,3}|f_{+}(q^2)|^2\sin\theta^2_\ell\,,
\label{eq:dGamma}
\end{eqnarray}
where ${p^*}$ is the magnitude of the momentum of the final state hadron
in the $D$ rest frame.


The form factor is traditionally parametrized with an explicit pole 
and a sum of effective poles:
\begin{eqnarray}
f_+(q^2) & = & \frac{f(0)}{1-\alpha}
\left(\frac{1}{1- q^2/m^2_{\rm pole}}\right)\ +\ 
\sum_{k=1}^{N}\frac{\rho_k}{1- q^2/(\gamma_k\,m^2_{\rm pole})}\,,
\label{eqn:expansion}
\end{eqnarray}
where $\rho_k$ and $\gamma_k$ are expansion parameters. The parameter
$m_{{\rm pole}}$ is the mass of the lowest-lying $c\bar{q}$ resonance
with the appropriate quantum numbers; this is expected to provide the
largest contribution to the form factor for the $c\ra q$ transition.  
For example, for $D\to\pi$ transitions the dominant resonance is
expected to be $D^*$, and thus $m^{}_{\rm pole}=m^{}_{D^*}$.

\subsubsection{Simple Pole}

Equation~(\ref{eqn:expansion}) can be simplified by neglecting the 
sum over effective poles, leaving only the explicit vector meson pole. 
This approximation is referred to as ``nearest pole dominance'' or 
``vector-meson dominance.''  The resulting parameterization is
\begin{eqnarray}
  f_+(q^2) & = & \frac{f_+(0)}{(1-q^2/m^2_{\rm pole})}\,. 
\label{SimplePole}
\end{eqnarray}
However, values of $m_{{\rm pole}}$ that give a good fit to the data 
do not agree with the expected vector meson masses~\cite{Hill:2006ub}. 
To address this problem, the ``modified pole'' or Becirevic-Kaidalov~(BK) 
parameterization~\cite{Becirevic:1999kt} was introduced.
This parametrization assumes that gluon 
hard-scattering contributions ($\delta$) are near zero, and scaling
violations ($\beta$) are near unity~\cite{Hill:2006ub}:
\begin{eqnarray}
1 + 1\slash \beta - \delta & \equiv & 
\frac{\left(M_D^2 - m_{P}^2\right)}{f_+(0)}\ 
\left.\frac{df_+}{dq^2}\right|_{q^2=0}\ \approx\ 2\,.
\end{eqnarray}
The parameterization takes the form
\begin{eqnarray}
f_+(q^2) & = & \frac{f_+(0)}{(1-q^2/m^2_{\rm pole})}
\left(1-\alpha^{}_{\rm BK}\frac{q^2}{m^2_{\rm pole}}\right)\,.
\end{eqnarray}
To be consistent with $1 + 1\slash \beta - \delta\approx 2$, the 
parameter $\alpha^{}_{\rm BK}$ should be near the value~1.75.

This parameterization has been used by several experiments to 
determine form factor parameters.
Measured values of $m^{}_{\rm pole}$ and $\alpha^{}_{\rm BK}$ are
listed Tables~\ref{kPseudoPole} and~\ref{piPseudoPole} for
$D\to K\ell\nu$ and $D\to\pi\ell\nu$ decays, respectively.
Both tables show $\alpha^{}_{BK}$ to be substantially lower than
the expected value of~$\sim$\,1.75.


\begin{table}[htbp]
\caption{Results for $m_{\rm pole}$ and $\alpha_{\rm BK}$ from various
  experiments for $D^0\to K^-\ell^+\nu$ and $D^+\to K_S\ell^+\nu$
  decays.   The last entry is a lattice QCD prediction.
\label{kPseudoPole}}
\begin{center}

\end{center}
\end{table}

\begin{table}[htbp]
\caption{Results for $m_{\rm pole}$ and
  $\alpha_{\rm BK}$ from various experiments for 
  $D^0\to \pi^-\ell^+\nu$ and $D^+\to \pi^0\ell^+\nu$ decays.  The last 
  entry is a lattice QCD prediction.
\label{piPseudoPole}}
\begin{center}
%
\end{center}
\end{table}

\subsubsection{$z$ Expansion}

Several groups have advocated an alternative series 
expansion around some value $q^2=t_0$ to parameterize 
$f^{}_+$~\cite{Boyd:1994tt,Boyd:1997qw,Arnesen:2005ez,Becher:2005bg}.
This expansion is given in terms of a complex parameter $z$,
which is the analytic continuation of $q^2$ into the
complex plane:
\begin{eqnarray}
z(q^2,t_0) & = & \frac{\sqrt{t_+ - q^2} - \sqrt{t_+ - t_0}}{\sqrt{t_+ - q^2}
	  + \sqrt{t_+ - t_0}}\,, 
\end{eqnarray}
where $t_\pm \equiv (M_D \pm m_h)^2$ and $t_0$ is the (arbitrary) $q^2$ 
value corresponding to $z=0$. The physical region corresponds to $|z|<1$.

The form factor is expressed as
\begin{eqnarray}
f_+(q^2) & = & \frac{1}{P(q^2)\,\phi(q^2,t_0)}\sum_{k=0}^\infty
a_k(t_0)[z(q^2,t_0)]^k\,,
\label{z_expansion}
\end{eqnarray}
where the $P(q^2)$ factor accommodates sub-threshold resonances via
\begin{eqnarray}
P(q^2) & \equiv & 
\begin{cases} 
1 & (D\to \pi) \\
z(q^2,M^2_{D^*_s}) & (D\to K)\,. 
\end{cases}
\end{eqnarray}
The ``outer'' function $\phi(t,t_0)$ can be any analytic function,
but a preferred choice (see, {\it e.g.}
Refs.~\cite{Boyd:1994tt,Boyd:1997qw,Bourrely:1980gp}) obtained
from the Operator Product Expansion (OPE) is
\begin{eqnarray}
\phi(q^2,t_0) & =  & \alpha 
\left(\sqrt{t_+ - q^2}+\sqrt{t_+ - t_0}\right) \times  \nonumber \\
 & & \hskip0.20in \frac{t_+ - q^2}{(t_+ - t_0)^{1/4}}\  
\frac{(\sqrt{t_+ - q^2}\ +\ \sqrt{t_+ - t_-})^{3/2}}
     {(\sqrt{t_+ - q^2}+\sqrt{t_+})^5}\,,
\label{eqn:outer}
\end{eqnarray}
with $\alpha = \sqrt{\pi m_c^2/3}$.
The OPE analysis provides a constraint upon the 
expansion coefficients, $\sum_{k=0}^{N}a_k^2 \leq 1$.
These coefficients receive $1/M$ corrections, and thus
the constraint is only approximate. However, the
expansion is expected to converge rapidly since 
$|z|<0.051\ (0.17)$ for $D\ra K$ ($D\ra\pi$) over 
the entire physical $q^2$ range, and Eq.~(\ref{z_expansion}) 
remains a useful parameterization.

The $z$-expansion formalism has been used by 
BaBar~\cite{Aubert:2006mc} and CLEO-c~\cite{Dobbs:2007sm}.
Their fits used the first three terms of the expansion,
and the results for the ratios $r_1\equiv a_1/a_0$ and $r_2\equiv a_2/a_0$ are 
listed in Table~\ref{piPseudoZ}.  The CLEO~III\cite{Huang:2004fra} results
listed are obtained by refitting their data using the full
covariance matrix. The BaBar correlation coefficient listed is 
obtained by refitting their published branching fraction using 
their published covariance matrix.
These measurements correspond to using the standard 
outer function $\phi(q^2,t_0)$ of Eq.~(\ref{eqn:outer}) and 
$t_0=t_+\left(1-\sqrt{1-t_-/t_+}\right)$. This choice of $t^{}_0$
constrains $|z|$ to be below a maximum value within the physical region.

\begin{table}[htbp]
\caption{Results for $r_1$ and $r_2$ from various experiments, for 
$D\to \pi\/K\ell\nu$. The correlation coefficient listed is 
for the total uncertainties (statistical $\oplus$ systematic) on 
$r^{}_1$ and~$r^{}_2$.}
\label{piPseudoZ}
\begin{center}
\begin{tabular}{cccccc}
\hline
\vspace*{-10pt} & \\
Expt.     & mode &  Ref.                         & $r_1$               & $r_2$               & $\rho$        \\
\hline
 \omit    & \omit         & \omit                & \omit               & \omit               & \omit         \\
 CLEO III & $D^0\to K^+$  & \cite{Huang:2004fra} & $0.2^{+3.6}_{-3.0}$ & $-89^{+104}_{-120}$ & -0.99         \\
 BaBar    & \omit         & \cite{Aubert:2006mc} & $-2.5\pm0.2\pm0.2$  & $0.6\pm6.\pm5.$     & -0.64         \\
 CLEO-c   & \omit         & \cite{Dobbs:2007sm}  & $-2.4\pm0.4\pm0.1$  & $21\pm11\pm2$       & -0.81         \\
 Average  & \omit         &  \omit               & $-2.3\pm0.23$       & $5.9\pm6.3$         & -0.74         \\ 
\hline
 CLEO-c   & $D^+\to K_S$  & \cite{Dobbs:2007sm}  & $-2.8\pm6\pm2$      & $32\pm18\pm4$       & -0.84         \\
 CLEO-c   & $D^0\to\pi^+$ & \cite{Dobbs:2007sm}  & $-2.1\pm7\pm3$      & $-1.2\pm4.8\pm1.7$  & -0.96         \\
 CLEO-c   & $D^+\to\pi^0$ & \cite{Dobbs:2007sm}  & $-0.2\pm1.5\pm4$    & $-9.8\pm9.1\pm2.1$  & -0.97         \\
\vspace*{-10pt} & \\
\hline
\end{tabular}
\end{center}
\end{table}

Table~\ref{piPseudoZ} also lists average values for $r_1$ and $r_2$  
obtained from a simultaneous fit to CLEO~III, BaBar, and CLEO-c 
branching fraction measurements. 
To account for final-state radiation in the BaBar
measurement, we allow a bias shift between the fit parameters for
the BaBar data and those for the other measurements
(a $\chi^2$ penalty is added to the fit for any deviation 
from BaBar's central value).
Table~\ref{piPseudoZ} shows satisfactory agreement between 
the parameters measured for $D^0$ and $D^+$ decays.  

\subsubsection{$D\ra V\ell\nu$ Decays}

When the final state hadron is a vector meson, the decay can proceed through
both vector and axial vector currents, and four form factors are needed.
The hadronic current is $H^{}_\mu = V^{}_\mu + A^{}_\mu$, 
where~\cite{Gilman:1989uy} 
\begin{eqnarray}
V_\mu & = & \left< V(p,\varepsilon) | \bar{q}\gamma^\mu c | D(p') \right> \ =\  
\frac{2V(q^2)}{M_D+m_h} 
\varepsilon_{\mu\nu\rho\sigma}\varepsilon^{*\nu}p^{\prime\rho}p^\sigma \\
 & & \nonumber\\
A_\mu & = & \left< V(p,\varepsilon) | -\bar{q}\gamma^\mu\gamma^5 c | D(p') \right> 
 \ =\  -i\,(M_D+m_h)A_1(q^2)\varepsilon^*_\mu \nonumber \\
 & & \hskip2.10in 
  +\ i \frac{A_2(q^2)}{M_D+m_h}(\varepsilon^*\cdot q)(p' + p)_\mu \nonumber \\
 & & \hskip2.30in 
+\ i\,\frac{2m_h}{q^2}\left(A_3(q^2)-A_0(q^2)\right)[\varepsilon^*\cdot (p' + p)] q_\mu\,.
\end{eqnarray}
In this expression, $m_h$ is the daughter meson mass and
\begin{eqnarray}
A_3(q^2) & = & \frac{M_D + m_h}{2m_h}A_1(q^2)\ -\ \frac{M_D - m_h}{2m_h}A_2(q^2)\,.
\end{eqnarray}
Kinematics require that $A_3(0) = A_0(0)$.
The differential partial width is
\begin{eqnarray}
\frac{d\Gamma(D \to V\ell\bar\nu_\ell)}{dq^2\, d\cos\theta_\ell} & = & 
  \frac{G_F^2\,|V_{cq}|^2}{128\pi^3M_D^2}\,p^*\,q^2 \times \nonumber \\
 & &  
\left[\frac{(1-\cos\theta_\ell)^2}{2}|H_-|^2\ +\  
\frac{(1+\cos\theta_\ell)^2}{2}|H_+|^2\ +\ \sin^2\theta_\ell|H_0|^2\right]\,,
\end{eqnarray}
where $H^{}_\pm$ and $H^{}_0$ are helicity amplitudes given by
\begin{eqnarray}
H_\pm & = & \frac{1}{M_D + m_h}\left[(M_B+m_h)^2A_1(q^2)\ \mp\ 
      2M^{}_D\,p^* V(q^2)\right] \\
 & & \nonumber \\
H_0 & = & \frac{1}{|q|}\frac{M_B^2}{2m_h(M_D + m_h)}\ \times\ \nonumber \\
 & & \hskip0.01in \left[
    \left(1- \frac{m_h^2 - q^2}{M_D^2}\right)(M_D^2 + m_h^2)A_1(q^2) 
    \ -\ 4{p^*}^2 A_2(q^2) \right]\,.
\label{HelDef}
\end{eqnarray}
The left-handed nature of the quark current manifests itself as
$|H_-|>|H_+|$. The differential decay rate for $D\ra V\ell\nu$ 
followed by the vector meson decaying into two pseudoscalars is

\begin{eqnarray}
\frac{d\Gamma(D\ra V\ell\nu, V\ra P_1P_2)}{dq^2 d\cos\theta_V d\cos\theta_\ell d\chi} 
 &  = & \frac{3G_F^2}{2048\pi^4}
       |V_{cq}|^2 \frac{p^*(q^2)q^2}{M_D^2} {\cal B}(V\to P_1P_2)\ \times \nonumber \\ 
 & & \hskip0.10in \big\{ (1 + \cos\theta_\ell)^2 \sin^2\theta_V |H_+(q^2)|^2 \nonumber \\
 & & \hskip0.20in +\ (1 - \cos\theta_\ell)^2 \sin^2\theta_V |H_-(q^2)|^2 \nonumber \\
 & & \hskip0.30in +\ 4\sin^2\theta_\ell\cos^2\theta_V|H_0(q^2)|^2 \nonumber \\
 & & \hskip0.40in +\ 4\sin\theta_\ell (1 + \cos\theta_\ell) 
             \sin\theta_V \cos\theta_V \cos\chi H_+(q^2) H_0(q^2) \nonumber \\
 & & \hskip0.50in -\ 4\sin\theta_\ell (1 - \cos\theta_\ell) 
          \sin\theta_V \cos\theta_V \cos\chi H_-(q^2) H_0(q^2) \nonumber \\
 & & \hskip0.60in -\ 2\sin^2\theta_\ell \sin^2\theta_V 
                \cos 2\chi H_+(q^2) H_-(q^2) \big\}\,,
\label{eq:dGammaVector}
\end{eqnarray}
where the angles $\theta^{}_\ell$, $\theta^{}_V$, and $\chi$ are defined
in Fig.~\ref{DecayAngles}. 

\begin{figure}[htbp]
  \begin{center}
\includegraphics[width=3.50in]{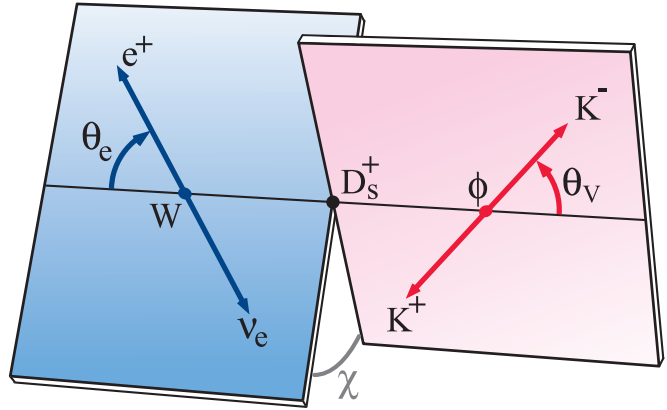}
  \end{center}
  \caption{
    Decay angles $\theta_V$, $\theta_\ell$ 
    and $\chi$. Note that the angle $\chi$ between the decay
    planes is defined in the $D$-meson reference frame, whereas
    the angles $\theta^{}_V$ and $\theta^{}_\ell$ are defined
    in the $V$ meson and $W$ reference frames, respectively.}
  \label{DecayAngles}
\end{figure}

Assuming that the simple pole form of Eq.~(\ref{SimplePole}) describes 
the $q^2$-dependence of the form factors, the 
distribution of Eq.~(\ref{eq:dGammaVector}) will depend only on the parameters
\begin{eqnarray}
r_V \equiv V(0) / A_1(0), & &  r_2 \equiv A_2(0) / A_1(0) \label{rVr2_eq}\,.
\end{eqnarray}
Table \ref{Table1} lists measurements of $r_V$ and $r_2$ from several
experiments. The average results from $D^+\ra\overline{K}^{*0}\ell^+\nu$
decays are also given. The measurements are plotted in
Fig.~\ref{fig:r2rv} which shows that they are all consistent with one
another.

\begin{table}[htbp]
\caption{Results for $r_V$ and $r_2$ from various experiments.
\label{Table1}}
\begin{center}
\begin{tabular}{cccc}
\hline
\vspace*{-10pt} & \\
Experiment & Ref. & $r_V$ & $r_2$ \\
\vspace*{-10pt} & \\
\hline
\vspace*{-10pt} & \\
$D^+\to \overline{K}^{*0}l^+\nu$ & \omit & \omit & \omit         \\
E691         & \cite{Anjos:1990pn}     & 2.0$\pm$  0.6$\pm$  0.3  & 0.0$\pm$  0.5$\pm$  0.2    \\
E653         & \cite{Kodama:1992tn}     & 2.00$\pm$ 0.33$\pm$ 0.16 & 0.82$\pm$ 0.22$\pm$ 0.11   \\
E687         & \cite{Frabetti:1993jq}     & 1.74$\pm$ 0.27$\pm$ 0.28 & 0.78$\pm$ 0.18$\pm$ 0.11   \\
E791 (e)     & \cite{Aitala:1997cm}    & 1.90$\pm$ 0.11$\pm$ 0.09 & 0.71$\pm$ 0.08$\pm$ 0.09   \\
E791 ($\mu$) & \cite{Aitala:1998ey}    & 1.84$\pm$0.11$\pm$0.09   & 0.75$\pm$0.08$\pm$0.09     \\
Beatrice     & \cite{Adamovich:1998ia} & 1.45$\pm$ 0.23$\pm$ 0.07 & 1.00$\pm$ 0.15$\pm$ 0.03   \\
FOCUS        & \cite{Link:2002wg}   & 1.504$\pm$0.057$\pm$0.039& 0.875$\pm$0.049$\pm$0.064  \\
Average      & \omit              & 1.62$\pm$0.055           & 0.83$\pm$0.054             \\
\hline
$D^0\to \overline{K}^0\pi^-\mu^+\nu$ & \omit & \omit & \omit         \\
FOCUS        & \cite{Link:2004uk}    & 1.706$\pm$0.677$\pm$0.342& 0.912$\pm$0.370$\pm$0.104 \\
\hline
$D_s^+ \to \phi\,e^+ \nu$ &\omit  &\omit     & \omit                  \\
BaBar        & \cite{Aubert:2006ru}    & 1.636$\pm$0.067$\pm$0.038& 0.705$\pm$0.056$\pm$0.029 \\
\hline
$D^0, D^+\to \rho\,e \nu$ & \omit  & \omit    & \omit                 \\
CLEO         & \cite{Mahlke:2007uf}    & 1.40$\pm$0.25$\pm$0.03   & 0.57$\pm$0.18$\pm$0.06    \\
\hline
\end{tabular}
\end{center}
\end{table}

\begin{figure}[htbp]
  \begin{center}
    \includegraphics[width=6.5in,angle=0]{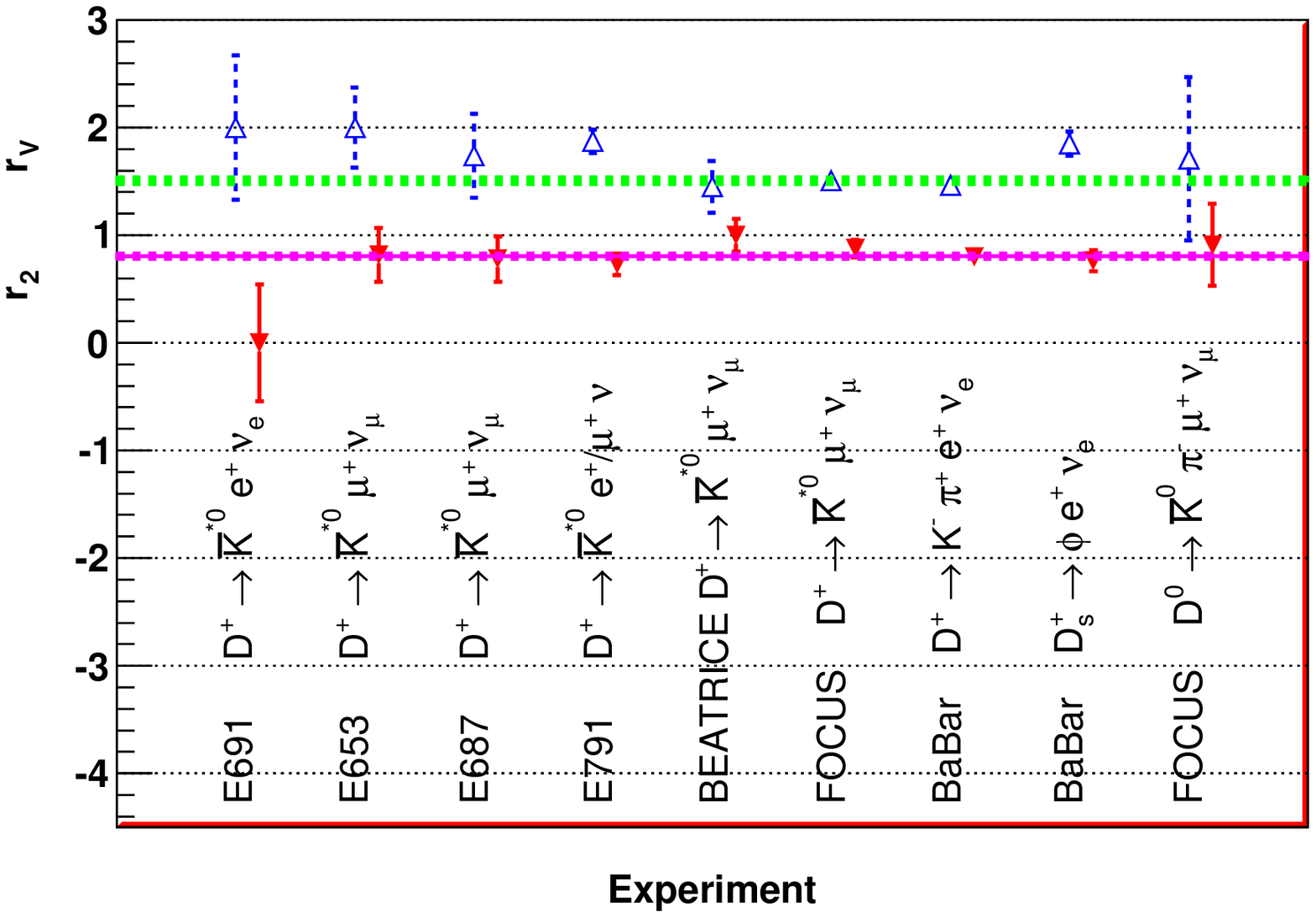}
  \end{center}
\vskip0.10in
  \caption{A comparison of $r_2$ and $r_V$ values 
    from various experiments. The first seven measurements are for $D^+
    \to K^-\pi^+ l^+\nu_l$ decays. Also shown as a line with
    1-$\sigma$ limits is the average of these. The last two points are
    $D_s^+$ decays and Cabibbo-suppressed $D$ decays. 
  \label{fig:r2rv}}
\end{figure}

\subsubsection{$S$-Wave Component}

In 2002 FOCUS reported~\cite{Link:2002ev} an asymmetry in
the observed $\cos(\theta_V)$ distribution. This is interpreted as
evidence for an $S$-wave component in the decay amplitude as follows. 
Since $H_0$ typically dominates over $H_{\pm}$, the distribution given 
by Eq.~(\ref{eq:dGammaVector}) is, after integration over $\chi$,
roughly proportional to $\cos^2\theta_V$. 
Inclusion of a constant $S$-wave amplitude of the form $A\,e^{i\delta}$ 
leads to an interference term proportional to 
$|A H_0 \sin\theta_\ell \cos\theta_V|$; this term causes an asymmetry 
in $\cos(\theta_V)$.
When FOCUS fit their data including this $S$-wave amplitude, 
they obtained $A = 0.330 \pm 0.022 \pm 0.015$ GeV$^{-1}$ and 
$\delta = 0.68 \pm 0.07 \pm 0.05$ \cite{Link:2002wg}. 

More recently, both BaBar~\cite{Aubert:2008rs} and 
CLEO-c~\cite{Ecklund:2009fia} have also found evidence 
for an $f^{}_0$ component in semileptonic $D^{}_s$ decays.

\subsubsection{Model-independent Form Factor Measurement}

Subsequently the CLEO-c collaboration extracted the form factors
$H_+(q^2)$, $H_-(q^2)$, and $H_0(q^2)$ in a model-independent fashion
directly as functions of $q^2$\cite{Briere:2010zc} and also determined the
$S$-wave form factor $h_0(q^2)$ via the interference term, despite the
fact that the $K\pi$ mass distribution appears dominated by the vector
$K^*(892)$ state. Their results are shown in Figs.~\ref{fig:cleoc_H} and
\ref{fig:cleoc_h0}.  Plots in Fig.~\ref{fig:cleoc_H} clearly show that
$H_0(q^2)$ dominates over essentially the full range of $q^2$, but
especially at low $q^2$. They also show that the transverse form factor
$H_t(q^2)$ (which can be related to $A_3(q^2)$ is small (compared to
Lattice Gauge Theory calculations) and suggest that the form factor
ratio $r_3 \equiv A_3(0) / A_1(0)$ is large and negative.

The product $H_0(q^2)\times h_0(q^2)$ is shown in
Fig.~\ref{fig:cleoc_h0} and clearly indicates the existence of
$h_0(q^2)$, although it seems to fall faster with $q^2$ than $H_0(q^2)$.
The other plots in that figure show that $D$- and $F$-wave versions of
the $S$-wave $h_0(q^2)$ are not significant.

\begin{figure}[htb]
  \begin{center}
    \vskip0.20in
    \includegraphics[width=4.75in,angle=0.]{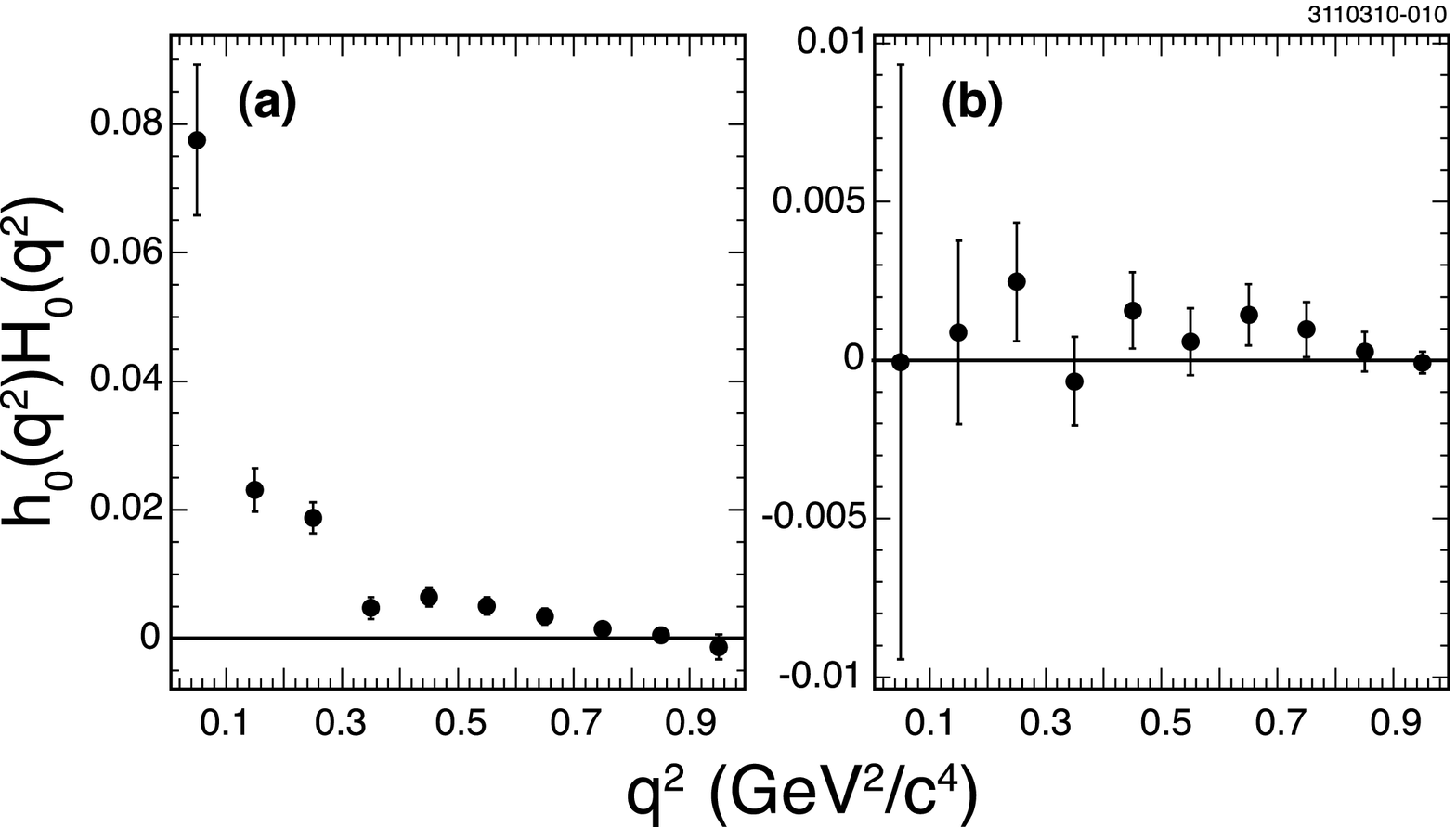}
  \end{center}
  \begin{center}
    \vskip0.20in
    \includegraphics[width=4.75in,angle=0.]{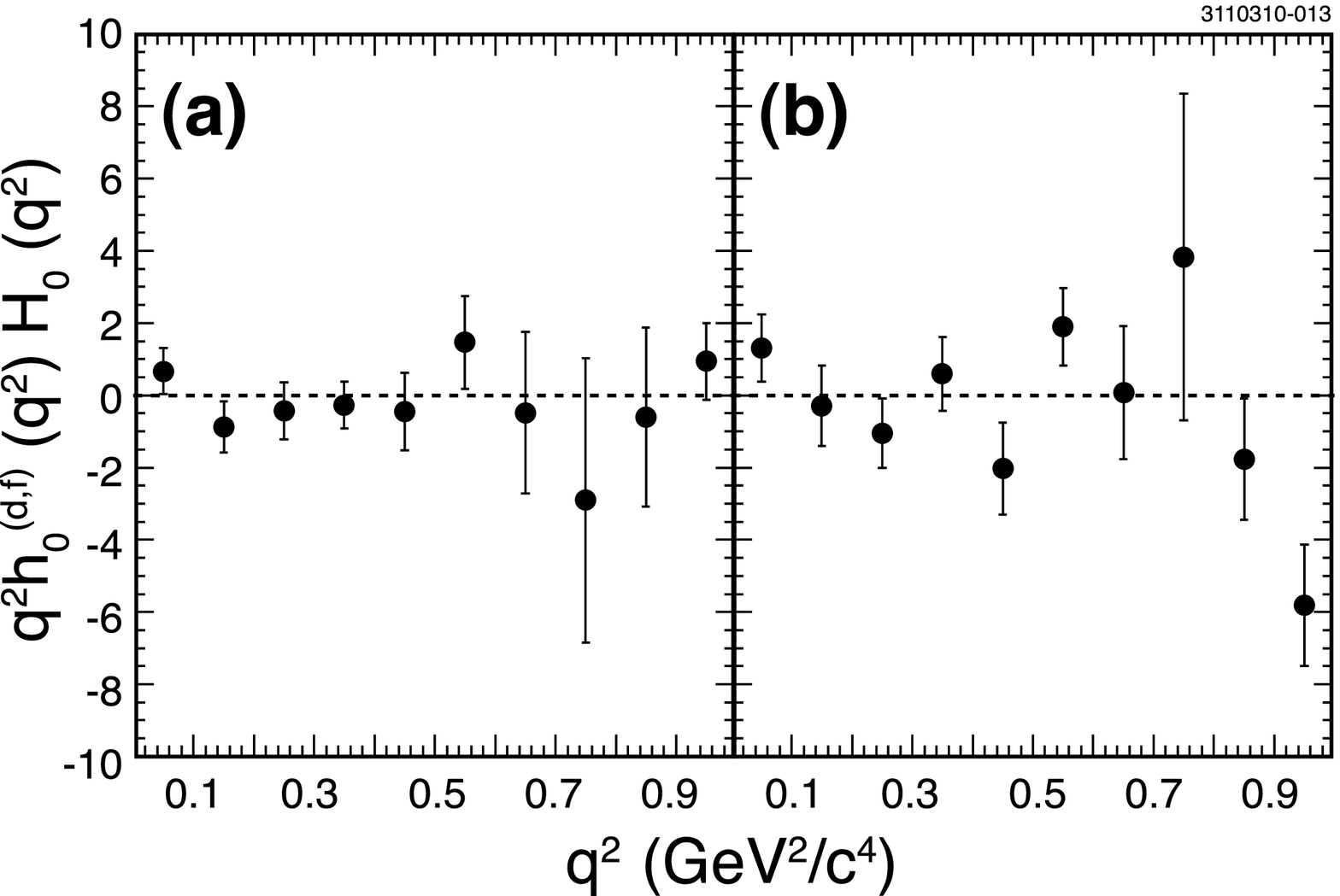}
  \end{center}
\vskip-0.20in
  \caption{Model-independent form factors $h_0(q^2)$ measured by 
    CLEO-c\cite{Briere:2010zc}.
  \label{fig:cleoc_h0}}
\end{figure}

\begin{figure}[htb]
  \begin{center}
    \vskip0.20in
    \includegraphics[width=4.75in,angle=0.]{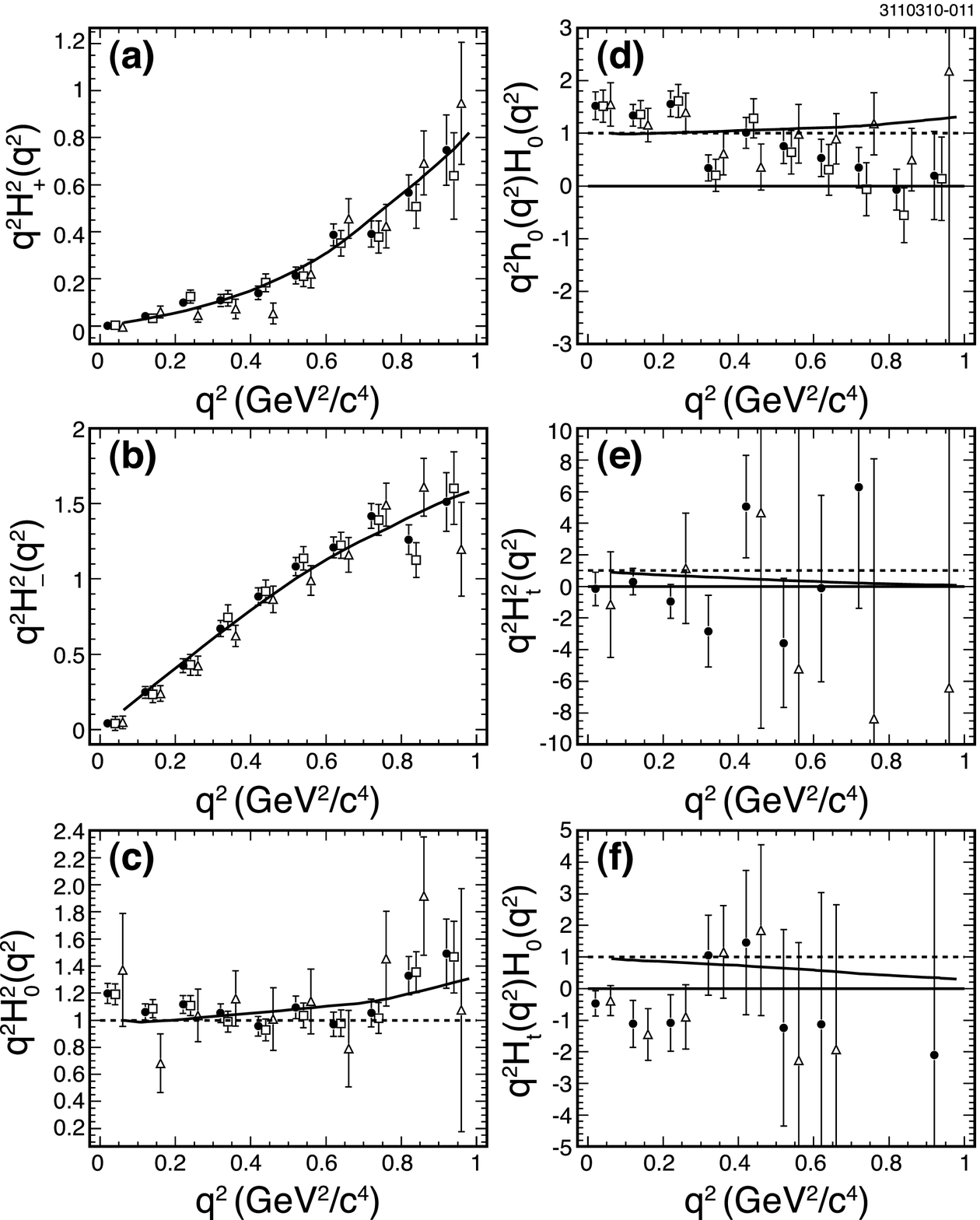}
  \end{center}
\vskip-0.20in
  \caption{Model-independent form factors $H(q^2)$ measured by 
    CLEO-c\cite{Briere:2010zc}.
  \label{fig:cleoc_H}}
\end{figure}

\clearpage
\subsection{\emph{CP} Asymmetries}

\emph{CP} violation occurs if the decay rate for a particle differs 
from that of its \emph{CP}-conjugate\cite{Bigi:2000yz}. 
In general there are two classes of \emph{CP} violation, termed
{\it indirect\/} and {\it direct\/}\cite{Nir:1999mg}. Indirect \emph{CP} 
violation refers to $\Delta C\!=\!2$ processes and 
arises in $D^0$ decays due to $D^0$-$\dbar$ mixing. 
It can occur as an asymmetry in the mixing itself, or it can 
result from interference between a decay 
amplitude arising via mixing and a non-mixed amplitude. 
Direct \emph{CP} violation refers to
$\Delta C\!=\!1$ processes and occurs in both charged and neutral 
$D$ decays. It results from interference between two different decay
amplitudes (e.g., a penguin and tree amplitude) that have
different weak (CKM) and strong phases\footnote{The weak 
phase difference will have opposite signs for $D\ra f$ and 
$\overline{D}\ra\bar{f}$ decays, while the strong phase difference 
will have the same sign. As a result, squaring the total amplitudes 
to obtain the decay rates gives interference terms having 
opposite sign, i.e., non-identical decay rates.}.
A difference in strong phases typically arises due to 
final-state interactions (FSI)\cite{Buccella:1994nf}. A difference
in weak phases arises from different CKM vertex couplings, as 
is often the case for spectator and penguin diagrams.

\vspace{0.8cm}
The \emph{CP} asymmetry is defined as the difference between 
$D$ and $\overline{D}$ partial widths divided by their sum:
\begin{eqnarray}  
A_{CP} & = & \frac{\Gamma(D)-\Gamma(\overline{D})}
{\Gamma(D)+\Gamma(\overline{D})}\,.
\end{eqnarray}
However, to take into account differences in production rates between 
$D$ and $\overline{D}$ (which would affect the number of respective 
decays observed), experiments usually normalize to a Cabibbo-favored 
mode. In this case there is the additional benefit that most corrections 
due to inefficiencies cancel out, reducing systematic uncertainties. An 
implicit assumption is that there is no measurable \emph{CP} 
violation in the Cabibbo-favored normalizing mode. 
The \emph{CP} asymmetry is calculated as
\begin{eqnarray}
A_{CP} & = & \frac{\eta(D)-\eta(\overline{D})}{\eta(D)+\eta(\overline{D})}\,,
\end{eqnarray}
where (considering, for example, $D^0 \to K^-K^+$)
\begin{eqnarray}
 \eta(D) & = & \frac{N(D^0 \rightarrow K^-K^+)}{N(D^0 \rightarrow K^-\pi^+)}\,, \\
 \eta(\overline{D}) & = & \frac{N(\dbar\rightarrow K^-K^+)}
{N(\dbar\rightarrow K^+\pi^-)}\,.
\end{eqnarray}
In the case of $D^+$ and $D^+_s$ decays, $A^{}_{CP}$ measures 
direct \emph{CP} violation; in the case of $D^0$ decays, $A^{}_{CP}$ 
measures direct and indirect \emph{CP} violation combined.
Values of $A^{}_{CP}$ for $D^+$, $D^0$ and $D_s^+$ decays are listed in
Tables~\ref{tab:cp_charged}, \ref{tab:cp_neutral} and \ref{tab:cp_ds} respectively.

\begin{table}
\renewcommand{\arraystretch}{1.4}
\caption{\cp\ asymmetries 
$A^{}_{CP}= [\Gamma(D^+)-\Gamma(D^-)]/[\Gamma(D^+)+\Gamma(D^-)]$
for $D^\pm$ decays.
\label{tab:cp_charged}}
\footnotesize
\begin{center}

\end{center} 
\end{table}

\begin{table}
\renewcommand{\arraystretch}{1.3}
\caption{\cp\ asymmetries 
$A^{}_{CP}=[\Gamma(D^0)-\Gamma(\dbar)]/[\Gamma(D^0)+\Gamma(\dbar)]$
for $D^0,\dbar$ decays.
\label{tab:cp_neutral}}
\footnotesize
\begin{center}
%
 \end{center} 
\end{table}

\begin{table}
\renewcommand{\arraystretch}{1.4}
\caption{\cp\ asymmetries 
$A^{}_{CP}= [\Gamma(D_s^+)-\Gamma(D_s^-)]/[\Gamma(D_s^+)+\Gamma(D_s^-)]$
for $D_s^\pm$ decays.
\label{tab:cp_ds}}
\footnotesize
\begin{center}
%
\end{center} 
\end{table}

\clearpage
\subsection{\emph{$T$}-violating Asymmetries}

$T$-violating asymmetries are measured using triple-product
correlations and assuming the validity of the $CPT$ theorem.
Triple-product correlations of the form 
$\vec{a}\cdot(\vec{b}\times\vec{c})$, 
where $a$, $b$, and $c$ are spins or momenta, are odd 
under time reversal~(\emph{T}).
For example, for $D^0 \to K^+K^-\pi^+\pi^-$ decays, 
$C_T \equiv \vec{p}^{}_{K^+}\cdot(\vec{p}_{\pi^+}\times \vec{p}_{\pi^-})$  
changes sign (i.e., is odd) under a \emph{T} transformation.
The corresponding quantity for $\dbar$ is
$\overline{C}_T \equiv 
      \vec{p}^{}_{K^-}\cdot(\vec{p}_{\pi^-}\times \vec{p}_{\pi^+})$.
Defining  
\begin{eqnarray}
 A_{T} & = &
    \frac{\Gamma(C_T>0)-\Gamma(C_T<0)}{\Gamma(C_T>0)+\Gamma(C_T<0)}
\end{eqnarray}
for $D^0$ decay and
\begin{eqnarray}
\overline{A}_{T} & = & 
   \frac{\Gamma(-\overline{C}_T>0)-\Gamma(-\overline{C}_T<0)}
                        {\Gamma(-\overline{C}_T>0)+\Gamma(-\overline{C}_T<0)}
\end{eqnarray} 
for $\dbar$ decay, in the absence of strong phases
either $A^{}_T\neq 0$ or $\overline{A}^{}_T\neq 0$ indicates
$T$ violation. In these expressions the $\Gamma$'s are partial widths. 
The asymmetry
\begin{eqnarray}
A^{}_{T\,{\rm viol}} & \equiv & \frac{A_{T}-\overline{A}_{T}}{2}
\end{eqnarray}
tests for $T$ violation even with nonzero strong phases (see 
Refs.~\cite{Golowich:1988ig,Bigi:2001sg,Bensalem:2002ys,Bensalem:2000hq,Bensalem:2002pz}).
Values of $A_{T\,{\rm viol}}$ for some $D^+$, $D^+_s$, and
$D^0$ decay modes are listed in Table~\ref{tab:t_viol}.

\begin{table}[h]
\renewcommand{\arraystretch}{1.4}
\caption{$T$-violating asymmetries 
$A^{}_{T\,{\rm viol}} = (A_{T}-\overline{A}_{T})/2$.
\label{tab:t_viol}}
\footnotesize
\begin{center}
\begin{tabular}{|l|c|c|c|} 
\hline
{\bf Mode} & {\bf Year} & {\bf Collaboration} & {\boldmath $A^{}_{T\,{\rm viol}}$} \\
\hline
{\boldmath $D^0 \to K^+K^-\pi^+\pi^-$} &
   2010 & BABAR~\cite{Sanchez:2010xj} &  $ +0.0010 \pm 0.0051 \pm 0.0044 $ \\
&  2005 & FOCUS~\cite{Link:2005th}  &  $ +0.010  \pm 0.057  \pm 0.037  $ \\
&       & COMBOS average          &  $ +0.0010 \pm 0.0067            $ \\  
\hline
{\boldmath $D^+ \to K^0_sK^+\pi^+\pi^-$} &
  2005 & FOCUS~\cite{Link:2005th}  &  $ +0.023 \pm  0.062  \pm 0.022  $ \\
\hline
{\boldmath $D^+_s \to K^0_sK^+\pi^+\pi^-$} &
  2005 & FOCUS~\cite{Link:2005th}  &  $ -0.036  \pm 0.067  \pm 0.023  $ \\
\hline                    
\end{tabular}
\end{center} 
\end{table}

\vskip0.30in
\begin{center}  ---------------  \end{center}
\vskip0.30in
In summary, Tables~\ref{tab:cp_charged}--\ref{tab:t_viol} show that
there is no evidence yet for \emph{CP} or $T$ violation in the charm 
sector. The most sensitive searches for \emph{CP} violation have 
reached a level of sensitivity well below~1\%.

\clearpage
\subsection{World Average for the \emph{$D^+_s$} Decay Constant~\fds}

The Heavy Flavor Averaging Group has used
Belle, BaBar, and CLEO measurements of ${\cal B}(D^+_s\ra\mu^+\nu)$ 
and ${\cal B}(D^+_s\ra\tau^+\nu)$ to calculate a world average
(WA) value for the $D^+_s$ decay constant~\fds.
The Belle results are from Ref.~\cite{Widhalm:2007ws},
the BaBar results are from Refs.~\cite{Aubert:2006sd,Lees:2010qj},
and the CLEO results are from 
Refs.~\cite{Alexander:2009ux,Onyisi:2009th,Naik:2009tk}.

The value for \fds\ is calculated via
\begin{eqnarray}
f^{}_{D^{}_s} & = & \frac{1}
{G^{}_F |V^{}_{cs}| m^{}_{\ell}
\biggl( 1-\frac{\displaystyle m^2_{\ell}}{\displaystyle m^2_{D^{}_s}}\biggr)}
\sqrt{\frac{8\pi\,{\cal B}(D^+_s\ra\ell^+\nu)}{m^{}_{D^{}_s} \tau^{}_{D^{}_s}}}\,,
\label{eqn:fds_inverted}
\end{eqnarray}
where, for ${\cal B}(D^+_s\ra\ell^+\nu)$, $\ell^+=\mu^+$ or $\ell^+=\tau^+$.
The error on \fds\ is calculated as follows:
values for variables on the right-hand-side of Eq.~(\ref{eqn:fds_inverted}) 
are sampled from Gaussian distributions having mean values equal to the 
central values and standard deviations equal to their respective 
errors. The resulting values of \fds\ are plotted, and the 
r.m.s. of the distribution is taken as the $\pm 1\sigma$ errors. 
The procedure is done separately for the WA values of 
${\cal B}(D^+_s\ra\mu^+\nu)$ and ${\cal B}(D^+_s\ra\tau^+\nu)$, 
and for the BaBar value~\cite{Aubert:2006sd} of 
${\cal B}(D^+_s\ra\mu^+\nu)$.

The BaBar result is treated separately because the 
signal yield is normalized to \dsphipi\ decays;
thus the measurement is
\begin{eqnarray}
\frac{\Gamma(D^+_s\ra\mu^+\nu)}{\Gamma(D^+_s\ra\phi\pi^+)}\Bigl|^{}_{\rm Babar}
& = & 0.143\,\pm 0.018{\rm\ (stat.)}\,\pm 0.006{\rm\ (syst.)}\,.
\label{eqn:dsratio_babar}
\end{eqnarray}
To obtain ${\cal B}(D^+_s\ra\mu^+\nu)$, one must multiply
by ${\cal B}(D^+_s\ra\phi\pi^+)$. However,  
for this analysis the $\phi$ is 
reconstructed via $\phi\ra K^+K^-$ with 
$|M^{}_{K^+K^-}-M^{}_\phi |\equiv\Delta M^{}_{KK}<5.5$\meve~\cite{Coleman:private}.
For the BaBar measurements of 
${\cal B}(D^+_s\ra\phi\pi^+)$~\cite{Aubert:2005xu,Aubert:2006nm},
a mass window $\Delta M^{}_{KK}<15$\meve\ was used.
Because the \dsphipi\ branching fraction depends on
$\Delta M^{}_{KK}$ (see Table II of Ref.~\cite{Alexander:2008cqa}),
we do not multiply together the BaBar results for
$\Gamma(D^+_s\ra\mu^+\nu)/\Gamma(D^+_s\ra\phi\pi^+)$ and 
${\cal B}(D^+_s\ra\phi\pi^+)$; instead we use the fact 
that CLEO has measured the branching fraction
${\cal B}(D^+_s\ra K^+K^-\pi^+)$ for $\Delta M^{}_{KK}=5{\rm\ MeV}$.
To multiply the BaBar result for 
$\Gamma(D^+_s\ra\mu^+\nu)/\Gamma(D^+_s\ra\phi\pi^+)$ by 
the CLEO result for ${\cal B}(D^+_s\ra K^+K^-\pi^+)$  
requires dividing Eq.~(\ref{eqn:dsratio_babar}) by 
${\cal B}(\phi\ra K^+K^-) = 0.491$ (as used in Ref.~\cite{Aubert:2006sd})
and subtracting in quadrature the 1.2\% uncertainty in 
${\cal B}(\phi\ra K^+K^-)$ from the systematic error. In
addition, for the result~(\ref{eqn:dsratio_babar}) BaBar 
subtracted off 
$D^+_s\ra f^{}_0(980)(K^+K^-)\pi^+$ background (48 events); 
as this process is included in the CLEO measurement, these 
events must be added back in to BaBar's $\phi\pi^+$ yield. 
The BaBar result then becomes
\begin{eqnarray}
\frac{\Gamma(D^+_s\ra\mu^+\nu)}{\Gamma(D^+_s\ra K^+K^-\pi^+)}
\biggr|^{}_{\Delta M^{}_{KK}=5.5{\rm\ MeV}} & = & 
0.285\,\pm 0.035{\rm\ (stat.)}\,\pm 0.011{\rm\ (syst.)}\,.
\label{eqn:ratio_adjusted}
\end{eqnarray}
Multiplying this by CLEO's measurement~\cite{Alexander:2008cqa}
\begin{eqnarray}
{\cal B}(D^+_s\ra K^+K^-\pi^+)
\Bigr|^{}_{\Delta M^{}_{KK}=5{\rm\ MeV}} & = &(1.69\,\pm 0.08\,\pm 0.06)\%
\end{eqnarray}
gives
\begin{eqnarray}
{\cal B}(D^+_s\ra\mu^+\nu)\Bigl|^{}_{\rm BaBar\ adjusted} &  = &
\Bigl(4.81\,\pm 0.63{\rm\ (stat.)}\,\pm 0.25{\rm\ (syst.)}\Bigr)\times 10^{-3}\,.
\end{eqnarray}

\vskip0.15in
\begin{center}\underline{\hskip0.80in}\end{center}
\vskip0.20in

In summary, three types of measurements are used to 
calculate the WA \fds:
\begin{enumerate}
\item the WA \dsmunu\ branching fraction, which is calculated 
from Belle and CLEO measurements (see Fig.~\ref{fig:charm_fds_mu});
\item the WA \dstaunu\ branching fraction, which is
calculated from CLEO and BaBar measurements (see 
Fig.~\ref{fig:charm_fds_tau}); and
\item the ratio $\Gamma(D^+_s\ra\mu^+\nu)/\Gamma(D^+_s\ra K^+K^-\pi^+)$
measured by BaBar, adjusted as described above~\cite{Chadha:note}.
\end{enumerate}
The WA \fds\ value is obtained by averaging 
the three results, accounting for correlations such as 
the values of $|V^{}_{cs}|$, $m^{}_{D^{}_s}$, and 
$\tau^{}_{D^{}_s}$ in Eq.~(\ref{eqn:fds_inverted}). The result
is shown in Fig.~\ref{fig:charm_fds}. The WA value is
\begin{eqnarray}
f^{}_{D^{}_s} & = & 254.6\,\pm 5.9 {\rm\ \ MeV},
\end{eqnarray}
where the statistical and systematic errors are combined.
This value can be compared to results from the two most precise 
lattice QCD calculations: it is $2.1\sigma$ higher than that 
from the HPQCD Collaboration 
($241\,\pm 3$\meve~\cite{Follana:2007uv}), and it is consistent 
with the less precise result from the Fermilab/MILC Collaboration 
($249\,\pm 11$\meve~\cite{Bernard:2009wr}).

\begin{figure}
\begin{center}
\includegraphics[width=5.0in]{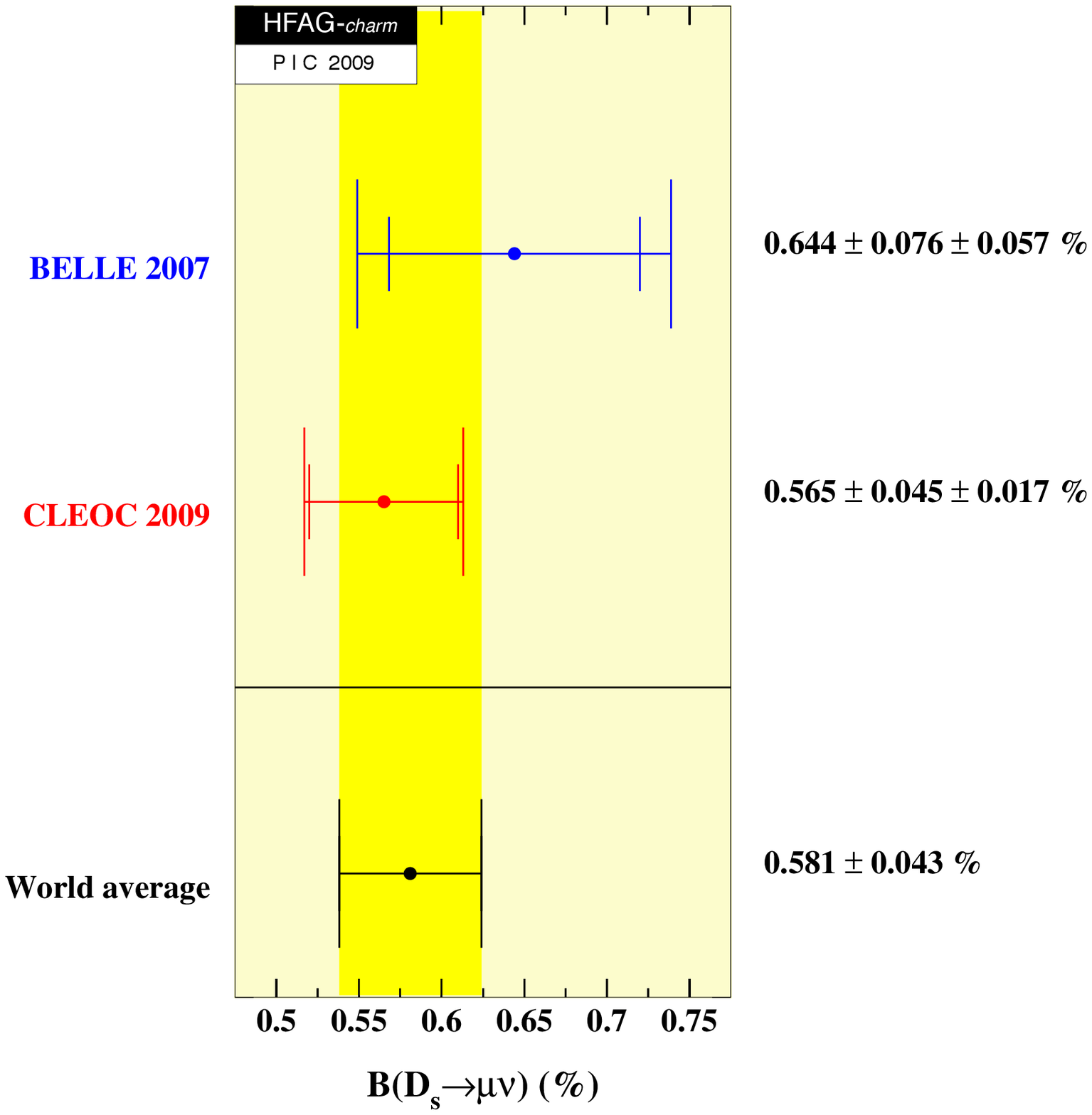}
\end{center}
\vskip-0.20in
\caption{\label{fig:charm_fds_mu}
WA value for ${\cal B}(D^+_s\ra\mu^+\nu)$, as
calculated from Refs.~\cite{Widhalm:2007ws,Alexander:2009ux}.
When two errors are listed, the first one is statistical and 
the second is systematic.}
\end{figure}

\begin{figure}
\begin{center}
\includegraphics[width=5.0in]{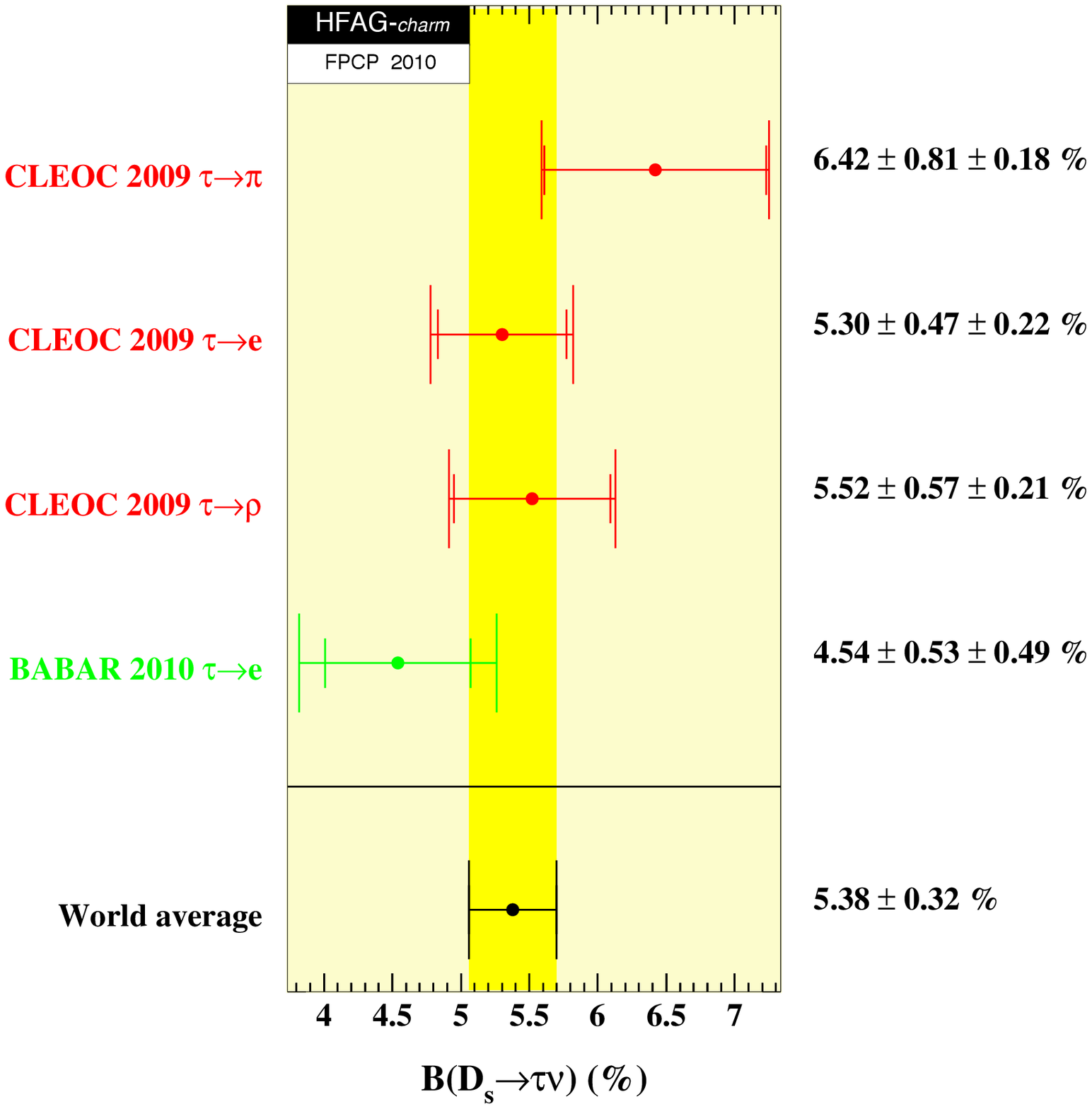}
\end{center}
\vskip-0.20in
\caption{\label{fig:charm_fds_tau}
WA value for ${\cal B}(D^+_s\ra\tau^+\nu)$, as calculated from 
Refs.~\cite{Alexander:2009ux,Onyisi:2009th,Naik:2009tk,Lees:2010qj}.
When two errors are listed, the first one is statistical and the
second is systematic.}
\end{figure}

\begin{figure}
\begin{center}
\includegraphics[width=5.0in]{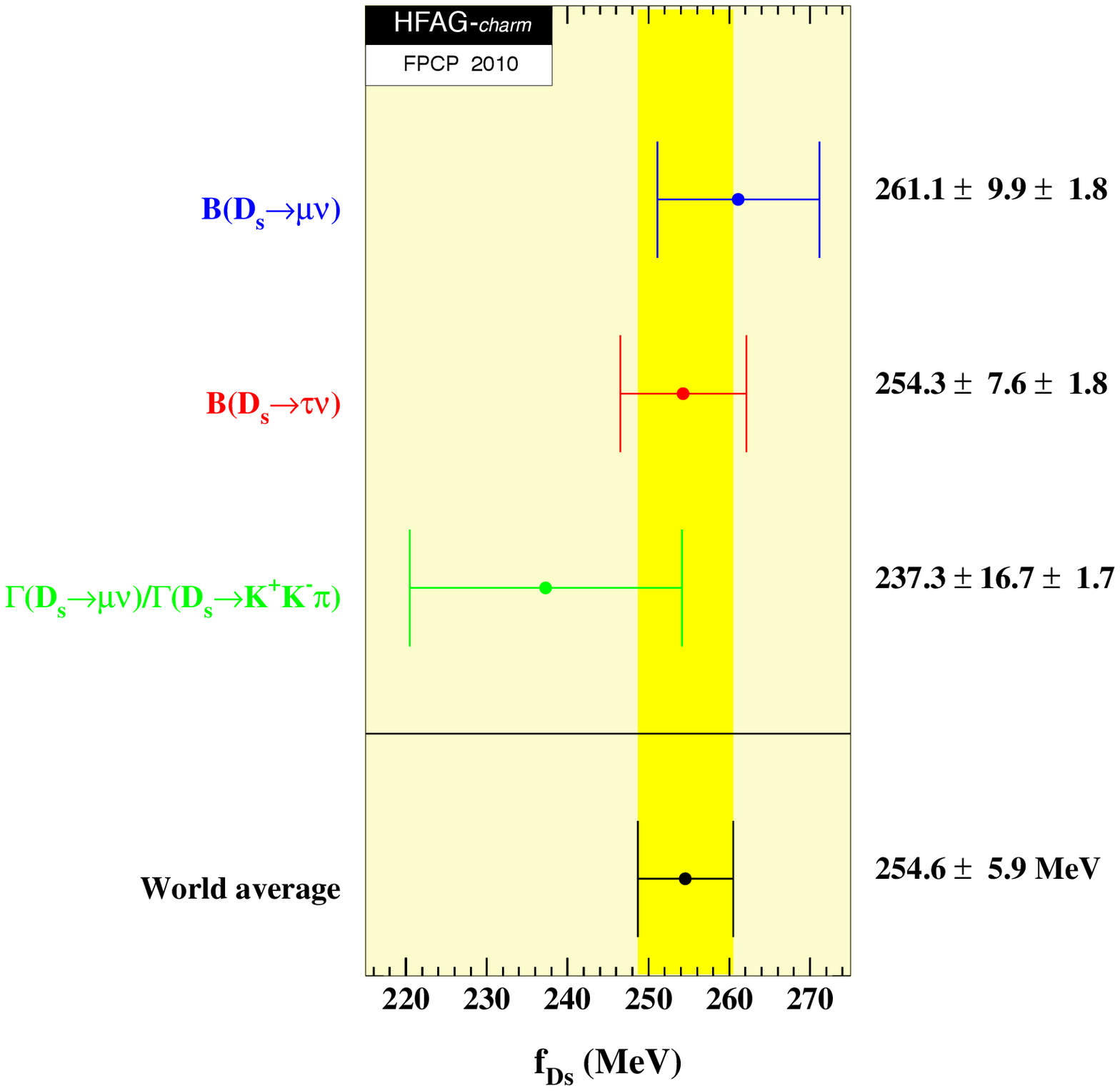}
\end{center}
\vskip-0.20in
\caption{\label{fig:charm_fds}
WA value for \fds. For each measurement, the first error listed 
is the total uncorrelated error, and the second error is the total
correlated error (mostly from $\tau^{}_{D^{}_s}$).}
\end{figure}

\clearpage

\subsection{Two-body Hadronic $D^0$ Decays and Final State Radiation}

Branching fractions measurements for $D^0\to K^-\pi^+$, $D^0\to \pi^+\pi^-$ 
and $D^0\to K^+ K^-$ have reached sufficient precision to allow averages 
with ${\cal O}(1\%)$ relative uncertainties. At these precisions, Final 
State Radiation (FSR) must be treated correctly and consistently across 
the input measurements for the accuracy of the averages to match the 
precision.  The sensitivity of measurements to FSR arises because of 
a tail in the distribution of radiated energy that extends to the 
kinematic limit.  The tail beyond $E_\gamma \approx 30$ MeV causes 
typical selection variables like the hadronic invariant mass to 
shift outside the selection range dictated by experimental 
resolution (see Fig.~\ref{fig:FSR_mass_shift}).  While the 
differential rate for the tail is small, the integrated rate 
amounts to several percent of the total $h^+ h^-(n\gamma)$ 
rate because of the tail's extent.  The tail therefore 
translates directly into a several percent loss in 
experimental efficiency.

All measurements that include an FSR correction have a correction 
based on use of 
PHOTOS~\cite{Barberio:1990ms,Barberio:1993qi,Golonka:2005pn,Golonka:2006tw} 
within the experiment's Monte Carlo simulation.  PHOTOS itself, however, 
has evolved, over the period spanning the set of measurements.  In 
particular, incorporation of interference between radiation off of 
the two separate mesons has proceeded in stages: it was first available 
for particle--antiparticle pairs in version 2.00 (1993), and extended 
to any two body, all charged, final states in version 2.02 (1999).  
The effects of interference are clearly visible 
(Figure~\ref{fig:FSR_mass_shift}), and cause a 
roughly 30\% increase in the integrated rate into 
the high energy photon tail.  To evaluate the FSR 
correction incorporated into a given measurement, 
we must therefore note whether any correction was 
made, the version of PHOTOS used in correction, 
and whether the interference terms in PHOTOS were 
turned on.  

\subsubsection{Branching Fraction Corrections}

\begin{figure}[bt]
\begin{center}
\includegraphics[width=0.5\textwidth,angle=-90.]{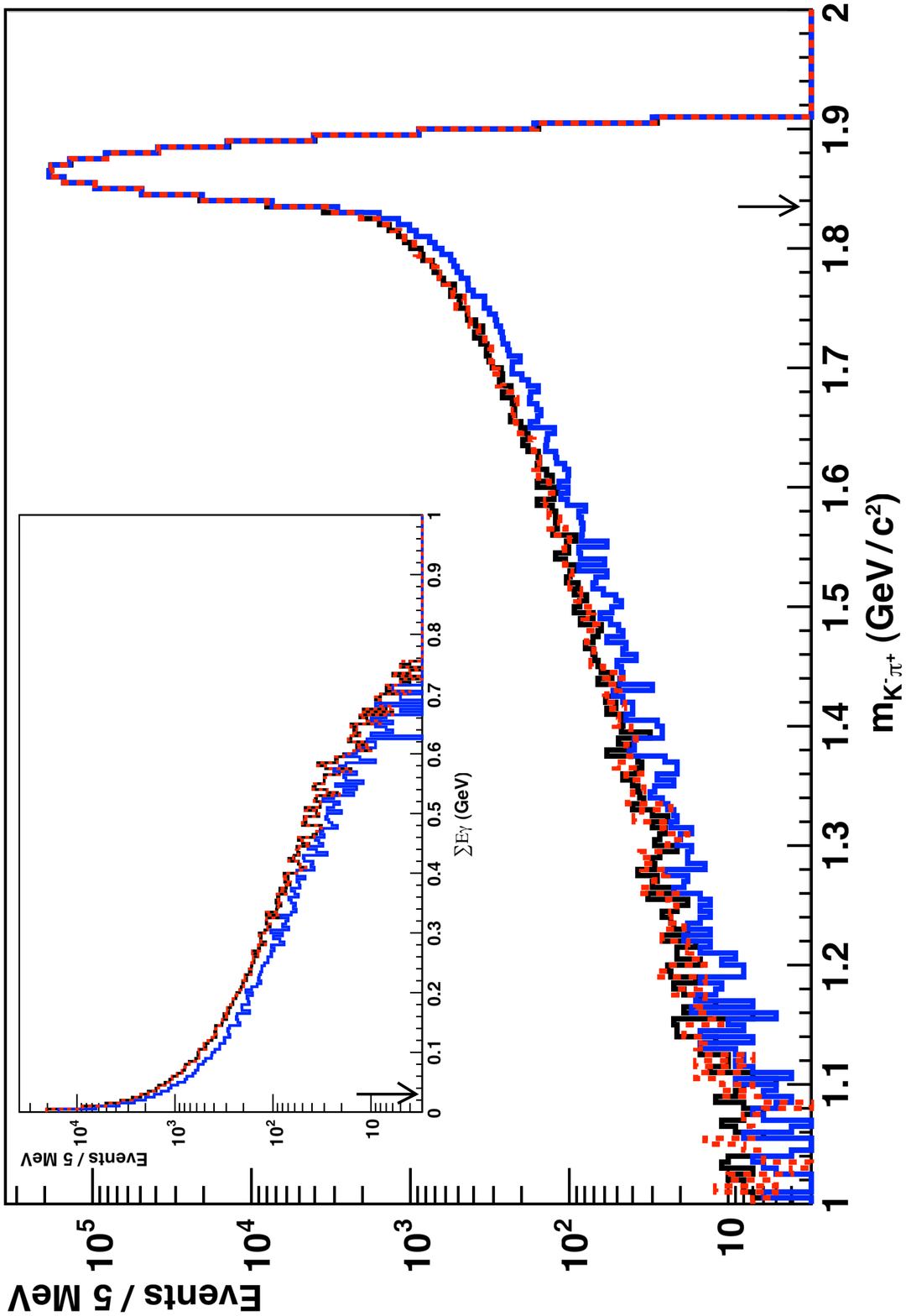}
\caption{The $K\pi$ invariant mass distribution for 
$D^0\to K^-\pi^+ (n\gamma)$ decays. The 3 curves correspond 
to three different configurations of PHOTOS for modeling FSR: 
version 2.02 without interference (blue), version 2.02 with 
interference (red dashed) and version 2.15 with interference (black).  
The true invariant mass has been smeared with a typical experimental 
resolution of 10 MeV${}/c^2$.  Inset: The corresponding spectrum of 
total energy radiated per event.  The arrow indicates the $E_\gamma$ 
value that begins to shift kinematic quantities outside of the range 
typically accepted in a measurement.}
\label{fig:FSR_mass_shift}
\end{center}
\end{figure}

Before averaging the measured branching fractions, the published 
results are updated, as necessary, to the FSR prediction of 
PHOTOS~2.15 with interference included.  The correction will 
always shift a branching fraction to a higher value: with no 
FSR correction or with no interference term in the correction, 
the experimental efficiency determination will be biased high, 
and therefore the branching fraction will be biased low.

Most of the branching fraction analyses used the kinematic quantity sensitive to FSR in the candidate selection criteria.  For the analyses at the $\psi(3770)$, the variable was $\Delta E$, the difference between the candidate $D^0$ energy and the beam energy ({\em e.g.}, $E_K + E_\pi - E_{\rm beam}$ for $D^0\to K^-\pi^+$).  In the remainder of the analyses, the relevant quantity was the reconstructed hadronic two-body mass $m_{h^+h^-}$.  To correct we need only to evaluate the fraction of decays that FSR moves outside of the range accepted for the analysis.  

The corrections were evaluated using an event generator (EvtGen \cite{Ryd:2005zz}) that incorporates PHOTOS to simulate the portions of the decay process most relevant to the correction.  We compared corrections determined both with and without smearing to account for experimental resolution.  The differences were negligible, typically of order of a 1\% of the correction itself.  The immunity of the correction to resolution effects comes about because most of the long FSR-induced tail in, for example, the $m_{h^+h^-}$ distribution resides well away from the selection boundaries.  The smearing from resolution, on the other hand, mainly affects the distribution of events right at the boundary.  

For measurements incorporating an FSR correction that did not include interference, we update by assessing the FSR-induced efficiency loss for both the PHOTOS version and configuration used in the analysis and our nominal version 2.15 with interference.  For measurements that published their sensitivity to FSR, our generator-level predictions for the original efficiency loss agreed to within a few percent (of the correction).  This agreement lends additional credence to the procedure.

Once the event loss from FSR in the most sensitive kinematic quantity is accounted for, the event loss from other quantities is very small.  Analyses using $D^*$ tags, for example, showed little sensitivity to FSR in the reconstructed $D^*-D^0$ mass difference: for example, in $m_{K^-\pi^+\pi^+}-m_{K^-\pi^+}$. Because the effect of FSR tends to cancel in the difference of the reconstructed masses, this difference showed a much smaller sensitivity than the two body mass even before a two body mass requirement. In the $\psi(3770)$ analyses, the beam-constrained mass distributions ($\sqrt{E_{\rm beam}^2 - |\vec{p}_K + \vec{p}_\pi|^2}$)  showed little further sensitivity.

\begin{figure}
\begin{center}
\includegraphics[width=0.24\textwidth,angle=-90.]{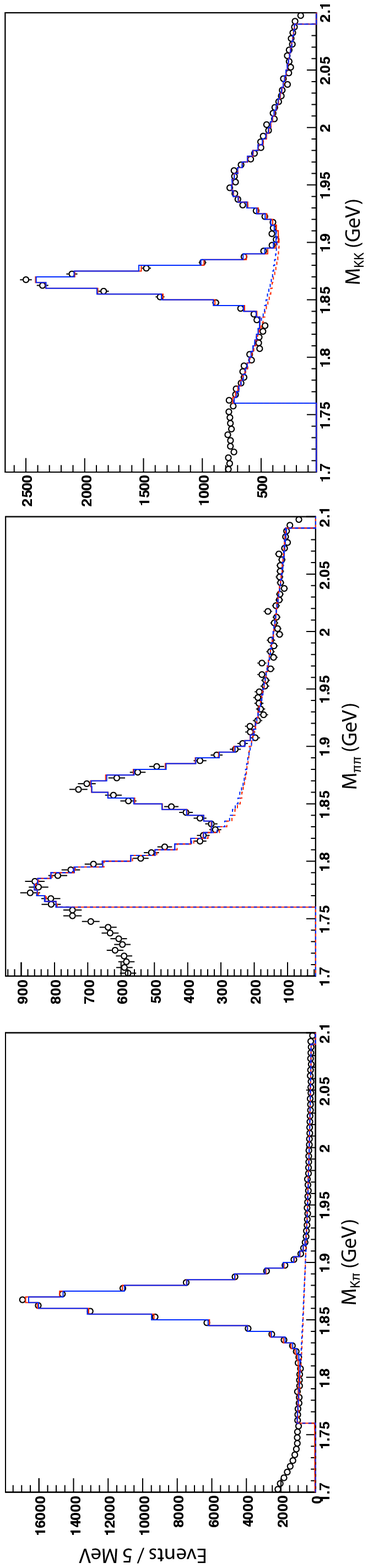}
\caption{FOCUS data (dots), original fits (blue) and 
toy MC parameterization (red) for $D^0\to K^-\pi^+$ (left) , 
$D^0\to \pi^+\pi^-$ (center) and $D^0\to \pi^+\pi^-$ (right).}
\label{fig:FocusFits}
\end{center}
\end{figure}

The FOCUS \cite{Link:2002hi} analysis of the branching ratios ${\cal B}(D^0\to \pi^+\pi^-)/{\cal B}(D^0\to K^-\pi^+)$ and ${\cal B}(D^0\to K^+ K^-)/{\cal B}(D^0\to K^-\pi^+)$ obtained yields using fits to the two body mass distributions.  FSR will both distort the low end of the signal mass peak, and will contribute a signal component to the low side tail used to estimate the background.  The fitting procedure is not sensitive to signal events out in the FSR tail, which would be counted as part of the background.

A more complex toy Monte Carlo procedure was required to analyze the effect of FSR on the fitted yields, which were published with no FSR corrections applied.  A detailed description of the procedure and results is available at the HFAG site \cite{charm_hh_webpage}, and a brief summary is provided here.  Determining the correction involved an iterative procedure in which samples of similar size to the FOCUS sample were generated and then fit using the FOCUS signal and background parameterizations.  The MC parameterizations were tuned based on differences between the fits to the toy MC data and the FOCUS fits, and the procedure was repeated. These steps were iterated until the fit parameters matched the original FOCUS parameters.  

\begin{table}
  \centering 
  \caption{The experimental measurements relating to ${\cal B}(D^0\to K^-\pi^+)$, ${\cal B}(D^0\to \pi^+\pi^-)$ and ${\cal B}(D^0\to K^+ K^-)$ after correcting to the common version and configuration of PHOTOS.  The uncertainties are statistical and total systematic, with the FSR-related systematic estimated in this procedure shown in parentheses.  Also listed are the percent shifts in the results from the correction, if any, applied here, as well as the original PHOTOS and interference configuration for each publication.}
  \label{tab:FSR_corrections}
\begin{tabular}{lccc}
\hline \hline
Experiment & result (rescaled) & correction [\%] & PHOTOS \\ \hline
\multicolumn{4}{l}{$D^{0} \to K^{-} \pi^{+}$} \\
      CLEO-c 07  (CC07) \cite{:2007zt}         & $3.891 \pm 0.035 \pm 0.065(27)\%$ & --   & 2.15/Yes \\      
      BaBar 07   (BB07) \cite{Aubert:2007wn}   & $4.035 \pm 0.037 \pm 0.074(24)\%$ & 0.69 & 2.02/No \\
      CLEO II 98 (CL98) \cite{Artuso:1997mc}   & $3.920 \pm 0.154 \pm 0.168(32)\%$ & 2.80 & none \\
      ALEPH 97   (AL97) \cite{Barate:1997mm}   & $3.930 \pm 0.091 \pm 0.125(32)\%$ & 0.79 & 2.0/No \\
      ARGUS 94   (AR94) \cite{Albrecht:1994nb} & $3.490 \pm 0.123 \pm 0.288(24)\%$ & 2.33 & none \\
      CLEO II 93 (CL93) \cite{Akerib:1993pm}   & $3.960 \pm 0.080 \pm 0.171(15)\%$ & 0.38 & 2.0/No \\
      ALEPH 91   (AL91) \cite{Decamp:1991jw}   & $3.730 \pm 0.351 \pm 0.455(34)\%$ & 3.12 & none \\
\multicolumn{4}{l}{$D^{0} \to \pi^{+}\pi^{-} / D^{0} \to K^{-} \pi^{+}$} \\
      CLEO-c 05  (CC05) \cite{Rubin:2005py}    & $0.0363 \pm 0.0010 \pm 0.0008(01)$ & 0.25 & 2.02/No \\
      CDF 05     (CD05) \cite{Acosta:2004ts}   & $0.03594 \pm 0.00054 \pm 0.00043(15)$ & -- & 2.15/Yes \\
      FOCUS 02   (FO02) \cite{Link:2002hi}     & $0.0364 \pm 0.0012 \pm 0.0006(02)$ & 3.10 & none \\
\multicolumn{4}{l}{$D^{0} \to K^{+}K^{-} / D^{0} \to K^{-} \pi^{+}$} \\
      CDF 05 \cite{Acosta:2004ts}    & $0.0992 \pm 0.0011 \pm 0.0012(01)$ & -- & 2.15/Yes \\
      FOCUS 02 \cite{Link:2002hi}     & $0.0982 \pm 0.0014 \pm 0.0014(01)$ & -1.12 & none \\
\multicolumn{4}{l}{$D^{0} \to K^{+}K^{-}$} \\
      CLEO-c 08  (CC08) \cite{:2008nr}       & $0.411 \pm 0.008 \pm 0.009\%$ & 0.64 & 2.02/No \\ \hline
\end{tabular}
\end{table}

The toy MC samples for the first iteration were based on the generator-level distribution of $m_{K^-\pi^+}$, $m_{\pi^+\pi^-}$ and $m_{K^+K^-}$, including the effects of FSR, smeared according to the original FOCUS resolution function, and on backgrounds thrown using the parameterization from the final FOCUS fits.  For each iteration, 400 to 1600 individual data-sized samples were thrown and fit. The means of the parameters from these fits determined the corrections to the generator parameters for the following iteration.  The ratio between the number of signal events generated and the final signal yield provides the required FSR correction in the final iteration.  Only a few iterations were required in each mode.  Figure ~\ref{fig:FocusFits} shows the FOCUS data, the published FOCUS fits, and the final toy MC parameterizations.  The toy MC provides an excellent description of the data.

The corrections obtained to the individual FOCUS yields were $1.0298\pm 0.0001$ for $K^-\pi^+$, $1.062 \pm 0.001$ for $\pi^+\pi^-$, and $1.0183 \pm 0.0003$ for $K^+K^-$.  These corrections tend to cancel in the branching ratios, leading to corrections of 1.031 to  ${\cal B}(D^0\to \pi^+\pi^-)/{\cal B}(D^0\to K^-\pi^+)$, and 0.9888 for ${\cal B}(D^0\to K^+ K^-)/{\cal B}(D^0\to K^-\pi^+)$.

Table~\ref{tab:FSR_corrections} summarizes the corrected branching fractions.  The published FSR-related modeling uncertainties have been replaced by with a new, common, estimate based on the assumption that the dominant uncertainty in the FSR corrections come from the fact that the mesons are treated like structureless particles. No contributions from structure-dependent terms in the decay process (eg. radiation off individual quarks) are included in PHOTOS. Internal studies done by various experiments have indicated that in $K\pi$ decay, the PHOTOS corrections agree with data at the 20-30\% level. We therefore attribute a 25 uncertainty to the FSR prediction from potential structure-dependent contributions. For the other two modes, the only difference in structure is the final state valence quark content. While radiative corrections typically come in with a $1/M$ dependence, one would expect the additional contribution from the structure terms to come in on time scales shorter than the hadronization time scale. In this case, you might expect LambdaQCD to be the relevant scale, rather than the quark masses, and therefore that the amplitude is the same for the three modes. In treating the correlations among the measurements this is what we assume. We also assume that the PHOTOS amplitudes and any missing structure amplitudes are relatively real with constructive interference.  The uncertainties largely cancel in the branching fraction ratios. For the final average branching fractions, the FSR uncertainty on $K\pi$ dominates. Note that because of the relative sizes of FSR in the different modes, the $\pi\pi/K\pi$ branching ratio uncertainty from FSR is positively correlated with that for $K\pi$ branching, while the $KK/K\pi$ branching ratio FSR uncertainty is negatively correlated.

The ${\cal B}(D^0\to K^-\pi^+)$ measurement of reference~\cite{Coan:1997ye}, the  
${\cal B}(D^0\to \pi^+\pi^-)/{\cal B}(D^0\to K^-\pi^+)$ measurements of references~\cite{Aitala:1997ff} 
and~\cite{Csorna:2001ww} and the ${\cal B}(D^0\to K^+ K^-)/{\cal B}(D^0\to K^-\pi^+)$ measurement
of reference~\cite{Csorna:2001ww} are excluded from the branching fraction averages presented here.
The measurements appear not to have incorporated any FSR corrections, and insufficient information
is available to determine the 2-3\% corrections that would be required.

\begin{sidewaystable}[p]
  \centering 
  \caption{The correlation matrix corresponding to the covariance matrix from the sum of statistical,
  systematic and FSR covariances.}\label{tab:correlations}
  \small

\end{sidewaystable}

\subsubsection{Average Branching Fractions}

\begin{figure}
\begin{center}
\includegraphics[width=0.55\textwidth,angle=-90.]{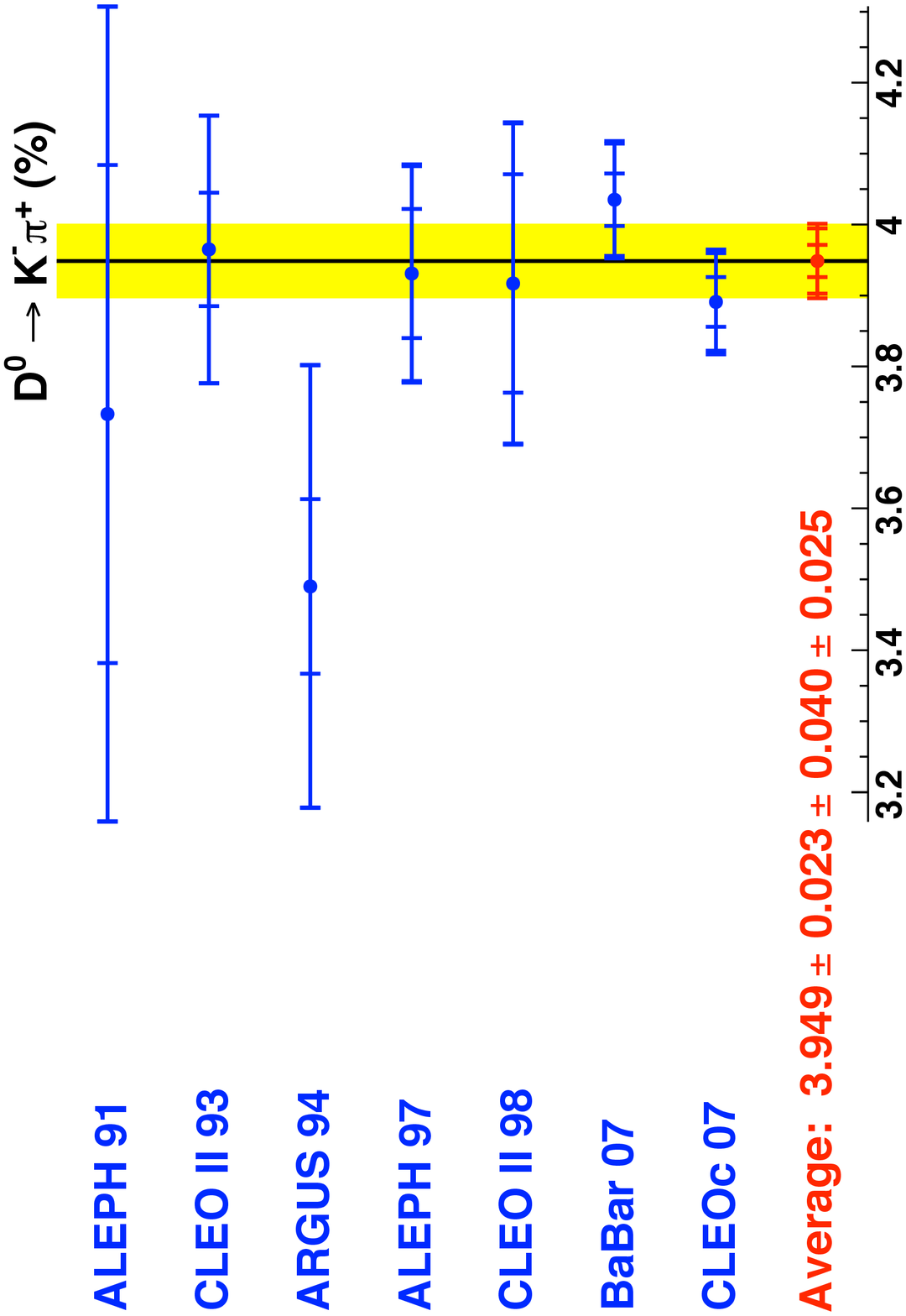}
\caption{Comparison of measurements of 
${\cal B}(D^0\to K^-\pi^+)$ (blue) with the average 
branching fraction obtained here (red, and yellow band).}
\label{D0bfs}
\end{center}
\end{figure}

The average branching fractions for $D^0\to K^-\pi^+$, $D^0\to \pi^+\pi^-$ and $D^0\to K^+ K^-$ are obtained from
a single $\chi^2$ minimization procedure,  in which the three branching fractions are floating parameters.  The
central values derive from a fit in which the covariance matrix is the sum of the covariance matrices for the
statistical, systematic (excluding FSR) and FSR uncertainties.  The statistical uncertainties are obtained from
a fit using only the statistical covariance matrix.  The systematic uncertainties are obtained from the
quadrature uncertainties from a fit with statistical-only and statistical+systematic covariance matrices, and
the FSR uncertainties on the averages from the quadrature differences in the uncertanties obtained from the
nominal fit and a fit excluding the FSR uncertainties.

In forming the covariance matrix for the FSR uncertainties, the FSR
uncertainties are treated as fully correlated (or anti-correlated) as described above.  For the
systematic covariance matrix, ALEPH's systematic uncertainties in the $\theta_{D^*}$ parameter are treated
as fully correlated between the ALEPH 97
and ALEPH 91 measurements.  Similarly, the tracking efficiency uncertainties in the CLEO II 98 and the
CLEO II 93 measurements are treated as fully
correlated.  Finally, the CLEO-c 07 $D^0\to K^-\pi^+$ measurement and the CLEO-c 08 $D^0\to K^+ K^-$
measurements have a significant statistical correlation.  The 2007 hadronic branching fraction analysis
derives the number of $N_{D^0\bar{D}^0}$ pairs produced in CLEO-c, and that quantity is statistically
correlated with the $D^0\to K^-\pi^+$ branching fraction in that analysis ($\rho=0.65$).  The 2008 
$K^+K^-$ analysis in turn uses that value of $N_{D^0\bar{D}^0}$ as the normalization for its branching
fraction.  Table~\ref{tab:correlations} presents the correlation matrix for the nominal fit (stat.+syst.+FR).

The averaging procedure results in a final $\chi^2$ of 8.5  for 13-3 degrees of freedom.  The branching
fractions obtained are
\begin{eqnarray*}
  {\cal B}(D^0\to K^-\pi^+)   & = & 3.949 \pm 0.023 \pm 0.040 \pm 0.025 \\
  {\cal B}(D^0\to \pi^+\pi^-) & = & 0.143 \pm 0.002 \pm 0.002 \pm 0.001 \\
  {\cal B}(D^0\to K^+ K^-)    & = & 0.394 \pm 0.004 \pm 0.005 \pm 0.002. \\
\end{eqnarray*}
The uncertainties, estimated as described above, are statistical, systematic (excluding FSR), and
FSR modeling.  The correlation coefficients from the fit using the total uncertainties are
\begin{center}
\begin{tabular}{llll}
               & $K^-\pi^+$ & $\pi^+\pi^-$ & $K^+ K^-$ \\
$K^-\pi^+$     &  1.00 & 0.72 & 0.74  \\
$\pi^+\pi^-$   &  0.72 & 1.00 & 0.53  \\
$K^+ K^-$      &  0.74 & 0.53 & 1.00  \\
\end{tabular}
\end{center}

\begin{table}[b]
  \centering 
  \caption{Evolution of the $D^0\to K^-\pi^+$ branching fraction from a fit with no FSR corrections or correlations (similar  to the average in the PDG 2009 update \cite{Amsler:2008zzb}) to the nominal fit presented here.}\label{tab:fit_evolution}
\begin{tabular}{cccll}
\hline\hline
Modes &  description                       & ${\cal B}(D^0\to K^-\pi^+)$ (\%)           & $\chi^2$ / (d.o.f.) \\
fit        &                               &                                       & \\ \hline
$K^-\pi^+$ & PDG summer 2009 equivalent    & $3.913 \pm 0.022 \pm 0.043 $          & 6.0 / (8-1)\\
$K^-\pi^+$ & drop Ref.~\cite{Coan:1997ye}  & $3.921 \pm 0.023 \pm 0.044$           & 4.8 / (7-1)\\
$K^-\pi^+$ & add FSR corrections           & $3.940 \pm 0.023 \pm 0.041 \pm 0.015$ & 4.0 / (7-1)\\
$K^-\pi^+$ & add FSR correlations          & $3.940 \pm 0.023 \pm 0.041 \pm 0.025$ & 4.2 / (7-1)\\
all        & CDF + FOCUS only              & $3.940 \pm 0.023 \pm 0.041 \pm 0.025$  & 4.5 /(12-3) \\
all        & add CLEO-c $K^+ K^-$          & $3.949 \pm 0.023 \pm 0.040 \pm 0.025$  & 8.5 /(13-3)\\
\hline
\end{tabular}
\end{table}

As the $\chi^2$ would suggest and Fig.~\ref{D0bfs}, the average value for ${\cal B}(D^0\to K^-\pi^+)$ and
the input branching fractions agree very well.  With the estimated uncertainty in the FSR modeling used here,
the FSR uncertainty dominates the statistical uncertainty in the average, suggesting that experimental
work in the near future should focus on verification of FSR with 
$E_\gamma \simge 100$ MeV.

The ${\cal B}(D^0\to K^-\pi^+)$ average obtained here is approximately one statistical standard deviation higher than the
2009 PDG update \cite{Amsler:2008zzb}.  Table~\ref{tab:fit_evolution} shows the evolution from a
fit similar to the PDG's (no FSR corrections or correlations, reference~\cite{Coan:1997ye}
excluded) to the average presented here.  There are three main contributions to the difference,
which only coincidentally all shift the result upwards.  The branching fraction in reference~\cite{Coan:1997ye} happens
to be on the low side, and its exclusion shifts the result by $+0.008\%$.  The FSR corrections are expected
to shift the result upwards, and indeed contribute a shift of $+0.019\%$.  Finally, including the CLEO-c
absolute $D^0\to K^+ K^-$ branching fraction contributes the final shift of $+0.009\%$.  As Fig.~\ref{fig:D0KK}
shows, the $K^+ K^-$ branching fractions inferred from the combining the CDF and FOCUS branching ratios and
the average $K^-\pi^+$ branching fraction (excluding the CLEO-c $K^+ K^-$ result) are both lower than
the CLEO-c absolute measurement.  The fit, therefore, exerts an upward pressure on the $K^-\pi^+$ result
to improve the agreement in the $K^+ K^-$ sector.

\begin{figure}
\begin{center}
\includegraphics[width=0.45\textwidth,angle=-90.]{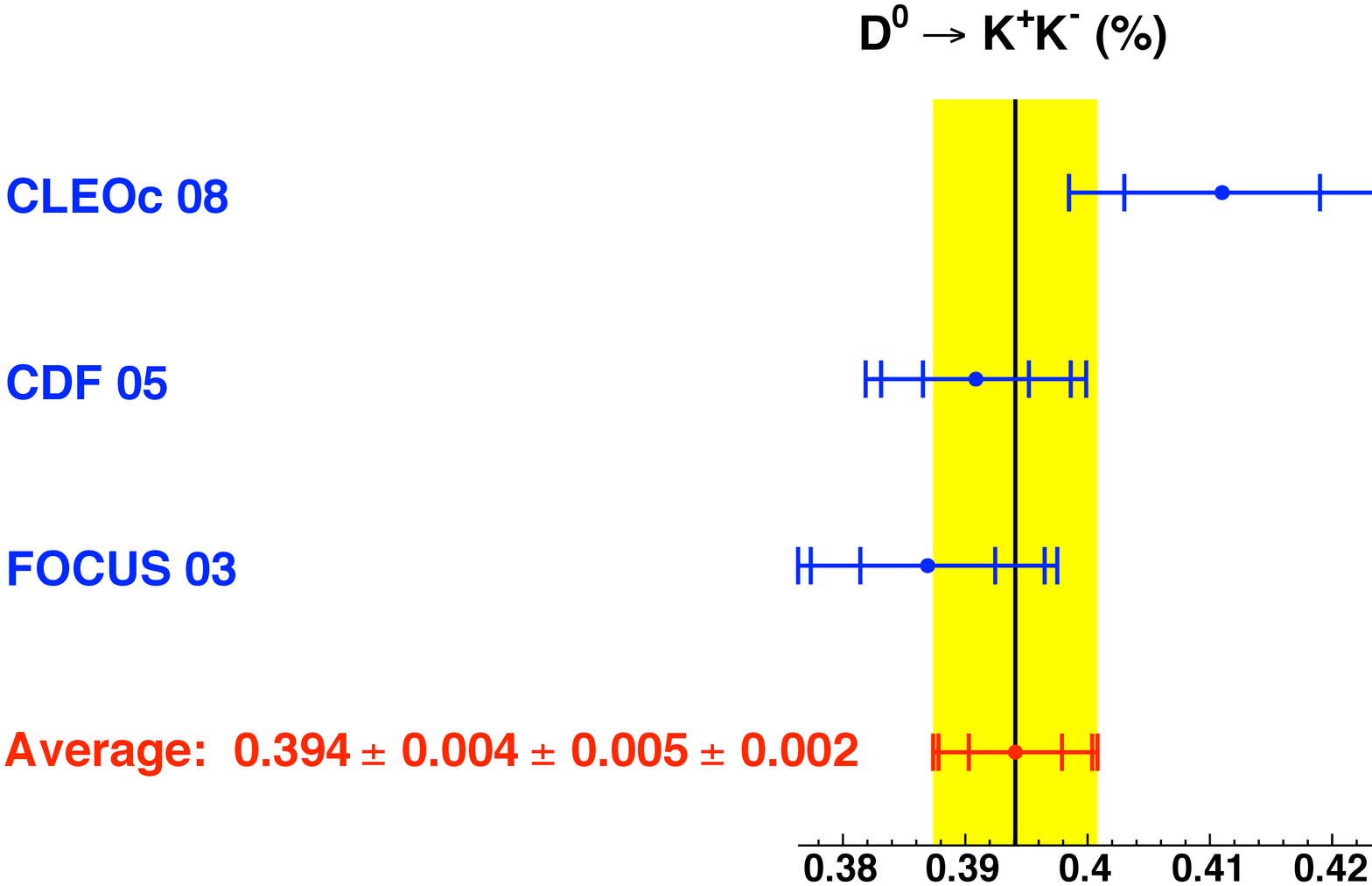}
\caption{Comparison of the absolute CLEO-c ${\cal B}(D^0\to K^+K^-)$ 
measurement, the CDF and FOCUS branching ratio measurements scaled 
by the ${\cal B}(D^0\to K^-\pi^+)$ branching fraction, and this 
average (red point, yellow band).}
\label{fig:D0KK}
\end{center}
\end{figure}

\clearpage
\mysection{$\tau$ lepton Properties}

\label{sec:tau}

The aim of this chapter is to provide average values of the properties of the $\tau$ lepton.
The mass of the $\tau$ lepton is presented in section~\ref{subsec:Tau_Mass}, and
the branching fractions of the decay modes are presented in section~\ref{subsec:Tau_BR}.
Using these average values, we present tests of charged current lepton universality in
section~\ref{subsec:Tau_LU} and obtain estimates for \Vus, the relative weak coupling 
between up and strange quarks, in section~\ref{subsec:Tau_Vus}.
We summarize the status of searches for lepton flavor violating decays of the $\tau$ lepton in section~\ref{subsec:Tau_LFV}.

\subsection{Mass of the $\tau$ lepton}
\label{subsec:Tau_Mass}

\begin{figure}[!hbtp]
\begin{center}
\includegraphics[height=.4\textheight,width=.44\textwidth]{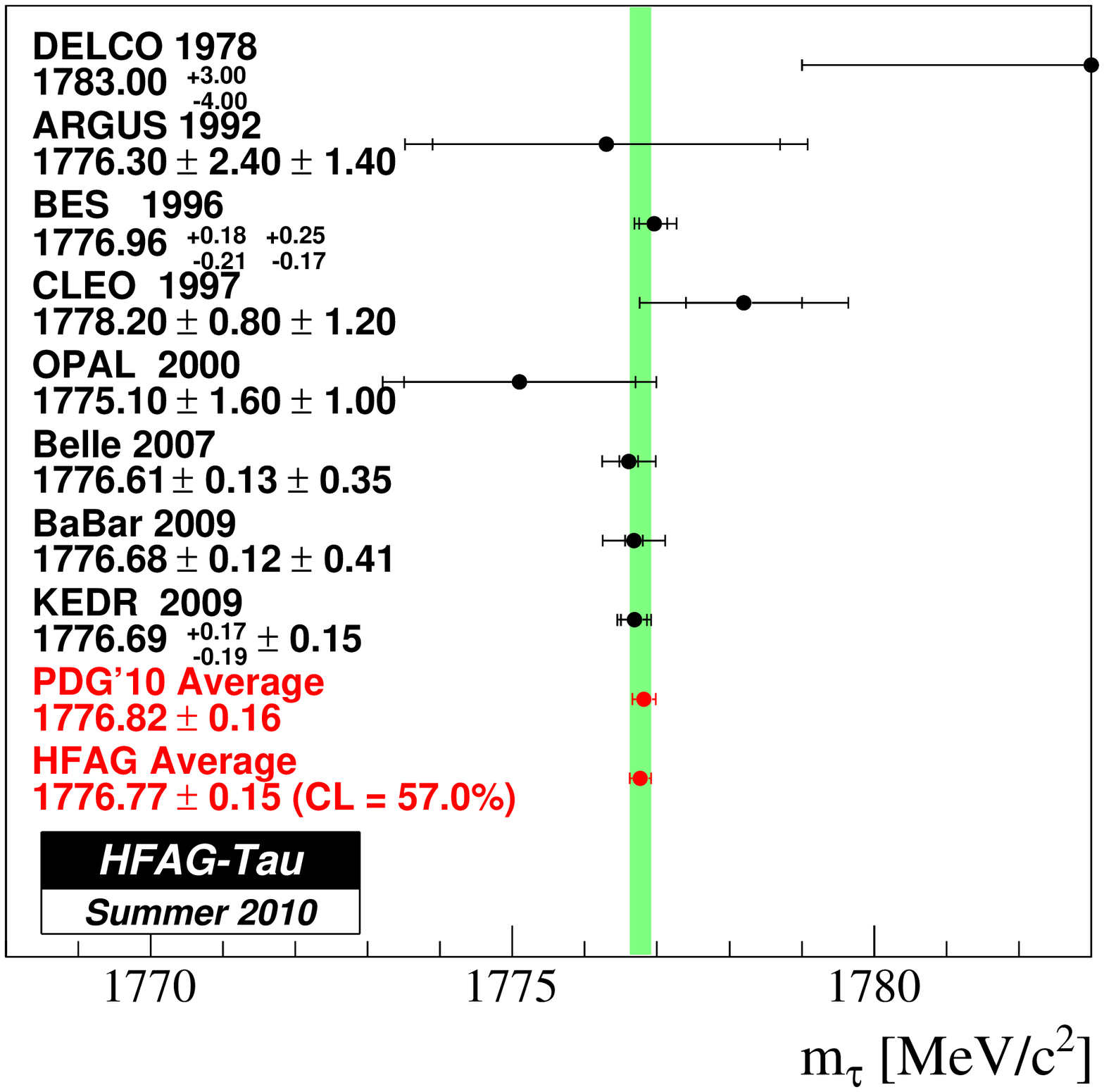}
\end{center}
\caption{Measurements and average value of $m_\tau$.}
\label{fig:Tau_Mass}
\end{figure}

The mass of the $\tau$ lepton has recently been measured by the BaBar and Belle experiments
using the end-point technique from the pseudo-mass distribution in $\tau^-\to\pi^-\pi^-\pi^+\nu_\tau$ decays,
as well as by the KEDR experiment from a study of the $\tau^+\tau^-$ cross-section around the production threshold.
In Figure~\ref{fig:Tau_Mass} we present the measurements and average values of $m_\tau$.


\subsection{$\tau$ Branching Fractions:}
\label{subsec:Tau_BR}

In this section we present the measurements and average values of the 
$\tau$ branching fractions~\footnote{Charge conjugate $\tau$ decays are implied throughout.},
including those which have been recently measured by the B-Factories. 
We take into account correlations between measurements,
arising from common dependence on the $\tau-$pair cross-section~\cite{Banerjee:2007is}
and the assumed knowledge of the branching fractions for the
background modes. For measurements from the same experiment,
we treat the dependence on detector-specific systematics
as sources of correlated systematic uncertainties.

We report here results from single-quantity averages,
which includes correlations between the B-Factories,
as well as the results from a global fit,
which includes correlations between the different branching fractions.
We label results from the former as ``HFAG Average'',
and the latter as ``HFAG Fit'' in the following figures.

For the ``HFAG Fit'', we use 131 measurements from non-B-Factory experiments, 
which includes the set of 124 measurements used in the global fit performed by
the PDG~\cite{PDG_2010}. The measurements from non-B-Factories include
     37 measurements from ALEPH,
      2 measurements from ARGUS,
      1 measurement from CELLO,
     36 measurements from CLEO,
      6 measurements from CLEO3,
     14 measurements from DELPHI,
      2 measurements from HRS,
     11 measurements from L3,
     19 measurements from OPAL, and
      3 measurements from TPC.

All of these measurements can be expressed as a linear function
of the form $(\sum_i \alpha_i P_i )\over(\sum_j \beta_j P_j )$ 
of few selected branching fractions $(P_i)$, which are labelled as
base modes. The base modes are chosen such that they sum up to unity. 

For the 124 measurements used in the PDG global fit, there are 31 base modes. 
These results and  their corresponding references are listed in Ref.~\cite{PDG_2010}.
We first augment this set with 4 additional base modes of
${\cal{B}}(\tau^- \to K^- \pi^0 \eta \nu_\tau)$, 
${\cal{B}}(\tau^- \to \bar{K}^0 \pi^- \eta  \nu_\tau)$, 
${\cal{B}}(\tau^- \to \bar{K}^0 \pi^- 2\pi^0 \nu_\tau)$
and
${\cal{B}}(\tau^- \to \bar{K}^0 h^- h^- h^+ \nu_\tau)$,
because 2 of these modes (containing $\eta$) have been recently measured by the
B-Factories with significant precision.
We also include 
2 measurements from CLEO for the modes containing $\eta$,
2 measurements from ALEPH for the other modes,
and 1 measurement from OPAL of ${\cal{B}}(\tau^- \to \bar{K}^0 \pi^- >= 1 \pi^0 \nu_\tau)$.
We further include 1 measurement of ${\cal{B}}(\tau^- \to \pi^- \pi^- \pi^+ \eta  \nu_\tau)$ from CLEO,
1 measurement of ${\cal{B}}(\tau^- \to  K^- \omega \nu_\tau)$ from CLEO3,
and replace 1 base node of ${\cal{B}}(\tau^- \to  h^- \omega \nu_\tau)$ with 2 base nodes
containing measurements of
${\cal{B}}(\tau^- \to  \pi^- \omega \nu_\tau)$ and ${\cal{B}}(\tau^- \to  K^- \omega \nu_\tau)$.
This leads us to a global fit to a set of 131 measurements with 36 base nodes.

Finally we include the following 22 measurements from the B-Factories:
\begin{itemize}
\item 12 measurements from the BaBar collaboration:
\begin{eqnarray*}
{{\cal{B}}(\tau^- \to \mu^- \bar{\nu}_\mu \nu_\tau)}/{{\cal{B}}(\tau^- \to e^- \bar{\nu}_e \nu_\tau)} & =  & (0.9796\, \pm\, 0.0016\, \pm\, 0.0036)~\cite{Aubert:2009qj},\\
{{\cal{B}}(\tau^- \to \pi^- \nu_\tau)}/{{\cal{B}}(\tau^- \to e^- \bar{\nu}_e \nu_\tau)} & =  & (0.5945\, \pm\, 0.0014\, \pm\, 0.0061)~\cite{Aubert:2009qj},\\
{{\cal{B}}(\tau^- \to  K^-  \nu_\tau)}/{{\cal{B}}(\tau^- \to e^- \bar{\nu}_e \nu_\tau)} & =  & (0.03882\, \pm\, 0.00032\, \pm\, 0.00057)~\cite{Aubert:2009qj},\\
{\cal{B}}(\tau^- \to K^- \pi^0 \nu_\tau) & =  & (0.416\, \pm\, 0.003\, \pm\, 0.018)\%~\cite{Aubert:2007jh}\\
{\cal{B}}(\tau^- \to \bar{K}^0 \pi^- \nu_\tau) & =  & (0.840\, \pm\, 0.004\, \pm\, 0.023)\%~\cite{Aubert:2008an}\\
{\cal{B}}(\tau^- \to \bar{K}^0 \pi^- \pi^0 \nu_\tau) & =  & (0.342\, \pm\, 0.006\, \pm\, 0.015)\%~\cite{Paramesvaran:2009ec}\\
{\cal{B}}(\tau^- \to \pi^- \pi^- \pi^+ \nu_\tau ~(\mathrm{ex.~}K^0)) & = & (8.834 \, \pm\, 0.007\, \pm\, 0.127)\%~\cite{Aubert:2007mh}\\
{\cal{B}}(\tau^- \to  K^-  \pi^- \pi^+ \nu_\tau ~(\mathrm{ex.~}K^0)) & = & (0.273\, \pm\, 0.002\, \pm\, 0.009)\%~\cite{Aubert:2007mh}\\
{\cal{B}}(\tau^- \to  K^-  \pi^-  K^+ \nu_\tau)  & = & (0.1346\, \pm\, 0.0010\, \pm\, 0.0036)\%~\cite{Aubert:2007mh}\\
{\cal{B}}(\tau^- \to  K^-   K^-   K^+ \nu_\tau)  & = & (1.58\, \pm\, 0.13\, \pm\, 0.12)\times 10^{-5}~\cite{Aubert:2007mh}\\
{\cal{B}}(\tau^- \to  3h^- 2h^+ \nu_\tau ~(\mathrm{ex.~}K^0))  & = & (8.56\, \pm\, 0.05\, \pm\, 0.42)\times 10^{-4}~\cite{Aubert:2005waa}\\
{\cal{B}}(\tau^- \to  2\pi^- \pi^+ \eta \nu_\tau ~(\mathrm{ex.~}K^0))  & = & (1.60\, \pm\, 0.05\, \pm\, 0.11)\times 10^{-4}~\cite{Aubert:2008nj}
\end{eqnarray*}
\item 10 measurements from the Belle collaboration:
\begin{eqnarray*}
{\cal{B}}(\tau^- \to h^- \pi^0 \nu_\tau) & =  & (25.67\, \pm\, 0.01\, \pm\, 0.39)\%~\cite{Fujikawa:2008ma}\\
{\cal{B}}(\tau^- \to \bar{K}^0 \pi^- \nu_\tau) & = & (0.808\, \pm\, 0.004\, \pm\, 0.026)\%~\cite{Epifanov:2007rf}\\
{\cal{B}}(\tau^- \to \pi^- \pi^- \pi^+ \nu_\tau ~(\mathrm{ex.~}K^0)) & = & (8.420\, \pm\, 0.003\, {~}^{+0.260}_{-0.250})\%~\cite{Lee:2010tc}\\
{\cal{B}}(\tau^- \to  K^-  \pi^- \pi^+ \nu_\tau ~(\mathrm{ex.~}K^0)) & = & (0.330\, \pm\, 0.001\, {~}^{+0.016}_{-0.017})\%~\cite{Lee:2010tc}\\
{\cal{B}}(\tau^- \to  K^-  \pi^-  K^+ \nu_\tau)  & = & (0.155\, \pm\, 0.001\, {~}^{+0.006}_{-0.005})\%~\cite{Lee:2010tc}\\
{\cal{B}}(\tau^- \to  K^-   K^-   K^+ \nu_\tau)  & = & (3.29\, \pm\, 0.17\, {~}^{+0.19}_{-0.20})\times 10^{-5}~\cite{Lee:2010tc}\\
{\cal{B}}(\tau^- \to  \pi^- \pi^0 \eta \nu_\tau)  & = & (1.35\, \pm\, 0.03\, \pm\, 0.07)\times 10^{-3}~\cite{Inami:2008ar}\\
{\cal{B}}(\tau^- \to  K^- \eta \nu_\tau)  & = & (1.58\, \pm\, 0.05\, \pm\, 0.09)\times 10^{-4}~\cite{Inami:2008ar}\\
{\cal{B}}(\tau^- \to  K^- \pi^0 \eta \nu_\tau)  & = & (0.46\, \pm\, 0.11\, \pm\, 0.04)\times 10^{-4}~\cite{Inami:2008ar}\\
{\cal{B}}(\tau^- \to  \bar{K}^0 \pi^- \eta \nu_\tau)  & = & (0.88\, \pm\, 0.14\, \pm\, 0.04)\times 10^{-4}~\cite{Inami:2008ar}
\end{eqnarray*}
\end{itemize}
where the uncertainties are statistical and systematic, respectively.

We add to the list of base nodes one additional measurement of 
${\cal{B}}( K^- \phi \nu_\tau (\phi \to KK))$ = 
${\cal{B}}( K^- K^+ K^- \nu_\tau) \times ({\cal{B}}(\phi \to K^+K^-) + {\cal{B}}(\phi \to K^0_SK^0_L))$,
which leads us to a global fit to a set of 153 measurements with 37 base nodes.

We try to take into account the correlations between measurements,
and avoid applying the PDG-style scale factors to all our measurements.
However, two of the measurements from the B-Factories have
significant discrepancy with respect to each other.
These are measurements of ${\cal{B}}({\tau^- \to K^-K^+K^-\nu_\tau})$
from BaBar and Belle experiments, which are more than $5 \sigma$ apart.
We scale the errors from these measurements by a scale factor
of 5.44 obtained from results of single-quantity ``HFAG Average'',
using the same prescription as the PDG collaboration.

As far as possible, we try to include the measurements quoted in the
original publications. For example, ALEPH presents results with
the correlation matrix between measurements of hadronic modes in Ref.~\cite{Schael:2005am}.
Their paper also quotes derived measurements for the pionic modes,
after subtracting the kaonic contributions as measured by other experiments.
PDG interpretes these published correlations between hadronic modes
as correlations between the pionic modes, and uses the measurements
of pionic modes. Our reasoning is that the B-Factories can measure
the kaonic modes with better precision, thanks to the excellent
particle-identification system in our respective experiments.
Thus the estimates for the kaonic contribution subtracted from
the hadronic modes should be revisited based on data from B-Factories.

We interpret the ALEPH data as measurements for the hadronic modes
and treat their measured correlation matrix as between the following decay modes :
$\tau^-\to e^-\bar{\nu}_e\nu_\tau$,
$\tau^-\to\mu^-\bar{\nu}_\mu\nu_\tau$,
$\tau^-\to h^-\nu_\tau$,
$\tau^-\to h^- \pi^0 \nu_\tau$,
$\tau^-\to h^- 2\pi^0 \nu_\tau$,
$\tau^-\to h^- 3\pi^0 \nu_\tau$,
$\tau^-\to h^- 4\pi^0 \nu_\tau$,
$\tau^-\to 3h^-\nu_\tau$,
$\tau^-\to 3h^- \pi^0 \nu_\tau$,
$\tau^-\to 3h^- 2\pi^0 \nu_\tau$,
$\tau^-\to 3h^- 3\pi^0 \nu_\tau$,
$\tau^-\to 5h^-\nu_\tau$ and
$\tau^-\to 5h^- \pi^0 \nu_\tau$,
where $h^-$ = $\pi^-$ or $^-K$,
as in the original publication.
Since the ALEPH measurements of these branching fractions 
have been constrained to add up to unity,
we exclude the weakest measurement of
$\tau^-\to 3h^- 3\pi^0 \nu_\tau$ in our global fit,
as in the PDG global fit.

If the unitarity constraint is dropped from the global fit to 124
measurements, sum of the 31 base modes fall short from unity by
1.0 $\sigma$ (0.9 $\sigma$) in the scenario when ALEPH
correlation matrix is modified (un-modified). From the global
fit to 153 measurements including those from B-Factories, 
sum of the 37 base modes fall short from unity by 1.6 $\sigma$
(1.9 $\sigma$) when ALEPH correlation matrix is modified
(un-modified). 

A summary of quality of these global fits are presented in Table~\ref{tab:TauGlobalFit_summary}
for  the constrained and unconstrained cases
with data from non-B-Factories and including those from B-Factories.
The results of the global fit to 131 and 153 measurements 
are presented in Tables~\ref{tab:TauGlobalFit_unconstrained} and~\ref{tab:TauGlobalFit}
for the constrained and unconstrained cases, respectively.

\begin{table}[!hbtp]
\begin{center}

\caption{Summary of global fits for branching fractions from unconstrained and constrained fits to data from non-B-Factories and including those from B-Factories.}
\label{tab:TauGlobalFit_summary}
\end{center}
\end{table}

\begin{table}[!hbtp]
\begin{center}
%
\caption{Results for branching fractions (in \%) from unconstrained fit to data from non-B-Factories and including those from B-Factories.}
\label{tab:TauGlobalFit_unconstrained}
\end{center}
\end{table}

\begin{table}[!hbtp]
\begin{center}
%
\caption{Results for branching fractions (in \%) from unitarity constrained fit to data from non-B-Factories and including those from B-Factories.}
\label{tab:TauGlobalFit}
\end{center}
\end{table}

In the following, we present results of branching fractions
of modes separated according to one or three or five charged tracks
(``prongs'') in the final state, or decays containing $K^0$, $\eta$, $K^{*0}$:

\begin{itemize}
\item 1-prong decays with 0 or 1 $\piz$:

The measurements and average values of 
${\cal{B}}({\tau^-\to\mu^-\bar{\nu}_\mu\nu_\tau})$,
${\cal{B}}({\tau^-\to\pi^-\nu_\tau})$,
${\cal{B}}({\tau^-\to K^- \nu_\tau})$,
${\cal{B}}({\tau^-\to h^-\nu_\tau})$,
where $h^-$ = $\pi^-$ or $K^-$, 
are presented in Figure~\ref{fig:TauTo1Prong0Piz}, 
and those of
${\cal{B}}({\tau^-\to\pi^-\pi^0\nu_\tau})$,
${\cal{B}}({\tau^-\to K^- \pi^0\nu_\tau})$ and
${\cal{B}}({\tau^-\to h^- \pi^0\nu_\tau})$ 
decays are presented in Figure~\ref{fig:TauTo1Prong1Piz}.

\begin{figure}[!hbtp]
\begin{center}
\includegraphics[height=.42\textheight,width=.49\textwidth]{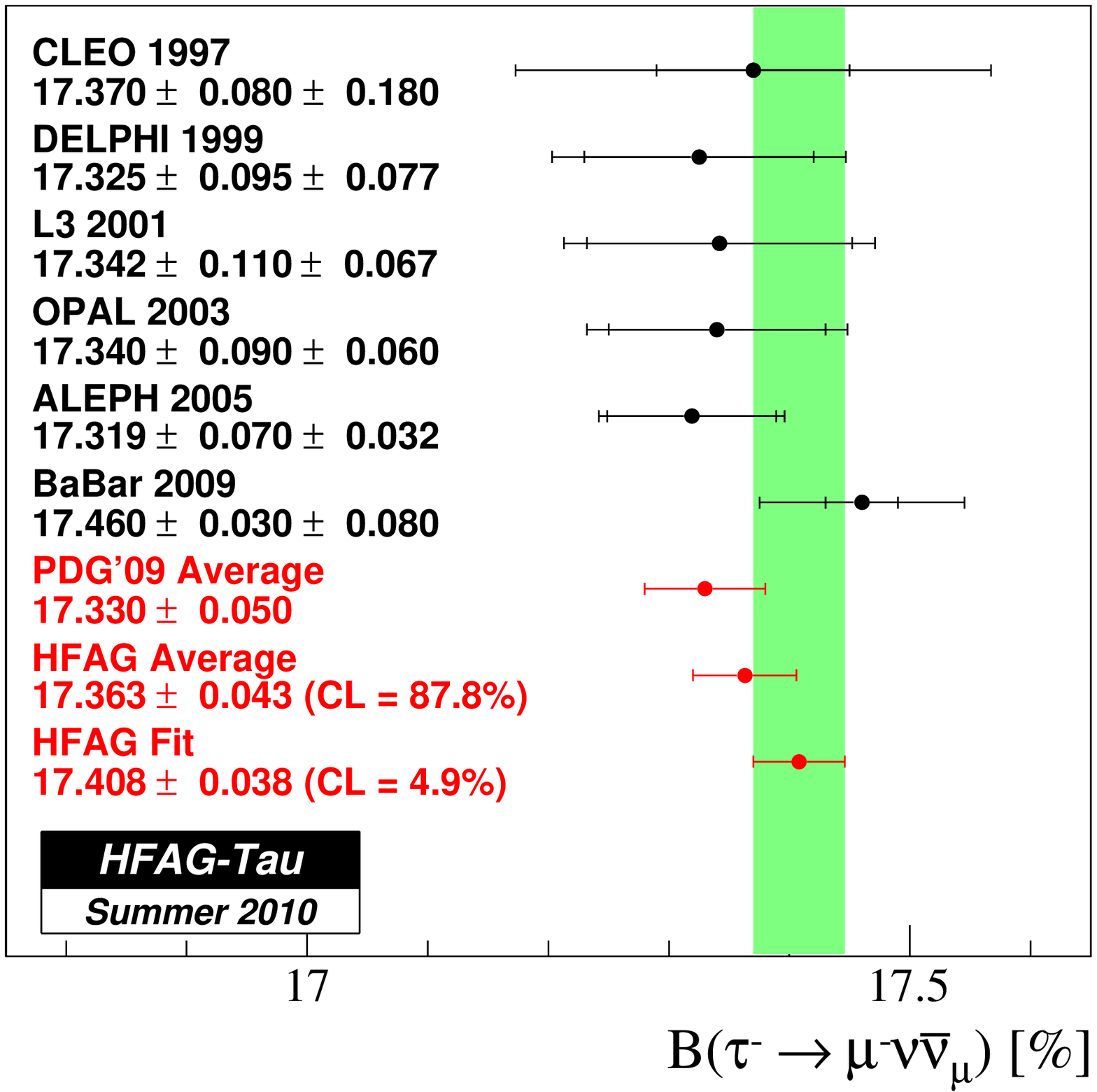}
\includegraphics[height=.42\textheight,width=.49\textwidth]{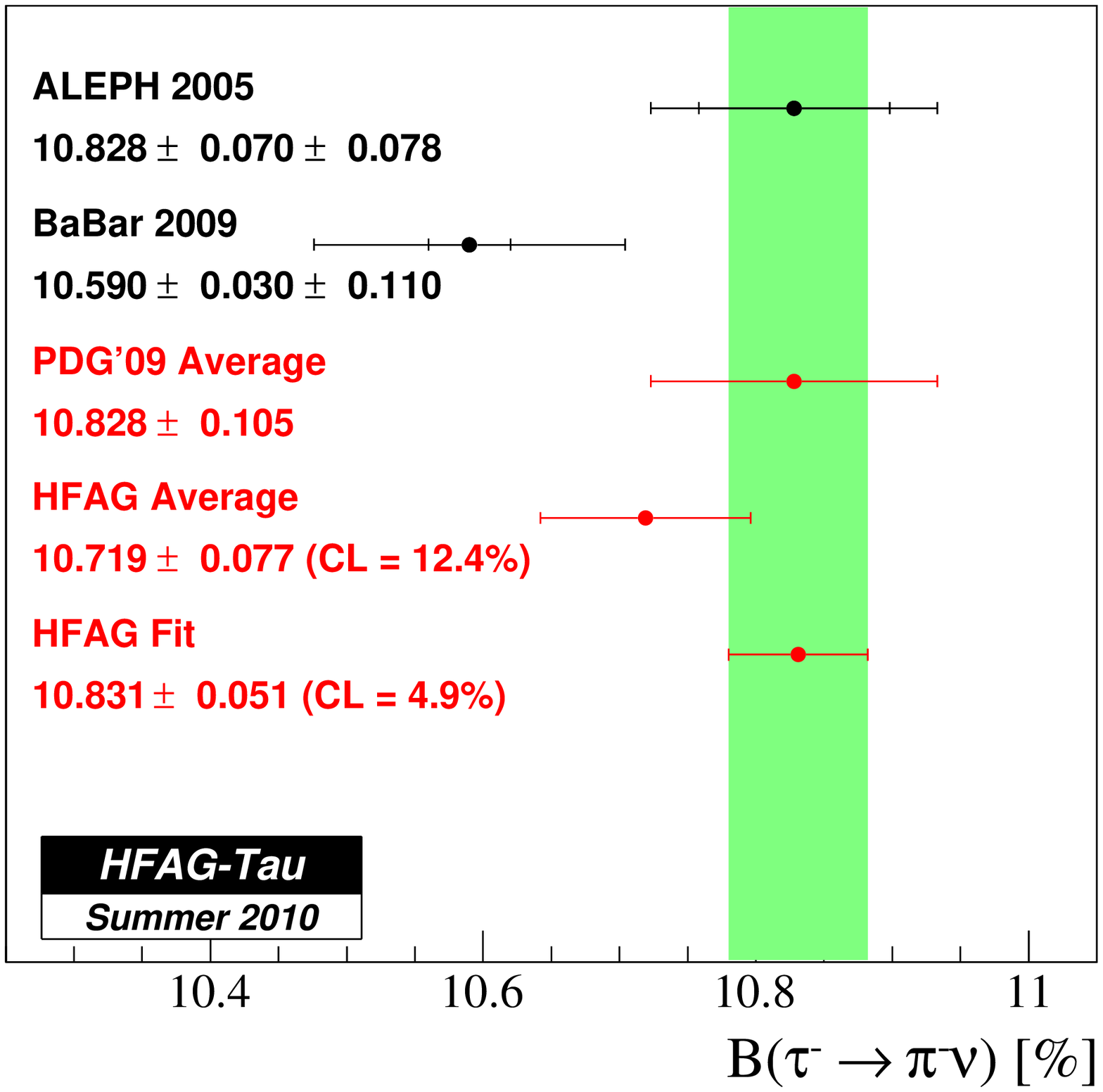}
\includegraphics[height=.42\textheight,width=.49\textwidth]{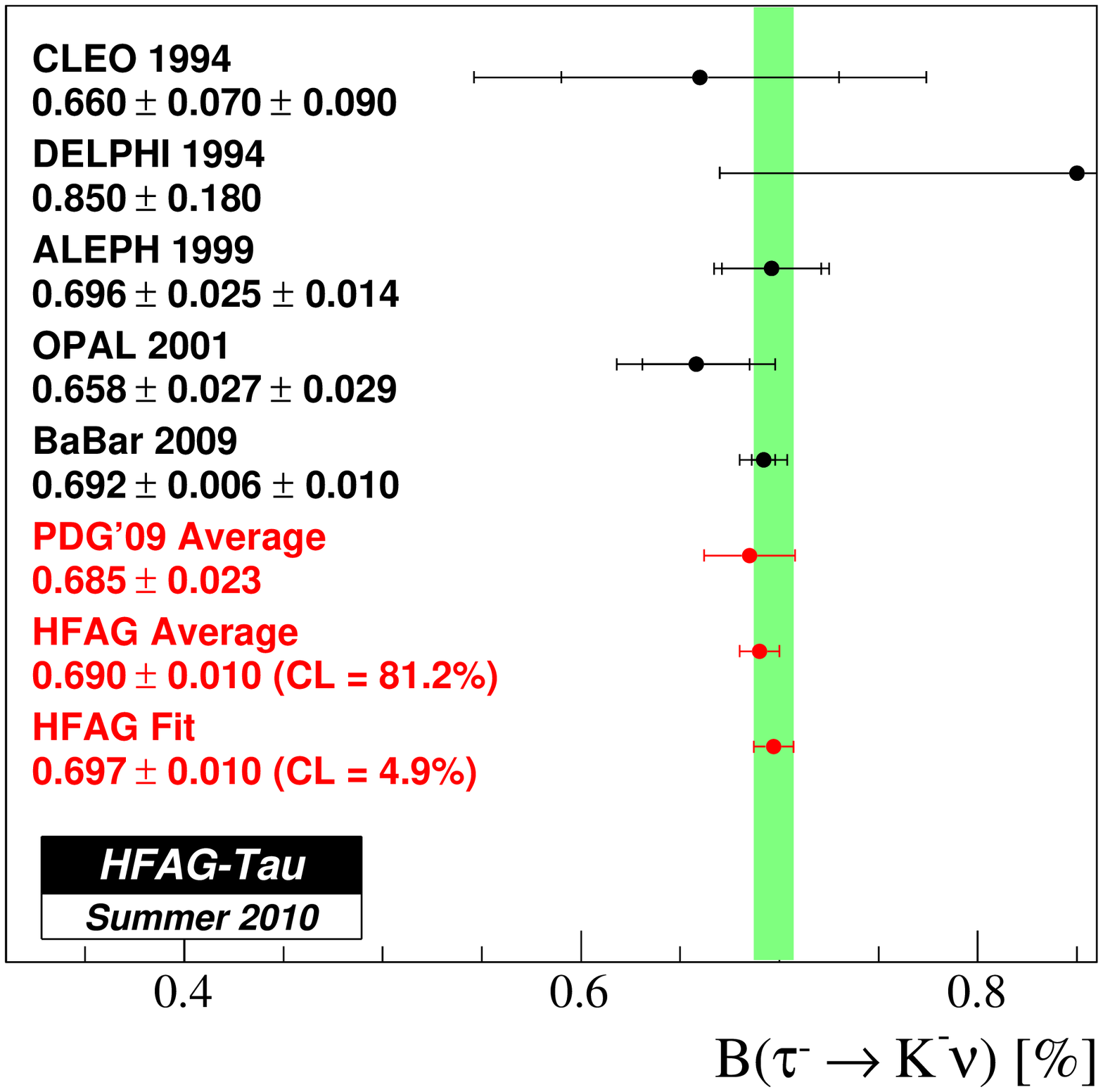}
\includegraphics[height=.42\textheight,width=.49\textwidth]{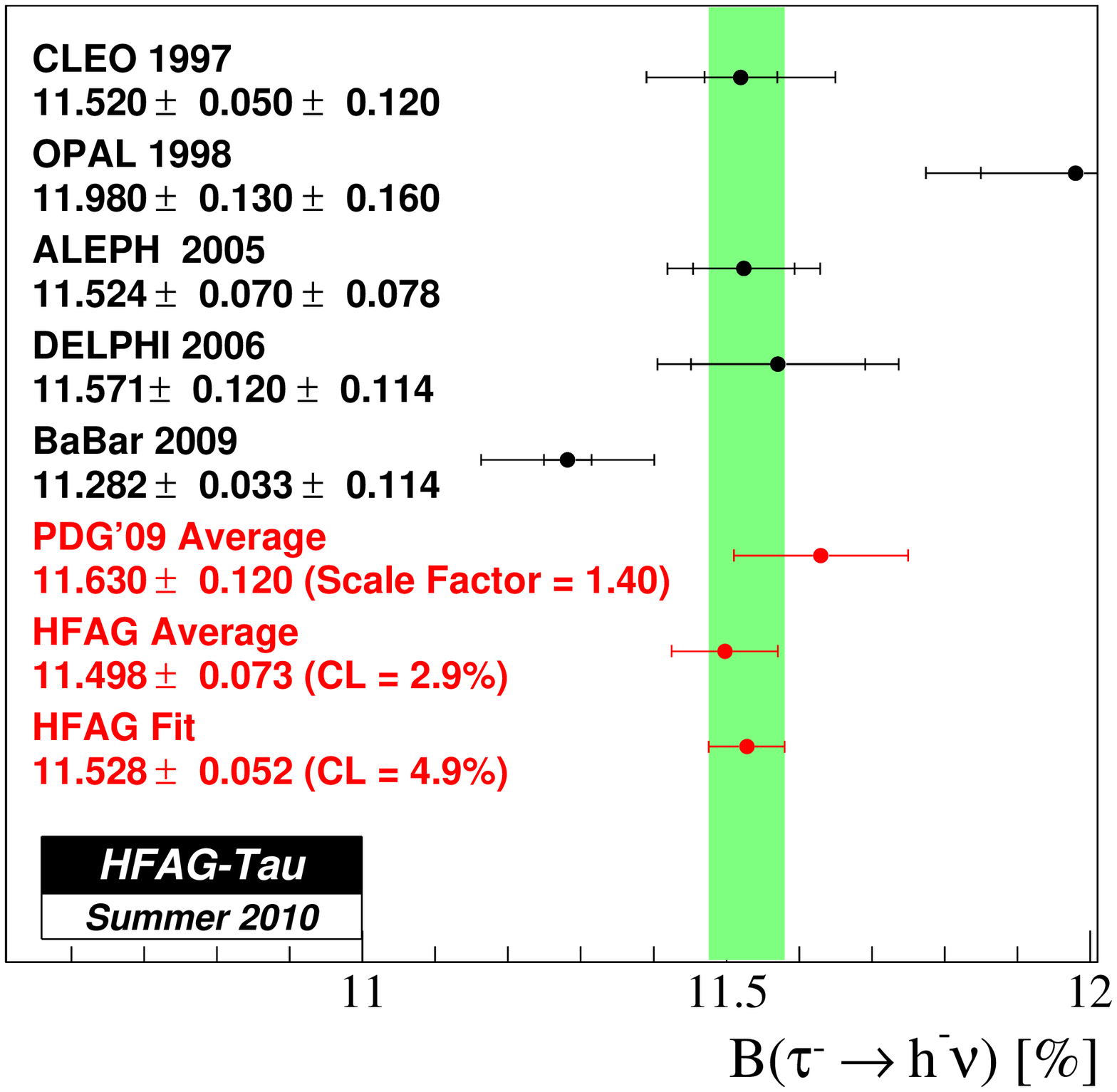}
\end{center}
\caption{Measurements and average values of 1-prong decays with 0 $\piz$.}
\label{fig:TauTo1Prong0Piz}
\end{figure}

\begin{figure}[!hbtp]
\begin{center}
\includegraphics[height=.42\textheight,width=.49\textwidth]{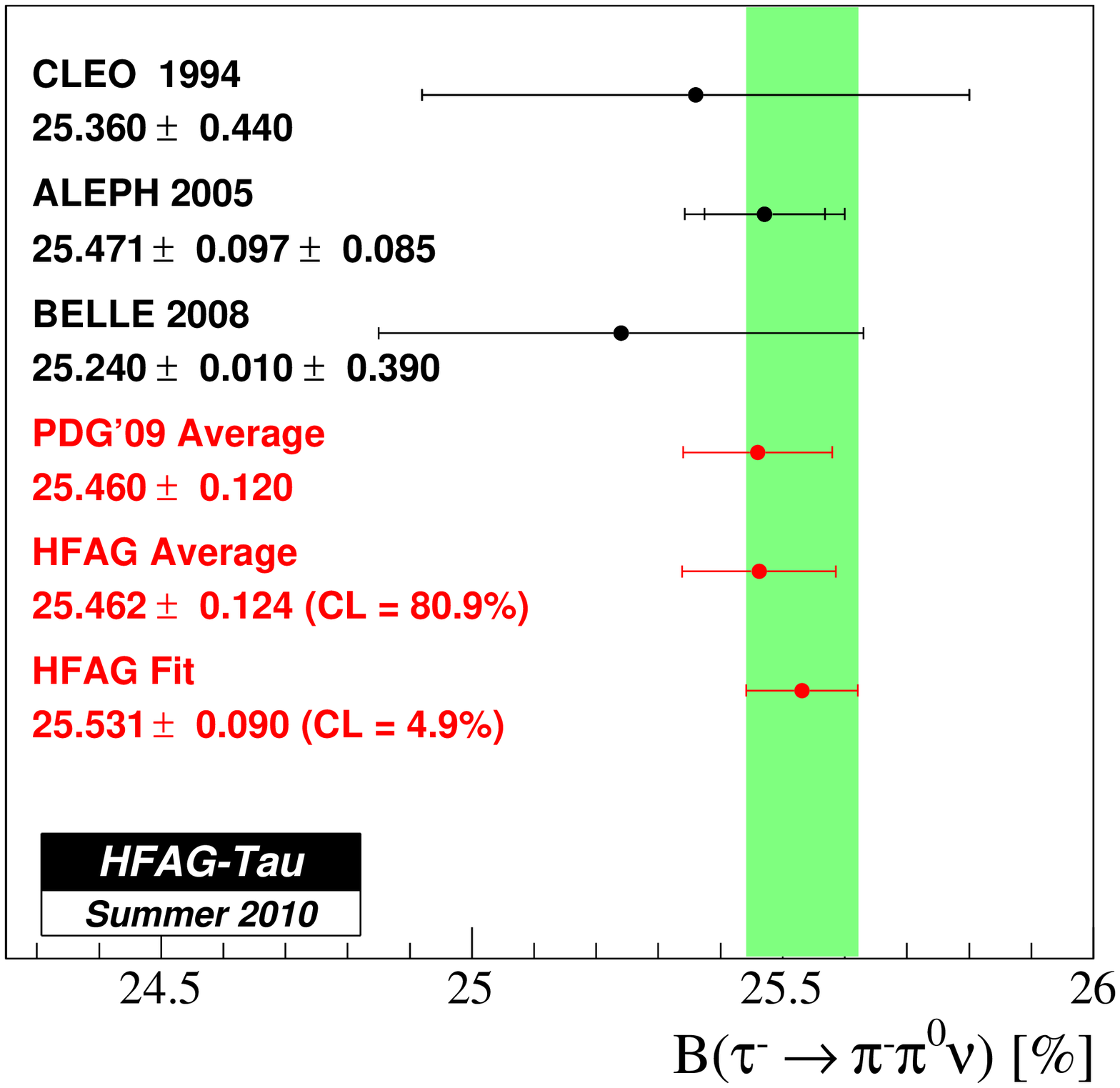}
\includegraphics[height=.42\textheight,width=.49\textwidth]{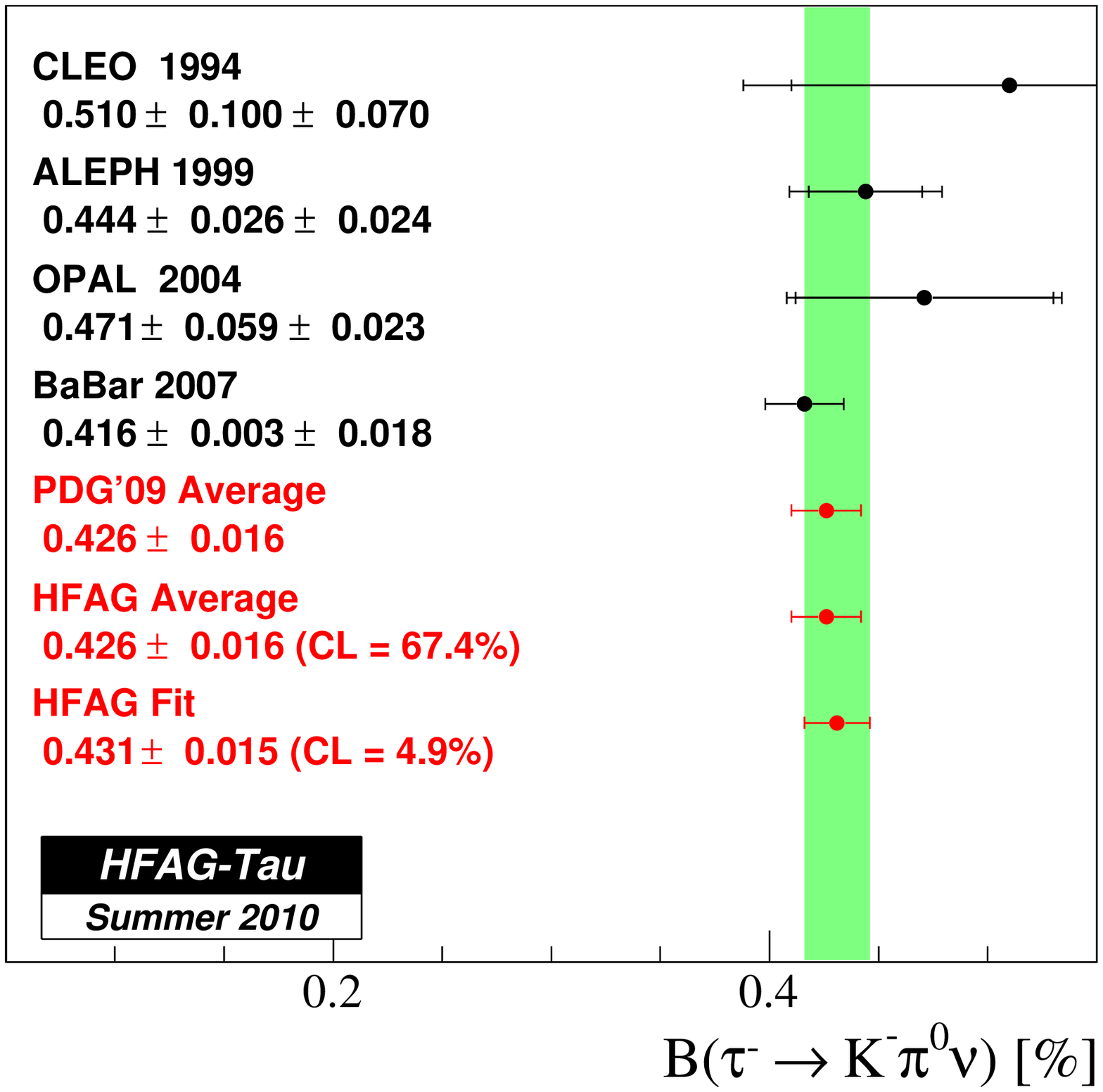}
\includegraphics[height=.42\textheight,width=.49\textwidth]{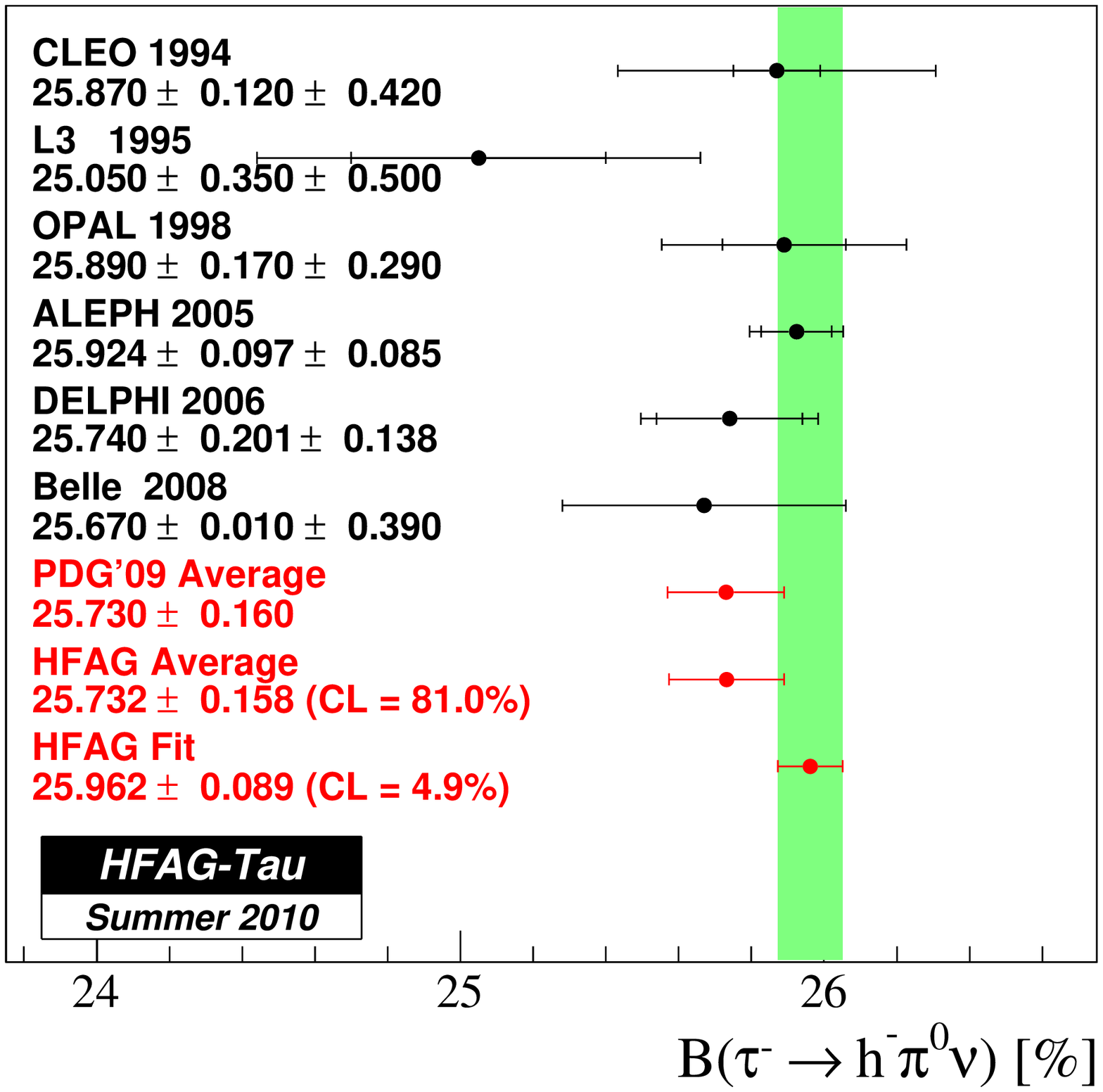}
\end{center}
\caption{Measurements and average values of 1-prong decays with 1 $\piz$.}
\label{fig:TauTo1Prong1Piz}
\end{figure}

\item 3-prong decays with 0 $\piz$, 0 $K^0$:

The measurements and average values of 
${\cal{B}}({\tau^-\to h^-h^-h^+\nu_\tau}$ ${(\mathrm{ex.~} K^0)})$,
${\cal{B}}({\tau^-\to \pi^-\pi^-\pi^+\nu_\tau}$ ${(\mathrm{ex.~} K^0)})$,
${\cal{B}}({\tau^-\to \pi^-K^-\pi^+\nu_\tau}$ ${(\mathrm{ex.~} K^0)})$,
${\cal{B}}({\tau^-\to \pi^-K^-K^+\nu_\tau})$ and
${\cal{B}}({\tau^-\to K^-K^-K^+\nu_\tau})$,
where $h^-$ = $\pi^-$ or $K^-$,
are presented in Figures~\ref{fig:TauToHmHmHpNu_1} and \ref{fig:TauToHmHmHpNu_2}.

The measurements of ${\cal{B}}({\tau^-\to K^-\phi\nu_\tau})$ are also presented in Figure~\ref{fig:TauToHmHmHpNu_2},
along with results from the single-quantity averaging procedure.
While the BaBar measurement uses the same data set as in 
the measurement of ${\cal{B}}({\tau^-\to K^-K^-K^+\nu_\tau})$,
Belle measurement uses a different data set for 
${\cal{B}}({\tau^-\to K^-\phi\nu_\tau})$ measurement
than used for their ${\cal{B}}({\tau^-\to K^-K^-K^+\nu_\tau})$ measurement.
To avoid redundancy, we do not use measurements of ${\cal{B}}({\tau^-\to K^-\phi\nu_\tau})$ in the global fit,
and no results for this mode from ``HFAG Fit'' are quoted in Figure~\ref{fig:TauToHmHmHpNu_2}.

The BaBar experiments also reports ${\cal{B}}({\tau^-\to \pi^-\phi\nu_\tau}) = (3.42\, \pm\, 0.55\, \pm\, 0.25) \times 10^{-5}$~\cite{Aubert:2007mh}.
Since it is the only measurement for this channel, no averaging has been performed.

\begin{figure}[!hbtp]
\begin{center}
\includegraphics[height=.42\textheight,width=.49\textwidth]{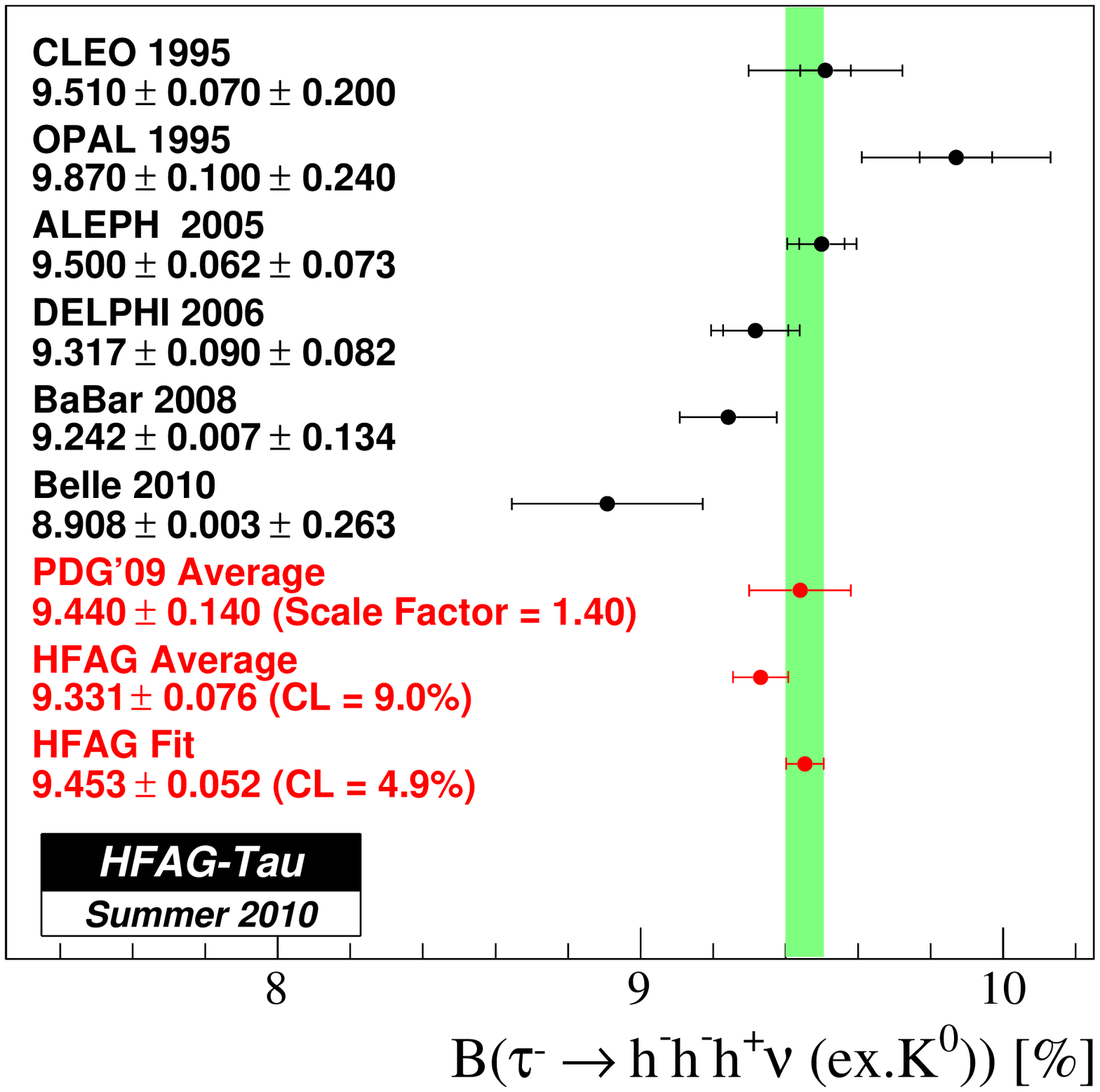}
\includegraphics[height=.42\textheight,width=.49\textwidth]{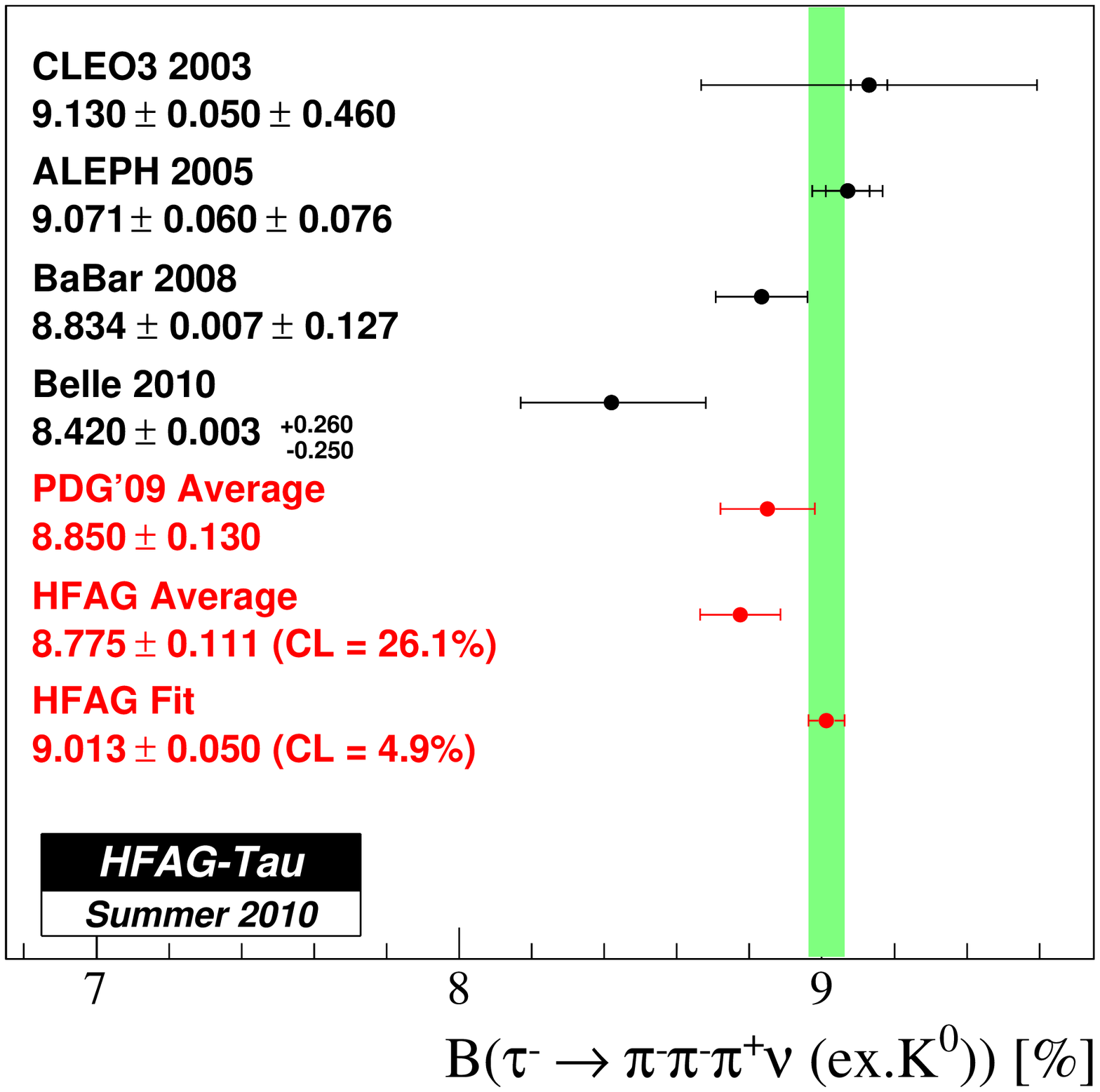}
\includegraphics[height=.42\textheight,width=.49\textwidth]{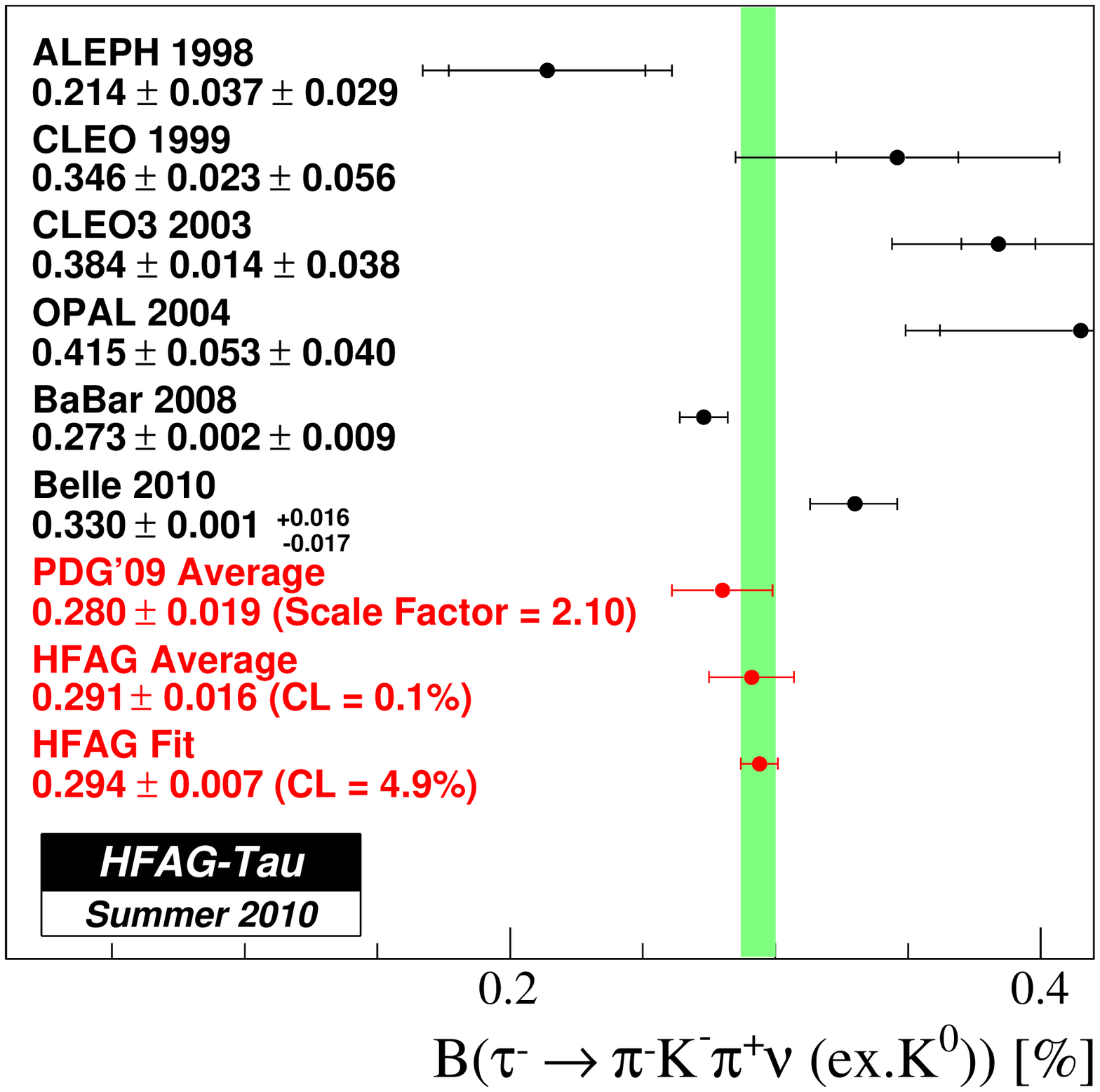}
\includegraphics[height=.42\textheight,width=.49\textwidth]{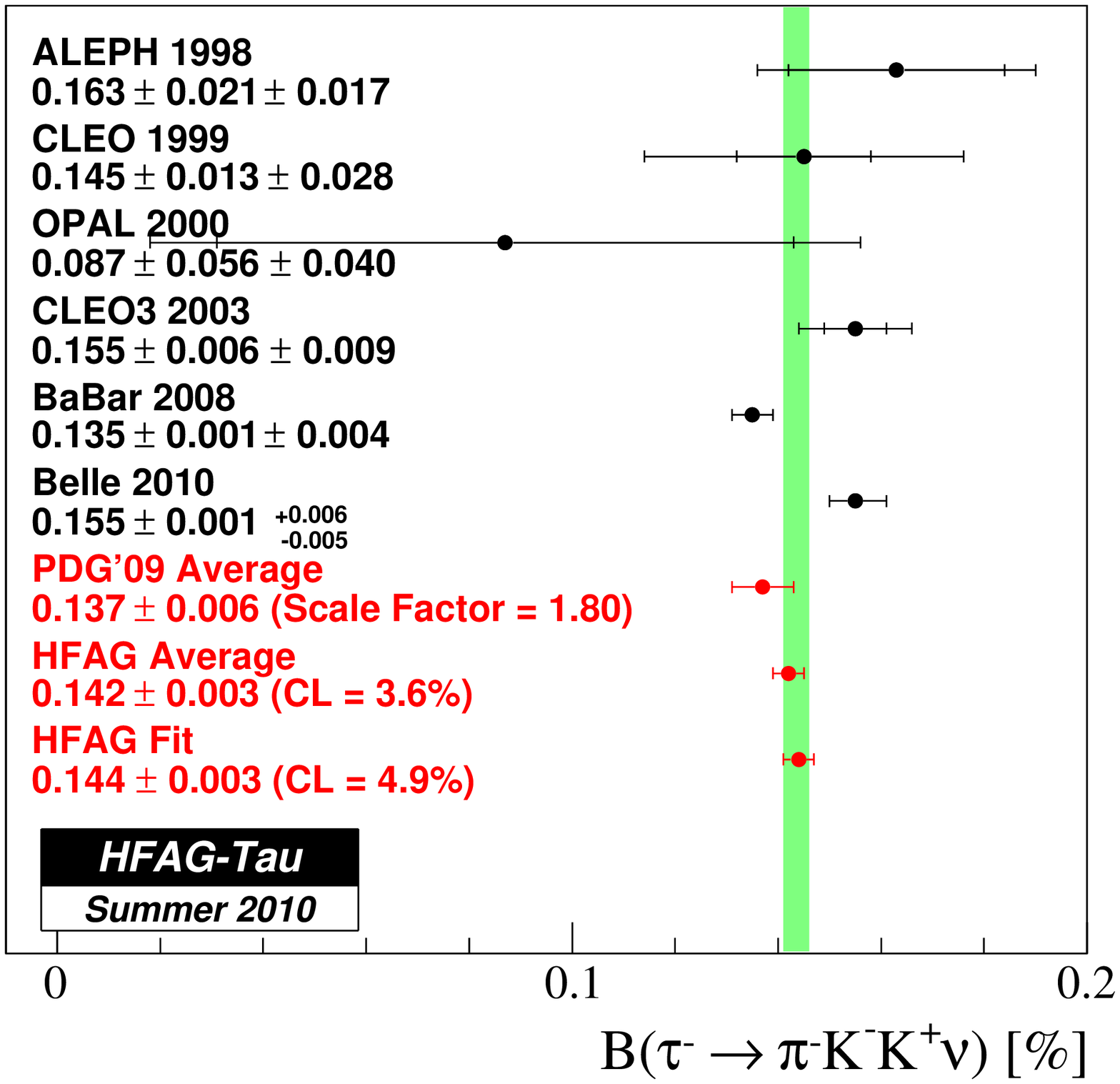}
\end{center}
\caption{Measurements and average values of 3-prong decays with 0 $\piz$, 0 $\bar{K}^0$.
 ALEPH quotes
 ${\cal{B}}({\tau^-\to h^-h^-h^+\nu_\tau}$ ${(\mathrm{ex.~} K^0,\omega)})$
= $(9.469 \pm\, 0.062 \pm\, 0.073)\%$ and
 ${\cal{B}}({\tau^-\to \pi^-\pi^-\pi^+\nu_\tau}$ ${(\mathrm{ex.~} K^0,\omega)})$
= $(9.041 \pm\, 0.060 \pm\, 0.076)\%$.
We add 
${\cal{B}}(\tau^- \to  h^- \omega \nu_\tau) \times {\cal{B}}(\omega \to \pi^- \pi^+)$ = $0.031\%$ and
${\cal{B}}(\tau^- \to  \pi^- \omega \nu_\tau) \times {\cal{B}}(\omega \to \pi^- \pi^+)$ = $0.030\%$, respectively,
to get the values quoted here.}
\label{fig:TauToHmHmHpNu_1}
\end{figure}

\begin{figure}[!hbtp]
\begin{center}
\includegraphics[height=.42\textheight,width=.49\textwidth]{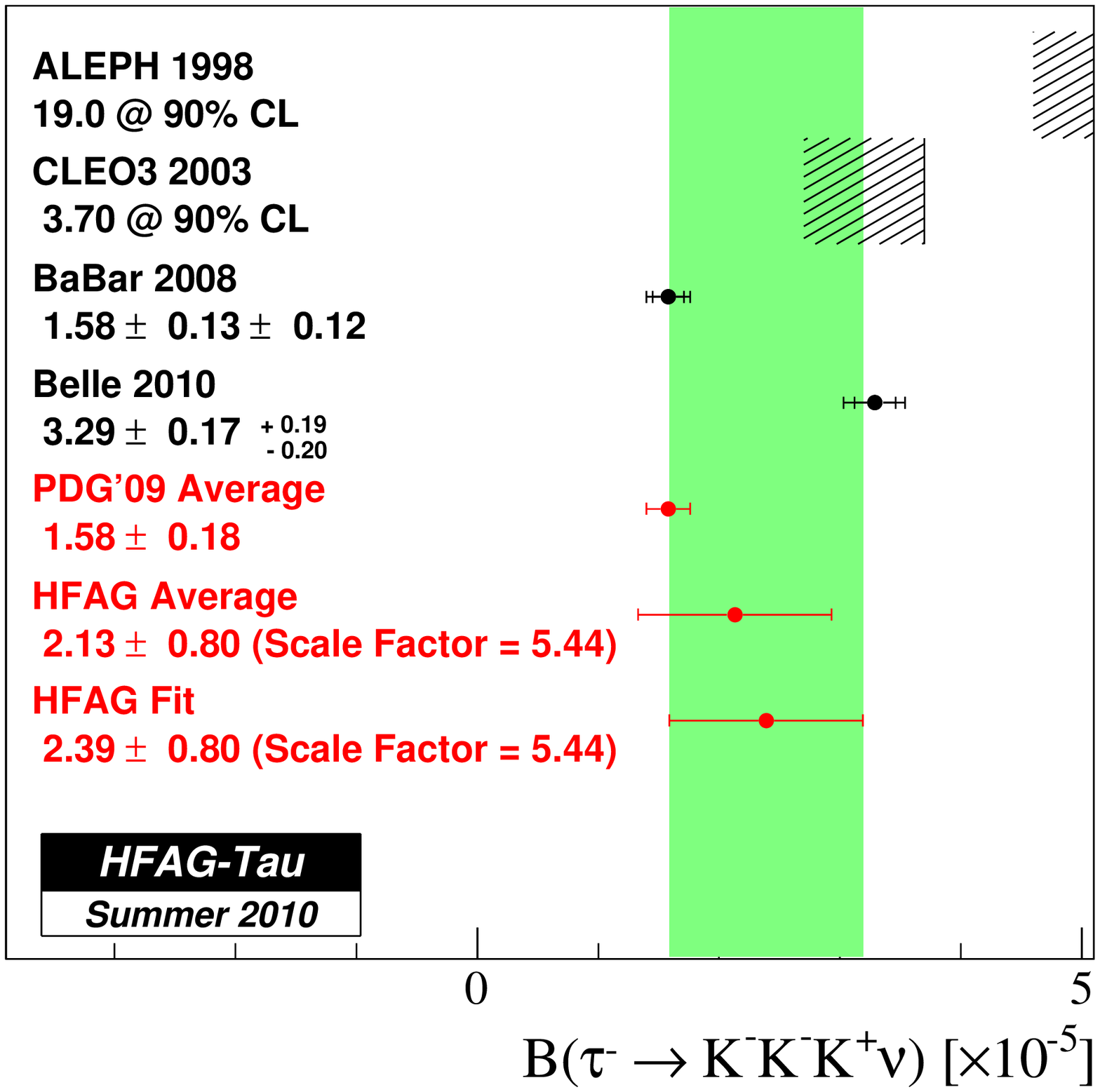}
\includegraphics[height=.42\textheight,width=.49\textwidth]{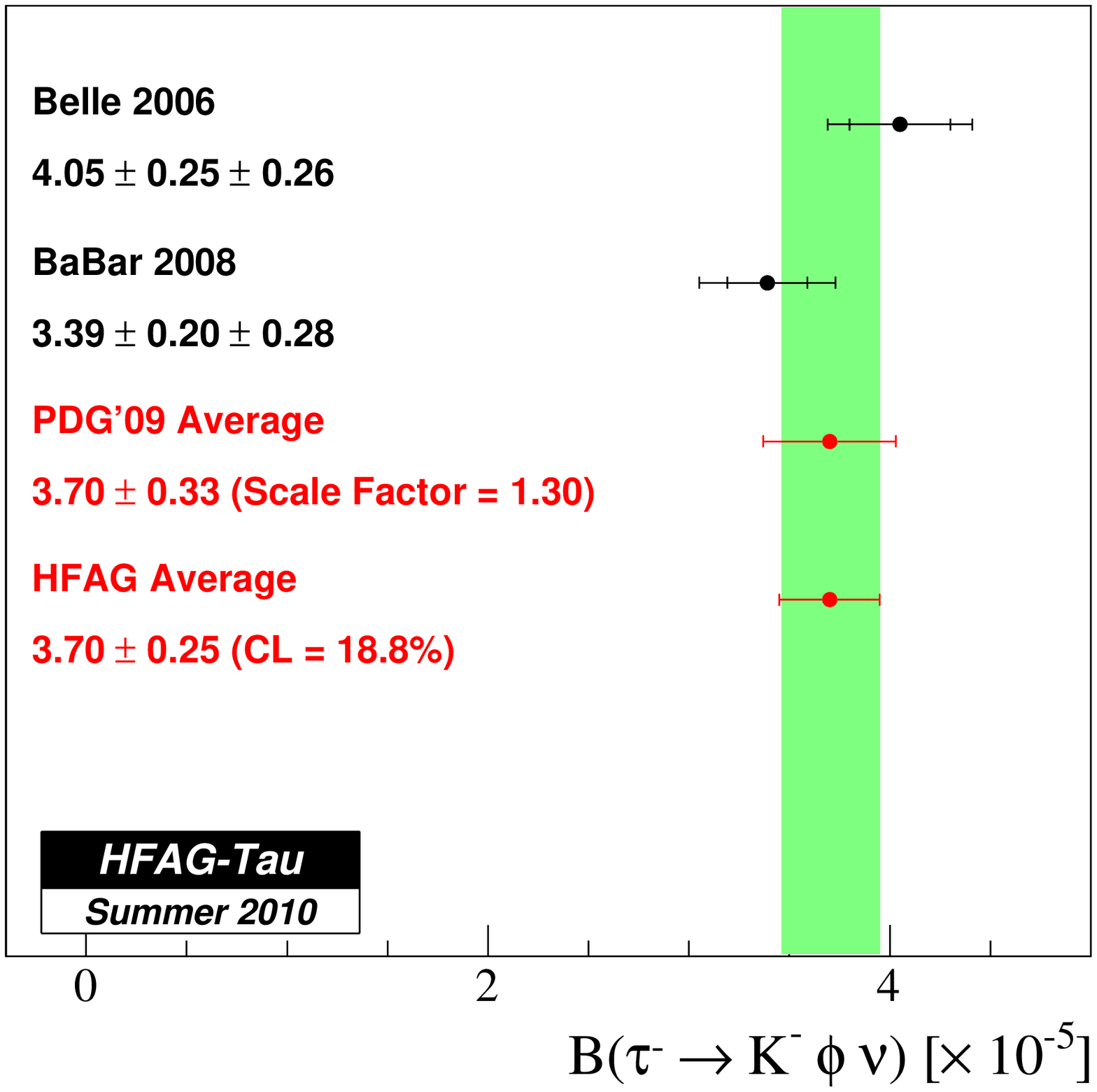}
\end{center}
\caption{Measurements and average values of 
${\cal{B}}({\tau^-\to K^-K^-K^+\nu_\tau})$ 
and 
${\cal{B}}({\tau^-\to K^-\phi\nu_\tau})$.}
\label{fig:TauToHmHmHpNu_2}
\end{figure}

\item 5-prong decays with 0 $\piz$, 0 $K^0$:

The measurements and average values of 
${\cal{B}}({\tau^-\to h^-h^-h^-h^+h^+\nu_\tau}$ ${(\mathrm{ex.~} K^0)})$
and 
${\cal{B}}({\tau^-\to \pi^-}$ ${f_1(1285)\nu_\tau})$
are presented in Figure~\ref{fig:TauTo5Prongs}.
The $f_1(1285)$ content is determined from $2\pi^-2\pi^+$ as well as $\pi^-\pi^+\eta$ final states~\cite{Aubert:2008nj}.
The average value of ${\cal{B}}({\tau^-\to \pi^-}$ ${f_1(1285)\nu_\tau})$ 
from the single-quantity averaging procedure is also quoted
in Figure~\ref{fig:TauTo5Prongs}.

\begin{figure}[!hbtp]
\begin{center}
\includegraphics[height=.42\textheight,width=.49\textwidth]{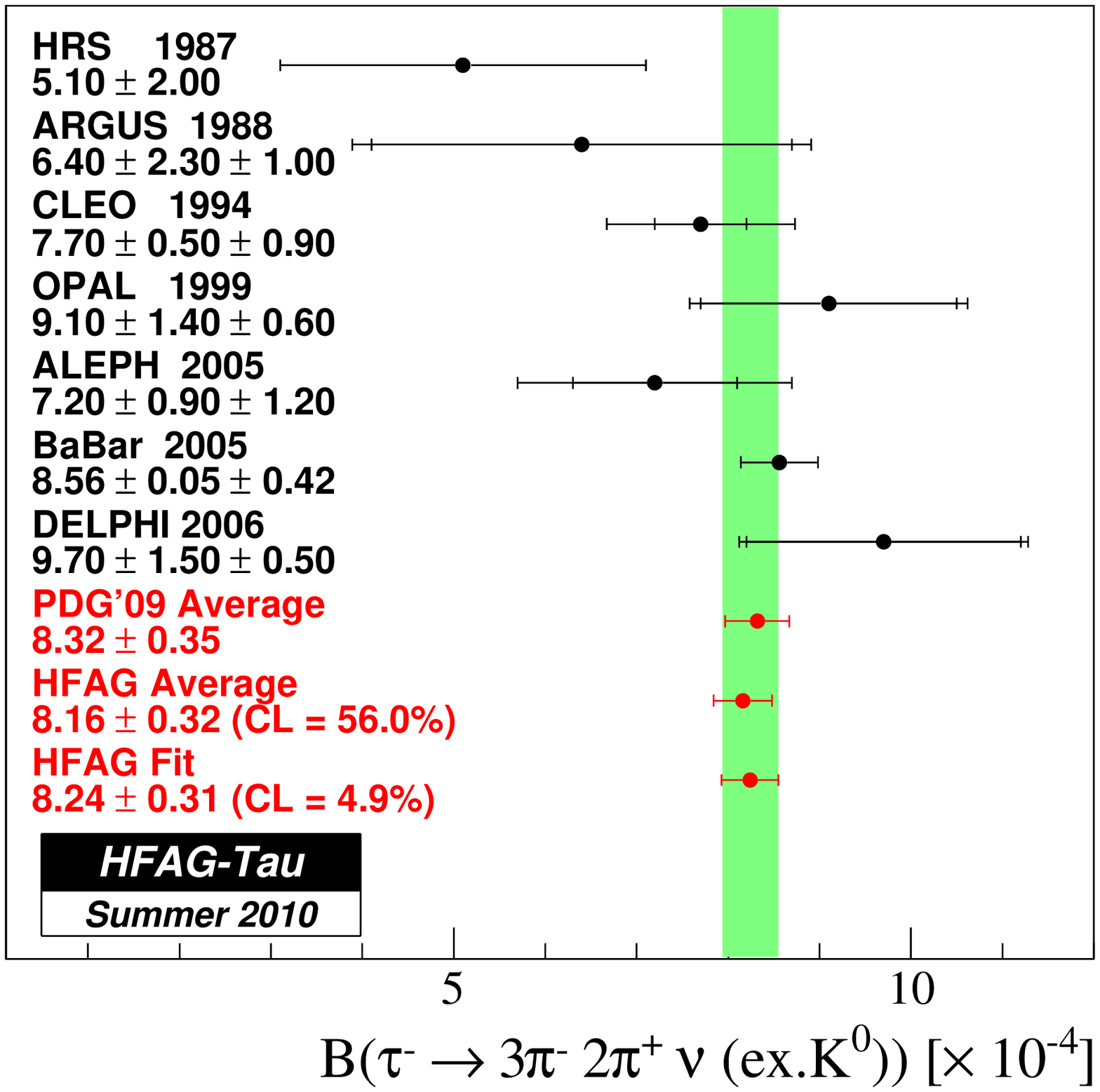}
\includegraphics[height=.42\textheight,width=.49\textwidth]{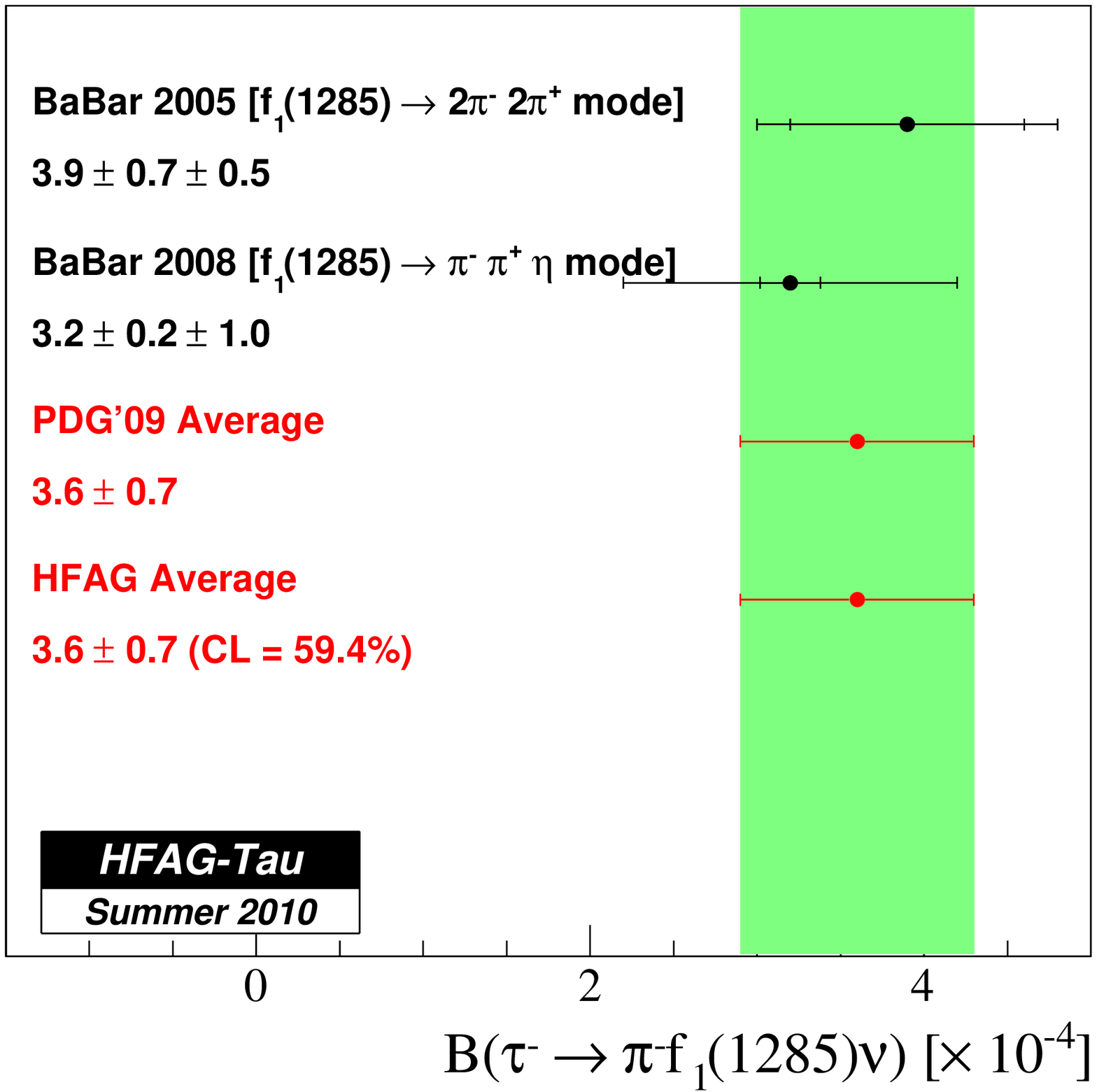}
\end{center}
\caption{Measurements and average values of 
${\cal{B}}({\tau^-\to h^-h^-h^-h^+h^+\nu_\tau ~(\mathrm{ex.~} {K}^0)})$
and
${\cal{B}}({\tau^-\to  \pi^-f_1(1285)\nu_\tau})$.}
\label{fig:TauTo5Prongs}
\end{figure}

\item decays with $K^0$:

The measurements and average values of
${\cal{B}}({\tau^-\to \pi^-\bar{K}^0\nu_\tau})$ and 
${\cal{B}}({\tau^-\to \pi^-\pi^0\bar{K}^0\nu_\tau})$ 
are presented in Figure~\ref{fig:TauToPimKzbX}.

\begin{figure}[!hbtp]
\begin{center}
\includegraphics[height=.42\textheight,width=.49\textwidth]{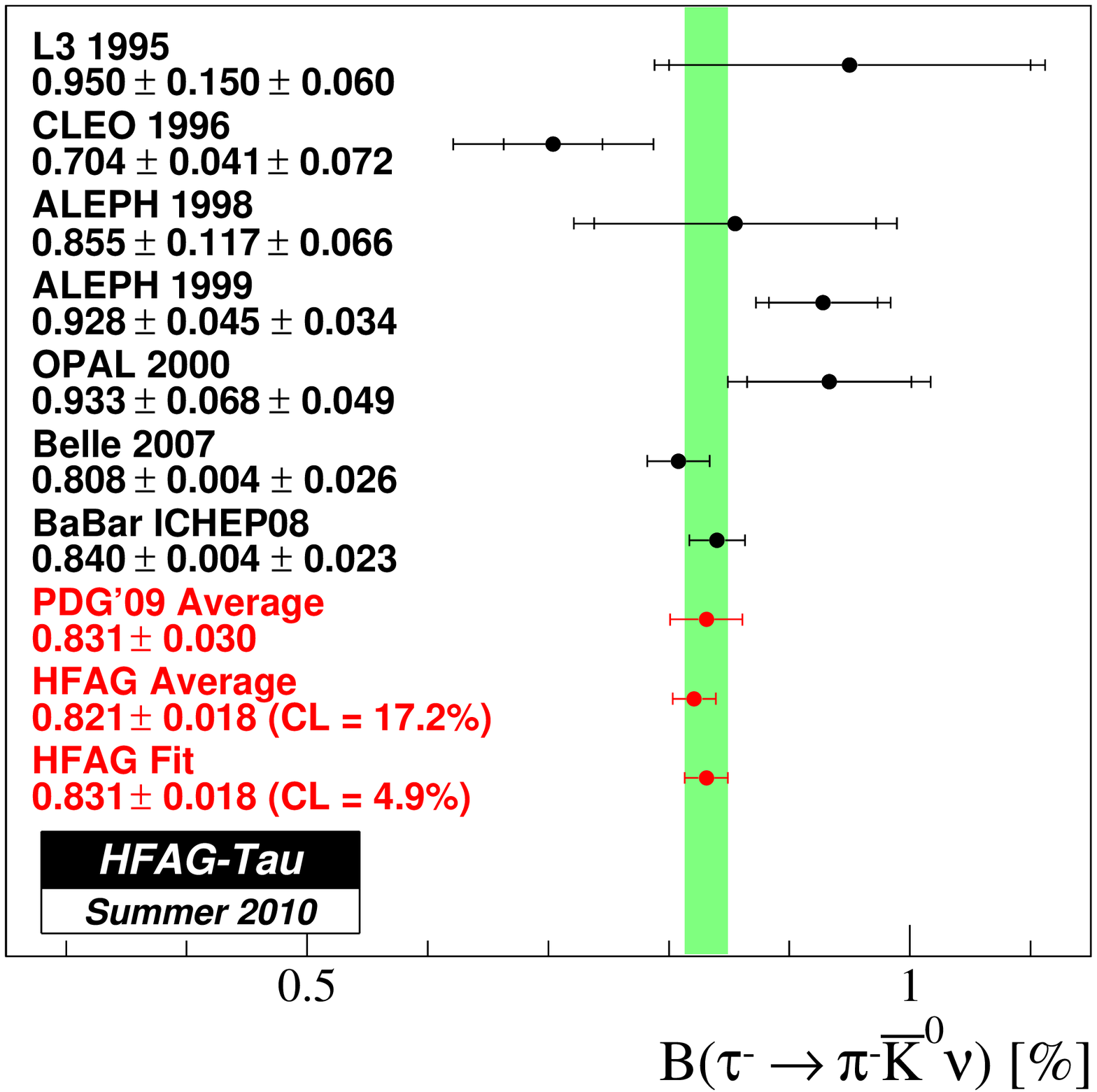}
\includegraphics[height=.42\textheight,width=.49\textwidth]{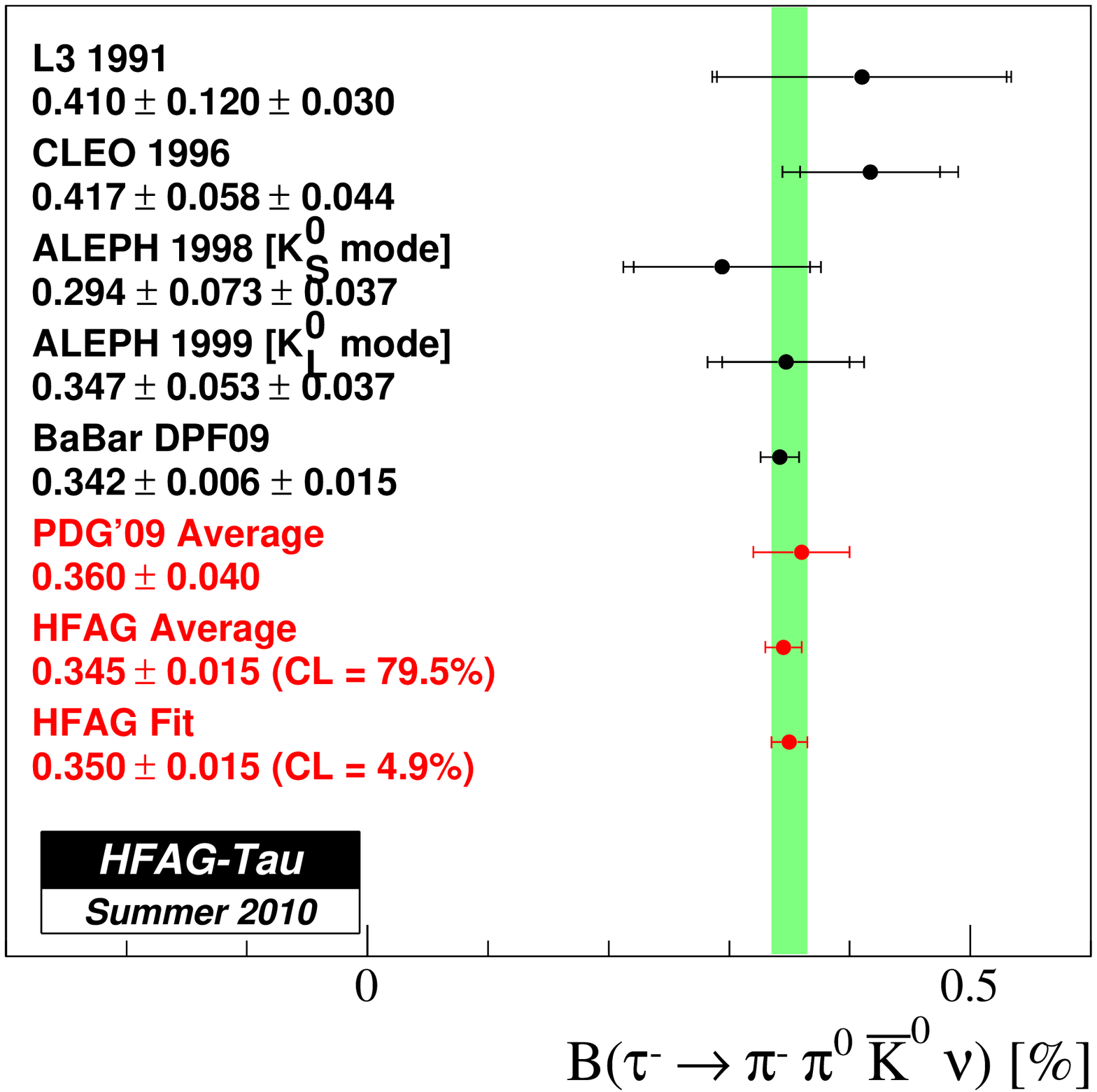}
\end{center}
\caption{Measurements and average values of 
${\cal{B}}({\tau^-\to \pi^-\bar{K}^0\nu_\tau})$ 
and 
${\cal{B}}({\tau^-\to \pi^-\pi^0\bar{K}^0\nu_\tau})$.}
\label{fig:TauToPimKzbX}
\end{figure}

\item decays with $\eta$:

The measurements and average values of 
${\cal{B}}({\tau^-\to K^-\eta\nu_\tau})$, 
${\cal{B}}({\tau^-\to \pi^-\bar{K}^0 \eta\nu_\tau})$ 
and ${\cal{B}}({\tau^-\to}$ ${K^-\pi^0\eta\nu_\tau})$ 
are presented in Figure~\ref{fig:TauToHmEtaX}.
The $K^{*-}$ content is determined from $\tau^-\to \pi^-K_S^0\eta\nu_\tau$ and $\tau^-\to K^-\pi^0\eta\nu_\tau$ decay modes.
The measurements and average values for ${\cal{B}}({\tau^-\to K^{*-}\eta\nu_\tau})$ 
are also presented in Figure~\ref{fig:TauToHmEtaX}.

\begin{figure}[!hbtp]
\begin{center}
\includegraphics[height=.42\textheight,width=.49\textwidth]{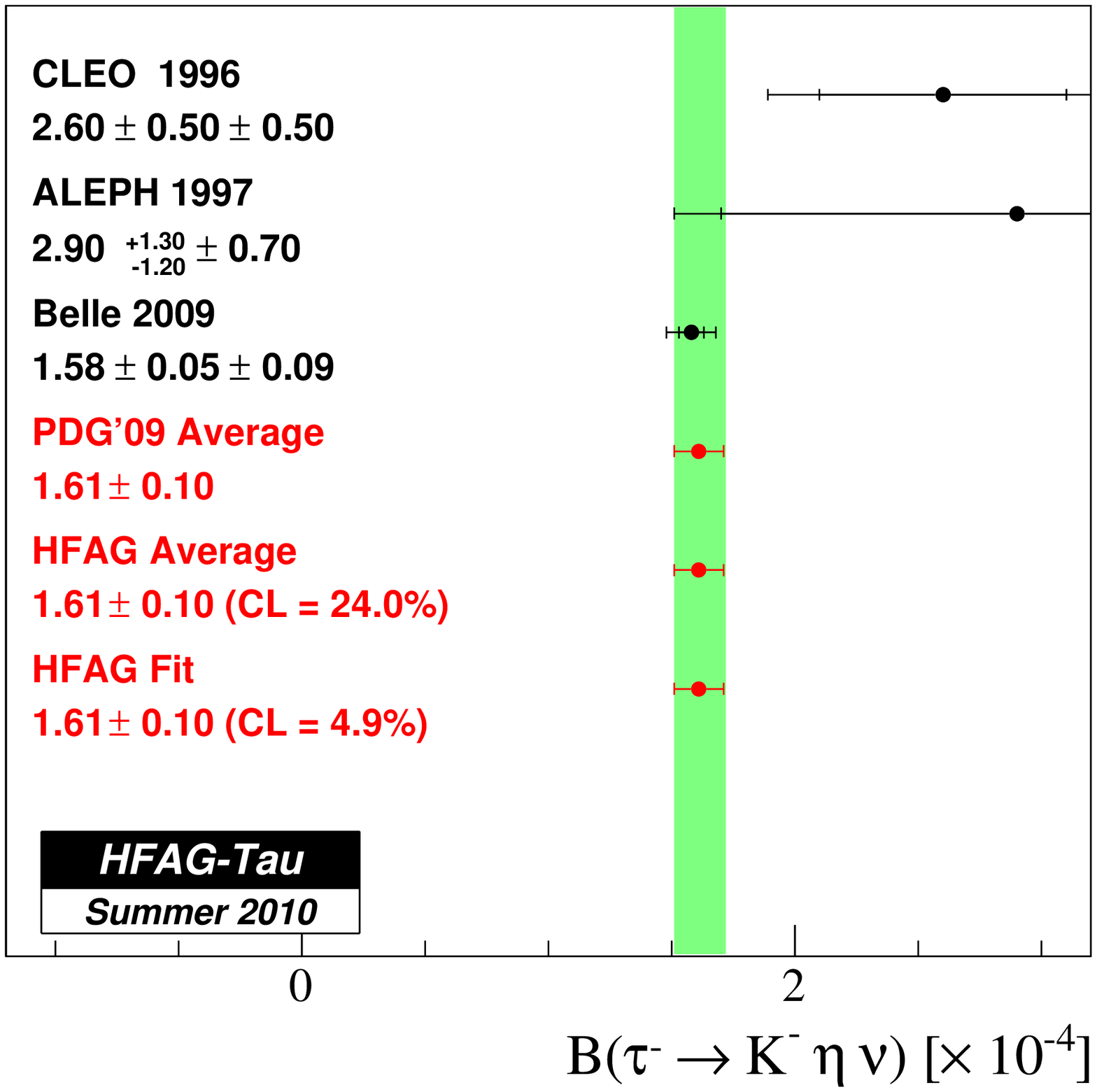}
\includegraphics[height=.42\textheight,width=.49\textwidth]{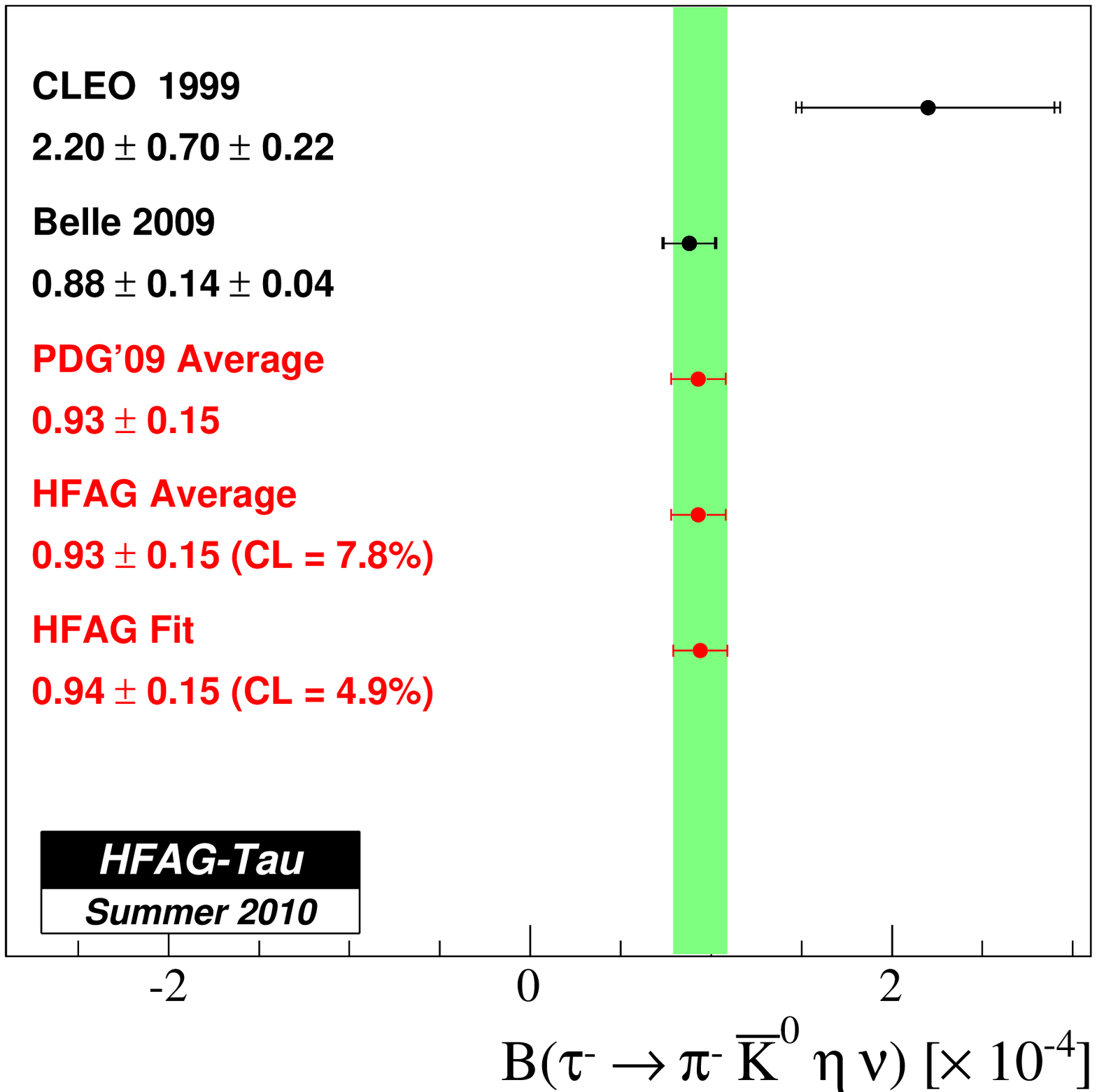}
\includegraphics[height=.42\textheight,width=.49\textwidth]{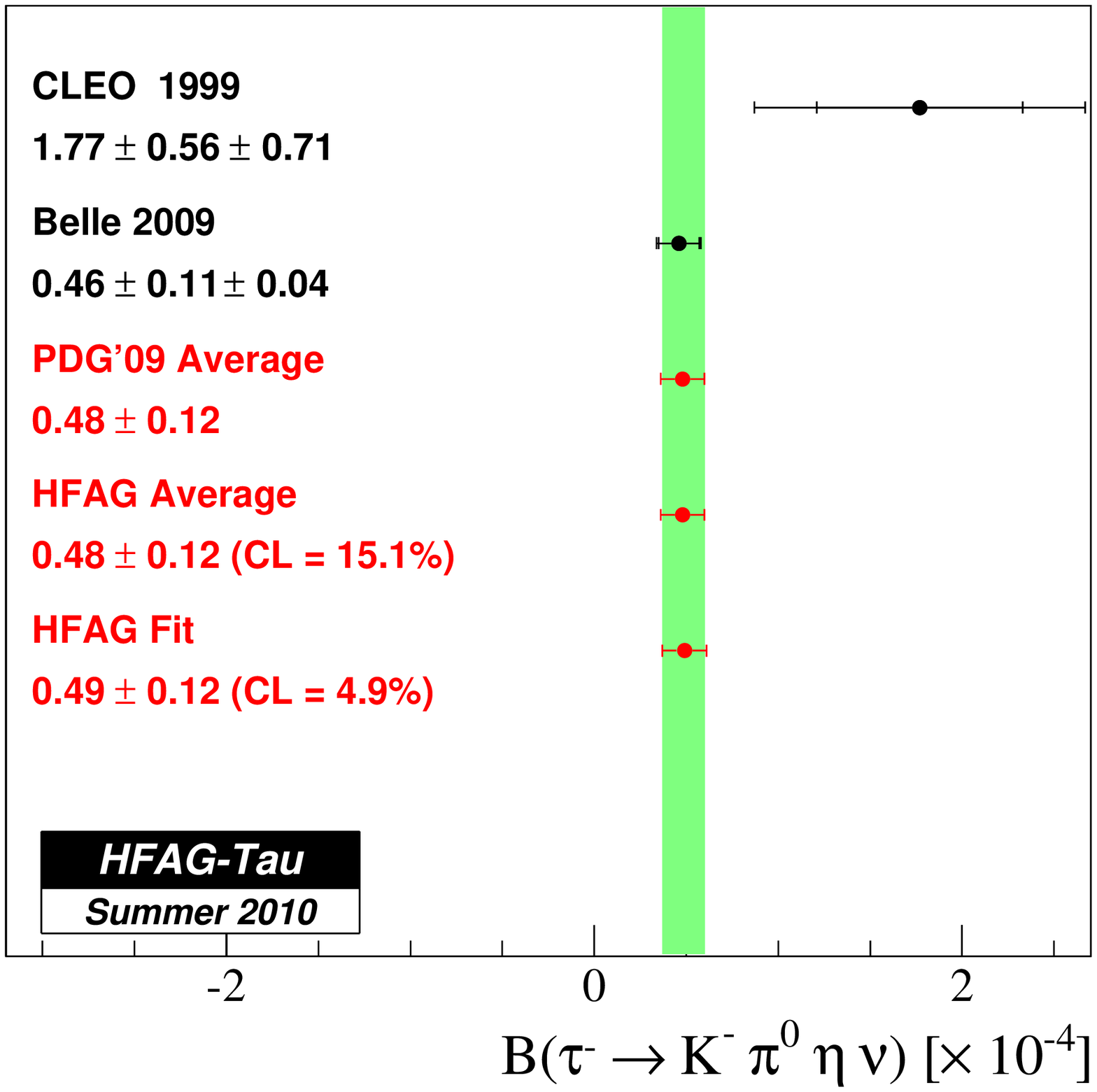}
\includegraphics[height=.42\textheight,width=.49\textwidth]{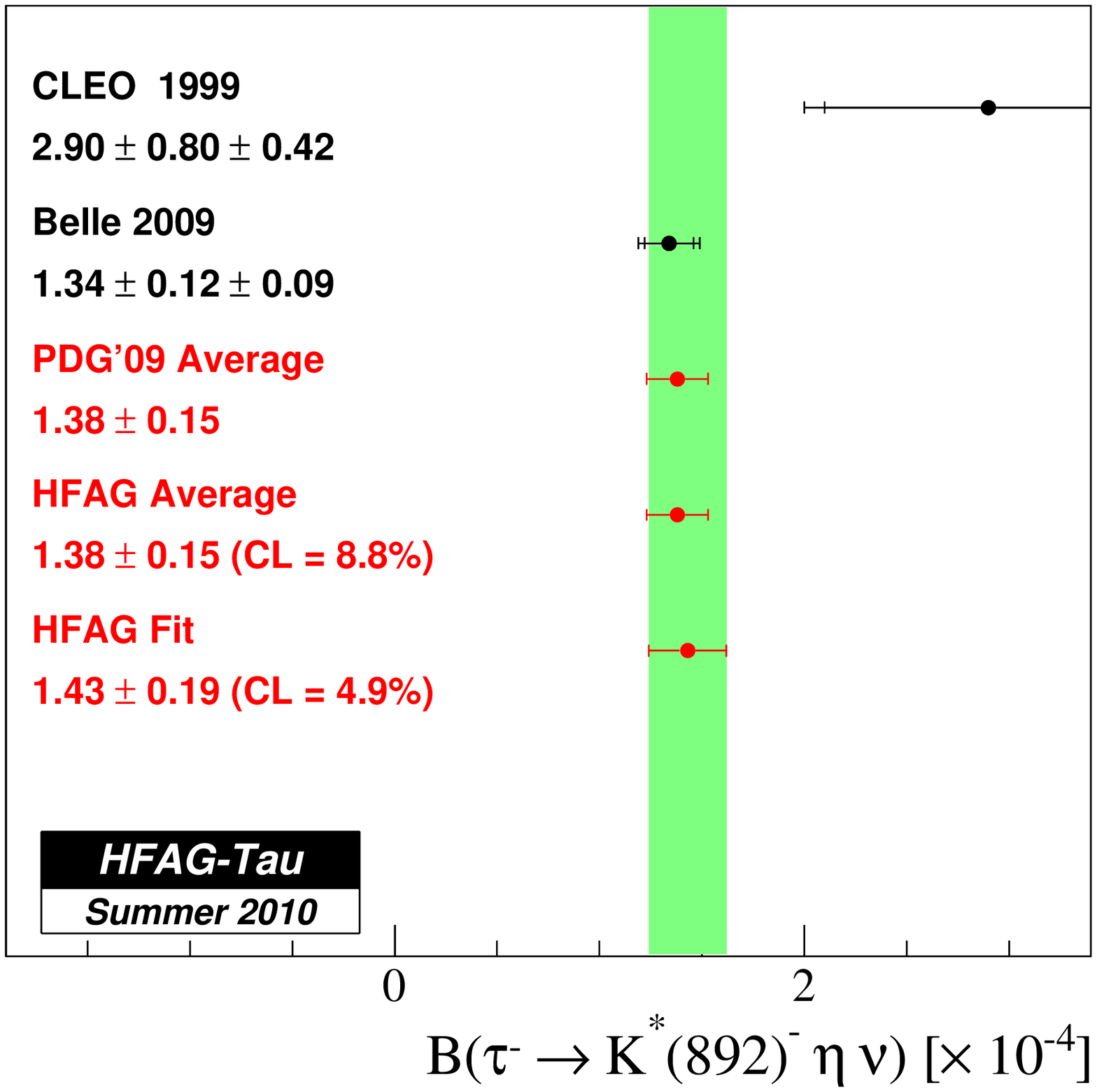}
\end{center}
\caption{Measurements and average values of 
${\cal{B}}({\tau^-\to K^-\eta\nu_\tau})$, 
${\cal{B}}({\tau^-\to \pi^-\bar{K}^0\eta\nu_\tau})$, 
${\cal{B}}({\tau^-\to K^-\pi^0\eta\nu_\tau})$, 
and 
${\cal{B}}({\tau^-\to  K^{*-}\eta\nu_\tau})$.}
\label{fig:TauToHmEtaX}
\end{figure}

The measurements and average values of 
${\cal{B}}({\tau^-\to \pi^-\pi^0\eta\nu_\tau})$ and 
${\cal{B}}({\tau^-\to \pi^-\pi^-\pi^+\eta\nu_\tau}$ ${(\mathrm{ex.~}K^0)})$ 
are presented in Figure~\ref{fig:TauToPimEtaX}.

\begin{figure}[!hbtp]
\begin{center}
\includegraphics[height=.42\textheight,width=.49\textwidth]{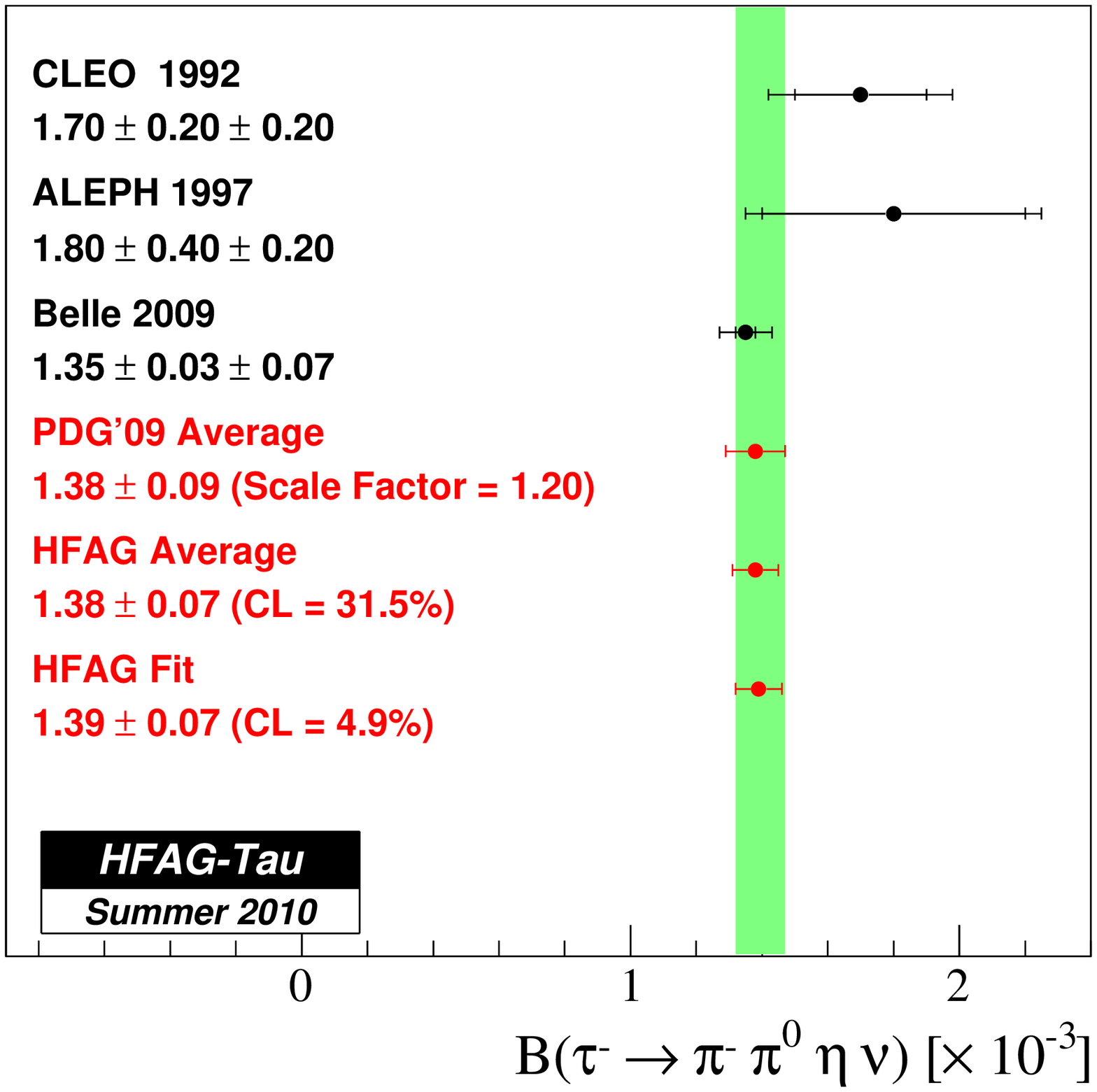}
\includegraphics[height=.42\textheight,width=.49\textwidth]{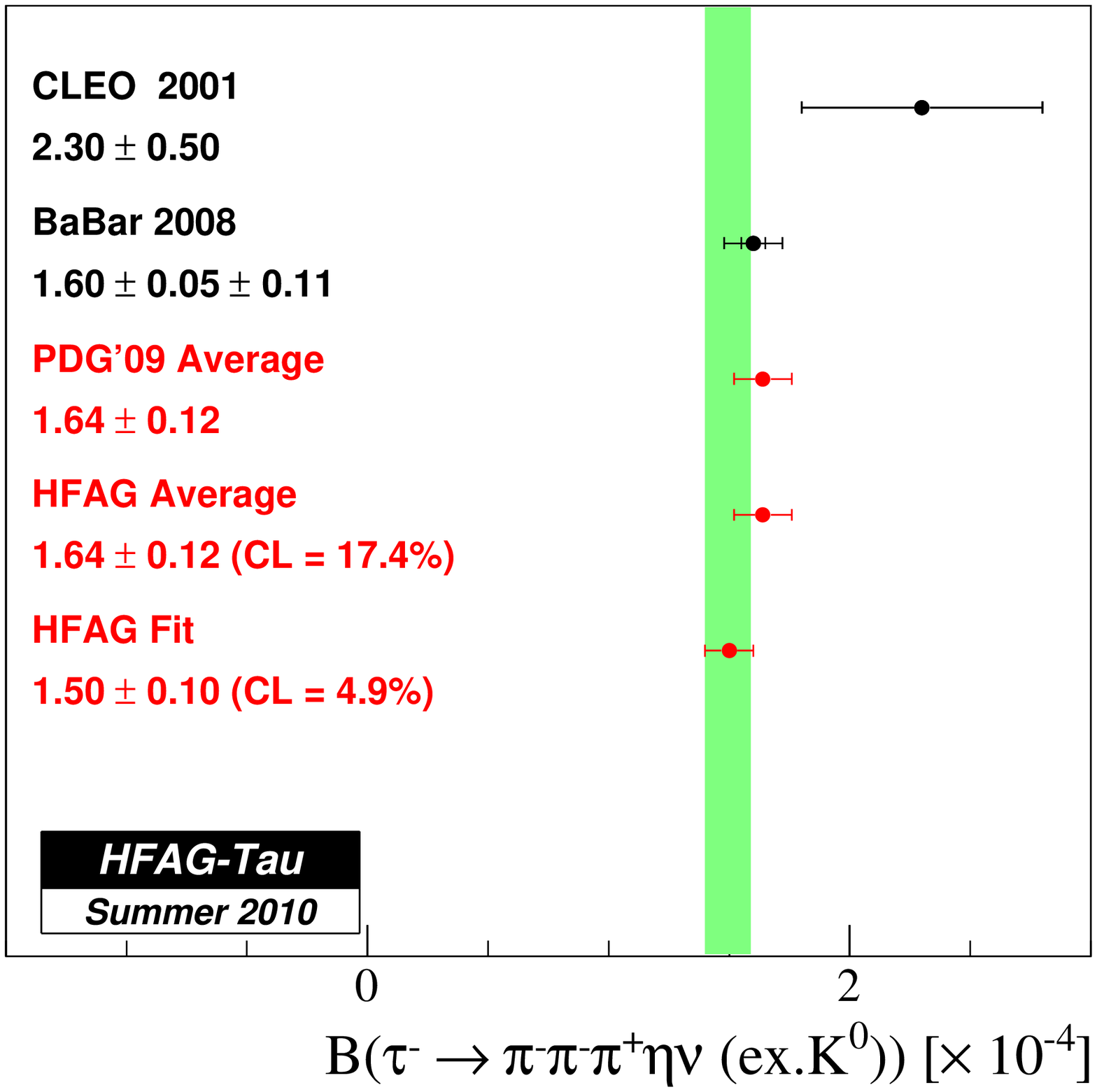}
\end{center}
\caption{Measurements and average values of 
${\cal{B}}({\tau^-\to \pi^-\pi^0\eta\nu_\tau})$ 
and
${\cal{B}}({\tau^-\to \pi^-\pi^-\pi^+\eta\nu_\tau~(\mathrm{ex.~}K^0)})$.}
\label{fig:TauToPimEtaX}
\end{figure}

\item decays with $K^{*0}$:

The measurements and average values for ${\cal{B}}({\tau^-\to K^-K^{*0}\nu_\tau})$ are presented in Figure~\ref{fig:TauToKstarEtaX}.
The Belle experiments also reports ${\cal{B}}({\tau^-\to K^-K^{*0}\piz\nu_\tau}) = (2.39\, \pm\, 0.46\, \pm\, 0.26) \times 10^{-5}$~\cite{Adachi:2008uy},
which is the first measurement for this mode.

\begin{figure}[!hbtp]
\begin{center}
\includegraphics[height=.42\textheight,width=.49\textwidth]{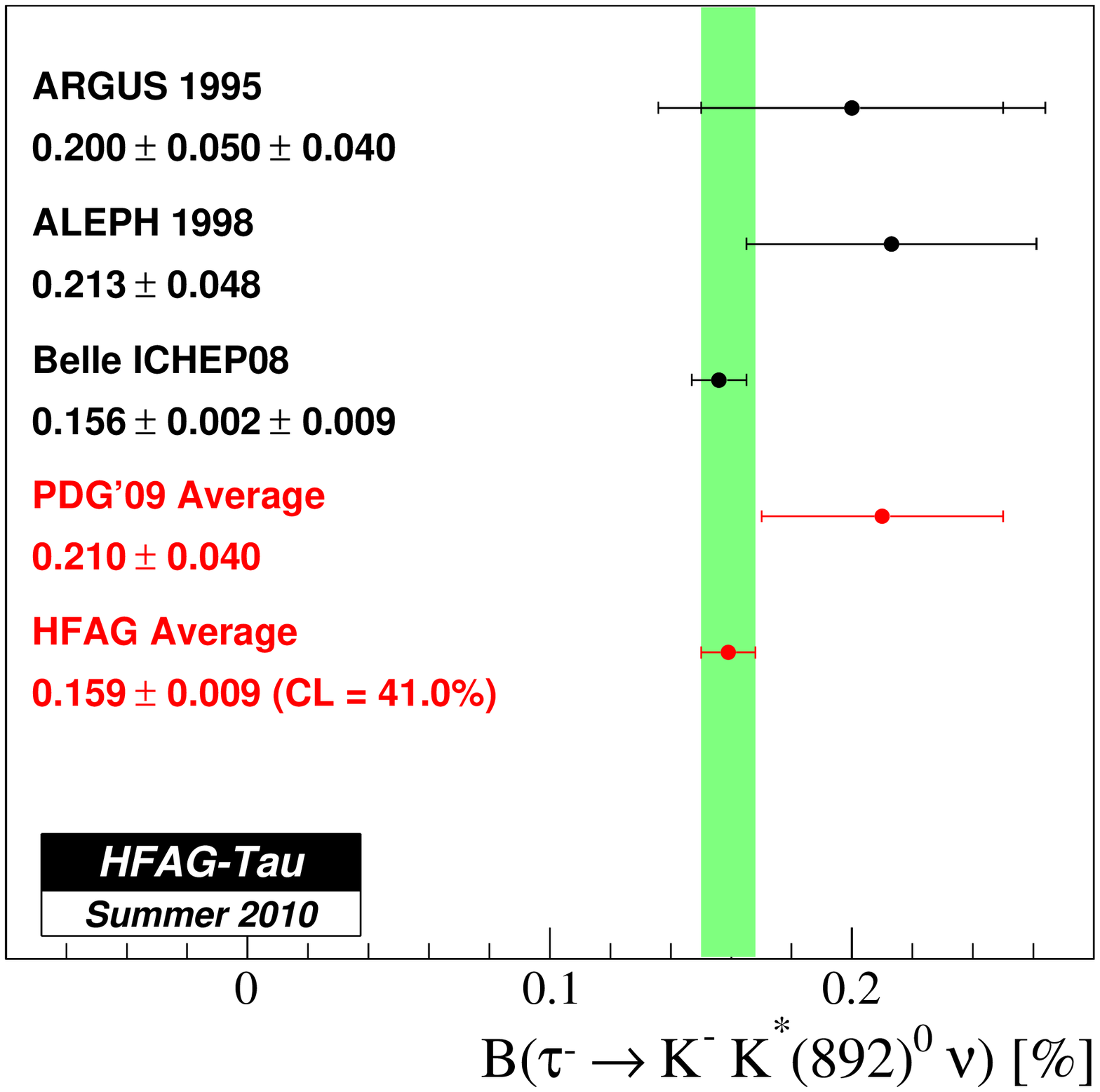}
\end{center}
\caption{Measurements and average values of ${\cal{B}}({\tau^-\to K^-K^{*0}\nu_\tau})$.}
\label{fig:TauToKstarEtaX}
\end{figure}

\end{itemize}


\subsection{Tests of Lepton Universality}
\label{subsec:Tau_LU}

Tests of $\mu-e$ universality can be expressed as
\begin{eqnarray}
\left( \frac{g_\mu}{g_e} \right)^2 = 
\frac
{{\cal{B}}({\tau^- \to \mu^- \overline{\nu}_\mu \nu_\tau})}
{{\cal{B}}({\tau^- \to e^- \overline{\nu}_e \nu_\tau})}
\frac{f(m_e^2/m_\tau^2)}{f(m_\mu^2/m_\tau^2)}~\label{emuuniv},
\end{eqnarray}
\noindent where $f(x) = 1-8x+8x^3-x^4-12x^2\log{x}$, 
assuming that the neutrino masses are negligible~\cite{Tsai:1971vv}.

From the unitarity constrained fit, we obtain
$
{{\cal{B}}({\tau^- \to \mu^- \overline{\nu}_\mu \nu_\tau})}/
{{\cal{B}}({\tau^- \to e^- \overline{\nu}_e  \nu_\tau})}$
= 
$0.9762\, \pm\,   0.0028$,
which includes a correlation co-efficient of $18.33\%$
between the branching fractions.
This yields a value of $\left( \frac{g_\mu}{g_e} \right)$ = 
$1.0019\, \pm\, 0.0014$,
which is consistent with the SM value.

These predictions from $\tau$ decays are more precise than the other determinations:
\begin{itemize}
\item We average the measurements of 
$
{{\cal{B}}(\pi \to e \nu_e (\gamma))}/
{{\cal{B}}(\pi \to \mu \nu_\mu (\gamma))}$
= $(1.2265$ $\pm$ $0.0034$ $\mathrm{(stat)}$ $\pm$ $0.0044$ $\mathrm{(syst)})$ $\times$ $10^{-4}$ from TRIUMF~\cite{Britton:1992pg} and
= $(1.2346$ $\pm$ $0.0035$ $\mathrm{(stat)}$ $\pm$ $0.0036$ $\mathrm{(syst)})$ $\times$ $10^{-4}$ from PSI~\cite{Czapek:1993kc},
to obtain a value of $(1.2310\, \pm\, 0.0037) \times 10^{-4}$.
Comparing this with the prediction of $(1.2352\, \pm\, 0.0001) \times 10^{-4}$ from
recent theoretical calculations~\cite{Cirigliano:2007ga},
we obtain a value of $\left( \frac{g_\mu}{g_e} \right)$ = $1.0017\, \pm\, 0.0015$.
\item The ratio ${{\cal{B}}(K \to e \nu_e (\gamma))}/{{\cal{B}}(K \to \mu \nu_\mu (\gamma))}$
has recently been measured very precisely by the KLOE~\cite{Ambrosino:2009rv} and the NA62~\cite{Goudzovski:2010uk} collaborations.
Using the new world average value of $(2.487\, \pm\, 0.012) \times 10^{-5}$ from Ref.~\cite{Goudzovski:BEACH2010:Preliminary},
and the predicted value of $(2.477\, \pm\, 0.001) \times 10^{-5}$ from Ref.~\cite{Cirigliano:2007ga},
we obtain $\left( \frac{g_\mu}{g_e} \right)$ = $0.9980\, \pm\, 0.0025$.
\item From the report of the FlaviaNet Working Group on Kaon Decays~\cite{Antonelli:2010yf},
we obtain $\left( \frac{g_\mu}{g_e} \right)$ = $1.0010\, \pm\, 0.0025$
using measurements of
${{\cal{B}}(K \to \pi \mu \overline{\nu})}/{{\cal{B}}(K \to \pi e \overline{\nu})}$.
\item From the report of the LEP Electroweak Working Group~\cite{Alcaraz:2006mx},
we obtain $\left( \frac{g_\mu}{g_e} \right)$ =  $0.997\, \pm\, 0.010$ 
using measurements of
${{\cal{B}}(W \to \mu \overline{\nu}_\mu)}/{{\cal{B}}(W \to e \overline{\nu}_e)}$.
\end{itemize}

Tau-muon universality is tested with
\begin{eqnarray}
\left( \frac{g_{\tau}}{g_{\mu}} \right)^2 = 
      \frac{{\cal{B}}({\tau \to h \nu_\tau})}{{\cal{B}}({h \to\mu \overline{\nu}_\mu})}
      \frac{2m_h m^2_{\mu}\tau_h}{(1+\delta_{h})m^3_{\tau}\tau_{\tau}} 
      \left( \frac{1-m^2_{\mu}/m^2_h}{1-m^2_h/m^2_{\tau}} \right)^2, ~\label{mutauuniv}
\end{eqnarray}
\noindent where $h$ = $\pi$ or $K$ and the radiative corrections are
$\delta_{\pi} = (0.16\, \pm\, 0.14)\%$ and
$\delta_{K} = (0.90\, \pm\, 0.22)\%$~\cite{Marciano:1993sh,Decker:1994ea,Decker:1994dd}.
Using the world averaged mass and lifetime values and meson decay rates~\cite{PDG_2010}
and our unitarity constrained fit, we determine 
$\left( \frac{g_{\tau}}{g_{\mu}} \right)$ =
$0.9966\, \pm\, 0.0030$ $(0.9860\, \pm\, 0.0073)$
from the pionic and kaonic branching fractions,
where the correlation co-efficient between these values are $13.10\%$.
Combining these results, we obtain 
$\left( \frac{g_{\tau}}{g_{\mu}} \right)$ = $0.9954\, \pm\, 0.0029$,
which is $1.6~\sigma$ below the SM expectation.

We also test lepton universality between $\tau$ and $\mu$ ($e$),
by comparing the averaged electronic (muonic) branching fractions of
the $\tau$ lepton with the predicted branching fractions from measurements
of the $\tau$ and $\mu$ lifetimes and their respective masses~\cite{PDG_2010}, 
using known electroweak and radiative corrections~\cite{Marciano:1993sh}.
This gives 
$\left( \frac{g_{\tau}}{g_{\mu}} \right)$ = $1.0011\, \pm\, 0.0021$
and
$\left( \frac{g_{\tau}}{g_{e}} \right)$ = $1.0030\, \pm\, 0.0021$.
The correlation co-efficient between the determination of
$\left( \frac{g_{\tau}}{g_{\mu}} \right)$ from electronic branching
fraction with the ones obtained from pionic and kaonic branching fractions
are $48.16\%$ and  $21.82\%$, respectively.
Averaging these three values, we obtain
$\left( \frac{g_{\tau}}{g_{\mu}} \right)$ = $1.0001\, \pm\, 0.0020$,
which is consistent with the SM value.
In Figure~\ref{fig:TauLU}, we compare these above determinations 
with each other and with the values obtained from W decays~\cite{Alcaraz:2006mx}.

\begin{figure}[!hbtp]
\begin{center}
\begin{minipage}{.42\textwidth}
\includegraphics[height=.32\textheight]{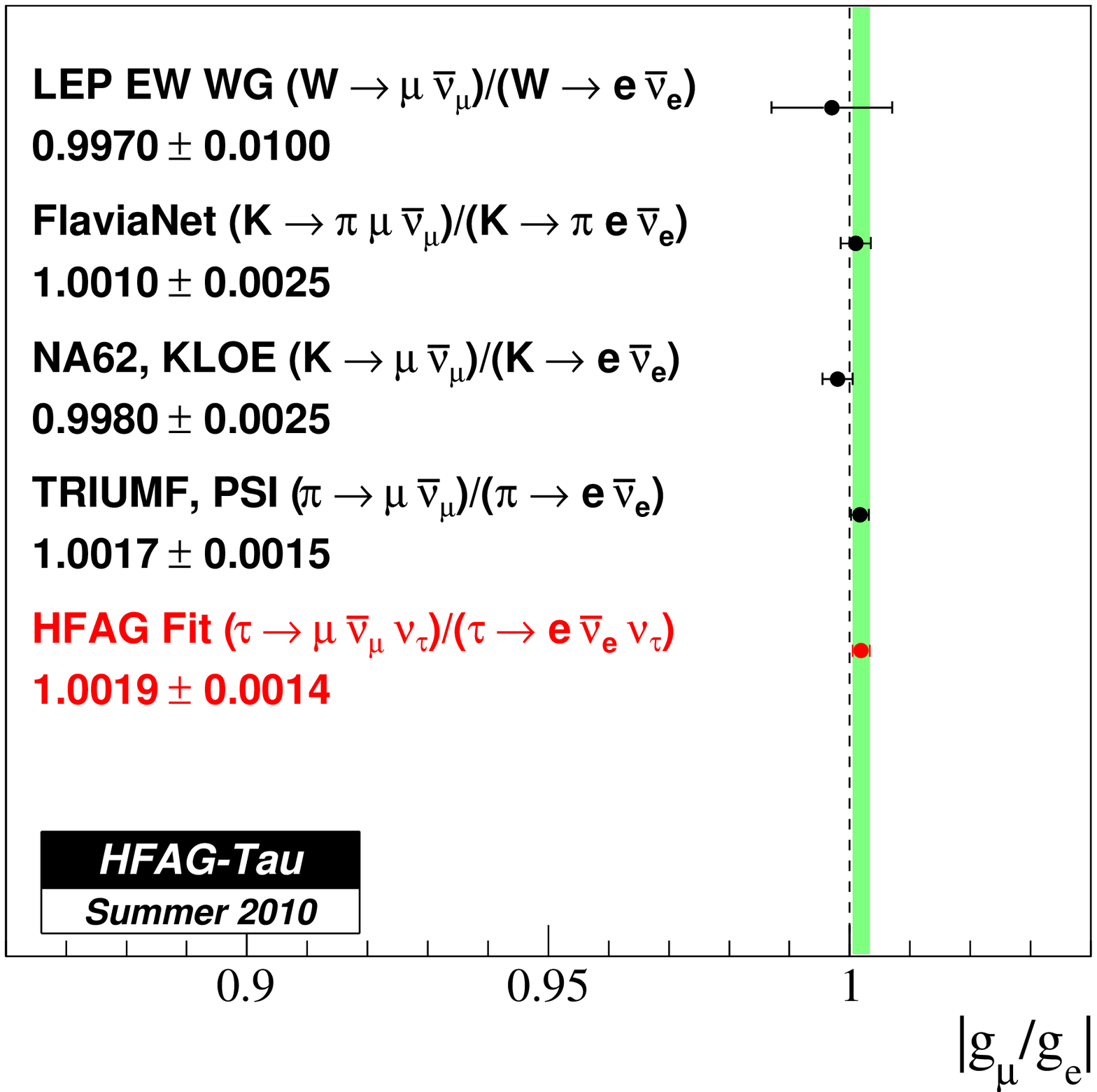}
\includegraphics[height=.32\textheight]{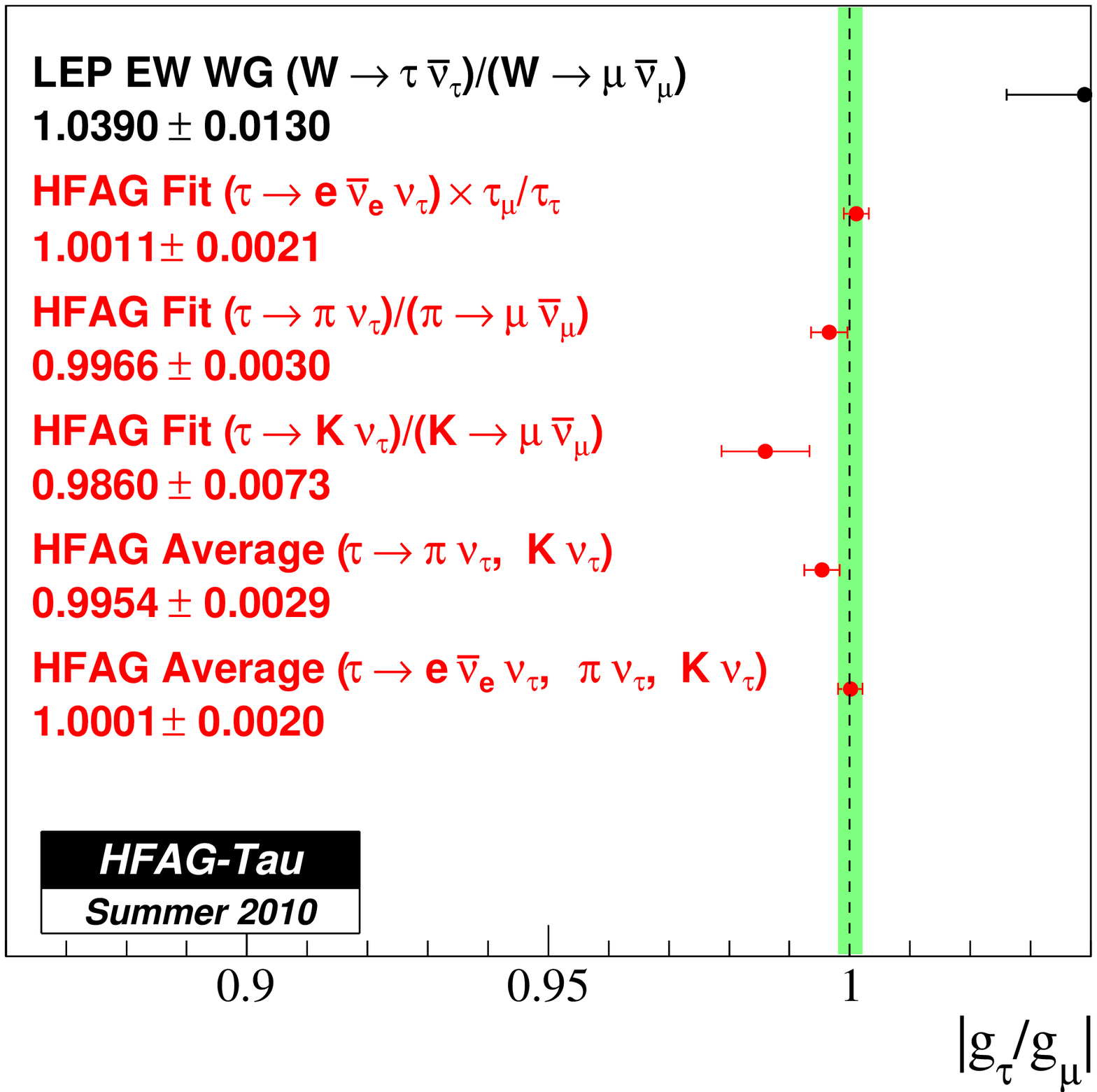}
\includegraphics[height=.32\textheight]{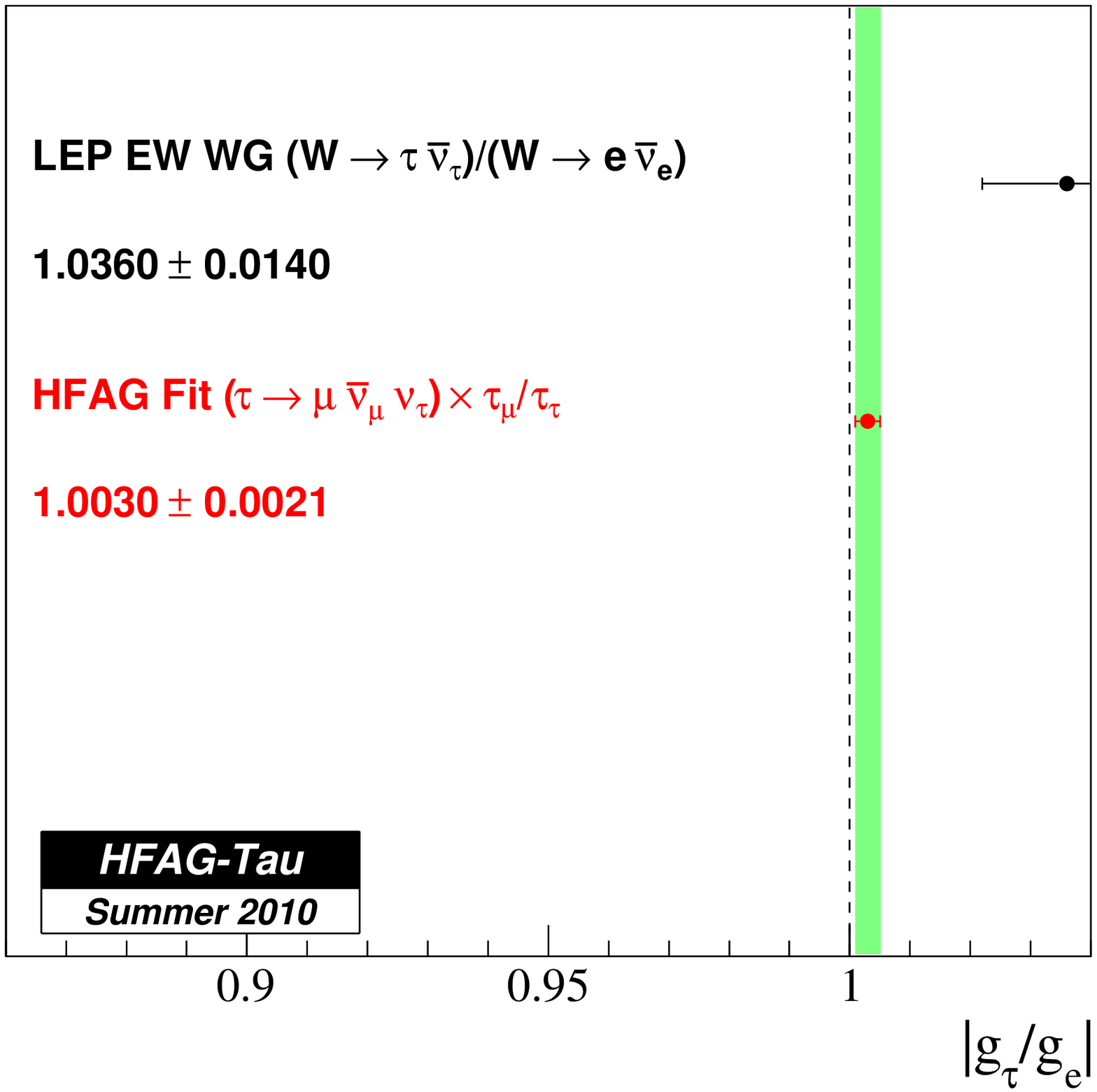}
\end{minipage}
\end{center}
\caption{Measurements of lepton universality from W, kaon, pion and tau decays.}
\label{fig:TauLU}
\end{figure}


\subsection{Measurement of $|V_{us}|$}
\label{subsec:Tau_Vus}

\newcommand{\Vudvalue} {\ensuremath{0.97425\, \pm\, 0.00022}\xspace}

We describe 3 extractions for $|V_{us}|$ using
${\cal{B}}({\tau^-\to K^-\nu_\tau})$, 
$
{{\cal{B}}({\tau^- \to K^-\nu_\tau})}/
{{\cal{B}}({\tau^- \to \pi^-\nu_\tau})}$,
and inclusive sum of $\tau$ branching fractions having net strangeness of unity in the final state:

\begin{itemize}
\item
We use the value of kaon decay constant $f_K = 157\, \pm\, 2 \mev$~\cite{Follana:2007uv}, 
and our value of 
\begin{eqnarray*}
{\cal{B}}({\tau^-\to K^-\nu_\tau}) = \frac{G^2_F f^2_K \Vus^2 m^3_{\tau} \tau_{\tau}}{16\pi\hbar} \left (1 - \frac{m_K^2}{m_\tau^2} \right )^2 S_{EW},
\end{eqnarray*}
where $S_{EW} = 1.0201\, \pm\, 0.0003$~\cite{Erler:2002mv},
to determine $\Vus= 0.2204\, \pm\, 0.0032$
from results of the unitarity constrained fit.
This value is consistent with the estimate of $\Vus = 0.2255\, \pm\, 0.0010$
obtained using the unitarity constraint on the first row of the CKM matrix.

\item
We  use $f_K/f_\pi = 1.189\, \pm\, 0.007$~\cite{Follana:2007uv}, 
$\Vud = \Vudvalue$~\cite{Hardy:2008gy}, and
the long-distance correction $\delta_{LD} = (0.03\, \pm\, 0.44)\%$,  
estimated~\cite{Banerjee:2008hg} 
using corrections to $\tau\to h\nu_\tau$ and $h \to
\mu\nu_\mu$~\cite{Marciano:1993sh,Decker:1994ea,Decker:1994dd,Marciano:2004uf},
for the ratio
\begin{eqnarray*}
\frac{{\cal{B}}({\tau^- \to K^-\nu_\tau})}{{\cal{B}}({\tau^- \to \pi^-\nu_\tau})} 
&=& 
\frac{f_K^2 |V_{us}|^2}{f_\pi^2 |V_{ud}|^2}
 \frac{\left( 1 -  \frac{m_K^2}{m_\tau^2} \right)^2}{\left( 1 -  \frac{m_\pi^2}{m_\tau^2} \right)^2} (1+\delta_{LD}),
\end{eqnarray*}
where short-distance electro-weak corrections cancel in this ratio.

From the unitarity constrained fit, we obtain
${{\cal{B}}({\tau^- \to K^-\nu_\tau})}/{{\cal{B}}({\tau^- \to \pi^-\nu_\tau})}$
$=$ $ 0.0644\, \pm\, 0.0009$,
which includes a correlation co-efficient of $-0.49\%$
between the branching fractions.
This yields $\Vus = 0.2238\, \pm\, 0.0022$,
which is also consistent with value of $\Vus$ from CKM unitarity prediction.
 
\item
The total hadronic width of the \mtau normalized to the electronic branching fraction,
$R_{\rm{had}} = {\cal{B}}_{\rm{had}}/{\cal{B}}_{\rm{e}}$,
can be written as $R_{\rm{had}} = R_{\rm{non-strange}} + R_{\rm{strange}}$.
We can then measure
\begin{eqnarray}
|V_{us}| &=& \sqrt{R_{\rm{strange}}/\left[\frac{R_{\text{non-strange}}}{|V_{ud}|^2} -  \delta R_{\text{theory}}\right]}.
\end{eqnarray}

Here, we use $\Vud = \Vudvalue$~\cite{Hardy:2008gy}, and $\delta R_{\text{theory}} = 0.240\, \pm\, 0.032$~\cite{Gamiz:2006xx} 
obtained with the updated average value of $m_{s}(2\gev) = 94\, \pm\, 6~\mev$~\cite{Jamin:2006tj},
which contributes to an error of $0.0010$ on $|V_{us}|$.
We note that this error is equivalent to  half the difference between calculations of $|V_{us}|$ 
obtained using fixed order perturbation theory (FOPT) and contour improved perturbation theory (CIPT) calculations of $\delta R_{\text{theory}}$~\cite{Maltman:2010hb}, 
and twice as large as the theoretical error proposed in Ref.~\cite{Gamiz:2007qs}.

As in Ref.~\cite{Davier:2005xq}, we improve upon the estimate of
electronic branching fraction by averaging its direct measurement 
with its estimates of $(17.899\, \pm\, 0.040)\%$ and $(17.794\, \pm\, 0.062)\%$
obtained from the averaged values of muonic branching fractions and 
the averaged value of the lifetime of the 
\mtau lepton = $(290.6\, \pm\, 1.0) \times 10^{-15} ~\mathrm{s}$~\cite{PDG_2010},
assuming lepton universality and taking into account the correlation between the leptonic branching fractions.
This gives a more precise estimate for the electronic branching fraction: ${\cal{B}}^{\rm{uni}}_{\rm{e}}$ = $ (17.852\, \pm\, 0.027)\%$.

Assuming lepton universality, the total hadronic branching fraction 
can be written as: ${\cal{B}}_{\rm{had}} = 1 - 1.972558 ~ {\cal{B}}^{\rm{uni}}_{\rm{e}}$,
which gives a value for the total \mtau hadronic width normalized to 
the electronic branching fraction as $R_{\rm{had}} = 3.6291\, \pm\, 0.0086$.

The non-strange width is $R_{\rm{non-strange}} = R_{\rm{had}} - R_{\rm{strange}}$,
where the estimate for the strange width $R_{\rm{strange}}$ = $0.1613\, \pm\, 0.0028$
is obtained from the sum of the strange branching fractions 
with the unitarity constrained fit as listed in Table~\ref{tab:TauGlobalFit}.
This gives a value of  $\Vus$ $=$ $0.2174\, \pm\, 0.0022$,
which is $3.3~\sigma$ lower than the CKM unitarity prediction.

A similar estimation using results from the unconstrained fit to the branching fractions gives 
$\Vus$ $=$ $0.2166\, \pm\, 0.0023$, which is $3.6~\sigma$ lower than the CKM unitarity prediction.
%
%
Since the sum of base modes from our unconstrained fit is less than unity by $1.6~\sigma$, 
instead of using ${\cal{B}}_{\rm{non-strange}} = 1 - {\cal{B}}_{\rm{leptonic}} - {\cal{B}}_{\rm{strange}}$, 
we also evaluate $\Vus$ from the sum of the averaged non-strange branching fractions. 
This gives $\Vus$ $=$ $0.2169\, \pm\, 0.0023$, which is $3.5~\sigma$ lower than the CKM unitarity prediction.

\end{itemize}

\begin{figure}[!hbtp]
\begin{center}
\includegraphics[height=.42\textheight,width=.49\textwidth]{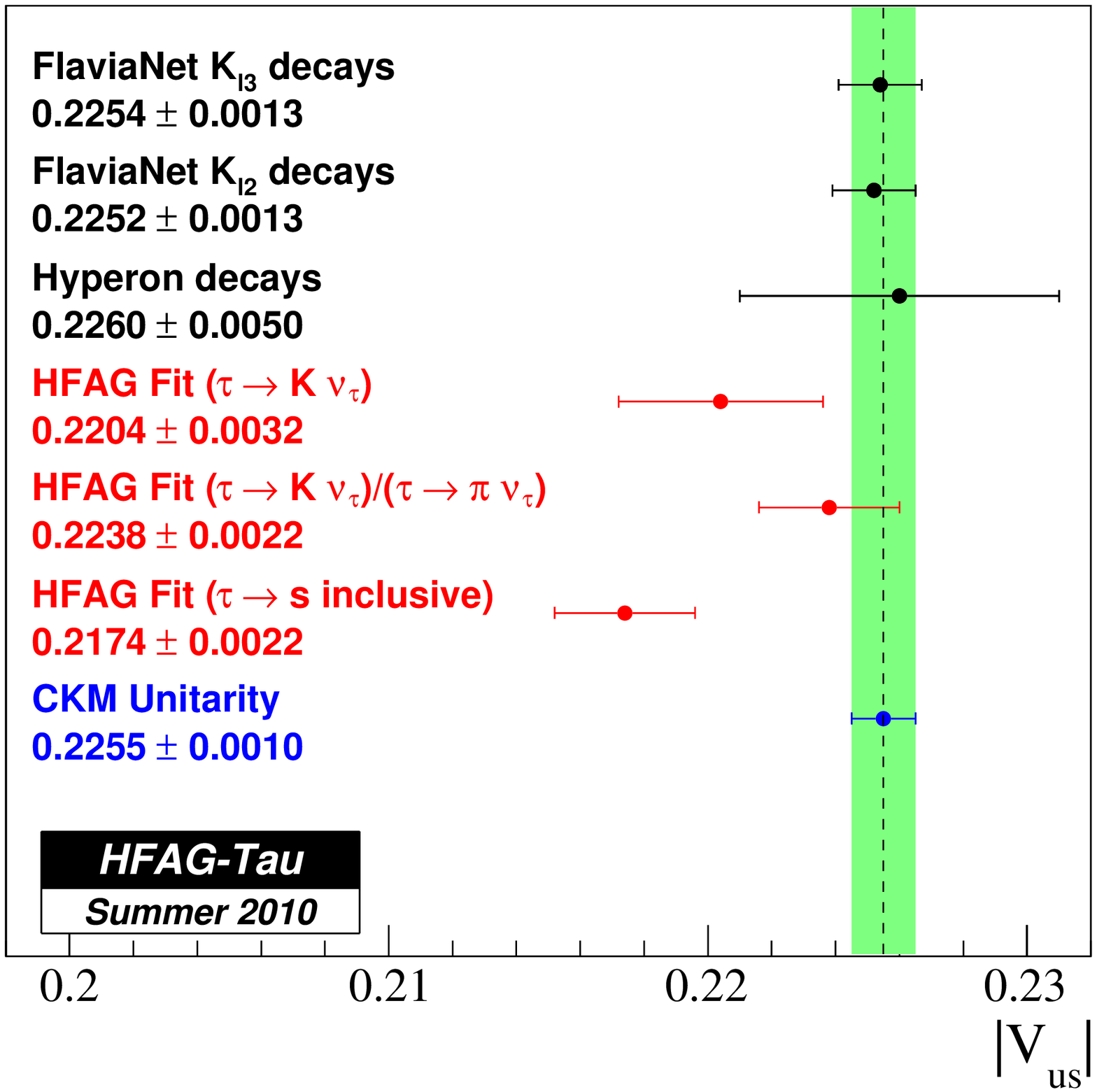}
\end{center}
\caption{Measurements of $\Vus$ from kaon, hyperon and tau decays.}
\label{fig:Vus}
\end{figure}
 
Summary of these $\Vus$ values are plotted in Figure~\ref{fig:Vus}, 
where we also include values from kaon decays obtained from Ref.~\cite{Antonelli:2010yf} 
and from hyperon decays obtained from Ref.~\cite{Jamin:MoriondEW2007:Preliminary}.


\subsection{Search for lepton flavor violation in $\tau$ decays}
\label{subsec:Tau_LFV}

\begin{figure}[!hbtp]
\begin{center}
\includegraphics[width=1.2\textwidth,angle=270]{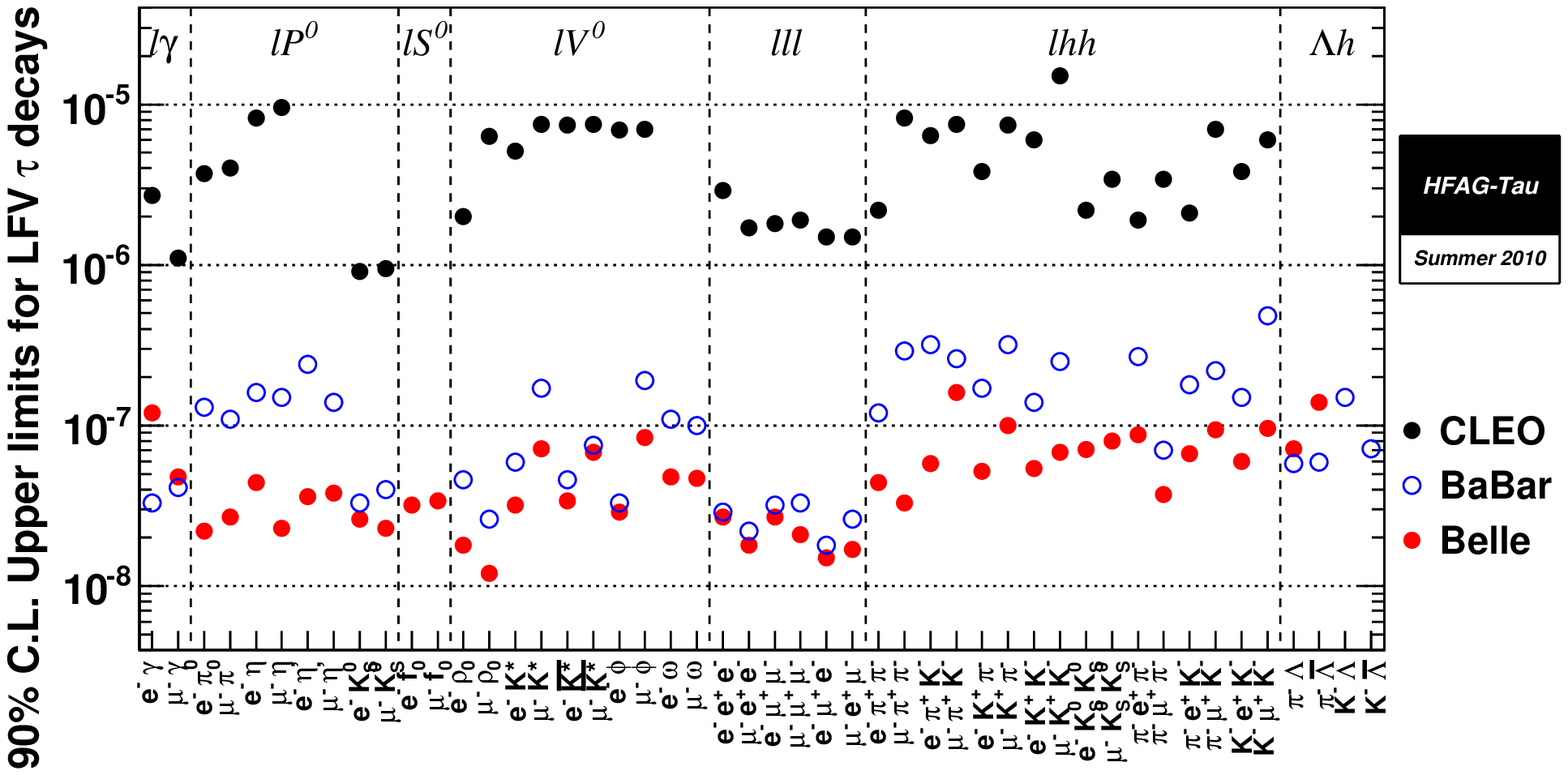}
\end{center}
\caption{Status of searches for lepton flavor violation in $\tau$ decays.}
\label{fig:Tau_LFV}
\end{figure} 

The status of searches for lepton flavor violation in $\tau$ decays is 
summarized in Figure~\ref{fig:Tau_LFV}. A table of these results and 
the corresponding references are provided on the HFAG web site
\vskip0.15in\hskip0.75in
{\tt http://www.slac.stanford.edu/xorg/hfag/tau/HFAG-TAU-LFV.htm}.


\def\babar{\mbox{\slshape B\kern-0.1em{\smaller A}\kern-0.1em
    B\kern-0.1em{\smaller A\kern-0.2em R}}\xspace}

\section{Summary}
\label{sec:summary}

This article provides updated world averages for 
$b$-hadron properties at least through the end of 2009. Some results
that appeared in the spring of 2010 are also included.
A small selection of highlights of the results described in Sections
\ref{sec:life_mix}-\ref{sec:tau} is given in 
Tables~\ref{tab_summary1} and~\ref{tab_summary2}.

\begin{table}
\caption{Selected world averages at the end of 2009
from Chapters~\ref{sec:life_mix} and~\ref{sec:cp_uta}.}
\label{tab_summary1}
\renewcommand{\arraystretch}{1.15}
\begin{center}

\end{center}
\end{table}
\begin{table}
\caption{Selected world averages at the end of 2009
from Chapters~\ref{sec:slbdecays}--\ref{sec:tau}.}
\label{tab_summary2}
\renewcommand{\arraystretch}{1.15}
\begin{center}
%
\end{center}
\end{table}

Concerning lifetime and mixing averages,
the most significant changes in the past two years
are due to new results from the Tevatron experiments,
mainly measurements of $b$-baryon lifetimes and searches
for \cp\ violation in \Bs mixing.  
While \dzero has measured a like-sign dimuon asymmetry
deviating by $3.2\,\sigma$ from the SM, 
the latest results on the  
\cp-violation phase in $\Bs\to J/\psi\phi$
no longer show any hint of New Physics.
On the other hand, averaging procedures
for the $b$-hadron production fractions have been
improved: for the first time we obtain a set of
fractions based on Tevatron measurements only, and we
also extract the fraction of $\Upsilon(5S)$ decays to \Bs pairs 
taking into account decays without open-bottom mesons.  

The measurement of $\sin 2\beta \equiv \sin 2\phi_1$ from $b \to
c\bar{c}s$ transitions such as $\Bz \to \jpsi\KS$ has reached $<4\%$
precision: $\sin 2\beta \equiv \sin 2\phi_1 = 0.673 \pm 0.023$.
Measurements of the same parameter using different quark-level processes
provide a consistency test of the Standard Model and allow insight into
possible new physics.  Recent improvements include the use of
time-dependent Dalitz plot analyses of $\Bz \to \KS\Kp\Km$ and $\Bz \to
\KS\pip\pim$ to obtain \CP\ violation parameters for $\phi\KS$,
$f_0(980)\KS$ and $\rho\KS$.  All results among hadronic $b \to s$
penguin dominated decays are currently consistent with the Standard
Model expectations.  Among measurements related to the Unitarity
Triangle angle $\alpha \equiv \phi_2$, updates of the parameters of the
$\rho\rho$ system now allow constraints at the level of $\approx
6^\circ$.  Knowledge of the third angle $\gamma \equiv \phi_3$ also
continues to improve.  Notwithstanding the well-known statistical issues
in extracting the value of the angle itself, the world average values of
the parameters in $B \to DK$ decays now show a significant direct \CP\
violation effect.





Concerning $D^0$-$\dbar$ mixing, three experiments 
have now found evidence for this phenomenon: Belle, \babar, and CDF.
These measurements and others (made by Belle, \babar, CLEO, FNAL E791, FNAL E831)
are combined to yield World Average (WA) values for mixing parameters $x$ and $y$, 
and for \cpv\ parameters $|q/p|$ and $\phi$. From this fit, the no-mixing 
point $x\!=\!y\!=\!0$ is excluded at $10.2\sigma$. The parameter $x$ differs 
from zero by $2.5\sigma$, and $y$ differs from zero by $5.7\sigma$. Mixing 
at this level is presumably dominated by long-distance processes, which are 
difficult to calculate. Thus, it may be difficult to identify new 
physics from mixing alone. The WA value for the observable \ycp\ 
is positive, which indicates that the \cp-even state is 
shorter-lived as in the $K^0$-$\kbar$ system. However, $x$ also 
appears to be positive, which implies that the \cp-even state is 
heavier; this is unlike in the $K^0$-$\kbar$ system. 
There is no evidence yet for \cpv\ (either direct or indirect) 
in the $D^0$-$\dbar$ system.

\section{Acknowledgments}

We are grateful for the strong support of the \belle, 
\babar, CLEO, CDF, and \dzero\ collaborations, without whom 
this compilation of results and world averages would not have 
been possible. The success of these experiments in turn would 
not have been possible without the excellent operations of the 
KEKB, PEP-II, CESR, and Tevatron accelerators, and fruitful 
collaborations between the accelerator groups and the experiments.

\clearpage


\bibliographystyle{HFAGutphys}
\raggedright
\bibliography{EndOfYear09.bib,life_mix.bib,cp_uta.bib,slbdecays/slb_ref.bib,rare/rare_refs.bib,charm/charm_refs.bib,tau/tau_refs.bib}{}

\begin{mcitethebibliography}{100}
\bibitem{CDFnote9203:2008_cont}
 we consider that these new results supersede the ones from the 3-year old
  CDF note 7386~\cite{CDFnote7386:2005}, although the lifetime analysis of one
  of the modes ($B_s \to D_s\pi\pi\pi$) has not been updated\relax
\mciteBstWouldAddEndPuncttrue
\mciteSetBstMidEndSepPunct{\mcitedefaultmidpunct}
{\mcitedefaultendpunct}{\mcitedefaultseppunct}\relax
\EndOfBibitem
\end{mcitethebibliography}

\end{document}